\documentclass{jpp}

\usepackage{graphicx}
\usepackage{epstopdf, epsfig}
\usepackage{color}

\shorttitle{Kinetic stability of Chapman-Enskog plasmas }
\shortauthor{A. F. A. Bott, S. C. Cowley and A. A. Schekochihin}

\title{Kinetic stability of Chapman-Enskog plasmas}

\author{Archie F. A. Bott\aff{1,2,3}
  \corresp{\email{archie.bott@physics.ox.ac.uk}},
  S. C. Cowley\aff{4},
  \and A. A. Schekochihin\aff{1,5} }

\affiliation{\aff{1}Department of Physics, University of Oxford,
Parks Road, Oxford OX1 3PU, UK
\aff{2} Trinity College, Broad St, Oxford OX1 3BH, UK
\aff{3} Department of Astrophysical Sciences, University of Princeton, 4 Ivy Ln, Princeton, NJ 08544, USA
\aff{4} Princeton Plasma Physics Laboratory, 100 Stellarator Road, Princeton, New Jersey 08540,  USA
\aff{5} Merton College, Merton St, Oxford OX1 4JD, UK}

\begin{document}

\maketitle

\abstract{In this paper, we investigate the kinetic stability of classical, collisional plasma – that is, plasma in which the mean-free-path $\lambda$ of constituent particles is short compared to the length scale $L$ over which fields and bulk motions in the plasma vary macroscopically, and the collision time is short compared to the 
evolution time. Fluid equations are typically used to describe such plasmas, since their distribution functions are close to being Maxwellian. The small deviations from the Maxwellian distribution are calculated via the Chapman-Enskog (CE) expansion in $\lambda/L \ll 1$, and determine macroscopic momentum and heat fluxes in the plasma. Such a calculation is only valid if the underlying CE distribution function is stable at collisionless length scales and/or time scales. We find that at sufficiently high plasma $\beta$, the CE distribution function can be subject to numerous microinstabilities across a wide range of scales. For a particular form of the CE distribution function arising in strongly magnetised plasma (viz., plasma in which the Larmor periods of particles are much smaller than collision times), we provide a detailed analytic characterisation of all significant microinstabilities, including peak growth rates and their associated wavenumbers. Of specific note is the discovery of several new microinstabilities, including one at sub-electron-Larmor scales (the `whisper instability') whose growth rate in certain parameter regimes is large compared to other instabilities. Our approach enables us to construct the kinetic stability maps of classical, two-species collisional plasma in terms of $\lambda$, the electron inertial scale $d_e$ and the plasma $\beta$. This work is of general consequence in emphasising the fact that high-$\beta$ collisional plasmas can be kinetically unstable; for strongly magnetised CE plasmas, the condition for instability is $\beta \gtrsim L/\lambda$. In this situation, the determination of transport coefficients via the standard CE approach is not valid.}

\tableofcontents

\section{Introduction} \label{sec:introduction}

Answering the question of when a plasma can be described adequately
by fluid equations is fundamental for a comprehensive understanding of plasma dynamics.  
It is well known that some physical effects in plasmas -- for example, Landau damping -- 
specifically require a fully kinetic description in terms of distribution functions of the plasma's constituent particles~\citep{L46}. 
However, for many other plasma processes, a detailed description of the underlying 
particle distribution provides little additional understanding of the essential physics governing that process. 
Characterising such processes with fluid equations, which describe the evolution of macroscopic physical quantities 
such as density, fluid velocity and temperature, 
often simplifies the description and therefore aids understanding. 
Fluid equations are also easier to solve numerically than kinetic equations:
the latter reside in six-dimensional phase space (and time), with three 
additional dimensions -- the velocity space -- when compared to the former. 
The underlying difficulty associated with determining when a plasma is a fluid 
is finding a closed set of equations in the macroscopic plasma variables. The derivation of fluid equations from the Maxwell-Vlasov-Landau equations 
governing the evolution of the plasma's distribution functions is carried out by taking moments 
(that is, integrating the governing equations and their outer products with velocity $\boldsymbol{v}$ over velocity space). However, the resulting equations
 are not closed: the evolution equation of the zeroth-order moment (density) requires
knowledge of the evolution of the first-order moment, the evolution equation for the first-order moment 
needs the second-order moment, and so on. For plasma fluid equations to be 
able to describe the evolution of a plasma without reference to that plasma's 
underlying distribution functions, a closure hypothesis or an approximation relating higher-order 
moments to lower ones is required. 

For a collisional plasma -- i.e., one in which the mean free paths $\lambda_{s}$ 
and collision times $\tau_s$ of the ions and electrons ($s = i, e$) are much smaller than
the typical length scale $L$ and time scale $\tau_L$ on which macroscopic properties of the plasma 
change -- there is a procedure for achieving such a closure: the \textit{Chapman-Enskog (CE)
expansion}~\citep{CC70, E1917,C88}. It is assumed that in a collisional plasma, 
the small perturbations of the distribution functions away from a Maxwellian equilibrium have typical size $\epsilon \sim \lambda_{s}/L \sim \tau_s/\tau_L \ll 1$ (assuming sonic motions, and $\lambda_i \sim \lambda_e$). 
Since the perturbation is small, its form can be determined 
explicitly by performing an asymptotic expansion of the Maxwell-Vlasov-Landau equations.  
Once the underlying distribution is known, the relevant moments can be 
calculated -- in particular, the momentum and heat fluxes are the second- and third-order moments of the $\textit{O}(\epsilon)$ non-Maxwellian component of the distribution 
function.
The CE expansion applied to a two-species magnetised plasma was worked out by~\citet{B65}. 
Subsequent studies have refined and extended various aspects of his calculation~\citep{E84,MT84,EH86,HKC94,SC04}. 
In this paper, we will refer to the distribution functions associated with the CE 
expansion as CE distribution functions, and plasmas 
with particle distribution functions given by CE distribution functions as CE 
plasmas. 

However, the theory constructed as outlined above is incomplete. For the CE 
expansion to provide an adequate fluid closure, the resulting distribution functions 
must be stable to all kinetic instabilities with 
length scales shorter than the longest mean free path, and timescales shorter than the 
macroscopic plasma timescale $\tau_L$. Such instabilities (if present) are known as 
microinstabilities. We emphasise that these microinstabilities should be distinguished 
conceptually from instabilities describable by the closed set of plasma-fluid equations: 
for example, Rayleigh-Taylor~\citep{Ray1883,Tay50,TMMM85,Kull91}, magnetorotational~\citep{BH91a,BH91b}, magnetoviscous~\citep{QDH02,Balb2004,IB05}, or magnetothermal/heat-flux-driven buoyancy instabilities~\citep{Balb00,Balb01,Q08,K11}. Kinetic microinstabilities should also be distinguished 
from the small-scale instabilities that arise in solving higher-order $(\textit{O}(\epsilon^2)$) fluid equations obtained
from the CE asymptotic expansion~\citep[for neutral fluids, these are called the \textit{Burnett 
equations} -- see][]{GVU08}.
Such instabilities are not physical because they arise at scales 
where the equations themselves do not apply~\citep{B82}.
Fluid instabilities do not call into question the 
validity of the fluid equations themselves; in contrast, if 
microinstabilities occur, the plasma-fluid equations obtained through the closure 
hypothesis are physically invalid, irrespective of their own stability. 

Microinstabilities have been studied in depth for a 
wide range of classical plasmas by many authors; see, for example,~\citet{D83},~\citet{G93}, and~\citet{H12} for three different general perspectives on microinstability theory.  Although it can be shown that a Maxwellian distribution is always immune to such instabilities~\citep{B58, CL14}, 
anisotropic distribution functions are often not~\citep{F62,K62,KMQ68}. A notable example is the Weibel instability, 
which occurs in counter-streaming unmagnetised plasmas~\citep{W59,F59}. The linear theory of such instabilities is generally well known~\citep[for modern reviews, 
see][]{LSP09,ILS12}.  Microinstabilities in magnetised plasma have also been comprehensively studied.
The ion firehose and mirror instabilities 
are known to occur in plasmas with sufficient ion-pressure anisotropy and large enough plasma~$\beta$~\citep{CKW58,P58,VS61,H69,H81,H07}, 
while electron-pressure anisotropy can also result in microinstabilities of various types~\citep{KP66,HV70,GM85}. 

A number of authors have noted that microinstabilities, if present, will have a significant effect on the macroscopic transport properties of plasmas~\citep{K64,SCKHS05,SCKR08,MSK16,RQV16,KCKS16,KSCS17,RDRS18,DPRR20}. 
Typically (although not always), once the small-scale magnetic and electric fields 
associated with microinstabilities have grown, they will start to scatter particles, 
which in turn will alter the plasma's distribution functions. This has micro- and macroscopic consequences for plasma behaviour.
From the microscopic perspective, it changes the course of the evolution of the microinstabilities 
themselves -- by, e.g., reducing the anisotropy of the underlying particle distribution functions~\citep{HTDS14,RQV18}.  
From the macroscopic perspective, the changes to the distribution functions will alter both heat and 
momentum fluxes in the plasma (which, as previously mentioned, are determined by non-Maxwellian terms in the distribution 
function). In this picture, a plasma subject to microinstabilities in some sense generates its own 
effective anomalous collisionality~\citep{SCKR08,MS14,KSS14,SKQS17,KSSQ20}.  
The typical values of the altered fluxes attained must depend on the saturated state of microinstabilities~\citep{ACRR10}. 
Exploring the mechanisms leading to saturation of both unmagnetised, Weibel-type instabilities~\citep[e.g., ][]{DHHW72,LWG79,CPBM98,CCC02,K05,PA11,RGDB15} 
and magnetised instabilities~\citep[e.g.,][]{KPS07,PSBO08,RSRC11,RQV15,RSC15} 
continues to be an active research area. Simulation results~\citep{HKPS09,KSS14,GSN14,RQV16,MSK16,GSN18,BAKQS21}
support the claim that the saturation amplitude of 
such microinstabilities is typically such that the plasma maintains itself close to 
marginality of the relevant instability. 

Do these kinetic instabilities afflict the CE distribution function? Naively, it might 
be assumed not, since it is `almost' Maxwellian. However, it turns out that, provided the plasma $\beta$ is sufficiently high, small 
distortions from a Maxwellian can be sufficient to lead to instability. 
Instabilities of a CE distribution function in an unmagnetised plasma were first explored 
by~\citet{K64}, who considered a collisional electron plasma  (mean free path $\lambda_e$)
with macroscopic variations in density, temperature and velocity (scale {$\sim$}$L$). He showed that 
the CE distribution function in such a plasma would have two non-Maxwellian terms of order $\lambda_e/L$ -- an antisymmetric term associated
with heat flux, and another term associated with velocity shear -- and 
that the latter term would result in the so-called \textit{transverse instability}. \citet{K64} 
also claimed that this instability would lead to a significant change in the 
plasma viscosity, and other transport coefficients. \citet{A70B,A70} further 
developed the theory of the transverse instability, including a quasi-linear 
theory resulting in isotropisation of the underlying electron distribution 
function. 

The stability of the CE distribution function was later considered by \citet{RL78}. They found that in an initially unmagnetised two-species plasma supporting 
a fluid-scale electron-temperature gradient (scale $L_T$, no flow shear), the second-order terms (in $\lambda/L_T$) in the electron distribution function 
could result in the formation of unstable waves, with typical real frequencies $\varpi \propto \lambda_{e}/L_T$, and growth rates $\gamma_{\rm RL} \propto 
\left(\lambda_{e}/L_T\right)^2$. Similarly to \citet{K64}, they argued that the presence of such instabilities would 
suppress the macroscopic heat flux in the plasma (which in a collisional plasma is carried predominantly by 
electrons). This particular instability has also been proposed as an explanation for the origin of the cosmic magnetic field~\citep{OH03}. 
Subsequent authors have explored further the idea that non-Maxwellian components of the electron distribution function required to support a 
macroscopic heat flux can lead to kinetic instability.  \citet{LE92} considered the 
effect of introducing a uniform, macroscopic magnetic field into the same problem, and found 
that a faster instability feeding off first-order heat-flux terms in the CE distribution function -- the \textit{whistler instability} -- arose at the electron Larmor scale, with $\gamma_{\rm whistler,T} \propto \lambda_{e}/L_T$. A 
quasi-linear theory of this instability was 
subsequently constructed by~\citet{PE98}. Both \citet{LE92} and \citet{PE98} proposed that the instability at saturation would result in a suppressed heat flux~\citep[see also][]{GL00}. 
More recently, the whistler instability has been studied in simulations of high-$\beta$ plasma-- 
with two groups independently finding both the onset of instability at electron 
scales, and evidence of a suppression of heat flux~\citep{RDRS16,RDRS18,KSCS17,R18b}. 
\citet{DPRR20} constructed a theoretical model for whistler-regulated heat transport based on a set of 
reasonable assumptions that were motivated by these prior simulations. 

The possibility of microinstabilities associated with the ion CE distribution 
function was also considered by~\citet{SCKHS05}, who found that
weakly collisional, magnetised plasma undergoing subsonic, turbulent shearing motions
 can be linearly unstable to firehose and mirror instabilities at 
sufficiently high $\beta_i$ (where $\beta_i$ is the ion plasma beta). This is because the shearing motions give rise 
to an ion pressure anisotropy $\Delta_i \sim \lambda_i^2/L_V^2$, where $L_V$ is the length scale associated with the shearing 
motions.
For $|\Delta_i| \gtrsim \beta_i^{-1}$, the mirror and firehose instability thresholds can 
be crossed (the mirror instability is trigged by sufficiently positive pressure anisotropy,
the firehose instability by negative pressure anisotropy). 
Beyond its threshold, the maximum firehose instability growth rate $\gamma_{\rm fire}$ was found to 
satisfy $\gamma_{\rm fire} \propto |\Delta_i+2/\beta_i|^{1/2}$, whilst for the mirror instability, the maximum 
growth rate was $\gamma_{\rm mirr} \propto \Delta_i-1/\beta_i$. Such destabilisation of shearing motions was confirmed numerically by~\citet{KSS14}, followed by many others~\citep[e.g., ][]{RQV15,RQV16,RQV18,MSK16}.  

In this paper, we examine the criteria for the CE distribution function to be stable to 
microinstabilities at collisionless scales -- i.e., at $k \lambda_s \gg 1$ (where $k$ is the microinstability wavenumber), and $\gamma \tau_L \gg 1$. 
In a two-species plasma with a fixed mass ratio $\mu_e \equiv m_e/m_i$ and a charge $Z$ that is not very large, these criteria turn out to be relationships between 
three dimensionless parameters: $\lambda/L$, $d_e/L$, and $\beta$, where $\lambda \equiv \lambda_e = 
\lambda_i$ is the mean free path for both ions and electrons, and $d_e$ is the electron inertial scale. 
The first criterion (which we refer to as the $\beta$\textit{-stabilisation condition}) 
is that the ratio $\lambda/L$ be much smaller than the reciprocal of the plasma 
$\beta$, viz. $\lambda \beta/L \ll 1$. This condition arises because the 
microinstabilities discussed in this paper are stabilised (usually by Lorentz forces) at sufficiently low 
$\beta$. The second criterion (the \textit{collisional-stabilisation condition}) is that 
the characteristic wavenumber $k_{\rm{peak}}$ of the fastest-growing 
microinstability in the absence of collisional effects be comparable to (or smaller than) the 
reciprocal of the mean-free-path: $k_{\rm{peak}} \lambda \lesssim 1$. 
Unlike the $\beta$-stabilisation condition, we do not justify this 
condition rigorously, because our calculations are only valid for wavenumbers $k$ such that $k \lambda \gg 
1$; thus, we cannot say anything definitive about the $k \lambda \lesssim 1$ 
regime. We do, however, show that another, more restrictive 
stabilisation condition that one might naively expect to exist on account of collisions -- that 
microinstabilities cannot occur if their growth rate $\gamma$ is smaller than 
the collision frequency (viz., $\gamma \tau_s \lesssim 1$) -- does not, in fact, apply
to the most significant microinstabilities in CE plasma. 
There are good physical reasons to believe that the CE distribution 
function is stable against collisionless microinstabilities if the collisional-stabilisation 
condition $k_{\rm{peak}} \lambda \lesssim 1$ is satisfied: not least that the typical growth time of the fastest microinstability in CE plasma (calculated neglecting collisional damping of microinstabilities) becomes comparable to the macroscopic evolution time scale 
$\tau_L$.  We thus assume the validity of 
the collisional-stabilisation condition throughout this paper. 
How $k_{\rm{peak}}$ relates to the other 
physical parameters is in general somewhat complicated; however, typically 
the collisional-stabilisation condition can be written as a lower bound on the ratio $d_e/L$. For example, 
in the limit of very high~$\beta$, it is $d_e/L > (m_e/m_i)^{-1/6} 
(\lambda/L)^{2/3}$ (see section \ref{poseps_stab}). 

If both the $\beta$-stabilisation and collisional-stabilisation conditions are violated, we demonstrate that
CE plasma will be subject to at least one microinstability, and quite possibly
multiple microinstabilities across a wide range of scales. Some of these microinstabilities are thresholdless -- that is, without including collisional 
effects, they will occur for CE distributions
departing from a Maxwellian distribution by an asymptotically small amount. Note that 
all significant microinstabilities associated with the CE distribution function are `low frequency': 
their growth rate $\gamma$ satisfies $\gamma \ll k v_{\mathrm{th}s}$, where $k$ is 
the typical wavenumber of the instability, and $v_{\mathrm{th}s}$ the thermal velocity of 
the particles of species $s$. This property enables a small anisotropy of the 
distribution function to create forces capable of 
driving microinstabilities (see section \ref{linear_stab_method}). 

In this paper, we characterise all significant microinstabilities that 
arise at different values of $\lambda/L$, $\beta$, and $d_e/L$ for a particular 
form of the CE distribution function appropriate for a strongly magnetised plasma -- that is, a plasma
where the Larmor radii of ions and electrons are much smaller than the 
corresponding mean free paths of these particles. We treat this particular case because of its importance to astrophysical  
systems, which almost always possess macroscopic magnetic fields of sufficient 
strength to magnetise their constituent particles~\citep{SC06}. 
Our characterisation of microinstabilities 
focuses on providing the maximum microinstability growth rates, as well as the wavenumbers 
at which this growth occurs. We find that there exist two general classes of 
microinstabilities: those driven by the non-Maxwellian component of the CE 
distribution associated with temperature gradients, and those driven by the 
non-Maxwellian component associated with bulk velocity gradients (`shear'). We refer to 
these two non-Maxwellian terms (which exist for both the ion and electron CE distribution functions) 
as the \textit{CE temperature-gradient terms} and the \textit{CE shear terms} 
respectively. Microinstabilities driven by the CE temperature-gradient terms are 
called the CE temperature-gradient-driven (CET) microinstabilities, while those driven by 
the CE shear terms are the CE shear-driven (CES) microinstabilities. 

As expected, within this general microinstability classification scheme, 
we recover a number of previously identified microinstabilities, 
including the (electron-shear-driven) transverse instability (which we discuss in sections \ref{pospres_electron_trans} and \ref{negpres_subelectron_obliquetrans}), the whistler instability (section \ref{pospres_electron_EC}), the electron mirror instability (section \ref{pospres_electron_oblique}),
the electron firehose instability (sections \ref{negpress_electronfire_prl} and \ref{negpres_electron_oblique}), the 
ordinary-mode instability (section \ref{negpres_subelectron_ord}), the
(electron-temperature-gradient-driven) whistler heat-flux instability (sections \ref{electron_heatflux_instab_whistl} and \ref{electron_heatflux_instab_whistl_obl}), and the (ion-shear-driven) mirror (section \ref{pospress_ion_mirror}) and firehose (sections  \ref{negpres_fire}, \ref{negpres_fire_par}, \ref{negpres_fire_oblique}, \ref{negpres_fire_critline}, and \ref{negpres_fire_subion}) instabilities. 
We also find four microinstabilities that, to our knowledge, have not been previously discovered: two ion-temperature-gradient-driven ones at ion Larmor scales -- the 
\textit{slow-hydromagnetic-wave instability} (section \ref{ion_heatflux_instab_slowwave}) and the \textit{long-wavelength kinetic-Alfv\'en wave instability} (section \ref{CET_KAW_instab})
-- and two electron-shear-driven ones -- the \textit{electron-scale-transition (EST) instability} (section \ref{negpres_elec_EST}) and the \textit{whisper instability} (section \ref{negpres_subelectron_phantom})
-- at electron-Larmor and sub-electron-Larmor scales, respectively. 
Of these microinstabilities, the whisper instability seems to be of particular 
significance: it has an extremely large growth rate in certain 
parameter regimes, and is associated with a new high-$\beta$ wave in a Maxwellian plasma, which also appears to have previously escaped attention.  
For convenience, a complete index of microinstabilities discussed in this paper is given in table 
\ref{tab:microinstabs_index}, while the peak growth rates of these microinstabilities 
and the scales at which they occur (for a hydrogen CE plasma) are given in table \ref{tab:microinstabs_prop}. 
\begin{table}
\centering
{\renewcommand{\arraystretch}{1.01}
\renewcommand{\tabcolsep}{0.15cm}
\begin{tabular}[c]{c|c|c|c}
Microinstability name & Section(s) & \renewcommand{\arraystretch}{1} \begin{tabular}{@{}c@{}} Other names \\ occurring in literature \end{tabular}  & Driving CE term   \\[0.8em]
\hline
\renewcommand{\arraystretch}{1.01}
Mirror instability & \ref{pospress_ion_mirror} & -- & Ion-velocity shear \\[0.8em]
Firehose instability & \renewcommand{\arraystretch}{1} \begin{tabular}{@{}c@{}}  \ref{negpres_fire}, \ref{negpres_fire_par}, \\ \ref{negpres_fire_oblique}, \ref{negpres_fire_critline}, \\ \ref{negpres_fire_subion} \end{tabular} & \renewcommand{\arraystretch}{1} \begin{tabular}{@{}c@{}} Garden-hose \\ instability  \end{tabular}  & Ion-velocity shear  \\[0.8em]
\renewcommand{\arraystretch}{1.01}
\renewcommand{\arraystretch}{1} \begin{tabular}{@{}c@{}} Slow-hydromagnetic- \\  wave instability* \end{tabular} & \ref{ion_heatflux_instab_slowwave} & -- & \renewcommand{\arraystretch}{1} \begin{tabular}{@{}c@{}} Ion-temperature \\ gradient \end{tabular}  \\[0.8em]
\renewcommand{\arraystretch}{1} \begin{tabular}{@{}c@{}} Long-wavelength kinetic Alfv\'en \\  wave (KAW) instability* \end{tabular} & \ref{CET_KAW_instab} & -- & \renewcommand{\arraystretch}{1} \begin{tabular}{@{}c@{}} Ion-temperature \\ gradient \end{tabular}  \\[0.8em]
\renewcommand{\arraystretch}{1.01}
\renewcommand{\arraystretch}{1} \begin{tabular}{@{}c@{}} CES whistler \\ instability \end{tabular}  & \ref{pospres_electron_EC} & \renewcommand{\arraystretch}{1} \begin{tabular}{@{}c@{}} Electron-cyclotron \\ (whistler) instability  \end{tabular} & \renewcommand{\arraystretch}{1} \begin{tabular}{@{}c@{}} Electron-velocity \\ shear \end{tabular} \\[0.8em]
\renewcommand{\arraystretch}{1.01}
\renewcommand{\arraystretch}{1} \begin{tabular}{@{}c@{}} Electron \\ mirror instability  \end{tabular} & \ref{pospres_electron_oblique} & \renewcommand{\arraystretch}{1} \begin{tabular}{@{}c@{}} KAW, field-swelling \\ instability \end{tabular} & \renewcommand{\arraystretch}{1} \begin{tabular}{@{}c@{}} Electron-/ion- \\ velocity shear  \end{tabular} \\[0.8em]
\renewcommand{\arraystretch}{1} \begin{tabular}{@{}c@{}} Electron \\ firehose instability  \end{tabular} & \ref{negpress_electronfire_prl}, \ref{negpres_electron_oblique} & KAW instability & \renewcommand{\arraystretch}{1} \begin{tabular}{@{}c@{}} Electron-/ion- \\ velocity shear  \end{tabular} \\[0.8em]
\renewcommand{\arraystretch}{1} \begin{tabular}{@{}c@{}} Electron-scale-transition \\ (EST) instability* \end{tabular} & \ref{negpres_elec_EST} & -- & \renewcommand{\arraystretch}{1} \begin{tabular}{@{}c@{}} Electron-velocity \\ shear  \end{tabular} \\[0.8em]
\renewcommand{\arraystretch}{1} Whisper instability* & \ref{negpres_subelectron_phantom} & -- & \renewcommand{\arraystretch}{1} \begin{tabular}{@{}c@{}} Electron-velocity \\ shear  \end{tabular} \\[0.8em]
\renewcommand{\arraystretch}{1.01}
Transverse instability & \ref{pospres_electron_trans}, \ref{negpres_subelectron_obliquetrans} &  \renewcommand{\arraystretch}{1} \begin{tabular}{@{}c@{}} Small-anisotropy \\ Weibel instability  \end{tabular} & \renewcommand{\arraystretch}{1} \begin{tabular}{@{}c@{}} Electron-velocity \\ shear  \end{tabular} \\[0.8em]
\renewcommand{\arraystretch}{1} \begin{tabular}{@{}c@{}} Ordinary-mode \\ instability \end{tabular}  & \ref{negpres_subelectron_ord} & -- & \renewcommand{\arraystretch}{1} \begin{tabular}{@{}c@{}} Electron-velocity \\ shear  \end{tabular} \\[0.8em]
\renewcommand{\arraystretch}{1} \begin{tabular}{@{}c@{}} CET whistler \\ instability \end{tabular}  & \ref{electron_heatflux_instab_whistl}, \ref{electron_heatflux_instab_whistl_obl} & \renewcommand{\arraystretch}{1} \begin{tabular}{@{}c@{}} Whistler heat \\ flux instability  \end{tabular} & \renewcommand{\arraystretch}{1} \begin{tabular}{@{}c@{}} Electron-temp. \\ gradient \end{tabular}
\renewcommand{\arraystretch}{1}
\end{tabular}} 
\caption{\textbf{Index of microinstabilities.} The microinstabilities listed here are those discussed in the main text, with the relevant sections indicated. 
We also indicate whether 
these microinstabilities are driven by macroscopic electron/ion temperature gradients associated with the CE 
distribution function, or by macroscopic electron/ion velocity gradients (shears): see section \ref{sec:CE_dist_func} for a discussion of this classification. Newly identified microinstabilities are indicated with an asterisk.}
\label{tab:microinstabs_index}
\end{table}
\begin{table}
\centering
{\renewcommand{\arraystretch}{1}
\renewcommand{\tabcolsep}{0.15cm}
\begin{tabular}[c]{c|c|c|c}
\renewcommand{\arraystretch}{1} \begin{tabular}{@{}c@{}} Microinstability \\ name \end{tabular} & Wavenumber scale & \renewcommand{\arraystretch}{1} \begin{tabular}{@{}c@{}} Growth rate \\ ($\times \Omega_e$) \end{tabular} & $\beta$-threshold  \\
\hline
\renewcommand{\arraystretch}{1} \begin{tabular}{@{}c@{}} Mirror \\ instability \end{tabular}  & $k_{\|} \rho_i \lesssim k_{\perp} \rho_i \sim 1 $ & $\mu_e \epsilon$ & $\epsilon \beta \sim  1$ \\
 & & & \\[-0.8em]
\renewcommand{\arraystretch}{1} \begin{tabular}{@{}c@{}} Parallel firehose \\ instability \end{tabular} & $k_{\|} \rho_i \sim k_{\perp} \rho_i \sim \epsilon^{1/2}$, $ k_{\bot} \ll \epsilon^{1/4} k_{\|}$ & $\mu_e \epsilon$ & $\epsilon \beta \sim 1 $ \\ 
 & & & \\[-0.8em]
\renewcommand{\arraystretch}{1} \begin{tabular}{@{}c@{}} Oblique firehose \\ instability \end{tabular}  & $ k_{\|} \rho_i \sim \epsilon^{1/4}$,  $k_{\bot} \sim k_{\|}$ & $\mu_e \epsilon^{3/4}$ & $\epsilon \beta \sim  1 $ \\
 & & & \\[-0.8em]
\renewcommand{\arraystretch}{1} \begin{tabular}{@{}c@{}} Critical-line firehose \\ instability ($\epsilon \gtrsim 10^{-6}$) \end{tabular} & $k_{\|} \rho_i \approx \sqrt{3/2} \, k_{\perp} \rho_i < 1 $ & $\mu_e \epsilon^{1/2}$ & $\epsilon \beta \sim  1 $ \\
 & & & \\[-0.8em]
 \renewcommand{\arraystretch}{1} \begin{tabular}{@{}c@{}} Critical-line firehose \\ instability ($\epsilon \ll 10^{-6}$) \end{tabular} & $k_{\|} \rho_i \approx \sqrt{3/2} \, k_{\perp} \rho_i \sim \epsilon^{1/12} $ & $\mu_e \epsilon^{7/12}$ & $\epsilon \beta \sim  1 $ \\
 & & & \\[-0.8em]
\renewcommand{\arraystretch}{1} \begin{tabular}{@{}c@{}} Slow-hydro.-wave \\ instability \end{tabular}  & $k_{\perp} \rho_i \lesssim k_{\|} \rho_i \sim 1 $ & $ \mu_e^{5/4} \epsilon$ & $\epsilon \beta \sim \mu_e^{-1/4}$ \\
 & & & \\[-0.8em]
 \renewcommand{\arraystretch}{1} \begin{tabular}{@{}c@{}} Long wavelength \\ KAW instability \end{tabular}  & $k_{\|} \rho_i < k_{\bot} \rho_i  \sim 1 $ & $\mu_e^{5/4} \epsilon k_{\|}/k_{\perp}$ & $\epsilon \beta \sim  \mu_e^{-1/4}$ \\
 & & & \\[-0.8em]
   \renewcommand{\arraystretch}{1} \begin{tabular}{@{}c@{}} CES whistler \\ instability \end{tabular}  & $k_{\bot} \rho_e \lesssim k_{\|} \rho_e \sim 1 $ & $\mu_e^{1/2} \epsilon$ & $\epsilon \beta \sim  \mu_e^{-1/2}$ \\
 & & & \\[-0.8em]
 \renewcommand{\arraystretch}{1} \begin{tabular}{@{}c@{}} Electron mirror \\ instability \end{tabular}  & $k_{\|} \rho_e \lesssim k_{\perp} \rho_e \sim 1 $ & $\mu_e^{1/2} \epsilon$ & $\epsilon \beta \sim  \mu_e^{-1/2}$ \\
 & & & \\[-0.8em]
 \renewcommand{\arraystretch}{1} \begin{tabular}{@{}c@{}} Parallel electron \\ firehose instability \end{tabular}  & $k_{\bot} \rho_e \lesssim k_{\|} \rho_e \sim 1 $ & $\mu_e \epsilon$ & $\epsilon \beta \sim  \mu_e^{-1/2}$ \\
 & & & \\[-0.8em]
\renewcommand{\arraystretch}{1} \begin{tabular}{@{}c@{}} Oblique electron \\ firehose instability \end{tabular}  & $k_{\|} \rho_e \lesssim k_{\perp} \rho_e \sim 1 $ & $\mu_e^{1/2} \epsilon$ & $\epsilon \beta \sim  \mu_e^{-1/2}$ \\
  & & & \\[-0.8em]
\renewcommand{\arraystretch}{1} \begin{tabular}{@{}c@{}} EST instability \\  ($\epsilon \beta^{5/7} \lesssim \mu_e^{-1/2}$) \end{tabular}  & $k_{\|} \rho_e < 1 \lesssim k_{\perp} \rho_e \sim \epsilon^{1/2} \beta^{1/2} \mu_e^{1/4}$ & $\mu_e^{5/4} \epsilon^{5/2} \beta^{3/2} $ & $\epsilon \beta \sim \mu_e^{-1/2}$   \\
  & & & \\[-0.8em]
  \renewcommand{\arraystretch}{1} \begin{tabular}{@{}c@{}} EST instability \\  ($\epsilon \beta^{5/7} \gtrsim \mu_e^{-1/2}$) \end{tabular}  & $k_{\|} \rho_e < 1 \ll k_{\perp} \rho_e \sim \epsilon^{1/5} \mu_e^{1/10}$ & $\mu_e^{1/5} \epsilon^{2/5} $ & $\epsilon \beta \sim \mu_e^{-1/2}$ \\
  & & & \\[-0.8em]
 \renewcommand{\arraystretch}{1} \begin{tabular}{@{}c@{}} Whisper \\ instability \end{tabular}  & $k_{\|} \rho_e < 1 \ll k_{\perp} \rho_e \sim \epsilon^{1/2} \beta^{1/2} \mu_e^{1/4}$ & $\mu_e^{3/8} \epsilon^{3/4} \beta^{1/4} $ & $\epsilon \beta \sim \beta^{2/7} \mu_e^{-1/2}$  \\
 & & & \\[-0.8em]
 \renewcommand{\arraystretch}{1} \begin{tabular}{@{}c@{}} Parallel transverse \\ instability \end{tabular}  &$k_{\perp} \rho_e \lesssim k_{\|} \rho_e \sim \epsilon^{1/2} \beta^{1/2} \mu_e^{1/4}$ & $\mu_e^{3/4} \epsilon^{3/2} \beta^{1/2}$ & $\epsilon \beta \sim \mu_e^{-1/2}$ \\
  & & & \\[-0.8em]
\renewcommand{\arraystretch}{1} \begin{tabular}{@{}c@{}} Oblique transverse \\ instability \end{tabular}  &  $1 \lesssim k_{\|} \rho_e \lesssim k_{\bot} \rho_e \sim \epsilon^{1/2} \beta^{1/2} \mu_e^{1/4}$ & $\mu_e^{3/4} \epsilon^{3/2} \beta^{1/2}$ & $\epsilon \beta \sim \mu_e^{-1/2}$ \\
  & & & \\[-0.8em]
\renewcommand{\arraystretch}{1} \begin{tabular}{@{}c@{}} Ordinary-mode \\ instability \end{tabular}  & $k_{\|} = 0$, $k_{\perp} \rho_e \sim \epsilon^{1/2} \beta^{1/2} \mu_e^{1/4}$ &  $\mu_e^{3/4} \epsilon^{3/2} \beta^{1/2}$ &  $\epsilon \beta \sim \beta^{2/3} \mu_e^{-1/2} $ \\
 & & & \\[-0.8em]
 \renewcommand{\arraystretch}{1} \begin{tabular}{@{}c@{}} CET whistler \\ instability \end{tabular}  & $k_{\perp} \rho_e \lesssim k_{\|} \rho_e \sim \epsilon^{1/5} \beta^{1/5} \mu_e^{1/20} $ & $\mu_e^{1/4} \epsilon$ & $\epsilon \beta \sim  \mu_e^{-1/4}$ 
\end{tabular}} 
\caption{\textbf{Properties of microinstabilities.} Typical wavenumbers and maximum growth rates of microinstabilities in strongly magnetised hydrogen CE plasma, and their $\beta$-stabilisation thresholds. Here, $\mu_e = m_e/m_i$.  
We assume scalings (\ref{brag_multiscale}) to relate the magnitude of CE temperature-gradient-driven and CE shear-driven microinstabilities. These scalings lead to the non-Maxwellian component of the ion distribution function having magnitude {$\sim$}$\epsilon = \mathrm{Ma} \, \lambda/L_V$, where $\lambda = \lambda_e = \lambda_i$, $L_V$ is the length scale of the CE plasma's bulk fluid motions in the direction parallel to the guide magnetic field [see (\ref{parallel_vel_scale}\textit{d})], and $\mathrm{Ma}$ is the Mach number of those bulk motions. 
The quoted wavenumbers and growth rates apply when the $\beta$-stabilisation 
threshold is exceeded by an order-unity or much larger factor.}
\label{tab:microinstabs_prop}
\end{table} 
There do exist microinstabilities in CE plasma that are not represented in tables \ref{tab:microinstabs_index}
and \ref{tab:microinstabs_prop}; however, we claim that the instabilities discussed in this paper are the most significant, on account of their large growth rates and/or low 
$\beta$-stabilisation thresholds compared to the unrepresented ones. 

\begin{figure}
\centerline{\includegraphics[width=0.99\textwidth]{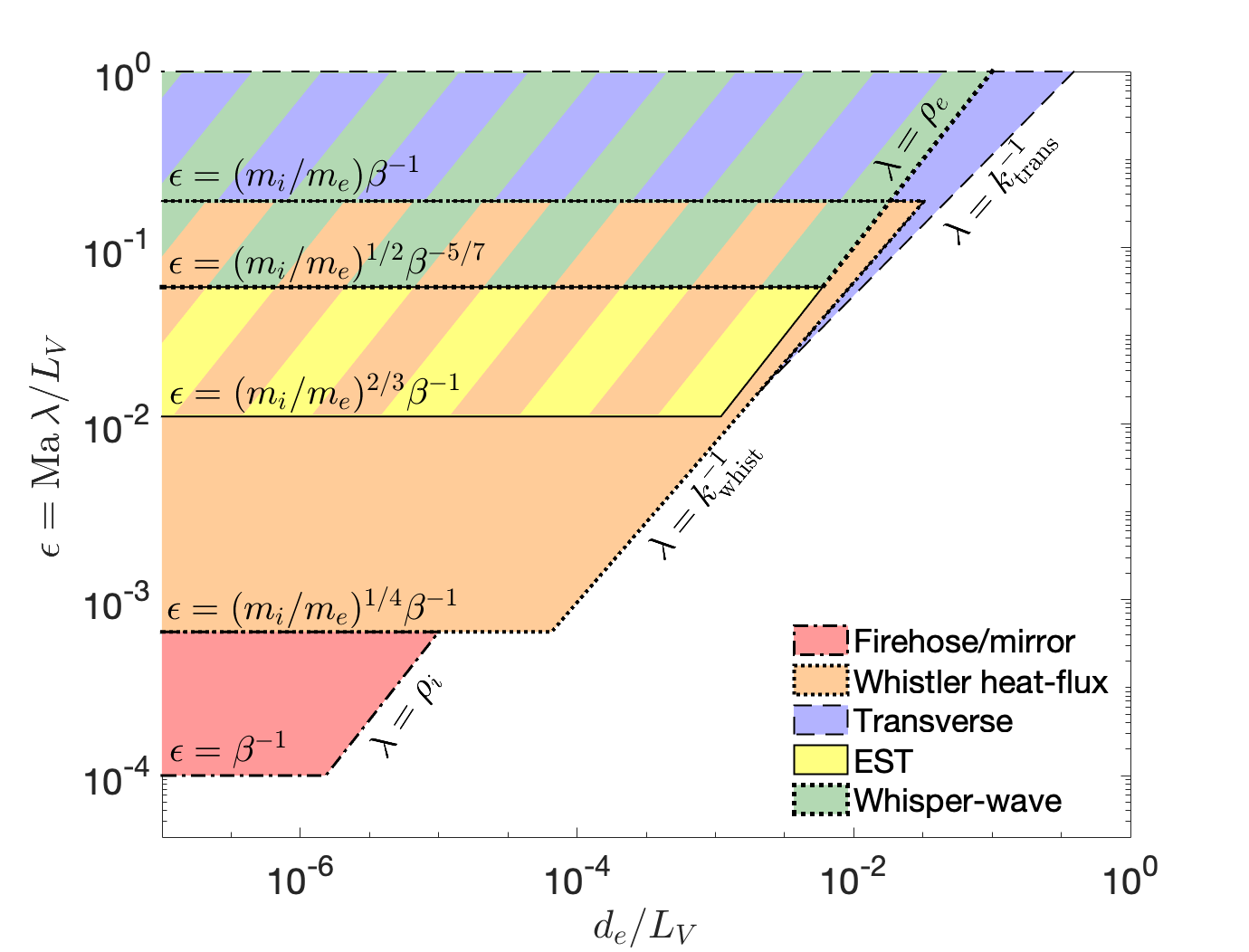}}
\caption{\textit{Stability map for the CE distribution function}. Idealised illustration of the stability 
of strongly magnetised, classical, collisional hydrogen plasma to microinstabilities
for different (non-dimensionalised) values of the mean free path $\lambda = \lambda_e = \lambda_i$ and the electron inertial scale~$d_e$.
Here, the length scale $L_V$ to which $\lambda$ and $d_e$ are normalised is 
the length scale of the CE plasma's bulk fluid motions in the direction parallel to the guide magnetic field [see (\ref{parallel_vel_scale}\textit{d})]; 
we assume scalings (\ref{brag_multiscale}) 
to relate the magnitude of CE temperature-gradient-driven and CE shear-driven 
microinstabilities, so the CE expansion parameter is $\epsilon = \mathrm{Ma} \, \lambda/L_V$ (see the caption of table \ref{tab:microinstabs_prop} for definitions).
The white region of the ($d_e/L_V$,$\mathrm{Ma} \, \lambda/L_V$) stability map is stable; the coloured 
regions are not. In the unstable regions, the fastest-growing microinstability is 
indicated by colour according to the figure's legend; in the regions where multiple microinstabilities could be 
operating simulataneously, multiple colours have been employed. The plasma beta $\beta$
here was taken to be $\beta = 10^{4}$, and the Mach number $\mathrm{Ma} = 1$.
} \label{stabilitymapintro}
\end{figure}

Having systematically identified all significant microinstabilities, we can construct `stability maps' of
strongly magnetised CE plasma using ``phase diagrams'' over a two-dimensional ($\lambda/L$, $d_e/L$) 
parameter space at a fixed $\beta$. An example of such a map (for a hydrogen plasma with equal ion and electron temperatures) is shown 
in figure \ref{stabilitymapintro}. 
The entire region of the ($\lambda/L$, $d_e/L$) space depicted in figure \ref{stabilitymapintro} could naively be 
characterised as pertaining to classical, collisional plasma, and thus 
describable by fluid equations, with transport coefficients given by standard 
CE theory. However, there is a significant region of the 
parameter space (which is demarcated by boundaries corresponding to the $\beta$-stabilisation and collisional-stabilisation conditions)
that is unstable to microinstabilities.  In fact, in strongly magnetised plasma,
the collisional-stabilisation condition is never satisfied, because there exist 
microinstabilities whose characteristic length scales are the ion and 
electron Larmor radii, respectively; this being the case, only the $\beta$-stabilisation
condition guarantees kinetic stability. 

The effect of microinstabilities being present in CE plasma would be to change the non-Maxwellian components of the 
distribution function, and therefore to alter the CE-prescribed resistivity, thermal conductivity 
and/or viscosity. Identifying the dominant microinstability or microinstabilities in such plasmas (as is done in figure \ref{stabilitymapintro} for a hydrogen plasma) is 
then necessary for calculating the true transport coefficients, which are
likely determined by the effective collisionality associated with the saturated state of the dominant microinstability rather than 
by Coulomb collisions. Although such calculations are not undertaken in this paper, it seems 
possible that the modified transport coefficients could be determined 
self-consistently in terms of macroscopic plasma properties such as temperature gradients or velocity shears.  
We note that the calculation presented here assumes that the CE distribution function is determined without 
the microinstabilities and thus is only correct when the plasma is stable. Therefore, strictly speaking, the only conclusion one can make when
the CE plasma is unstable is that the naive CE values of transport coefficients should not be taken as 
correct.

We emphasise that kinetic instability of CE plasmas is a phenomenon of practical importance as well as   
academic interest. We illustrate this in tables \ref{tab:physicalparams_inputs} and 
\ref{tab:physicalparams}, where the possibility of microinstabilities is considered  
for a selection of physical systems composed of classical, 
collisional plasma.
\begin{table}
\centering
{\renewcommand{\arraystretch}{1.3}
\renewcommand{\tabcolsep}{0.15cm}
\begin{tabular}[c]{c|c|c|c|c|c}
Environment & $T_{e}$ (eV) & $T_{i}$ (eV) & $n_{e}$ (cm$^{-3}$)& $B$ (G) & $L$ (cm) \\
\hline
Warm intergalatic medium (WIGM) & $10^{2}$ & $10^{2}$ & $10^{-5}$ & $10^{-8}$ & $3 \times 10^{24}$ \\
Intracluster medium (ICM) & $10^{4}$ & $10^{4}$ & $10^{-2}$ & $10^{-5}$ & $3 \times 10^{23}$  
\\
IGM post reionisation  & $1$ & $1$ & $10^{-6}$ & $10^{-19}$ & $3 \times 10^{24}$  
\\
Solar photosphere & $1$ & $1$ & $10^{17}$ & $500$ & $10^{7}$  \\
Solar chromosphere & $1$ & $1$ & $10^{12}$ & $10$ & $10^{7}$  \\
ICF hot spot (NIF) & $5 \times 10^3$ & $5 \times 10^3$ & $10^{25}$ & $10^7$ & $2 \times 10^{-3}$ 
\\
Laser-ablated plasma (long pulse) & $10^3$ & $5 \times 10^2$ & $4 \times 10^{21}$ & $10^6$ & $10^{-2}$ 
\\
NIF `TDYNO' laser-plasma & $10^3$ & $10^3$ & $5 \times 10^{20}$ & $10^6$ & $10^{-2}$ 
\end{tabular}} 
\caption{\textbf{Plasma parameters for some physical systems composed of classical, collisional plasma.} 
The values of temperature and density of the WIGM given here are from~\citet{NME08}, while those of the ICM come from~\citet{F94}. The estimates of the typical magnetic-field strengths and scale lengths for both the WIGM and the ICM are from~\citet{RKCD08}. For simplicity, we have assumed equal ion and 
electron temperatures; however, we acknowledge that there is some uncertainty as to the validity of this assumption~\citep[see, e.g.,][]{YFH05}. \citet{BL01} is the source of estimates for the IGM post reionisation. Estimates for typical solar parameters are from~\citet{WTS14} 
and~\citet{S12}. The values of ion temperature and electron density  
for ICF hot spots are from~\citet{H14}, who reported DT experiments carried out on the National Ignition 
Facility (NIF); the estimates of magnetic-field strength, electron temperature and scale 
length come from numerical simulations of the same 
experiment~\citep{W17}. The parameters for the laser-ablated CH plasma are from an 
experiment on the OMEGA laser facility, with a 1 ns, 500 J pulse with a $0.351 \, \mu \mathrm{m}$ wavelength~\citep{L06}; 
we assume that the measured fields are found in front of the critical-density surface when 
estimating the density. The `TDYNO' laser-plasma is a turbulent CH plasma that was
produced as part of a recent laboratory astrophysics experiment on the NIF which found evidence of suppressed 
heat conduction~\citep[][see main text]{M21}.
 Naturally, the systems described here often support a range of density, temperatures 
and magnetic fields, so the values provided should be 
understood as representative, but negotiable.}
\label{tab:physicalparams_inputs}
\hfill \break
\centering
{\renewcommand{\arraystretch}{1.3}
\renewcommand{\tabcolsep}{0.15cm}
\begin{tabular}[c]{c|c|c|c|c|c|c|c|c}
Environment & $\lambda_e/L$ & $\lambda_i/L$ & $d_e/L$ & $\beta$ & $\beta \lambda_e/L$ & $\rho_e/\lambda_e$ & $\rho_i/\lambda_i$ & $k_{\rm{peak}} \lambda_e$\\
\hline
WIGM & $2 \times 10^{-3}$ & $2 \times 10^{-3}$ & $ 2 \times 10^{-17}$ & $10^{4}$ & 20 & $ 10^{-12}$ & $10^{-11}$ & $10^{12}$ \\
ICM & $ 10^{-2}$ &  $ 10^{-2}$ & $10^{-17}$ & $10^{2}$ & $1$ & $10^{-14}$ & 
$10^{-13}$ & $10^{14}$ \\
Reion. IGM & $ 10^{-7}$ &  $ 10^{-7}$ & $10^{-16}$ & $10^{22}$ & $10^{15}$ & $0.5$ & 
$20$ & $10^{5}$ \\
Photosphere & $ 6 \times 10^{-12} $ &  $6 \times 10^{-12}$  & $2 \times 10^{-10}$ & $30$ & $10^{-10}$ & $110$ & 
$4 \times 10^{3}$ & $10^{-4}$ \\
Chromosphere & $ 2 \times 10^{-7}$ &  $2 \times 10^{-7}$ & $5 \times 10^{-8}$ & $1$ & $10^{-7}$ & $0.2$ & 
$6$ & $0.2$ \\
ICF hot spot & $ 0.3$ &  $ 0.2$ & $4 \times 10^{-5}$ & $4 \times 10^{6}$ & $10^6$ & $0.1$ & 
$10$ & $10^3$ \\
Laser-abl. pl. & $ 7 \times 10^{-3}$ &  $ 3 \times 10^{-4}$ & $8 \times 10^{-4}$ & $200$ & $1$ & $0.4$ & 
$800$ & $2.5$ \\
NIF TDYNO & $ 6 \times 10^{-2}$ &  $2 \times 10^{-2}$ & $2 \times 10^{-3}$ & $45$ & $2.5$ & $0.1$ & 
$200$ & $10$
\end{tabular}} 
\caption{\textbf {Derived plasma parameters for systems composed of classical, collisional plasma.}
All parameters are 
calculated using~\citet{H94}, except for $k_{\mathrm{peak}} \lambda_e$. This is calculated by 
considering all possible instabilities, and then finding the magnitude of $k_{\mathrm{peak}} \lambda_e$
for the fastest-growing instability satisfying $k_{\mathrm{peak}} \lambda_e \gtrsim 1$.  Depending on the values of other 
parameters, the fastest-growing instability varies between systems; in the WIGM, ICM, laser-ablation and TDYNO plasmas, the whistler heat-flux 
instability is the fastest-growing one, while in the reionised IGM or ICF hot spots, the transverse instability is.} 
\label{tab:physicalparams}
\end{table} 
We find that, while there exist some systems where CE plasmas are immune to  
microinstabilities -- for example, the photosphere and chromosphere -- there 
are many other astrophysical plasma systems that are likely susceptible to them. Similar considerations apply to a 
range of laser plasmas, including plasmas generated in inertial-confinement-fusion 
and laboratory-astrophysics
experiments. Indeed, a recent experiment carried out on the National Ignition Facility (NIF) -- part of a wider
programme of work exploring magnetic-field amplification in turbulent 
laser-plasmas~\citep{T18,B21a,B21b,B23} -- found evidence for the existence of large-amplitude local temperature fluctuations over a range of scales, a finding that was inconsistent with Spitzer thermal conduction~\citep{M21}. This claim was corroborated by MHD simulations (with the code FLASH) of the experiment that modelled thermal conduction either using the Spitzer model, or no explicit thermal conduction model: the latter simulations were found to be much closer to the actual experimental data. Because the plasma created in the NIF experiment is also anticipated by our theory to be susceptible to CE microinstabilities, observations of a discrepancy with CE-derived transport coefficients are tantalising. We note that the idea of microinstabilities emerging in both collisional 
astrophysical plasmas and laser plasmas is not a new one: see, e.g.~\citet{SCKHS05} or~\citet{HT15}
in the former context; in the latter,~\citet{EB87} or~\citet{BKWM20}. 
However, to our knowledge there does not exist 
a systematic treatment of the general kinetic stability of 
CE plasmas. This is the gap that this paper attempts to fill. 

This paper has the following structure. In section \ref{ProbOut}, we discuss kinetic and fluid descriptions of 
classical plasma. We then describe the CE expansion in collisional plasma:
we work out the CE distribution function arising in a two-species strongly magnetised plasma, 
evaluate the friction forces, heat and momentum fluxes necessary to 
construct a closed set of plasma-fluid equations, and systematically estimate the size of the non-Maxwellian components of this 
distribution. Next, we discuss qualitatively 
the existence and nature of microinstabilities potentially arising in CE plasma, before 
presenting the methodology that we later use to perform the full linear, kinetic stability 
calculation. We provide an overview of this methodology in section \ref{linear_stab_overview}, 
and then a much more detailed exposition of it in section \ref{linear_stab_method}: in particular, 
we describe in the latter how a simple form of the dispersion relation for the fastest microinstabilities can be
obtained by considering the low-frequency limit $\gamma \ll k v_{\mathrm{th}s}$ of the hot-plasma dispersion relation, 
and how this simplified dispersion relation can be solved analytically. 
Readers who are uninterested in the technical details of this calculation are 
encouraged to pass over section \ref{linear_stab_method}; knowledge of its 
contents is not a pre-requisite for subsequent sections. 
In sections \ref{Results} and \ref{Results_shearingterm}, we construct stability maps (analogous to figure \ref{stabilitymapintro}) showing the parameter ranges in which the 
CE distribution function is stable, to CET and CES microinstabilities, respectively. The parameters are $\beta$ and $\lambda/L$, and we construct separate 
stability maps for CET and CES microinstabilities in order to take into account the fact that $L$ is in general not the same in the situations where these two types of microinstabilities occur. 
In section \ref{Results}, we also discuss the
significant CET microinstabilities that can occur (or not) at different values $\lambda/L$ and 
$\beta$, and provide simple analytic characterisations of them; in section \ref{Results_shearingterm}, we do the same for significant CES 
microinstabilities. Finally, in section \ref{Discussion}, we discuss the general implications of these 
instabilities for classical, collisional plasmas, and consider future research directions. 
Throughout this paper, most lengthy calculations are 
exiled to appendices; a glossary of mathematical notation is given in appendix 
\ref{Glossary}. 

\section{Problem setup} \label{ProbOut}

\subsection{Kinetic versus fluid description of classical plasma}

The evolution of classical plasma is most generally described by kinetic theory, via the solution 
of Maxwell-Vlasov-Landau equations for the distribution functions
of constituent particles. More specifically, in a kinetic description of a plasma, the distribution function $f_s(\boldsymbol{r},\boldsymbol{v},t)$ of the
particle of species $s$ satisfies
\begin{equation}
  \frac{\partial f_{s}}{\partial t} + \boldsymbol{v} \bcdot \bnabla f_{s} + \frac{Z_s e}{m_s} \left(\boldsymbol{E}+ \frac{\boldsymbol{v} \times \boldsymbol{B}}{c}\right) \bcdot \frac{\partial f_s}{\partial \boldsymbol{v}}= 
  \sum_{s'} \mathfrak{C}(f_s,f_{s'}) , \label{MaxVlasLan}
\end{equation}
where $t$ is time, $\boldsymbol{r}$ spatial position, $\boldsymbol{v}$ the velocity, $e$ the elementary 
charge, $Z_s e$ the charge and $m_s$ the mass of species $s$, $\boldsymbol{E}$ the  
electric field, $\boldsymbol{B}$ the magnetic field, $c$ the speed of 
light, and $\mathfrak{C}(f_s,f_{s'})$ the collision operator for interactions between species $s$ and $s'$. Equation (\ref{MaxVlasLan}) is coupled to 
Maxwell's equations: 
\begin{subeqnarray}
   \bnabla \bcdot \boldsymbol{E} & = &
   4 \upi \sum_{s} Z_s e \int \mathrm{d}^3 \boldsymbol{v} \, f_s,\\ [3pt]
      \bnabla \bcdot \boldsymbol{B} & = & 0,\\[3pt] 
  \bnabla \times \boldsymbol{E} & = &
   - \frac{1}{c} \frac{\p \boldsymbol{B}}{\p t},\\[3pt]
 \bnabla \times \boldsymbol{B} & = &
    \frac{1}{c} \frac{\p \boldsymbol{E}}{\p t} + \frac{4 \upi}{c} \sum_{s} Z_s e \int \mathrm{d}^3 \boldsymbol{v} \, \boldsymbol{v} \, f_s 
    . \label{Maxwell}
\end{subeqnarray}
Together, (\ref{MaxVlasLan}) and (\ref{Maxwell}) form a closed set of governing equations. 

The density $n_s$, bulk fluid velocity $\boldsymbol{V}_{s}$ and 
temperature $T_s$ of species $s$ can be formally defined in terms of moments of the 
distribution function:
\begin{subeqnarray}
  n_{s} & \equiv & \int \mathrm{d}^3 \boldsymbol{v} \, f_s,\\[3pt] 
  \boldsymbol{V}_{s} & \equiv & \frac{1}{n_s} \int \mathrm{d}^3 \boldsymbol{v} \, \boldsymbol{v} \, f_s,\\[3pt] 
  T_{s} & \equiv & \frac{1}{n_s}  \int \mathrm{d}^3 \boldsymbol{v} \, \frac{1}{3} m_s |\boldsymbol{v}-\boldsymbol{V}_s|^2 \, 
  f_s. \label{dens_vel_temp_def}
\end{subeqnarray}
Governing ``fluid'' equations are then derived by integrating (\ref{MaxVlasLan}) or outer products of (\ref{MaxVlasLan}) 
and the velocity variable $\boldsymbol{v}$ with respect to $\boldsymbol{v}$:
\begin{subeqnarray}
&&  \frac{\mathrm{D} n_s}{\mathrm{D} t}\bigg{|}_s + n_s \bnabla \cdot \boldsymbol{V}_s  = 0 , \\[3pt]
&& m_s n_s {\mathrm{D} \boldsymbol{V}_s \over \mathrm{D} t} \bigg{|}_s = -\bnabla p_s - \bnabla \cdot \boldsymbol{\pi}_s +Z_s e n_s \left(\boldsymbol{E} + \frac{\boldsymbol{V}_s \times \boldsymbol{B}}{c}\right) + \boldsymbol{R}_s , \\ 
&&  \frac{3}{2} n_s {\mathrm{D} T_s \over \mathrm{D} t} \bigg{|}_s + p_s \bnabla \cdot \boldsymbol{V}_s = - \bnabla \cdot \boldsymbol{q}_s - \boldsymbol{\pi}_s:\bnabla \boldsymbol{V}_s + \mathcal{Q}_s , \label{fluideqns}
\end{subeqnarray}
where 
\begin{equation}
\frac{\mathrm{D}}{\mathrm{D} t}\bigg{|}_s \equiv \frac{\p }{\p  t} + \boldsymbol{V}_s \bcdot \bnabla 
\end{equation}
is the convective derivative with respect to the fluid motions of species $s$, $p_s$ the pressure, $\boldsymbol{\pi}_s$ the viscosity tensor, and $\boldsymbol{q}_s$ the heat flux of species $s$, $\boldsymbol{R}_s$ 
the friction force on this species due to collisional interactions with 
other species, and $\mathcal{Q}_s$ the 
heating rate due to inter-species collisions. The latter quantities are formally defined in terms of 
the distribution function as follows:
\begin{subeqnarray}
  p_{s} & \equiv & \int \mathrm{d}^3 \boldsymbol{v} \, \frac{1}{3} m_s |\boldsymbol{v}-\boldsymbol{V}_s|^2 \, f_s = n_s T_s ,\\[3pt] 
 \boldsymbol{\pi}_s & \equiv & - p_s \mathsfbi{I} + \int \mathrm{d}^3 \boldsymbol{v} \, m_s \left(\boldsymbol{v}-\boldsymbol{V}_s\right) \left(\boldsymbol{v}-\boldsymbol{V}_s\right) \, f_s , \\[3pt] 
  \boldsymbol{q}_{s} & \equiv & \int \mathrm{d}^3 \boldsymbol{v} \, \frac{1}{2} m_s |\boldsymbol{v}-\boldsymbol{V}_s|^2 \left(\boldsymbol{v}-\boldsymbol{V}_s\right)  \, f_s, 
  \\[3pt]
  \boldsymbol{R}_s & \equiv & \sum_{s'} \int \mathrm{d}^3 \boldsymbol{v} \, m_s \boldsymbol{v} \, \mathfrak{C}(f_s,f_{s'}), \\[3pt]
  \mathcal{Q}_s & \equiv & -\boldsymbol{R}_s \bcdot \boldsymbol{V}_s + \sum_{s'} \int \mathrm{d}^3 \boldsymbol{v} \, \frac{1}{2} m_s |\boldsymbol{v}|^2 \, 
  \mathfrak{C}(f_s,f_{s'}). \label{fluxes_fluid}
\end{subeqnarray}

The distribution function only appears in Maxwell's equations via 
its zeroth and first moments; namely, Gauss' law (\ref{Maxwell}\textit{a}) and the Maxwell-Amp\`ere law (\ref{Maxwell}\textit{d}) can be 
written as
\begin{subeqnarray}
   \bnabla \bcdot \boldsymbol{E} & = &
   4 \upi \sum_{s} Z_s e n_s, \\ [3pt]
 \bnabla \times \boldsymbol{B} & = &
    \frac{1}{c} \frac{\p \boldsymbol{E}}{\p t} + \frac{4 \upi}{c} \sum_{s} Z_s e n_s \boldsymbol{V}_s.  \label{Maxwell_fluid}
\end{subeqnarray}
Unlike the kinetic description, the fluid equations (\ref{fluideqns}) combined with Maxwell's equations (\ref{Maxwell}\textit{b}), (\ref{Maxwell}\textit{d}), (\ref{Maxwell_fluid}\textit{a}), and (\ref{Maxwell_fluid}\textit{b}) are not 
a closed system: knowledge of the distribution function, not just of $n_s$, $\boldsymbol{V}_s$ or 
$T_s$, is required to calculate momentum and heat fluxes, as well as the friction force or 
heating. 

As discussed in the Introduction, solving fluid equations 
as opposed to kinetic equations is advantageous in many cases of interest. Since the dimensionality of the kinetic system 
is greater (a six-dimensional phase space vs. three-dimensional position
space), solving the kinetic system introduces both significant 
numerical and conceptual complexity. However, the system of fluid equations (\ref{fluideqns}) is only usable 
if some type of closure can be introduced to calculate 
$ \boldsymbol{\pi}_s$, $\boldsymbol{q}_{s}$, $\boldsymbol{R}_s$ and $\mathcal{Q}_s$ 
in terms of $n_s$, $\boldsymbol{V}_s$ and $T_s$. For classical plasmas, such a 
closure is generally not possible, except in the case of strongly
collisional plasmas. 

\subsection{The Chapman-Enskog (CE) expansion}

\subsubsection{The CE distribution functions} \label{sec:CE_dist_func}

For a classical, collisional plasma -- i.e., a plasma where the mean free path $\lambda_s$ of particles of species $s$ 
satisfies $\lambda_s/L \ll 1$ for all $s$, $L$ being the length scale over which the macroscopic properties of the plasma vary -- a formal procedure exists for deriving a closed system of fluid equations from a 
kinetic description of the plasma. 
This procedure is the Chapman-Enskog (CE) expansion, which gives distribution 
functions that are close to, but not exactly, Maxwellian. 
We call them Chapman-Enskog (CE) distribution functions. 
The non-Maxwellian components of the CE distribution functions of particle species $s$ are proportional to $\lambda_s/L$, and 
must be present in order to support gradients of $n_s$, $\boldsymbol{V}_s$ and 
$T_s$ on $\textit{O}(L)$ length scales, because (\ref{fluxes_fluid}\textit{b}-\textit{e}) are all zero for a Maxwellian plasma.  

We consider a collisional electron-ion plasma (in which, by definition, $\mu_e \equiv m_e/m_i \ll 1$) with the property that 
all constituent particle species are strongly magnetised by the macroscopically varying magnetic field $\boldsymbol{B}$: 
that is, the Larmor radius $\rho_s \equiv m_s v_{\mathrm{th}s} c/|Z_s| e |\boldsymbol{B}|$ satisfies $\rho_s \ll \lambda_{s}$ both for the ions and for the electrons (here $v_{\mathrm{th}s} \equiv \sqrt{2 T_s/m_s}$ 
is the thermal speed of species $s$). Equivalently, a strongly magnetised plasma is one in which the Larmor frequency $\Omega_s \equiv e |Z_s|/m_s c$ satisifies $\Omega_s \tau_s \gg 1$, where $\tau_s$ is the collision time of species $s$. In such a plasma, the macroscopic variation of the fluid moments is locally anisotropic with respect to $\boldsymbol{B}$;  
$L$ is the typical length scale of variation in the direction 
locally parallel to $\boldsymbol{B}$. It can then be shown that, to first order of the Chapman-Enskog expansion in $\lambda_{s}/L \ll 1$,  
and to zeroth order in $\rho_s/\lambda_{s} \ll 1$, the CE distribution 
functions of the electrons and ions
are
\begin{subeqnarray}
f_{e}(\tilde{v}_{e\|},\tilde{v}_{e\bot}) & = &\frac{n_{e}}{v_{\mathrm{th}e}^3 \upi^{3/2}} \exp \left(-\tilde{v}_{e}^2\right) \nonumber \\
&& \times \Bigg\{1+\left[\eta_e^{T} A_e^T(\tilde{v}_e) + \eta_e^{R} A_e^R(\tilde{v}_e) + \eta_e^{u} A_e^u(\tilde{v}_e) \right] \tilde{v}_{e\|} \nonumber \\
&& \qquad \qquad \qquad + \epsilon_e C_e(\tilde{v}_e) \left(\tilde{v}_{e\|}^2- \frac{\tilde{v}_{e\bot}^2}{2}\right)\Bigg\} , \qquad \quad  
\\
f_{i}(\tilde{v}_{i\|},\tilde{v}_{i\bot}) & = &\frac{n_{i}}{v_{\mathrm{th}i}^3 \upi^{3/2}} \exp \left(-\tilde{v}_{i}^2\right) \nonumber \\
&& \qquad \times \left\{1+\eta_i A_i(\tilde{v}_i) \tilde{v}_{i\|} + \epsilon_i C_i(\tilde{v}_i) \left(\tilde{v}_{i\|}^2- \frac{\tilde{v}_{i\bot}^2}{2} \right)\right\} .  
\label{ChapEnskogFunc}
\end{subeqnarray}
Let us define the various symbols employed in (\ref{ChapEnskogFunc}), 
before discussing the origin of these expressions and their significance for formulating fluid equations (see section \ref{fluxfluids}). 

The particle velocity $\boldsymbol{v}$ (with the corresponding speed $v = |\boldsymbol{v}|$) is split into components parallel and perpendicular to the macroscopic magnetic field $\boldsymbol{B} = B \hat{\boldsymbol{z}}$ as $\boldsymbol{v} = v_{\|} \hat{\boldsymbol{z}} + \boldsymbol{v}_{\bot}$, and the perpendicular plane is in turn characterised by two vectors $\hat{\boldsymbol{x}}$ and $\hat{\boldsymbol{y}}$ chosen so that $\left\{\hat{\boldsymbol{x}},\hat{\boldsymbol{y}},\hat{\boldsymbol{z}}\right\}$ is an orthonormal basis. The perpendicular velocity is related to these basis vectors by the gyrophase angle $\phi$:
\begin{equation}
\boldsymbol{v}_{\bot} = v_{\bot} \left(\cos{\phi} \, \hat{\boldsymbol{x}} - \sin{\phi} \, \hat{\boldsymbol{y}} \right) .
\end{equation}
The non-dimensionalised peculiar velocity $\tilde{\boldsymbol{v}}_s$ in the rest frame of the ion fluid is defined by $\tilde{\boldsymbol{v}}_s \equiv (\boldsymbol{v}-\boldsymbol{V}_i)/v_{\mathrm{th}s}$, $\tilde{v}_s \equiv 
|\tilde{\boldsymbol{v}}_s|$, $\tilde{v}_{s\|} \equiv \hat{\boldsymbol{z}} \bcdot 
\tilde{\boldsymbol{v}}_s$, and $\tilde{v}_{s\bot} \equiv |\tilde{\boldsymbol{v}}_s-\tilde{v}_{s\|} 
\hat{\boldsymbol{z}}|$. The number densities satisfy the quasi-neutrality condition
\begin{equation}
 Z n_i = n_e \, ,
\end{equation}
where we have utilised $Z_{e} = -1$, and defined $Z \equiv Z_i$. We emphasise that $n_{s}$, 
$\left\{\hat{\boldsymbol{x}},\hat{\boldsymbol{y}},\hat{\boldsymbol{z}}\right\}$
and $v_{\mathrm{th}s}$ all vary over length scales $L$ in the plasma, but 
not on shorter scales (at least not in the direction locally parallel to $\boldsymbol{B}$). The functions $A_e^T(\tilde{v}_e)$, $A_e^R(\tilde{v}_e)$, $A_e^u(\tilde{v}_e)$, $C_e(\tilde{v}_e)$, $A_i(\tilde{v}_i)$ and 
$C_i(\tilde{v}_i)$ are isotropic functions. Their magnitude is $\textit{O}(1)$ when $\tilde{v}_e \sim 
1$ or $\tilde{v}_i \sim 1$, for electrons and ions respectively. Finally, the parameters $\eta_e^T$, $\eta_e^R$, $\eta_e^u$, $\eta_i$, $\epsilon_e$ and $\epsilon_i$ 
are defined as follows: 
\begin{subeqnarray}
  \eta_{e}^T & = & \lambda_{e} \nabla_{\|} \log T_e = \mathrm{sgn}(\nabla_{\|} \log T_e) \frac{\lambda_{e}}{L_T}  ,\\[3pt]
  \eta_{e}^R & = &  \lambda_{e} \frac{R_{e\|}}{p_e} ,\\[3pt]
   \eta_{e}^u & = &  \lambda_{e} \frac{m_e u_{ei\|}}{T_e \tau_{e}} ,\\[3pt]
     \eta_{i} & = & \lambda_{i} \nabla_{\|} \log T_i = \mathrm{sgn}(\nabla_{\|} \log T_i) \frac{\lambda_{i}}{L_{T_i}}, \\[3pt]
      \epsilon_{e} & = &
\frac{\lambda_{e}}{v_{\mathrm{th}e}} {\left(\hat{\boldsymbol{z}} \hat{\boldsymbol{z}} - \frac{1}{3}\mathsfbi{I} \right) \! \boldsymbol{:} \! \mathsfbi{W}_e} = \mathrm{sgn}\left[\left(\hat{\boldsymbol{z}} \hat{\boldsymbol{z}} - \frac{1}{3}\mathsfbi{I} \right) \! \boldsymbol{:} \! \mathsfbi{W}_e\right] \frac{V_e}{v_{\mathrm{th}e}} \frac{\lambda_e}{L_{V_e}} , 
  \\[3pt]
   \epsilon_{i} & = &
 \frac{\lambda_{i}}{v_{\mathrm{th}i}} {\left(\hat{\boldsymbol{z}} \hat{\boldsymbol{z}} - \frac{1}{3}\mathsfbi{I} \right) \! \boldsymbol{:} \! \mathsfbi{W}_i}{v_{\mathrm{th}i}} = \mathrm{sgn}\left[\left(\hat{\boldsymbol{z}} \hat{\boldsymbol{z}} - \frac{1}{3}\mathsfbi{I} \right) \! \boldsymbol{:} \! \mathsfbi{W}_i\right]\frac{V_i}{v_{\mathrm{th}i}} \frac{\lambda_i}{L_{V}} ,
  \label{smallparam}
\end{subeqnarray}
where $\lambda_{e}$ is the electron mean free path, $\lambda_{i}$ the ion mean-free-path, $\tau_e$ the electron collision time,  $R_{e\|} \equiv \hat{\boldsymbol{z}} \bcdot \boldsymbol{R}_{e}$ the parallel electron friction force, $\boldsymbol{u}_{ei} \equiv \boldsymbol{V}_e - \boldsymbol{V}_i$
the relative electron-ion velocity, $u_{ei\|} \equiv \hat{\boldsymbol{z}} \bcdot \boldsymbol{u}_{ei}$, 
\begin{equation}
   \mathsfbi{W}_s =
   \bnabla \boldsymbol{V}_s + \left(\bnabla \boldsymbol{V}_s\right)^{T} -\frac{2}{3}  \left(\bnabla \bcdot \boldsymbol{V}_s\right) \mathsfbi{I} 
   \label{ChapEnsTerms}
\end{equation}
the traceless rate-of-strain tensor of species $s$, $V_e$ ($V_i$) the bulk electron-(ion-)fluid speed, and 
\begin{subeqnarray}
L_T & \equiv & \left|\nabla_{\|} \log T_e \right|^{-1} \, , \\
L_{T_i} & \equiv & \left|\nabla_{\|} \log T_i \right|^{-1} \, , \\
L_{V_e} & \equiv & \frac{1}{V_e}\left|\left(\hat{\boldsymbol{z}} \hat{\boldsymbol{z}} - \frac{1}{3}\mathsfbi{I} \right) \! \boldsymbol{:} \! \mathsfbi{W}_e \right|^{-1} \, , \\
L_V & \equiv & \frac{1}{V_i} \left|\left(\hat{\boldsymbol{z}} \hat{\boldsymbol{z}} - \frac{1}{3}\mathsfbi{I} \right) \! \boldsymbol{:} \! \mathsfbi{W}_i \right|^{-1} \, , \label{parallel_vel_scale}
\end{subeqnarray}
are, respectively, the electron- and ion-temperature and the electron- and ion-flow length scales parallel to the 
background magnetic field.
The mean free paths are formally defined 
for a two-species plasma by 
\begin{subeqnarray}
\lambda_{e} & \equiv & {v_{\mathrm{th}e}}{\tau_{e}} , \\
\lambda_{i} & \equiv & {v_{\mathrm{th}i}}{\tau_{i}} , \label{meanfreepath_defs}
\end{subeqnarray}
and the collision times $\tau_e$ and $\tau_i$ are given in terms of macroscopic plasma parameters by
\begin{subeqnarray}
\tau_{e} & \equiv & \frac{3 m_e^{1/2} T_e^{3/2}}{4 \sqrt{2 \upi} Z_i^2 e^4 n_i \log{\Lambda_{\mathrm{CL}}}} , \\
\tau_{i} & \equiv &  \frac{3 m_i^{1/2} T_i^{3/2}}{4 \sqrt{2 \upi} Z_i^4 e^4 n_i \log{\Lambda_{\mathrm{CL}}}} 
, \label{colltimes}
\end{subeqnarray}
where $\log{\Lambda_{\mathrm{CL}}}$ is the Coulomb logarithm~\citep{B65}\footnote{Braginskii defined his ion collision time as equal to (\ref{colltimes}\textit{b}) multiplied by a factor of $\sqrt{2}$; for the sake of species equality, we remove this factor.}. In a 
collisional plasma, $\eta_e^T$, $\eta_e^R$, $\eta_e^u$, $\eta_i$, $\epsilon_e$ and $\epsilon_i$ 
are assumed small. We note that all these parameters can be either positive or 
negative, depending on the orientation of temperature and velocity 
gradients.

It is clear from their definitions that each of the non-Maxwellian terms associated with 
the parameters $\eta_e^T$, $\eta_e^R$, $\eta_e^u$, $\eta_i$, $\epsilon_e$ and $\epsilon_i$
is linked to a different macroscopic physical quantity. Thus, $\eta_e^T$ and $\eta_i$ 
are proportional to the electron- and ion-temperature gradients, respectively; we will therefore 
refer to the associated non-Maxwellian terms as the \textit{CE electron-temperature-gradient term} 
and the \textit{CE ion-temperature-gradient term}. We refer to the 
non-Maxwellian term proportional to $\eta_e^R$
as the \textit{CE electron-friction term}, to the non-Maxwellian term proportional to $\eta_e^u$
as the \textit{CE electron-ion-drift term}, and the non-Maxwellian terms 
proportional to $\epsilon_e$ and $\epsilon_i$ as the \textit{CE electron-shear term} 
and the \textit{CE ion-shear term}. We note that the friction and electron-ion-drift terms appear in the electron CE distribution function
but not the ion CE distribution function because of our choice to define all velocities in the ion-fluid rest frame.

The derivation of the CE distribution functions (\ref{ChapEnskogFunc}) for a two-species strongly magnetised plasma undergoing sonic motions (that is, $V_i \sim v_{\mathrm{th}i}$) from the kinetic equation~(\ref{MaxVlasLan}) was first 
completed by~\citet{B65} for arbitrary values of $\rho_s/\lambda_{s}$. We do 
not reproduce the full derivation in the main text, but, for the reader's convenience, we provide a derivation of $(\ref{ChapEnskogFunc})$ in appendix \ref{ChapEnskogDev_Append_comp}. 
The gist of the full derivation is to assume that the distribution function is close to a Maxwellian, 
with parameters that only evolve on a slow time scale $t' \sim t L/\lambda_e \sim t 
L/\lambda_i \gg t$. The kinetic equation (\ref{MaxVlasLan}) is then expanded 
and solved order by order in $ \lambda_e/L \sim \lambda_i/L \ll 1$, allowing for 
the calculation of the (small) non-Maxwellian components of the distribution 
function. The small parameters $\eta_e^T$, $\eta_e^R$, $\eta_e^u$, $\eta_i$, $\epsilon_e$ and 
$\epsilon_i$, as well as the isotropic functions $A_e^T(\tilde{v}_e)$, $A_e^R(\tilde{v}_e)$, $A_e^u(\tilde{v}_e)$, $C_e(\tilde{v}_e)$, $A_i(\tilde{v}_i)$ and 
$C_i(\tilde{v}_i)$ emerge during this calculation. The precise forms of these functions depend only on the collision operator assumed in the original 
 Maxwell-Vlasov-Landau system; in appendix \ref{ChapEnskogIsoFunc}, we provide a simple illustration 
 of this, by calculating these isotropic functions explicitly for 
Krook~\citep{BGK54} and Lorentz collision operators (Appendices \ref{ChapEnskogIsoFunc_Krook} and \ref{ChapEnskogIsoFunc_Lorentz}, respectively). 
For the full Landau collision operator, 
the equivalent calculation is more complicated, but can be performed (for example)
by expanding the isotropic functions in Sonine polynomials~\citep[see][]{HS05}. 

\subsubsection{Closure of fluid equations (\ref{fluideqns})} 
\label{fluxfluids}

Once the CE distribution function has been calculated, the desired fluid closure can be obtained by evaluating  
the heat fluxes, the friction forces, and the momentum fluxes~(\ref{fluxes_fluid}) associated with the non-Maxwellian 
components of the CE distribution functions. These 
calculations were carried out in full for arbitrary values of $\rho_s/\lambda_{s}$ by~\citet{B65}. 

We do not reproduce the full fluid
closure relations here; instead, we illustrate how the non-Maxwellian terms 
in the CE distribution functions (\ref{ChapEnskogFunc}) give rise to the friction force and heat fluxes 
parallel to the macroscopic magnetic field, as well as to the viscosity tensor. In a strongly 
magnetised two-species plasma (where $\rho_s \ll \lambda_s$), parallel friction forces and heat fluxes 
are typically much larger than their perpendicular or diamagnetic counterparts. 

\begin{itemize}
  
\item \underline{Heat fluxes}. Recalling (\ref{fluxes_fluid}\textit{c}), the parallel heat flux $q_{s\|} \equiv \hat{\boldsymbol{z}} \bcdot \boldsymbol{q}_{s}$ associated with 
 species $s$ is given by
\begin{equation}
  q_{s\|} = \frac{1}{2} \int \mathrm{d}^3\boldsymbol{v}_s' \, m_s \left|\boldsymbol{v}_s'\right|^2 v_{s\|}' \,  f_{s}
  , \label{par_heat_flux}
\end{equation}
where $\boldsymbol{v}_s' \equiv \boldsymbol{v}-\boldsymbol{V}_s$. Noting that the electron distribution function (\ref{ChapEnskogFunc}\textit{a}) is specified in the rest frame of the ions, not electrons, 
it is necessary first to calculate the electron distribution function in the electron rest frame 
before calculating the parallel electron heat flux. An expression for this quantity is given by (\ref{CE_dist_electronframe}) in appendix 
\ref{ChapEnskogDev_Append_comp} as part of our derivation of (\ref{ChapEnskogFunc}\textit{a}):
\begin{eqnarray}
 f_{e}(v_{e\|}',v_{e\bot}') & = & \frac{n_{e}}{v_{\mathrm{th}e}^3 \upi^{3/2}} \exp \left(-\frac{|\boldsymbol{v}_{e}'|^2}{v_{\mathrm{th}e}^2}\right) \nonumber \\
 && \times \Bigg\{1 + \bigg[\eta_e^T A_e^{T}\!\left(\frac{|\boldsymbol{v}_e'|}{v_{\mathrm{th}e}}\right) +\eta_e^R A_e^{R}\!\left(\frac{|\boldsymbol{v}_e'|}{v_{\mathrm{th}e}}\right) + \eta_e^u \left(A_e^{u}\!\left(\frac{|\boldsymbol{v}_e'|}{v_{\mathrm{th}e}}\right) -1 \right) \bigg] \frac{v_{e\|}'}{v_{\mathrm{th}e}} \nonumber \\
&& \qquad \qquad + \epsilon_e C_e\!\left(\frac{|\boldsymbol{v}_e'|}{v_{\mathrm{th}e}}\right) \left(\frac{v_{e\|}'^2}{v_{\mathrm{th}e}^2} - \frac{v_{e\bot}'^2}{2 v_{\mathrm{th}e}^2} \right) \Bigg\} \, ,\label{CE_dist_electronframe_maintext}
\end{eqnarray}
Now substituting (\ref{CE_dist_electronframe_maintext}) 
into (\ref{par_heat_flux}) (with $s = e$), we find that the 
parallel electron heat flux is
\begin{equation}
 q_{e\|} = - n_{e} T_e v_{\mathrm{th}e} \left[\mathcal{A}_e^T \eta_e^T + \mathcal{A}_e^R \eta_e^R + \left(\mathcal{A}_e^u - \frac{1}{2}\right) \eta_e^u\right], \label{heatflux_expression}
\end{equation}
where  
\begin{equation}
\mathcal{A}_e^{T,R,u} = -\frac{4}{3\sqrt{\upi}} \int_0^{\infty} \mathrm{d}\tilde{v}_e \, \tilde{v}_e^6 A_e^{T,R,u}(\tilde{v}_e)  \exp \left(-\tilde{v}_e^2\right) 
 .
\end{equation}
The minus signs in the definitions of $\mathcal{A}_e^{T,R,u}$ 
have been introduced so that $\mathcal{A}_e^{T,R,u} \geq 0$ for a typical 
collision operator (determining that these constants are indeed positive for any given 
collision operator is non-trivial, but it is a simple exercise to show this for a 
Krook collision operator, using the expressions for $A_e^T(\tilde{v}_e)$, $A_e^R(\tilde{v}_e)$, and $A_e^u(\tilde{v}_e)$ given in appendix \ref{ChapEnskogIsoFunc_Krook}).
Expression (\ref{heatflux_expression}) for the electron heat flux can be rewritten as
\begin{equation}
 q_{e\|} =  - \kappa_{e}^{\|} \nabla_{\|} T_e - \left[ \mathcal{A}_e^u - \frac{1}{2} - \frac{ \mathcal{A}_e^R}{\tilde{\mathcal{A}}_e^R}\left(\tilde{\mathcal{A}}_e^u-\frac{1}{2}\right)\right] 
n_e T_e u_{ei\|} \, , \label{brag_heatflux_elec}
\end{equation}
where the parallel electron heat conductivity is defined by
\begin{equation}
\kappa_{e}^{\|} = 2 \left(\mathcal{A}_e^T - \frac{\mathcal{A}_e^R}{\tilde{\mathcal{A}}_e^R} \tilde{\mathcal{A}}_e^T\right) \frac{n_e T_e \tau_e}{m_e} 
\, , \label{parallel_cond_electrons}
\end{equation}
and 
\begin{equation}
\tilde{\mathcal{A}}_e^{T,R,u} = -\frac{4}{3\sqrt{\upi}} \int_0^{\infty} \mathrm{d}\tilde{v}_e \, \tilde{v}_e^4 A_e^{T,R,u}(\tilde{v}_e)  \exp \left(-\tilde{v}_e^2\right) \, 
.
\end{equation}
Numerical evaluation of the coefficients $\mathcal{A}_e^{T,R,u}$ and $\tilde{\mathcal{A}}_e^{T,R,u}$ for the Landau collision operator 
gives~\citep{B65}
\begin{equation}
 q_{e\|} \simeq  - 3.16 \frac{n_e T_e \tau_e}{m_e}  \nabla_{\|} T_e + 0.71 n_e T_e u_{ei\|} \, 
 . 
\end{equation}

The ion heat flux can be calculated directly from (\ref{par_heat_flux}) ($s = i$) using (\ref{ChapEnskogFunc}\textit{b}):
\begin{equation}
 q_{i\|} =  - n_{i} T_i v_{\mathrm{th}i} \mathcal{A}_i \eta_i , \label{ion_par_heat_flux}
\end{equation}
where 
\begin{equation}
\mathcal{A}_i = -\frac{4}{3\sqrt{\upi}} \int_0^{\infty} \mathrm{d}\tilde{v}_i \, \tilde{v}_i^6 A_i(\tilde{v}_i)  \exp \left(-\tilde{v}_i^2\right) \, .   
\end{equation}
This becomes
\begin{equation}
 q_{i\|} =   - \kappa_{i}^{\|} \nabla_{\|} T_i , \label{brag_heatflux_ion}
\end{equation}
where the parallel ion heat conductivity is
\begin{equation}
\kappa_{i}^{\|} = -2 \mathcal{A}_i \frac{n_i T_i \tau_i}{m_i} \simeq - 3.9 \frac{n_i T_i \tau_i}{m_i} 
\, . \label{parallel_cond_ions}
\end{equation}
The last equality is for the Landau collision operator~\citep{B65}. 
Note that the absence of a term proportional to the
electron-ion-drift in the ion heat flux (\ref{ion_par_heat_flux}) is physically due to the smallness of 
the ion-electron collision operator~\citep{HS05}.
  
  \item \underline{Friction force}. 
  We evaluate the friction force by considering the electron-ion-drift associated with electron CE distribution function. 
  Namely, noting that 
\begin{equation}
 u_{ei\|} = \frac{v_{\mathrm{th}e}^4}{n_e} \int \mathrm{d}^3 \tilde{\boldsymbol{v}}_e \, \tilde{v}_{e\|} f_{e} ,
\end{equation}
it follows from (\ref{ChapEnskogFunc}\textit{a}) that 
\begin{equation}
 u_{ei\|} = v_{\mathrm{th}e} \left(\tilde{\mathcal{A}}_e^T \eta_e^T + \tilde{\mathcal{A}}_e^R \eta_e^R +  \tilde{\mathcal{A}}_e^u \eta_e^u \right) . 
 \label{electronparticleflux}
\end{equation}
This expression can in turn be used to relate the 
parallel electron-friction force $R_{e\|}$, defined in (\ref{fluxes_fluid}\textit{d}), to electron flows and temperature 
gradients:
\begin{equation}
R_{e\|} = -\left(\frac{2 \tilde{\mathcal{A}}_e^u + 1}{2 \tilde{\mathcal{A}}_e^R}\right) \frac{n_e m_e u_{ei\|}}{\tau_e} 
- \frac{\tilde{\mathcal{A}}_e^T}{\tilde{\mathcal{A}}_e^R} n_e \nabla_{\|} T_e \, . 
\label{Friction_force}
\end{equation}
Evaluating the coefficients $\tilde{\mathcal{A}}_e^T$, $\tilde{\mathcal{A}}_e^R$ and $\tilde{\mathcal{A}}_e^u$ for the full Landau collision operator, one 
finds~\citep{B65}
\begin{equation}
R_{e\|} \simeq -0.51 \frac{n_e m_e u_{ei\|}}{\tau_e} 
- 0.71 n_e \nabla_{\|} T_e \, .  
\end{equation}

\item \underline{Viscosity tensor}. 
For gyrotropic distributions such as the CE distribution functions~(\ref{ChapEnskogFunc}), the viscosity tensor $\boldsymbol{\pi}_s$ 
of species $s$ defined by (\ref{fluxes_fluid}\textit{b})
-- which is the momentum flux excluding the convective terms and isotropic pressure  -- is given by
\begin{equation}
\boldsymbol{\pi}_s = \left(p_{s\|} - p_{s\perp}\right) \left(\hat{\boldsymbol{z}} \hat{\boldsymbol{z}}-\frac{1}{3}\mathsfbi{I}\right)  , 
\label{viscosity_tens_gyro}
\end{equation}
where the parallel pressure $p_{s\|}$ and the perpendicular pressure $p_{s\bot}$ are 
defined by
\begin{subeqnarray}
p_{s\|} & \equiv & \int \mathrm{d}^3 \boldsymbol{v}_s' \, m_s |v_{s\|}'|^2 f_{s} = n_{s} T_{s} \left(1-\frac{2}{3}\epsilon_s \mathcal{C}_s\right) \, , \\
p_{s\bot} & \equiv & \frac{1}{2}\int \mathrm{d}^3 \boldsymbol{v}_{s}'  \, m_s |\boldsymbol{v}_{s\bot}'|^2 f_{s} = n_{s} T_{s} \left(1+\frac{1}{3}\epsilon_s \mathcal{C}_s\right)\, 
, \label{press_perpparr}
\end{subeqnarray}
with the last expressions having being obtained on substitution of the CE distribution function~(\ref{ChapEnskogFunc}), 
and 
\begin{equation}
\mathcal{C}_s = -\frac{8}{5\sqrt{\upi}} \int_0^{\infty} \mathrm{d}\tilde{v}_s \, \tilde{v}_s^6 C_s(\tilde{v}_s)  \exp \left(-\tilde{v}_s^2\right) \, . 
\label{pressureanisop_const}
\end{equation}
The sign of the constant $\mathcal{C}_s$ is again chosen so that $\mathcal{C}_s > 0$ for typical 
collision operators; for the Landau collision operator, $\mathcal{C}_e \simeq 1.1$ and $\mathcal{C}_i \simeq 1.44$~\citep{B65}.  
We note for reference that the parameter $\epsilon_s$ [see (\ref{smallparam}e-f)]
has a simple relationship to the pressure anisotropy of species $s$: utilising (\ref{press_perpparr}), 
one finds
\begin{equation}
\Delta_s \equiv \frac{p_{s\bot}-p_{s\|}}{p_{s}} = \mathcal{C}_s \epsilon_s \, . \label{pressureanisop}
\end{equation}

Using (\ref{press_perpparr}), the viscosity tensor (\ref{viscosity_tens_gyro}) can 
be written
\begin{equation}
\boldsymbol{\pi}_s = - \frac{\mu_{\mathrm{v}s}}{2} \left(\hat{\boldsymbol{z}} \hat{\boldsymbol{z}} - \frac{1}{3}\mathsfbi{I} \right)  {\left(\hat{\boldsymbol{z}} \hat{\boldsymbol{z}} - \frac{1}{3}\mathsfbi{I} \right) \! \boldsymbol{:} \! \mathsfbi{W}_s}
 , \label{brag_viscosity}
\end{equation}
where the dynamic viscosity of species $s$ is 
\begin{equation}
 \mu_{\mathrm{v}s} \equiv 2 \mathcal{C}_s n_s T_s \tau_s \, . \label{dynam_viscosity}
\end{equation}

  \item \underline{Thermal energy transfer between species}. It can be shown that for 
  the CE distribution functions $(\ref{ChapEnskogFunc})$, the rate of thermal 
   energy transfer from electrons to ions $\mathcal{Q}_e$ is simply
   \begin{equation}
    \mathcal{Q}_e = - \boldsymbol{R}_e \bcdot \boldsymbol{u}_{ei}  ,
   \end{equation}
   while the rate of thermal energy transfer from ions to electrons vanishes: $\mathcal{Q}_i \approx 
   0$. This is because the ion-electron collision rate is assumed small (by a factor of the mass ratio) 
   compared to the ion-ion collision rate when deriving 
   (\ref{ChapEnskogFunc}\textit{b}), and is thus neglected. \citet{B65} shows that, in 
   fact,
   there is a non-zero (but small) rate of transfer:
   \begin{equation}
   \mathcal{Q}_i = -\mathcal{Q}_e - \boldsymbol{R}_e \bcdot \boldsymbol{u}_{ei}  
   = \frac{3 n_e m_e}{m_i \tau_{e}} \left(T_e-T_i\right) \, .
   \end{equation}
   The time scale on which the ion and electron temperatures equilibrate is the ion-electron temperature equilibration time 
   \begin{equation}
   \tau_{ie}^{\rm eq} \equiv \frac{1}{2}  \mu_e^{-1/2} \tau_{i} 
   . \label{ion_elec_equil_time}
   \end{equation} 
\end{itemize}

In summary, the non-Maxwellian components of the CE distribution function are 
essential for a collisional plasma to be able to support fluxes of heat 
and momentum. More specifically,  (\ref{brag_heatflux_elec}) demonstrates that the
electron heat fluxes in a CE plasma are proportional to both temperature gradients and 
electron-ion drifts, and are carried by the electron-temperature-gradient, 
friction and electron-ion-drift terms of the CE distribution function. In contrast, the ion 
heat fluxes (\ref{brag_heatflux_ion}) are proportional only to ion temperature gradients (and carried by 
the CE ion-temperature-gradient term). Momentum fluxes~(\ref{brag_viscosity}) for electrons and ions 
are carried by the CE electron- and ion-shear terms, respectively, and are 
proportional to components of the rate-of-strain tensor. 

\subsubsection{Relative size of non-Maxwellian terms in the CE distribution function} 
\label{rel_size_CEterms}

In the case of magnetised, two-species plasma satisfying $T_i \sim 
T_e$, (\ref{smallparam}) can be used to estimate the size of 
the small parameters $\eta_e^T$, $\eta_e^R$, $\eta_e^u$, $\eta_i$, $\epsilon_e$ and 
$\epsilon_i$. Although these parameters are \textit{a priori} proportional to $\lambda_s/L$ for both ions and electrons, their 
precise magnitudes are, in fact, subtly different. Namely, the terms associated 
with $\eta_e^T$, $\eta_e^R$, $\eta_e^u$ and $\eta_i$ are gradients of the electron and ion temperatures and electron-ion relative parallel drift velocities, whereas terms associated with $\epsilon_e$ and $\epsilon_i$ 
involve gradients of the bulk flows [cf. (\ref{smallparam})] -- and these gradients do not necessarily occur on the same 
length scale. Recalling that the (electron) temperature
and the (ion) flow length scales parallel to the macroscopic magnetic field are defined by [cf. (\ref{parallel_vel_scale})]
\begin{subeqnarray}
L_T & = & \left|\nabla_{\|} \log T_e \right|^{-1} \, , \\
L_V & = & \frac{1}{V_i}\left|\left(\hat{\boldsymbol{z}} \hat{\boldsymbol{z}} - \frac{1}{3}\mathsfbi{I} \right) \! \boldsymbol{:} \! \mathsfbi{W}_i \right|^{-1} \, , \label{parallel_vel_scale_2nd}
\end{subeqnarray}
where $\mathsfbi{W}_i$ is the ion rate-of-strain tensor (\ref{ChapEnsTerms}), and assuming that $L_{T_i} = \left(\nabla_{\|} \log T_i \right)^{-1} \sim L_{T}$ (an assumption we will check \textit{a posteriori}), 
it follows from (\ref{smallparam}) that 
\begin{subeqnarray}
\eta_e^T & \sim &  \frac{\lambda_{e}}{L_T}, \\
\eta_e^R & \sim & \lambda_{e} \frac{R_{e\|}}{p_e} \sim \frac{\lambda_{e}}{L_T} \sim \eta_e^T, \\
\eta_e^u & \sim & \frac{u_{ei\|}}{v_{\mathrm{th}e}} \sim \frac{\lambda_{e}}{L_T} \sim \eta_e^T, \\
\eta_i & \sim & \frac{\lambda_{i}}{L_T} \sim \frac{1}{Z^2} \eta_e^T , \\
\epsilon_e & \sim &  \frac{V_i}{v_{\mathrm{th}e}} \frac{\lambda_{e}}{L_V} \sim \mathrm{Ma}\, \mu_e^{1/2} \frac{L_T}{L_V} \eta_e^T,\\
\epsilon_i & \sim & \frac{V_i}{v_{\mathrm{th}i}} \frac{\lambda_{i}}{L_V} \sim \mathrm{Ma} \, \frac{L_T}{Z^2 L_V} \eta_e^T, 
\label{smallparamscalings}
\end{subeqnarray}
where $\mathrm{Ma} \equiv V_i/v_{\mathrm{th}i}$ is the Mach number. Note that, to arrive at (\ref{smallparamscalings}\textit{b}), we assumed that $R_{e\|} \sim p_e/L_{T}$ and $u_{ei\|} \sim v_{\mathrm{th}e} \lambda_e/L_T$, justified by (\ref{Friction_force}) and (\ref{electronparticleflux}), respectively. The relative magnitudes of 
$\eta_e^T$, $\eta_e^R$, $\eta_e^u$, $\eta_i$, $\epsilon_e$ and 
$\epsilon_i$ therefore depend on the Mach number of the plasma, as well as on
the length scales $L_T$ and $L_V$. 

In the work of~\citet{B65}, who \textit{a priori} presumes all ``fluid'' quantities in the plasma to vary on just a single scale $L \sim L_T \sim L_V$, 
with sonic ordering $\mathrm{Ma} \lesssim 1$, determining the relative size of these parameters for a hydrogen plasma ($Z = 1$) is simple:
\begin{equation}
\epsilon_e \sim  \mu_e^{1/2} \epsilon_i \ll  \epsilon_i \sim  \eta_i \sim \eta_e^T \sim \eta_e^R \sim \eta_e^u 
 \, . \label{brag_onescale}
\end{equation} 
However, in most interesting applications, this single-scale ordering is incorrect. In a plasma with $\lambda_s/L \ll 1$ under Braginskii's ordering, 
motions on many scales naturally arise. The fluid Reynolds 
number in such a plasma is given by 
\begin{equation}
  \mathrm{Re} \equiv \frac{V_0 L_0}{\nu} \, ,
\end{equation}
where $V_0$ is the typical fluid velocity at the scale $L_0$ of driving motions and $\nu \equiv \mu_{\mathrm{v}i}/m_i n_i \sim v_{
\mathrm{th}i} \lambda_i$ 
is the kinematic viscosity [see (\ref{dynam_viscosity})]. Typically, this number is large:
\begin{equation}
  \mathrm{Re} \sim \frac{V_0}{v_{\mathrm{th}i}} \frac{L_0}{\lambda_{i}} \gtrsim \frac{1}{\epsilon_i} \gg 1 \, , 
  \label{Reynolds_num}
\end{equation}
where we have assumed $\mathrm{Ma}_0 \equiv V_0/v_{\mathrm{th}i} \lesssim 1$, in line with Braginskii's sonic ordering. Therefore, such a plasma will naturally become turbulent and exhibit motions across a range of scales. As a consequence, velocity and temperature fluctuations
on the smallest (fluid) scales must be considered, since the associated shears and temperature gradients are the largest. 
To estimate $\eta_e^T$, $\eta_e^R$, $\eta_e^u$, $\eta_i$, $\epsilon_e$ and 
$\epsilon_i$ accurately, we must determine the magnitude of these gradients.

First, let $\ell_{\nu}$ be the smallest scale on which the velocity varies due to turbulent motions (the Kolmogorov scale), with velocity fluctuations on scales $\ell \ll \ell_{\nu}$ being suppressed by viscous diffusion. 
Then it follows that $\mathrm{Re}_{\ell_{\nu}} \sim 1$, where $\mathrm{Re}_\ell \equiv V\!\left(\ell\right) \ell/\nu$ 
is the scale-dependent Reynolds number and $V\!\left(\ell\right)$ is the typical fluid velocity on scale 
$\ell$.
For Kolmogorov 
turbulence, 
\begin{equation}
\frac{V\!\left(\ell\right)}{V_0} \sim \left(\frac{\ell}{L_0}\right)^{1/3} \sim \left(\frac{\mathrm{Re}_\ell}{\mathrm{Re}} \right)^{1/4}, 
\end{equation}
and $\ell/L_0 \sim \left({\mathrm{Re}_\ell}/{\mathrm{Re}} \right)^{3/4}$, which gives $V\!\left(\ell\right)\!/\ell \sim (V_0/L_0) \left({\mathrm{Re}_\ell}/{\mathrm{Re}} 
\right)^{-1/2}$, and thus, from~(\ref{Reynolds_num}),
\begin{equation}
 \frac{V\!\left(\ell_\nu\right)}{\ell_\nu} \sim  \frac{V_0}{L_0} \left(\frac{\mathrm{Re}_{\ell_\nu}}{\mathrm{Re}} 
\right)^{-1/2} \sim \mathrm{Ma}_0^{1/2} \left(\frac{\lambda_{i}}{L_0}\right)^{-1/2} 
\frac{V_0}{L_0} \, .
\end{equation}
We therefore conclude that 
\begin{equation}
 L_V \sim \ell_{\nu} \frac{V_0}{V\!\left(\ell_\nu\right)} \sim L_0 \mathrm{Ma}_0^{-1/2} \left(\frac{\lambda_{i}}{L_0}\right)^{1/2} 
 . 
\end{equation}

Next, the smallest scale on which the electron temperature varies, $\ell_{\chi}$, 
is the scale below which temperature fluctuations are suppressed by 
thermal diffusion; it satisfies $\mathrm{Pe}_{\ell_{\chi}} \sim 1$, where $\mathrm{Pe}_\ell \equiv V\!\left(\ell\right) L/\chi$ 
is the scale-dependent P\'eclet number and $\chi \equiv 2 \kappa_{e}^{\|}/3 n_e \sim v_{\mathrm{th}e} \lambda_{e}$ is the (parallel) thermal diffusivity [see (\ref{parallel_cond_electrons})]. 
Because temperature is passively advected by the flow, the temperature fluctuation $T(\ell)$ at any scale $\ell > \ell_\chi$ 
obeys the same scaling as the bulk velocity: 
\begin{equation}
\frac{T\!\left(\ell\right)}{T(L_0)} \sim \frac{V\!\left(\ell\right)}{V_0} \sim \left(\frac{\mathrm{Pe}_\ell}{\mathrm{Pe}} \right)^{1/4}  \, . 
\end{equation}
In addition, the magnitude of temperature fluctuations at the driving scale is 
related to the mean temperature by the Mach number of the driving-scale motions, $T(L_0) \sim T_0 \mathrm{Ma}_0$, which then gives
\begin{equation}
\frac{T\!\left(\ell\right)}{T_0} \sim \mathrm{Ma}_0 \left(\frac{\mathrm{Pe}_\ell}{\mathrm{Pe}} \right)^{1/4}  \, ,   
\end{equation}
where $\mathrm{Pe} \equiv \mathrm{Pe}_{L_0}$. It follows from an analogous argument to that just given for the velocity fluctuations that
\begin{equation}
 \frac{T\!\left(\ell_\chi\right)}{\ell_\chi} \sim  \frac{T_0}{L_0} \mathrm{Ma}_0 \mathrm{Pe}^{1/2} 
 .
\end{equation} 
Under Braginskii's ordering, the Prandtl number of CE 
plasma is
\begin{equation}
   \mathrm{Pr} \equiv \frac{\nu}{\chi} = \frac{\mathrm{Pe}}{\mathrm{Re}} \sim \frac{v_{\mathrm{th}i} \lambda_{i}}{v_{\mathrm{th}e} \lambda_{e} } 
   \sim \mu_e^{1/2} \ll 1 \, ,
\end{equation}
and, therefore,
\begin{equation}
 L_T \sim \ell_{\chi} \frac{T_0}{T\!\left(\ell_\chi\right)} \sim L_0 \mu_e^{-1/4} \mathrm{Ma}_0^{-3/2} \left(\frac{\lambda_{i}}{L_0}\right)^{1/2} 
 . 
\end{equation}
Thus, $L_V \sim \mathrm{Ma}_0 \mu_e^{1/4} L_T \ll L_T$ under the assumed ordering.

Finally, we consider whether our \textit{a priori} assumption that $L_{T_i} \sim L_T$ is, in fact, justified. A sufficient condition for ion-temperature gradients 
to be the same as electron-temperature gradients is for the evolution time $\tau_{L}$ of all macroscopic motions to be much longer than the ion-electron temperature equilibration time 
$\tau_{ie}^{\rm eq}$ defined by (\ref{ion_elec_equil_time}). Since $\tau_{L} \gtrsim {\ell_\nu}/{V\!\left(\ell_\nu\right)}$, it follows that
\begin{equation}
\frac{\tau_{L}}{\tau_{ie}^{\rm eq}} \sim \left(\frac{m_i}{m_e}\right)^{1/2}\mathrm{Ma}_0^{3/2} \left(\frac{\lambda_i}{L_0}\right)^{1/2} \sim \epsilon_i \left(\frac{m_i}{m_e}\right)^{1/2} \, . \label{macroieequilrat}
\end{equation}
Thus, if $\epsilon_i \gg \mu_e^{1/2}$, we conclude that collisional equilibration of ion and electron temperatures might be too inefficient to regulate small-scale ion-temperature fluctuations, in which case it would follow that $L_{T_i} < L_{T}$. However, it has been previously demonstrated 
via numerical solution of the Vlasov-Fokker-Planck equation that the CE expansion procedure breaks down due to 
nonlocal transport effects if $\lambda_e/L$ is only moderately small~\citep{BEN81}; thus, the only regime in which there is not ion-electron equilibration over all scales is one where
the CE expansion is not valid anyway. In short, we conclude that assuming $L_{T_i} \sim L_T$ is reasonable.

Bringing these considerations together with (\ref{smallparamscalings}), we find that
\begin{subeqnarray}
    \eta_e^T & \sim & \mu_e^{1/4} \mathrm{Ma}_0 \frac{\lambda_i}{L_V} \sim
  \mathrm{Ma}_0^{3/2} \mu_e^{1/4} \left(\frac{\lambda_i}{L_0}\right)^{1/2} \sim \eta_e^{R} \sim \eta_e^{u} \sim \eta_i \, , \\ 
  \epsilon_e & \sim & \mu_e^{1/2} \mathrm{Ma}_0 \frac{\lambda_i}{L_V}\sim \mu_e^{1/2} \mathrm{Ma}_0^{3/2} \left(\frac{\lambda_i}{L_0}\right)^{1/2} \, ,   \\
  \epsilon_i & \sim & \mathrm{Ma}_0 \frac{\lambda_i}{L_V} \sim \mathrm{Ma}_0^{3/2} \left(\frac{\lambda_i}{L_0}\right)^{1/2} \, 
  . \label{brag_multiscale}
\end{subeqnarray}
Thus, we conclude that the largest distortions of the ion CE distribution are due to 
flow gradients, while temperature gradients cause the greatest distortions of the 
electron CE distribution function.

\subsection{Kinetic stability of classical, collisional plasma}

\subsubsection{Overview}

We have seen that the CE expansion provides a procedure for the calculation of 
the distribution functions arising in a classical, collisional plasma in terms of gradients of 
temperature, electron-ion drifts and bulk fluid velocities; these calculations in turn allow for the 
closure of the system (\ref{fluideqns}) of fluid equations. However, these same gradients are 
sources of free energy in the plasma, so they can lead to instabilities. Some 
of these instabilities will be `fluid', i.e., they are captured within the CE 
description and are features of the fluid dynamics of plasmas; others 
are kinetic (`microinstabilities'), and their existence implies that the CE expansion is, in fact, illegitimate. 
Our primary purpose in this paper is to determine when such microinstabilities do not occur in a strongly magnetised two-species plasma.
If, however, they do occur, we wish to determine their growth rates. 
We begin by making a few general qualitative comments concerning the existence and nature of these 
microinstabilities, before presenting the technical details of their derivation. 

\subsubsection{Existence of microinstabilities in classical, collisional plasma} 
\label{existmicrostab}

It might 
naively be assumed that a classical, collisional plasma is kinetically stable, on two 
grounds. The first of these is that the distribution function of such a plasma is `almost' Maxwellian, 
and thus stable. While it is certainly the case that a plasma whose constituent particles have 
Maxwellian distribution functions is kinetically stable~\citep{B58,CL14}, it is also known that a plasma with anisotropic particle distribution functions is typically 
not~\citep{F62,KMQ68,D83,G93}.  
The (small) non-Maxwellian component of the CE distribution 
function is anisotropic (as, e.g., was explicitly demonstrated by the calculation of pressure anisotropy in section \ref{fluxfluids}), and thus we cannot \textit{a priori} rule out 
microinstabilities associated with this anisotropy. 

The second naive reason for dismissing the possibility of microinstabilities in 
classical, collisional plasma is the potentially stabilising effect of collisional damping on microinstability growth 
rates. If collisional processes are sufficiently dominant to 
be responsible for the mediation of macroscopic momentum and heat 
fluxes in the plasma, it might be naively inferred that they would also suppress microinstabilities. This is, in fact, far from guaranteed, for the following reason. The 
characteristic scales of the microinstabilities are not fluid scales, but are rather 
intrinsic plasma length scales related to quantities such as the Larmor radius $\rho_s$ 
or the inertial scale $d_s$ of species $s$, or the Debye length $\lambda_{\mathrm{D}}$ 
-- quantities given in terms of macroscopic physical properties of plasma~by
\begin{subeqnarray}
\rho_s & = & \frac{m_s v_{\mathrm{th}s} c}{Z_s e |\boldsymbol{B}|} , \\
 d_s & \equiv & \left(\frac{4 \upi Z_s^2 e^2 n_s}{m_s c^2}\right)^{-1/2} = \rho_s \beta_s^{-1/2} , \\
\lambda_{\mathrm{D}} & \equiv & \left(\sum_s \frac{4 \upi Z_s^2 e^2 n_s}{T_s}\right)^{-1/2} = \left(\sum_s \frac{2 c}{d_s^2 v_{\mathrm{th}s}} \right)^{-1/2}, \label{skin_depth_def}
\end{subeqnarray}
where 
\begin{equation}
\beta_s \equiv  \frac{8 \upi n_s T_s}{B^2} \label{plasma_beta_def}
\end{equation}
is the plasma beta of species $s$. The crucial observation is then that the dynamics on characteristic 
microinstability scales may 
be collisionless. For a classical, collisional hydrogen plasma
 (where $\lambda \equiv \lambda_e \sim \lambda_i$ for $T_e \sim T_i$), the mean free path is 
 much larger than the Debye length: $\lambda/\lambda_{\mathrm{D}} \sim n_e \lambda_{\mathrm{D}}^3 \gg 
 1$; so there exists a range of wavenumbers $k$ on which microinstabilities 
 are both possible ($k \lambda_{\mathrm{D}} \lesssim 1$) and collisionless 
 ($k \lambda \gg 1$). For a strongly magnetised collisional 
 plasma, $\lambda_s \gg \rho_s$ for all species by definition; thus, any microinstability with a characteristic scale 
 comparable to the Larmor radius of any constituent particle will be 
 effectively collisionless. We note that such a range of collisionless wavenumbers only 
 exists in classical (viz., weakly coupled) plasmas; in strongly coupled plasmas, for which $\lambda \lesssim 
\lambda_{\mathrm{D}}$, all hypothetically possible microinstability wavenumber scales are collisional.   
Thus the phenomenon of microinstabilities in collisional plasmas is solely a concern for the classical 
regime.  
 
 \subsubsection{A simple example: the firehose instability in CE plasmas} \label{CEexam_firehose}
 
Perhaps the simplest example of a microinstability that can occur in CE
plasma is the firehose instability. This example was previously discussed 
by~\citet{SCKHS05}, but we nonetheless outline it here to illustrate the 
central concept of our paper. 

Consider bulk fluid motions of the plasma on length scales $L_V$ that are much smaller 
than the mean free path $\lambda_i$, but much larger than the ion Larmor radius 
$\rho_i$; the characteristic frequencies associated with these motions are assumed to be much smaller that the ion Larmor 
frequency $\Omega_i$, but much larger than the inverse of the ion collision time $\tau_{i}^{-1}$.  Under these assumptions, 
the following four statements can be shown to be true~\citep{SCKHS05}:
\begin{enumerate}
 \item The bulk velocities of the electron and ion species are approximately 
 equal: $\boldsymbol{V}_e \approx \boldsymbol{V}_i$. 
 \item The electric field in a frame co-moving with the ion fluid vanishes; transforming to the stationary frame of the system, 
 this gives
 \begin{equation}
 \boldsymbol{E} = -\frac{\boldsymbol{V}_i \times \boldsymbol{B}}{c} \, .  \label{idealMHDohmslaw}
 \end{equation}
 \item The contribution of the displacement current to the Maxwell-Amp\`ere law 
 (\ref{Maxwell}\textit{d}) is negligible, and so
 \begin{equation}
   e n_e \left(\boldsymbol{V}_i - \boldsymbol{V}_e \right) \approx \frac{c}{4 \upi} 
   \bnabla \times \boldsymbol{B} \, . 
   \end{equation}
 \item The electron and ion viscosity tensors both take the form 
 (\ref{viscosity_tens_gyro}), and the electron pressure anisotropy, defined by (\ref{pressureanisop}), is small compared to the ion pressure anisotropy: $\Delta_e \ll \Delta_i$.  
\end{enumerate}
It then follows directly from (\ref{fluideqns}\textit{b}), summed over both ion and electron species, that 
\begin{equation}
m_i n_i {\mathrm{D} \boldsymbol{V}_i \over \mathrm{D} t} \bigg{|}_i = -\bnabla \left(\frac{B^2}{8 \upi} + p_{e\perp} + p_{i\perp}\right) -  \bnabla \bcdot \left[ \hat{\boldsymbol{z}} \hat{\boldsymbol{z}} \left(p_{i\perp}-p_{i\|}\right) 
\right] + \frac{\boldsymbol{B} \bcdot \bnabla \boldsymbol{B}}{4 \upi}
\, . \label{momequation_ionscales}
\end{equation}
We remind the reader that $\hat{\boldsymbol{z}} = \boldsymbol{B}/B$, and emphasize that we have neglected 
the electron inertial term on the grounds that it is small compared to the ion inertial term:  
\begin{equation}
 m_e n_e {\mathrm{D} \boldsymbol{V}_e \over \mathrm{D} t} \bigg{|}_e \ll m_i n_i {\mathrm{D} \boldsymbol{V}_i \over \mathrm{D} t} \bigg{|}_i \, .  
\end{equation}
 The 
evolution of the magnetic field is described by the induction equation, 
\begin{equation}
{\mathrm{D} \boldsymbol{B} \over \mathrm{D} t} \bigg{|}_i = \boldsymbol{B} \bcdot 
\bnabla \boldsymbol{V}_{i} -  \boldsymbol{B} \bnabla \bcdot \boldsymbol{V}_{i} 
\, , \label{inductioneqn}
\end{equation}
which is derived by substituting (\ref{idealMHDohmslaw}) into Faraday's law
(\ref{Maxwell}\textit{c}). 

Now consider small-amplitude perturbations with respect to a particular macroscale state of the 
plasma
\begin{subeqnarray}
   \delta \boldsymbol{V}_{i} & = &
   \widehat{\delta \boldsymbol{V}}_{i\perp} \exp\left\{\mathrm{i}\left(\boldsymbol{k} \bcdot \boldsymbol{r} - \omega t\right)\right\}  ,\\[3pt]
  \delta \boldsymbol{B} & = &
   \widehat{\delta \boldsymbol{B}}_{\perp} \exp\left\{\mathrm{i}\left(\boldsymbol{k} \bcdot \boldsymbol{r} - \omega t\right)\right\} \label{FourierEBfieldsFirehose} 
   ,
\end{subeqnarray}
whose characteristic frequency $\omega$ is much greater than that of the plasma's bulk fluid motions (but is still much smaller than $\Omega_i$), 
whose wavevector $\boldsymbol{k} = k_{\|} \hat{\boldsymbol{z}}$ is parallel to 
$\boldsymbol{B}$, and assume also that the velocity and magnetic-field perturbations are 
perpendicular to $\boldsymbol{B}$. It is then easy to show that (\ref{momequation_ionscales}) and (\ref{inductioneqn}) 
become
\begin{subeqnarray}
- \mathrm{i} m_i n_i \omega  \widehat{\delta \boldsymbol{V}}_{i\perp}  & = & 
\mathrm{i} \left(\frac{B_0^2}{4 \upi} + p_{i\perp} - p_{i\|}\right) k_{\|} \frac{ \widehat{\delta \boldsymbol{B}}_{\perp} }{B} 
\, , \\
- \mathrm{i} \omega  \widehat{\delta \boldsymbol{B}}_{\perp} & = & 
\mathrm{i} B k_{\|} \widehat{\delta \boldsymbol{V}}_{i\perp} \, ,
\end{subeqnarray}
where $p_{i\perp}$ and $p_{i\|}$ are the perpendicular and parallel ion pressures 
associated with the macroscale state (which, on account of its comparatively slow evolution compared to the perturbation, can be 
regarded a quasi-equilibrium). Physically, the macroscale flow gives rise 
to different values of $p_{i\perp}$ and $p_{i\|}$, and thereby an ion pressure 
anisotropy $\Delta_i$, because it changes the strength $B$ of the macroscale magnetic 
field; thanks to the effective conservation of the first and second adiabatic 
moments of the ions on the evolution timescale of the macroscale flow~\citep{CGL56}, 
an increase (decrease) in $B$ results in an increase (decrease) in $p_{i\perp}$, 
and a decrease (increase) in $p_{i\|}$.
The dispersion relation for the perturbation is then
\begin{equation}
  \omega^2 = k_{\|}^2 v_{\mathrm{th}i}^2 \left(\frac{1}{\beta_i} + \frac{\Delta_i}{2}\right) \label{firehosegrowth_Sec2}
  \, ,
\end{equation}
where $\beta_i $, defined by (\ref{plasma_beta_def}), is the ion plasma beta. For a sufficiently negative ion pressure anisotropy, viz., $\Delta_i < -2/\beta_i$, the perturbation is unstable. This instability is known as the (parallel) firehose instability.  

The underlying physics of the parallel firehose instability has been discussed extensively 
elsewhere~\citep[see][and references therein; also see section \ref{negpres_fire}]{RSRC11}. 
Here, we simply note that the firehose instability arises in a magnetised plasma with sufficiently negative
pressure anisotropy as compared to the inverse of the ion plasma beta; because the ion CE distribution 
function has a small, non-zero pressure anisotropy, this statement applies to CE plasma at 
large $\beta_i$. We also observe that the product of the growth rate (\ref{firehosegrowth_Sec2}) of the firehose instability with the ion-ion collision time 
satisfies
\begin{equation}
\omega \tau_i \sim k_{\|} \lambda_i \left|\frac{1}{\beta_i} + \Delta_i \right|^{1/2} \sim 
\frac{1}{\beta_i} \frac{\lambda_i}{\rho_i} 
\, ,
\end{equation}
where we have assumed that $\Delta_i \lesssim 2 \beta_i^{-1}$, and employed the (non-trivial) result that the peak growth of the parallel firehose instability occurs at wavenumbers satisfying $k_{\|} \rho_i \sim \beta_i^{-1/2}$ (see sections \ref{negpres_fire} and 
\ref{negpres_fire_par}).  
Thus, if $\beta_i \ll \lambda_i/\rho_i$ 
-- a condition easily satisifed in weakly collisional astrophysical environments such as the ICM (see table \ref{tab:physicalparams})
-- it follows that $\omega \tau_i \gg 1$, and so collisional damping is unable to inhibit the parallel 
firehose in a CE plasma\footnote{In fact, the naive condition $\gamma \tau_i \lesssim 1$ 
is not sufficient to ensure collisional stabilisation of the firehose 
instability; the true stabilisation condition is instead $k_{\|} \lambda_i \lesssim 1$ 
(see section \ref{shortcomings_coll} for a discussion of this claim).}. This 
failure is directly attributable to its characteristic wavelength being at 
collisionless scales: the parallel wavenumber satisfies $k_{\|} \lambda_i \sim \beta_i^{-1/2} \lambda_i/\rho_i \gg 
1$. 

This simple example clearly illustrates
that microinstabilities are indeed possible in a classical, collisional plasma, 
for precisely the reasons given in section \ref{existmicrostab}.
 
 \subsubsection{Which microinstabilities are relevant} 
 \label{sec:CharacMicroQual}
 
Although the naive arguments described in section \ref{existmicrostab} do not imply kinetic 
stability of CE plasma, these same arguments do lead to significant restrictions 
on the type of microinstabilities that can arise. Namely, for some plasma modes, 
the small anisotropy of CE distribution functions is an insufficient free-energy 
source for overcoming the competing collisionless damping mechanisms that ensure stability for pure Maxwellian distribution functions -- e.g., Landau damping or 
cyclotron damping. For other plasma modes, the characteristic length scales 
are so large that collisional damping does suppress growth. In magnetised plasmas, there also 
exist cyclotron harmonic oscillations that, despite minimal damping, can only become unstable
for sufficiently large anisotropy of the particle distribution function: e.g., the electrostatic Harris instability~\citep{H59,HHK64}. Since the anisotropy threshold for such microinstabilities is typically $\Delta_s \gtrsim 1$~\citep{SH65}, they cannot operate in a CE plasma. 

We claim that there are only two classes of microinstabilities that can be triggered in a CE plasma. 
The first are \textit{quasi-cold plasma modes}: these are modes whose frequency is 
so large that resonant wave-particle interactions (Landau or cyclotron resonances) 
only occur with electrons whose speed greatly exceeds the electron thermal speed 
$v_{\mathrm{th}e}$. Collisionless damping of such modes is typically very weak, 
and thus small anisotropies of particle distribution functions can be sufficient to 
drive an instability. Well-known examples of a small non-Maxwellian part of
the distribution function giving rise to microinstabilities include the bump-on-tail instability associated with 
a fast beam of electrons~\citep[see section 3.3.3 of][]{D83}, 
or the whistler instability for small temperature anisotropies~\citep[see section 3.3.5 of][]{D83}. 
The existence of such instabilities for the CE distribution can be 
demonstrated explicitly: e.g., the peak growth rate of the bump-on-tail instability associated with the CE distribution function (`the CE bump-on-tail instability') is calculated
in appendix \ref{existence_electrostatic}. However, the growth rates $\gamma$ of such instabilities are 
exponentially small in $\lambda_e/L \ll 1$. This claim, which is explicitly 
proven for the CE bump-on-tail instability in appendix 
\ref{existence_electrostatic}, applies to all electrostatic instabilities (see appendix \ref{impossibility_electrostatic_slowgrowth}), and it can be argued that it also applies to all quasi-cold plasma modes (see 
appendix \ref{arg_stab_highfreq}). 
When combined with the constraint that the resonant wave-particle interactions 
required for such instabilities cannot occur if $\gamma \tau_{r} \lesssim 1$, where $\tau_{r}$ is the collision time of the resonant particles,
the exponential smallness of the growth rate suggests that such microinstabilities will not be 
significant provided $\lambda_e/L$ really is small. As discussed in section \ref{rel_size_CEterms}, plasmas in which $\lambda_e/L$ is only moderately small are not well modelled as CE plasmas anyway, and thus, for the rest of 
this paper, we will not study quasi-cold-plasma-mode instabilities.

The second class of allowed microinstabilities comprises modes that are electromagnetic and low-frequency in the sense that 
the complex frequency $\omega$ of the microinstability satisfies, for at least one particle species $s$,
\begin{equation}
    \frac{\omega}{k v_{\mathrm{th}s}} \sim \left(\frac{\lambda_s}{L}\right)^{\iota} \label{omegascale_A} 
    \ll 1 , 
\end{equation}
 where $\iota$ is some order-unity number. Low-frequency electromagnetic modes are 
in general only subject to weak Landau and cyclotron damping (of order $\omega/k v_{\mathrm{th}s} \ll 1 $ or less), and thus can become unstable for small distribution-function anisotropies. By contrast, electromagnetic 
modes satisfying $\omega \sim k v_{\mathrm{th}s}$ would typically generate strong 
inductive electric fields, which would in turn be subject to significant Landau or cyclotron 
damping, overwhelming any unstable tendency. 
The firehose instability introduced in section \ref{CEexam_firehose} is one example of this
 type of microinstability: it satisfies (\ref{omegascale_A}) with $\iota = 1/2$, provided its $\beta$-stabilisation 
 threshold is surpassed.

In this paper, we will focus on microinstabilities in this second class. 
Whilst small compared 
to the streaming rate $k v_{\mathrm{th}s}$ of species $s$, the growth 
rates satisfying (\ref{omegascale_A}) can still be significantly larger than the rate at which the plasma evolves 
on macroscopic scales, and thus invalidate the CE expansion. 
We do not in this paper present a rigorous proof that there are no 
microinstabilities of the CE distribution function which do not fall into either of the two classes considered 
above. However, there do exist more precise arguments supporting the latter claim than 
those based on physical intuition just presented; these are discussed further in sections \ref{disprel_simps_overview} and \ref{shortcomings_othermicro}. 

The microinstabilities satisfying  
(\ref{omegascale_A}) fall into two sub-classes. The first sub-class consists of microinstabilities driven by 
the CE temperature-gradient, CE electron-friction and CE electron-ion-drift terms in the CE distribution functions (\ref{ChapEnskogFunc}); we refer to 
these collectively as \textit{CE temperature-gradient-driven microinstabilities}, or CET microinstabilities,
on account of the parameters $\eta_e^R$ and $\eta_e^u$ scaling with temperature gradients (see section \ref{fluxfluids}). 
The second sub-class is microinstabilities driven by the CE shear terms, or 
\textit{CE shear-driven microinstabilities} (CES microinstabilities). This sub-classification is 
necessary for two reasons. First, the velocity-space anisotropy associated with the CE shear 
terms is different from other non-Maxwellian terms, and thus different types of 
microinstabilities can emerge for the two sub-classes. Secondly, as was discussed in section \ref{rel_size_CEterms} for the case
of CE plasma, the typical size of small parameters $\eta_e^T$, $\eta_e^R$, $\eta_e^u$ and $\eta_i$ 
is different from that of $\epsilon_e$ and $\epsilon_i$. In our initial 
overview of our calculations (section \ref{linear_stab_overview}) and in the more detailed discussion of our method (section \ref{linear_stab_method}), 
we will consider all microinstabilities driven by the 
non-Maxwellian terms of the CE distribution together; however, when it comes to 
presenting detailed results, we will consider CET and CES microinstabilities 
separately (sections \ref{Results} and \ref{Results_shearingterm}, 
respectively). 

\subsection{Linear stability calculation: overview} \label{linear_stab_overview}

\subsubsection{General dispersion relation} \label{HotPlasmaDispDis}

Our linear kinetic stability calculation proceeds as follows: we consider an electromagnetic perturbation with wavevector $\boldsymbol{k}$ and (complex) frequency $\omega$ of the form
\begin{subeqnarray}
   \delta \boldsymbol{E} & = &
   \widehat{\delta \boldsymbol{E}} \exp\left\{\mathrm{i}\left(\boldsymbol{k} \bcdot \boldsymbol{r} - \omega t\right)\right\}  ,\\[3pt]
  \delta \boldsymbol{B} & = &
   \widehat{\delta \boldsymbol{B}} \exp\left\{\mathrm{i}\left(\boldsymbol{k} \bcdot \boldsymbol{r} - \omega t\right)\right\} \label{FourierEBfields} 
   ,
\end{subeqnarray}
in a plasma with the equilibrium electron and ion distribution functions
given by (\ref{ChapEnskogFunc}\textit{a}) and (\ref{ChapEnskogFunc}\textit{b}), respectively. We 
assume that all macroscopic parameters in the CE distribution function are 
effectively constant on the time scales and length scales associated with 
microinstabilities: this is equivalent to assuming that $k \lambda_e, k \lambda_i \gg 1$
(where $k \equiv |\boldsymbol{k}|$ is the wavenumber of the perturbation), and $|\omega| \tau_L \gg 1$. 
To minimise confusion between quantities evolving on short, collisionless time scales, and those on long, fluid time scales, 
we relabel the equilibrium number density of species $s$ as $n_{s0}$, and the macroscopic magnetic field 
as $\boldsymbol{B}_0$ in subsequent calculations. For notational convenience, we define
\begin{equation}
\eta_e \equiv \eta_e^T ,
\end{equation}
and 
\begin{equation}
A_e(\tilde{v}_e) \equiv A_e^T(\tilde{v}_e) + \frac{\eta_e^R}{\eta_e^T} A_e^R(\tilde{v}_e) 
+ \frac{\eta_e^u}{\eta_e^T} A_e^u(\tilde{v}_e) 
\, ,
\end{equation}
which in turn allows for the equilibrium distribution function of species $s$ to be 
written as
\begin{equation}
f_{s0}(\tilde{v}_{s\|},\tilde{v}_{s\bot}) = \frac{n_{s0}}{v_{\mathrm{th}s}^3 \upi^{3/2}} \exp \left(-\tilde{v}_{s}^2\right) \left[1+\eta_s A_s(\tilde{v}_s) \tilde{v}_{s\|} + \epsilon_s C_s(\tilde{v}_s) \left(\tilde{v}_{s\|}^2- \frac{\tilde{v}_{s\bot}^2}{2} \right)\right] . \quad  
\label{ChapEnskogFunc_s}
\qquad \;
\end{equation}
Finally, without loss of generality, we can set $\boldsymbol{V}_{i} = 0$ 
by choosing to perform the kinetic calculation in the frame of the ions; thus, $\tilde{\boldsymbol{v}}_s = 
\boldsymbol{v}/v_{\mathrm{th}s}$. 

It is well known~\citep{S62,P17} that the electric field of all linear electromagnetic perturbations in a collisionless, magnetised 
plasma with equilibrium distribution function $f_{s0}$ 
must satisfy 
\begin{equation}
\left[\frac{c^2 k^2}{\omega^2} \left(\hat{\boldsymbol{k}}\hat{\boldsymbol{k}}-\mathsfbi{I}\right)+ \boldsymbol{\mathfrak{E}} \right] 
\bcdot \widehat{\delta \boldsymbol{E}} = 0 \, , \label{singulareigenvaleqnFull}
\end{equation}
where $\hat{\boldsymbol{k}} \equiv \boldsymbol{k}/k$ 
is the direction of the perturbation, 
\begin{equation}
  \boldsymbol{\mathfrak{E}} \equiv \mathsfbi{I} + \frac{4 \upi \mathrm{i}}{\omega} 
  \boldsymbol{\sigma} \label{dielecttensfull}
\end{equation}
the plasma dielectric tensor, and $\boldsymbol{\sigma}$ the plasma 
conductivity tensor. 
The hot-plasma dispersion relation is then given by
\begin{equation}
   \mbox{det}\left[\frac{c^2 k^2}{\omega^2} \left(\hat{\boldsymbol{k}}\hat{\boldsymbol{k}}-\mathsfbi{I}\right)+ \boldsymbol{\mathfrak{E}} \right]=0 
   .
   \label{hotplasmadisprel}
\end{equation}
The conductivity tensor in a hot, magnetised plasma is best displayed in an orthogonal coordinate system with
basis vectors $\left\{\hat{\boldsymbol{x}},\hat{\boldsymbol{y}},\hat{\boldsymbol{z}}\right\}$ defined in terms of $\boldsymbol{B}_0$ and 
$\boldsymbol{k}$:
\begin{equation}
     \hat{\boldsymbol{z}} \equiv \frac{\boldsymbol{B}_0}{B_0}, \quad 
     \hat{\boldsymbol{x}}  \equiv \frac{\boldsymbol{k}_{\perp}}{k_{\perp}} \equiv \frac{\boldsymbol{k}- k_{\|} \hat{\boldsymbol{z}}}{k_{\bot}} , 
     \quad
     \hat{\boldsymbol{y}}  \equiv \hat{\boldsymbol{z}} \times \hat{\boldsymbol{x}} \label{basisdef} 
     , \label{xyzcoordinatebasis}
\end{equation}
where $B_0 \equiv |\boldsymbol{B}_0|$, $k_{\|} \equiv \boldsymbol{k}\bcdot \hat{\boldsymbol{z}}$, and $k_{\bot} \equiv \left|\boldsymbol{k}_{\perp}\right|$.
In this notation, $\boldsymbol{k} = k_{\|} \hat{\boldsymbol{z}} + k_{\bot} 
\hat{\boldsymbol{x}}$. 
The conductivity tensor is then given by
\begin{eqnarray}
\boldsymbol{\sigma} = \sum_s \boldsymbol{\sigma} _s & = & - \frac{\mathrm{i}}{4 \upi \omega} \sum_s \omega_{\mathrm{p}s}^2 \bigg[ \frac{2}{\sqrt{\upi}} \frac{k_{\|}}{|k_{\|}|}  \int_{-\infty}^{\infty} \mathrm{d} \tilde{w}_{s\|} \, \tilde{w}_{s\|} \int_0^{\infty} \mathrm{d} \tilde{v}_{s\bot} \Lambda_s(\tilde{w}_{s\|},\tilde{v}_{s\bot}) \hat{\boldsymbol{z}} \hat{\boldsymbol{z}} \nonumber \\
&& \mbox{} + \tilde{\omega}_{s\|} \frac{2}{\sqrt{\upi}} \int_{C_L} \mathrm{d} \tilde{w}_{s\|} \int_0^{\infty} \mathrm{d} \tilde{v}_{s\bot} \tilde{v}_{s\bot}^2 \Xi_s(\tilde{w}_{s\|},\tilde{v}_{s\bot}) \sum_{n = -\infty}^{\infty} \frac{\mathsfbi{R}_{sn}}{\zeta_{sn} -\tilde{w}_{s\|}}  \bigg] \, , \qquad \label{conductivity}
\end{eqnarray}
where
\begin{equation}
  \omega_{\mathrm{p}s} \equiv \sqrt{\frac{4 \upi Z_s^2 e^2 n_{s0}}{m_s}} , \label{plasmafrequency_def}
\end{equation} 
\begin{equation}
\tilde{w}_{s\|} \equiv \frac{k_{\|} \tilde{v}_{s\|}}{|k_{\|}|},   \label{w_pl_def}
\end{equation}
\begin{equation}
  \tilde{\rho}_s \equiv \frac{m_s c v_{\mathrm{th}s}}{Z_{s} e B_0} =  \frac{|Z_{s}|}{Z_{s}} \rho_s,
\end{equation}
\begin{equation}
  \tilde{\omega}_{s\|} \equiv \frac{\omega}{|k_{\|}| v_{\mathrm{th}s}} , \label{normfrequency}
\end{equation}
\begin{equation}
  \zeta_{sn} \equiv \tilde{\omega}_{s\|} - \frac{n}{|k_{\|}| \tilde{\rho}_s} , \label{zeta_def}
\end{equation}
\begin{equation}
  \tilde{f}_{s0}(\tilde{v}_{s\|},\tilde{v}_{s\bot}) \equiv \frac{\upi^{3/2} v_{\mathrm{th}s}^3}{n_{s0}} f_{s0}\left(\frac{k_{\|}}{|k_{\|}|} v_{\mathrm{th}s} \tilde{w}_{s\|},v_{\mathrm{th}s} \tilde{v}_{s\bot}\right)
  , \label{renorm_distfunc_def}
\end{equation}
\begin{equation}
  \Lambda_s(\tilde{w}_{s\|},\tilde{v}_{s\bot}) \equiv \tilde{v}_{s\bot} \frac{\p \tilde{f}_{s0}}{\p \tilde{w}_{s\|}}-\tilde{w}_{s\|} \frac{\p \tilde{f}_{s0}}{\p \tilde{v}_{s\bot}} 
  , \label{anisotropyfunc}
\end{equation}
\begin{equation}
  \Xi_s(\tilde{w}_{s\|},\tilde{v}_{s\bot}) \equiv \frac{\p \tilde{f}_{s0}}{\p \tilde{v}_{s\bot}}
  + \frac{\Lambda_s(\tilde{w}_{s\|},\tilde{v}_{s\bot})}{\tilde{\omega}_{s\|}} , \label{IntgradCond}
\end{equation}
\normalsize
\begin{subeqnarray}
 (\mathsfbi{R}_{sn} )_{xx} & \equiv & \frac{n^2 J_n(k_{\bot} \tilde{\rho}_s \tilde{v}_{s\bot})^2}{k_{\bot}^2 \tilde{\rho}_s^2 \tilde{v}_{s\bot}^2} , \\
 (\mathsfbi{R}_{sn} )_{xy} & \equiv & \frac{\mathrm{i} n J_n(k_{\bot} \tilde{\rho}_s \tilde{v}_{s\bot}) J_n'(k_{\bot} \tilde{\rho}_s \tilde{v}_{s\bot})}{k_{\bot} \tilde{\rho}_s \tilde{v}_{s\bot}} , \\
 (\mathsfbi{R}_{sn} )_{xz} & \equiv & \frac{n J_n(k_{\bot} \tilde{\rho}_s \tilde{v}_{s\bot})^2}{k_{\bot} \tilde{\rho}_s \tilde{v}_{s\bot}} \frac{k_{\|} \tilde{w}_{s\|}}{|k_{\|}| \tilde{v}_{s\bot}}  ,\\
 (\mathsfbi{R}_{sn} )_{yx} & \equiv & - (\mathsfbi{R}_{sn} )_{xy} ,\\
 (\mathsfbi{R}_{sn} )_{yy} & \equiv & J_n'(k_{\bot} \tilde{\rho}_s \tilde{v}_{s\bot})^2 ,\\
 (\mathsfbi{R}_{sn} )_{yz} & \equiv & -\mathrm{i} n J_n(k_{\bot} \tilde{\rho}_s \tilde{v}_{s\bot}) J_n'(k_{\bot} \tilde{\rho}_s \tilde{v}_{s\bot}) \frac{k_{\|} \tilde{w}_{s\|}}{|k_{\|}| \tilde{v}_{s\bot}} , \\
 (\mathsfbi{R}_{sn} )_{zx} & \equiv & (\mathsfbi{R}_{sn} )_{xz} ,\\
 (\mathsfbi{R}_{sn} )_{zy} & \equiv & -(\mathsfbi{R}_{sn} )_{yz} ,\\
 (\mathsfbi{R}_{sn} )_{zz} & \equiv & \frac{\tilde{w}_{s\|}^2}{\tilde{v}_{s\bot}^2} J_n(k_{\bot} \tilde{\rho}_s \tilde{v}_{s\bot})^2 . \label{defMsn}
\end{subeqnarray}
Here $(\mathsfbi{R}_{sn})_{xy} = \hat{\boldsymbol{x}} \bcdot \mathsfbi{R}_{sn} \bcdot \hat{\boldsymbol{y}}$, and similarly 
for other components of $\mathsfbi{R}_{sn}$. For the reader's convenience, a summary of the derivation of the hot-plasma 
dispersion relation is given in appendix \ref{HotPlasmDispDeriva}.

We note that the dielectric and conductivity tensors have the following symmetries:
\begin{equation}
  \mathfrak{E}_{yx} = - \mathfrak{E}_{xy} \, , \quad  \mathfrak{E}_{zx} = \mathfrak{E}_{xz} \, , 
  \quad 
  \mathfrak{E}_{zy} = - \mathfrak{E}_{yz} \, , \label{dielectricsimsgen}
\end{equation}
\begin{equation}
  \sigma_{yx} = - \sigma_{xy} \, , \quad \sigma_{zx} = \sigma_{xz} \, , \quad \sigma_{zy} = - \sigma_{yz} \, , \label{conductsimsgen}
\end{equation}
where, for tensors with no species subscript, we use the notation $ \mathfrak{E}_{xy} \equiv \hat{\boldsymbol{x}} \bcdot  \boldsymbol{\mathfrak{E}} \bcdot 
\hat{\boldsymbol{y}}$. We also observe that if $f_{s0}(v_{\|},v_{\bot})$ 
 is an even function with respect to $v_{\|}$, then, for $k_{\|} > 0$,
 \begin{subeqnarray}
  \sigma_{xx}(-k_{\|}) & = & \sigma_{xx}(k_{\|}) \, , \\
  \sigma_{xy}(-k_{\|}) & = & \sigma_{xy}(k_{\|}) \, , \\
  \sigma_{xz}(-k_{\|}) & = & -\sigma_{xz}(k_{\|}) \, , \\
  \sigma_{yy}(-k_{\|}) & = & \sigma_{yy}(k_{\|}) \, , \\
  \sigma_{yz}(-k_{\|}) & = & -\sigma_{yz}(k_{\|}) \, , \\
  \sigma_{zz}(-k_{\|}) & = & \sigma_{zz}(k_{\|}) \, ,\label{conductnegwav_even}
\end{subeqnarray}
with the remaining components of the conductivity tensor given by equations 
(\ref{conductsimsgen}). If $f_{s0}(v_{\|},v_{\bot})$ 
 is an odd function with respect to $v_{\|}$, then 
 \begin{subeqnarray}
  \sigma_{xx}(-k_{\|}) & = & -\sigma_{xx}(k_{\|}) \, , \\
  \sigma_{xy}(-k_{\|}) & = & -\sigma_{xy}(k_{\|}) \, , \\
  \sigma_{xz}(-k_{\|}) & = &\sigma_{xz}(k_{\|}) \, , \\
  \sigma_{yy}(-k_{\|}) & = &-\sigma_{yy}(k_{\|}) \, , \\
  \sigma_{yz}(-k_{\|}) & = & \sigma_{yz}(k_{\|}) \, , \\
  \sigma_{zz}(-k_{\|}) & = &-\sigma_{zz}(k_{\|}) \, .\label{conductnegwav_odd}
\end{subeqnarray}
These symmetries can be used to determine completely the behaviour of perturbations with $k_{\|} < 0$ 
directly from perturbations with  $k_{\|} > 0$, without any additional 
calculations. Thus, unless stated otherwise, from this point on, we assume $k_{\|} > 
0$, and thus $\tilde{w}_{s\|} = \tilde{v}_{s\|}$ [see (\ref{w_pl_def})].

\subsubsection{Simplifications of dispersion relation: overview of our approach} \label{disprel_simps_overview}

The full hot-plasma dispersion relation (\ref{hotplasmadisprel}) is a transcendental equation, and 
thus, for general distribution functions, the growth rates of perturbations can only be determined numerically; 
this hinders the systematic investigation of stability over wide-ranging parameter regimes. 
However, adopting a few simplifications both to the form of the CE distribution 
functions (\ref{ChapEnskogFunc_s}) and to the type of microinstabilities being 
considered (see section \ref{sec:CharacMicroQual}) turns out to be advantageous when attempting a systematic study. It enables us to obtain simple analytical results 
for microinstability growth rates and characteristic wavenumbers, 
as well as greatly reducing the numerical cost of evaluating these quantities.
The former allows us to make straightforward comparisons between
microinstabilities, while the latter facilitates the calculation of stability plots over a wide range of parameters 
without requiring intensive computational resources.

First, we choose a Krook collision operator, with constant collision time $\tau_s$ for each species~$s$~\citep{BGK54}, when evaluating the isotropic 
functions $A_e^T(\tilde{v}_e)$, $A_e^R(\tilde{v}_e)$, $A_e^u(\tilde{v}_e)$, 
$A_i(\tilde{v}_i)$, $C_e(\tilde{v}_e)$, and $C_i(\tilde{v}_i)$ in (\ref{ChapEnskogFunc_s}). 
As was explained in section \ref{sec:CE_dist_func}, these functions are determined 
by the collision operator. 
While the full Landau collision operator might seem to be the most appropriate choice, the conductivity tensor $\boldsymbol{\sigma}$ defined 
by (\ref{conductivity}) cannot be written 
in terms of standard mathematical functions if this choice is made. Instead, the relevant integrals must be done numerically. 
If a simplified collision operator is assumed, $\boldsymbol{\sigma}$ 
can be evaluated analytically with only a moderate amount of algebra. In 
appendix \ref{ChapEnskogIsoFunc_Krook}, we show that for the Krook collision operator, 
\begin{subeqnarray}
  A_e^T(\tilde{v}_e) & = & -\left(\tilde{v}_e^2-\frac{5}{2}\right) \, , \\
  A_e^R(\tilde{v}_e) & = & -1 \, , \\
  A_e^u(\tilde{v}_e) & = & 0 \, , \\
  A_i(\tilde{v}_e) & = & -\left(\tilde{v}_i^2-\frac{5}{2}\right) \, , \\
  C_e(\tilde{v}_e) & = & -1 \, , \\
  C_i(\tilde{v}_i) & = & -1 \, ,
  \label{CETKrook_isofunc}
\end{subeqnarray}
where it is assumed that $\tilde{v}_e, \tilde{v}_i \ll \eta_e^{-1/3}, \epsilon_i^{-1/2}$ in order that the 
CE distribution functions retain positive signs (the vanishing of the CE electron-ion-drift term is discussed in appendix \ref{ChapEnskogIsoFunc_Krook}). 
Adopting the Krook collision operator has the additional advantage of allowing
a simple prescription for collisional damping of microinstabilities to be introduced self-consistently
into our stability calculation (see section \ref{shortcomings_coll} for further discussion of this).    

Secondly, as discussed in section \ref{sec:CharacMicroQual}, the most important microinstabilities associated with the CE 
distribution function are low-frequency, i.e., they 
satisfy (\ref{omegascale_A}). Therefore, instead of solving the full hot-plasma dispersion 
relation, we can obtain 
a less complicated algebraic dispersion relation. We also always consider 
electromagnetic rather than electrostatic perturbations. This is because it can 
be shown for a CE plasma that purely electrostatic microinstabilities are limited 
to the quasi-cold plasma modes (see appendix 
\ref{Electrostatic}). Describing how the simplified dispersion relation for 
low-frequency, electromagnetic perturbations is obtained from the full hot-plasma 
dispersion relation requires a rather lengthy exposition, and necessitates the introduction of a substantial amount of additional mathematical 
notation. In addition to this, certain shortcomings of this approach warrant an extended discussion. 
Readers who are interested these details will find them in the next section (section \ref{linear_stab_method}). 
Readers who are instead keen to see the results of the stability calculations as soon as possible are encouraged to jump to 
sections \ref{Results} and \ref{Results_shearingterm}.  

\subsection{Linear stability calculation: detailed methodology} \label{linear_stab_method}

\subsubsection{Low-frequency condition in a magnetised plasma}

Before applying to the hot-plasma dispersion relation
(\ref{hotplasmadisprel}) the simplifications discussed in section \ref{disprel_simps_overview}, we refine the low-frequency condition (\ref{omegascale_A})
based on the specific form (\ref{conductivity}) of the conductivity tensor for a magnetised plasma. 
It is clear that the equilibrium distribution function only affects the conductivity tensor via the functions $\Lambda_s(\tilde{v}_{s\|},\tilde{v}_{s\bot})$ 
and $\Xi_s(\tilde{v}_{s\|},\tilde{v}_{s\bot})$ [see (\ref{anisotropyfunc}) and (\ref{IntgradCond})]. For a distribution function of the form (\ref{ChapEnskogFunc_s}), 
it can be shown that
\begin{equation}
 \Lambda_s(\tilde{v}_{s\|},\tilde{v}_{s\bot})= - \tilde{v}_{s\bot} \exp \left(-\tilde{v}_{s}^2\right) \left[\eta_s A_s(\tilde{v}_s) - 3 \epsilon_s C_s(\tilde{v}_s)  \tilde{v}_{s\|} \right] 
  , \label{anisopparameter}
\end{equation}
and 
\begin{eqnarray}
 \Xi_s(\tilde{v}_{s\|},\tilde{v}_{s\bot}) & = &  - \tilde{v}_{s\bot} \exp \left(-\tilde{v}_{s}^2\right) \Bigg[ 2 + 2 \tilde{v}_{s\|} \eta_s A_s(\tilde{v}_s) - \frac{\tilde{v}_{s\|}}{\tilde{v}_s} \eta_s A_s'(\tilde{v}_s) 
\nonumber \\
&& \quad + 2 \epsilon_s C_s(\tilde{v}_s) \left(\tilde{v}_{s\|}^2-\frac{\tilde{v}_{s\bot}^2}{2}+\frac{1}{2}\right) 
- \frac{1}{\tilde{v}_s} \left(\tilde{v}_{s\|}^2-\frac{\tilde{v}_{s\bot}^2}{2}\right) 
\epsilon_s C_s'(\tilde{v}_s) \nonumber \\
&& \quad \quad + \frac{\eta_s}{ \tilde{\omega}_{s\|}} A_s(\tilde{v}_s) - 3 \frac{\epsilon_s}{ \tilde{\omega}_{s\|}} C_s(\tilde{v}_s)  \tilde{v}_{s\|} \Bigg] 
\, ,\label{Xi_CE}
\end{eqnarray}
where the first term in the square brackets in (\ref{Xi_CE}) originates from the Maxwellian part of the distribution function. 
A comparison of the size of the second, third, fourth, and fifth terms with the first
indicates that for $\tilde{v}_s \sim 1$ -- for which $\Xi_s$ attains its largest characteristic values -- the non-Maxwellian terms of the CE distribution function only provide a small, $\textit{O}(\eta_e, \epsilon_e)$ 
contribution, and thus the conductivity is only altered slightly. 
However, considering the sixth and seventh terms in the square brackets in (\ref{Xi_CE}) 
(which are only present thanks to the anisotropy of the CE distribution function), it 
is clear that the non-Maxwellian contribution to the conductivity tensor can be 
significant for $\tilde{v}_s \sim 1$ provided the frequency (\ref{normfrequency})
satisfies one of
\begin{equation}
    \tilde{\omega}_{s\|} \sim \eta_s \ll 1  \quad \mathrm{or} \quad \tilde{\omega}_{s\|} \sim \epsilon_s \ll 1 \, . \label{omegascale}
\end{equation}
Thus, the relevant low-frequency condition in a magnetised plasma involves the 
parallel particle streaming rate $k_{\|} v_{\mathrm{th}s}$. 

There do exist certain 
caveats to the claim that it is necessary for microinstabilities of CE 
plasma to satisfy (\ref{omegascale}); we defer detailed statement and discussion of these caveats -- as well as of other potential shortcomings of our approach -- to sections 
\ref{shortcomings_twospecies}, \ref{shortcomings_coll} and
\ref{shortcomings_othermicro}. 

\subsubsection{Simplification I: non-relativistic electromagnetic fluctuations} \label{disprel_simps_I}

The requirement that the mode be electromagnetic, combined with the fact we are 
interested in non-relativistic fluctuations ($\omega \ll k c$) enables our first simplification. We see from (\ref{hotplasmadisprel}) that for any 
perturbation of interest, the dielectric tensor must satisfy $\|\boldsymbol{\mathfrak{E}}\| \gtrsim k^2 
c^2/\omega^2 \gg 1$ (where $\| \cdot \|$ is the Euclidean tensor norm); therefore, it simplifies to
\begin{equation}
  \boldsymbol{\mathfrak{E}} \approx \frac{4 \upi \mathrm{i}}{\omega} 
  \boldsymbol{\sigma} \, . \label{dielectric_nonrelsimp}
\end{equation}
This amounts to ignoring the displacement current in the 
Amp\`ere-Maxwell law, leaving Amp\`ere's original equation. For convenience of exposition, we 
denote the contribution of each species $s$ to (\ref{dielectric_nonrelsimp}) by
\begin{equation}
  \boldsymbol{\mathfrak{E}}_s \equiv \frac{4 \upi \mathrm{i}}{\omega} 
  \boldsymbol{\sigma}_s \, . \label{dielectric_species_s}
\end{equation}

\subsubsection{Simplification II: expansion of dielectric tensor in $\omega \ll k_{\|} v_{\mathrm{th}s}$} 
\label{disprel_simps_II}

The next simplification involves an expansion of the matrices $\boldsymbol{\mathfrak{E}}_s$ in 
the small parameters $\tilde{\omega}_{s\|} \sim \eta_s \sim \epsilon_s \ll 1$. The general principle of the expansion is as follows. We first divide the matrix $\boldsymbol{\mathfrak{E}}_s$ [see (\ref{dielecttensfull}), (\ref{conductivity}), and (\ref{dielectric_species_s})] into the Maxwellian contribution $\mathsfbi{M}_s$ and the non-Maxwellian one $\mathsfbi{P}_s$:
\begin{equation}
\boldsymbol{\mathfrak{E}}_s = \frac{\omega_{\mathrm{p}s}^2}{\omega^2} \left(\mathsfbi{M}_s + \mathsfbi{P}_s \right) , \label{Maxnonmaxsep_s}
\end{equation}
where the $\omega_{\mathrm{p}s}^2/\omega^2$ factor is introduced for later convenience. Next, we note 
that for a Maxwellian distribution, $\Lambda_s(\tilde{v}_{s\|},\tilde{v}_{s\bot}) = 
0$ [see (\ref{anisotropyfunc})], whereas $\Lambda_s \sim \epsilon_{s}, \eta_s$ for the non-Maxwellian component of the CE distribution function. Thus, from (\ref{conductivity}) 
considered under the ordering $k \rho_s \sim 1$, $\mathsfbi{M}_s = \textit{O}(\tilde{\omega}_{s\|})$
as $\tilde{\omega}_{s\|} \rightarrow 0$, while $\mathsfbi{P}_s = \textit{O}(\eta_s, 
\epsilon_s)$. The expansion of $\mathsfbi{M}_s$ and $\mathsfbi{P}_s$ in $\tilde{\omega}_{s\|}$ 
is, therefore,
 \begin{subeqnarray}
   \mathsfbi{M}_s\!\left(\tilde{\omega}_{s\|},\boldsymbol{k}\right) & \equiv & \tilde{\omega}_{s\|} \mathsfbi{M}_s^{(0)}\!\left(\boldsymbol{k}  \right)  + \tilde{\omega}_{s\|}^2 \mathsfbi{M}_s^{(1)}\!\left(\boldsymbol{k} \right) + ... 
   \, , \\
   \mathsfbi{P}_s\!\left(\tilde{\omega}_{s\|},\boldsymbol{k}\right)  & \equiv & \mathsfbi{P}_s^{(0)}\!\left(\boldsymbol{k}\right) + \tilde{\omega}_{s\|} \mathsfbi{P}_s^{(1)}\!\left(\boldsymbol{k}\right) + ... \, \label{MandPexpansion_arbspecies}
   .
  \end{subeqnarray}
  where the matrices $\mathsfbi{M}_s^{(0)}$ and 
  $\mathsfbi{M}_s^{(1)}$ are $\textit{O}(1)$ functions of $\boldsymbol{k}$ only, and $\mathsfbi{P}_s^{(0)}$ 
  and $\mathsfbi{P}_s^{(1)}$ are $\textit{O}(\eta_s, 
\epsilon_s)$. We then expand $\boldsymbol{\mathfrak{E}}_s$ as follows:
  \begin{equation}
    \boldsymbol{\mathfrak{E}}_s = \tilde{\omega}_{s\|} \boldsymbol{\mathfrak{E}}_s^{(0)} + \tilde{\omega}_{s\|}^2 \boldsymbol{\mathfrak{E}}_s^{(1)}
    + ... \, , \label{dielectric_expand}
  \end{equation}
  where
   \begin{subeqnarray}
  \boldsymbol{\mathfrak{E}}_s^{(0)} & \equiv & \frac{\omega_{\mathrm{p}s}^2}{\omega^2} \left[ \mathsfbi{M}_s^{(0)}\!\left(\boldsymbol{k} \right)  + \frac{1}{\tilde{\omega}_{s\|}} \mathsfbi{P}_s^{(0)}\!\left(\boldsymbol{k}\right) \right] 
   \, ,  \\
   \boldsymbol{\mathfrak{E}}_s^{(1)}  & \equiv & \frac{\omega_{\mathrm{p}s}^2}{\omega^2} \left[ \mathsfbi{M}_s^{(1)}\!\left(\boldsymbol{k} \right) + \frac{1}{\tilde{\omega}_{s\|}}  \mathsfbi{P}_s^{(1)}\!\left(\boldsymbol{k}\right) \right] \, 
   . \label{Dielectric_0}
  \end{subeqnarray}
  
  \subsubsection{Additional symmetries of low-frequency dielectric tensor $ \boldsymbol{\mathfrak{E}}_s^{(0)} $} \label{disprel_simps_addition_simp}
  
 The tensor $\boldsymbol{\mathfrak{E}}_s^{(0)}$ defined by (\ref{Dielectric_0}\textit{a}) has some rather convenient additional symmetries, which lead to significant 
 simplification of the dispersion relation.
In appendix \ref{DispLowFreq} we show that in combination with the general symmetries 
(\ref{dielectricsimsgen}), which apply to $\boldsymbol{\mathfrak{E}}_s^{(0)}$ in addition to $\boldsymbol{\mathfrak{E}}$,
for any distribution function of particle species $s$ with a 
small anisotropy, 
\begin{subeqnarray}
  (\boldsymbol{\mathfrak{E}}_s^{(0)})_{xz} & = & - \frac{k_{\bot}}{k_{\|}} (\boldsymbol{\mathfrak{E}}_s^{(0)})_{xx} \, , \\
   (\boldsymbol{\mathfrak{E}}_s^{(0)})_{yz} & = & \frac{k_{\bot}}{k_{\|}} (\boldsymbol{\mathfrak{E}}_s^{(0)})_{xy} \, , \\
   (\boldsymbol{\mathfrak{E}}_s^{(0)})_{zz} & = & \frac{k_{\bot}^2}{k_{\|}^2} (\boldsymbol{\mathfrak{E}}_s^{(0)})_{xx} \, . 
  \label{dielectricsymsB}
\end{subeqnarray}
These symmetries have the consequence that
\begin{equation}
  \hat{\boldsymbol{k}} \bcdot \boldsymbol{\mathfrak{E}}_s^{(0)} = 
  \boldsymbol{\mathfrak{E}}_s^{(0)}
  \bcdot \hat{\boldsymbol{k}}
  = 0 \, . \label{dielectric_zeroth_order}
\end{equation}
As a result of this identity, it is convenient to calculate the components of $\boldsymbol{\mathfrak{E}}_s^{(0)}$ (and $\boldsymbol{\mathfrak{E}}_s$)
in the coordinate basis $\{\boldsymbol{e}_1,\boldsymbol{e}_2,\boldsymbol{e}_3\}$
defined by
 \begin{equation}
   \boldsymbol{e}_1 \equiv \hat{\boldsymbol{y}} \times \hat{\boldsymbol{k}} 
   \, , \quad
   \boldsymbol{e}_2 \equiv \hat{\boldsymbol{y}} \, , \quad
   \boldsymbol{e}_3 \equiv \hat{\boldsymbol{k}} \, . \label{e_coordinatebasis_def}
   \end{equation}
Carrying out this calculation (see appendix \ref{DispLowFreq}), we find
\begin{subeqnarray}
  (\boldsymbol{\mathfrak{E}}_s^{(0)})_{11} & =  & \frac{k^2}{k_{\|}^2} (\boldsymbol{\mathfrak{E}}_s^{(0)})_{xx}  \, , \\
   (\boldsymbol{\mathfrak{E}}_s^{(0)})_{12} & = & -  (\boldsymbol{\mathfrak{E}}_s^{(0)})_{21} = \frac{k}{k_{\|}}  (\boldsymbol{\mathfrak{E}}_s^{(0)})_{xy}  \, , \\
  (\boldsymbol{\mathfrak{E}}_s^{(0)})_{22} & = &  (\boldsymbol{\mathfrak{E}}_s^{(0)})_{yy} \, , \\
  (\boldsymbol{\mathfrak{E}}_s^{(0)})_{13} & = &  (\boldsymbol{\mathfrak{E}}_s^{(0)})_{31}  =  (\boldsymbol{\mathfrak{E}}_s^{(0)})_{23} =  (\boldsymbol{\mathfrak{E}}_s^{(0)})_{32} =  (\boldsymbol{\mathfrak{E}}_s^{(0)})_{33} = 0 \, ,
     \label{dielectric123}
\end{subeqnarray}
where $(\boldsymbol{\mathfrak{E}}_s^{(0)})_{ij}$ is the $(i,j)$-th component of $\boldsymbol{\mathfrak{E}}_s^{(0)}$ in the basis $\{\boldsymbol{e}_1,\boldsymbol{e}_2,\boldsymbol{e}_3 
\}$. We conclude that, if $k \rho_s \sim 1$ and $\tilde{\omega}_{s\|}  \ll 1$, the components of $\boldsymbol{\mathfrak{E}}_s$ 
satisfy
\begin{equation} 
   (\boldsymbol{\mathfrak{E}}_s)_{13} \sim (\boldsymbol{\mathfrak{E}}_s)_{23} \sim (\boldsymbol{\mathfrak{E}}_s)_{33} \sim \tilde{\omega}_{s\|} (\boldsymbol{\mathfrak{E}}_s)_{11} \sim \tilde{\omega}_{s\|}  (\boldsymbol{\mathfrak{E}}_s)_{12} \sim \tilde{\omega}_{s\|}  (\boldsymbol{\mathfrak{E}}_s)_{22} \, .\label{dielectric3_smallterm}
\end{equation}
These components can be written in terms of the components of 
 $\boldsymbol{\mathfrak{E}}_s$ in the 
 $\{\hat{\boldsymbol{x}},\hat{\boldsymbol{y}},\hat{\boldsymbol{z}}\}$ coordinate 
 frame [see (\ref{xyzcoordinatebasis})] via a coordinate transformation; 
 the resulting expressions are rather bulky, so we do not reproduce them here -- they are detailed in appendix 
 \ref{ChapEnskogFunc_Append_comp}.
  
\subsubsection{Consequences for dispersion relation} \label{consequences}
 
On account of the additional symmetries described in the previous section, a 
simplified dispersion relation for low-frequency modes can be derived in place of the full 
hot-plasma dispersion relation (\ref{hotplasmadisprel}). However, depending on the frequency and characteristic wavelengths of modes, this derivation 
has a subtlety because of the large discrepancy between ion and electron masses. 
In, e.g., a two-species plasma with $\mu_e = m_e/m_i \ll 1$ (and ion charge 
$Z$), we have
\begin{equation}
\frac{\tilde{\omega}_{e\|}}{\tilde{\omega}_{i\|}} = \sqrt{\mu_e \tau} \, 
, \label{ratio_tildeomegas}
\end{equation}
where $\tau = T_i/T_e$. If $\tau \sim 1$ [as would be expected in a collisional plasma on macroscopic evolution time scales $\tau_L$ greater than the ion-electron temperature equilibration time $\tau_{ie}^{\rm eq}$ -- cf. (\ref{macroieequilrat})], then $\tilde{\omega}_{i\|} \sim {\mu_e}^{-1/2} \tilde{\omega}_{e\|} \gg 
\tilde{\omega}_{e\|}$. Thus, in general, $\tilde{\omega}_{i\|} \not \sim \tilde{\omega}_{e\|}$, and any dispersion relation will in principle depend on an additional (small) dimensionless parameter $\mu_e$. This introduces various complications to the simplified dispersion relation's derivation, most significant of which being that, since $\rho_e = Z \mu_e^{1/2} \tau^{-1/2} \rho_i \ll \rho_i$ (for $Z \gtrsim 1$), to assume the ordering $k \rho_s \sim 1$ 
for both ions and electrons is inconsistent (see section \ref{shortcomings_twospecies}). 

To avoid the description of our approach being obscured by these complications, we consider a special case at first: we adopt the ordering $k \rho_e \sim 1$ in a two-species plasma and assume that $\tilde{\omega}_{i\|} \sim \mu_e^{-1/2} \tilde{\omega}_{e\|} \ll 1$. In this case, $\tilde{\omega}_{i\|} \|\boldsymbol{\mathfrak{E}}_i^{(0)}\| \sim \mu_e^{1/2} Z \tau^{-1/2} \tilde{\omega}_{e\|} \|\boldsymbol{\mathfrak{E}}_e^{(0)}\| \ll \tilde{\omega}_{e\|} \|\boldsymbol{\mathfrak{E}}_e^{(0)}\|$, and so the dielectric tensor $\boldsymbol{\mathfrak{E}}$ is given by
  \begin{equation}
    \boldsymbol{\mathfrak{E}} = \tilde{\omega}_{e\|} \boldsymbol{\mathfrak{E}}^{(0)} + \tilde{\omega}_{e\|}^2 \boldsymbol{\mathfrak{E}}^{(1)}
    + ... \, , \label{dielectric_expand_multispecies}
  \end{equation}
  where
   \begin{subeqnarray}
  \boldsymbol{\mathfrak{E}}^{(0)} & \equiv & \boldsymbol{\mathfrak{E}}_e^{(0)} + \frac{\tilde{\omega}_{i\|}}{\tilde{\omega}_{e\|}} \boldsymbol{\mathfrak{E}}_i^{(0)} \approx \boldsymbol{\mathfrak{E}}_e^{(0)}
   \, ,  \\
   \boldsymbol{\mathfrak{E}}_s^{(1)}  & \equiv & \boldsymbol{\mathfrak{E}}_e^{(1)} + \frac{\tilde{\omega}_{i\|}^2}{\tilde{\omega}_{e\|}^2} \boldsymbol{\mathfrak{E}}_i^{(1)} \, 
   . \label{dielectric_0_multispecies}
  \end{subeqnarray}
 Thus, to leading order in the $\tilde{\omega}_{e\|} \ll 1$ expansion, only the electron species contributes to the dielectric tensor for electron-Larmor-scale modes. We revisit the derivation of simplified dispersion relations for CE microinstabilities more generally in section \ref{shortcomings_twospecies}.   

To derive the simplified dispersion relation for electron-Larmor-scale modes, we start by considering the  
component of (\ref{singulareigenvaleqnFull}) for the electric field that is parallel to the wavevector $\hat{\boldsymbol{k}}$, 
\begin{equation}
 \hat{\boldsymbol{k}} \bcdot \boldsymbol{\mathfrak{E}} \bcdot \widehat{\delta \boldsymbol{E}}  
 = 0 \, , \label{singulareigenvaleqnFull_33}
\end{equation}
and then substitute the expanded form (\ref{dielectric_expand_multispecies}) of the 
dielectric tensor (with $s = e$). The orthogonality of $\boldsymbol{\mathfrak{E}}_e^{(0)}$ to $\hat{\boldsymbol{k}}$ -- viz., (\ref{dielectric_zeroth_order})
-- implies that (\ref{singulareigenvaleqnFull_33}) becomes
\begin{equation}
\hat{\boldsymbol{k}} \bcdot \boldsymbol{\mathfrak{E}}^{(1)} \bcdot \widehat{\delta \boldsymbol{E}}  
 =\mathfrak{E}_{33}^{(1)} \hat{\boldsymbol{k}} \bcdot \widehat{\delta \boldsymbol{E}} + \hat{\boldsymbol{k}} \bcdot \boldsymbol{\mathfrak{E}}^{(1)} \bcdot \widehat{\delta 
  \boldsymbol{E}}_T =\textit{O}(\tilde{\omega}_{e\|} |\widehat{\delta 
\boldsymbol{E}}|) \, , \label{consistencycond}
\end{equation}
where the transverse electric field is 
defined by $\widehat{\delta \boldsymbol{E}}_{T} \equiv \widehat{\delta \boldsymbol{E}} \bcdot \left(\mathsfbi{I} -  \hat{\boldsymbol{k}} \hat{\boldsymbol{k}} 
\right)$.
In appendix \ref{lowfrequency_electrostatic}, we show that for $\tilde{\omega}_{e\|}, \tilde{\omega}_{i\|} \ll 1$, 
\begin{equation}
\mathfrak{E}_{33}^{(1)} \approx \frac{\omega_{\mathrm{p}e}^2}{\omega^2} \frac{2 k_{\|}^2}{k^2} (1+Z\tau^{-1}) \left[1+ \textit{O}(\eta_e, \epsilon_e)\right] 
\, .
\end{equation}
Since this is strictly positive, we can rewrite (\ref{consistencycond}) to 
give the electrostatic field in terms of the transverse electric field:
\begin{equation}
  \hat{\boldsymbol{k}} \bcdot \widehat{\delta \boldsymbol{E}} = -\left(\mathfrak{E}_{33}^{(1)} \right)^{-1} \left(\hat{\boldsymbol{k}} \bcdot \boldsymbol{\mathfrak{E}}^{(1)} \bcdot \widehat{\delta 
  \boldsymbol{E}}_T \right)
  \, . \label{electrostaticcomp}
\end{equation}
We conclude that $|\hat{\boldsymbol{k}} \bcdot \widehat{\delta \boldsymbol{E}}| \sim |\widehat{\delta 
  \boldsymbol{E}}_T|$ for all low-frequency perturbations with $k_{\|} \sim k$; a corollary of this result is that there can be no low-frequency 
purely electrostatic perturbations (see appendix 
\ref{low_freq_electrostatic_stab} for an alternative demonstration of this).

We can now derive the dispersion relation from the other two components of (\ref{singulareigenvaleqnFull}),
\begin{equation}
\left[\frac{c^2 k^2}{\omega^2} \left(\hat{\boldsymbol{k}}\hat{\boldsymbol{k}}-\mathsfbi{I}\right)+  \left(\hat{\boldsymbol{k}}\hat{\boldsymbol{k}}-\mathsfbi{I}\right) \bcdot \boldsymbol{\mathfrak{E}} \right] 
\bcdot \widehat{\delta \boldsymbol{E}} = 0 \, , \label{singulareigenvaleqnFull_transverse}
\end{equation}
by (again) substituting the expanded dielectric tensor (\ref{dielectric_expand_multispecies}) into (\ref{singulareigenvaleqnFull_transverse}):
 \begin{equation}
   \left[\tilde{\omega}_{e\|} \boldsymbol{\mathfrak{E}}^{(0)} + \frac{c^2 k^2}{\omega^2} \left(\hat{\boldsymbol{k}}\hat{\boldsymbol{k}}-\mathsfbi{I}\right) \right] 
\bcdot \widehat{\delta \boldsymbol{E}}_T = - \left(\hat{\boldsymbol{k}}\hat{\boldsymbol{k}}-\mathsfbi{I}\right) \bcdot \left( \boldsymbol{\mathfrak{E}} - \tilde{\omega}_{e\|} \boldsymbol{\mathfrak{E}}^{(0)} \right) \bcdot \widehat{\delta \boldsymbol{E}} \, , \label{perpdieletricA}
\end{equation}
where we have used the identity
\begin{equation}
\boldsymbol{\mathfrak{E}}^{(0)} =  \left(\hat{\boldsymbol{k}}\hat{\boldsymbol{k}}-\mathsfbi{I}\right) \bcdot  \boldsymbol{\mathfrak{E}}^{(0)} 
\bcdot  \left(\hat{\boldsymbol{k}}\hat{\boldsymbol{k}}-\mathsfbi{I}\right) \, ,
\end{equation}
and ordered $k^2 c^2/\omega^2 \sim \tilde{\omega}_{e\|} 
\|\boldsymbol{\mathfrak{E}}^{(0)}\|$. The ratio of the right-hand side of (\ref{perpdieletricA})
to the left-hand side is $\textit{O}(\tilde{\omega}_{e\|})$; we thus conclude 
that, to leading order in the $\tilde{\omega}_{e\|} \ll 1$ expansion,
 \begin{equation}
   \left[\tilde{\omega}_{e\|} \boldsymbol{\mathfrak{E}}_e^{(0)} + \frac{c^2 k^2}{\omega^2} \left(\hat{\boldsymbol{k}}\hat{\boldsymbol{k}}-\mathsfbi{I}\right) \right] 
\bcdot \widehat{\delta \boldsymbol{E}}_T = 0 \, , \label{perpdieletric}
\end{equation}
and the dispersion relation is approximately 
\begin{equation}
\left[\tilde{\omega}_{e\|} (\boldsymbol{\mathfrak{E}}_e^{(0)})_{11}-\frac{k^2 c^2}{\omega^2}\right]\left[\tilde{\omega}_{e\|} (\boldsymbol{\mathfrak{E}}_e^{(0)})_{22}-\frac{k^2 c^2}{\omega^2}\right]+\left[\tilde{\omega}_{e\|} (\boldsymbol{\mathfrak{E}}_e^{(0)})_{12}\right]^2
 = 0 \, . \label{disprel_simp_1}
\end{equation}
Finally, writing the dielectric tensor in terms of $\mathsfbi{M}_{e}$ and 
$\mathsfbi{P}_{e}$ as defined by (\ref{Maxnonmaxsep_s}\textit{a}), we find
 \begin{eqnarray}
   \left[\tilde{\omega}_{e\|} (\mathsfbi{M}_{e}^{(0)})_{11} + (\mathsfbi{P}_{e}^{(0)})_{11}- k^2 d_e^2\right]\left[\tilde{\omega}_{e\|} (\mathsfbi{M}_{e}^{(0)})_{22} + (\mathsfbi{P}_{e}^{(0)})_{22} - k^2 d_e^2\right] 
   \nonumber \\
  + \left[\tilde{\omega}_{e\|} (\mathsfbi{M}_{e}^{(0)})_{12} + (\mathsfbi{P}_{e}^{(0)})_{12}\right]^2 = 0 \, 
  ,
  \qquad \label{disprel_simp_2}
 \end{eqnarray}
 where $d_e = c/\omega_{\mathrm{p}e}$ is the electron inertial scale [see (\ref{skin_depth_def}\textit{b})]. This can be re-written as a quadratic equation in $\omega$ -- and thus, expressions for the complex 
 frequency of any low-frequency perturbation can be found for any given positive wavenumber. 
We note that the electron inertial scale is related to the electron Larmor 
radius by $d_e = \rho_e \beta_e^{-1/2}$; therefore, our expansion scheme is only 
consistent with the low-frequency assumption (\ref{omegascale}) under our assumed ordering, 
$\tilde{\omega}_{e\|} \sim \beta_e^{-1}$, when $\beta_e \gg 1$.
 
We note that one only needs to know $\boldsymbol{\mathfrak{E}}_e^{(0)}$ in order 
to obtain the dispersion relation of low-frequency perturbations and the transverse component of the 
electric field, whereas to determine the electrostatic component of the electric 
field (and other quantities, such as the density perturbation -- see appendix \ref{CES_density_modes}), one must go to higher order in the $\tilde{\omega}_{e\|} \ll 1$ 
expansion. Since we are primarily interested in microinstability growth rates and  
wavenumber scales, we will not explicitly calculate the 
electrostatic fields associated with perturbations using 
(\ref{electrostaticcomp}), and thus can avoid the rather laborious calculation of 
$\boldsymbol{\mathfrak{E}}^{(1)}$ for CE distribution functions. 
We do, however, in appendix \ref{Maxwellresponse_low123} derive an explicit expression for 
$\boldsymbol{\mathfrak{E}}^{(1)}$ for a plasma with Maxwellian distribution 
functions for all particle species; 
this in turn allows us to relate the electrostatic 
electric field to the transverse field
for such a plasma (see appendix \ref{electrostaticcompcalc}).  

For the sake of completeness, we also observe that if the non-Maxwellian part 
of the CE distribution function is even with respect to 
$v_{\|}$, the transformation rules (\ref{conductnegwav_even}) combined with (\ref{dielectric123}) imply 
that a perturbation with a negative parallel wavenumber $k_{\|}$  
will obey exactly the same dispersion relation as a perturbation for a
positive parallel wavenumber, viz., for $k_{\|} > 0$, 
\begin{equation}
 \mathsfbi{P}_e^{(0)}\!\left(-k_{\|}, k_{\bot}\right) = \mathsfbi{P}_e^{(0)}\!\left(k_{\|}, k_{\bot} \right) \, . 
\end{equation} 
If instead the non-Maxwellian part 
is odd, then, for $k_{\|} > 0$, 
\begin{equation}
 \mathsfbi{P}_e^{(0)}\!\left(-k_{\|}, k_{\bot}\right) = -\mathsfbi{P}_e^{(0)}\!\left(k_{\|}, k_{\bot} 
\right) \, .
\end{equation}
The dispersion relation for perturbations with $k_{\|} < 0$ can, therefore, be recovered by considering  
perturbations with $k_{\|} > 0$, but under the substitution $\mathsfbi{P}_{e}^{(0)} 
\rightarrow -\mathsfbi{P}_{e}^{(0)}$. Thus, we can 
characterise all unstable perturbations under the assumption that $k_{\|} > 0$. 

In all subsequent calculations, we require the Maxwellian part
$\mathsfbi{M}_e^{(0)}$ of the dielectric tensor. The elements of the matrix $\mathsfbi{M}_s^{(0)}$ of species $s$ are as follows:
\begin{subeqnarray}
  (\mathsfbi{M}_{s}^{(0)})_{11} & = & \mathrm{i} \frac{k^2}{k_{\|}^2}  F\!\left(k_{\|} \tilde{\rho}_s,k_{\bot} \tilde{\rho}_s\right) \, , \\
  (\mathsfbi{M}_{s}^{(0)})_{12} & = & -\mathrm{i} \frac{k}{k_{\|}} G\!\left(k_{\|} \tilde{\rho}_s,k_{\bot} \tilde{\rho}_s\right) \, , \\
  (\mathsfbi{M}_{s}^{(0)})_{21} & = & \mathrm{i} \frac{k}{k_{\|}} G\!\left(k_{\|} \tilde{\rho}_s,k_{\bot} \tilde{\rho}_s\right) \, , \\  
  (\mathsfbi{M}_{s}^{(0)})_{22} & = & \mathrm{i} H\!\left(k_{\|} \tilde{\rho}_s,k_{\bot} \tilde{\rho}_s\right) \, , \label{Mmaxcomp}
\end{subeqnarray}
where the functions $F\!\left(x,y\right)$, $G\!\left(x,y\right)$ and $H\!\left(x,y\right)$ are 
\begin{subeqnarray}
  F\!\left(x,y\right) & \equiv & \frac{4 \sqrt{\upi}}{y^2} \exp{\left(-\frac{y^2}{2}\right)} \sum_{m=1}^{\infty} m^2 I_m\!\left(\frac{y^2}{2}\right)\exp{\left(-\frac{m^2}{x^2}\right)} \, , \\
  G\!\left(x,y\right) & \equiv & \exp{\left(-\frac{y^2}{2}\right)} \sum_{m = -\infty}^{\infty} m \, \Real{\; Z\!\left(\frac{m}{x}\right)} \left[ I_m'\!\left(\frac{y^2}{2}\right)- I_m\!\left(\frac{y^2}{2}\right)\right] \, , \\
  H\!\left(x,y\right) & \equiv &  F\!\left(x,y\right) + \sqrt{\upi} y^2 \exp{\left(-\frac{y^2}{2}\right)} \sum_{m = -\infty}^{\infty} \left[ I_m\!\left(\frac{y^2}{2}\right)- I_m'\!\left(\frac{y^2}{2}\right)\right] \exp{\left(-\frac{m^2}{x^2}\right)} \, ,  
  \qquad \qquad \label{specialfuncMax}
\end{subeqnarray}
$I_m\!\left(\alpha\right)$ is the $m$-th modified Bessel function, and 
\begin{equation}
  Z\!\left(z\right) = \frac{1}{\sqrt{\upi}} \int_{C_L} \frac{\mathrm{d}u \exp{\left(-u^2\right)}}{u-z} \label{Plasma_disp_func_def}
\end{equation}
is the plasma dispersion function ($C_L$ is the Landau contour)~\citep{F61}.  The derivation of 
these results from the full dielectric tensor (which is calculated in appendix \ref{Maxwellresponse_gen}) for a plasma whose constituent particles all have Maxwellian distributions is presented in 
Appendices \ref{Maxwellresponse_lowxyz}  (expansion in the $\{\hat{\boldsymbol{x}},\hat{\boldsymbol{y}},\hat{\boldsymbol{z}}\}$ basis) and \ref{Maxwellresponse_low123} (expansion in the $\{\boldsymbol{e}_1,\boldsymbol{e}_2,\boldsymbol{e}_3 
\}$ basis). 

\subsubsection{Effect of multiple species on dispersion-relation derivations} \label{shortcomings_twospecies}
 
We now relax the assumptions adopted in section \ref{consequences} that the low-frequency modes of interest are on electron Larmor scales, and discuss how we derive simplified dispersion relations for (low-frequency) CE microinstabilities more generally. 

First, it is unnecessarily restrictive to assume that, for all CE microinstabilities, $\tilde{\omega}_{s\|} \ll 1$
for all particle species. There are some instabilities for which $\tilde{\omega}_{e\|} \sim \eta_e \sim \epsilon_e \ll 1$  
while $\tilde{\omega}_{i\|} \gtrsim 1$. Recalling the orderings $\tilde{\omega}_{e\|} \sim \beta_e^{-1}$ and $k \rho_e \sim 1$ that were adopted for the electron-Larmor-scale instabilities described in section \ref{consequences}, it follows that $\tilde{\omega}_{i\|} \gtrsim 1$ whenever $\beta_e \lesssim \tau^{-1/2} \mu_e^{-1/2}$; in other words, electron-Larmor-scale CE microinstabilities in plasmas with $\beta_e$ that is not too large will satisfy $\tilde{\omega}_{i\|} \gtrsim 1$. Therefore, we cannot naively apply our low-frequency approximation to both $\boldsymbol{\mathfrak{E}}_e$ and $\boldsymbol{\mathfrak{E}}_i$ 
in all cases of interest. We will remain cognisant of this in the calculations 
that follow -- a concrete example of $\tilde{\omega}_{i\|} \gtrsim 1$ will 
be considered in section \ref{electron_heatflux_instab_whistl}.

Secondly, because of the large separation between electron and ion Larmor scales, it is 
necessary to consider whether the approximation $\mathsfbi{M}_s\!\left(\tilde{\omega}_{s\|},\boldsymbol{k}\right) \approx \tilde{\omega}_{s\|} \mathsfbi{M}_s^{(0)}\!\left(\boldsymbol{k}\right)$
remains valid for parallel or perpendicular wavenumbers much larger or smaller than the inverse Larmor radii of 
each species. We show in appendix \ref{ordering} that the leading-order term in the $\tilde{\omega}_{s\|} \ll 1$ expansion remains larger than higher-order terms for all $k_{\|} \rho_s \gtrsim 1$ (as, indeed, was implicitly assumed in section \ref{consequences}). 
However, for $k_{\|} \rho_s$ sufficiently small, the same statement does not hold for all components of 
$\mathsfbi{M}_s$. More specifically, it is shown in the same appendix that the dominant contribution to $\mathsfbi{M}_s\!\left(\boldsymbol{k}\right)$
when $k_{\|} \rho_s \ll 1$ instead comes from the quadratic term $\tilde{\omega}_{s\|}^2 \mathsfbi{M}_s^{(1)}\!\left(\boldsymbol{k}\right)$ (rather than any higher-order 
term). Thus, in general, our simplified dispersion relation for low-frequency modes in a two-species plasma has the form of 
a quartic in $\omega$, rather than a quadratic, if $k_{\|} \rho_s \ll 1$ for at least 
the electron species. Physically, the reason why a
quadratic dispersion relation is no longer a reasonable approximation
is the existence of more than two low-frequency modes in a two-species Maxwellian plasma in certain
wavenumber regimes. 
For example, for quasi-parallel modes with characteristic parallel wavenumbers satisfying $k_{\|} \rho_i \ll 1$, 
there are four low-frequency modes (see section \ref{negpres_fire}).
Nevertheless, in other situations, the components of $\mathsfbi{M}_s\!\left(\boldsymbol{k}\right)$ 
for which the $\mathsfbi{M}_s\!\left(\tilde{\omega}_{s\|},\boldsymbol{k}\right) \approx \tilde{\omega}_{s\|} \mathsfbi{M}_s^{(0)}\!\left(\boldsymbol{k}\right)$ 
approximation breaks down are not important, on account of their small size 
compared with terms in the dispersion relation associated with other Maxwellian 
components. In this case, the original quadratic dispersion relation is 
sufficient. An explicit wavenumber regime in which this is realised is $k_{\|} \rho_e \sim k_{\perp} \rho_e \ll 
1$ but $k \rho_i \gg 1$ -- see sections \ref{pospres_electron_oblique} and \ref{negpres_electron_oblique}. 

Taking these multiple-species effects into account, the reasons behind the decision made in section \ref{sec:CharacMicroQual} to 
consider the CES microinstabilities separately from the CET microinstabilities 
come into plain focus. First, the characteristic sizes of the CE 
electron-temperature-gradient and ion-temperature-gradient terms are comparable ($\eta_i \sim 
\eta_e$), while the CE ion-shear term is much larger than the CE electron-shear 
term: $\epsilon_i \sim \mu_e^{-1/2} \epsilon_e$. This has the consequence that the natural orderings of $\tilde{\omega}_{e\|}$ and $\tilde{\omega}_{i\|}$ with respect to other parameters are different for CES and CET 
microinstabilities. Secondly, the fact that the velocity-space anisotropy 
associated with the CE temperature-gradient terms differs from the CE shear terms 
-- which excite microinstabilities with different characteristic wavevectors -- 
means that the form of the dispersion relations of CET and CES 
microinstabilities are distinct. More specifically, the dispersion relation for CET microinstabilities at both electron and ion scales 
can always be simplified to a quadratic equation in $\omega$; in contrast, for CES microinstabilities, 
the dispersion relation cannot in general be reduced to anything simpler than a quartic. 

\subsubsection{Modelling collisional effects on CE microinstabilities} \label{shortcomings_coll}
 
  As proposed thus far, our method for characterising microinstabilities in a CE plasma does not include explicitly the effect of collisions on the 
  microinstabilities themselves. In principle, this can be worked out by introducing a collision operator into the linearised 
  Maxwell-Vlasov-Landau equation from which the hot-plasma dispersion relation (\ref{hotplasmadisprel}) 
  is derived. Indeed, if a Krook collision operator is assumed (as was done in section \ref{disprel_simps_overview} when determining 
  the precise form of the CE distribution functions of ions and electrons), the 
  resulting modification of the hot-plasma dispersion relation is quite 
  simple: the conductivity tensor (\ref{conductivity}) remains the same, but 
 with the substitution
 \begin{equation}
 \tilde{\omega}_{s\|} \rightarrow {\hat{\omega}}_{s\|} \equiv \tilde{\omega}_{s\|} + 
 \frac{\mathrm{i}}{k_{\|} \lambda_s} \, , \label{collresconduct}
 \end{equation} in the resonant denominators (see appendix \ref{HotPlasmDispDeriva}). 
  As for how this affects the simplifications to the dispersion relation outlined in section \ref{disprel_simps_II}, 
  the expansion parameter in the dielectric tensor's expansion (\ref{dielectric_expand}) 
  is altered, becoming ${\hat{\omega}}_{s\|} \ll 1$ (as opposed to $\tilde{\omega}_{s\|} \ll 
  1$); in other words, $\|\boldsymbol{\mathfrak{E}}_s^{(1)}\|/\|\boldsymbol{\mathfrak{E}}_s^{(0)}\| \sim {\hat{\omega}}_{s\|}$. 
  
  The latter result leads to an seemingly counterintuitive conclusion: collisions typically fail to 
  stabilise low-frequency instabilities in CE plasma if $\omega \tau_s \lesssim 
  1$ (where $\tau_s$ is the collision time of 
  species $s$) but $k_{\|} v_{\mathrm{th}i} \tau_s = k_{\|} \lambda_s \gg 1$. This 
  is because the simplified dispersion relation (\ref{disprel_simp_2}) 
  only involves leading-order terms in the expanded dielectric tensor. These 
  terms are independent of ${\hat{\omega}}_{s\|}$, and thus 
  the growth rate of any microinstability that is adequately described by (\ref{disprel_simp_2}) 
  does not depend on the size of $\omega \tau_s$.  
 For these microinstabilities, the effect of collisions only becomes relevant if 
    \begin{equation}
   k_{\|} \lambda_s \lesssim 1 \, . \label{colldamp}
  \end{equation}
  This is inconsistent with the assumptions $k \lambda_e \gg 1$, $k \lambda_i \gg 1$ 
  made when setting up our calculation in section \ref{HotPlasmaDispDis}. 
  Thus, the only regime where collisions can reasonably be included in our calculation is one
  where they are typically not important. An exception to this rule arises when 
  two-species plasma effects mean that the first-order terms in the ${\hat{\omega}}_{s\|} \ll 1$ 
  expansion are needed for a correct characterisation of the growth rate of certain microinstabilities (see section \ref{shortcomings_twospecies}); for 
  these instabilities, we include the effect of collisions using 
  (\ref{collresconduct}). 
  
  Although our calculation is not formally valid when (\ref{colldamp}) holds,  
  so we cannot show explicitly that growth ceases, this condition nonetheless represents a sensible criterion 
  for suppression of microinstabilities by collisional damping. Physically, it 
  signifies that collisions are strong enough to scatter a particle 
  before it has streamed across a typical wavelength of fluctuation excited by a microinstability.
  This collisional scattering prevents particles from being resonant, which in turn would suppress the growth of many different microinstabilities. 
  However, we acknowledge that there exist microinstabilities that do not involve resonant-particle populations 
  (e.g., the firehose instability -- see sections \ref{CEexam_firehose} and \ref{negpres_fire}), 
  and thus it cannot be rigorously concluded from our work that all microinstabilities are suppressed when (\ref{colldamp}) 
  applies. 
  
  Yet even without an actual proof of collisional stabilisation, there is another 
  reason implying that (\ref{colldamp}) is a reasonable threshold for 
  microinstabilities: the characteristic growth time of microinstabilities at 
  wavenumbers satisfying (\ref{colldamp}) is comparable the evolution 
  time $\tau_L$ of macroscopic motions in the plasma. To illustrate this idea, we 
  consider the ordering (\ref{omegascale}) relating the complex frequency of 
  microinstabilities to the small parameter $\epsilon_s$ for CES (CE shear-driven) microinstabilities, and use it to 
  estimate
  \begin{equation}
  \omega \tau_L \sim \epsilon_s k_{\|} v_{\mathrm{th}s} \tau_L \lesssim \epsilon_s \frac{L_V}{\lambda_s} 
  \frac{v_{\mathrm{th}s}}{V} , \label{growthtime}
  \end{equation}
  where $V \sim L_V/\tau_L$ is the characteristic ion bulk-flow velocity. Considering 
  orderings~(\ref{brag_multiscale}), it follows that $\epsilon_e \sim \mu_e^{1/2} 
  \epsilon_i$, and so
  \begin{equation}
   \epsilon_i \frac{v_{\mathrm{th}i}}{V} \sim \epsilon_e \frac{v_{\mathrm{th}e}}{V} 
   \sim \frac{\lambda_e}{L_V} \sim \frac{\lambda_i}{L_V} \, .
  \end{equation}
Then (\ref{growthtime}) becomes
  \begin{equation}
   \omega \tau_L \lesssim  1 , 
  \end{equation}
  implying (as claimed) that the CES microinstability growth rate is smaller than the fluid turnover rate $\tau_L^{-1}$. Spelled out clearly, this means that the underlying quasiequilibrium state changes before going unstable. 
  Similar arguments can be applied to CET (CE temperature-gradient-driven) microinstabilities. 
  
  Thus, (\ref{colldamp}) represents a lower bound on the 
  characteristic wavenumbers at which microinstabilities can operate. We shall 
  therefore assume throughout the rest of this paper that microinstabilities 
  are suppressed (or rendered irrelevant) if they satisfy (\ref{colldamp}). 
  
\subsubsection{Caveats: microinstabilities in CE plasma where $\omega/k_{\|} v_{\mathrm{th}s} \not \sim \eta_s, \epsilon_s$} 
\label{shortcomings_othermicro}

As mentioned in section \ref{disprel_simps_overview}, there are a number of important caveats to the claim that 
the ordering (\ref{omegascale}) must be satisfied by microinstabilities in a CE plasma. 

The first of these is that our 
comparison of non-Maxwellian with the Maxwellian terms in expression (\ref{Xi_CE}) for $\Xi_s$ is in 
essence a pointwise comparison at characteristic values of $\tilde{v}_s$ for which $\Xi_s$ attains its largest typical magnitude. However, 
$\Xi_s$ affects the components of the conductivity tensor via 
the velocity integral of its product with a complicated function of frequency and wavenumber [see (\ref{conductivity})]. 
Thus, it does not necessarily follow that
the ratio of the integrated responses of the Maxwellian and non-Maxwellian 
contributions to the conductivity tensor is the same as the pointwise ratio of the respective contributions to 
$\Xi_s$. In some circumstances, this can result in the Maxwellian part 
being smaller than anticipated, leading to faster microinstabilities. 
An example of this phenomenon was given in section \ref{shortcomings_twospecies}: for $k_{\|} \rho_s \ll 1$, 
the characteristic magnitude of 
the Maxwellian contribution to some components of the dielectric tensor is 
$\textit{O}(\tilde{\omega}_{s\|}^2)$, as compared with the naive estimate 
$\textit{O}(\tilde{\omega}_{s\|})$. This leads to certain CES microinstabilities (for example, the CE ion-shear-driven firehose instability -- section \ref{negpres_fire}) 
satisfying a modified low-frequency condition
\begin{equation}
    \tilde{\omega}_{s\|} \sim \epsilon_s^{1/2} \ll 1 . \label{omegascale_v2}
\end{equation}
A similar phenomenon affects the limit $k_{\|} \rightarrow 0$ for fixed $k_{\bot}$, 
in which case it can be shown that the Maxwellian contribution to $\sigma_{zz}$ is 
$\textit{O}(k_{\|}/k_{\bot})$; this leads to
a CES microinstability (the CE electron-shear-driven ordinary-mode instability -- see section \ref{negpres_subelectron_ord}) satisfying a modified ordering 
\begin{equation}
   \frac{\omega}{k_{\bot} v_{\mathrm{th}s}}\sim \epsilon_s \ll 1 . \label{omegascale_v3}
\end{equation}   

The second caveat is that for some plasma modes, the particles predominantly responsible for 
collisionless damping or growth are suprathermal, i.e., $\tilde{v}_s \gg 1$. 
Then the previous comparison of terms in (\ref{Xi_CE}) is not applicable. Modes 
of this sort are the quasi-cold plasma modes discussed in section 
\ref{sec:CharacMicroQual} and appendix \ref{Electrostatic}. They can be unstable, but always with a growth rate 
that is exponentially small in $\eta_s$ and $ \epsilon_s$. 

In spite of these two caveats, we proceed by considering the full hot-plasma dispersion relation (\ref{hotplasmadisprel}) 
in the low-frequency limit $\omega \ll k_{\|} v_{\mathrm{th}s}$. This approach enables the treatment of all microinstabilities 
satisfying condition
\begin{equation}
    \tilde{\omega}_{s\|} \sim \eta_s^{\iota_\eta} , \epsilon_s^{\iota_\epsilon} \ll 1 , \label{omegascale_new}
\end{equation}
where ${\iota_\eta}$ and ${\iota_\epsilon}$ are any fractional powers. Similarly to the discussion in section 
\ref{sec:CharacMicroQual}, we claim that the microinstabilities satisfying the low-frequency condition (\ref{omegascale_new}) 
are likely to be the most rapid of all possible microinstabilities in CE plasma. 
A formal justification of this claim relies on the argument -- presented in appendix \ref{arg_stab_highfreq} 
-- that for all plasma modes satisfying $\omega \gtrsim k_{\|} v_{\mathrm{th}s}$ 
and $|\Real{\; \omega}| \gg |\Imag{\; \omega}|$, the growth rate is 
exponentially small in $\eta_s$ and $\epsilon_s$. By definition, this class of 
modes includes the quasi-cold modes. In a plasma where $\epsilon_s, \eta_s \ll 
1$, the growth rates of such microinstabilities will be exponentially small, and 
thus of little significance. The only situation that we are aware of in which the low-frequency 
condition (\ref{omegascale_new}) is not appropriate is the aforementioned 
CES ordinary-mode instability; a separate treatment of it involving the 
full hot-plasma dispersion relation is provided in appendix 
\ref{derivation_ordinarymode}.

\section{CET (Chapman-Enskog, temperature-gradient-driven) microinstabilities} \label{Results}

\subsection{Form of CE distribution function} \label{heatfluxrterm}

We consider first the non-Maxwellian terms of the CE distribution function arising from temperature gradients and electron-ion drifts. 
Neglecting bulk-flow gradients [viz., setting $\epsilon_s = 0$ for both species -- see (\ref{smallparam}\textit{e},\textit{f})], the CE distribution functions (\ref{ChapEnskogFunc_s}) for the electrons and ions become
\begin{subeqnarray}
f_{e0}(\tilde{v}_{e\|},\tilde{v}_{e\bot}) & = & \frac{n_{e0}}{v_{\mathrm{th}e}^3 \upi^{3/2}} \exp \left(-\tilde{v}_{e}^2\right) \bigg\{1-\tilde{v}_{\|e} \left[\eta_e^T \left(\tilde{v}_e^2-\frac{5}{2}\right)+\eta_e^R  \right] \bigg\} , 
\qquad \\
f_{i0}(\tilde{v}_{i\|},\tilde{v}_{i\bot}) & = & \frac{n_{i0}}{v_{\mathrm{th}i}^3 \upi^{3/2}} \exp \left(-\tilde{v}_{i}^2\right) \bigg\{1-\eta_i  \tilde{v}_{\|i} \left(\tilde{v}_i^2-\frac{5}{2}\right) \bigg\} , 
\label{CEheatflux}
\end{subeqnarray}
where we have written out explicitly the electron-temperature-gradient [$\eta_e^T$, $\eta_i$ -- see (\ref{smallparam}\textit{a},\textit{d})] and electron-friction [$\eta_e^R$ -- see (\ref{smallparam}\textit{b})] 
terms under the assumption that the Maxwell-Vlasov-Landau system from which these CE distribution functions were derived is governed 
by a Krook collision operator.  We remind the reader that the electron-ion-drift term 
[$\eta_e^u$ -- see (\ref{smallparam}\textit{c})]
disappears for this choice of collision operator. We also observe that the 
non-Maxwellian part of the distribution functions (\ref{CEheatflux}) have 
odd parity; thus, any unstable mode with $k_{\|} > 0$ has a corresponding 
unstable mode with $k_{\|} < 0$ and the signs of $\eta_e^T$, $\eta_e^R$, and $\eta_i$ 
reversed (see section \ref{consequences}, last paragraph). 

The precise methodology that we
employ to calculate the growth rates of CET microinstabilities is described in 
appendix~\ref{CET_method}; here, we focus on the results of those calculations. 
In section \ref{heatflux_stab}, we will present the overview of the CET 
stability landscape, while the microinstabilities referred to there will be 
treated analytically in section \ref{CETclass}. 

\subsection{Stability} \label{heatflux_stab}

We determine the stability (or otherwise) of the CE distribution functions of the form (\ref{CEheatflux}\textit{a}) and (\ref{CEheatflux}\textit{b}) 
for different values of $\eta_e^T$, $\eta_e^R$, and $\eta_i$, the electron inertial scale 
$d_e$, the electron-temperature scale length $L_T = |\nabla_{\|} \log{T_e}|^{-1}$, and for fixed electron and ion plasma betas ($\beta_e$ and $\beta_i$, respectively).  
Stability calculations are carried out for particular combinations of values of $\eta_e^T$, $\eta_e^R$, $\eta_i$, $d_e$, $L_T$, $\beta_e$ and $\beta_i$ by solving for the maximum 
microinstability growth rate across all wavevectors (see appendix \ref{CET_method} for explanation of how this is done), 
and determining whether this growth rate is positive for the microinstabilities whose wavelength is smaller
than the Coulomb mean free paths (a condition necessary for our calculation to be valid).

\begin{figure}
\centerline{\includegraphics[width=0.99\textwidth]{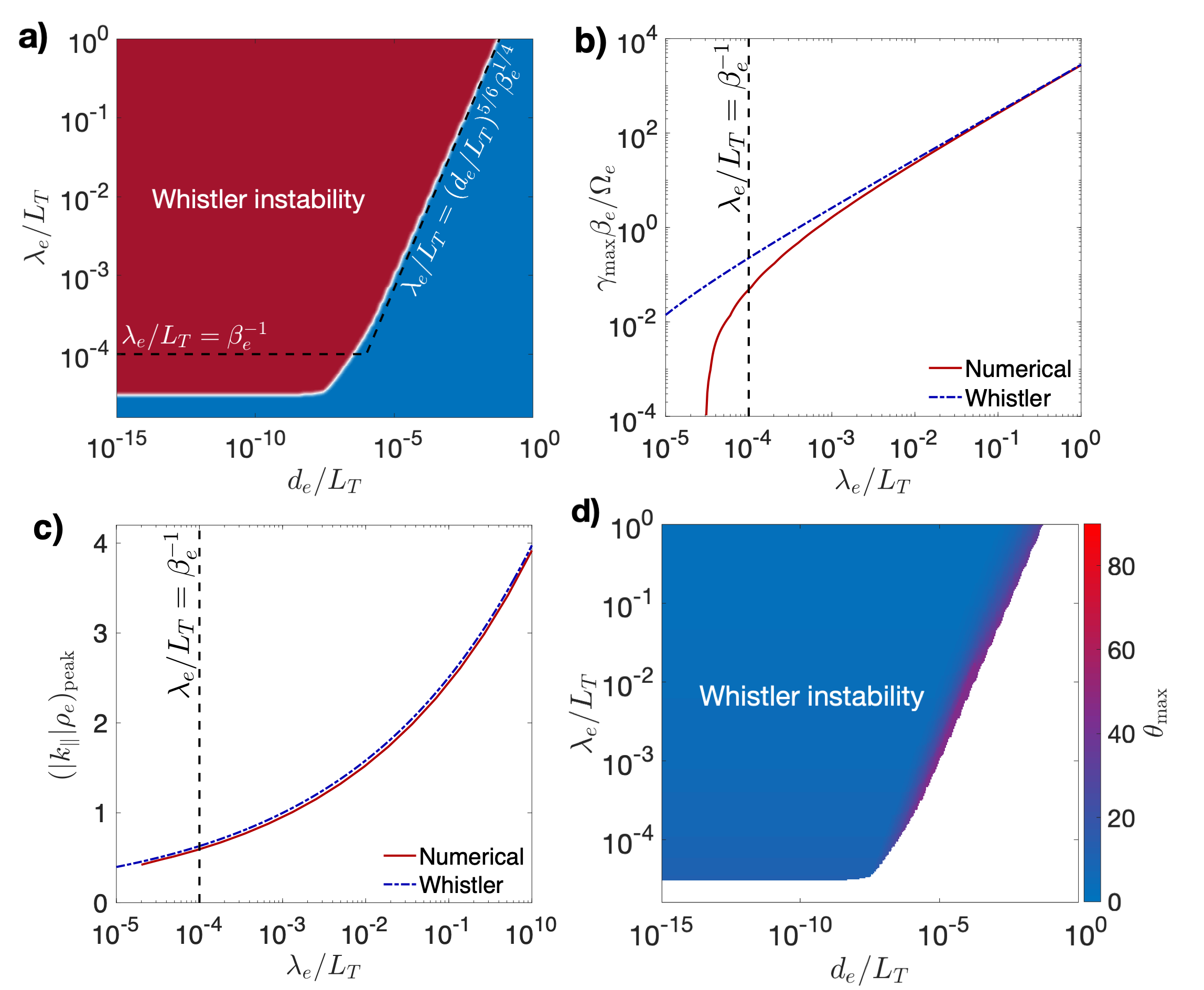}}
\caption{\textit{CE-distribution-function stability map for CET microinstabilities}. Exploration of the stability of the CE distribution functions
 (\ref{CEheatflux}\textit{a}) and (\ref{CEheatflux}\textit{b}) for different values of small parameters $\eta_e^T$, $\eta_e^R$, and $\eta_i$, and the ratio of the electron inertial scale $d_e$ to the temperature scale length $L_T$, in a temperature-equilibrated hydrogen plasma. In this 
 plot, we chose $\eta_e^R = 0$ and $\eta_e^T = \eta_i $, and then show $\lambda_e/L_T = |\eta_e^T|$ with equal logarithmic spacing in the range $\left[10^{-5},10^{0}\right]$; $d_e/L_T$ is chosen with equal logarithmic spacing in the range 
 $\left[10^{-15},10^{0}\right]$. The total size of the grid is $400^2$. For reasons of efficiency, we calculate growth rates on a $40^2$ grid in wavenumber space with
 logarithmic spacing for both parallel and perpendicular wavenumbers. In this plot, $\beta_e = \beta_i = 10^4$. \textbf{a)} Stable (blue) and unstable (red) regions of $\left(d_e/L_T,\lambda_e/L_T\right)$ phase space. The theoretically anticipated collisional cutoff [right -- see (\ref{colldampB})] and $\beta$-stabilisation threshold (bottom) 
 of the CET whistler instability are shown as black dashed lines. \textbf{b)} Maximum normalised microinstability growth rate (red) versus $\lambda_e/L_T$ for a fixed electron inertial scale $d_e/L_T = 10^{-15}$, 
 along with analytically predicted maximum growth rate for the CET whistler instability in the limit $\lambda_e \beta_e/L_T  \gg 1$ [blue, see (\ref{maxgrowthrate_heatflux})]. 
 \textbf{c)} Parallel wavenumber of fastest-growing microinstability (red) versus $\lambda_e/L_T$ for a fixed electron inertial scale $d_e/L_T = 10^{-15}$, 
 along with the same quantity analytically predicted for the CET whistler instability in the limit $\lambda_e \beta_e/L_T \gg 1$ [blue, see (\ref{maxgrowthrate_heatflux_wavenumber})]
  \textbf{d)} Wavevector angle $\theta \equiv \tan^{-1}{(k_{\perp}/k_{\|})}$ of the fastest-growing instability over $\left(d_e/L_T,\lambda_e/L_T\right)$ parameter space. \label{Figure3}}
\end{figure}

The results of one such stability
calculation -- for a temperature-equilibrated hydrogen plasma ($\eta_e^T = \eta_i$, $\beta_i = \beta_e$) -- are presented in figure \ref{Figure3}. 
In spite of the five-dimensional ($\eta_e^T,\eta_e^R,d_e,L_T,\beta_e$) parameter space that seemingly needs to be explored, 
we can, in fact, convey the most salient information concerning the stability of the 
CE distribution functions (\ref{CEheatflux}) using plots over a two-dimensional $\left(d_e/L_T,\lambda_e/L_T\right)$ 
parameter space at a fixed $\beta_e$ [where we remind the reader that $\lambda_e/L_T = |\eta_e^T|$ -- see (\ref{smallparam}\textit{a})]. This reduction in phase-space dimensionality is 
possible for two reasons. First, it transpires that the CE electron-friction 
term of the form given in (\ref{CEheatflux}\textit{a}) does not drive any 
microinstabilities, bur merely modifies the real frequency of 
perturbations with respect to their Maxwellian frequencies (this is proven in appendix \ref{heatfluxrterm_di_res}). Thus, 
we can set $\eta_e^R = 0$ without qualitatively altering the stability properties of the CE distribution functions  ~(\ref{CEheatflux}). Secondly, none of the salient stability thresholds applying to
CET microinstabilities depends on $d_e$ and $L_T$ separately: 
one is a function of $d_e/L_T$, while another is independent of both 
quantities. 

Figure \ref{Figure3}a shows the regions of instability and stability of the CE 
distribution function~(\ref{CEheatflux}) over the $\left(d_e/L_T,\lambda_e/L_T\right)$ parameter space. 
The unstable region is bracketed by two thresholds. For $d_e/L_T$ below a 
critical value $(d_e/L_T)_{\rm c0}$, stability is independent of $d_e/L_T$, and  
only depends on the relative magnitude of $\lambda_e/L_T$ and $\beta_e$: 
CET microinstabilities are quenched if $\lambda_e \beta_e/L_T \ll 1$. For 
$d_e/L_T \gtrsim (d_e/L_T)_{\rm c0}$, and $\lambda_e \beta_e/L_T \gtrsim 1$, 
stability is attained at fixed $\lambda_e/L_T$ for $d_e/L_T > (d_e/L_T)_{\rm c}$, where $(d_e/L_T)_{\rm c}$ increases monotonically 
with $\lambda_e/L_T$. If $\lambda_e \beta_e/L_T \gtrsim 1$ and $d_e/L_T \lesssim 
(d_e/L_T)_{\rm c}$, then the CE 
distribution function~(\ref{CEheatflux}) is unstable. 

The fastest-growing CET microinstability 
is the \emph{whistler (heat-flux) instability}: whistler waves driven unstable by the small anisotropy of the CE electron-temperature-gradient term (see section 
\ref{electron_heatflux_instab_whistl}). 
That this instability with wavevector parallel to the magnetic field is indeed the 
dominant microinstability is most easily ascertained by comparing simple analytic 
expressions for its peak growth rate and wavevector to the equivalent quantities 
recorded when performing the general stability calculation (see figures
\ref{Figure3}b, \ref{Figure3}c and \ref{Figure3}d). The maximum microinstability growth rate matches 
the analytic result (\ref{maxgrowthrate_heatflux}) for the CET whistler instability in the limit $\lambda_e \beta_e/L_T
\gg 1$, while the parallel wavenumber $(|k_{\|}| \rho_e)_{\rm peak}$ of the fastest-growing mode
is extremely well described by (\ref{maxgrowthrate_heatflux_wavenumber}).
In addition, figure \ref{Figure3}d demonstrates that the parallel instability is indeed the fastest. The CET whistler 
instability has been considered previously by a number of authors (see references in section \ref{electron_heatflux_instab_whistl}); we note that these prior studies of this instability suggest that, nonlinearly, oblique CET whistler modes may be the more important ones, even though linearly the parallel modes are the fastest growing (see section \ref{electron_heatflux_instab_whistl_obl}). 

The two thresholds demarcating the unstable region can then be associated with stabilisation conditions of the CET whistler instability, 
each with a simple physical interpretation. The first condition is 
the $\beta$-stabilisation condition of the whistler instability. It is shown in 
section \ref{electron_heatflux_instab_whistl} that when $\lambda_e \beta_e/L_T \ll 1$, 
cyclotron damping on whistler modes is sufficiently strong that only quasi-parallel modes with 
parallel wavenumbers $k_{\|} \rho_e \lesssim (\lambda_e \beta_e/L_T)^{1/3} \ll 1$ can be 
destabilised by the anisotropy of the CE distribution function, and that the peak growth 
rate $\gamma_{\rm whistler,T}$ of these unstable modes is exponentially small in $\lambda_e \beta_e/L_T$
compared to the electron Larmor frequency [see (\ref{whistlermode_smallk})]: 
$\gamma_{\rm whistler,T}/\Omega_e \sim \lambda_e \exp{[-(\lambda_e \beta_e/2 L_T)^{-2/3}]}/L_T $. This means that 
if $\lambda_e \beta_e/L_T$ is reduced below unity, the growth rate of the CET whistler instability decreases 
dramatically, and thus the instability is unable to operate effectively on timescales shorter than those over which the CE
plasma is evolving macroscopically. 

The second condition is collisional stabilisation of the CET whistler 
instability. Naively, it might be expected that two conditions must be satisfied 
in order for the microinstability to operate: that its growth rate must satisfy $\gamma_{\rm whistler,T} \tau_{e} \gg 
1$, and its characteristic wavenumber $k \lambda_{e} \gg 1$ [see (\ref{colldamp})]. Noting that for the 
CET whistler instability [cf. (\ref{maxgrowthrate_heatflux})], 
\begin{equation}
  \frac{\gamma_{\rm whistler,T} \tau_{e}}{k \lambda_{e}} = \frac{\gamma_{\rm whistler,T}}{k v_{\mathrm{th}e}} \sim 
  \frac{\lambda_e}{L_T} \left(  \frac{\lambda_e \beta_e}{L_T} \right)^{-1/5} \ll 1 \, , 
\end{equation}
it follows that the former condition is more restrictive. Written as a condition 
on $d_e/L_T$ in terms of $\lambda_e/L_T$ [and using $\gamma_{\rm whistler,T} \sim \lambda_e \Omega_e/L_T$ -- see (\ref{maxgrowthrate_heatflux})], $\gamma_{\rm whistler,T} \tau_{e} \gg 
1$ becomes
\begin{equation}
\frac{d_e}{L_T} \ll \beta_e^{-5/2} \left(  \frac{\lambda_e \beta_e}{L_T} \right)^2 \, ,
\end{equation}
while the condition $k \lambda_{e} \gg 1$ on the instability wavenumber $k_{\|} \rho_e \sim (\lambda_e \beta_e/L_T)^{1/5}$ [see (\ref{maxgrowthrate_heatflux_wavenumber})] leads~to
\begin{equation}
\frac{d_e}{L_T} \ll \left(\frac{d_e}{L_T}\right)_{\rm c} \equiv \beta_e^{-3/2} \left(\frac{\lambda_e \beta_e}{L_T}\right)^{6/5} \, . \label{colldampB}
\end{equation}
It is the latter that agrees well with the true result, as shown in figure 
\ref{Figure3}a, implying that $(d_e/L_T)_{\rm c0} = \beta_e^{-3/2}$. The (arguably surprising) result that the CET whistler instability can operate even if 
$\gamma_{\rm whistler,T} \tau_{e} \lesssim 1$ is, in fact, a generic feature of 
low-frequency (viz., $\omega \ll k v_{\mathrm{th}e}$) plasma instabilities (see section \ref{shortcomings_coll}).  
The physical instability mechanism underlying such modes can be sustained provided the time taken for thermal 
particles (in this case, electrons) to cross the mode's wavelength is much shorter 
than the collision time, irrespective of the mode's own frequency -- in other words, $\tau_e k v_{\mathrm{th}e} = k \lambda_e \gg 1$. 
We point out that the collisional-stabilisation condition of the CET whistler 
instability can \emph{never} be satisfied in a strongly magnetised plasma if $\lambda_e \beta_e/L_T \gtrsim 1$: this is 
because its wavenumber $k$ satisfies $k^{-1} \lesssim \rho_e \ll \lambda_e$. 

Whilst it is the fastest-growing one (assuming $\eta_e^T \sim \eta_i$), the CET whistler instability is not the only 
CET microinstability of interest. There are two other instabilities driven by the CET ion-temperature gradient 
term, neither of which has previously been identified, to our knowledge: the \emph{slow (hydromagnetic) wave 
instability} (see section \ref{ion_heatflux_instab_slowwave}), and the \emph{long-wavelength kinetic-Alfv\'en wave instability} (see section \ref{CET_KAW_instab}). 
The former, whose characteristic wavenumber scale satisfies $k \rho_i \sim 1$, has a larger characteristic growth rate $\gamma_{\rm SW} \sim \lambda_i \Omega_i/L_{T_i}$ 
(where $L_{T_i} = |\nabla_{\|} \log{T_i}|^{-1}$ is the 
scale length of the ion temperature gradient). 
Similarly to the CET whistler instability, the CET slow-wave
instability has $\beta$-stabilisation and collisional-stabilisation conditions $\lambda_i \beta_i/L_{T_i} \ll   
1$ and $\lambda_i \lesssim \rho_i$, respectively. Thus, unless $\lambda_i \beta_i/L_{T_i} > \lambda_e \beta_e/L_{T_e}$ (a condition equivalent to $\tau^3 L_{T_e}/L_{T_i} > Z^3$, where $\tau = T_i/T_e$), the CET 
slow-wave instability only operates when the CET whistler wave instability does, but on larger, ion rather than electron, scales. 
Nevertheless, the CET slow-wave instability is worth noting because,
on account of being an ion instability, it should
continue to operate even if the electron-scale CET whistler instability modifies 
the underlying electron distribution function. The slow-wave instability will then be responsible for modifying the ion distribution function. We are not aware of any work on the CET slow-wave instability and, thus, on its effect on ion heat conduction.  

Readers who are 
interested in knowing more about the properties and growth rates of CET microinstabilities are 
encouraged to continue section \ref{CETclass}; those who are focused on the wider 
question of the kinetic stability of the CE distribution function should jump 
ahead to section \ref{Results_shearingterm}. 

\subsection{CET microinstability classification} \label{CETclass}

\subsubsection{Parallel whistler (heat-flux) instability}  \label{electron_heatflux_instab_whistl}

The CET whistler instability, which has been studied previously by a
number of authors~\citep{LE92,PE98,GL00,RDRS16,KSCS17,RDRS18,R18b,SLYP19,KVC19,DPRR20}, %
is driven by parallel electron heat fluxes. These heat fluxes introduce the asymmetry to the CE electron 
distribution function (i.e., the electron-temperature-gradient term), which, 
if it is sufficiently large, can overcome electron cyclotron damping of (electromagnetic) whistler waves 
and render them unstable. The instability is mediated by gyroresonant 
wave-particle interactions that allow whistlers to drain free energy from electrons with parallel 
velocities $v_{\|} = \pm \Omega_e/k_{\|}$. For a positive, parallel electron heat 
flux, which is driven by an anti-parallel temperature gradient ($\nabla_{\|} T_e < 
0$, so $\eta_e^T < 0$), it is only whistlers with a positive parallel wavenumber that are 
unstable. Whistler waves 
with both parallel and oblique wavevectors with respect to the magnetic field  
can be destabilised, although the parallel modes are the fastest-growing ones.  

\begin{figure}
\centerline{\includegraphics[width=0.99\textwidth]{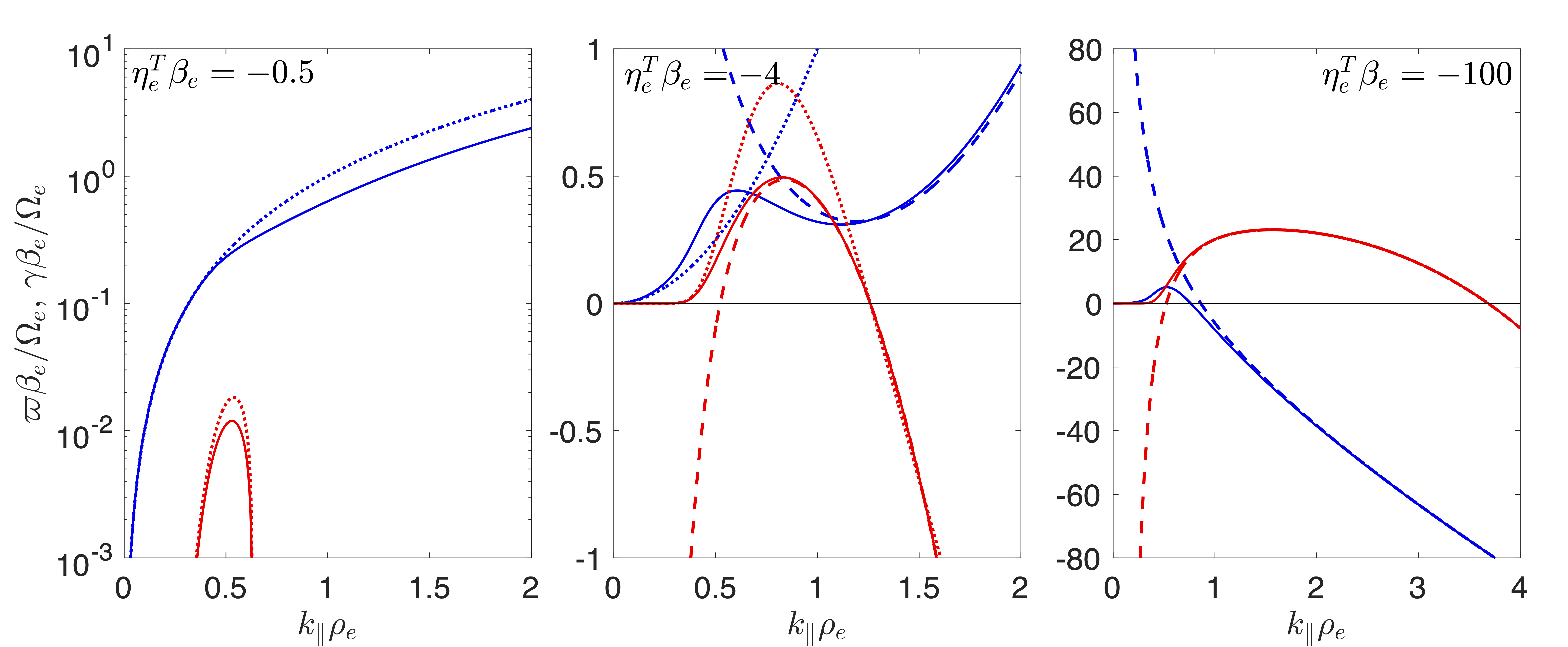}}
\caption{\textit{Parallel CET whistler instability}.  Dispersion curves of unstable whistler modes, whose instability is driven by the electron-temperature-gradient term in the CE distribution function~(\ref{CEheatflux}\textit{a}), for wavevectors that are co-parallel with the background 
magnetic field (viz., $\boldsymbol{k} = k_{\|} \hat{\boldsymbol{z}}$). 
The frequency (solid blue) and growth rates (solid red) of the modes are calculated using 
(\ref{whistlerwave_growthrate_allkpl}\textit{a}) and 
(\ref{whistlerwave_growthrate_allkpl}\textit{b}), respectively. 
The resulting frequencies and growth rates, when normalised as $\gamma \beta_e/\Omega_e$, are functions of the dimensionless quantity $\eta_e^{T} \beta_e$; we show the dispersion curves for three different values of $\eta_e^{T} \beta_e$. 
The approximations (\ref{whistlerwave}\textit{a}) and (\ref{whistlerwave}\textit{b}) for 
the frequency (dotted blue) and growth rate (dotted red) in the limit $k_{\|} \rho_e \ll 1$ are also plotted, 
as are the approximations (\ref{whistlerwave_growthrate_lgekpl}\textit{a}) and (\ref{whistlerwave_growthrate_lgekpl}\textit{b}) 
for the frequency (dashed blue) and growth rate (dashed red) in the limit $k_{\|} \rho_e \gg 
1$. 
 \label{Figure_newCETwhist}}
\end{figure}

The CET whistler instability is most simply characterised analytically for parallel wavenumbers (i.e., $k = k_{\|}$). Then, it can be shown [see appendix \ref{derivation_parwhistlerheatflux},  and also~\citet{LE92} and~\citet{RDRS16}]
that the real frequency $\varpi$ and growth rate $\gamma$ at arbitrary $k_{\|} > 0$ 
are given by
\begin{subeqnarray}
  \frac{\varpi}{\Omega_e} & = &  \eta_e^T \left(\frac{k_\| \rho_e}{4} - \frac{1}{2 k_\| \rho_e}\right) 
- \frac{\left(\eta_e^T/2 + k_\|^3 \rho_e^3/\beta_e\right) \Real{\; Z\!\left({1}/{k_\| \rho_e}\right)} }{\left[\Real{\; Z\!\left({1}/{k_\| \rho_e}\right)}\right]^2 + \upi \exp{\left(-{2}/{k_\|^2 \rho_e^2}\right)}} 
\, ,  \\
\frac{\gamma}{\Omega_e} & = & 
-\frac{\sqrt{\upi}\left(\eta_e^T/2 + k_\|^3 \rho_e^3/\beta_e\right) }{\left[\Real{\; Z\!\left({1}/{k_\| \rho_e}\right)}\right]^2 \exp{\left({1}/{k_\|^2 \rho_e^2}\right)}+ \upi \exp{\left(-{1}/{k_\|^2 \rho_e^2}\right)}} 
\, . \label{whistlerwave_growthrate_allkpl}
\end{subeqnarray}
For $\eta_e^T > 0$, $\gamma < 0$, but if $\eta_e^T < 0$, then $\gamma$ is  
non-negative for $k_{\|} \rho_e \leq \left(\eta_e^T \beta_e/2\right)^{1/3}$. 
The dispersion curves $\varpi = \varpi(k_{\|})$ and $\gamma = \gamma(k_\|)$ of 
unstable whistler waves with parallel wavevectors for three different values of $|\eta_e^T| 
\beta_e$ are plotted in figure~\ref{Figure_newCETwhist} using the above formulae.
For $|\eta_e^T| \beta_e \gtrsim 1$, the range of unstable parallel wavenumbers, $\Delta k_{\|}$, is comparable to  
the characteristic wavenumber of the instability: $\Delta k_{\|} \sim k_{\|} \sim 
\rho_e^{-1}$. 

The expressions (\ref{whistlerwave_growthrate_allkpl}\textit{a}) and (\ref{whistlerwave_growthrate_allkpl}\textit{b}) can be simplified in two subsidiary limits, which in turn allows for the derivation of analytic expressions for the maximum 
growth rate of the instability and the (parallel) wavenumber at which that growth rate is realised. 

First, adopting the ordering $k_{\|} \rho_e \sim \left(\eta_e^T 
\beta_e\right)^{1/3} \ll 1$ under which the destabilising $\eta_e^T$ terms and the stabilising electron FLR terms are the same order, we find 
\begin{subeqnarray}
\varpi& \approx & \frac{k_\|^2 \rho_e^2}{\beta_e} \Omega_e \, , \\
\gamma & \approx & 
-\frac{\sqrt{\upi}}{k_\|^2 \rho_e^2} \left( \frac{\eta_e^T}{2} + \frac{k_\|^3 \rho_e^3}{\beta_e}\right) 
\exp{\left(-\frac{1}{k_\|^2 \rho_e^2}\right)} \Omega_e
\, . \label{whistlerwave}
\end{subeqnarray}
The frequency corresponds to that of a whistler wave in the $k_{\|} \rho_e \ll 1$ limit~\citep{BHXP13}. The fastest growth, which occurs at the wavenumber
\begin{equation}
  k_{\|} \rho_e \approx \left(\frac{|\eta_e^T|  \beta_e}{2}\right)^{1/3} -\frac{|\eta_e^T|  \beta_e}{4} 
  \, ,
\end{equation}
is exponentially slow in $|\eta_e^T|  \beta_e \ll 1$: 
\begin{equation}
  \gamma_{\rm max} \approx \frac{3 \sqrt{\upi}}{4} |\eta_e^T| \exp{\left[-\frac{2^{2/3}}{\left(|\eta_e^T|  \beta_e\right)^{2/3}}-1\right]} \Omega_e
  \, . \label{whistlermode_smallk}
\end{equation}

Next, considering the opposite limit $k_\| \rho_e \gg 1$, we obtain
\begin{subeqnarray}
\varpi& \approx & \left[\eta_e^T \beta_e\left(\frac{1}{4} k_\| \rho_e-\frac{\upi-2}{2 \upi k_\| \rho_e}\right)+ \frac{2}{\upi} k_\|^2 \rho_e^2 \right] \frac{\Omega_e}{\beta_e}  \, , \\
\gamma & \approx & 
-\frac{1}{\sqrt{\upi}} \left[\eta_e^T \beta_e\left(\frac{1}{2}-\frac{4-\upi}{2 \upi k_{\|}^2 \rho_e^2}\right) + k_{\|}^3 \rho_e^3 \right] \frac{\Omega_e}{\beta_e}
\, . \label{whistlerwave_growthrate_lgekpl}
\end{subeqnarray}
We then find that the maximum growth rate of the parallel mode is given by
\begin{eqnarray}
  \gamma_{\rm max} & \approx & \frac{|\eta_e^{T}|}{\sqrt{\upi}} \left\{1- \left[\frac{1}{\sqrt{\upi}}\left(\frac{4}{\upi}-1\right)\right]^{3/5} \left[\left(\frac{3}{2}\right)^{2/5}-\left(\frac{2}{3}\right)^{3/5}\right]\left(|\eta_e^{T}|\beta_e \right)^{-2/5} \right\}
  \Omega_e \quad \nonumber \\
  & \approx & 0.56 |\eta_e^{T}| \left[1-0.13 \left(|\eta_e^{T}|\beta_e \right)^{-2/5} \right] \Omega_e \, , \label{maxgrowthrate_heatflux}
\end{eqnarray}
at the parallel wavenumber 
\begin{equation}
  k_{\|} \rho_e = \left[\frac{2}{3\sqrt{\upi}}\left(\frac{4}{\upi}-1\right)\right]^{1/5} \left(|\eta_e^{T}|\beta_e \right)^{1/5} 
  \approx 0. 63  \left(|\eta_e^{T}|\beta_e \right)^{1/5}  \, . \label{maxgrowthrate_heatflux_wavenumber}
\end{equation}
In addition, we see that the real frequency of 
modes with $k_\| \rho_e \lesssim \left(|\eta_e^T| 
\beta_e/{2}\right)^{1/3}$ is larger than the growth rate of the mode:
$\varpi\sim k_\| \rho_e \gamma \gg \gamma$. Thus, these modes oscillate more rapidly than they grow. 

The approximate expressions for (\ref{whistlerwave}) and (\ref{whistlerwave_growthrate_lgekpl}) are valid in the limits $|\eta_e^T| \beta_e \ll 1$ and $|\eta_e^T|  \beta_e \gg 1$, 
respectively,
and are plotted in figure~\ref{Figure_newCETwhist} alongside the exact results (\ref{whistlerwave_growthrate_allkpl}). Of particular note is the accuracy of the approximate expression (\ref{whistlerwave_growthrate_lgekpl}\textit{b}) 
for the growth rate when $k_{\|} \rho_e \gtrsim 0.6$; this suggests that (\ref{maxgrowthrate_heatflux})
is a reasonable estimate of the peak growth rate for $|\eta_e^T| \beta_e \gtrsim 
1$.

\subsubsection{Oblique whistler (heat-flux) instability}  \label{electron_heatflux_instab_whistl_obl}

Analytical expressions for the frequency and growth rate of unstable 
modes with an oblique wavevector at an angle to the magnetic field are more complicated than 
the analogous expressions for parallel modes. In appendix \ref{electron_heatflux_instab}, we show that there are two low-frequency oblique modes, whose complex 
frequencies $\omega$ are given by 
\begin{equation}
  \omega = \frac{\Omega_e}{\beta_e} k_{\|} \rho_e \frac{-B_{\mathrm{T}} \pm \sqrt{B_{\mathrm{T}}^2 + 4A_{\mathrm{T}}C_{\mathrm{T}}}}{2 A_{\mathrm{T}}} 
  \, , \label{heatflux_elec_frequency}
\end{equation}
where the coefficients $A_{\mathrm{T}} = A_{\mathrm{T}}(k_{\|} \rho_e,k_{\perp} \rho_e,\eta_e^T \beta_e)$, $B_{\mathrm{T}} = B_{\mathrm{T}}(k_{\|} \rho_e,k_{\perp} 
\rho_e,\eta_e^T \beta_e)$, and $C_{\mathrm{T}} = C_{\mathrm{T}}(k_{\|} \rho_e,k_{\perp} \rho_e,\eta_e^T \beta_e)$ 
are composed of the sums and products of the special 
functions defined in (\ref{specialfuncMax}), and also other special functions
defined in appendix \ref{heatfluxterm}. 
For a given wavenumber, we can use (\ref{heatflux_elec_frequency}) to calculate the growth rates 
of any unstable oblique modes -- and, in particular, demonstrate that positive growth rates 
are present for certain values of $\eta_e^T$. When they do exist, (\ref{heatflux_elec_frequency}) suggests that they will have the typical size 
$\gamma \sim \Omega_e/\beta_e \sim \eta_e^T \Omega_e$ when $k \rho_e \sim 1$
and $\eta_e^T \beta_e \sim 1$. 

\begin{figure}
\centerline{\includegraphics[width=0.99\textwidth]{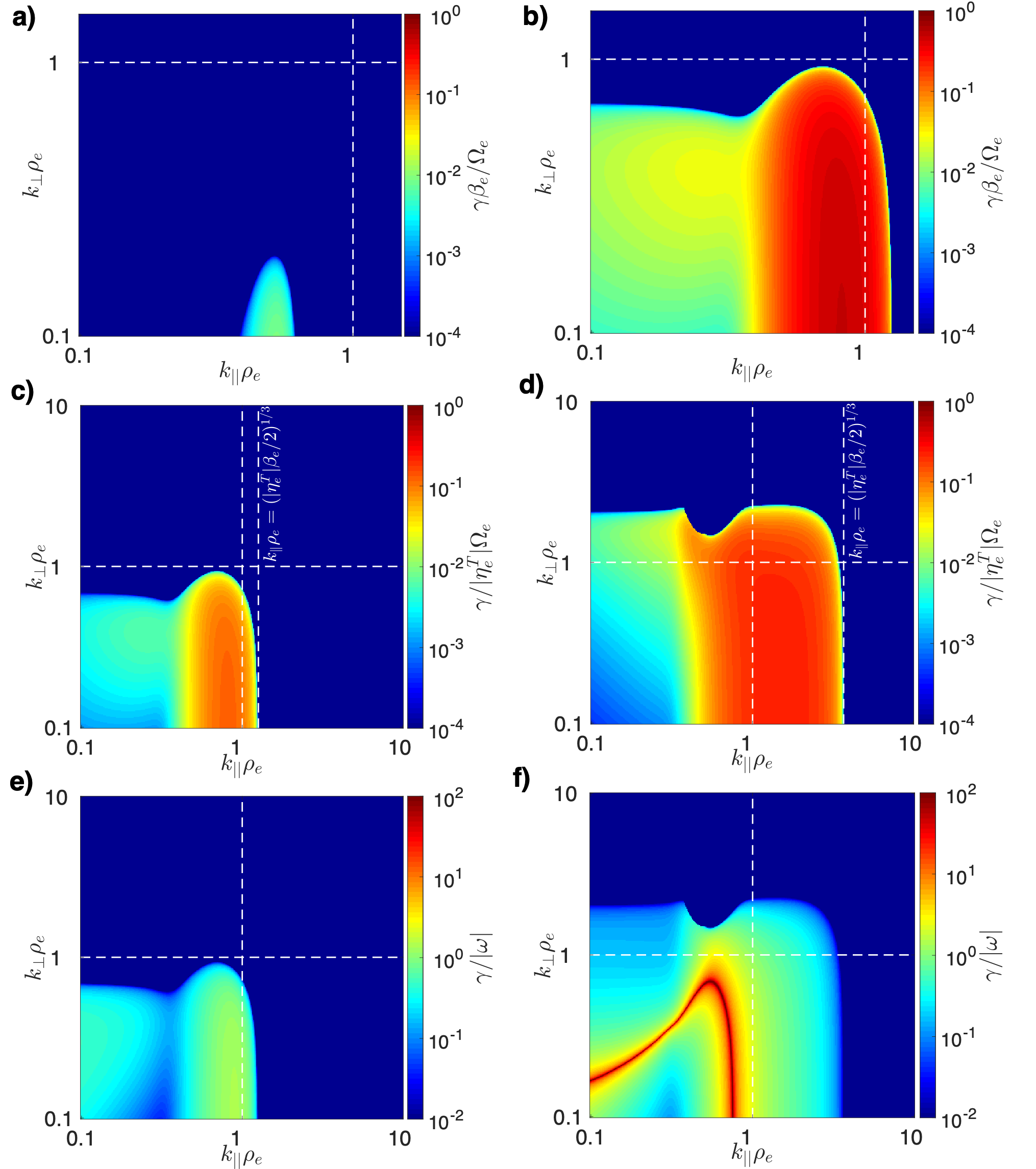}}
\caption{\textit{Oblique CET whistler instabilities}. Maximum positive growth rates of unstable whistler modes whose instability is driven by the electron-temperature-gradient term in CE distribution function 
(\ref{CEheatflux}\textit{a}), at arbitrary wavevectors with respect to the background 
magnetic field. 
The growth rates of the modes are calculated by taking the imaginary part of (\ref{heatflux_elec_frequency}), where coefficients $A_{\mathrm{T}}$, $B_{\mathrm{T}}$ and $C_{\mathrm{T}}$ are known functions of the wavevector. 
The growth rates are calculated on a $400^2$ grid, with equal logarithmic spacing in both perpendicular and parallel directions between the minimum and maximum wavenumbers.
The resulting growth rates, when normalised as $\gamma \beta_e/\Omega_e$, are functions of the dimensionless quantity $\eta_e^{T} \beta_e$. \textbf{a)} $\eta_e^{T} \beta_e = -0.5$. 
\textbf{b)} $\eta_e^{T} \beta_e = -4$. \textbf{c)} Same as b) but with normalisation $\gamma/|\eta_e^{T}| \Omega_e$.
\textbf{d)} Same as c), but with $\eta_e^{T} \beta_e = -100$.
 \textbf{e)} Ratio of growth rate to absolute value of real frequency for unstable modes for $\eta_e^{T} \beta_e = -4$.  \textbf{f)} Same as e), but with $\eta_e^{T} \beta_e = -100$.  \label{Figure1}}
\end{figure}

For $\eta_e^T > 0$, we find that both modes (\ref{heatflux_elec_frequency}) are damped; for $\eta_e^T < 0$, 
one mode is damped for all wavenumbers, but the other is not. 
Figure \ref{Figure1} shows the maximum (positive) growth rate $\gamma$ (normalised to $\Omega_e/\beta_e$) of this mode at a fixed value of 
$\eta_e^T$, for a range of $\beta_e$. The growth rate is calculated by evaluating 
the imaginary part of~(\ref{heatflux_elec_frequency}) at a given wavenumber. 
For $-\eta_e^T < 1/\beta_e$, the mode of interest is
damped for most wavenumbers, except for a small region of wavenumbers quasi-parallel to the magnetic field: in this region, there is a very small growth 
rate $\gamma \ll \Omega_e/\beta_e$ (figure \ref{Figure1}a). This finding is 
consistent with the exponentially small growth rates found for the parallel 
whistler modes [see (\ref{whistlermode_smallk})]. When $-\eta_e^T \sim 1/\beta_e$, there is a 
marked change in behaviour: a larger region of unstable modes appears, with $\gamma \sim 
\Omega_e/\beta_e$, at wavenumbers $k \rho_e \sim 1$ (figures \ref{Figure1}b and c). The growth rate is the largest 
for parallel modes -- but there also exist oblique modes with $k_{\bot} \lesssim k_{\|}$ whose growth rate is close to the peak growth rate. 
For example, for $\eta_e^T \beta_e = -4$, we find that the growth rate of the fastest-growing mode with a wavevector angle $\theta = 10^{\circ}$
is only ${\sim}2\%$ smaller than the fastest-growing parallel mode; for a 
wavevector angle $\theta = 20^{\circ}$, the reduction is by ${\sim}6\%$; and for 
$\theta = 30^{\circ}$, the reduction is by ${\sim}20\%$. Finally, if $-\eta_e^T \gg 1/\beta_e$, there exists a 
extended region of unstable modes, with $1 \lesssim k \rho_e \lesssim \left|\eta_e^T \beta_e\right|^{1/3}$, and $\gamma \sim |\eta_e^T \Omega_e|$ (figure \ref{Figure1}d). Again, the peak growth rate is at $k_{\perp} = 0$, 
but oblique modes also have a significant growth rate (for unstable modes with $\theta = 30^{\circ}$, the reduction in the largest growth rate compared to the fastest-growing parallel mode is only by ${\sim}4\%$). 
Most of the unstable modes have a non-zero real frequency: 
for $-\eta_e^T \sim 1/\beta_e$, $\omega \sim \gamma$ (figure \ref{Figure1}e), while for $-\eta_e^T \gg 1/\beta_e$, $\omega \gg \gamma$ 
for $k \rho_e \gg 1$ (figure \ref{Figure1}f). Note, however, that in the latter 
case there exists a band of wavenumbers at which there is no real frequency. 

In summary, we have (re-)established that the fastest-growing modes of the CET 
whistler instability are parallel to the magnetic field; however, we have shown semi-analytically (a novel result of this work) 
that the growth of oblique perturbations can be almost as large. This result is of 
some significance, because it has been argued that oblique whistler modes are necessary 
for the instability to scatter heat-carrying electrons efficiently~\citep[see, e.g.,][]{KSCS17}. It was proposed previously that such modes could arise 
from modifications to the CET 
electron-temperature-gradient terms induced by the unstable parallel whistler 
modes rendering the oblique modes the fastest-growing ones; our calculations 
suggest that it would only a require a small change to the CET whistler growth rates 
for this to be realised.  

As a further aside, we observe that 
in a plasma with sufficiently high plasma $\beta_e$, these oblique modes are in fact closer in nature to 
kinetic Alfv\'en waves (KAWs) than to whistler waves. Whistler waves are 
characterised as having effectively immobile ions ($\omega \gg k_{\bot} v_{\mathrm{th}i}$), while KAWs have warm ions ($\omega \ll k_{\bot} v_{\mathrm{th}i}$); as a 
consequence, whistler waves have a negligible density perturbation ($\delta n_e \ll Z e n_{e} \varphi/T_i$, where $\varphi$ is the electrostatic potential associated with the wave), while KAWs 
do not: $\delta n_e \approx Z e n_{e} {\varphi}/T_i$~\citep{BHXP13}. In a $\beta_e \sim 1$ plasma for $k_{\bot} \gtrsim k_{\|}$, the real frequency of whistler 
modes satisfies $\omega/k_{\bot} v_{\mathrm{th}i}  \sim k_{\|} \rho_i/\beta_e \sim  k_{\|} \rho_i$; thus, we conclude 
from our above considerations that the two waves must operate in different 
regions of wavenumber space, viz., $k_{\|} \rho_i \ll 1$, $k_{\bot} \rho_i > 1$ for KAWs, and $k_{\|} \rho_i \gg 1$ for 
whistlers. However, for $\beta_e \gtrsim \mu_e^{-1/2}$ (where $\mu_e = m_e/m_i$) and $k_{\bot} \sim k_{\|} \gg \rho_i^{-1}$, the  
frequency of whistler waves is too low for $\omega \gg k_{\bot} v_{\mathrm{th}i}$ 
to be satisfied whilst also maintaining $k_{\|} \rho_e \ll 1$. Instead, the 
ions participate in the wave mechanism, and $\delta n_e \approx -Z e n_{e} {\varphi}/T_i$ (see appendix \ref{KAWs_densityperturb}). 

For further discussion of the 
physics of the whistler instability (as well as its nonlinear evolution), see~\citet{KSCS17} and the other references given at the beginning of section \ref{electron_heatflux_instab_whistl}. 

\subsubsection{Slow-(hydromagnetic)-wave instability}  \label{ion_heatflux_instab_slowwave}

Although parallel ion heat fluxes in a classical, collisional plasma are typically much 
weaker than electron heat fluxes, they can still act as a free-energy source for 
instabilities, by introducing anisotropy to the ion distribution function (\ref{CEheatflux}\textit{b}) (i.e., the CE ion-temperature-gradient 
term). Furthermore, anisotropy in the ion distribution function can enable the 
instability of plasma modes that are not destabilised by the CE 
electron-temperature-gradient term. This exact situation is realised in the CET 
slow-hydromagnetic-wave instability, in which a sufficiently large
CET ion-temperature-gradient term counteracts the effect of ion cyclotron damping on 
slow hydromagnetic waves. The slow hydromagnetic wave (or slow wave)~\citep{R71,FK79} is the left-hand-polarised quasi-parallel electromagnetic mode in high-$\beta$ plasma; it exists for parallel wavenumbers~$k_{\|}$ that satisfy $\beta_i^{-1/2} \ll k_{\|} \rho_i \lesssim 1$, and 
has a characteristic frequency $\omega \approx 2 \Omega_i/\beta_i$. 
To the authors' knowledge, no instability of the slow wave due to the ion heat flux has previously been 
reported. The instability's mechanism is analogous to the CET whistler 
instability: the slow waves drain energy from ions with parallel velocities $v_{\|} = \pm \Omega_i/k_{\|}$ via gyroresonant wave-particle interactions. 
For an anti-parallel ion temperature gradient (i.e., $\nabla_{\|} T_i < 0$, so $\eta_i < 
0$), slow waves propagating down the temperature gradient are destabilised, 
while those propagating up the temperature gradient are not.

\begin{figure}
\centerline{\includegraphics[width=0.99\textwidth]{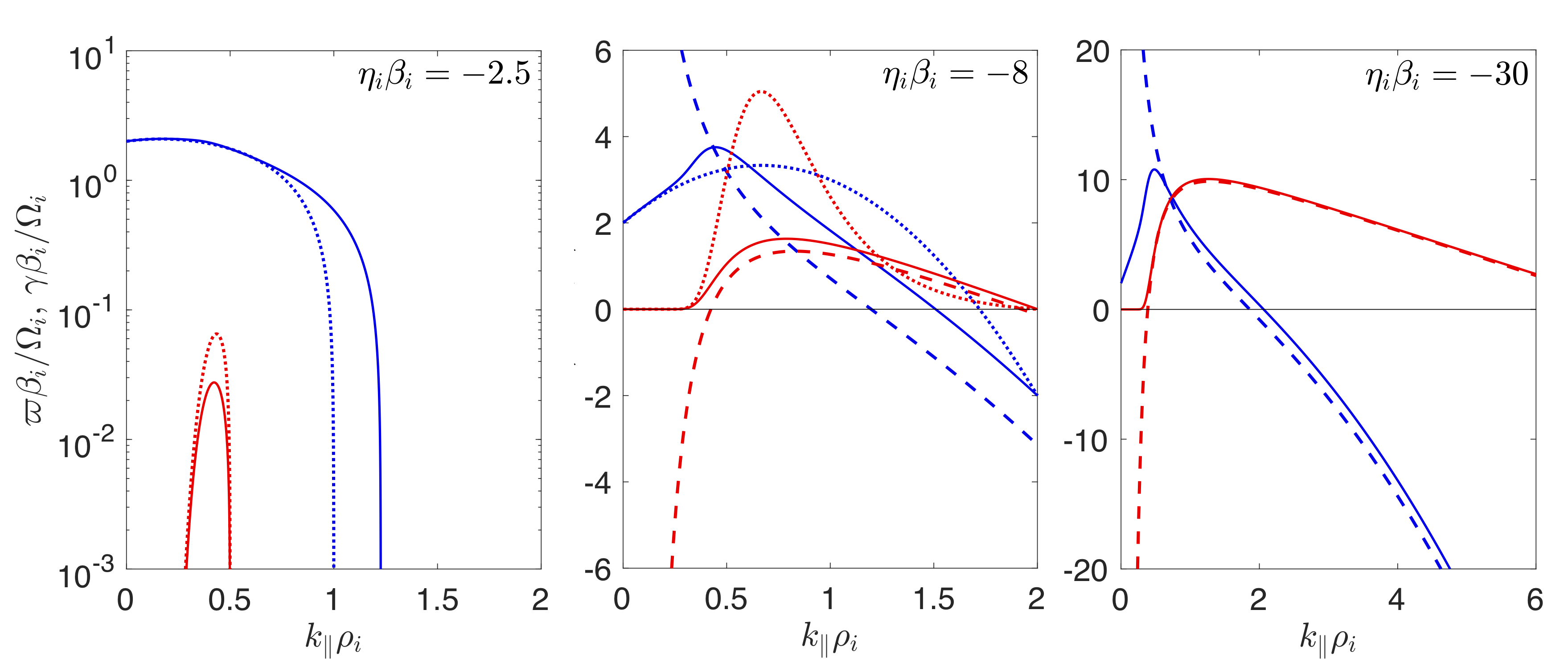}}
\caption{\textit{Parallel CET slow-hydromagnetic-wave instability}.  Dispersion curves of slow hydromagnetic waves whose instability is driven by the ion-temperature-gradient term in the CE distribution function 
(\ref{CEheatflux}\textit{b}), for wavevectors co-parallel with the background 
magnetic field (viz., $\boldsymbol{k} = k_{\|} \hat{\boldsymbol{z}}$). 
The frequency (solid blue) and growth rates (solid red) of the modes are calculated using  
(\ref{slowwavegen}\textit{a}) and 
(\ref{slowwavegen}\textit{b}), respectively. 
The resulting frequencies and growth rates, when normalised as $\gamma \beta_i/\Omega_i$, are functions of the dimensionless quantity $\eta_i \beta_i$; we show the dispersion curves for three different values of $\eta_i \beta_i$. 
The approximations (\ref{slowmode_smallk_freq}) and (\ref{slowmode_smallk_growth}) for 
the frequency (dotted blue) and growth rate (dotted red) in the limit $k_{\|} \rho_i \ll 1$ are also plotted, 
as are the approximations (\ref{slowwave_growthrate_lgekpl}\textit{a}) and (\ref{slowwave_growthrate_lgekpl}\textit{b}) 
for the frequency (dashed blue) and growth rate (dashed red) in the limit $k_{\|} \rho_i \gg 
1$. 
 \label{Figure_newCETion}}
\end{figure}

As before, the slow-wave instability is most easily characterised in the subsidiary limit $k_{\bot} \rho_i \rightarrow 0$ ($k = k_{\|}$).  
Under the ordering $k_{\|} \rho_i \sim 1$, the real frequency $\varpi$ and growth rate $\gamma$ are given by (see appendix \ref{derivation_slowwave})
\begin{subeqnarray}
\frac{\varpi}{\Omega_i} & = &  \eta_i \left(\frac{k_{\|} \rho_i}{4}-\frac{1}{2 k_\| \rho_i}\right) 
- \frac{k_{\|}^2 \rho_i^2 \left[\Real{\; Z\!\left({1}/{k_\| \rho_i}\right)} + k_{\|} \rho_i\right] \left(\eta_i/4 + k_\| \rho_i/\beta_i\right) }{\left[\Real{\; Z\!\left({1}/{k_\| \rho_i}\right)} + k_{\|} \rho_i\right]^2 + \upi \exp{\left(-{2}/{k_\|^2 \rho_i^2}\right)}} 
\, , \quad \\
\frac{\gamma}{\Omega_i} & = & -
\frac{\sqrt{\upi} k_{\|}^2 \rho_i^2 \left(\eta_i/4 + k_\| \rho_i/\beta_i\right) }{\left[\Real{\; Z\!\left({1}/{k_\| \rho_i}\right)} + k_{\|} \rho_i\right]^2 \exp{\left({1}/{k_\|^2 \rho_i^2}\right)}+ \upi \exp{\left(-{1}/{k_\|^2 \rho_i^2}\right)}} 
\, .  \label{slowwavegen}
\end{subeqnarray}
The CET electron-temperature-gradient term does not appear because its 
contributions to the frequency and growth rate are much smaller than the 
equivalent contributions of the CET ion-temperature-gradient term at $k_{\|} \rho_i \sim 1$. 
Plots of $\varpi = \varpi(k_{\|})$ and $\gamma = \gamma(k_{\|})$ for different 
values of $\eta_i \beta_i < 0$ are shown in figure \ref{Figure_newCETion}. 

As with the CET whistler instability, we can derive simple expressions for the 
peak growth rate (and the wavenumber associated with that growth rate) in 
subsidiary limits. 

First, ordering $k_{\|} \rho_i \sim \eta_i \beta_i/4 \ll 1$ so that the destabilising $\eta_i$ terms and the stabilising ion FLR terms are the same order, 
we find that the real frequency (\ref{slowwavegen}\textit{a}) becomes 
\begin{equation}
 \varpi \approx \frac{2 \Omega_i}{\beta_i} \left(1-\frac{1}{4} k_{\|} \rho_i \eta_i \beta_i -\frac{3}{2} k_{\|}^2 \rho_i^2 \right) \, , 
 \label{slowmode_smallk_freq}
\end{equation}
which is precisely that of the slow hydromagnetic wave, with first-order FLR corrections included~\citep{FK79}. 
For $\eta_i < 0$ and $k_{\|} \rho_i < |\eta_i| \beta_i/4$, the growth rate (\ref{slowwavegen}\textit{b}) 
is positive:
\begin{equation}
\gamma \approx - \frac{4 \sqrt{\upi}}{k_\|^4 \rho_i^4} \left(\frac{\eta_i}{4}  + \frac{k_{\|} \rho_i}{\beta_i}\right)\exp{\left(-\frac{1}{k_\|^2 \rho_i^2}\right)} 
\Omega_i \, .   \label{slowmode_smallk_growth}
\end{equation}
The maximum growth rate (which is exponentially small in $\eta_i \beta_i/4 \ll 1$) is 
\begin{equation}
  \gamma_{\rm max} \approx \frac{8 \sqrt{\upi}}{|\eta_i| \beta_i^2} \exp{\left(-\frac{16}{|\eta_i|^2  \beta_i^{2}}-1\right)} 
  \Omega_i
  \, , \label{slowmode_smallk}
\end{equation}
achieved at the parallel wavenumber
\begin{equation}
  k_{\|} \rho_i \approx \frac{|\eta_i|  \beta_i}{4} -\frac{|\eta_i|^3  \beta_i^3}{128} 
  \, .
\end{equation}

In the opposite limit, $k_{\|} \rho_i \sim \left(|\eta_i| \beta_i/4\right)^{1/3} \gg 1$, we obtain 
\begin{subeqnarray}
\varpi& \approx & -\left(\eta_i \beta_i \frac{1-\upi/4}{k_\| \rho_i} -k_\|^2 \rho_i^2 \right) \frac{\Omega_i}{\beta_i}  \, , \\
\gamma & \approx & 
-\sqrt{\upi} \left[\frac{\eta_i}{4} \beta_i\left(1-\frac{\upi-3}{k_{\|}^2 \rho_i^2}\right) + k_{\|} \rho_i \right] \frac{\Omega_i}{\beta_i}
\, . \label{slowwave_growthrate_lgekpl}
\end{subeqnarray}
The maximum positive growth rate is 
\begin{equation}
  \gamma_{\rm max} \approx \frac{\sqrt{\upi}}{4} \left\{1- 3\left[4 \left(\upi-3\right)\right]^{1/3} \left(|\eta_i|\beta_i \right)^{-2/3} \right\} 
   |\eta_i|
  \Omega_i  \approx 0.44 \left[1-2.48 \left(|\eta_i|\beta_i \right)^{-2/3} \right]  |\eta_i| \Omega_i \, , \,  \label{maxgrowthrate_slowwave}
\end{equation}
realised for $\eta_i < 0$ at the parallel wavenumber
\begin{equation}
k_{\|} \rho_i \approx \left(\frac{\upi -3}{2}\right) ^{1/3} \left(|\eta_i|\beta_i\right)^{1/3} 
\approx 0.41 \left(|\eta_i|\beta_i\right)^{1/3} 
\, .
\end{equation}
We note that, in contrast to the CET whistler instability, the real frequency of the 
fastest-growing unstable mode is smaller than its growth rate: $\omega_{\rm peak}/\gamma_{\rm max} \approx 0.36 
(|\eta_i|\beta_i)^{-1/3}$. 

The approximate expressions (\ref{slowmode_smallk_freq}), (\ref{slowmode_smallk_growth}),
(\ref{slowwave_growthrate_lgekpl}\textit{a}),
and (\ref{slowwave_growthrate_lgekpl}\textit{b}) for the frequency and growth rate in the 
limits $k_{\|} \rho_i \ll 1$ and $k_{\|} \rho_i \gg 1$, are plotted in figure \ref{Figure_newCETion}, along with the exact results (\ref{slowwavegen}).

\begin{figure}
\centerline{\includegraphics[width=0.99\textwidth]{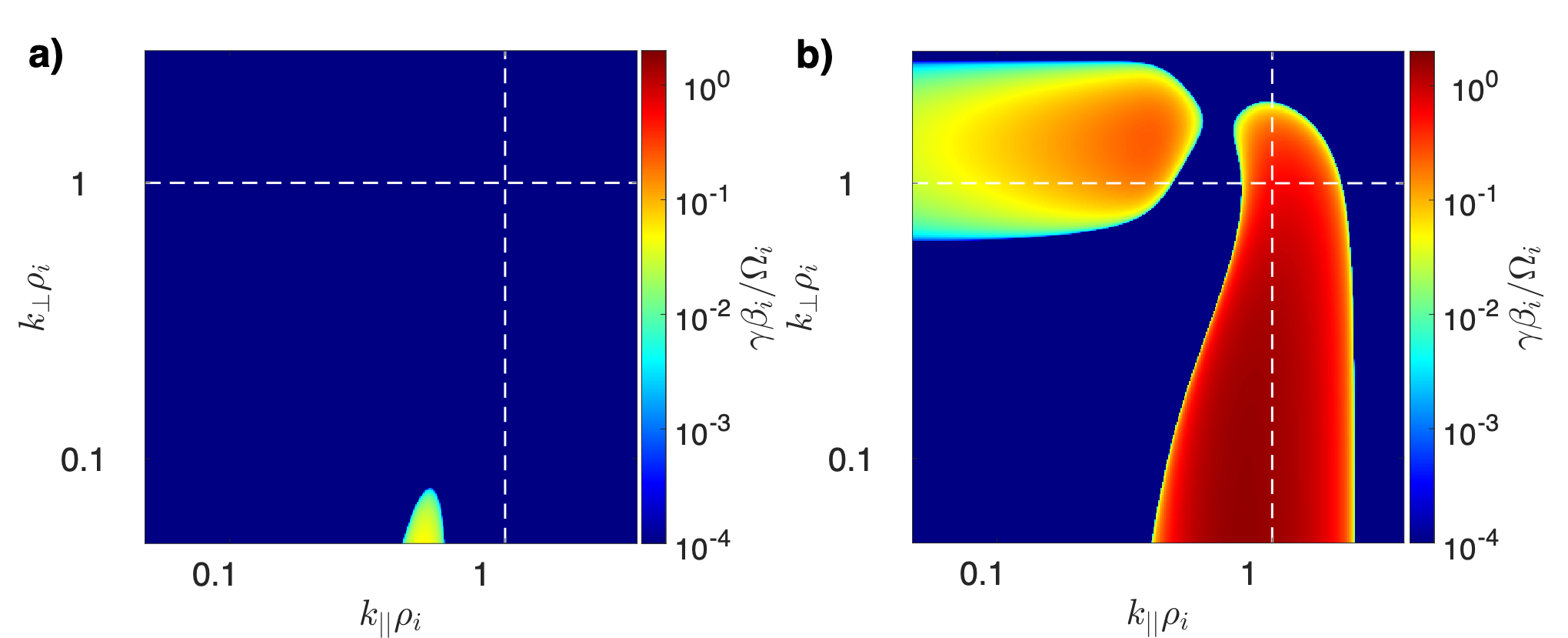}}
\caption{\textit{Oblique CET ion-Larmor-scale instabilities}. 
Maximum positive growth rates of unstable ion-Larmor-scale modes whose instability is driven by the CE ion-temperature-gradient term in the CE distribution function 
(\ref{CEheatflux}\textit{b}), at arbitrary wavevectors with respect to the background 
magnetic field. 
The growth rates of all modes are calculated by taking the imaginary part of (\ref{heatflux_ion_frequency}), with coefficients $\tilde{A}_{\mathrm{T}}$, $\tilde{B}_{\mathrm{T}}$ and $\tilde{C}_{\mathrm{T}}$ being known functions of the wavevector (see appendix \ref{ion_heatflux_instab}). The growth rates are calculated on a $400^2$ grid, with logarithmic spacing in both perpendicular and parallel directions between the minimum and maximum wavenumber magnitudes.
The resulting growth rates, when normalised as $\gamma \beta_i/\Omega_i$, are functions of $\eta_i \beta_i$. \textbf{a)} $\eta_i \beta_i = -2.5$. 
\textbf{b)} $\eta_i \beta_i = -8$. The unstable $k_{\|} \rho_i \ll k_{\perp} \rho_i \sim 1$ modes appearing in b) are dealt with in section \ref{CET_KAW_instab}. \label{Figure2}}
\end{figure}

As with the CET whistler instability, a general expression for the complex frequency of 
oblique ion CET instabilities can be derived in the form (see appendix \ref{ion_heatflux_instab}):
\begin{equation}
  \omega = \frac{\Omega_i}{\beta_i} k_{\|} |\rho_i| \frac{-\tilde{B}_{\mathrm{T}} \pm \sqrt{\tilde{B}_{\mathrm{T}}^2 + 4\tilde{A}_{\mathrm{T}}\tilde{C}_{\mathrm{T}}}}{2 \tilde{A}_{\mathrm{T}}} 
  \, , \label{heatflux_ion_frequency}
\end{equation}
where $\tilde{A}_{\mathrm{T}} = \tilde{A}_{\mathrm{T}}(k_{\|} \rho_i,k_{\perp} \rho_i,\eta_i \beta_i)$, $\tilde{B}_{\mathrm{T}} = \tilde{B}_{\mathrm{T}}(k_{\|} \rho_i,k_{\perp} 
\rho_i,\eta_i \beta_i)$, and $\tilde{C}_{\mathrm{T}} = \tilde{C}_{\mathrm{T}}(k_{\|} \rho_i,k_{\perp} \rho_i,\eta_i \beta_i)$ 
are again sums and products of various special mathematical functions defined in (\ref{specialfuncMax}).  Investigating such modes by evaluating (\ref{heatflux_ion_frequency}) numerically
for a range of wavenumbers (see figure \ref{Figure2}), we find that, for $\eta_i < 0$, there is one mode that is always damped and one 
that can be unstable. For $-\eta_i \lesssim 4/\beta_i$, the unstable modes are 
restricted to quasi-parallel modes (see figure \ref{Figure2}a); for $-\eta_i \gtrsim 4/\beta_i$, there is a much broader spectrum of 
unstable modes (including oblique ones). The positive growth rates of the unstable mode are shown in figure 
\ref{Figure2}b for $\eta_i \beta_i = -8$. The typical growth rate $\gamma$ satisfies $\gamma \sim 
\Omega_i/\beta_i \sim \eta_i \Omega_i$, as anticipated from (\ref{heatflux_ion_frequency}). 
We also observe in figure \ref{Figure2}b the existence of an unstable mode at 
quasi-perpendicular wavenumbers, which is discussed in section \ref{CET_KAW_instab}. 

In summary, an ion temperature gradient can destabilise ion-Larmor-scale, slow hydromagnetic 
waves via a similar mechanism to an electron temperature gradient 
destabilising electron-Larmor-scale whistler waves. If $\beta_i \gg L_{T_i}/\lambda_i$, the 
characteristic growth rate of these modes is $\gamma \sim \lambda_i 
\Omega_i/L_{T_i}$. 
Unstable modes whose wavevector is parallel to $\boldsymbol{B}_0$ grow most 
rapidly, although the growth rate of (moderately) oblique modes is only somewhat smaller.
While the CET whistler instability is faster growing than the CET slow-wave 
instability, both modes grow much more quickly than characteristic hydrodynamic 
time scales in a strongly magnetised plasma. In any conceivable 
saturation mechanism, the electron mode will adjust the electron heat flux, and 
the ion mode the ion heat flux. Thus, it seems likely that 
understanding the evolution (and ultimately, the saturation) of both 
instabilities would be necessary to model correctly the heat transport in a classical, 
collisional plasma that falls foul of the $\beta$-stabilisation condition. 

\subsubsection{Long-wavelength kinetic-Alfv\'en-wave instability} \label{CET_KAW_instab}

The instability observed in figure \ref{Figure2}b at wavevectors satisfying $k_{\|} \rho_i \ll k_{\perp} \rho_i \sim 1$ 
is different in nature to the slow-hydromagnetic-wave instability:
it is an ion-temperature-gradient-driven instability of long-wavelength KAWs. 
Like the CET slow-wave instability, it operates on account of resonant wave-particle 
interactions that allow free energy to be drained from the anisotropy of the ion 
distribution function, which itself arises from the ion temperature gradient. However, 
the gyroresonances $v_{\|} \approx \pm \Omega_i/k_{\|}$ operate 
inefficiently for modes with $k_{\|} \rho_i \ll 
1$ in a CE plasma, because there are comparatively few particles with $v_{\|} \gg v_{\mathrm{th}i}$; 
the dominant resonance is instead the Landau resonance $v_{\|} = \omega/k_{\|}$. More specifically, 
KAWs with $k_{\perp} \rho_i \gtrsim 1$, which are usually subject to strong Landau and Barnes damping (that is, the 
damping rate of the waves is comparable to their real frequency),
can be destabilised if the (ion) plasma beta is sufficiently large: $\beta_i \gtrsim L_{T_i}/\lambda_i$. 
In figure \ref{Figure2}b, the peak growth rate of the CET KAW instability 
is smaller than that of the CET slow-hydromagnetic-wave instability by an 
order of magnitude; as will be shown below, this is, in fact, a generic feature of the instability. 

Similarly to quasi-parallel unstable modes, quasi-perpendicular ones such as unstable KAWs can be 
characterised analytically, allowing for a simple 
identification of unstable modes and their peak growth rates. 
It can be shown (see appendix \ref{derivation_CETKAW}) that, in the limit $k_{\|} \rho_i \ll 
1$, $k_\perp \rho_i \sim 1$, the complex frequency of the low-frequency ($\omega \ll k_{\|} v_{\mathrm{th}i}$) modes in a plasma whose 
ion distribution function is (\ref{CEheatflux}\textit{b}) is
\begin{eqnarray}
\frac{\omega}{k_{\|} v_{\mathrm{th}i}} & = &  \frac{\eta_i \mathcal{G}_i}{2 \left(1-\mathcal{F}_i\right)}  + \frac{k_{\perp} \rho_i}{\beta_i  \left(1-\mathcal{F}_i\right)^{2}} \Bigg[ -\frac{\mathrm{i}\sqrt{\upi}}{2} k_{\perp} \rho_i \left(\mathcal{F}_i + \sqrt{\frac{\mu_e Z^2}{\tau}}\right)
\nonumber \\
& & \pm \sqrt{1-\frac{\upi}{4} \frac{k_{\perp}^2 \rho_i^2}{\beta_i} \bigg(\mathcal{F}_i + \sqrt{\frac{\mu_e Z^2}{\tau}}\bigg)^2 - \frac{\mathrm{i} \sqrt{\upi} \eta_i \beta_i}{4} \frac{2 \mathcal{G}_i - \mathcal{F}_i\left(1 - \mathcal{F}_i\right)}{1-\mathcal{F}_i} } \Bigg] \, , 
\label{heatfluxKAWfreq}
\end{eqnarray}
where $\mathcal{F}_i \equiv \mathcal{F}(k_{\perp} \rho_i)$, $\mathcal{G}_i \equiv \mathcal{G}(k_{\perp} 
\rho_i)$, and 
\begin{eqnarray}
\mathcal{F}(\alpha) & \equiv & \exp{\left(-\frac{\alpha^2}{2}\right)} \left[I_{0}\left(\frac{\alpha^2}{2}\right) - I_{1}\left(\frac{\alpha^2}{2}\right)\right] \, 
, \\
\mathcal{G}(\alpha) & \equiv & 2 \alpha^2 \mathcal{F}(\alpha)-\exp{\left(-\frac{\alpha^2}{2}\right)} I_{1}\left(\frac{\alpha^2}{2}\right) \, 
. \label{specialfunction_quasiperp}
\end{eqnarray}
In a Maxwellian plasma (i.e., when $\eta_i = 0$), (\ref{heatfluxKAWfreq}) becomes
\begin{eqnarray}
\frac{\omega}{k_{\|} v_{\mathrm{th}i}} & = &  \frac{1}{\left(1-\mathcal{F}_i\right)^{2}} \Bigg[ -\frac{\mathrm{i}\sqrt{\upi}}{2} \frac{k_{\perp}^2 \rho_i^2}{\beta_i} \left(\mathcal{F}_i + \sqrt{\frac{\mu_e Z^2}{\tau}}\right) 
\nonumber \\
&& \pm \sqrt{\frac{k_{\perp}^2 \rho_i^2}{\beta_i^2}-\frac{\upi}{4} \frac{k_{\perp}^4 \rho_i^4}{\beta_i^2} \bigg(\mathcal{F}_i + \sqrt{\frac{\mu_e Z^2}{\tau}}\bigg)^2} \Bigg] \, .
\end{eqnarray}
In the subsidiary limit $k_{\perp} \rho_i \gg 1$, we recover 
$\omega \approx \pm k_{\|} v_{\mathrm{th}i} k_{\perp} \rho_i/\beta_i$, which is the 
well-known dispersion relation of a KAW~\citep{SCDHHQ09,BHXP13,KAKS18}. 

\begin{figure}
\centerline{\includegraphics[width=0.99\textwidth]{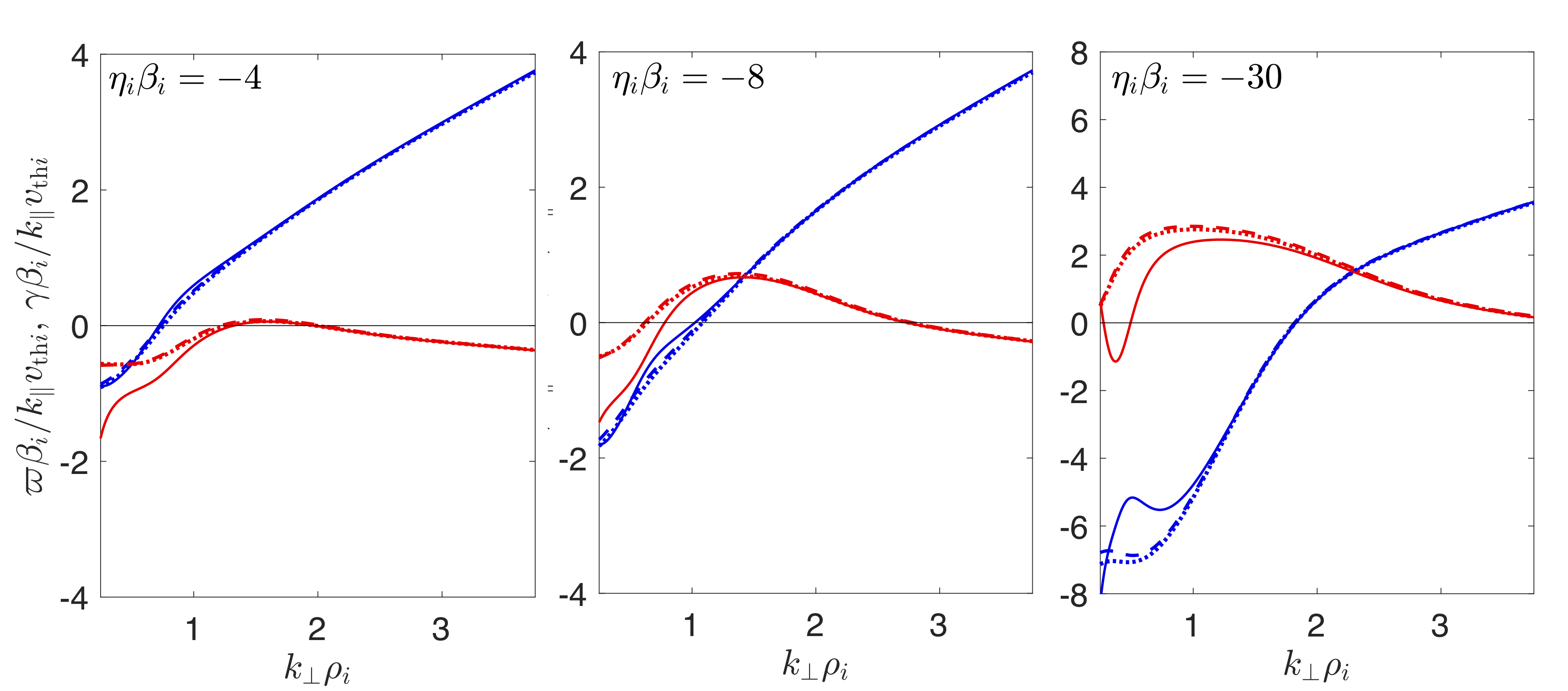}}
\caption{\textit{Quasi-perpendicular CET KAW instability}.  Dispersion curves of unstable KAWs whose instability is driven by the ion-temperature-gradient term in the CE distribution function~(\ref{CEheatflux}\textit{b}), for wavevectors that are almost perpendicular to the background 
magnetic field (viz., $k_{\perp} \gg k_{\|}$). 
The frequency (blue) and growth rates (red) of unstable modes  
are calculated at (small) fixed values of $k_{\|} \rho_i$ from the real and imaginary parts of
(\ref{heatflux_ion_frequency}); the solid curves are calculated for $k_{\|} \rho_i = 0.35$, while the 
dashed curves are for $k_{\|} \rho_i = 0.05$.
The resulting frequencies and growth rates, when normalised as $\gamma \beta_i/k_{\|} v_{\mathrm{th}i}$, 
are functions of the dimensionless quantity $\eta_i \beta_i$; we show the dispersion curves for three different values of $\eta_i \beta_i$. 
The frequency (dotted blue) and growth rate (dotted red) in the limit $k_{\|} \rho_i \ll 1$, which are 
calculated by taking the real and imaginary parts of (\ref{heatfluxKAWfreq}), are also plotted. 
 \label{Figure_newCETKAW}}
\end{figure}

For $\eta_i \neq 0$, we find that, for modes with a positive propagation direction with 
respect to the background magnetic field (viz., $k_{\|} > 0$), there 
is an instability provided
\begin{equation}
  \eta_i \lesssim -3.14 \left(1 + 6.5 \sqrt{\frac{\mu_e Z^2}{\tau}}\right)
  \beta_i^{-1}
  \, ,
\end{equation}
with the perpendicular wavenumber $k_{\perp} \rho_i$ of the fastest-growing unstable mode 
at fixed $k_{\|}$ just beyond this threshold being approximately given by
\begin{equation}
k_{\perp} \rho_i \approx 1.77 \left(1 - 3.4 \sqrt{\frac{\mu_e Z^2}{\tau}}\right) 
\, .
\end{equation}
Figure \ref{Figure_newCETKAW} shows the real frequency and growth rate of such modes at three 
different (negative) values of $\eta_i \beta_i$. 
As $\eta_i$ is 
decreased beyond the threshold, modes over an increasingly large range of perpendicular 
wavenumbers are destabilised at both super- and sub-ion Larmor scales. Indeed, in 
the limit $|\eta_i| \beta_i \gg 1$, the peak growth rate $\gamma_{\rm max}$ (for a fixed $k_{\|}$)
occurs at a perpendicular wavenumber $k_{\perp} \rho_i < 1$, which decreases as $|\eta_i| \beta_i$ increases.
Such modes are, in fact, no longer well described physically as KAWs; their analogues in a Maxwellian 
plasma are Barnes-damped, non-propagating slow modes. 

Although it is possible to characterise analytically the peak growth rate of the 
unstable modes (and the perpendicular wavenumber at which such growth is attained) in the limit $k_{\|} \rho_i \ll 1$ 
by analysing (\ref{heatfluxKAWfreq}), such estimates do not capture accurately 
the behaviour of the fastest-growing modes across all wavevectors, because these 
fastest-growing modes occur at finite values of $k_{\|} \rho_i$; at such values, the 
dependence of the frequency and growth rate on $k_{\perp} \rho_i$ departs somewhat from (\ref{heatfluxKAWfreq}) (see figure \ref{Figure_newCETKAW}). Instead, we 
find numerically that, for $\eta_i \beta_i \lesssim -6$,
\begin{equation}
  \gamma_{\rm max} \approx 0.025 |\eta_{i}| \Omega_i \quad \mathrm{at}  \quad (k_{\|} 
  \rho_i)_{\rm peak} \approx 0.35 \, ,
\end{equation}
independent of the specific value of either $\eta_i$ or $\beta_i$. 
For values of $k_{\|} \rho_i$ that are larger than $(k_{\|} \rho_i)_{\rm peak}$, the instability is 
quenched. It is clear that, in comparison to the slow-hydromagnetic wave instability, the 
growth rate of the fastest-growing perpendicular modes is small [see (\ref{slowwave_growthrate_lgekpl})]. This 
difference can be attributed to the fact that, for unstable modes in 
the limit $|\eta_i| \beta_i \gg 1$, $\gamma_{\rm max} \sim |\eta_i| k_{\|} \rho_i 
\Omega_i$ and the value of $k_{\|} \rho_i$ at which maximum growth is achieved 
is still rather small compared to unity. We conclude that the instability of  
slow hydromagnetic waves that are driven by an ion temperature gradient is likely to be more 
significant than the analogous instability of quasi-perpendicular/KAW 
modes. 

\section{CES (Chapman-Enskog, shear-driven) microinstabilities} \label{Results_shearingterm}

\subsection{Form of CE distribution function} \label{Results_shearingterm_theory}

Next, we consider the non-Maxwellian terms of the CE distribution arising from 
bulk-flow gradients. If we set $\eta_s = 0$ for both ions and electrons (viz., neglecting both temperature gradients and electron-ion drifts), 
the CE distribution functions (\ref{ChapEnskogFunc}) for both species become
\begin{equation}
f_{s0}(v_{\|},v_{\bot}) = \frac{n_{s0}}{v_{\mathrm{th}s}^3 \upi^{3/2}} \exp \left(-\tilde{v}_{s}^2\right) \left[1 - \epsilon_s \left(\frac{v_{\|}^2}{v_{\mathrm{th}s}^2}- \frac{v_{\bot}^2}{2 v_{\mathrm{th}s}^2} \right)\right] ,  
\label{CEsheardistfuncexpression}
\end{equation}
where we have again chosen the isotropic functions $C_s(\tilde{v}_s)$ 
to be the ones that arise from the Krook collision operator (see section \ref{disprel_simps_overview}).
We note that for this choice of collision operator, the constant $\mathcal{C}_s$ defined by (\ref{pressureanisop_const}) is  
$\mathcal{C}_s \approx 3/2$, and so the relationship (\ref{pressureanisop}) between the CE distribution functions' pressure anisotropy $\Delta_s$ 
and the shear parameter $\epsilon_s$ becomes
\begin{equation}
\Delta_s = \frac{3}{2} \epsilon_s \, .  \label{pressanisop_krook}
\end{equation}
We also observe that the CE shear terms have even parity with respect to the 
parallel velocity $v_{\|}$, and thus for any unstable mode with positive parallel wavenumber $k_{\|} > 0$, there is
a corresponding unstable mode with $k_{\|} < 0$.  This conclusion has the consequence 
that the sign of $\epsilon_s$ [which is the same as the sign of $\left(\hat{\boldsymbol{z}} \hat{\boldsymbol{z}} - \mathsfbi{I}/3 \right) \! \boldsymbol{:} \! \mathsfbi{W}_s$, where $\mathsfbi{W}_s$ is the rate-of-strain tensor of species $s$ -- see (\ref{ChapEnsTerms})]
has a significant effect on possible types of CES microinstabilities. Thus, we 
must consider the cases $\epsilon_s > 0$ (positive pressure anisotropy, $\Delta_s > 0$) 
and $\epsilon_s < 0$ (negative pressure anisotropy, $\Delta_s < 0$) separately. For easier comparison to 
previous work by other authors, we will sometimes substitute $\epsilon_s =2 \Delta_s/3$, and work in terms of $\Delta_s$.  

As with the discussion of CET microinstabilities in section \ref{Results}, in the main text, we 
only present the main findings of our calculations: namely, the overview of the CES stability landscape (section \ref{poseps_stab}), and
the analytical characterisation of CES microinstabilities with $\epsilon_s > 0$ (section \ref{CES_pos_micro}) and $\epsilon_s < 0$ (section \ref{CES_neg_aniso}).
The methodology underlying the calculations of growth rates of CES microinstabilities 
is presented in appendix \ref{CES_method_append}. 

\subsection{Stability} \label{poseps_stab}

\begin{figure}
\centerline{\includegraphics[width=0.99\textwidth]{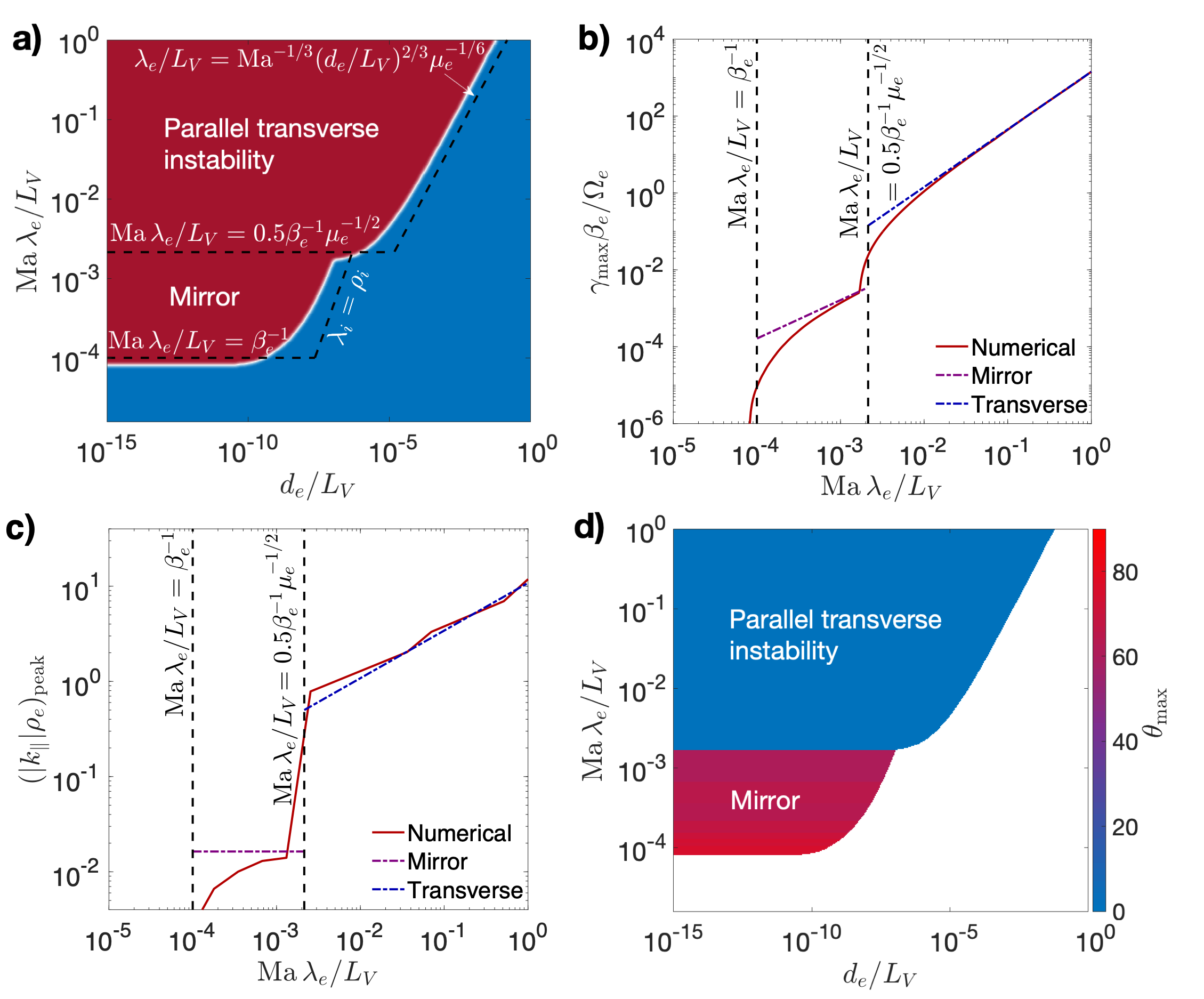}}
\caption{\textit{CE-distribution-function stability map for CES microinstabilities driven by positive pressure anisotropy}. Exploration of the stability of the ion and electron CE distribution functions (\ref{CEsheardistfuncexpression}) 
for different positive values of small parameters $\epsilon_e$ and $\epsilon_i$ (viz., electron or ion pressure anisotropies), and the ratio of the electron inertial scale $d_e$ to the velocity scale length $L_V,$ in a temperature-equilibrated hydrogen plasma. In this 
 plot, we chose $\epsilon_e = \mu_e^{1/2} \epsilon_i$, and then show $\mathrm{Ma} \, \lambda_e/L_V = |\epsilon_i|$ with equal logarithmic spacing in the range $\left[10^{-5},10^{0}\right]$; $d_e/L_V$ is chosen with equal logarithmic spacing in the range 
 $\left[10^{-15},10^{0}\right]$. The total size of the grid is $400^2$. For reasons of efficiency, we calculate growth rates on a $40^2$ grid in wavenumber space with
 logarithmic spacing for both parallel and perpendicular wavenumbers. In this plot, $\beta_e = \beta_i = 10^4$, and $\mathrm{Ma} = 1$. \textbf{a)} Stable (blue) and unstable (red) regions of $\left(d_e/L_V,\mathrm{Ma} \, \lambda_e/L_V\right)$ phase space. The theoretically anticipated collisional 
 cutoffs [right -- see (\ref{de_LVc})] and $\beta$-stabilisation thresholds (horizontal dashed lines) for the CES mirror and parallel transverse instabilities, respectively, are also shown. \textbf{b)} 
 Maximum normalised microinstability growth rate (red) versus $\mathrm{Ma} \, \lambda_e/L_V$ for a fixed electron inertial scale $d_e/L_V = 10^{-15}$, 
 along with the maximum growth rate for the mirror instability (purple) in the limit $\mathrm{Ma} \, \lambda_e \beta_e/L_V  \gg 1$ [see (\ref{mirrorgrowth})], 
 and for the parallel transverse instability 
 in the limit $\mathrm{Ma} \, \lambda_e \beta_e/L_V  \gg \mu_e^{-1/2}$ [see (\ref{transverse_maxgrowth}), with $\theta = 
 0^{\circ}$].
 \textbf{c)} Parallel wavenumber of the fastest-growing microinstability (red) versus $\mathrm{Ma} \, \lambda_e/L_V$ for a fixed electron inertial scale $d_e/L_V = 10^{-15}$, 
 along with the same quantity analytically predicted for the mirror instability (purple) in the limit $\mathrm{Ma} \, \lambda_e \beta_e/L_V \gg 1$ [see 
 (\ref{mirrorgrowth_kval})], and for the parallel transverse instability (blue) in the limit $\mathrm{Ma} \, \lambda_e \beta_e/L_V  \gg \mu_e^{-1/2}$ [see 
 (\ref{transverse_maxgrowth_kpl_pp_val}), with $\theta = 0^{\circ}$].
  \textbf{d)} Wavevector angle $\theta \equiv \tan^{-1}{(k_{\|}/k_{\perp})}$ of the fastest-growing instability over the $\left(d_e/L_V,\mathrm{Ma} \, \lambda_e \beta_e/L_V\right)$ parameter space. \label{Figure9}}
\end{figure}

\begin{figure}
\centerline{\includegraphics[width=0.99\textwidth]{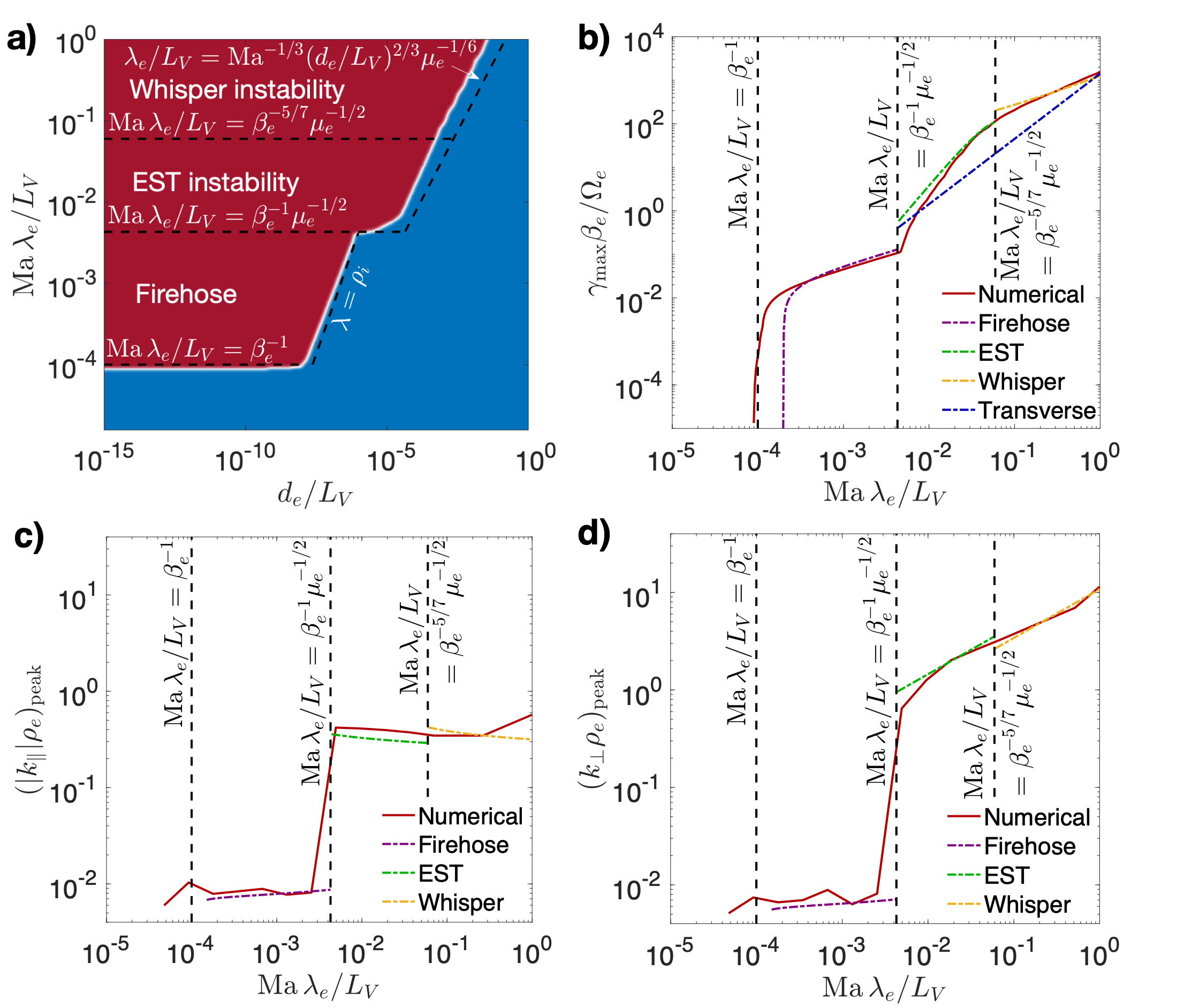}}
\caption{\textit{CE-distribution-function stability map for CES microinstabilities driven by negative pressure anisotropy}. Same as figure \ref{Figure9}, but for negative values of the small parameters $\epsilon_e$ and $\epsilon_i$. \textbf{a)} Stable (blue) and unstable (red) regions of $\left(d_e/L_V,\mathrm{Ma} \, \lambda_e/L_V\right)$ phase space. The theoretically anticipated collisional 
 cutoffs [right -- see (\ref{de_LVc})] for the CES firehose and oblique transverse instabilities, respectively, and the $\beta$-stabilisation thresholds (horizontal dashed lines)
 for the CES firehose, CES electron-scale-transition (EST) and whisper instabilities are also shown. \textbf{b)} 
 Maximum normalised microinstability growth rate (red) versus $\mathrm{Ma} \, \lambda_e/L_V$ for a fixed electron inertial scale $d_e/L_V = 10^{-15}$, 
 along with analytically predicted maximum growth rate for the firehose instability (purple) [see (\ref{firehose_growth_critline_cyclo})], 
for the EST instability (green) in the limit $\mu_e^{-1/2} \beta_e^{-5/7} \gg \mathrm{Ma} \, \lambda_e/L_V \gg \mu_e^{-1/2} \beta_e^{-1}$ [see (\ref{EST_maxgrowth})]
for the whisper instability (yellow) in the limit $\mu_e^{-1/2} \beta_e^{-1/3} \gg \mathrm{Ma} \, \lambda_e/L_V \gg \mu_e^{-1/2} 
\beta_e^{-5/7}$ [see (\ref{whiswavinstab_gamma})], and for the oblique transverse instability (blue) in the limit $\mathrm{Ma} \, \lambda_e/L_V \gg \mu_e^{-1/2} \beta_e^{-1}$
 [see (\ref{oblique_transverseinstab_peakgrowth})].
 \textbf{c)} Same as b), but for the parallel wavenumber of the fastest-growing microinstability. The analytical predictions of this quantity for the firehose instability (purple) [see (\ref{firehosepeakgrowthrate_cyclo_kval})], 
for the EST instability (green) [see (\ref{EST_maxgrowth_kval}\textit{b})],
and for the whisper instability (yellow) [see (\ref{whiswavinstab_k}\textit{b})], respectively, are also shown. \textbf{d)} Same as b), but for the perpendicular wavenumber of the fastest-growing microinstability. The analytical predictions of this quantity for the firehose instability (purple) [see (\ref{firehosepeakgrowthrate_cyclo_kval})], 
for the EST instability (green) [see (\ref{EST_maxgrowth_kval}\textit{a})],
and for the whisper instability (yellow) [see (\ref{whiswavinstab_k}\textit{a})], are also shown. \label{Figure10}}
\end{figure}

 The stability of 
 CE distribution functions of the form (\ref{CEsheardistfuncexpression}) is determined as a function of 
 the parameters $\epsilon_i$, $\epsilon_e$, $d_e$, $\beta_e$, $\beta_i$, and the velocity scale length $L_V = |\left(\hat{\boldsymbol{z}} \hat{\boldsymbol{z}} - \frac{1}{3}\mathsfbi{I} \right) \! \boldsymbol{:} \! \mathsfbi{W}_i/V_i|^{-1}$ by assessing whether the maximum microinstability growth rate across all wavelengths smaller than $\lambda_e$ and $\lambda_i$
 is negative or positive (see appendix \ref{CES_method_append} for the methodology underpinning this calculation).   
 As with the temperature-gradient-driven instabilities, we report the 
 results of stability calculations that pertain to a 
 temperature-equilibrated hydrogen plasma; that is, the particular case in which $\beta_i = \beta_e$ and $\epsilon_e = \mu_e^{1/2} 
 \epsilon_i$ [where we recall that the characteristic magnitude of 
 the CE electron velocity-shear term in such a plasma is smaller than the analogous CE ion velocity-shear 
 term by a factor of $\mu_e^{1/2} = (m_e/m_i)^{1/2}$]. 
 Because $\epsilon_i$ can take both positive and negative values (see section 
 \ref{Results_shearingterm_theory}), we do one stability calculation for each 
 case; the results of these two calculations are shown in figures \ref{Figure9} and \ref{Figure10}, 
 respectively. 
The key characteristics of the stability of the CE distribution function 
(\ref{CEsheardistfuncexpression}) for ions and electrons can be shown using 
plots over a two-dimensional $(d_e/L_V,\mathrm{Ma} \, \lambda_e/L_V)$ parameter space at fixed $\beta_e$ 
and $\mathrm{Ma}$ -- we remind the reader that $\mathrm{Ma} \, \lambda_e/L_V = |\epsilon_i|$, and that 
the Mach number $\mathrm{Ma}$ is assumed to satisfy $\mathrm{Ma} \lesssim 1$  -- as opposed to the five-dimensional $(\epsilon_i,d_e,L_V,\beta_e,\mathrm{Ma})$ parameter space that might naively be anticipated, because the two relevant stability thresholds are not independent 
functions of $d_e$, $\mathrm{Ma}$, and $L_V$. 

The regions of stability presented in figure \ref{Figure9}a for $\epsilon_i > 0$ (viz., for shear flows that drive
positive pressure anisotropy) and in figure \ref{Figure10}a for $\epsilon_i < 0$ (viz., for shear flows driving 
negative pressure anisotropy), respectively, are broadly similar
to the region of stability for CET microinstabilities described in section 
\ref{heatflux_stab} (and shown in figure \ref{Figure3}a), but with one crucial difference. Once again, for $d_e/L_V$ less than a critical value $\left(d_e/L_V\right)_{\rm c0}$, stability is 
independent of $d_e/L_V$, and there are no instabilities for $\mathrm{Ma} \, \lambda_e \beta_e/L_V \ll 
1$; for $d_e/L_V \gtrsim \left(d_e/L_V\right)_{\rm c0}$ and $\mathrm{Ma} \, \lambda_e \beta_e/L_V 
> 1$, stability is guaranteed if (and only if) $d_e/L_V > \left(d_e/L_V\right)_{\rm c}$ at fixed $\mathrm{Ma} \, \lambda_e/L_V$, where $\left(d_e/L_V\right)_{\rm c}$
is a monotonically increasing function of $\mathrm{Ma} \, \lambda_e/L_V$. 
As before, these two bounding thresholds correspond to the 
$\beta$-stabilisation conditions and collisional stabilisation conditions, 
respectively, of CES microinstabilities. However, the dependence of $\left(d_e/L_V\right)_{\rm c}$ on 
$\mathrm{Ma} \, \lambda_e/L_V$ is more complicated than the analogous relationship 
between $\left(d_e/L_T\right)_{\rm c}$ and 
$\mathrm{Ma} \, \lambda_e/L_T$ that was presented in figure \ref{Figure3}a. Namely,
if $\mathrm{Ma} \, \lambda_e/L_T \gtrsim \beta_e^{-1} \mu_e^{-1/2}$, then
$\left(d_e/L_V\right)_{\rm c}$ suddenly shifts towards a larger value, with the 
subsequent (power-law) relationship between $\left(d_e/L_V\right)_{\rm c}$ and $\mathrm{Ma} \, \lambda_e/L_V$
being distinct from the analogous relationship when $\mathrm{Ma} \, \lambda_e/L_T \lesssim \beta_e^{-1} 
\mu_e^{-1/2}$. This behaviour is the result of a feature of the unstable region that is present for CES but not CET microinstabilities: different instabilities being dominant in different regions of 
the $(d_e/L_V,\mathrm{Ma} \, \lambda_e/L_V)$ parameter space. 
As we will see, this arises because CES microinstabilities on ion scales have less stringent $\beta$-stabilisation thresholds
than those on electron scales. Although their regions of stability are 
qualitatively similar, the types of  
microinstabilities that arise when $\epsilon_i > 0$ or $\epsilon_i < 0$ are quite different, so we now discuss each case in turn. 

\subsubsection{Positive pressure anisotropy} \label{poseps_stab_pospress}

For $\epsilon_i > 0$ and $0.5 \mu_e^{-1/2} \beta_e^{-1} \gtrsim \mathrm{Ma} \, \lambda_e/L_V \gg \beta_e^{-1}$, 
the fastest-growing CES microinstability is the \emph{mirror instability}: 
that is, a non-propagating, compressible slow mode on ion scales that is destabilised by positive 
ion pressure anisotropy. For $\mathrm{Ma} \, \lambda_e \beta_e/L_V \gtrsim 0.5
\mu_e^{-1/2}$, a faster-growing CES microinstability emerges on electron Larmor scales, driven 
by positive electron pressure anisotropy: the \emph{whistler (electron-cyclotron) 
instability}. For fixed $\beta_i$, 
the CES mirror instability can operate at smaller values of $\mathrm{Ma} \, \lambda_e/L_V$ 
than the CES whistler instability, because the mirror-instability threshold 
$\Delta_i \beta_i = 3 \mathrm{Ma} \, \lambda_e \beta_i/2 L_V \geq 1$ (see section \ref{pospress_ion_mirror})
is a less stringent condition on $\mathrm{Ma} \, \lambda_e/L_V$ for fixed $\beta_e$ than the threshold $\Delta_e \beta_e = 3 \mu_e^{1/2} \mathrm{Ma} \, \lambda_e \beta_i/2 L_V \gtrsim 
0.5$ of the CES whistler instability (see section \ref{pospres_electron_EC}). 
On the other hand, once $\mathrm{Ma} \, \lambda_e \beta_e/L_V \gtrsim 0.5 \mu_e^{-1/2}$,
the maximum growth rate of the CES mirror instability $\gamma_{\rm mirr} \sim \Delta_i \Omega_i$ 
is much smaller than that of the CES whistler instability: $\gamma_{\rm whistler,S} \sim \Delta_e \Omega_e \sim \mu_e^{-1/2} \Delta_i \Omega_i  
\gg \Delta_i \Omega_i$. For $\mathrm{Ma} \, \lambda_e \beta_e/L_V \gg 
\mu_e^{-1/2}$, in addition to unstable whistler modes, modes on sub-electron-Larmor scales
are also destabilised: this is the \emph{parallel transverse instability}, a microinstability that is
essentially unmagnetised ($k \rho_i \gg 1$) in character. 
When it can operate, the CES parallel transverse instability has a much larger growth rate than the unstable electron-Larmor-scale whistler waves, $\gamma_{\rm trans} \sim \Delta_e \left(\Delta_e \beta\right)^{1/2} \Omega_e \gg 
\gamma_{\rm whist} \sim \Delta_e \Omega_e$, so if $\mathrm{Ma} \, \lambda_e \beta_e/L_V \gg 
\mu_e^{-1/2}$, the transverse instability dominates. 

Numerical evidence for the dominance of the CES mirror instability when $\mu_e^{-1/2} \gg \mathrm{Ma} \, \lambda_e/L_V \gg 1$, and then the CES parallel transverse 
instability when $\mathrm{Ma} \, \lambda_e/L_V \gg \mu_e^{-1/2}$, can be produced by isolating the maximum growth rate, the parallel wavenumber and the wavevector angle associated with peak 
growth for the unstable regions of the $(d_e/L_V,\mathrm{Ma} \, \lambda_e/L_V)$ parameter space. Figure 
\ref{Figure9}b shows that, for fixed $d_e/L_V$ and a range of $\mathrm{Ma} \, \lambda_e/L_V$, 
the peak microinstability growth rate is a reasonable match for that 
of the mirror instability [viz., (\ref{mirrorgrowth})] for $0.5 \mu_e^{-1/2} \beta_e^{-1} \gtrsim \mathrm{Ma} \, \lambda_e/L_V \gg \beta_e^{-1}$, 
and a good match for the parallel transverse instability [viz., (\ref{transverse_maxgrowth})] for $\mathrm{Ma} \, \lambda_e/L_V \gtrsim \mu_e^{-1/2} \beta_e^{-1}$. 
Figure \ref{Figure9}c demonstrates that, for $\mu_e^{-1/2} \beta_e^{-1} \gtrsim \mathrm{Ma} \, \lambda_e/L_V \gg \beta_e^{-1}$, the (non-dimensionalised) parallel wavenumber $(k_{\|} \rho_e)_{\rm peak}$ of peak growth satisfies $(k_{\|} \rho_e)_{\rm peak} \sim \mu_e^{-1/2}$, in agreement with the expected parallel wavenumber of the fastest-growing mirror modes [see (\ref{mirrorgrowth_kval})]. At $\mathrm{Ma} \, \lambda_e/L_V \sim \mu_e^{-1/2} 
\beta_e^{-1}$, there is a dramatic shift in $(k_{\|} \rho_e)_{\rm peak}$ to a value $(k_{\|} \rho_e)_{\rm peak} \gtrsim 1$ that agrees with the expected parallel wavenumber of the parallel transverse instability [see (\ref{transverse_maxgrowth_kpl_pp_val})]. As for the peak-growth wavevector angle (figure \ref{Figure9}d), for $\beta_e^{-1} \lesssim \mathrm{Ma} \, \lambda_e/L_V \lesssim \mu_e^{-1/2} 
\beta_e^{-1}$, the dominant instability is oblique (as would be expected for the mirror 
instability), while for $\mathrm{Ma} \, \lambda_e/L_V \gtrsim 0.5 \mu_e^{-1/2} \beta_e^{-1}$, it is parallel (implying that the CES whistler/parallel transverse instability dominates). We conclude that the mirror instability 
is indeed dominant when $0.5 \mu_e^{-1/2} \beta_e^{-1} \gtrsim \mathrm{Ma} \, \lambda_e/L_V \gg \beta_e^{-1}$, 
and the parallel transverse instability when $\mathrm{Ma} \, \lambda_e/L_V \gg \mu_e^{-1/2} \beta_e^{-1}$. 

\subsubsection{Negative pressure anisotropy} \label{poseps_stab_negpress}

Now considering the case when $\epsilon_i < 0$, i.e., the case of negative 
pressure anisotropy, the only CES microinstability that operates when 
$\mu_e^{-1/2} \beta_e^{-1} \gtrsim \mathrm{Ma} \, \lambda_e/L_V \gg \beta_e^{-1}$
is the \emph{firehose instability}: the destabilisation of Alfv\'en waves by ion pressure 
anisotropies $\Delta_i \lesssim -1/\beta_i$\footnote{In the limit of wavelengths much larger than the ion Larmor radius, the firehose instability 
threshold is well known to be $\Delta_i = (\Delta_i)_{\rm c} < -2/\beta_i$. However, for plasmas whose ion species have either a CE distribution function or a bi-Maxwellian distribution,  
the instability threshold for oblique ion-Larmor-scale firehose modes is somewhat less stringent: see section 
\ref{negpres_fire}.}. 
If $\mathrm{Ma} \, \lambda_e/L_V \gtrsim \mu_e^{-1/2} \beta_e^{-1}$, several 
electron-scale CES microinstabilities arise, all of which tend to have larger growth rates than the firehose 
instability. The first of these to develop (at $\mathrm{Ma} \, \lambda_e/L_V \sim \mu_e^{-1/2} \beta_e^{-1}$) 
is the \emph{oblique electron firehose instability}: the destabilisation of oblique kinetic-Alfv\'en waves by 
negative electron pressure anisotropy. For
$\mu_e^{-1/2} \beta_e^{-1} \lesssim  \mathrm{Ma} \, \lambda_e/L_V \lesssim  \mu_e^{-1/2} 
\beta_e^{-5/7}$, the \emph{electron-scale-transition (EST) instability} begins to operate; this is a non-propagating 
quasi-perpendicular mode on electron Larmor scales ($k_{\perp} \rho_e \sim 1 \gg k_{\|} \rho_e$), which, 
while damped in a Maxwellian plasma, is unstable for sufficiently negative 
electron pressure anisotropies, and grows more rapidly than the oblique electron firehose instability. For 
$\mu_e^{-1/2} \beta_e^{-5/7} \lesssim \mathrm{Ma} \, \lambda_e/L_V \lesssim \mu_e^{-1/2} 
\beta_e^{-1/3}$, the EST instability is surpassed by the 
\emph{whisper instability}: the instability of a newly discovered propagating wave in a Maxwellian plasma (a \emph{whisper wave}) whose 
perpendicular wavelength is on sub-electron-Larmor scales ($k_{\perp} \rho_e \gg 1$), but 
whose parallel wavelength is above the electron-Larmor scale ($k_{\|} \rho_e < 
1$). Finally, when $\mathrm{Ma} \, \lambda_e/L_V \gtrsim \mu_e^{-1/2} 
\beta_e^{-1/3}$, the \emph{oblique transverse instability} comes to predominate; 
unlike either the oblique electron firehose, the EST, or whisper instabilities,
it is unmagnetised in nature (like its parallel relative). Of these four 
instabilities, the oblique electron firehose and transverse instabilities have been 
identified previously (see references in sections \ref{negpres_electron_oblique} and \ref{negpres_subelectron_obliquetrans}, respectively), but not the EST or whisper instabilities. 

We support these claims (in an analogous manner to the $\epsilon_i > 0$ case) by calculating the growth rate of the dominant microinstabilities for given points 
in the $(d_e/L_V,\mathrm{Ma} \, \lambda_e/L_V)$ parameter space.
 Figure \ref{Figure10}b shows the maximum growth rate for a 
fixed value of $d_e/L_V$. For $\mu_e^{-1/2} \beta_e^{-1} \gtrsim \mathrm{Ma} \, \lambda_e/L_V \gg \beta_e^{-1}$, 
the peak growth rate follows the analytical prediction for the ion firehose instability, $\gamma_{\rm fire} \sim |\Delta_i|^{1/2} \Omega_i/\sqrt{\log{1/|\Delta_i|}}$,
when $\Delta_i \ll -2/\beta_i$ [see (\ref{firehose_growth_critline_cyclo})]. 
For $\mathrm{Ma} \, \lambda_e/L_V \gtrsim \mu_e^{-1/2} \beta_e^{-1}$, the peak growth rate becomes much greater than   
$\gamma_{\rm fire}$; for $\beta_e^{-5/7} \gtrsim \mu_e^{1/2} \mathrm{Ma} \, \lambda_e/L_V \gg 
\beta_e^{-1}$, it instead matches that of the EST instability, $\gamma_{\mathrm{EST}} \sim |\Delta_e| \left(|\Delta_e|
\beta_e\right)^{3/2}
 \Omega_e/\sqrt{\log{|\Delta_e| \beta_e}}$ [see (\ref{EST_maxgrowth})], where we remind the reader that 
 $|\Delta_e| = 3 \mu_e^{1/2} \mathrm{Ma} \, \lambda_e/2 L_V$. For $\mu_e^{1/2} \mathrm{Ma} \, \lambda_e/L_V \gg 
\beta_e^{-5/7}$, the observed growth rate agrees with an analytical prediction for the whisper instability, 
$\gamma_{\mathrm{whisp}} \sim |\Delta_e|^{1/2} \left(|\Delta_e| \beta_e\right)^{1/4}
 \Omega_e/\sqrt{\log{|\Delta_e| \beta_e}}$ [see (\ref{whiswavinstab_gamma})].
Finally, because of the value of $\beta_e$ chosen for this numerical example, the condition  
$\mathrm{Ma} \, \lambda_e/L_V \gtrsim \mu_e^{-1/2} \beta_e^{-1/3}$ under which 
the oblique transverse instability dominates is never met for $\mathrm{Ma} \, \lambda_e/L_V \ll 
1$, and thus the numerically measured growth rate of the dominant CES microinstability is larger than the tranverse instability's peak growth rate $\gamma_{\rm trans} \sim |\Delta_e| \left(|\Delta_e| \beta_e\right)^{1/2} \Omega_e$ [see (\ref{oblique_transverseinstab_peakgrowth})] for 
the entire range of $\mathrm{Ma} \, \lambda_e/L_V$ that we show in figure \ref{Figure10}b, (blue line) . 
 
A further confirmation that the most important microinstabilities are those that we have explicitly identified is obtained by
 calculating the parallel 
and perpendicular wavenumbers associated with 
the dominant microinstability. Figures \ref{Figure10}c and \ref{Figure10}d show that, for $\beta_e^{-1} \ll \mathrm{Ma} \, \lambda_e/L_V \ll \mu_e^{-1/2} 
\beta_e^{-1}$, $(k_\| \rho_e)_{\rm{peak}} \sim (k_\perp \rho_e)_{\rm{peak}} \sim 
\mu_e^{1/2}$. These values of $(k_\| \rho_e)_{\rm{peak}}$ are consistent with the properties of the fastest-growing
unstable firehose modes (see sections \ref{negpres_fire} and \ref{negpres_fire_critline}), whose parallel wavenumber (approximately) satisfies $(k_\| \rho_i)_{\rm{peak}} \sim 1/\sqrt{\log{1/|\Delta_i|}}$ when $\Delta_i \ll -2/\beta_{i}$ [see (\ref{firehosepeakgrowthrate_cyclo_kval})], and whose wavevector angle is 
$\theta_{\mathrm{peak}} \approx 39^{\rm{o}}$. 
At $\mathrm{Ma} \, \lambda_e/L_V \sim \mu_e^{-1/2} \beta_e^{-1}$,
the magnitudes of the parallel and perpendicular wavenumbers changes abruptly, to $(k_\| \rho_e)_{\rm{peak}} \sim (k_\perp \rho_e)_{\rm{peak}} 
\sim 1$; this is in line with expectations from the onset of the oblique electron firehose instability 
when $|\Delta_e| \beta_e \sim 1$. For $\mathrm{Ma} \, \lambda_e/L_V \gg 
\beta_e^{-1}$ ($|\Delta_e| \beta_e \gg 1$), the parallel scale of the fastest-growing mode  
remains above electron Larmor scales [$(k_\| \rho_e)_{\rm{peak}}  < 
1$], while $(k_\perp \rho_e)_{\rm{peak}}$ increases monotonically 
above unity. Both findings match theoretical expectations 
concerning the evolution of the parallel and perpendicular wavenumbers of the EST and whisper instabilities 
as functions of increasing $|\Delta_e| \beta_e$, and analytic formulae for these 
quantities are in reasonable agreement with the numerical results (see sections \ref{negpres_elec_EST} and  \ref{negpres_subelectron_phantom}). 

\subsubsection{Collisional stabilisation} \label{poseps_stab_collstab}

For both $\epsilon_i > 0$ and $\epsilon_i < 0$, the shift in 
$\left(d_e/L_V\right)_{\rm c}$ at $\mathrm{Ma} \, \lambda_e/L_V \sim \mu_e^{-1/2} \beta_e^{-1}$ observed in 
figures \ref{Figure9}a and \ref{Figure10}a can be explained in terms of the ion-scale and electron-scale microinstabilities 
having distinct collisional-stabilisation conditions of the form (\ref{colldamp}) (viz., $k \lambda_e \sim k \lambda_i \lesssim 1$), with the condition on the 
ion-scale instabilities being more restrictive. 
The wavenumbers $k_{\rm mirr}$ and $k_{\rm fire}$ at which 
maximal growth of the ion mirror and firehose instabilities occurs satisfy $k_{\rm mirr} \rho_i \sim 1$ and $k_{\rm fire} \rho_i \lesssim 1$, respectively, for 
$\mathrm{Ma} \, \lambda_e \beta_e/L_V \gg 1$,
leading to the collisional-stabilisation condition
\begin{equation}
\frac{\lambda_e}{L_V} \lesssim \frac{\rho_i}{L_V} \sim \mu_e^{-1/2} \beta_e^{1/2} \frac{d_e}{L_V} \,  .  \label{mirror_coll_stab}
\end{equation}
For the electron-scale microinstabilities, the parallel and the oblique transverse 
instabilities have the largest (common) wavenumber of all such instabilities
that operate when $\epsilon_i > 0$ and $\epsilon_i < 0$, respectively, and 
so provide the most demanding collisional-stabilisation conditions. 
For both transverse instabilities, the wavenumber at which peak growth occurs for the 
satisfies $k_{\rm trans}  \rho_e \sim (\mu_e^{1/2} \mathrm{Ma} \, \lambda_e \beta_e/L_V)^{1/2}$ 
[see (\ref{transverse_maxgrowth_kval})], which in turn can be rearranged to give 
the collisional-stabilisation condition 
\begin{equation}
 \frac{\lambda_e}{L_V} \lesssim  \mathrm{Ma}^{-1/3} \mu_e^{-1/6} \left(\frac{d_e}{L_V}\right)^{2/3} \, .  \label{transverse_coll_stab}
\end{equation}
Bringing these results together, we find
\begin{equation}
    \left(\frac{d_e}{L_V}\right)_{\rm c} = \left\{
      \begin{array}{ll}
      \mu_e^{1/2} \beta_e^{-1/2} {\lambda_e}/{L_V} , & \beta_e^{-1} \ll \mathrm{Ma} \, {\lambda_e}/{L_V} < \mu_e^{-1/2} \beta_e^{-1},  \\[2pt]
      \mu_e^{1/4} \mathrm{Ma}^{1/2} \left({\lambda_e}/{L_V}\right)^{3/2},  & \mathrm{Ma} \, {\lambda_e}/{L_V} \gtrsim \mu_e^{-1/2} 
      \beta_e^{-1},
      \end{array} \right. \label{de_LVc}
\end{equation}
with $\left(d_e/L_V\right)_{\rm c0} = \mu_e^{1/2} \beta_e^{-3/2}$.
This matches 
asymptotically the numerical results shown in figures \ref{Figure9}a and \ref{Figure10}a.
These findings confirm that, 
once again, the relevant collisional-stabilisation condition for the microinstabilities with
wavenumber $k$ is $k \lambda_e  = k \lambda_i\ll 1$
[viz., (\ref{colldamp})], as opposed to the more restrictive conditions $\gamma \tau_i \gg 1$ and $\gamma \tau_e \gg 1$ on the CES ion-scale and electron-scale instabilities, respectively.
Similary to the collisional-stabilisation condition on the CET whistler instability (see section \ref{heatflux_stab}), we note that the collisional-stabilisation condition on any of these 
microinstabilities can \emph{never} actually be satisfied in a strongly magnetised plasma, because $k \lambda_i \gtrsim \lambda_i/\rho_i \gg 
1$ for the ion-scale instabilities, and $k \lambda_e \gtrsim \lambda_e/\rho_e \gg 1$ for the electron-scale instabilities.

\subsubsection{Outline of the rest of this section} \label{poseps_stab_outlinesec} 

Further discussion about the properties and growth rates of CES microinstabilities 
with $\epsilon_s > 0$ (viz., those driven by positive pressure anisotropy) can be found in 
section \ref{CES_pos_micro}, with the mirror, whistler and transverse 
instabilities discussed in sections \ref{pospress_ion_mirror}, \ref{pospres_electron_EC} and 
\ref{pospres_electron_trans}, respectively. In addition to these, 
there is another instability (the \emph{electron mirror instability}) that can be driven by positive pressure 
anisotropy of CE distribution functions that we note in passing: it consists in KAWs driven unstable by the CE electron-shear term, and to some extent by the ion-shear term (section \ref{pospres_electron_oblique}). 
The electron mirror instability does not appear 
to be the fastest-growing CES microinstability anywhere in the $(d_e/L_V,\mathrm{Ma} \, \lambda_e/L_V)$ parameter 
space; since the instability is subdominant to two other electron-scale instabilities (the whistler and transverse 
instabilities), this would seem to imply that the instability is comparatively less important. 

CES microinstabilities with $\epsilon_s < 0$ (viz., those driven by negative pressure 
anisotropy) are explored in section \ref{CES_neg_aniso}. The firehose 
instability is overviewed in section 
\ref{negpres_fire}, with then four subclasses of the instability (parallel, oblique, critical-line, and sub-ion-Larmor-scale) considered in sections \ref{negpres_fire_par}, \ref{negpres_fire_oblique}, \ref{negpres_fire_critline}, and\ref{negpres_fire_subion}. The oblique electron firehose instability is discussed in section 
\ref{negpres_electron_oblique}, the EST instability in section 
\ref{negpres_elec_EST}, the oblique transverse instability in section
\ref{negpres_subelectron_obliquetrans}, and the whisper instability in section 
\ref{negpres_subelectron_phantom}. We identify two additional CES microinstabilities 
which are never the fastest-growing microinstability in any unstable region:
the parallel electron firehose instability (section 
\ref{negpress_electronfire_prl}), which (in spite of its name) has a different 
underlying physical mechanism than the oblique electron firehose, and the 
ordinary-mode instability (section \ref{negpres_subelectron_ord}), which only 
operates at very high $\beta_e$ ($\beta_e \gtrsim |\Delta_e|^{-3}$), and is only characteristically 
distinct from the oblique transverse instability in a regime in which 
it is slower growing. 

Readers who do not wish to dwell on specific CES
microinstabilities should proceed directly to section \ref{Discussion}.

\subsection{CES microinstability classification: positive pressure anisotropy ($\epsilon_i > 0$)} \label{CES_pos_micro}

\subsubsection{Mirror instability}  \label{pospress_ion_mirror}

The CES mirror instability consists in the destabilisation of compressive slow modes 
by a sufficiently large positive 
ion pressure anisotropy associated with the ion-shear term of the ion CE
distribution function. In a high-$\beta$ plasma with Maxwellian ion and electron distribution functions, the slow mode -- which is one of the two plasma 
modes which exist at oblique wavevector angles $\theta \gtrsim \beta_i^{-1/4}$ (the other being the shear Alfv\'en 
wave), and consists of a perturbation to the magnetic field's strength -- is non-propagating, being subject to strong Barnes' (equivalently, transit-time) 
damping~\citep{B66}. This damping is the result of Landau-resonant interactions
between the slow mode and co-moving ions with $v_{\|} = \omega/k_{\|}$; since, 
for a distribution function that decreases monotonically with $v_{\|} > 0$, there 
are more ions with $v_{\|} < \omega/k_{\|}$ than with $v_{\|} > \omega/k_{\|}$, 
there is a net transfer of free energy from the slow modes to the ions (as a particle acceleration 
process, this is sometimes called betatron acceleration). However, in a plasma
with $\Delta_i > 0$, there is an increase in the relative number of ions with large pitch angles 
in the troughs of the slow mode's magnetic-field strength perturbation, 
giving rise to excess perpendicular pressure. When $\Delta_i > 1/\beta_i$, this excess pressure overbalances the
magnetic pressure, leading to the mirror instability. In CE plasma with $0 < \Delta_i \beta_i -1 \ll 1$, only quasi-perpendicular long-wavelength
mirror modes ($k_{\|} \rho_i \ll k_{\perp} \rho_i \ll 1$) are destabilised; for 
larger values of $\Delta_i$, a broad range of slow modes (including ion-Larmor-scale ones) 
become unstable. Chronologically, the earliest discussions of the 
mirror instability in pressure-anisotropic plasmas are due to~\citet{P58} and~\citet{H69}. 
\citet{SK93} provide a detailed and lucid discussion of the linear physics of the mirror instability~\citep[see also][]{KSCAC15}; various analytical~\citep{PSBO08,RSC15} and 
numerical~\citep{HKPS09,KSS14,RQV15,MSK16}
studies investigating its nonlinear evolution have also been carried out. 

The CES mirror instability can be characterised analytically -- and simple expressions 
derived for the maximum growth rate and the wavevector at which that growth is attained -- in the limit of 
marginal instability. First, we define the threshold parameter $\Gamma_i \equiv \beta_i \Delta - 1$, 
where $\Delta \equiv \Delta_i + \Delta_e = (1+\mu_e^{1/2})\Delta_i$, 
and assume that $\Gamma_i \ll 1$. It can then be shown (see appendix \ref{derivation_mirror}) that under the orderings
\begin{equation}
  k_{\|} \rho_i  \sim k_{\bot}^2 \rho_i^2 \sim \Gamma_i \ll 1 \, , \quad  \frac{\gamma}{\Omega_i} \sim \frac{\Gamma_i^2}{\beta_i} 
  \ll 1
  \, ,
  \label{mirror_order_marg}
\end{equation}
the mirror modes have a growth rate given by
 \begin{equation}
  \frac{\gamma}{\Omega_i} = \frac{k_{\|} \rho_i}{\sqrt{\upi} \beta_i}  \left(\Gamma_i -\frac{3}{2} \frac{k_{\|}^2}{k_{\bot}^2}-\frac{3}{4} k_{\bot}^2 \rho_i^2 \right) 
  \, . \label{mirrorgrowthrate_marg}
 \end{equation}
  This is the same result as the growth rate of the mirror instability in a 
  bi-Maxwellian plasma, with 
 (the anticipated) threshold $\Gamma_i > 0$~\citep{H07}. 
The peak growth rate $\gamma_{\mathrm{max}}$ 
 is then given by
 \begin{equation}
 \gamma_{\mathrm{max}} = \frac{ \Gamma_i^2}{6 \sqrt{2 \upi} \beta_i} \Omega_i \, ,
 \end{equation}
achieved at the wavenumber 
\begin{equation}
(k_{\|} \rho_i)_{\mathrm{peak}} = 
\frac{\Gamma_i}{3\sqrt{2}}    \, , \quad (k_{\bot} \rho_i)_{\mathrm{peak}} = \frac{\Gamma_i^{1/2}}{\sqrt{3}}\, . \label{mirror_marg_kval}
\end{equation}
This recovers the results of \citet{H07}. 
 
\begin{figure}
\centerline{\includegraphics[width=0.99\textwidth]{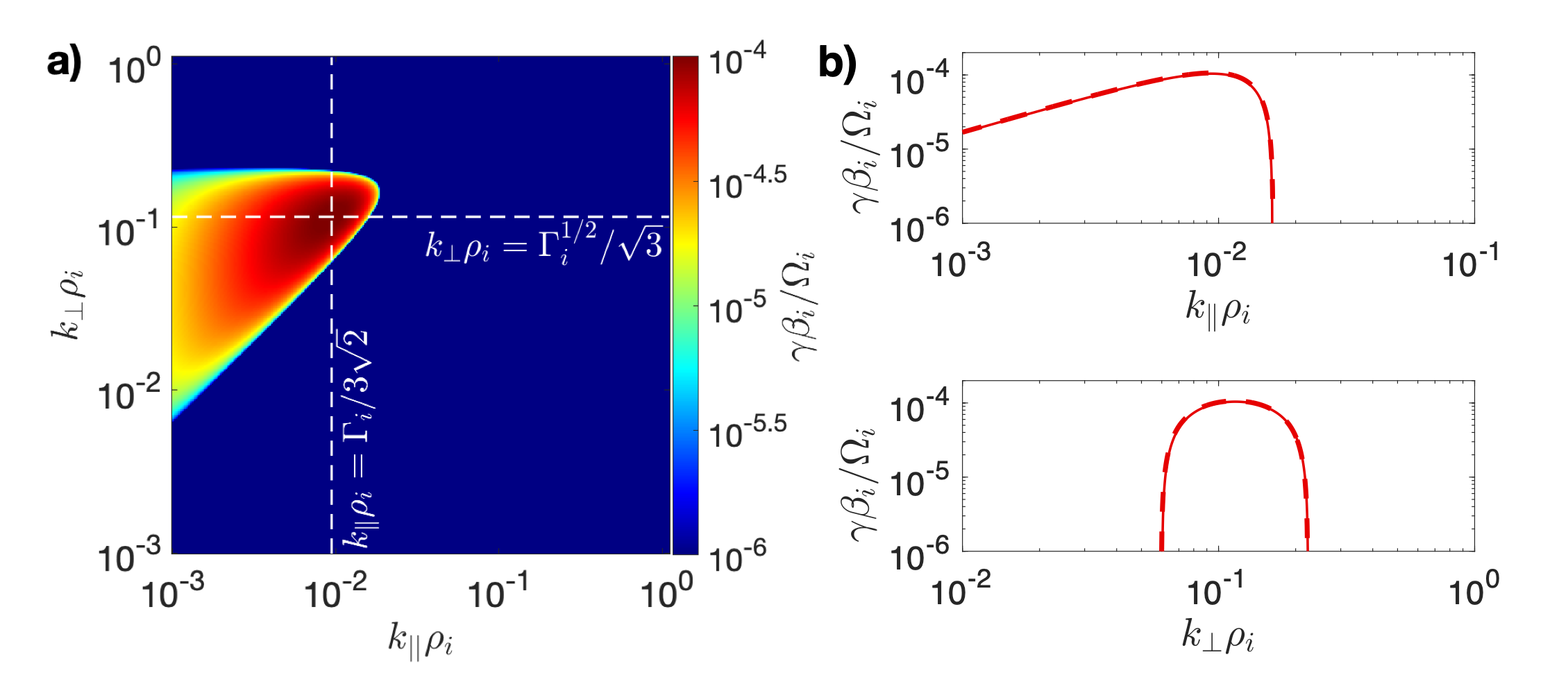}}
\caption{\textit{Mirror instability at $\Gamma_i = \Delta \beta_i -1 \ll 1$}. \textbf{a)} Growth rates of unstable mirror modes resulting from the CE ion-shear term in the CE distribution function (\ref{CEsheardistfuncexpression}) for $\Gamma_i = 0.04 \ll 1$ ($\Delta \beta_i = 1.04$).
The growth rates of all modes are calculated using the approach outlined in appendix \ref{CES_method_append_quartic}. The growth rates are calculated on a $400^2$ grid, with logarithmic spacing 
in both perpendicular and parallel directions between the minimum and maximum wavenumber magnitudes.
The resulting growth rates, when normalised as $\gamma \beta_i/\Omega_i$, are functions of the dimensionless quantity $\Delta \beta_i$. 
The dashed white lines indicate the analytical prediction (\ref{mirror_marg_kval}) 
for the wavenumber at which peak growth is achieved. 
\textbf{b)} The mirror mode's growth rate (solid line) as a function of $k_{\|} \rho_i$ with $k_{\bot} \rho_i = \Gamma_i^{1/2}/\sqrt{3}$ (top), and as a function of $k_{\perp} \rho_i$ with $k_{\|} \rho_i = \Gamma_i/3\sqrt{2}$ (bottom). The 
dashed lines show the analytical prediction (\ref{mirrorgrowthrate_marg}) for these quantities. \label{newFig_mirror_marg}}
\end{figure}

Figure \ref{newFig_mirror_marg} illustrates the accuracy of the above predictions for $\gamma$ (and therefore $\gamma_{\rm max}$), 
$(k_{\|} \rho_i)_{\mathrm{peak}}$ and $(k_{\bot} \rho_i)_{\mathrm{peak}}$ by 
comparing them with the equivalent values obtained numerically using the general method outlined in appendix  
\ref{CES_method_append} for a particular value of $\Gamma_i \ll 1$.
The wavenumber dependence of the numerically determined growth rate
(see figure \ref{newFig_mirror_marg}a) corroborates that, close to marginality, the unstable 
mirror modes are quasi-perpendicular; more quantitatively, the values of 
$k_{\|} \rho_i$ and $k_{\bot} \rho_i$
at which peak growth is obtained numerically match (\ref{mirror_marg_kval}). 
Furthermore, the growth rate (\ref{mirrorgrowthrate_marg}) 
agrees well with the numerical result when plotted as a function of $k_{\|} \rho_i$ with fixed $k_{\bot} \rho_i$,
and also as a function of $k_{\bot} \rho_i$ with fixed $k_{\|} \rho_i$ (figure 
\ref{newFig_mirror_marg}b).

\begin{figure}
\centerline{\includegraphics[width=0.99\textwidth]{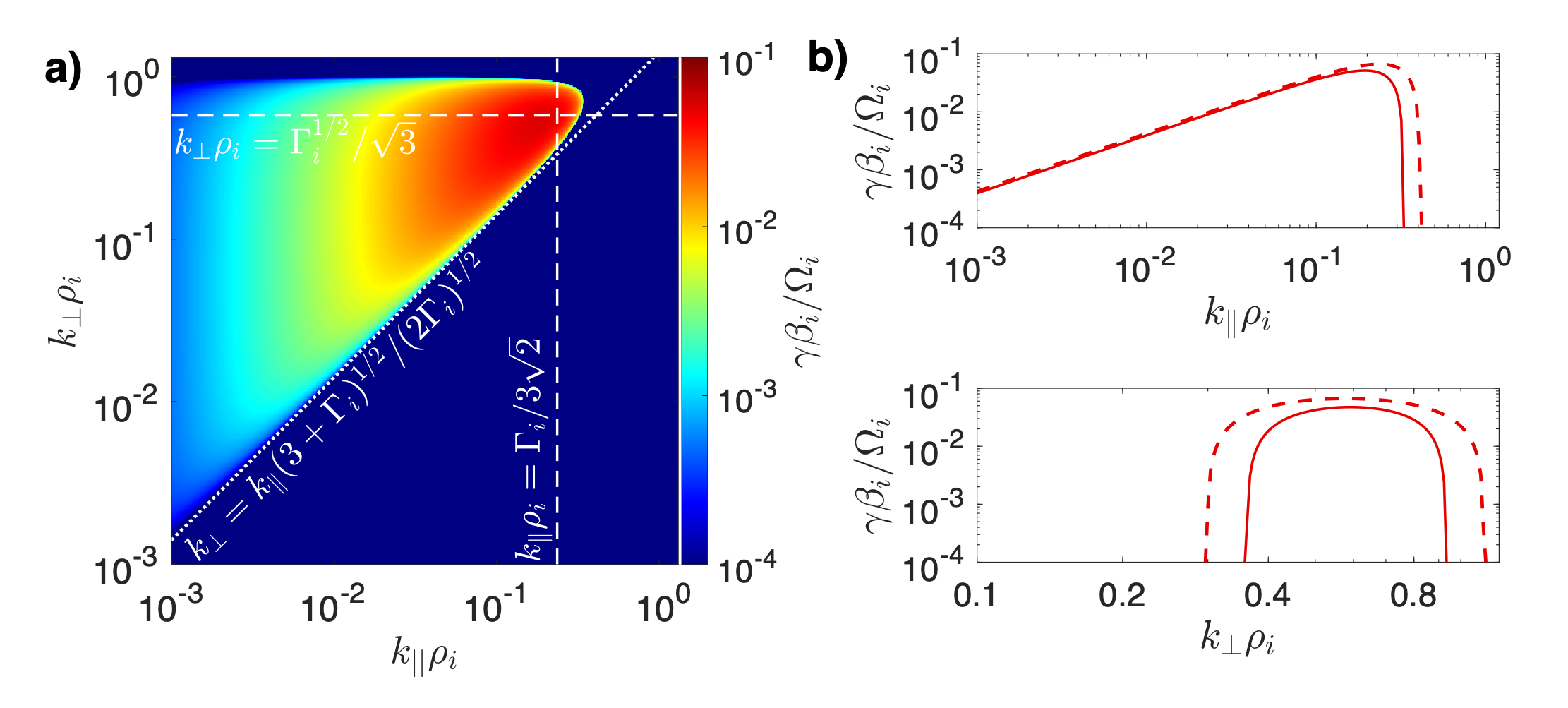}}
\caption{\textit{Mirror instability at $\Gamma_i = \Delta \beta_i -1 \sim 1$}. \textbf{a)} Growth rates of unstable mirror modes resulting from the CE ion-shear term in the CE distribution function (\ref{CEsheardistfuncexpression}) for $\Gamma_i = 1$ ($\Delta \beta_i = 2$).
The growth rates of all modes are calculated in the same way as in figure \ref{newFig_mirror_marg}. The dashed white lines indicate the analytic prediction (\ref{mirror_marg_kval}) 
for the parallel/perpendicular wavenumber at which peak growth is achieved, while the dotted line
indicates the analytical prediction (\ref{mirror_angleinstab}) for the perpendicular
wavenumber above which 
long-wavelength ($k_{\|} \rho_i \lesssim k_{\perp} \rho_i \ll 1$) mirror modes 
become unstable. 
\textbf{b)} The mirror mode's growth rate (solid line) as a function of $k_{\|} \rho_i$ with $k_{\bot} \rho_i = \Gamma_i^{1/2}/\sqrt{3}$ (top), and as a function of $k_{\perp} \rho_i$ with $k_{\|} \rho_i = \Gamma_i/3\sqrt{2}$ (bottom). The 
dashed lines show the analytical prediction (\ref{mirrorgrowthrate_marg}) for this quantity.  \label{newFig_mirror_unity}}
\end{figure}

In contrast, for finite $\Gamma_i \gtrsim 1$, simple expressions for $\gamma_{\rm max}$, 
$(k_{\|} \rho_i)_{\mathrm{peak}}$, and $(k_{\bot} \rho_i)_{\mathrm{peak}}$ are 
challenging to derive analytically. Our numerical calculations indicate that, when $\Gamma_i \sim 
1$, a broad range of (purely growing) oblique modes becomes unstable, with 
 maximum growth rate $\gamma_{\rm max} \sim \Omega_i/\beta_i \sim \Delta
\Omega_i$ attained when $k_{\|} \rho_i \lesssim k_{\perp} \rho_i \sim 1$ 
 (figure \ref{newFig_mirror_unity}a).
Therefore, asymptotic expansions that treat $k_{\bot} \rho_i$ and $k_{\|} \rho_i$ 
as small or large cannot be used to derive simplified expressions for the growth 
rate of the fastest-growing mirror modes. While the expressions (\ref{mirror_marg_kval}) for the wavenumber of peak growth derived
in the case of near-marginality remain 
qualitatively correct, they are no longer quantitatively accurate; the same 
conclusion applies to the expression (\ref{mirrorgrowthrate_marg}) for the growth rate 
when $k_{\|} \rho_i \sim k_{\perp} \rho_i \sim 1$ (figure  
\ref{newFig_mirror_unity}b). That being said, an expression similar to 
(\ref{mirrorgrowthrate_marg}) can be derived (see appendix \ref{derivation_mirror}) for long-wavelength unstable mirror 
modes
that satisfy the ordering 
\begin{equation}
  k_{\|} \rho_i  \sim k_{\bot} \rho_i  \ll 1 \, , \quad  \frac{\gamma}{\Omega_i} \sim \frac{k_{\|} \rho_i}{\beta_i} \sim {\Delta} k_{\|} \rho_i
  \ll 1
  \, .
  \label{mirror_order}
\end{equation}
This expression is 
 \begin{equation}
  \frac{\gamma}{\Omega_i} = \frac{k_{\|} \rho_i}{\sqrt{\upi} \beta_i}  \left(\Gamma_i -\frac{\Gamma_i + 3}{2} \frac{k_{\|}^2}{k_{\bot}^2}\right) 
  \, . \label{mirrorgrowthrate}
 \end{equation}
 It implies that all such modes with 
 \begin{equation}
 k_{\bot} > \left(\frac{3+\Gamma_i}{2 \Gamma_i}\right)^{1/2} k_{\|} \label{mirror_angleinstab}
 \end{equation} 
 will be unstable, a prediction that is consistent with the unstable region 
 observed in figure \ref{newFig_mirror_unity}a. 

\begin{figure}
\centerline{\includegraphics[width=0.99\textwidth]{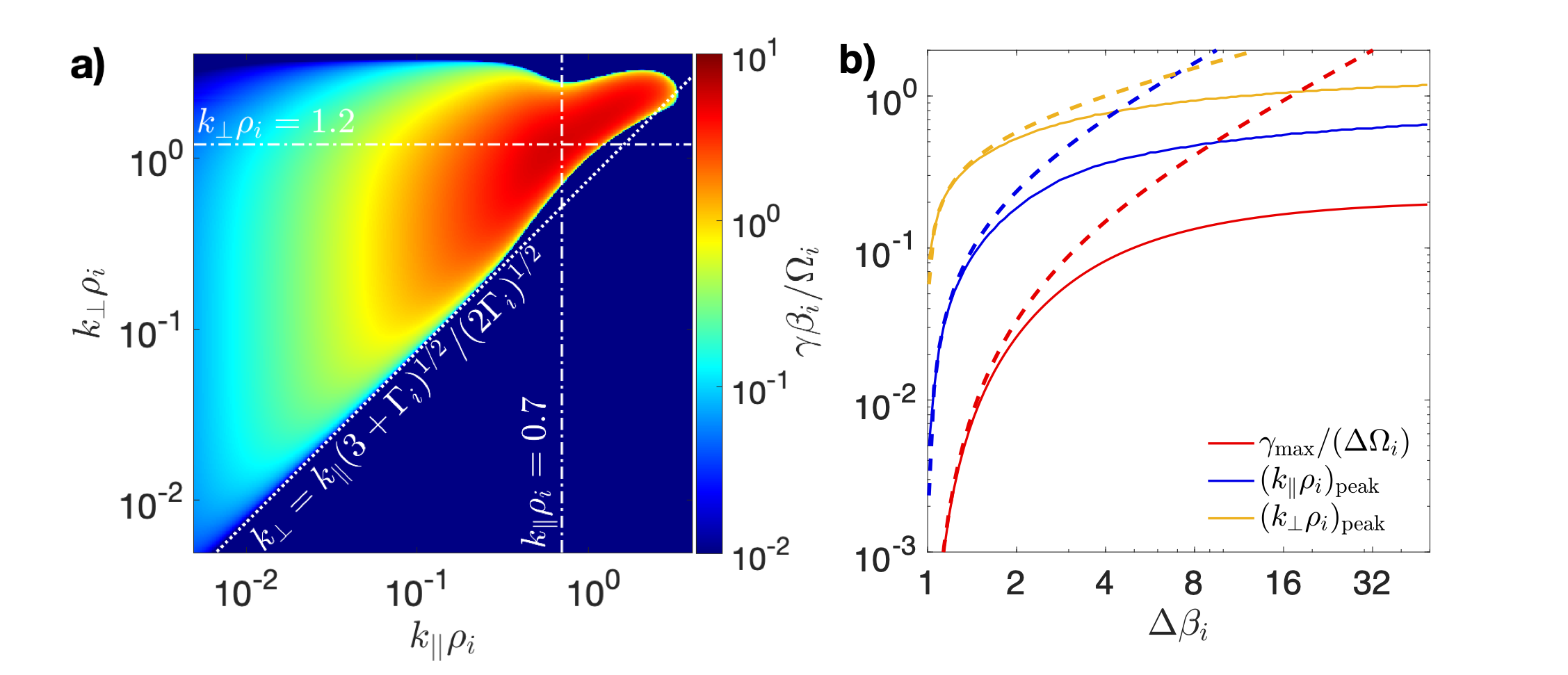}}
\caption{\textit{Mirror instability at $\Gamma_i = \Delta \beta_i \gg 1$}. \textbf{a)} Growth rates of unstable mirror modes resulting from the CE ion-shear term in the CE distribution function (\ref{CEsheardistfuncexpression}) for $\Gamma_i = 29 \gg 1$ ($\Delta \beta_i = 30$).
The growth rates of all modes are calculated in the same way as in figure \ref{newFig_mirror_marg}. 
The dot-dashed white lines indicate the parallel/perpendicular wavenumbers (\ref{mirrorgrowth_kval}) at which peak growth is achieved, 
while the dotted line indicates the analytical prediction (\ref{mirror_angleinstab}) for the perpendicular
wavenumber above which 
long-wavelength ($k_{\|} \rho_i \lesssim k_{\perp} \rho_i \ll 1$) mirror modes 
become unstable. 
\textbf{b)} Normalised maximum positive growth rate $\gamma_{\rm max}/\Delta \Omega_i$ (solid red line) of the unstable mirror mode as
a function of $\Delta \beta_i$ along with the parallel (solid blue line) and perpendicular (solid yellow line) wavenumbers, $(k_{\|} \rho_i)_{\mathrm{peak}}$ and $(k_{\perp} \rho_i)_{\mathrm{peak}}$ respectively, at
which that growth is attained. The analytical prediction (\ref{mirrorgrowthrate_marg}) of $\gamma_{\rm max}$ for marginally unstable modes,
as well as the analogous predictions (\ref{mirror_marg_kval}) for $(k_{\|} \rho_i)_{\mathrm{peak}}$ and $(k_{\perp} \rho_i)_{\mathrm{peak}}$, are shown as dashed lines. \label{newFig_mirror_lge}}
\end{figure}

When $\Gamma_i \gg 1$, but $\Gamma_i < (m_i/m_e)^{1/2}$, 
the region of $(k_{\|},k_{\bot})$ space in which mirror modes are unstable is qualitatively similar 
to the $\Gamma_i \sim 1$ case, albeit more extended (figure 
\ref{newFig_mirror_lge}a). 
We find that in this limit, the maximum growth rate 
$\gamma_{\mathrm{max}}$ becomes  
directly proportional to $\Delta$ (see figure 
\ref{newFig_mirror_lge}b), in contrast to the marginal case (\ref{mirrorgrowthrate_marg}):
\begin{equation}
  \gamma_{\mathrm{max}} \approx 0.2 \Delta \Omega_i \, . \label{mirrorgrowth}
\end{equation}
This growth is attained at parallel and perpendicular wavenumbers
\begin{equation}
  (k_{\bot} \rho_i)_{\mathrm{peak}} \approx 1.2 \, , \quad (k_{\|} \rho_i)_{\mathrm{peak}} \approx 
0.7 \, , \label{mirrorgrowth_kval}
\end{equation}
which depend only weakly on $\Delta \beta_i$.

Some understanding of these results can be derived by considering the dispersion 
relation of mirror modes on sub-ion Larmor scales. Adopting the ordering
\begin{equation}
 k_{\|} \rho_i \sim k_{\perp} \rho_i \sim (\Delta_i \beta_i)^{1/2} \gg 1 , \quad 
\frac{\gamma}{\Omega_i} \sim \Delta_i \, , \label{order_subionmirror}
\end{equation}
while assuming that $\Delta_i \beta_i \ll \mu_e^{-1/2}$, 
one finds (see appendix \ref{derivation_mirror}) that
\begin{equation}
\frac{\gamma}{\Omega_i} \approx \frac{k_\|}{k} \sqrt{\left(\frac{k^2 \rho_i^2}{\beta_i} - \Delta_i \frac{k_{\|}^2-k_{\bot}^2}{k^2} \right) \left(\Delta_i \frac{k_{\|}^2}{k^2} - \frac{k^2 \rho_i^2}{\beta_i} \right)} \, . 
 \label{ionmirror_subionscale}
\end{equation}
This can be re-written in terms of the wavevector angle $\theta = \tan^{-1}{(k_{\perp}/k_{\|})}$ as
\begin{equation}
\frac{\gamma}{\Omega_i} \approx \cos{\theta} \sqrt{\left[\frac{k^2 \rho_i^2}{\beta_i} - \Delta_i \left(\cos^2{\theta}-\sin^2{\theta}\right) \right] \left(\Delta_i \cos^2{\theta} - \frac{k^2 \rho_i^2}{\beta_i} \right)} \, . 
 \label{ionmirror_subionscale_theta}
\end{equation}
Analysing this expression leads to three conclusions. First, for $\theta > 45^{\circ}$, there is an
instability at all wavenumbers satisfying $k \rho_i < (\Delta_i \beta_i)^{1/2} 
\cos{\theta}$, explaining the expansion of the unstable region of $(k_{\|},k_{\bot})$-space 
with increasing $\Delta_i \beta_i$. For $\theta \leq 45^{\circ}$, growth only occurs over a more limited range 
of wavenumbers $\sqrt{\cos^2{\theta}-\sin^2{\theta}} < k \rho_i/(\Delta_i \beta_i)^{1/2} <  
\cos{\theta}$. Secondly, growth in this limit is maximised when $k \rho_i \ll (\Delta_i 
\beta_i)^{1/2}$, with the maximal growth rate
\begin{equation}
 \gamma_{\rm max} = \frac{1}{3 \sqrt{3}} \Delta_i \Omega_i  \approx 0.19 \Delta_i \Omega_i \label{mirrorgrowth_largek}
\end{equation}
attained at $\cos{\theta} = 1/\sqrt{3}$ ($\theta \approx 55^{\circ}$). This expression for $\gamma_{\rm max}$ 
is (surprisingly) close to the numerically measured peak growth rate 
(\ref{mirrorgrowth}). For $k \rho_i \sim (\Delta_i 
\beta_i)^{1/2}$, the maximum growth rate is smaller than 
(\ref{mirrorgrowth_largek}) by an order-unity factor. Finally, when $k \rho_i \gg (\Delta_i \beta_i)^{1/2}$, viz., 
in a wavenumber regime where there are no unstable mirror modes, (\ref{ionmirror_subionscale})
becomes imaginary, implying that the modes have a real frequency given by
\begin{equation}
\omega \approx \pm k_\| k_{\bot} \rho_e \frac{\Omega_e}{\beta_i} \, . \label{KAWdisprel_mirror}
\end{equation} 
This is the dispersion relation of kinetic Alfv\'en waves (KAWs) in a high-$\beta$ 
plasma\footnote{We note that (\ref{KAWdisprel_mirror}) is also the same dispersion relation as that of oblique whistler waves~\citep[see, e.g.,][]{GM15}. 
However, as was discussed in section \ref{electron_heatflux_instab_whistl}, in a 
high-$\beta$ plasma ($\beta_e \gg \mu_e^{-1/2}$), the small frequency ($\omega \ll k_{\|} v_{\mathrm{th}i}$) of 
perturbations prohibits all but parallel perturbations from not interacting significantly with 
the ions, and thus we believe that the modes are more accurately identified as KAWs.}. 
In short, at $\Delta_i \beta_i \gg 1$, KAWs are also destabilised by positive ion pressure 
anisotropy in addition to longer-wavelength mirror modes. We note that KAWs can also be destabilised by positive electron anisotropy, but the characteristic wavelength of such modes is preferentially comparable to electron Larmor scales (see section \ref{pospres_electron_oblique}). 

\subsubsection{Whistler instability} \label{pospres_electron_EC}

The CES whistler instability arises when  
the free energy associated with positive electron-pressure anisotropy $\Delta_e$ of the
electron CE distribution function destabilises whistler waves, overwhelming both the electron cyclotron damping (which 
is the dominant stabilisation mechanism for whistler 
waves with $k_{\|} \rho_e \sim 1$) and the Landau damping due to the ion species (the domininant 
stabilisation mechanism for waves with $k_{\|} \rho_e \ll 1$). 
In the special case of static ions, electron cyclotron damping can
be overcome by a positive electron-pressure anisotropy of any magnitude for whistler waves
with sufficiently long wavelengths. 
Retaining mobile ions, the instability operates only if $\Delta_e$
exceeds a threshold of order $(\Delta_e)_{\rm c} \sim \beta_e^{-1}$.
When $\Delta_e > (\Delta_e)_{\rm c}$, 
gyroresonant interactions between electrons with $v_{\|} = \pm \Omega_e/k_{\|}$
and whistler waves allow for free energy to pass from the former to the latter, 
and so an increasingly broad spectrum of unstable parallel and oblique modes emerges on 
electron Larmor scales. 
The analogue of this instability in a bi-Maxwellian plasma
was found by~\citet{KP66}, and it has since been studied numerically in moderately high-$\beta$ plasma ($\beta_e \sim 1$-$10$) by
several authors~\citep[e.g.,][]{GW96,GSN14,RQV16}. 

\begin{figure}
\centerline{\includegraphics[width=0.99\textwidth]{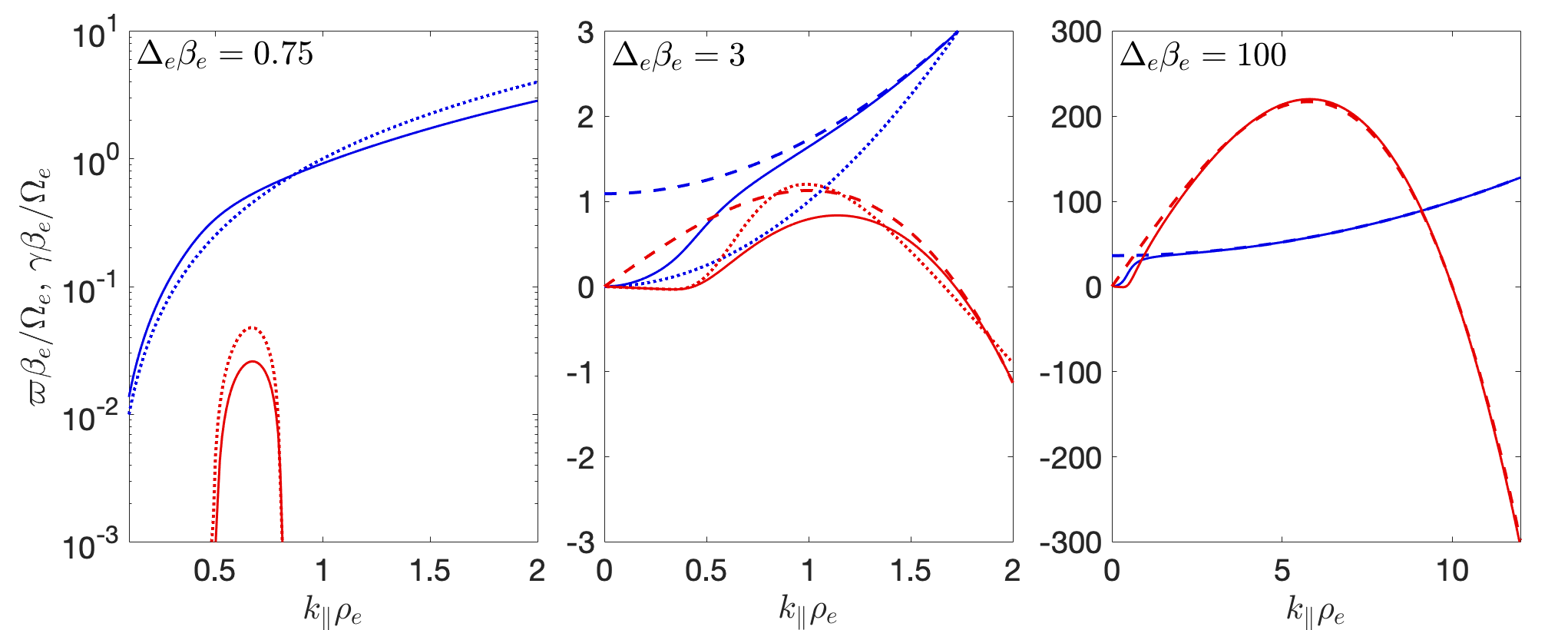}}
\caption{\textit{Parallel CES whistler instability}.  Dispersion curves of unstable whistler modes whose instability is driven by the electron-shear term in CE distribution function 
(\ref{CEsheardistfuncexpression}), for wavevectors that are co-parallel with the background 
magnetic field (viz., $\boldsymbol{k} = k_{\|} \hat{\boldsymbol{z}}$). 
The frequency (solid blue) and growth rate (solid red) of the modes are calculated using 
(\ref{electronweibelgrowthrate}\textit{a}) and 
(\ref{electronweibelgrowthrate}\textit{b}), respectively. 
The resulting frequencies and growth rates, when normalised as $\gamma \beta_e/\Omega_e$, are functions of the dimensionless quantity $\Delta_e \beta_e$; we show the dispersion curves for three different values of $\Delta_e \beta_e$. 
The approximations (\ref{CESwhistlerinstabgrowthrate_smallk}\textit{a}) and (\ref{CESwhistlerinstabgrowthrate_smallk}\textit{b}) for 
the frequency (dotted blue) and growth rate (dotted red) in the limit $k_{\|} \rho_e \ll 1$ are also plotted, 
as are the approximations (\ref{electronweibelgrowthrate_largek}\textit{a}) and (\ref{electronweibelgrowthrate_largek}\textit{b}) 
for the frequency (dashed blue) and growth rate (dashed red) in the limit $k_{\|} \rho_e \gg 
1$. 
 \label{Figure_newCESwhist}}
\end{figure}

Similarly to the CET whistler instability, the simplest characterisation
of the CES whistler instability is for unstable parallel whistler modes 
(viz., $k \approx k_{\|}$). Assuming that these modes satisfy the 
orderings
\begin{equation}
  \tilde{\omega}_{e\|} = \frac{\omega}{k_{\|} v_{\mathrm{th}e}} \sim \Delta_e \sim \frac{1}{\beta_e} \, , k_{\|} \rho_e \sim 1 , 
  \label{electronWeibel_order}
\end{equation}  
it can be shown (see appendix \ref{derivation_parwhistler}) that their real frequency $\varpi$ and growth rate $\gamma$
satisfy
\begin{subeqnarray}
\frac{\varpi \beta_e}{\Omega_e} & = &  \pm \Delta_e \beta_e
\pm \frac{k_{\|} \rho_e \left[\Delta_e \beta_e \left(1+\mu_e^{1/2}\right)- k_\|^2 \rho_e^2\right] \Real{\; Z\!\left({1}/{k_\| \rho_e}\right)} }{\left[\Real{\; Z\!\left({1}/{k_\| \rho_e}\right)}\right]^2 + \upi \exp{\left(-{2}/{k_\|^2 \rho_e^2}\right)}} 
 , \\
\frac{\gamma \beta_e}{\Omega_e}  & = & 
\frac{k_{\|} \rho_e \left[\exp{\left(-{1}/{k_\|^2 \rho_e^2}\right)}+\mu_e^{1/2}\right]\left(\Delta_e \beta_e- k_\|^2 \rho_e^2\right) +\mu_e^{1/2} \Delta_e \beta_e \Real{\; Z\!\left({1}/{k_\| \rho_e}\right)} }{\left[\Real{\; Z\!\left({1}/{k_\| \rho_e}\right)}\right]^2/\sqrt{\upi} + \sqrt{\upi} \exp{\left(-{2}/{k_\|^2 \rho_e^2}\right)}} 
 ,  \qquad \quad \, \label{electronweibelgrowthrate}
\end{subeqnarray}
where the terms proportional to $\mu_e^{1/2}$ are associated with the ion species\footnote{Formally, these terms are $\textit{O}(\mu_e^{1/2})$ under our assumed ordering,
and so should be dropped. However, because of the exponential dependence of the other damping/growth terms on $k_{\|} \rho_e$, 
these terms play an important role for moderate values of $k_{\|} \rho_e$, viz. $\mu_e^{1/2} \exp{\left({1}/{k_\|^2 \rho_e^2}\right)} \geq 1$ for $k_\| \rho_e \leq \sqrt{2}/\sqrt{\log{m_i/m_e}} \approx 
0.5$, so we retain them.}. In the limit $\mu_e \rightarrow 0$,  
formally there is always instability provided $\Delta_e \beta_e > 0$; however, 
for a hydrogen plasma ($\mu_e \approx 1/1836$), it can be shown numerically that the 
numerator of (\ref{electronweibelgrowthrate}\textit{b}) only becomes positive (over a narrow interval 
of parallel wavenumbers around $k_{\|} \rho_e \approx 0.60$) for $\Delta_e \beta_e > 
0.56$. The
dispersion curves $\varpi(k_{\|})$ and $\gamma(k_{\|})$ of the unstable whistler waves in 
a hydrogen plasma for three different values of $\Delta_e \beta_e$ 
that are above the necessary value for instability are shown in 
figure \ref{Figure_newCESwhist}. 
When $\Delta_e \beta_e \gtrsim 1$, the growth rate is postive 
for a range $\Delta k_{\|} \sim \rho_e^{-1}$ around $k_\| \rho_e \sim 
1$, attaining a characteristic magnitude $\gamma \sim \varpi \sim \Omega_e/\beta_e$. 

As before, we characterise the growth rate for various 
values of $\Delta_e \beta_e$ by taking subsidiary limits. 
First, for $\Delta_e \beta_e \ll 1$, a necessary (though not always sufficient) condition  
for positive growth is $k_\| \rho_e < (\Delta_e \beta_e)^{1/2} \ll 1$. 
We therefore expand (\ref{electronweibelgrowthrate}) in $k_\| \rho_e \sim (\Delta_e \beta_e)^{1/2} \ll 
1$, finding that
\begin{subeqnarray}
\varpi& \approx & \frac{k_\|^2 \rho_e^2}{\beta_e} \Omega_e
\, , \\
\gamma & \approx & 
\frac{\sqrt{\upi}}{k_\| \rho_e} \left\{\exp{\left(-\frac{1}{k_\|^2 \rho_e^2}\right)} \left(\Delta_e - \frac{k_\|^2 \rho_e^2}{\beta_e}\right) -\mu_e^{1/2} \frac{k_\|^2 \rho_e^2}{\beta_e} \right\} \Omega_e . \qquad  \label{CESwhistlerinstabgrowthrate_smallk}
\end{subeqnarray}
Similarly to what we showed in section \ref{electron_heatflux_instab_whistl} for the CET whistler instability, 
we have once again found unstable 
whistler waves. For comparison's sake, the approximate expressions (\ref{CESwhistlerinstabgrowthrate_smallk}) 
are plotted in figure \ref{Figure_newCESwhist} in addition 
to their exact analogues (\ref{electronweibelgrowthrate}); it is clear that 
there is reasonable agreement for a moderately small value of $\Delta_e 
\beta_e$, but that the approximations become less accurate for  $k_\| \rho_e \gtrsim 
0.5$ and $\Delta_e \beta_e > 1$. 

In the limit $\mu_e \rightarrow 0$, the expression (\ref{CESwhistlerinstabgrowthrate_smallk}\textit{b}) 
for the growth rate is very similar to that of the whistler (electron-cyclotron) instability in a plasma with a bi-Maxwellian 
distribution and positive electron pressure anisotropy~\citep{D83}. In this case, whistler modes with $k_\| \rho_e < (\Delta_e \beta_e)^{1/2}$
are always unstable, although the growth rate 
of such modes is exponentially small in $\Delta_e \beta_e \ll 1$ as compared to 
the frequency (\ref{CESwhistlerinstabgrowthrate_smallk}\textit{a}), and so $\gamma \ll \varpi \sim \Omega_e/\beta_e$. 
By contrast, with small but finite $\mu_e = m_e/m_i$, it can be shown analytically that, 
for (\ref{CESwhistlerinstabgrowthrate_smallk}\textit{b}) to be positive, $\Delta_e > (\Delta_e)_{\rm c}$, where 
\begin{eqnarray}
(\Delta_e)_{\rm c} & = & \frac{1}{\beta_e W_{\rm Lam}\left[\mu_e^{-1/2}\exp{(-1)}\right]} 
\nonumber\\
& \approx &  \frac{1}{\beta_e} \frac{1}{\log{(\mu_e^{-1/2})}-1-\log{[\log{(\mu_e^{-1/2})}-1}]}  \, . 
\label{CETwhist_thresh}
\end{eqnarray}
Here, $W_{\rm Lam}(x)$ denotes the Lambert W function~\citep{CGHJK96}. 
Unstable modes first develop around $(k_{\|} \rho_e)_c = (\Delta_e)_{\rm c}^{1/2} /[(\Delta_e)_{\rm c}+1/\beta_e]^{1/2}$. In a hydrogen plasma, this gives $(\Delta_e)_{\rm c} \approx 0.49/\beta_e$ and $(k_{\|} \rho_e)_c \approx 
0.57$, which are similar to the instability threshold and wavenumber, respectively, determined numerically if $\gamma$ is computed for  
arbitrary values of $k_{\|} \rho_e$; the small discrepancy is 
due to the finite value of $k_{\|}  \rho_e$ at which instability 
first emerges. Formally, $(\Delta_e)_{\rm c} \rightarrow 0$ as $\mu_e \rightarrow 0$, but 
the limit converges only logarithmically in $\mu_e$, suggesting that in an 
actual plasma, the CES whistler instability will generically have a threshold at a finite value of $\Delta_e \beta_e$. 

Let us now turn to the opposite subsidiary limit $\Delta_e \beta_e \gg 1$. We find from (\ref{electronweibelgrowthrate}\textit{b}) that maximal growth 
occurs at $k_\| \rho_e \sim (\Delta_e \beta)^{1/2} \gg 1$:
\begin{subeqnarray}
\varpi& \approx & \frac{1}{\upi}\left[\Delta_e \left(\upi-2\right)+ \frac{k_\|^2 \rho_e^2}{\beta_e}\right] \Omega_e
\, , \\
\gamma & \approx & 
\frac{k_\| \rho_e}{\sqrt{\upi}} \left(\Delta_e - \frac{k_\|^2 \rho_e^2}{\beta_e}\right) \Omega_e \, .  \label{electronweibelgrowthrate_largek}
\end{subeqnarray}
Alongside $k_\| \rho_e \ll 1$ approximations, these approximations are plotted in figure 
\ref{Figure_newCESwhist}, and agree well with the numerical results for $\Delta_e \beta_e \gtrsim 3$ 
and $k_\| \rho_e \gtrsim 2$. The maximum growth rate 
\begin{equation}
\gamma_{\mathrm{max}} = \frac{2}{3 \sqrt{3 \upi}} \Delta_e (\Delta_e \beta_e)^{1/2} \Omega_e \approx 0.22 \Delta_e (\Delta_e \beta_e)^{1/2} \Omega_e \label{CESwhistler_maxgrowth}
\end{equation}
is attained at the parallel wavenumber 
\begin{equation}
(k_\| \rho_e)_{\mathrm{peak}} = (\Delta_e \beta_e/3)^{1/2}. \label{CESwhistler_maxgrowth_kval}
\end{equation}
A notable feature of the CES whistler instability in this subsidary limit is that the fastest-growing 
modes are on sub-electron-Larmor scales; thus, such modes are arguably better 
conceptualised not as whistler modes, but as unstable, unmagnetised plasma modes (see section \ref{pospres_electron_trans}).

Similarly to the CET whistler instability, analytical expressions for the frequency and growth rate of 
unstable modes that have an oblique wavevector angle are much less simple 
that the analogous expressions for parallel whistler modes. It can be shown (see appendix \ref{formfordispshear}) that 
the complex frequency of such modes is given by 
\begin{equation}
  \omega = \frac{\Omega_e}{\beta_e} k_{\|} \rho_e \frac{- \mathrm{i} B_{\mathrm{S}} \pm \sqrt{-B_{\mathrm{S}}^2 + 4A_{\mathrm{S}}C_{\mathrm{S}}}}{2 A_{\mathrm{S}}} 
  \, , \label{shear_frequency}
\end{equation}
where the functions $A_{\mathrm{S}} = A_{\mathrm{S}}(k_{\|} \rho_e,k_{\perp} \rho_e,\Delta_e \beta_e)$, $B_{\mathrm{S}} = B_{\mathrm{S}}(k_{\|} \rho_e,k_{\perp} 
\rho_e,\Delta_e \beta_e)$, and $C_{\mathrm{S}} = C_{\mathrm{S}}(k_{\|} \rho_e,k_{\perp} \rho_e,\Delta_e \beta_e)$ 
are composed of the sums and products of special mathematical functions. When $\Delta_e \beta_e \sim 1$,
(\ref{shear_frequency}) implies that if there is an instability, its growth rate will be of order $\gamma \sim 
\Omega_e/\beta_e$ at $k_{\|} \rho_e, k_{\perp} \rho_e \sim 1$. 

\begin{figure}
\centerline{\includegraphics[width=0.99\textwidth]{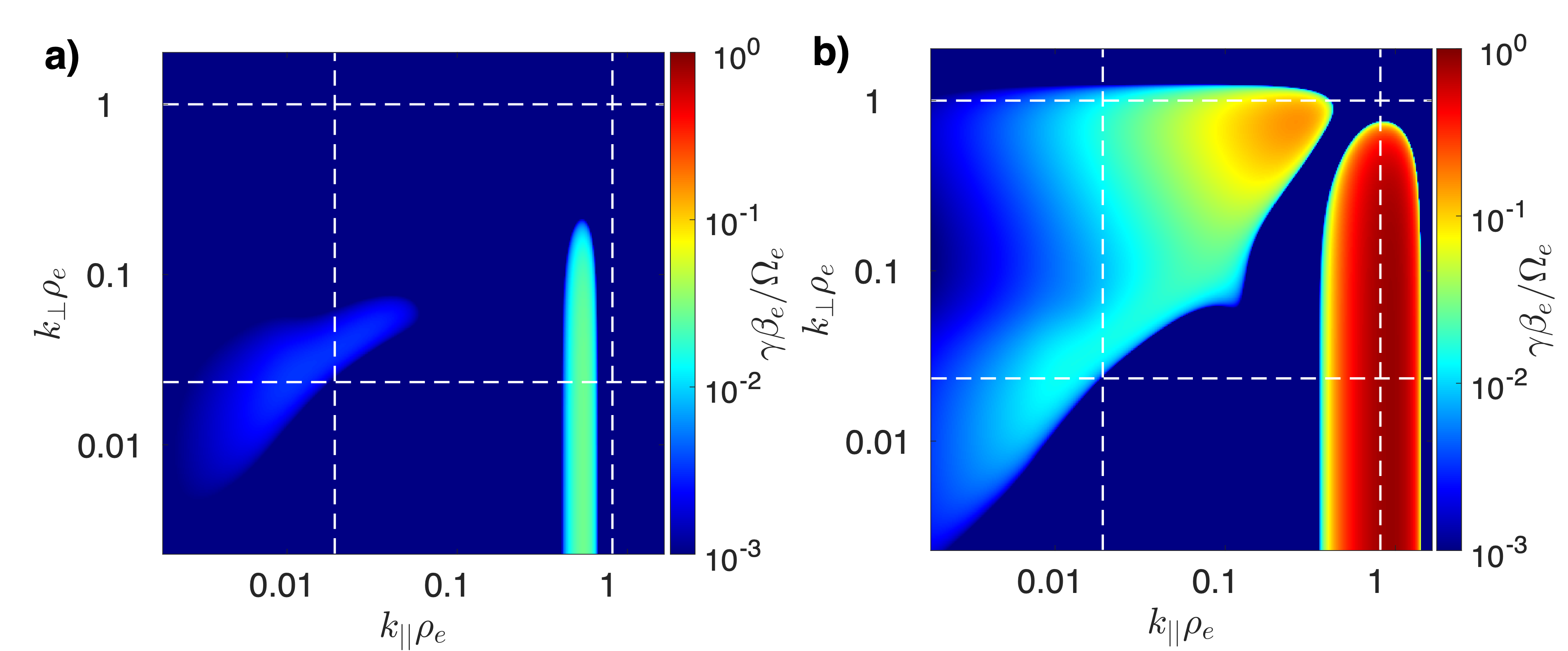}}
\caption{\textit{Oblique unstable modes at $\Delta_e \beta_e \sim 1$}: \textbf{a)} $\Delta_e \beta_e = 0.75$. 
\textbf{b)} $\Delta_e \beta_e = 3$. Maximum positive growth rates of linear perturbations 
resulting from CE ion- and electron-shear terms in the CE distribution function~(\ref{CEsheardistfuncexpression}) for $\Delta_e \beta_e \sim 1$.
Here, a temperature-equilibrated hydrogen plasma is considered, viz. $\Delta_e = \mu_e^{1/2} \Delta_i$, and $\beta_i = \beta_e$.  
The growth rates of all modes are calculated using the approach outlined in appendix \ref{CES_method_append_quartic}. The growth rates are 
calculated on a $400^2$ grid, with logarithmic spacing between wavenumbers in 
both perpendicular and parallel directions.
The resulting growth rates, when normalised as $\gamma \beta_e/\Omega_e$, are functions of $\Delta_e \beta_e$, or, equivalently, $\epsilon_e \beta_e$. The vertical dashed lines indicate $k_{\|} \rho_i = 1$ and $k_{\|} \rho_e = 
1$, respectively, while the horizontal ones indicate $k_{\perp} \rho_i = 1$ and $k_{\perp} \rho_e = 
1$. \label{Figure5}}
\end{figure}

To confirm this expectation, in figure \ref{Figure5} we plot the maximum growth rate (obtained numerically) of oblique modes across the $(k_{\|},k_{\perp})$-plane 
for two of the values of $\Delta_e \beta_e$ used in figure \ref{Figure_newCESwhist}. 
For $\Delta_e \beta_e$ not far beyond the threshold of the CES whistler 
instability (figure \ref{Figure5}a), the unstable modes are quasi-parallel
and have growth rates $\gamma \ll \Omega_e/\beta_e$
(cf. figure \ref{Figure_newCESwhist}, left panel). 
For $\Delta_e \beta_e \gtrsim 
1$, a broader spectrum of wavenumbers becomes unstable (figure 
\ref{Figure5}b). The parallel mode remains the fastest growing in this case; 
however, oblique modes with $k_{\perp} \lesssim k_{\|}/2$ also have growth rates of comparable magnitude: e.g., the fastest-growing mode with wavevector angle $\theta = 10^{\circ}$ has $\gamma_{\rm max}/\gamma_{\rm max}(k_{\perp} = 0) \approx 
0.93$, and for a wavevector angle $\theta = 10^{\circ}$, $\gamma_{\rm max}/\gamma_{\rm max}(k_{\perp} = 0) \approx 
0.76$. For more oblique angles, the growth rate is reduced significantly: e.g., for $\theta = 30^{\circ}$, $\gamma_{\rm max}/\gamma_{\rm max}(k_{\perp} = 0) \approx 
0.22$. Thus, we conclude that a spectrum of oblique modes in addition to parallel ones is 
indeed destabilised, with $\gamma \sim \Omega_e/\beta_e \lesssim \gamma(k_{\perp} = 0)$. 

We note that, in addition to oblique CES whistler modes, whose characteristic wavenumber domain is $k_{\perp} \rho_e \lesssim k_{\|} \rho_i \sim 1$, we observe two other 
unstable modes in figure \ref{Figure5}a with different characteristic values of $k_{\|}$ and $k_{\perp}$. The first of 
these, which exists on ion scales, is the CES mirror instability, which we already discussed in section 
\ref{pospress_ion_mirror}. The second is the CES electron mirror instability -- we shall consider this instability 
in section~\ref{pospres_electron_oblique}. 

\subsubsection{Parallel transverse instability} \label{pospres_electron_trans}

As was shown in section \ref{poseps_stab}, in the limit $\Delta_e \beta_e \gg 1$, the fastest-growing 
CES microinstability is essentially unmagnetised, and is a variant of the so-called transverse 
instability~\citep{K62,K64,A70}. This instability is also sometimes referred to as the resonant (electron) Weibel instability, 
or the Weibel instability at small anisotropy~\citep{W59,F59}. Both the linear theory of 
this instability and its physical mechanism have been explored extensively for 
bi-Maxwellian plasmas~\citep[see, e.g.][]{LSP09,ILS12}, and various
studies (both analytical and numerical) of its nonlinear evolution have also been performed~\citep{A70B,DHHW72,LWG79,CPBM98,CCC02,K05,PA11,RGDB15}. 
For the small anisotropy case that is relevant to CE 
plasma, the mechanism of the instability is somewhat subtle, involving both
non-resonant and Landau-resonant wave-particle interactions. In a Maxwellian plasma, transverse 
modes are non-propagating 
and Landau-damped by electrons 
with velocities $v \approx \omega/k_{\|}$. However, this damping can be 
reversed by the free energy associated with positive electron-pressure anisotropy at 
wavenumbers that satisfy $k d_e \lesssim \Delta_e^{1/2}$; the 
electron Landau damping increases more rapidly with $k$ than the instability's drive, which in turn sets 
the wavenumber at which peak growth occurs. The requirement for the corresponding scale to be well below 
the electron Larmor scale -- and thus for the plasma to be quasi-unmagnetised with respect to the transverse modes --
sets the restriction $\Delta_e \beta_e \gg 1$ on the instability's operation. 
In general, transverse modes whose wavevectors are co-parallel to the velocity-space direction along
which the temperature is smallest are the fastest growing; in the case of a CE electron distribution function of the form (\ref{CEsheardistfuncexpression}) with $\Delta_e > 0$, these modes' wavevectors
are parallel to the magnetic field. However, a broad spectrum of 
oblique transverse modes is also destabilised when $\Delta_e > 0$. 

To characterise the transverse instability's growth analytically, we first assume $\Delta_e \beta_e \gg 
1$, and then take directly the unmagnetised limit of the full CES dispersion relation 
(see appendix \ref{derivation_transverse}) under the orderings
\begin{equation}
k_{\bot} \rho_e \sim k_{\|} \rho_e \sim \left(\Delta_e \beta_e\right)^{1/2} \gg 1 \, , \quad  \tilde{\omega}_{e\|} = \frac{\omega}{k_{\|} v_{\mathrm{th}e}}  \sim \Delta_e \, .
\end{equation}
We obtain two non-propagating modes (real frequency $\varpi = 0$) that have growth rates
 \begin{subeqnarray}
 \gamma_1 & = & \frac{k v_{\mathrm{th}e}}{\sqrt{\upi}} \left(\Delta_e \frac{k_{\|}^2-k_{\bot}^2}{k^2} - \frac{k^2 \rho_e^2}{\beta_e}\right) \, , 
 \\
  \gamma_2 & = & \frac{k v_{\mathrm{th}e}}{\sqrt{\upi}} \left(\Delta_e \frac{k_{\|}^2}{k^2} - \frac{k^2 \rho_e^2}{\beta_e}\right) \, . \label{transverse_oblique_growthrate}
 \end{subeqnarray}
For $\Delta_e > 0$, the growth rate of the second mode is always positive 
and larger than that of the first mode; the first mode only has a positive growth rate provided $k_{\bot} < k_{\|}$. Now 
taking the subsidiary limit $k_{\|} \rho_e \gg k_{\bot} \rho_e \gg 1$, we find that 
both roots have the same growth rate:
\begin{equation} 
\gamma \approx \frac{k_{\|} v_{\mathrm{th}e}}{\sqrt{\upi}} \left(\Delta_e - \frac{k_{\|}^2 \rho_e^2}{\beta}\right) 
\, , \label{partransverse_growthrate}
\end{equation}
which is identical to (\ref{electronweibelgrowthrate_largek}\textit{b}). We note by comparison with (\ref{electronweibelgrowthrate_largek}\textit{a}) that 
the unmagnetised limit fails to recover the non-zero real frequencies of the $k_{\|} \rho_e \gg 1$
whistler modes; this is because the 
ratio of these modes' real frequency $\varpi$ to their growth rate $\gamma$ is $\varpi/\gamma \sim 1/k_{\|} 
\rho_e \ll 1$.

The maximum growth rate $\gamma_{\mathrm{max}}$ of the second mode (\ref{transverse_oblique_growthrate}\textit{b}) for an oblique wavevector with angle~$\theta$ is 
\begin{equation}
\gamma_{\mathrm{max}} = \frac{2}{3 \sqrt{3 \upi}} \cos^{3}{\theta} \, \Delta_e (\Delta_e \beta_e)^{1/2} \Omega_e  ,\label{transverse_maxgrowth}
\end{equation}
attained at the (total) wavenumber 
\begin{equation}
(k \rho_e)_{\mathrm{peak}} = \cos{\theta} \, (\Delta_e \beta_e/3)^{1/2}. \label{transverse_maxgrowth_kval}
\end{equation}
The parallel and perpendicular wavenumbers of this maximum growth are then
\begin{equation}
(k_{\|} \rho_e)_{\mathrm{peak}} = \cos^2{\theta} \, (\Delta_e \beta_e/3)^{1/2} , \quad (k_{\bot} \rho_e)_{\mathrm{peak}} = \cos{\theta} \sin{\theta} \, (\Delta_e \beta_e/3)^{1/2} . \label{transverse_maxgrowth_kpl_pp_val}
\end{equation}
In the special case of parallel modes ($\theta = 0^{\circ}$), this recovers the peak growth rate (\ref{CESwhistler_maxgrowth})
of the CES whistler instability at $k_{\|} $ in the limit $\Delta_e \beta_e \gg 1$. 

\begin{figure}
\centerline{\includegraphics[width=0.99\textwidth]{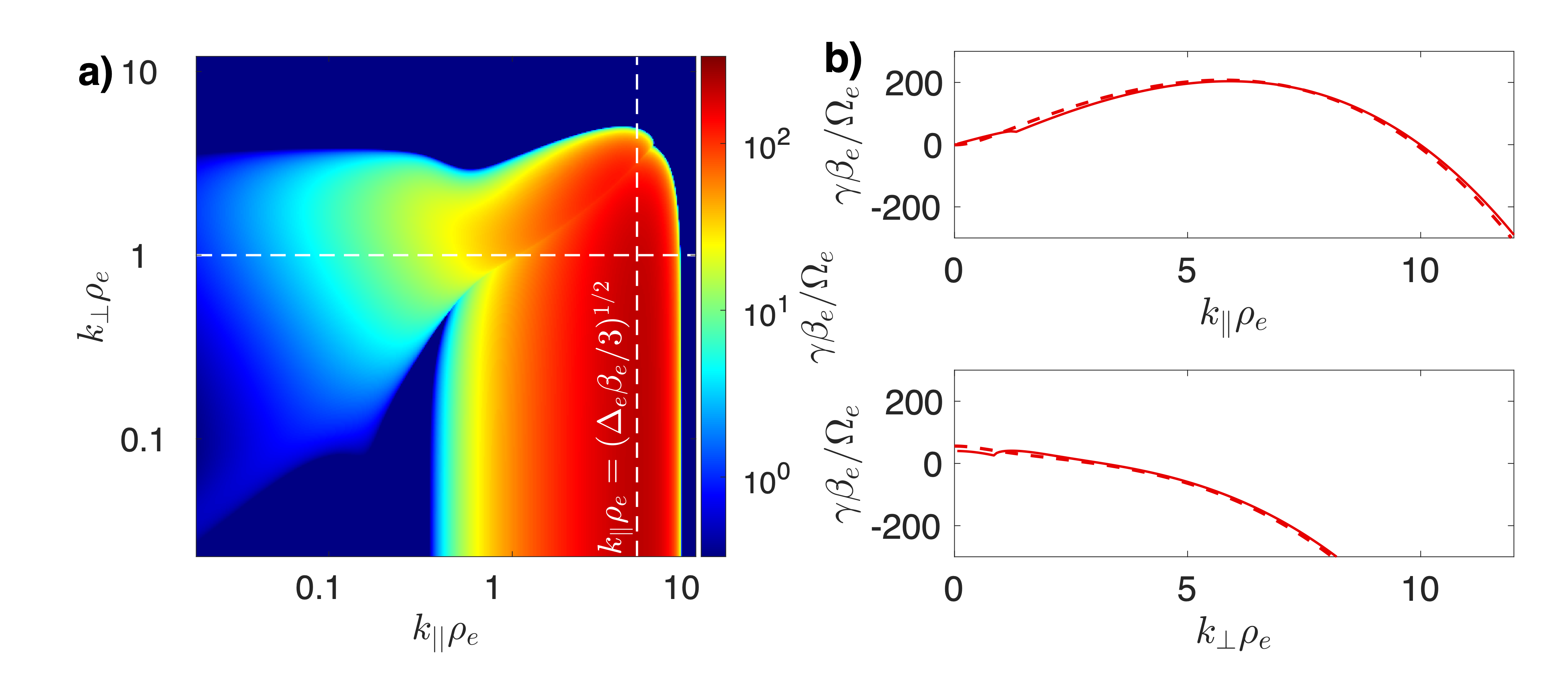}}
\caption{\textit{Oblique unstable modes at $\Delta_e \beta_e \gg 1$}. \textbf{a)} Maximum positive growth rates of linear perturbations 
resulting from CE ion- and electron-shear terms in the CE distribution function~(\ref{CEsheardistfuncexpression}) for $\Delta_e \beta = 100$.
Here, a temperature-equilibrated hydrogen plasma is considered, viz., $\Delta_e = \mu_e^{1/2} \Delta_i$ and $\beta_i = \beta_e$.  
The growth rates of all modes are calculated in the same way as figure \ref{Figure5}. The vertical dashed line 
indicates the value of $k_{\|} \rho_e$ at which maximum growth of the parallel transverse instability is attained [see (\ref{transverse_maxgrowth_kpl_pp_val})], while the horizontal one indicates $k_{\perp} \rho_e = 
1$. 
\textbf{b)} The transverse mode's growth rate (solid line) as a function of $k_{\|} \rho_e$ with $k_{\bot} \rho_e = 1$ (top), and as a function of $k_{\perp} \rho_e$ with $k_{\|} \rho_e = 
(\Delta_e \beta_e/3)^{1/2}$ (bottom). The 
dashed lines show the analytical prediction (\ref{transverse_oblique_growthrate}\textit{b}) for this quantity. \label{Figure_oblique_subelec}}
\end{figure}

In figure \ref{Figure_oblique_subelec}, we demonstrate that the fastest-growing unstable modes in the limit $\Delta_e \beta_e \gg 1$ are indeed transverse 
ones. This figure shows the numerically determined growth rate
as a function of $k_{\|}$ and $k_{\perp})$, for a particular large value of $\Delta_e \beta_e$. 
A broad range of sub-electron-Larmor scale modes are unstable (figure \ref{Figure_oblique_subelec}a), with the 
parallel wavenumber of the fastest-growing ones closely agreeing with the 
analytical prediction (\ref{transverse_maxgrowth_kpl_pp_val}). The analytical expression (\ref{transverse_oblique_growthrate}\textit{b}) for 
the transverse instability's growth rate also agrees well with the numerical result as a function 
of both $k_{\|}$ and $k_{\perp}$ (figure \ref{Figure_oblique_subelec}b).

\subsubsection{Electron mirror instability} \label{pospres_electron_oblique}

The oblique microinstability evident in figure \ref{Figure5}b at sub-ion-Larmor scales 
is the CES electron mirror instability: the destablisation of KAWs 
by excess perpendicular electron pressure (viz., $\Delta_e > 0$) associated with the CE electron-shear term. 
The instability~\citep[which has also been referred to as the field-swelling instability -- see][]{BC84} is 
perhaps confusingly named, given that its physical mechanism is rather 
different to that of the (ion-scale) mirror instability: non-resonant 
interactions between the anisotropic distribution of electrons and the KAWs causes the restoring force
underpinning the latter's characteristic oscillation to be negated if $\Delta_e > 1/\beta_e$.  
The electron mirror instability has been extensively explored in $\beta_e \sim 1$ 
plasma~\citep[see][and references therein]{HS18}; 
in plasmas with $\beta_e \gg 1$, it has been analytically characterised and its 
physical mechanism elucidated in the quasi-perpendicular ($k_{\|} \ll k_{\perp}$) limit
of gyrokinetics~\citep{KAKS18}. Here, we find that once its marginality
condition ($\Delta_e = 1/\beta_e$) is surpassed sufficiently, oblique modes
with $k_{\|} \lesssim k_{\perp}$ are also destabilised. 

\begin{figure}
\centerline{\includegraphics[width=0.99\textwidth]{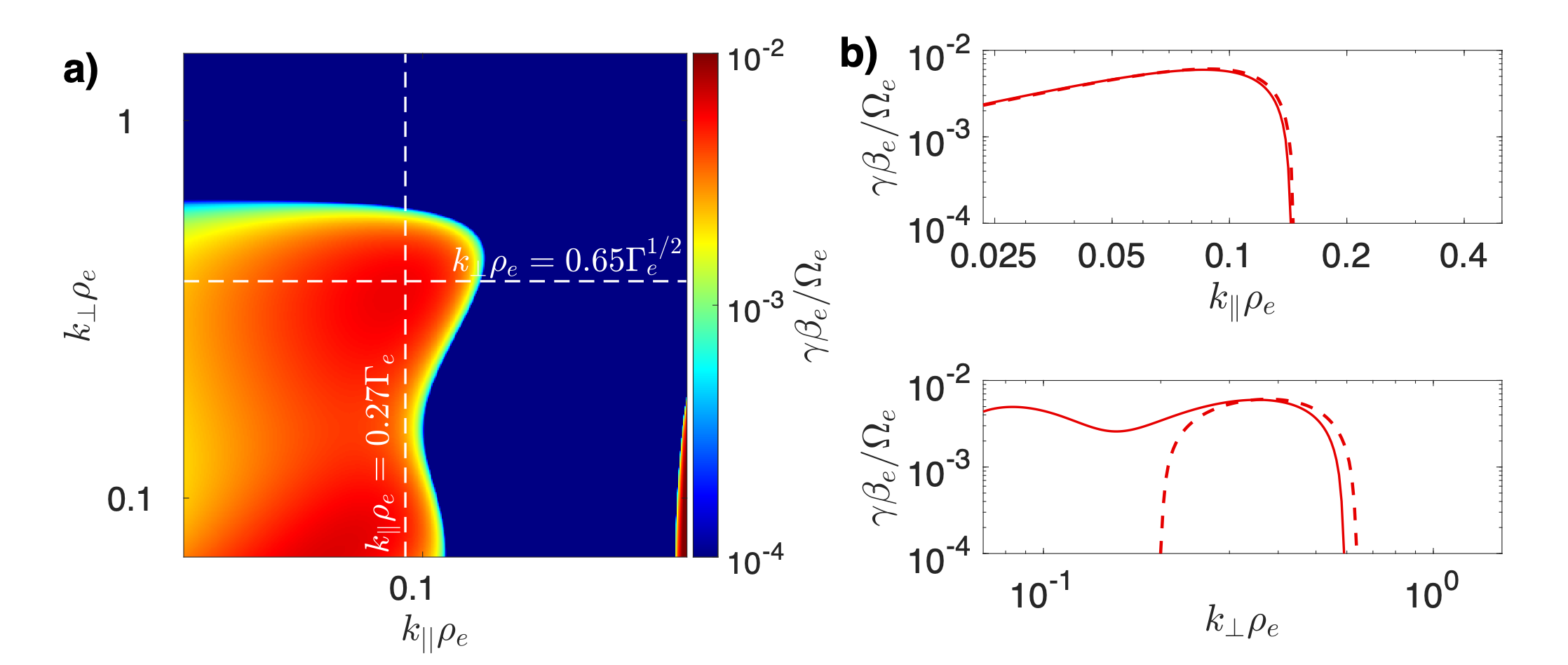}}
\caption{\textit{Electron mirror instability at $\Gamma_e = \Delta_e \beta_e -1 \ll 1$}. \textbf{a)} Growth rates of unstable electron mirror modes associated with the CE distribution function (\ref{CEsheardistfuncexpression}) for $\Gamma_e= 1/3$ ($\Delta_e \beta_e = 4/3$).
The growth rates of all modes are calculated in the same way as figure \ref{Figure5}. 
The dashed white lines indicate the analytical prediction (\ref{CESelectronmirror_peakgrowth_kval}) 
for the parallel/perpendicular wavenumber at which peak growth is achieved. 
\textbf{b)} Plot of the electron mirror mode's growth rate (solid line) as a function of $k_{\|} \rho_e$ with $k_{\bot} \rho_e = 0.65 \Gamma_e^{1/2}$ (top), and as a function of $k_{\perp} \rho_e$ with $k_{\|} \rho_e = 0.27 \Gamma_e$ (bottom). The 
dashed lines show the analytical prediction (\ref{electronmirror_growthrate}) for this quantity. \label{newFig_elecmirror_marg}}
\end{figure}

As with the mirror instability, a simple analytic characterisation of the CES electron mirror instability can be 
performed in the case of marginal instability. We define 
the marginality parameter $\Gamma_e \equiv \Delta_e \beta_e -1$, and adopt the
ordering
\begin{equation}
k_{\perp}^2 \rho_e^2 \sim  k_{\|} \rho_e \sim \tilde{\omega}_{e\|} \beta_e \sim \Gamma_e  
\ll 1 , 
\label{KAW_ords_marg}
\end{equation}
with the additional assumption that $\Gamma_e \gg \mu_e^{1/2}$ in order that 
the effect of ion pressure anisotropy can be neglected. Then, it can be 
shown (see appendix \ref{derivation_elecmirror}) that the growth rate is
\begin{equation}
 \frac{\gamma}{\Omega_e} = \frac{k_{\|} \rho_e}{\beta_e} \left[-\frac{3\sqrt{\upi}}{4} k_{\perp}^2 \rho_e^2+ \sqrt{\frac{3}{2} \Gamma_e k_{\perp}^2 \rho_e^2 -\frac{9}{4} k_{\|}^2 \rho_e^2 + \frac{9}{16}\left(\upi-2\right) k_{\perp}^4 \rho_e^4 } 
 \right]
 . \label{electronmirror_growthrate}
\end{equation}
It follows that the maximum growth rate is 
\begin{equation}
  \gamma_{\rm max} = \frac{\left[\upi-8+\sqrt{\upi\left(16+\upi\right)}\right]^{3/2}}{48\left(\upi-2\right)} \left[ \sqrt{\frac{\upi+4+\sqrt{\upi\left(16+\upi\right)}}{\upi-8+\sqrt{\upi\left(16+\upi\right)}}}-\sqrt{\frac{\upi}{\upi-2}}\right] \approx 0.055 \frac{\Gamma_e^2}{\beta_e} , \label{CESelectronmirror_peakgrowthrate}
\end{equation}
attained at
\begin{subeqnarray}
(k_\| \rho_e)_{\mathrm{peak}} & = & \sqrt{\frac{\upi-8+\sqrt{\upi\left(16+\upi\right)}}{36\left(\upi-2\right)}} \Gamma_e \approx 0.27 \Gamma_e, \\
(k_\perp \rho_e)_{\mathrm{peak}} & = & \sqrt{\frac{\upi-8+\sqrt{\upi\left(16+\upi\right)}}{6\left(\upi-2\right)}} \Gamma_e^{1/2} \approx 0.65 \Gamma_e^{1/2}.  \label{CESelectronmirror_peakgrowth_kval}
\end{subeqnarray}
Figure \ref{newFig_elecmirror_marg} demonstrates that these predictions are accurate
by comparing them to numerical results for a particular (small) value of $\Gamma_e$. 
More specifically, figure \ref{newFig_elecmirror_marg}a shows that the location in the 
$(k_{\|},k_{\perp})$ plane at which the maximum growth of the electron mirror instability is attained 
closely matches the analytical prediction 
(\ref{CESelectronmirror_peakgrowth_kval}), while figure \ref{newFig_elecmirror_marg}b confirms that the 
wavenumber dependence of the growth rate agrees with (\ref{electronmirror_growthrate}) for $k_{\perp} \rho_e \gtrsim 
\mu_e^{1/4}$. We note that, in addition to the electron mirror, another instability operating at smaller characteristic values
of $k_{\perp} \rho_e$ is evident in figure \ref{newFig_elecmirror_marg}. These are the $k_{\perp} \rho_i \gtrsim 1$ mirror modes 
driven unstable by the CE ion-shear term that were discussed in section 
\ref{pospress_ion_mirror}; for $1 \ll k \rho_i \ll \mu_e^{-1/4}$, the ion-pressure anisotropy associated 
with the CE ion-shear terms remains a greater free-energy source for KAW instabilities 
than the CE electron-shear term, even when $\Delta_e > 1/\beta_e$. 

\begin{figure}
\centerline{\includegraphics[width=0.99\textwidth]{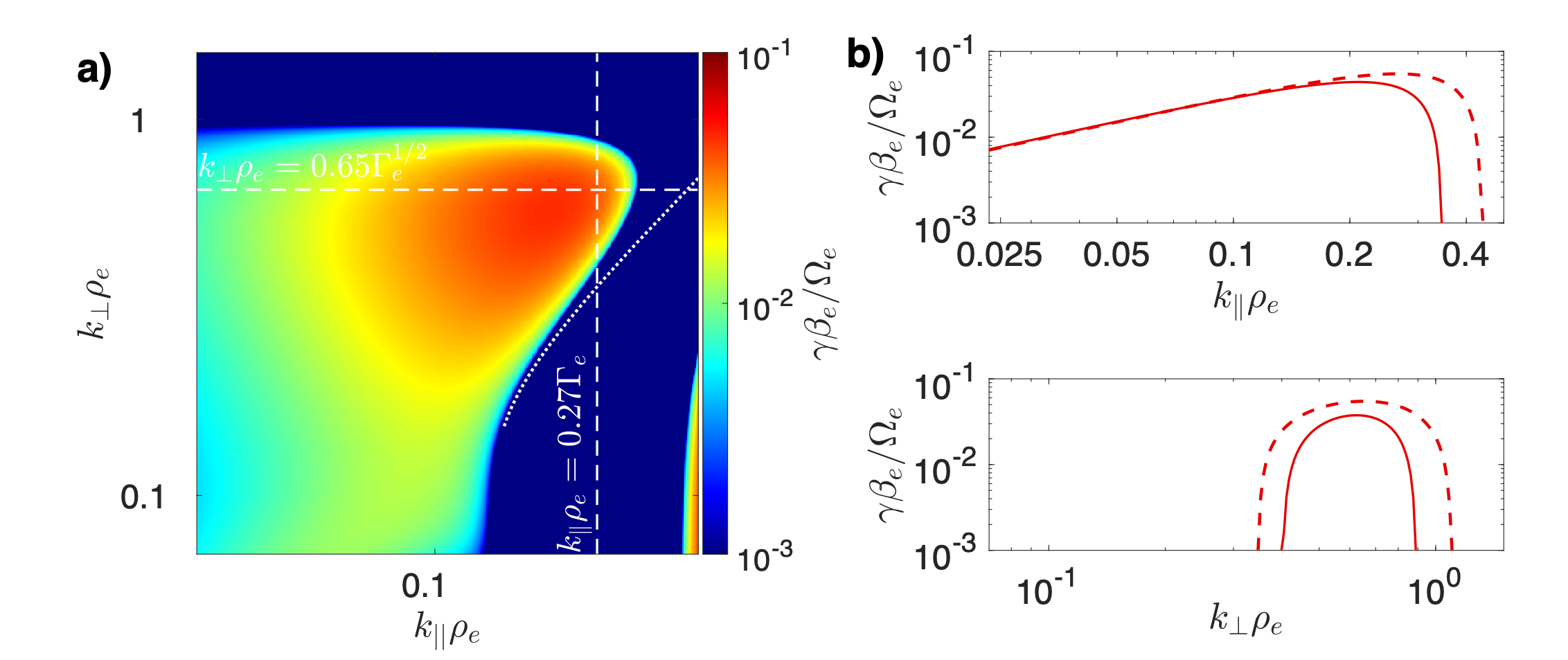}}
\caption{\textit{Electron mirror instability at $\Gamma_e = \Delta_e \beta_e -1 \sim 1$}. \textbf{a)} Growth rates of unstable electron mirror modes associated with the CE distribution function (\ref{CEsheardistfuncexpression}) for $\Gamma_e = 1$ ($\Delta_e \beta_e = 2$).
The growth rates of all modes are calculated in the same way as figure \ref{Figure5}.
The dashed white lines indicate the analytical prediction (\ref{CESelectronmirror_peakgrowth_kval}) 
for the parallel/perpendicular wavenumber at which peak growth is achieved, while the dotted line
indicates the analytical prediction (\ref{elecmirror_instabbound}) for the total
wavenumber below which oblique
long-wavelength ($k_{\|} \rho_e < k_{\perp} \rho_e \ll 1$) electron mirror modes 
become unstable. 
\textbf{b)} The electron mirror mode's growth rate (solid line) as a function of $k_{\|} \rho_e$ with $k_{\bot} \rho_e = 0.65 \Gamma_e^{1/2}$ (top), and as a function of $k_{\perp} \rho_e$ with $k_{\|} \rho_e = 0.27 \Gamma_e$ (bottom). The 
dashed lines show the analytical prediction (\ref{electronmirror_growthrate}) for this quantity.  \label{newFig_elecmirror_unity}}
\end{figure}

\begin{figure}
\centerline{\includegraphics[width=0.99\textwidth]{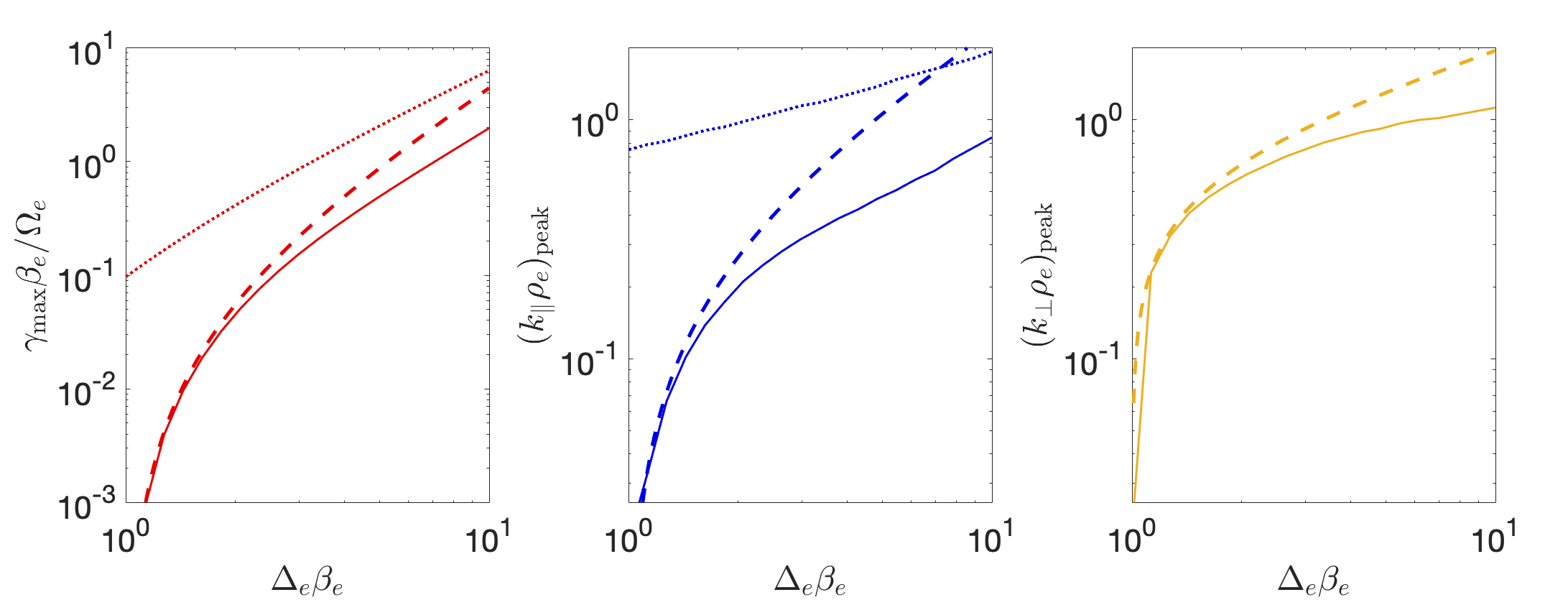}}
\caption{\textit{The maximum growth of the electron mirror instability}. The maximum normalised growth rate $\gamma_e \beta_e/\Omega_e$ (left panel, solid red line) of unstable electron mirror modes as
a function of $\Delta_e \beta_e$, as well as parallel (middle panel, solid blue line) and perpendicular (right panel, solid yellow line) wavenumbers, $(k_{\|} \rho_e)_{\mathrm{peak}}$ and $(k_{\perp} \rho_e)_{\mathrm{peak}}$, respectively, at
which that growth is attained. The analytical prediction (\ref{CESelectronmirror_peakgrowthrate}) of $\gamma_{\rm max}$ for marginally unstable electron mirror modes,
as well as the analogous predictions (\ref{CESelectronmirror_peakgrowth_kval}) for $(k_{\|} \rho_e)_{\mathrm{peak}}$ and $(k_{\perp} \rho_e)_{\mathrm{peak}}$, are shown as dashed 
lines. The dotted lines are the maximum growth rate and (parallel) wavenumber of peak growth 
for the CET parallel whistler instability (see section \ref{pospres_electron_EC}) as functions of $\Delta_e \beta_e$. 
\label{newFig_elecmirror_param}}
\end{figure}

For $\Gamma_e \gtrsim 1$, our near-marginal theory anticipates that peak growth 
occurs at electron Larmor scales ($k_{\|} \rho_e \lesssim k_{\|} \rho_e \sim 1$), with $\gamma_{\rm max} \sim 
\Omega_e/\beta_e$. These expectations are indeed realised numerically, as shown in figure \ref{newFig_elecmirror_unity} (see also 
figure \ref{Figure5}). 
The expression (\ref{electronmirror_growthrate}) for the growth rate as a function of wavenumber that was derived in the case of $\Gamma_e \ll 1$ remains qualitatively -- but not quantitatively -- 
accurate (see figure \ref{newFig_elecmirror_unity}b). Figure \ref{newFig_elecmirror_param} shows that a similar conclusion holds for the expression (\ref{CESelectronmirror_peakgrowthrate})
for the peak growth rate, and also for the expressions (\ref{CESelectronmirror_peakgrowth_kval}\textit{a}) and (\ref{CESelectronmirror_peakgrowth_kval}\textit{b}) of the parallel and perpendicular wavenumbers at 
which that growth occurs. 

To confirm our prior claim in section \ref{poseps_stab} that the CES parallel whistler 
instability is faster growing than the electron mirror instability, we show the 
former's numerically computed growth rate on figure \ref{newFig_elecmirror_param} (left panel); as
it approaches the asymptotic value (\ref{CESwhistler_maxgrowth}) that is valid in the 
limit $\Delta_e \beta_e \gg 1$, we observe that the electron mirror's growth 
rate is a factor of ${\sim}3$ smaller (cf. figure \ref{Figure_oblique_subelec}a). The 
parallel wavenumber at which peak growth of the whistler instability occurs is 
also larger than the analogous quantity for the electron mirror by an order-unity factor. 

While we cannot derive a simple analytic expression for the growth rate of the dominant
electron mirror modes when $\Gamma_e \gtrsim 1$, we can calculate this quantity for long-wavelength (viz., $k \rho_e \ll 1$) modes. For this calculation, 
we assume that $k \rho_e \sim \mu_e^{1/4} \ll 1$, $k_{\bot} \sim k_{\|}$, and  
the ordering
\begin{equation}
\tilde{\omega}_{e\|}  = \frac{\omega}{k_{\|} v_{\mathrm{th}e}}  \sim \frac{k \rho_e}{\beta_e}  \sim |\Delta_e| k \rho_e \, . 
\label{KAW_ords}
\end{equation}
 Under these assumptions, we obtain (see appendix \ref{derivation_elecmirror}) two modes whose complex frequencies  
 $\omega$ are given by 
\begin{eqnarray}
\omega & \approx & \pm k_\| \rho_e \Omega_e \Bigg\{\left[\frac{1}{\beta_e} + \Delta_e\left(\frac{1}{2}- \mu_e^{1/2} \frac{k_{\|}^2 \rho_e^2- k_{\bot}^2 \rho_e^2}{k^4 \rho_e^4}\right) \right] \nonumber\\
 && \qquad \qquad \times \left[ \frac{k^2 \rho_e^2}{\beta_e} - \Delta_e \left(k_{\bot}^2 \rho_e^2 + \mu_e^{1/2} \frac{k_{\|}^2}{k^2} - \frac{1}{2} k_{\|}^2 \rho_e^2 \right) \right]\Bigg\}^{1/2} . 
 \label{obliqueinstab_freq_gen}
\end{eqnarray}
The terms proportional to $\mu_e^{1/2} \Delta_e$ are associated
with the CE ion-shear term, which plays a non-negligible role for $k \rho_e \lesssim 
\mu_e^{1/4}$. In the subsidiary limit $k \rho_e \ll \mu_e^{1/4}$, (\ref{obliqueinstab_freq_gen})
becomes the dispersion relation (\ref{mirrorgrowth_largek}) obtained in section \ref{pospress_ion_mirror} for 
unstable mirror modes in the limit $\Delta_i \beta_i \gg 1$.
In the opposite subsidiary limit $k \rho_e \gg \mu_e^{1/4}$ (but $k \rho_e \ll 1$), 
(\ref{obliqueinstab_freq_gen}) simplifies~to
\begin{eqnarray}
\omega & \approx & \pm k_\| \rho_e \Omega_e \sqrt{\left(\frac{1}{\beta_e} + \frac{\Delta_e}{2}\right)\left[\frac{k^2 \rho_e^2}{\beta_e} - \Delta_e \left(k_{\bot}^2 \rho_e^2 - \frac{1}{2} k_{\|}^2 \rho_e^2\right) \right]} \, . 
 \label{obliqueinstab_freq_elec}
\end{eqnarray}
For $k_{\|} \ll k_{\perp}$, this recovers the high-$\beta$ limit of the 
dispersion relation for unstable KAWs previously derived in the gyrokinetic 
calculations of~\citet{KAKS18}; our calculations show that this dispersion 
relation also applies to oblique ($k_{\|} \lesssim k_{\perp}$) electron mirror modes. For $\Delta_e > 0$, we (as expected) have an unstable root if and only if
\begin{equation}
 \Delta_e > \frac{1}{\beta_e} , \label{threshold_electronmirror}
\end{equation}
with the unstable mode's growth rate being
\begin{eqnarray}
\gamma & \approx & k_\| \rho_e \Omega_e \sqrt{\left(\frac{1}{\beta_e} + \frac{\Delta_e}{2}\right)\left[\Delta_e \left(k_{\bot}^2 \rho_e^2 - \frac{1}{2} k_{\|}^2 \rho_e^2\right) - \frac{k^2 \rho_e^2}{\beta_e} \right]} \, . 
 \label{obliqueinstab_growth_elec}
\end{eqnarray}

We can now provide an analytical demonstration that 
a broad spectrum of electron mirror modes is unstable if $\Gamma_e \gtrsim 1$. 
It follows directly from (\ref{obliqueinstab_freq_gen}) that instability arises for all 
modes with $k_{\perp} > k_{\|}$ if the following constraint on the total wavenumber $k$ is satisfied:
\begin{equation}
k \rho_i < \sqrt{\frac{2 \mu_e^{1/2} \left(\Gamma_e+1\right) \cos^2{\theta}}{\left(\Gamma_e+3\right) \cos^2{\theta}-2 \Gamma_e \sin^2{\theta}}} 
\, , \label{elecmirror_instabbound}
\end{equation}
where $\theta = \tan^{-1}{(k_{\perp}/k_{\|})}$ is, as normal, the 
wavevector angle. The validity of this bound is illustrated in figure \ref{newFig_elecmirror_unity}a.
(\ref{elecmirror_instabbound}) is particularly simple to interpret in the
subsidiary limit $k \rho_e \gg \mu_e^{1/4}$, yielding a lower bound on $\theta$ alone:
\begin{equation}
\theta > \tan^{-1}{\sqrt{\frac{\Gamma_e+3}{2 \Gamma_e}}} . \label{elecmirror_angleinstab}
\end{equation}
For $\Gamma_e \ll 1$ (but $\Gamma_e > 0$), this implies that the only unstable electron mirror modes are 
quasi-perpendicular, as anticipated from our calculations pertaining to the marginal state of the 
instability. On the other hand, for $\Gamma_e \gtrsim 1$, modes with a 
wide range of wavevector angles will  
be destabilised. 

\subsection{CES microinstability classification: negative pressure anisotropy $(\epsilon_i < 0)$} \label{CES_neg_aniso}

\subsubsection{Firehose instability} \label{negpres_fire}

The best-known instability to be triggered by either negative ion or electron pressure anisotropy
associated with the CE ion- and electron-shear terms, respectively, is the CES firehose 
instability. The linear theory of the firehose (or garden-hose) instability in high-$\beta$ plasma, the first
studies of which were completed over half a century ago~\citep{R56,P58,CKW58,VS61},
has previously been explored in the contexts of plasmas with bi-Maxwellian distributions~\citep[e.g.,][]{KS67,DV68,YWA93,HM00}, 
CE distributions~\citep[e.g.,][]{SCKHS05}, and even characterisations that are 
independent of the ion distribution function~\citep[e.g.,][]{ACRR10,KSCAC15}. 
Its physical mechanism is well established: negative pressure anisotropies 
reduce the elasticity of magnetic-field lines that gives rise to Alfv\'en waves, and can completely 
reverse it when $\Delta_i$ is negative enough. The long-wavelength `fluid' 
firehose instability (whose mechanism is independent of the particular ion distribution 
function) is non-resonant in nature; however, resonant damping mechanisms such 
as Barnes damping or cyclotron damping play an important role in regulating  
the growth of modes on scales comparable to the ion-Larmor scale, and thereby 
set the scale of peak firehose growth. Beyond linear theory, nonlinear analytical 
studies of the parallel firehose instability in high-$\beta$ plasma have been 
completed~\citep[e.g.,][]{RSRC11}, as well as numerical 
ones~\citep[e.g.,][]{KSS14,MSK16,RQV18}.

While there is much in common between firehose modes
across all wavevector angles, there are certain differences that, 
on account of their significance for determining the fastest-growing firehose mode, are 
important to highlight. Based on these differences, firehose modes can be 
categorised into three different types: \emph{quasi-parallel}, 
\emph{oblique}, and \emph{critical-line} firehose modes. Quasi-parallel 
firehose modes, which are destabilised left-handed and/or 
right-handed high-$\beta$ Alfv\'en waves~\citep{KS67,DV68}, exist inside a 
narrow cone of wavevector angles $\theta \lesssim \beta_i^{-1/4}$~\citep{A13}. 
The peak wavenumber of their growth ($k_{\|} \rho_i \sim |\Delta_i + 2/\beta_i|^{1/2}$) is determined by gyroviscosity, an FLR 
effect~\citep{ACRR10}. 
For $\theta \gtrsim \beta_i^{-1/4}$, the characteristic low-frequency (viz., $\omega \ll \Omega_i$) waves that exist above ion-Larmor-scales in high-$\beta$ 
plasma are shear-Alfv\'en waves and (compressible) slow modes; the former 
remains susceptible to firehose instability, but, on account of its FLR coupling to 
the slow mode, its instability proceeds quite differently at sufficiently 
small wavenumbers ($k \rho_i \gtrsim |\Delta_i + 2/\beta_i|^{1/2}$),
with peak growth occurring at smaller scales ($k_{\|} \rho_i \sim |\Delta_i + 2/\beta_i|^{1/4} \ll 1$). Finally, 
along a `critical line' in the $(k_{\|},k_{\perp})$ plane ($k_{\perp} \approx \sqrt{2/3} k_{\|}$, $\theta \approx 39^{\circ}$), the FLR coupling 
between the slow mode and shear-Alfv\'en wave becomes anomalously weak 
due to two opposing FLR effects cancelling each other out. This results in
much weaker collisionless damping on critical-line firehose modes, and so they 
can exist on scales that are close to (though, as we prove here for the first time, not strictly at) the ion-Larmor 
scale. Thus critical-line firehose modes are generically the fastest-growing ones in 
high-$\beta$ plasma~\citep{SCKHS05}. 

\begin{figure}
\centerline{\includegraphics[width=0.99\textwidth]{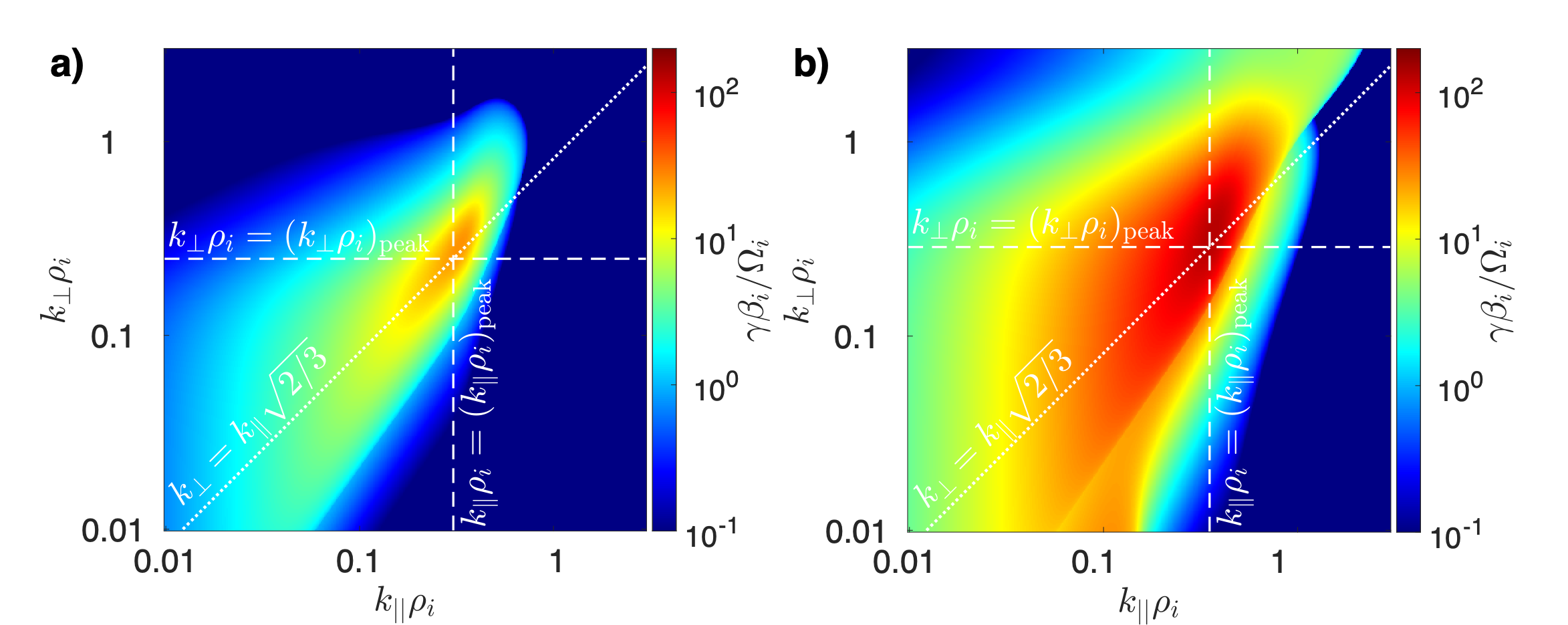}}
\caption{\textit{CES firehose instability when $\Delta_i \beta_i+2 \lesssim -1$}. Maximum positive growth rates of linear perturbations resulting from the CE ion-shear term in the CE distribution function (\ref{CEsheardistfuncexpression}) with $\Delta_i$ negative enough to surpass the long-wavelength firehose-instability threshold $\Delta_i = -2/\beta_i$
by at least an order-unity factor. 
The growth rates of all modes are calculated using the approach outlined in appendix \ref{CES_method_append_quartic}. The growth rates are calculated on a $400^2$ grid, with logarithmic spacing in both perpendicular and parallel directions between wavenumbers.
The resulting growth rates are normalised as $\gamma \beta_i/\Omega_i$ are functions of two dimensionless parameters: $\Delta_i \beta_i$ and $\beta_i$.
The dashed white lines indicate the analytical predictions (\ref{firehosepeakgrowthrate_cyclo_kval}) 
for the parallel/perpendicular wavenumber at which peak growth is achieved, while the dotted line
indicates the critical line $k_{\perp} = k_{\|} \sqrt{2/3}$ along which the firehose growth rate is predicted to be maximal. 
\textbf{a)} $\Delta_i \beta_i = -3$. 
\textbf{b)} $\Delta_i \beta_i = -30$. In both cases, $\beta_i = 10^{4}$. \label{newFig_firehose_2D}}
\end{figure}

We support this claim with figure \ref{newFig_firehose_2D}, which shows the 
maximum growth rate of the firehose-unstable modes as a function of both $k_{\|}$ and 
$k_{\perp}$ for two different (unstable) values of $\Delta_i \beta_i$ (and with the same value of $\beta_i$ as was used
to calculate the stability maps presented in section \ref{poseps_stab}). 
Both examples confirm that, although a broad spectrum of unstable parallel and oblique firehose modes emerge when $\Delta_i \beta_i +2 \lesssim -1$, it is the critical-line firehose modes 
that are the fastest growing. 

\begin{figure}
\centerline{\includegraphics[width=0.99\textwidth]{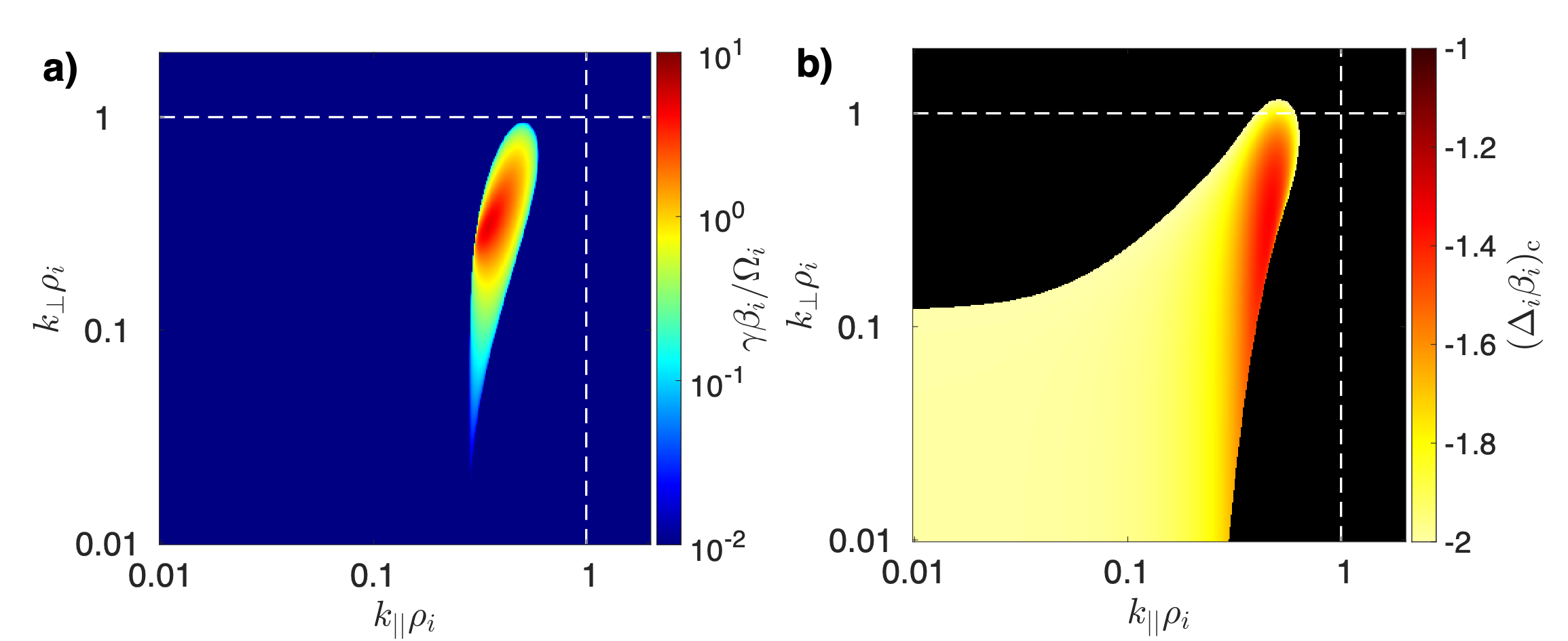}}
\caption{\textit{Onset of the CES firehose instability}. \textbf{a)} Maximum positive growth rates of linear perturbations resulting from the CE ion-shear term in the CE distribution function (\ref{CEsheardistfuncexpression}) with $\beta_i = 10^{4}$ and $\Delta_i = 
-1.7/\beta_i$ (which is below the long-wavelength firehose instability threshold $\Delta_i = 
-2/\beta_i$). The growth rates of all modes are calculated in the same way as figure~\ref{newFig_firehose_2D}. 
\textbf{b)} Threshold value $(\Delta_i \beta_i)_{\rm c}$  of $\Delta_i \beta_i$ at which modes with parallel and perpendicular wavenumber $k_{\|}$ and $k_{\perp}$, respectively, become firehose 
unstable. Regions of $(k_{\|},k_{\perp})$ that are shaded black are stable. \label{newFig_firehose_marginal}}
\end{figure}

The value of $\Delta_i$ required to trigger the CES firehose instability is, as 
with the case of the firehose instability in a plasma with a bi-Maxwellian ion distribution, dependent on the scale of the unstable 
firehose modes. For long-wavelength firehose modes (i.e. those with $k \rho_i \ll 
1$), the threshold is $\Delta_i < (\Delta_i)_{\rm c} = -2/\beta_i$; it can be shown that this result is independent of the 
particular form of the ion distribution function~\citep{ACRR10}.
However, our numerical solutions for the wavenumber-dependent growth rate of firehose modes in CE plasma 
when $\Delta_i > -2/\beta_i$ (see figure \ref{newFig_firehose_marginal}a) suggest that oblique ion-Larmor-scale firehose modes
can be destabilised at less negative 
pressure anisotropies. 
This is consistent with the findings of 
previous studies of the oblique firehose in $\beta \sim 1$ plasma~\citep{HM00,HT08,AF16}, although
this finding has not until now been comprehensively studied in plasma with $\beta \gg 1$.  
We can, in fact, calculate the threshold semi-analytically 
for the CES firehose instability as a function of wavenumber (see appendix 
\ref{thresholdcal_quadratic}); the results, which are shown in figure~\ref{newFig_firehose_marginal}b show that oblique 
firehose modes with $k_{\|} \rho_i \approx 0.45$, $k_{\perp} \rho_i \approx 0.3$ become unstable 
when $\Delta_i \approx -1.35/\beta_i$.  
The reduced threshold of ion-Larmor-scale firehose modes, which
can be shown to depend only on fourth- and higher-order moments of the ion 
distribution function, is considered in greater depth in Bott \textit{et al.} (2023, in prep.).

The growth of the three different sub-categories of unstable CES firehose modes (quasi-parallel, oblique, and critical-line firehoses)
can be described analytically. However, the relative orderings of $\tilde{\omega}_{i\|}$, $k_{\|} \rho_i$, $k_{\bot} 
 \rho_i$, $\beta_i$ and $|\Delta_i|$ for these sub-categories are different, so it is necessary to treat them separately. 

   \subsubsection{Quasi-parallel firehose instability} \label{negpres_fire_par}
   
\begin{figure}
\centerline{\includegraphics[width=0.99\textwidth]{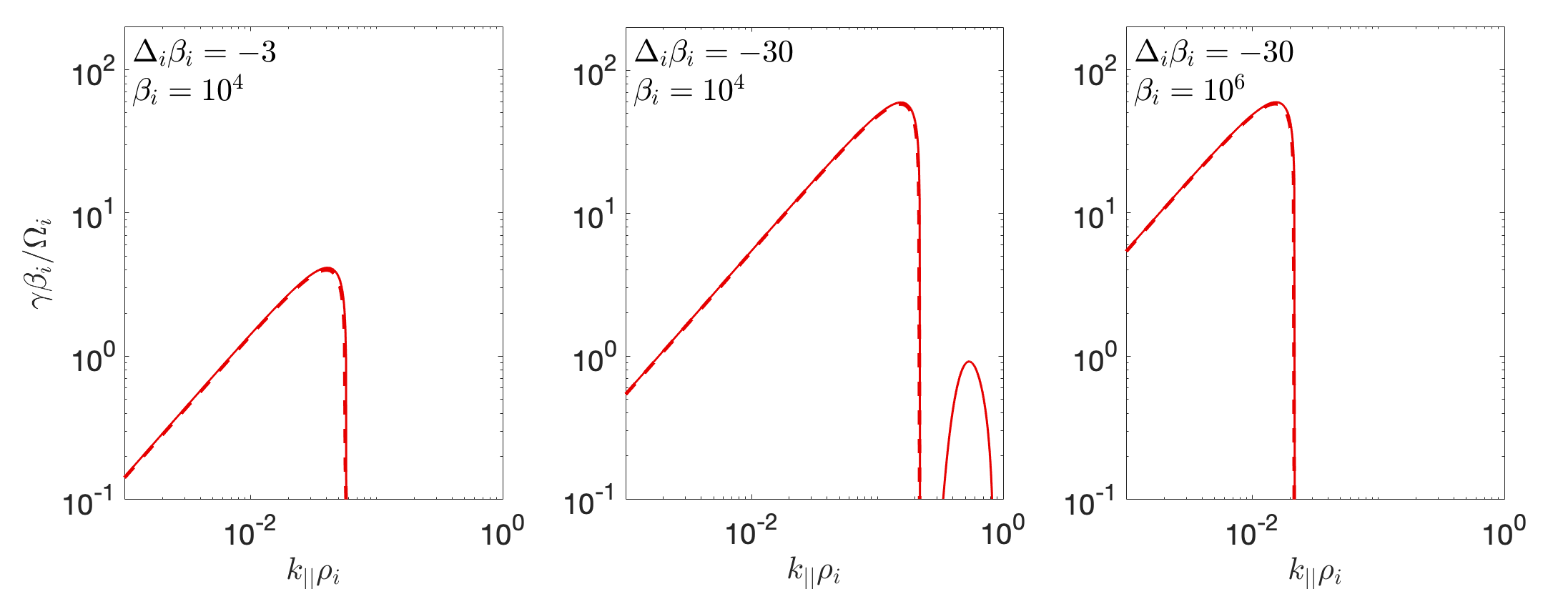}}
\caption{\textit{Parallel CES firehose instability}. Growth rates of Alfv\'en waves whose instability is driven by the CE ion-shear term in the CE distribution function (\ref{CEsheardistfuncexpression}), 
for wavevectors co-parallel with the background 
magnetic field (viz., $\boldsymbol{k} = k_{\|} \hat{\boldsymbol{z}}$). 
The growth rates (solid lines) of all modes are calculated in the same way as figure~\ref{newFig_firehose_2D}.
We show the growth rates for a selection of different values of $\Delta_i \beta_i$ 
and $\beta_i$.
The approximation (\ref{growthrate_firehose_parallel}) for 
the growth rate (dashed red) in the limit $k_{\|} \rho_i \ll 1$ is also plotted. 
 \label{newFig_firehose_parallel}}
\end{figure}

   The relevant orderings of parameters in for quasi-parallel firehose modes is
   \begin{equation}
     \tilde{\omega}_{i\|}  = \frac{\omega}{k_{\|} v_{\mathrm{th}i}}  \sim \beta_i^{-1/2} \sim |\Delta_i|^{1/2} \sim 
     k_{\|} \rho_i \, , \label{quasiparallelord}
   \end{equation}
   with the additional small wavenumber-angle condition
   \begin{equation}
   k_{\bot} \rho_i \ll  \beta_i^{-1/4} k_{\|} \rho_i \sim \beta_i^{-3/4}.
   \end{equation}
    Under the
   ordering (\ref{quasiparallelord}), we find (see appendix \ref{derivation_firehose_par}) that there are four modes with complex 
 frequencies given by
 \begin{equation}
   \frac{\omega}{\Omega_i} = \pm k_{\|} \rho_i \left(\frac{1}{4} k_{\|} \rho_i \pm \sqrt{\frac{1}{16} k_{\|}^2 \rho_i^2 + \frac{1}{\beta_i}+\frac{\Delta_i}{2}} \right) \, , \label{growthrate_firehose_parallel}
 \end{equation}
 where the $\pm$ signs can be chosen independently. This is the standard 
 parallel firehose dispersion relation~\citep{KS67,DV68,ACRR10}. 
 To (re-)identify the modes that are destabilised by the negative ion-pressure anisotropy, we set $\Delta_i = 0$: 
 the resulting dispersion relation agrees with \citet{FK79}, recovering the dispersion relation of Alfv\'en waves in the limit $k_\| \rho_i \ll \beta_i^{-1/2}$ [see see their eqn. (19)]
 and the dispersion relation of the slow and fast hydromagnetic waves in the limit $k_\| \rho_i \gg \beta_i^{-1/2}$ [see see their eqn. (20)].
The growth rate of the unstable parallel firehose modes that follows from (\ref{growthrate_firehose_parallel}) 
is shown in figure \ref{newFig_firehose_parallel} for several different values of $\Delta_i$ and $\beta_i$; the results closely match the analogous result 
determined numerically\footnote{An inquisitive reader might wonder why 
the numerical solution suggests that, in addition to the long-wavelength parallel firehose modes, parallel ion-Larmor scale
modes are also unstable in some cases (see figure \ref{newFig_firehose_parallel}, middle panel), albeit with a 
much smaller growth rate. This instability is the \emph{CES resonant parallel firehose instability}, so named
because of its mediation via gyroresonant interactions beween ions and ion-Larmor-scale 
modes~\citep{YWA93}. 
In a $\beta_i \sim 1$ plasma, this instability can have a growth rate comparable to (or even larger than) the  
longer-wavelength non-resonant firehose modes; however, because of the exponential 
dependence of the resonant parallel firehose instability's growth rate on $|\Delta_i|^{-1} \sim \beta_i$, 
the instability is generically much weaker than the non-resonant firehose in plasma with $\beta_i \gg 1$ (see Bott \textit{et al.}, in prep.).
In the language of section \ref{sec:CharacMicroQual}, resonant parallel firehose modes are quasi-cold in CE plasma. We therefore do not consider this instability further in this paper.}. 

$\quad$ For non-zero $\Delta_i$ and fixed $k_\| \rho_i$, (\ref{growthrate_firehose_parallel}) implies that we have instability provided
 \begin{equation}
   |\Delta_i| >  \frac{2}{\beta_i} + \frac{1}{8} k_\|^2 \rho_i^2 \, . \label{firehose_thres}
 \end{equation}
The fastest-growing mode 
\begin{equation}
 \frac{\gamma_{\mathrm{max}}}{\Omega_i} = \left|\frac{2}{\beta_i}+\Delta_i\right|
\end{equation}
occurs at the characteristic wavenumber  
\begin{equation}
 (k_\| \rho_i)_{\mathrm{peak}} = 2 \left|\frac{2}{\beta_i}+\Delta_i\right|^{1/2} \, .
\end{equation}
For $k_\| \rho_i > 2 \sqrt{2} \left|2 \beta_i^{-1}+\Delta_i\right|^{1/2}$, the 
unstable mode is stabilised. This agrees with previous analytical characterisations of the firehose instability~\citep{RSRC11}. 

 \subsubsection{Oblique firehose instability} \label{negpres_fire_oblique}

In this case, we order
   \begin{equation}
     \tilde{\omega}_{i\|} \sim \frac{1}{\beta_i^{1/2}} \sim |\Delta_i|^{1/2} \sim 
     k_{\|}^2 \rho_i^2 \sim k_{\bot}^2 \rho_i^2 \, . \label{obliqueord}
   \end{equation}
  Aside from the finite propagation angle of oblique modes, the key difference between the oblique and quasiparallel cases is the larger magnitude 
   of the typical wavenumber $k \rho_i \sim \beta_i^{-1/4}$. The unstable oblique 
   firehose modes have the complex frequency (see appendix \ref{derivation_firehose_oblique}) 
 \begin{eqnarray}
 \frac{\omega}{\Omega_i} & = & -k_{\|} \rho_i \Bigg[\frac{\mathrm{i}}{8 \sqrt{\upi} k_{\bot}^2 \rho_i^2} \left(k_{\|}^2 \rho_i^2  - \frac{3}{2} k_{\bot}^2 \rho_i^2 \right)^2 \nonumber \\
 && \quad \pm \sqrt{\frac{1}{\beta_i}+\frac{\Delta_i}{2} - \frac{1}{64 \upi k_{\bot}^4 \rho_i^4} \left(k_{\|}^2 \rho_i^2  - \frac{3}{2} k_{\bot}^2 \rho_i^2 \right)^4} \Bigg]
 \, . \label{growthrate_firehose_oblique}
 \end{eqnarray}
 Setting $|\Delta_i| = 0$, and considering the subsidiary limit $k \rho_i \ll \beta_i^{-1/4}$, 
 we recover the dispersion relation of the shear Alfv\'en mode~\citep{FK79}.

 $\quad$Similarly to the quasi-parallel firehose instability, the instability condition 
 is still
  \begin{equation}
   \Delta_i < -\frac{2}{\beta_i} \, . \label{firehose_thres_oblique} 
 \end{equation}
 If this condition is met, the maximum growth rate of the instability is 
  \begin{equation}
  \frac{\gamma_{\mathrm{max}}}{\Omega_i} \approx \left(\frac{8 \upi}{27}\right)^{1/4} \left|\frac{2}{\beta_i}+\Delta_i\right|^{3/4} \tan{\theta}
  \left[1-\frac{3}{2} \tan^2{\theta}\right]^{-1} \, , \label{firehose_growth_oblique}
 \end{equation}  
 and is attained at (parallel) wavenumber
  \begin{equation}
  (k_{\|} \rho_i)_{\mathrm{peak}} \approx \left(\frac{32 \upi}{3}\right)^{1/4} \left|\frac{2}{\beta_i}+\Delta_i\right|^{1/4} \tan{\theta}
  \left[1-\frac{3}{2}  \tan^2{\theta} \right]^{-1} \, , \label{obliqfire_parwav}
 \end{equation}  
 where $\theta = \tan^{-1}(k_{\perp}/k_\|)$ is (again) the wavevector angle with 
 respect to the magnetic field. 
In contrast to the quasi-parallel case, if the condition (\ref{firehose_thres_oblique}) is met, the instability persists for all wavenumbers 
 satisfying $k \rho_i \lesssim 1$, albeit with an decreasing growth rate beyond the parallel wavenumber 
 given by (\ref{obliqfire_parwav}). We notice that along the critical line $k_{\bot} = k_{\|}\sqrt{2/3}$ ($\theta \approx 39^{\circ}$), the maximum 
growth rate (\ref{firehose_growth_oblique}) of the oblique firehose diverges. 
This divergence is mathematically the result 
of failing to take into account higher-order terms in the $k \rho_i \ll 1$ 
expansion, but, as was discussed earlier in this section, it is indicative of a physical effect (viz., much faster growth of 
firehose modes with $k_{\bot} = k_{\|}\sqrt{2/3}$). 
 
\begin{figure}
\centerline{\includegraphics[width=0.99\textwidth]{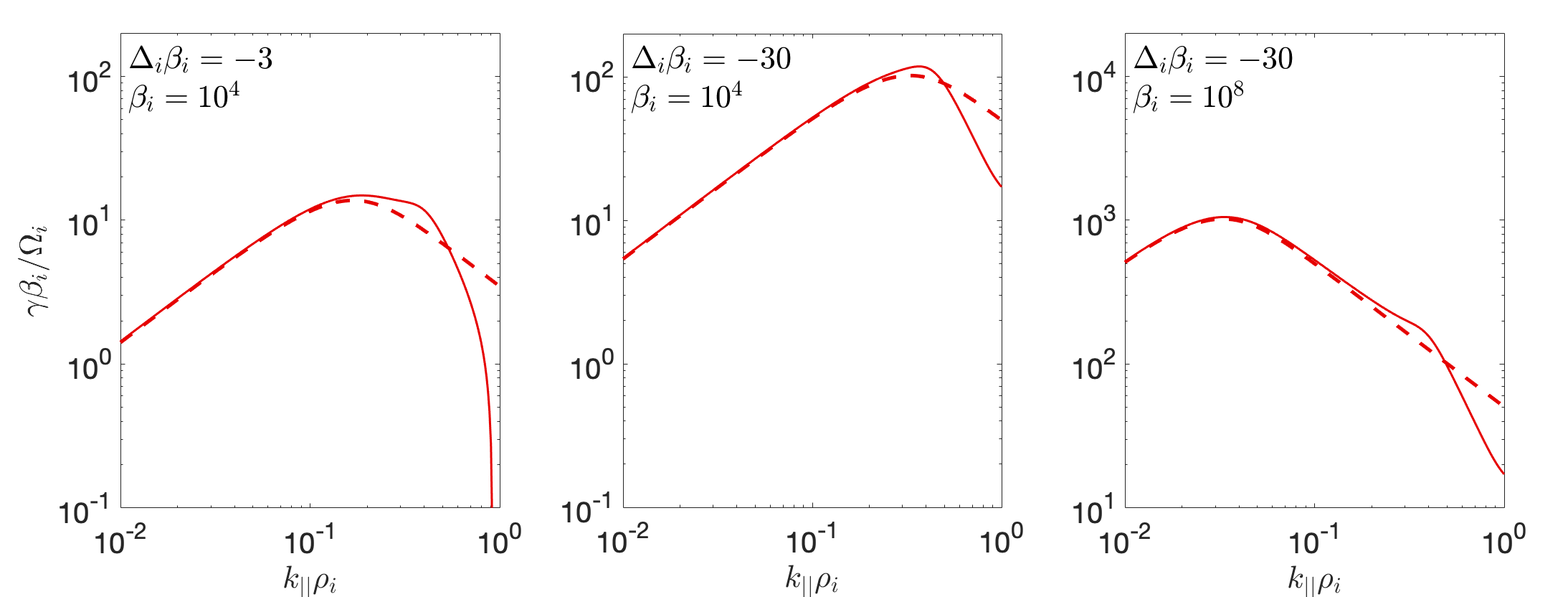}}
\caption{\textit{Oblique CES firehose instability}. Growth rates of the shear Alfv\'en mode whose instability is driven by the CE ion-shear term in the CE distribution function (\ref{CEsheardistfuncexpression}), 
for wavevectors at an angle $\theta = 60^{\circ}$ with the background 
magnetic field (viz., $k_{\perp} = \sqrt{3} k_{\|}$). 
The growth rates (solid lines) of all modes are calculated in the same way as figure~\ref{newFig_firehose_2D}. We show the growth rates for a selection of different values of $\Delta_i \beta_i$ 
and $\beta_i$.
The approximation (\ref{growthrate_firehose_oblique}) for 
the growth rate (dashed red) in the limit $k_{\|} \rho_i \ll 1$ is also plotted.
 \label{newFig_firehose_oblique}}
\end{figure}

 $\quad$The degree to which the growth rate of unstable modes determined from (\ref{growthrate_firehose_oblique})
 follows a numerical solution for a particular choice of $\theta$ is demonstrated in figure 
\ref{newFig_firehose_oblique}.
The agreement is reasonable, although an increasingly large discrepancy develops 
as $k \rho_i$ approaches unity due to FLR effects. 

 \subsubsection{Critical-line firehose instability} \label{negpres_fire_critline} 

In this third and final case, we set $k_{\bot} = k_{\|} 
\sqrt{2/3}$. The FLR coupling between the shear Alfv\'en mode and the 
Barnes'-damped slow-mode then vanishes to leading order in $k \rho_i \ll 1$, and 
next order FLR terms must be considered. Depending on the value of $\beta_i$, we 
find two sub-cases.  

First, for $\beta_i \sim \Delta_i^{-1} \gg 10^{6}$ -- a numerical bound that we will justify \textit{a posteriori} following our 
calculations -- the FLR term responsible for 
setting the wavenumber of the fastest-growing mode is the second-order correction 
to the FLR coupling between the shear Alfv\'en and slow modes. The appropriate 
ordering to adopt then depends on the relative magnitude of $\Delta_i$ and $\beta_{i}^{-1}$. For $\Delta_i \beta_i + 2 \lesssim -1$, 
we use the ordering
\begin{equation}
  \tilde{\omega}_{i\|} \sim \frac{1}{\beta_i^{1/2}} \sim |\Delta_i|^{1/2} \sim 
     k_{\|}^6 \rho_i^6 \, . \label{ordering_firehose_critline_Barnes}
\end{equation}
In this case, we find (see appendix \ref{derivation_firehose_critline}) that the frequency of the two shear Alfv\'en modes 
is given by
  \begin{equation}
 \frac{\omega}{\Omega_i} = -k_{\|} \rho_i \Bigg[\frac{ 6889 \mathrm{i} k_{\|}^6 \rho_i^6}{27648 \sqrt{\upi}}
 \pm \sqrt{\left(\frac{1}{\beta_i}+\frac{\Delta_i}{2}\right) - \frac{6889^2}{27648^2 \upi } k_{\|}^{12} \rho_i^{12} } \Bigg]
 \, . \label{growthrate_firehose_critline_Barnes}
 \end{equation}
 The wavelength at which the growth rate is maximised
 scales with an extraordinarily low power of $\left|2\beta_i^{-1}+\Delta_i\right|$: 
 \begin{equation}
  (k_{\|} \rho_i)_{\mathrm{peak}} \approx \frac{2^{19/12} 3^{1/2}\upi^{1/12}}{83^{1/3} 35^{1/12}} \left|\frac{2}{\beta_i}+\Delta_i\right|^{1/12} \approx 0.97 \left|\frac{2}{\beta_i}+\Delta_i\right|^{1/12} \, , 
  \label{firehosepeakgrowthrate_kval}
 \end{equation}  
 with associated maximum growth rate
 \begin{equation}
  \frac{\gamma_{\mathrm{max}}}{\Omega_i} \approx  \frac{2^{13/12} 3^{1/2}\upi^{1/12}}{83^{1/3} 35^{1/12}} \left|\frac{2}{\beta_i}+\Delta_i\right|^{7/12} \approx 0.58 \left|\frac{2}{\beta_i}+\Delta_i\right|^{7/12} \, . \label{firehose_growth_critline}
 \end{equation}  
 
 As discussed in section \ref{negpres_fire}, the instability threshold for critical-line firehose modes is not
 (\ref{firehose_thres_oblique}), but is a less stringent value. We can demonstrate this analytically by showing that, 
 for $\Delta_i \simeq -2/\beta_i$, critical-line firehose modes are still unstable. 
 Adopting the ordering
 \begin{equation}
  \tilde{\omega}_{i\|} \sim \frac{1}{\beta_i^{3/5}} \sim 
     k_{\|}^6 \rho_i^6 \, , \label{ordering_firehose_critline_Barnes_marg}
\end{equation}
it follows (see appendix \ref{derivation_firehose_critline}) that the growth rate of the critical-line firehose modes is
   \begin{equation}
 \frac{\gamma}{\Omega_i} = -k_{\|} \rho_i \Bigg[\frac{ 6889 k_{\|}^6 \rho_i^6}{27648 \sqrt{\upi}}
 \pm \sqrt{\frac{5}{4 \beta_i} k_{\|}^2 \rho_i^2 + \frac{6889^2}{27648^2 \upi } k_{\|}^{12} \rho_i^{12} } \Bigg]
 \, . \label{growthrate_firehose_margcritline_Barnes}
 \end{equation}
  The maximum growth rate of such modes is then given by
  \begin{equation}
  \frac{\gamma_{\mathrm{max}}}{\Omega_i} \approx  \frac{2^{3} 5^{7/10} 3^{3/2} \upi^{1/5}}{83^{4/5} 7^{7/10}} \beta_i^{-7/10}  \approx 1.2 \beta_i^{-7/10}  \label{firehose_growth_critline_Barnes_marg}
 \end{equation}  
obtained at parallel wavenumber
 \begin{equation}
  (k_{\|} \rho_i)_{\mathrm{peak}}  \approx \frac{2 5^{1/10} 3^{1/2} \upi^{1/10}}{83^{2/5} 7^{1/10}} \beta_i^{-1/10} \approx 0.64 \beta_i^{-1/10}   \, . \label{firehosepeakgrowthrate_Barnes_kval_marg}
 \end{equation}
 
 When $\beta_i \sim \Delta_i^{-1} \ll 10^{6}$ the fastest-growing critical-line 
 firehose modes have a sufficiently large wavenumber that the effect of FLR coupling between
 shear Alfv\'en and slow modes is sub-dominant to the effect of cyclotron damping. 
 Assuming that $\Delta_i \beta_i + 2 \lesssim -1$ and adopting the ordering
 \begin{equation}
  \tilde{\omega}_{i\|} \sim \frac{1}{\beta_i^{1/2}} \sim |\Delta_i|^{1/2} \, , 
  \quad k_{\|} \rho_i \sim \frac{1}{\sqrt{\log{1/\left|\beta_i^{-1}+\Delta_i/2\right|}}}
  , \label{ordering_firehose_critline_cyclo}
\end{equation}
we show in appendix \ref{derivation_firehose_critline} that the frequency of the shear Alfv\'en modes becomes 
   \begin{equation}
 \frac{\omega}{\Omega_i} = - \frac{\mathrm{i} \sqrt{\upi}}{2 k_{\|} \rho_i} \exp{\left(-\frac{1}{k_\|^2 \rho_i^2}\right)}
 \pm k_{\|} \rho_i \sqrt{\left(\frac{1}{\beta_i}+\frac{\Delta_i}{2}\right) - \frac{\pi}{4 k_{\|}^4 \rho_i^4} \exp{\left(-\frac{1}{k_\|^2 \rho_i^2}\right)} } 
 \, .
 \label{growthrate_firehose_critline_cyclotron}
 \end{equation}
In this case, the maximum growth rate
 \begin{equation}
  \frac{\gamma_{\mathrm{max}}}{\Omega_i} \approx (k_{\|} \rho_i)_{\mathrm{peak}} \left|\frac{1}{\beta_i}+\frac{\Delta_i}{2}\right|^{1/2} \label{firehose_growth_critline_cyclo}
 \end{equation}  
 is attained at
 \begin{equation}
  (k_{\|} \rho_i)_{\mathrm{peak}}  \approx  \frac{\sqrt{2}}{\sqrt{\log{1/\left|\beta_i^{-1}+\Delta_i/2\right|}}} \left[1-\frac{4 \log{\left(\log{1/\sqrt{\left|\beta_i^{-1}+\Delta_i/2\right|}}\right)}}{\log{1/\left|\beta_i^{-1}+\Delta_i/2\right|}}\right] \, . \label{firehosepeakgrowthrate_cyclo_kval}
 \end{equation}
 Figure \ref{newFig_firehose_2D} corroborates that the analytical approximation~(\ref{firehosepeakgrowthrate_cyclo_kval})
provides a reasonable estimate of the parallel wavenumber at which peak growth 
occurs. 

Similarly to the $\beta_i \gg 10^{6}$ regime, when $\beta_i \ll 10^{6}$, critical-line firehose modes still grow when $\Delta_i \approx -2/\beta_i$. Their growth rate as a function of wavenumber is given by 
  \begin{equation}
 \frac{\gamma}{\Omega_i} = - \frac{\sqrt{\upi}}{2 k_{\|} \rho_i} \exp{\left(-\frac{1}{k_\|^2 \rho_i^2}\right)}
 \pm k_{\|} \rho_i \sqrt{\frac{5}{4 \beta_i} k_{\|}^2 \rho_i^2 + \frac{\pi}{4 k_{\|}^4 \rho_i^4} \exp{\left(-\frac{1}{k_\|^2 \rho_i^2}\right)} } 
 \, .
 \label{growthrate_firehose_critline_cyclo_marg}
 \end{equation}
 The maximum of (\ref{growthrate_firehose_critline_cyclo_marg}), 
  \begin{equation}
  \frac{\gamma_{\mathrm{max}}}{\Omega_i} \approx \frac{\sqrt{5}}{2} (k_{\|} \rho_i)_{\mathrm{peak}}^2 \beta_i^{-1/2}  , \label{firehose_growth_critline_cyclo_marg}
 \end{equation}  
 is achieved at
 \begin{equation}
  (k_{\|} \rho_i)_{\mathrm{peak}}  \approx \frac{\sqrt{2}}{\sqrt{\log{\left(\upi \beta_i/20\right)}}} \left\{1 - \frac{3\log{\left[\log{\left(\upi \beta_i/20\right)}/2\right]}}{\log{\left(\upi \beta_i/20\right)}}\right\}  \, . \label{firehosepeakgrowthrate_cyclo_kval_marg}
 \end{equation}

\begin{figure}
\centerline{\includegraphics[width=0.99\textwidth]{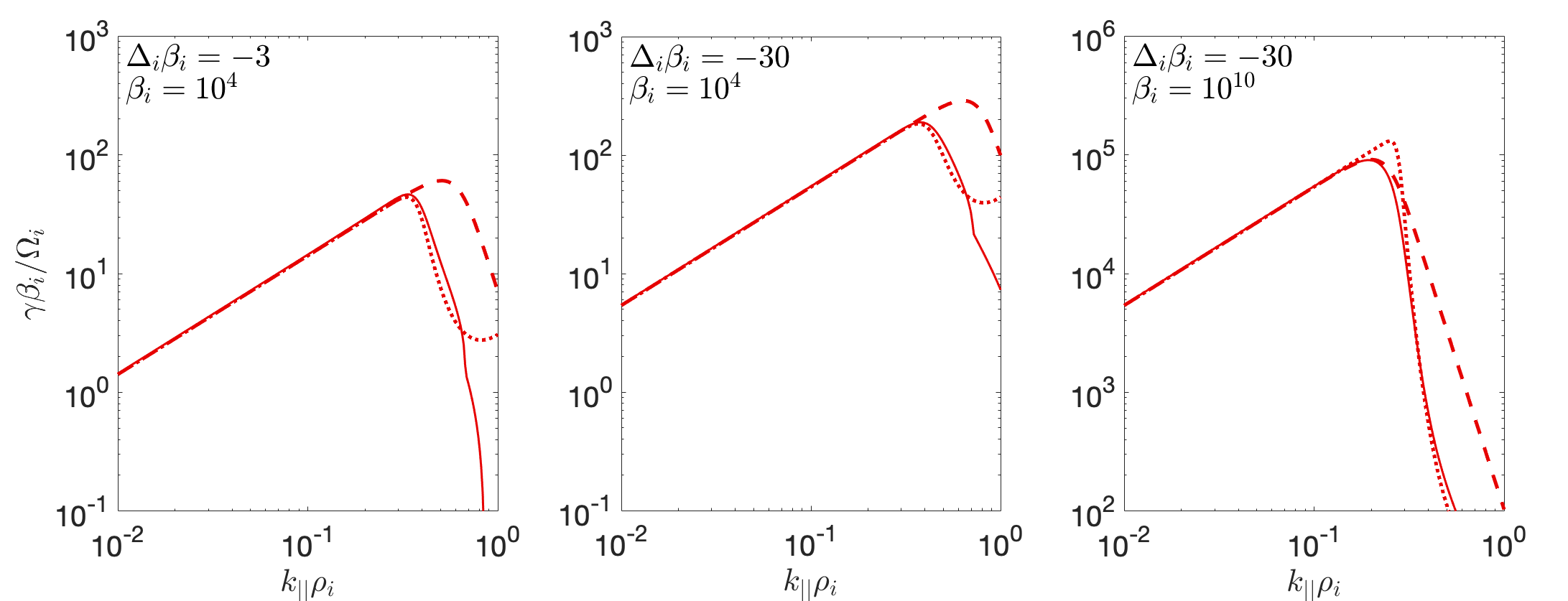}}
\caption{\textit{Critical-line CES firehose instability}. Growth rates of shear Alfv\'en modes whose instability is driven by the CE ion-shear term in the CE distribution function (\ref{CEsheardistfuncexpression}), 
for wavevectors at an angle $\theta \approx 39^{\circ}$ with the background 
magnetic field (viz., $k_{\perp} = \sqrt{2/3} k_{\|}$). 
The growth rates (solid lines) of all modes are calculated in the same way as figure~\ref{newFig_firehose_2D}. We show the growth rates for a selection of different values of $\Delta_i \beta_i$ 
and $\beta_i$.
The approximations (\ref{growthrate_firehose_critline_Barnes}) and (\ref{growthrate_firehose_critline_cyclotron}) for 
the growth rate (dashed and dotted red, respectively) in the limit $k_{\|} \rho_i \ll 1$ are also plotted. 
 \label{newFig_firehose_critline}}
\end{figure}

 By comparing the expressions (\ref{growthrate_firehose_critline_Barnes}) 
 and (\ref{growthrate_firehose_critline_cyclotron})
 for the complex frequency of shear Alfv\'en modes -- specifically, the ratio of the final terms -- the dependence on $\beta_i$ (equivalently, $\Delta_i$) of the relative importance  
 of FLR slow-mode 
 coupling and cyclotron damping can be determined. This ratio is
 ${\sim}0.16 k_{\|}^8 \rho_i^8 \exp{(-{1}/{k_\|^2 
 \rho_i^2})}$, with equality being achieved when $k_{\|} \rho_i \approx 0.3$. 
 Using (\ref{firehosepeakgrowthrate_kval}) to estimating the value of $\left|2\beta_i^{-1}+\Delta_i\right|$ at 
 which this value of $k_\| \rho_i$ is achieved, we find that $\left|2\beta_i^{-1}+\Delta_i\right| \approx 8 \times 
 10^{-7}$. Assuming $|\Delta_i \beta_i^{-1} + 2| \sim 1$, we conclude that, for $\beta_i \lesssim 
 10^{6}$, cyclotron damping will determine the wavenumber cutoff, with this transition value of $\beta_i$ proportional to the value of $|\Delta_i| \beta_i$. This  
 estimate can be validated numerically by comparing (\ref{growthrate_firehose_critline_Barnes}) 
 and (\ref{growthrate_firehose_critline_cyclotron})
 with the numerically determined growth rate (see figure 
 \ref{newFig_firehose_critline}).
We indeed find that, for $\beta_i \sim \Delta_i^{-1} \ll 10^6$, the effect of cyclotron damping sets the 
wavenumber of peak growth, while FLR slow-mode coupling does so for $\beta_i \sim \Delta_i^{-1} \gg 
10^6$. In both cases, the superior of the two analytic approximations closely matches the 
numerical growth rate. 
 
These results suggest that, for very large $\beta_i$, the wavenumber of the maximum growth of the firehose instability 
 satisfies $k \rho_i \ll 1$, rather than $k \rho_i \sim 1$. This result 
 might seem to contradict previous authors who claim to have found numerical evidence 
 that the fastest growth rates of the firehose instability occur at $k \rho_i \sim 
 1$~\citep{YWA93,SCKHS05,KSS14}; however, given the logarithmic dependence of the characteristic wavenumber (\ref{firehosepeakgrowthrate_cyclo_kval}), 
 we conclude that it would take simulations at very high $\beta_i$ to be able to distinguish 
 between $k \rho_i \sim 1$ and $k \rho_i \sim \beta_i^{-1/12} \ll 1$. In addition, the 
 results presented in figure \ref{newFig_firehose_marginal}b indicate 
 that firehose modes with $k \rho_i \sim 1$ have a less stringent  
 instability threshold on $\Delta_i$ than (\ref{firehose_thres_oblique}), 
providing an opportunity for such modes to grow significantly before 
longer-wavelength modes can do so. In short, it seems reasonable to 
assume for all practical purposes that the dominant firehose modes occur at $k \rho_i \sim 1$, provided $\beta_i$ is not extremely 
 large.

 \subsubsection{Sub-ion-Larmor-scale firehose instability} \label{negpres_fire_subion}

Figure \ref{newFig_firehose_2D}b also suggests that, once $|\Delta_i| \beta_i \gg 1$, 
firehose modes on sub-ion-Larmor scales develop -- albeit with a smaller growth rate
than the critical-line ones. Similarly to sub-ion-Larmor-scale mirror modes (see the end of 
section \ref{pospress_ion_mirror}), we can characterise these modes analytically by adopting the ordering
\begin{equation}
 k_{\|} \rho_i \sim k_{\perp} \rho_i \sim (|\Delta_i| \beta_i)^{1/2} \gg 1 , \quad 
\frac{\gamma}{\Omega_i} \sim \Delta_i \, .  
\end{equation}
If we also assume that $|\Delta_i| \beta_i \ll \mu_e^{-1/2}$, 
it is shown in appendix \ref{derivation_mirror} that the growth rate of these modes is given by 
\begin{eqnarray}
\frac{\gamma}{\Omega_i} & \approx & \frac{k_\|}{k} \sqrt{\left(-\Delta_i \frac{k_{\bot}^2-k_{\|}^2}{k^2} - \frac{k^2 \rho_i^2}{\beta_i} \right) \left(\frac{k^2 \rho_i^2}{\beta_i} - \Delta_i \frac{k_{\|}^2}{k^2}\right)}  \nonumber \\
& = & \cos{\theta} \sqrt{\left[-\Delta_i \left(\sin^2{\theta}-\cos^2{\theta}\right)- \frac{k^2 \rho_i^2}{\beta_i} \right] \left(\frac{k^2 \rho_i^2}{\beta_i} - \Delta_i \cos^2{\theta} \right)} \, . 
 \label{ionfirehose_subionscale_theta}
\end{eqnarray}
If $\Delta_i < 0$, we have an instability for all modes with $\theta > 45^{\circ}$
whose total wavenumber satisfies 
\begin{equation}
k \rho_i < \sqrt{|\Delta_i| \beta_i \left(\sin^2{\theta}-\cos^2{\theta}\right)}  
\, . \label{ionfirehose_subionscale_kbound}
\end{equation}
Analogously to the sub-ion-Larmor-scale mirror modes [cf. (\ref{mirrorgrowth_largek})], the growth is maximised when $k \rho_i\ll (|\Delta_i| 
\beta_i)^{1/2}$ and $\theta \approx 55^{\circ}$, with
\begin{equation}
 \gamma_{\rm max} = \frac{1}{3 \sqrt{3}} |\Delta_i| \Omega_i  \approx 0.19 |\Delta_i| \Omega_i \, . \label{firehosegrowth_largek}
\end{equation}
In contrast to the case of the mirror instability, this growth rate is asymptotically small in $\Delta_i \ll 1$ compared to the peak 
growth rate of the critical-line firehose modes [cf. (\ref{firehose_growth_critline}) and 
(\ref{firehosepeakgrowthrate_cyclo_kval})], and thus the instability of sub-ion-Larmor-scale firehose modes 
is always subdominant. For completeness, we note that, once $|\Delta_i| \beta_i \sim 
\mu_e^{-1/2}$, the electron-pressure anisotropy associated with the CE electron-shear term begins to play a comparable role to 
the ion-pressure anisotropy for modes with $k \rho_i \sim (|\Delta_i| 
\beta_i)^{1/2}$. In this case, the expression for the growth rate becomes
\begin{eqnarray}
\frac{\gamma}{\Omega_i} & \approx & \frac{k_\|}{k} \Bigg\{\left[-\Delta_i \frac{k_{\bot}^2-k_{\|}^2}{k^2} - k^2 \rho_i^2\left(\frac{1}{\beta_i} +\frac{\mu_e^{1/2} \Delta_i}{2}\right) \right] \nonumber \\
&& \qquad \times \left[\frac{k^2 \rho_i^2}{\beta_i} - \Delta_i \left(\mu_e^{1/2} k_{\perp}^2 \rho_i^2 -\frac{1}{2} \mu_e^{1/2} k_{\|}^2 \rho_i^2 + \frac{k_{\|}^2}{k^2}\right)\right]\Bigg\}^{1/2} \, .  
 \label{ionfirehose_subionscalewithelec}
\end{eqnarray}
The bound (\ref{ionfirehose_subionscale_kbound}) on the total wavenumber required for the instability of modes with $k_{\perp} > k_{\|}$ is then 
\begin{equation}
k \rho_i < \sqrt{\frac{|\Delta_i| \beta_i \left(\sin^2{\theta}-\cos^2{\theta}\right)}{1+\mu_e^{1/2} \Delta_i \beta_i/2}}  
\, . \label{ionfirehose_subionscale_kbound_withelec}
\end{equation}
Because the denominator tends to zero as $\Delta_i \rightarrow -2 \mu_e^{-1/2} 
\beta_i^{-1}$, the bound becomes increasingly weak, and so the
region of $(k_{\|},k_{\perp})$-space in which there is instability extends 
significantly towards electron Larmor scales. This extension precedes 
the onset of the oblique electron firehose instability (see section 
\ref{negpres_electron_oblique}). 

\subsubsection{Parallel electron firehose instability} \label{negpress_electronfire_prl}

The CES parallel electron firehose instability arises when the negative electron-pressure anisotropy ($\Delta_e < 0$)
associated with the CE electron-shear term becomes a sufficiently large free-energy
source to overcome the relatively weak collisionless damping mechanisms that act on
long-wavelength ($k_{\|} \rho_e \ll 1$) quasiparallel whistler 
waves by changing their handedness from right- to left-handed. More specifically, whistler waves with quasi-parallel wavevectors do not 
have a component of electric field parallel to $\boldsymbol{B}_0$, and so are 
not subject to electron Landau damping. 
Electron cyclotron damping does occur, but is very inefficient for $k_{\|} \rho_e \ll 
1$. The resonant interaction primarily responsible for damping is that between the whistler waves and Maxwellian ions in the CE plasma streaming along 
field lines with $v_{\|} \ll v_{\mathrm{th}i}$. When the handedness of the whistler waves changes, this interaction instead leads to the waves' growth. 
Because the resonant interaction driving the instability involves the plasma's ions,  
the CES parallel electron firehose instability has a rather small growth rate compared to other CES electron-scale microinstabilities, 
with growth disappearing entirely in the special case of cold ions. The 
parallel wavenumber of peak growth, which is a small but finite fraction of the electron Larmor scale, viz., $(k_{\|} \rho_e)_{\rm peak} \approx 0.4$ for $\Delta_e \lesssim -2/\beta_e$,
is set by electron cyclotron damping, which prevents shorter-wavelength modes 
from becoming unstable. The CES parallel 
electron firehose instability was first identified by~\citet{HV70} and
has been studied subsequently using theory and simulations in plasma 
with $\beta_e \sim 1$-$20$ by a number of authors~\citep[e.g.,][]{PB99,LH00,M02,GN03,CB08,CB10,RQV18}. 

\begin{figure}
\centerline{\includegraphics[width=0.99\textwidth]{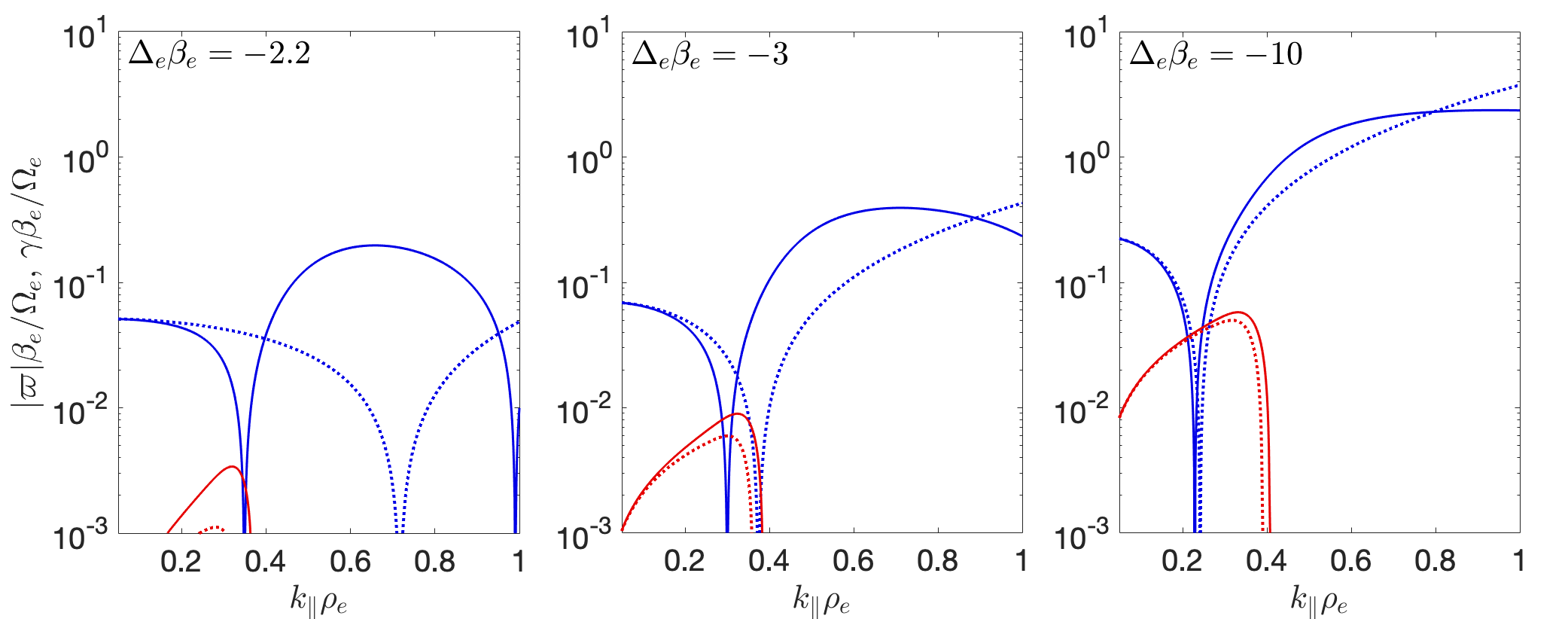}}
\caption{\textit{Parallel CES electron firehose instability}.  Dispersion curves of unstable whistler modes, whose instability is driven by the negative electron-pressure anisotropy associated with the electron-shear term in CE distribution function 
(\ref{CEsheardistfuncexpression}), for wavevectors that are co-parallel with the background 
magnetic field (viz., $\boldsymbol{k} = k_{\|} \hat{\boldsymbol{z}}$). 
The frequency (solid blue) and growth rates (solid red) of the modes are calculated using 
(\ref{electronweibelgrowthrate}\textit{a}) and 
(\ref{electronweibelgrowthrate}\textit{b}), respectively. 
The resulting frequencies and growth rates, when normalised as $\gamma \beta_e/\Omega_e$, are functions of the dimensionless quantity $\Delta_e \beta_e$; we show the dispersion
curves for three different values of $\Delta_e \beta_e$. 
The $k_{\|} \rho_e \ll 1$ approximations (\ref{elecfire_growthrate_kplsml}\textit{a}) for 
the frequency (dotted-blue) and (\ref{elecfire_growthrate_kplsml}\textit{b}) growth rate (dotted-red) are also plotted. 
\label{Figure_newCESparelecfire}}
\end{figure}

To characterise the parallel electron firehose instability analytically, we can simply use the 
expressions (\ref{electronweibelgrowthrate}\textit{a}) and (\ref{electronweibelgrowthrate}\textit{b}) given in section \ref{pospres_electron_EC} for the real frequency $\varpi$ 
and growth rate $\gamma$, respectively, of the parallel whistler waves that satisfy the ordering 
\begin{equation}
  \tilde{\omega}_{e\|} = \frac{\omega}{k_{\|} v_{\mathrm{th}e}} \sim \Delta_e \sim \frac{1}{\beta_e} \, , 
  \label{elecfireorder}
\end{equation} 
and have $k_{\|} \rho_e \sim 1$, but this time with $\Delta_e \beta_e < 0$. 
Plots of the dispersion curves $\varpi(k_{\|})$ and $\gamma(k_{\|})$ of CES parallel electron firehose modes 
are then shown  in figure 
\ref{Figure_newCESparelecfire} for a selection of different (negative) values of $\Delta_e \beta_e$.
In a hydrogen plasma, we find an instability for $\Delta_e < (\Delta_e)_{\rm c} \approx 
-1.7/\beta_e$. For $\Delta_e \lesssim -2/\beta_e$, 
modes with $k_{\|} \rho_e \lesssim 0.4$ become unstable. Figure~\ref{Figure_newCESparelecfire} also shows
that parallel electron firehose modes generically have a real frequency that is 
much greater than their growth rate ($\varpi \sim \Omega_e/\beta_e \gg \gamma$); however, this frequency changes sign at a 
wavenumber which, when $\Delta_e \lesssim -2/\beta_e$, is comparable
to the wavenumber $(k_{\|} \rho_e)_{\rm peak}$ at which peak growth occurs. 

These results can be elucidated by considering the expressions (\ref{electronweibelgrowthrate}) in the subsidiary limit
\begin{equation}
k_{\|} \rho_i \sim \frac{1}{\sqrt{\log{\left(2 \mu_e^{-1/2}|1+2/\Delta_e \beta_e|\right)}}} \ll 1 \, 
.
\end{equation}
Then (\ref{electronweibelgrowthrate}) simplifies to
\begin{subeqnarray}
\varpi & = &  \pm \left[\left(1+\frac{\Delta_e \beta_e}{2}\right) k_{\|}^2 \rho_e^2 - \mu_e^{1/2} \Delta_e \beta_e \right] \frac{\Omega_e}{\beta_e}, \\
\gamma  & = & 
\frac{\sqrt{\upi}}{k_\| \rho_e} \left[\Delta_e \exp{\left(-\frac{1}{k_\|^2 \rho_e^2}\right)}-\left(\frac{\Delta_e}{2} + \frac{1}{\beta_e}\right) \mu_e^{1/2} k_\|^2 \rho_e^2 \right] \Omega_e . \qquad  
\label{elecfire_growthrate_kplsml}
\end{subeqnarray}
These approximations are plotted alongside (\ref{electronweibelgrowthrate}) in figure 
\ref{Figure_newCESparelecfire}; the agreement is qualitative rather than quantitative for $\Delta_e \sim 
-2/\beta_e$, but becomes increasingly good as $\Delta_e$ is decreased further. 

Using these simplified expressions, we can derive approximate analytical 
expressions for the instability's threshold $(\Delta_e)_{\rm c}$, as well as its peak growth rate and the 
wavenumber at which that growth occurs. First considering the sign of (\ref{elecfire_growthrate_kplsml}), it is easy to show that there exists a 
range of wavenumbers $k_{\|}$ at which $\gamma > 0$ if and only if $\Delta_e < 
-2/\beta_e$, so $(\Delta_e)_{\rm c} \approx -2/\beta_e$. This is somewhat more stringent than the numerically observed 
threshold, a discrepancy attributable to FLR effects, not taken into account by the 
approximation (\ref{elecfire_growthrate_kplsml}\textit{b}). When $\Delta_e < 
-2/\beta_e$, it can be proven that the growth rate (\ref{elecfire_growthrate_kplsml}\textit{b}) is 
maximised at
\begin{equation}
 (k_{\|} \rho_e)_{\rm peak} \approx  \frac{1}{\sqrt{\log{\left(\mu_e^{-1/2}|1/2+1/\Delta_e \beta_e|\right)}}} 
 \left\{1-\frac{\log{\left[\sqrt{2}\log{\left(\mu_e^{-1/2}|1/2+1/\Delta_e \beta_e|\right)}\right]}}{\log{\left(\mu_e^{-1/2}|1/2+1/\Delta_e \beta_e|\right)}}\right\} 
 \, , \label{elecfirepar_kval}
\end{equation}
attaining the value
\begin{equation}
\gamma_{\rm max} = \sqrt{\upi} \mu_e^{1/2} (k_{\|} \rho_e)_{\rm peak} \left|\frac{\Delta_e}{2}+\frac{1}{\beta_e}\right| 
\Omega_e
\, . \label{elecfirepar_gammamax}
\end{equation} 

Comparing (\ref{elecfirepar_gammamax}) with the characteristic magnitude of $\varpi$ evaluated using (\ref{elecfire_growthrate_kplsml}\textit{a})
at $k_{\|} \rho_e = (k_{\|} \rho_e)_{\rm peak}$ (and assuming that $(k_{\|} \rho_e)_{\rm peak} \gtrsim \mu_e^{1/4}$), 
we conclude that $\gamma \lesssim \mu_e^{1/4} \varpi$, thereby explaining our 
previous observation that the growth rate of parallel electron firehose modes is generically 
much smaller than the real frequency of those modes. We can also show that the one exception to this 
occurs when $(k_{\|} \rho_e)_{\rm peak} \approx \mu_e^{1/4} [2 \Delta_e \beta_e/(1+2 \Delta_e 
\beta_e)]^{1/2}$, an approximate expression for the wavenumber below which $\varpi$
changes sign. As we will see, the characteristic growth rate of the CES parallel electron firehose 
is typically much smaller than its oblique relative in high-$\beta$ plasma (see section \ref{negpres_electron_oblique}),  
a conclusion that also applies in $\beta_e \sim 1$ plasmas with bi-Maxwellian 
distributions~\citep[see][]{LH00}.

\subsubsection{Oblique electron firehose instability} \label{negpres_electron_oblique}

In spite of its similar name, the CES oblique electron firehose instability is quite
distinct from its parallel cousin: it is a non-propagating mode than arises from the
destablisation of oblique KAWs by a sufficiently negative electron 
pressure anisotropy. The linear theory of the analogous instability in $\beta_e \sim 1$ plasma 
with bi-Maxwellian electrons was first presented by~\citet{LH00}, with a number 
of simulation studies of this instability having been conducted subsequently~\citep{GN03,CB08,CB10,RQV18}. 
The high-$\beta$ variant of the (linear) instability for general anisotropic electron distribution 
functions was studied in the $k_{\|} \ll k_{\perp}$ limit of gyrokinetics 
by~\citet{KAKS18}. In contrast to the findings of~\citet{GN03}, who showed that the oblique 
electron firehose instability in a bi-Maxwellian plasma at $\beta_e \sim 1$ involves 
gyroresonant wave-particle interactions between electrons and the unstable 
modes, instability of CES oblique electron firehose modes at $\beta_e \gg 1$ is essentially 
non-resonant, with sufficient large negative electron pressure anisotropies 
negating the restoring force that underpins the oscillation of high-$\beta$ 
KAWs. 

\begin{figure}
\centerline{\includegraphics[width=0.99\textwidth]{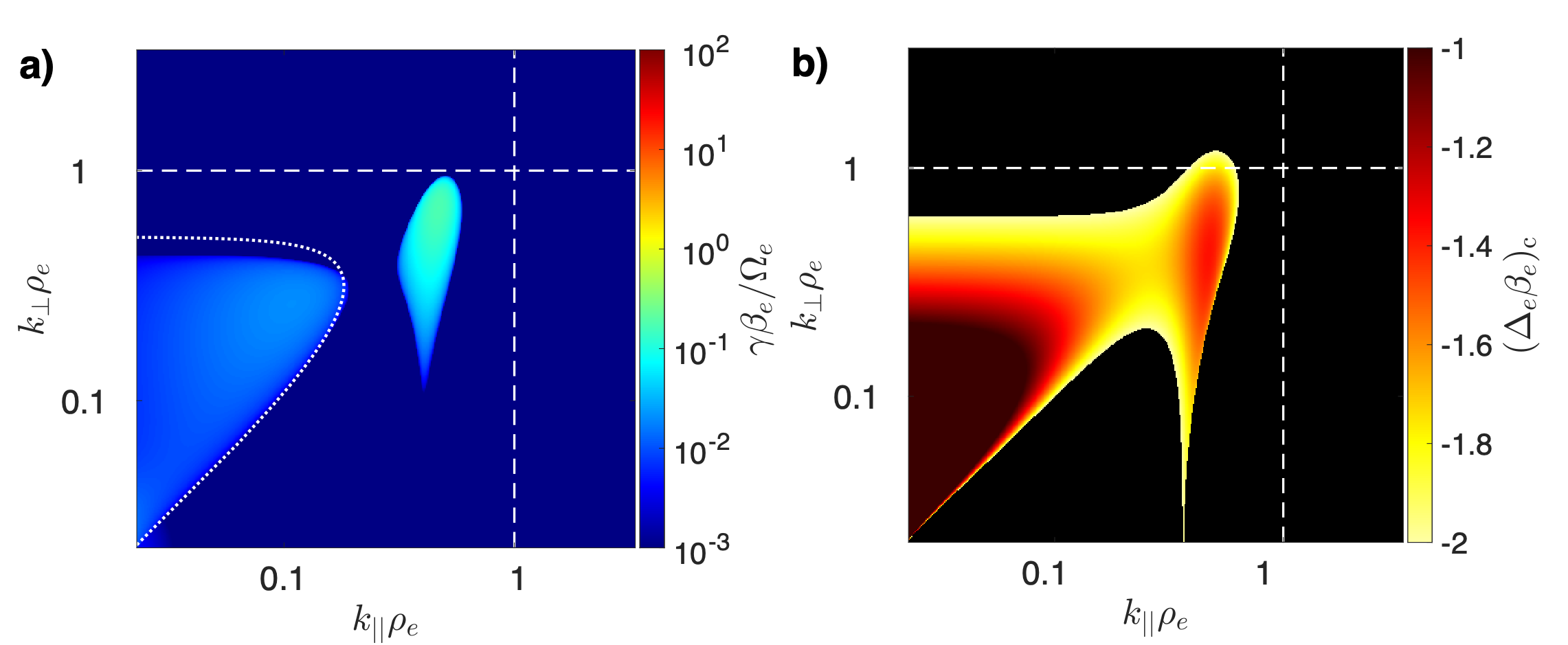}}
\caption{\textit{Onset of the CES oblique electron firehose instability}. \textbf{a)} Maximum positive growth rates of linear perturbations resulting from both the CE ion- and electron-shear term in the CE distribution function (\ref{CEsheardistfuncexpression}) with $\beta_i = 10^{4}$ and $\Delta_e = 
-1.7/\beta_e$ (which is above the long-wavelength oblique electron-firehose instability-threshold $\Delta_e = 
-2/\beta_e$). The growth rates of all modes are calculated in the same way as figure \ref{newFig_firehose_2D}. 
The resulting growth rates are normalised as $\gamma \beta_e/\Omega_e$ are functions of the dimensionless parameter $\Delta_e \beta_e$.
The dotted line denotes the instability boundary (\ref{ionfirehose_subionscale_kbound_withelec}) 
on KAWs driven unstable by ion pressure anisotropy of the CE ion-shear term. 
\textbf{b)} Threshold value of $\Delta_e \beta_e$ at which modes with parallel and perpendicular wavenumber $k_{\|}$ and $k_{\perp}$, respectively, become 
unstable. Regions of $(k_{\|},k_{\perp})$ that are shaded black are stable. \label{newFig_elecfirehose_marginal}}
\end{figure}

Similarly to the parallel electron firehose instability, the CES oblique 
electron firehose instability is triggered when $\Delta_e \lesssim -2/\beta_e$. 
The precise value of the threshold depends on the wavevector of the 
mode being destabilised. Analogously to the parallel electron firehose, long-wavelength oblique electron 
firehose modes are unstable when $\Delta_e < (\Delta_e)_{\rm c} = 
-2/\beta_e$. However, figure \ref{newFig_elecfirehose_marginal}a shows that 
there is positive growth of $k \rho_e \sim 1$ oblique electron firehose modes
for less negative values of $\Delta_e$, illustrating that 
the threshold is less stringent for such modes. 
This phenomenon is reminiscent of the ion firehose instability (see figure \ref{newFig_firehose_marginal}): ion-Larmor-scale oblique firehose modes 
also have a less stringent threshold than longer-wavelength modes. In addition 
to the $k \rho_e \sim 1$ modes, a region of unstable KAWs with characteristic wavenumbers 
$\mu_e^{1/2} \ll k \rho_e \ll \mu_e^{1/4}$, $k_{\bot} \sim k_{\|}$, is evident in figure \ref{newFig_elecfirehose_marginal}a. 
These modes, which were discussed at the end of section \ref{negpres_fire}, are 
destabilised by negative ion pressure anisotropy; the extent of this region closely matches the analytic prediction (\ref{ionfirehose_subionscale_kbound_withelec}).
Using a similar semi-analytic approach to that employed for the case of the ion firehose 
instability (see appendix \ref{thresholdcal_quadratic}), we can determine the 
approximate threshold for the oblique electron firehose instability as a function of $k_{\|} \rho_e$ 
and $k_{\perp} \rho_e$. The results are shown in figure 
\ref{newFig_elecfirehose_marginal}b; modes with $k_{\|} \rho_e \sim 0.5$, $k_{\perp} \rho_e \sim 0.4$
have the least stringent threshold ($\Delta_e \approx -1.4/\beta_e$). 

\begin{figure}
\centerline{\includegraphics[width=0.99\textwidth]{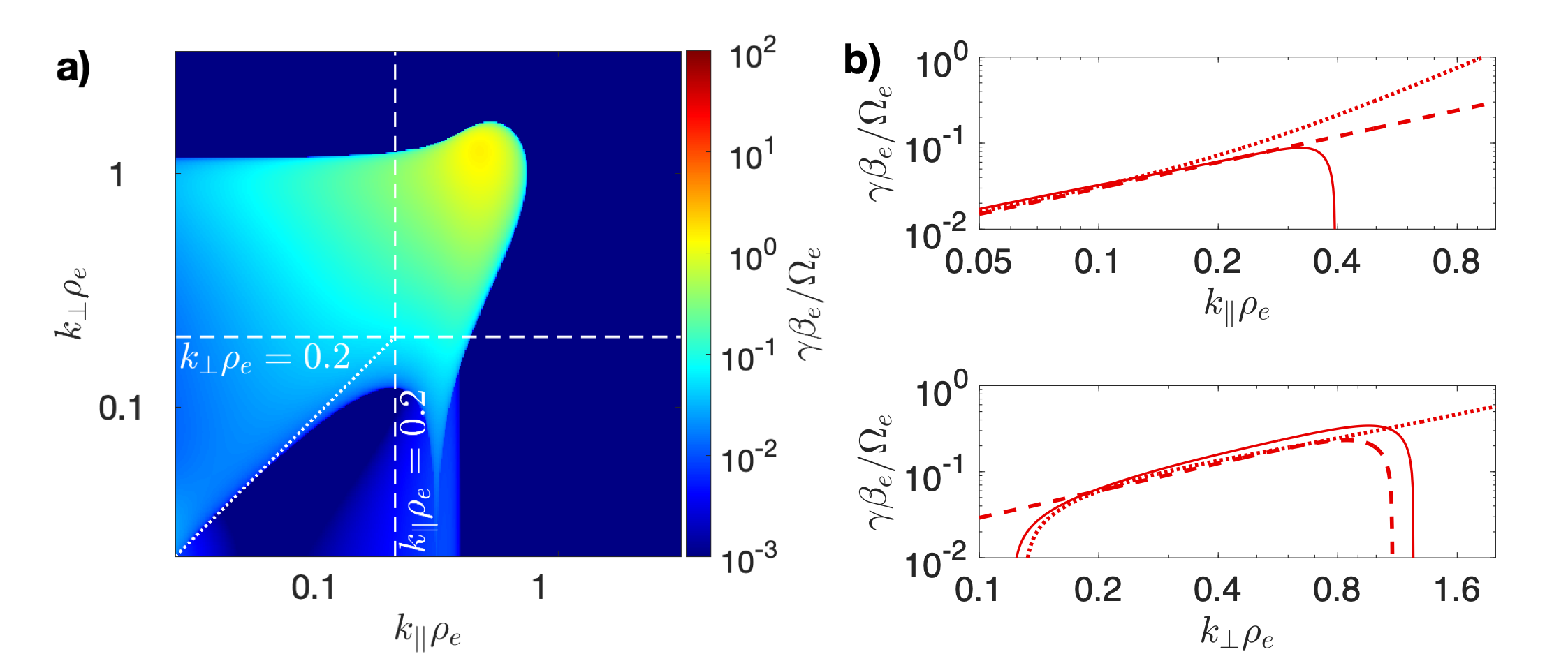}}
\caption{\textit{Oblique electron firehose instability at $\Delta_e \beta_e +2 \lesssim -1$}. \textbf{a)} Maximum positive growth rates of linear perturbations 
resulting from CE ion- and electron-shear terms in the CE distribution function (\ref{CEsheardistfuncexpression}) for $\Delta_e \beta_e = -3$.
Here, a temperature-equilibrated hydrogen plasma is considered, viz., $\Delta_e = \mu_e^{1/2} \Delta_i$, and $\beta_i = \beta_e$.  
The growth rates of all modes are calculated in the same way as figure \ref{newFig_elecfirehose_marginal}. 
\textbf{b)} Plots of the oblique electron firehose mode growth rate (solid line) as a function of $k_{\|} \rho_e$ with $k_{\bot} \rho_e = 0.2$ (top), and as a function of $k_{\perp} \rho_e$ with $k_{\|} \rho_e = 0.2$ (bottom). The 
dotted and dashed lines show the analytical predictions (\ref{obliqueinstab_freq_gen_elecfire}) and (\ref{ESTmodefreq}), respectively. \label{Figure_obliqueelecfire_unity}}
\end{figure}

Well into the unstable regime, i.e., when $\Delta_e\beta_e + 2 \lesssim -1$, electron firehose 
modes across a broad range of wavevectors are destabilised (see figure 
\ref{Figure_obliqueelecfire_unity}a).
The fastest-growing electron firehose modes are oblique and occur at electron Larmor 
scales ($k_{\perp} \rho_e \sim 1 > k_{\|} \rho_e$), with characteristic 
growth rate $\gamma \sim |\Delta_e| \Omega_e \sim \Omega_e/\beta_e$. 
This growth rate is much larger than the peak growth rate of the parallel electron 
firehose instability (\ref{elecfirepar_gammamax}).

Similarly to the electron mirror instability, a simple analytic expression for the growth rate of the fastest-growing 
electron firehose modes when $\Delta_e\beta_e + 2 \lesssim -1$ is challenging to 
establish. We can, however, characterise the growth of two particular classes of 
electron firehose modes analytically. 

The first of these are long-wavelength (viz., $k \rho_e \ll 1$)
electron firehose modes. For these, we adopt the same ordering (\ref{KAW_ords}) as was 
considered when characterising long-wavelength electron mirror modes: 
\begin{equation}
k_{\|} \rho_e \sim k_{\perp} \rho_e \sim \mu_e^{1/4} \ll 1 , \qquad \tilde{\omega}_{e\|} = \frac{\omega}{k_{\|} v_{\mathrm{th}e}}   \sim \frac{k \rho_e}{\beta_e}  \sim |\Delta_e| k \rho_e \, 
.
\label{KAW_ords_elecfire}
\end{equation}
We then obtain a closed-form expression [cf. (\ref{obliqueinstab_freq_gen}), and also (\ref{ionfirehose_subionscalewithelec})] for the complex 
frequencies of the electron firehose modes: 
\begin{eqnarray}
\omega & \approx & \pm k_\| \rho_e \Omega_e \Bigg\{\left[\frac{1}{\beta_e} + \Delta_e\left(\frac{1}{2}- \mu_e^{1/2} \frac{k_{\|}^2 \rho_e^2- k_{\bot}^2 \rho_e^2}{k^4 \rho_e^4}\right) \right] \nonumber\\
 && \qquad \qquad \times \left[ \frac{k^2 \rho_e^2}{\beta_e} - \Delta_e \left(k_{\bot}^2 \rho_e^2 + \mu_e^{1/2} \frac{k_{\|}^2}{k^2} - \frac{1}{2} k_{\|}^2 \rho_e^2 \right) \right]\Bigg\}^{1/2} . 
 \label{obliqueinstab_freq_gen_elecfire}
\end{eqnarray}
If $\Delta_e < -2/\beta_e$, the right-hand side of (\ref{obliqueinstab_freq_gen_elecfire})
is purely imaginary for $k_{\perp} > k_{\|}$, and so we have positive growth for all long-wavelength 
electron firehose modes with $\theta > 45^{\circ}$\footnote{In fact, this condition is stronger than necessary to 
guarantee instability -- but the exact condition is somewhat complicated, so we omit discussion of it.}.
This approximation should be compared with the numerically determined growth rate in 
figure \ref{Figure_obliqueelecfire_unity}b. 
If it is further assumed that 
$\mu_e^{1/4} \ll k \rho_e \ll 1$, $k_{\bot} \sim k_{\|}$, it is shown in section \ref{pospres_electron_oblique} that 
(\ref{obliqueinstab_freq_gen_elecfire}) simplifies to an analogue of
(\ref{obliqueinstab_freq_elec}), viz., 
\begin{eqnarray}
\omega & \approx & \pm k_\| \rho_e \Omega_e \sqrt{\left(\frac{1}{\beta_e} + \frac{\Delta_e}{2}\right)\left(k^2 \rho_e^2 \frac{1}{\beta_e} + \frac{\Delta_e}{2} \left[k_{\|}^2 \rho_e^2 -2 k_{\bot}^2 \rho_e^2 \right] \right)} \, . 
 \label{obliqueinstab_freq_elec_negpres}
\end{eqnarray}
This result is again in agreement with the gyrokinetic calculations of~\citet{KAKS18}. 
Extrapolating (\ref{obliqueinstab_freq_elec_negpres}) to $k_{\|} \rho_e \sim k_{\perp} \rho_e \sim 
1$, we recover that $\gamma \sim \Omega_e/\beta_e$ when $|\Delta_e \beta_e +2| \gtrsim 1$. 

A second sub-category of electron firehose modes that can be described 
analytically are quasi-perpendicular ones. For any fixed $k_{\|} \rho_e \ll 1$, the most rapidly growing modes are strongly anisotropic: they occur when the
perpendicular wavelength is comparable to the electron Larmor radius, 
$k_{\bot} \rho_e \sim 1$. These modes can therefore be elucidated 
analytically by considering their dispersion relation
under the ordering
\begin{equation}
\tilde{\omega}_{e\|} \sim |\Delta_e| \sim \frac{1}{\beta_e} \label{elec_fire_ord}
\end{equation}
in the wavenumber domain $\mu_e^{1/2} \ll k_{\|} \rho_e \ll k_{\bot} \rho_e \sim 
1$. We solve the dispersion relation (see appendix \ref{derivation_obliqueelecfirehose}) to find
 \begin{equation}
 \frac{\omega}{\Omega_e} =  \frac{k_{\|} \rho_e}{\mathcal{F}(k_{\perp} \rho_e)} \Bigg\{-\mathrm{i} \frac{\sqrt{\upi}}{2} \left[\frac{k_{\bot}^2 \rho_e^2}{\beta_e}+\Delta_e \mathcal{H}(k_{\perp} \rho_e)\ \right] \pm \sqrt{\mathfrak{D}\left(k_{\bot} \rho_e, \beta_e, \Delta_e\right)} 
 \Bigg\}
 \, , \label{ESTmodefreq}
 \end{equation}
 where the discriminant is 
  \begin{eqnarray}
\mathfrak{D}\left(k_{\bot} \rho_e, \beta_e, \Delta_e\right) & \equiv &  \left[\frac{k_{\bot}^2 \rho_e^2}{\beta_e}+\Delta_e \mathcal{H}(k_{\perp} \rho_e) \right] \nonumber \\
&& \times \left\{\frac{1}{\beta_e} \left(1-\frac{\upi}{4} k_{\bot}^2 \rho_e^2 \right) - \Delta_e \left[\frac{\upi}{4} \mathcal{H}(k_{\perp} \rho_e) + \mathcal{F}(k_{\perp} \rho_e) \right] \right\} \, , \label{ESTmodedisc}
 \end{eqnarray}
 and the two auxiliary functions are [cf. (\ref{specialfunction_quasiperp})]
\begin{eqnarray}
\mathcal{F}(\alpha) & = & \exp{\left(-\frac{\alpha^2}{2}\right)} \left[I_{0}\left(\frac{\alpha^2}{2}\right) - I_{1}\left(\frac{\alpha^2}{2}\right)\right] \, 
, \\
\mathcal{H}(\alpha) & \equiv & 1 - \exp \left(-\frac{\alpha^2}{2}\right) I_0\!\left(\frac{\alpha^2}{2}\right) \, 
. \label{EST_specfunc}
\end{eqnarray}
As a sanity check, we observe that in the subsidiary limit $k_{\bot} \rho_e \ll 1$, (\ref{ESTmodefreq}) 
 becomes
\begin{eqnarray}
 \omega  & \approx & \pm k_{\bot} k_{\|} \rho_e^2 \Omega_e \sqrt{\left(\frac{1}{\beta_e} + \frac{\Delta_e}{2}\right)\left(\frac{1}{\beta_e} -\Delta_e \right)} \, , 
 \label{obliqueinstab_freq_elec_negpres_EST}
\end{eqnarray}
returning us to the dispersion relation (\ref{obliqueinstab_freq_elec_negpres}) of unstable kinetic Alfv\'en 
waves taken in the limit $k_{\|} \ll k_{\bot}$.

In the case when $\Delta_e < -2 \beta_e^{-1}$, one of the modes described by (\ref{ESTmodefreq}) can be 
destabilised by sufficiently negative pressure anisotropy, and become purely 
growing. The wavenumbers susceptible to this instability are those 
satisfying
\begin{equation}
k_{\bot}^2 \rho_e^2 \left[1 - \exp \left(-\frac{k_{\bot}^2 \rho_e^2}{2}\right) I_0\!\left(\frac{k_{\bot}^2 
\rho_e^2}{2}\right)\right]^{-1} < |\Delta_e| \beta_e . \label{ESTunstab_wav}
\end{equation}
Provided $\Delta_e < -2 \beta_e^{-1}$ and $|\Delta_e| \beta_e \sim 1$, this gives a range of unstable perpendicular wavenumbers $k_{\bot} \rho_e \lesssim  
1$. That these wavenumbers are indeed unstable follows immediately from the observation 
that if (\ref{ESTunstab_wav}) holds, then the discriminant (\ref{ESTmodedisc}) 
satisfies
  \begin{eqnarray}
\mathfrak{D}\left(k_{\bot} \rho_e, \beta_e, \Delta_e\right) & = &  -\upi \left[\Delta_e \mathcal{H}(k_{\perp} \rho_e)-\frac{k_{\bot}^2 \rho_e^2}{\beta_e}\right] \left[\Delta_e \mathcal{H}(k_{\perp} \rho_e)-\frac{k_{\bot}^2 \rho_e^2}{\beta_e}+ \frac{1}{\beta_e}  +|\Delta_e| \mathcal{F}(k_{\perp} \rho_e)\right] \nonumber \\
& < & \upi \left[\Delta_e \mathcal{H}(k_{\perp} \rho_e)-\frac{k_{\bot}^2 \rho_e^2}{\beta_e}\right]^2 \, , \qquad \label{ESTmodedisc_B}
 \end{eqnarray}
from which it follows that the imaginary part of (\ref{ESTmodefreq}) for the `+' root is 
positive. When $|\Delta_e \beta_e + 2| \sim 1$, the characteristic growth rate of the instability is 
\begin{equation}
\gamma_{\mathrm{max}} \sim k_{\|} \rho_e |\Delta_e| 
\Omega_e \, ,  
\end{equation}
which is consistent with the numerical findings shown in figure 
\ref{Figure_obliqueelecfire_unity}a. Indeed, (\ref{ESTmodefreq})
agrees reasonably with the numerically determined growth rate for small values of $k_{\|} \rho_i$ 
(see figure \ref{Figure_obliqueelecfire_unity}b). 

One particularly interesting subsidiary limit of (\ref{obliqueinstab_freq_elec_negpres_EST}) is $|\Delta_e| \beta_e \gg 1$, in which it can be 
shown that, under the ordering $k_{\bot} \rho_e \sim (|\Delta_e| \beta_e)^{1/2} \gg 1$, the growth rate~is 
\begin{equation}
\gamma \approx \upi k_{\|} k_{\bot}^3 \rho_e^4 \left(|\Delta_e| - \frac{k_{\bot}^2 \rho_e^2}{\beta_e}\right) 
\Omega_e \, . \label{EST_growth_kpplge}
\end{equation} 
This implies that the perpendicular wavelength of peak growth transitions smoothly
to values below the electron Larmor radius as $|\Delta_e| \beta_e$ is increased 
beyond order-unity values. As we shall discuss in the next section, these unstable 
sub-electron-Larmor scale modes are best regarded as a distinct instability from 
the electron firehose, and so we introduce it properly in a new section. 

\subsubsection{Electron-scale-transition (EST) instability} \label{negpres_elec_EST}

When $|\Delta_e| \beta_e$ is increased significantly past unity, the fastest-growing microinstability changes character 
from that of a destabilised KAW, and instead 
becomes a destabilised non-propagating mode. 
The authors of this paper are not aware of this instability 
having been identified previously; we call it the \textit{electron-scale-transition (EST)} 
instability, on account of it providing a smooth transition between unstable 
KAWs with $k_{\bot} \rho_e \ll 1$, and microinstabilities on sub-electron scales ($k_{\bot} \rho_e \gtrsim 
1$). Unstable EST modes are quasi-perpendicular ($k_{\|} \rho_e < 1 \lesssim k_{\bot} \rho_e \lesssim \beta_e^{1/7}$), 
with the parallel wavenumber of the fastest-growing modes determined by 
a balance between the instability's drive and the electron-cyclotron damping 
that arises at sufficiently large $k_{\|} \rho_e$. 
In contrast to the oblique electron firehose instability, Landau-resonant electrons 
with $v_{\|} \approx \omega/k_{\|}$ also play a role in the EST instability's physical mechanism. 

To demonstrate that the EST modes are not unstable KAWs, we consider the expression (\ref{ESTmodefreq}) in a Maxwellian plasma (viz., $\Delta_e = 
0$). It is easy to show that in this case, $\mathfrak{D}\left(k_{\bot} \rho_e, \beta_e, \Delta_e\right) \leq 0$ 
if and only if
\begin{equation}
k_{\bot} \rho_e \geq \frac{2}{\sqrt{\upi}} \, .  \label{mink_EST}
\end{equation}
Thus, for sufficiently large values of $k_{\bot} \rho_e$, KAWs cease to be able to 
propagate, and we obtain two purely damped non-propagating modes. Thus, any 
microinstabilities for $\Delta_e < 0$ associated with these modes can no longer 
be considered to be unstable KAWs. Substituting (\ref{mink_EST}) into the 
threshold condition (\ref{ESTunstab_wav}), we estimate that EST modes first become 
unstable when $\Delta_e < (\Delta_e)_{\rm c} \approx -3/\beta_e$. 

\begin{figure}
\centerline{\includegraphics[width=0.99\textwidth]{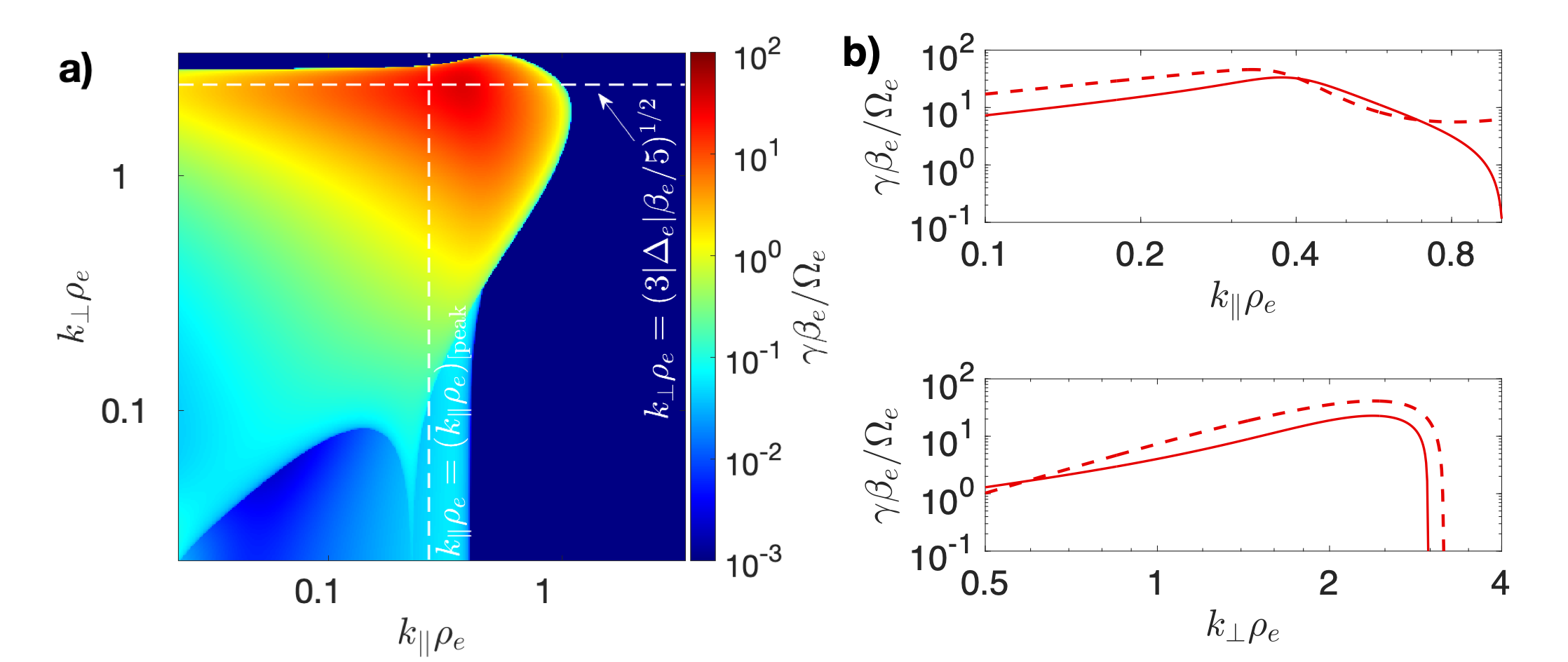}}
\caption{\textit{CES electron-scale-transition (EST) instability}. \textbf{a)} Maximum positive growth rates of linear perturbations 
resulting from CE ion- and electron-shear terms in the CE distribution function (\ref{CEsheardistfuncexpression}) for $\Delta_e \beta_e = -10$ and $\beta_e = 10^4$.
Here, a temperature-equilibrated hydrogen plasma is considered viz., $\Delta_e = \mu_e^{1/2} \Delta_i$, and $\beta_i = \beta_e$.  
The growth rates of all modes are calculated in the same way as figure \ref{newFig_elecfirehose_marginal}. \textbf{b)} Plot of the EST mode growth rate (solid line) as a function of $k_{\|} \rho_e$ with $k_{\bot} \rho_e = (3 |\Delta_e|\beta_e/5)^{1/2}$ (top), and as a function of $k_{\perp} \rho_e$ with $k_{\|} \rho_e = (k_{\|} \rho_e)_{\rm peak}$ (bottom), where $(k_{\|} \rho_e)_{\rm peak}$ is
given by (\ref{EST_maxgrowth_kval}\textit{b}). The 
dotted and dashed lines show the analytical prediction (\ref{EST_frequency}). \label{Figure_EST}}
\end{figure}

As $\Delta_e$ is decreased below $(\Delta_e)_{\rm c}$, 
the EST modes quickly acquire a faster growth rate than all the other CES 
microinstabilities that can operate for such values of $\Delta_e$. We illustrate 
this numerically in figure \ref{Figure_EST}a by showing the maximum growth rate 
of all CES microinstabilities as a function of $(k_{\|},k_{\perp})$ for a particular value of $\Delta_e < 0$. 
The EST modes with $k_{\|} \rho_e$, $k_{\perp} \rho_e > 1$ are the fastest growing, 
with $\gamma \gg \Omega_e/\beta_e$. 

In the limit $|\Delta_e| \beta_e \gg 1$ (but $|\Delta_e| \beta_e \ll \beta_e^{2/7}$), the maximum growth rate of the EST 
instability can be estimated analytically. Adopting the orderings 
\begin{equation}
 k_{\|} \rho_e \sim \frac{1}{\sqrt{\log{|\Delta_e| \beta_e}}} , \quad k_{\bot} \rho_e \sim (|\Delta_e| \beta_e)^{1/2} , \quad \frac{\omega}{k_\| v_{\mathrm{th}e}} 
\sim |\Delta_e|^{5/2} \beta_e^{3/2} , \label{ESTorder}
\end{equation}
it can be shown (see appendix \ref{derivation_EST}) that the EST mode has the growth rate
\begin{equation}
\frac{\gamma}{\Omega_e} = \upi k_{\|} k_{\bot}^3 \rho_e^4 \left(|\Delta_e| - \frac{k_{\bot}^2 \rho_e^2}{\beta_e}\right) 
\left\{1+\frac{\upi k_{\bot}^2 \rho_e^2}{k_{\|}^2 \rho_e^2} \left[4 \exp{\left(-\frac{1}{k_{\|}^2 \rho_e^2}\right)} +\sqrt{\upi} \mu_e^{1/2} k_{\|}^3 \rho_e^3 \right]\right\}^{-1} 
\, , \label{EST_frequency}
\end{equation}
where the term proportional to $\mu_e^{1/2}$ is associated with Landau damping on the ion species.  
 Taking the subsidiary limit $k_{\|} \rho_e \ll 1/\sqrt{\log{|\Delta_e| 
\beta_e}}$, we recover (\ref{EST_growth_kpplge}). The EST mode's growth rate is, therefore, anticipated to be positive provided 
$k_{\bot} \rho_e < \left(|\Delta_e| \beta_e\right)^{1/2}$. 
It can then be shown that (\ref{EST_frequency}) has the approximate maximum value
\begin{equation}
 \gamma_{\mathrm{max}} \approx \frac{6 \sqrt{3} \upi}{25 \sqrt{5}}  (k_{\|} \rho_e)_{\mathrm{peak}} \left[1-\frac{3 \upi^{3/2}}{5} \mu_e^{1/2} (k_{\|} \rho_e)_{\mathrm{peak}} |\Delta_e| \beta_e \right]|\Delta_e| \left(|\Delta_e| \beta_e\right)^{3/2}
 \Omega_e \, , \label{EST_maxgrowth}
\end{equation}
 at the wavenumbers
  \begin{subeqnarray}
(k_{\bot} \rho_e)_{\mathrm{peak}} & = & \left(\frac{3 |\Delta_e| \beta_e}{5}\right)^{1/2} , \\
 (k_{\|} \rho_e)_{\mathrm{peak}} & = & \frac{1}{\sqrt{\log{(24 \upi|\Delta_e| \beta_e/5)}}} \left[1-\frac{\log{\log{\left(24 \upi|\Delta_e| \beta_e/5\right)}}}{\log{24 \upi|\Delta_e| \beta/5}}\right] \, . \label{EST_maxgrowth_kval}
 \end{subeqnarray}
The growth rate (\ref{EST_frequency}) is plotted in figure 
\ref{Figure_EST}b along with the numerically determined growth rate; reasonable agreement is found. 

We note that, for perpendicular wavenumbers $k_{\perp} \rho_e \gtrsim 
\beta_e^{1/7}$, the characteristic quasi-perpendicular plasma modes in a Maxwellian plasma are not EST modes, but are instead whisper waves (see section \ref{negpres_subelectron_phantom}). 
Therefore, when $|\Delta_e| \beta_e \gtrsim \beta_e^{2/7}$ [see (\ref{ESTinvaliditybound})], the expressions (\ref{EST_maxgrowth})
and (\ref{EST_maxgrowth_kval}\textit{a}) for the EST mode's maximum growth rate and the 
perpendicular wavenumber at which that growth is attained are no longer valid. 
Instead, when $|\Delta_e| \beta_e \gtrsim \beta_e^{2/7}$, the fastest-growing EST modes (which coexist with faster-growing unstable whisper waves) are those close to the scale $k_{\perp} \rho_e \sim 
\Delta_e^{-1/5}$; extrapolating from (\ref{EST_frequency}), we find that $\gamma_{\rm max} \sim |\Delta_e|^{2/5} 
\Omega_e/\sqrt{\log{|\Delta_e| \beta_e}}$.

\subsubsection{Oblique transverse instability} \label{negpres_subelectron_obliquetrans}

\begin{figure}
\centerline{\includegraphics[width=0.99\textwidth]{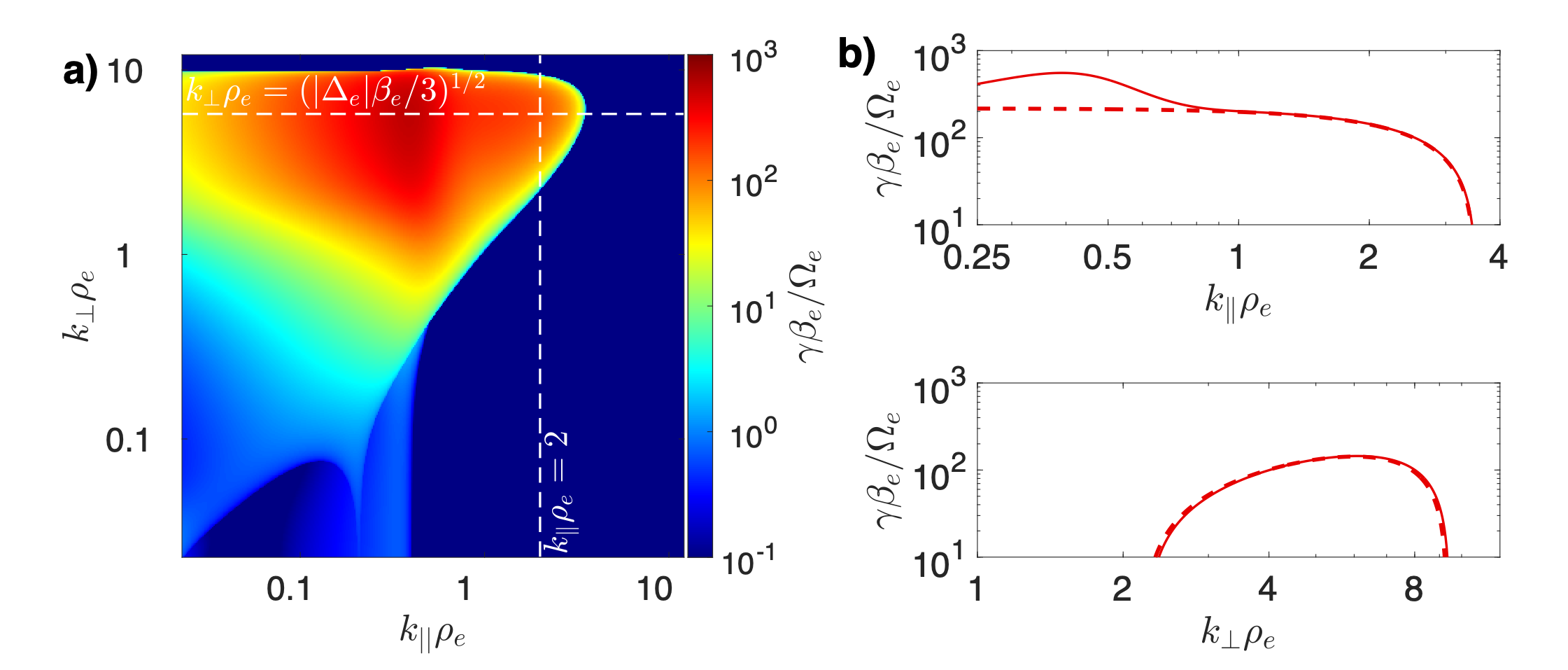}}
\caption{\textit{CES oblique transverse instability}. \textbf{a)} Maximum positive growth rates of linear perturbations 
resulting from CE ion- and electron-shear terms in the CE distribution function (\ref{CEsheardistfuncexpression}) for $\Delta_e \beta = -100$ and $\beta_e = 10^4$.
Here, a temperature-equilibrated hydrogen plasma is considered viz., $\Delta_e = \mu_e^{1/2} \Delta_i$, and $\beta_i = \beta_e$.  
The growth rates of all modes are calculated in the same way as figure \ref{newFig_elecfirehose_marginal}. \textbf{b)} Plot of the oblique transverse mode's growth rate (solid line) as a function of $k_{\|} \rho_e$ with $k_{\bot} \rho_e = (|\Delta_e|\beta_e/3)^{1/2}$ (top), and as a function of $k_{\perp} \rho_e$ with $k_{\|} \rho_e = 2$ (bottom). The 
dotted and dashed lines show the analytical prediction (\ref{EST_frequency}). \label{Figure_obltransverse}}
\end{figure}

The transverse instability (whose physical mechanism was discussed in section \ref{pospres_electron_trans}) can be excited for sufficiently large negative 
electron pressure ansotropies as well as positive ones; however, when $\Delta_e < 
0$, the fastest-growing modes are highly oblique with respect to the background magnetic field as opposed to 
parallel to it. In contrast to the $\Delta_e > 0$ case, the oblique transverse instability does 
not become the fastest-growing CES microinstability for all $\Delta_e \ll 
-\beta_e^{-1}$, only becoming so once its maximum growth rate exceeds the electron Larmor frequency (which 
occurs when $\Delta_e \lesssim -\beta_e^{-1/3}$). While $\Delta_e > 
-\beta_e^{-1/3}$, the fastest-growing oblique transverse modes, which have $k_{\perp} \rho_e \sim (|\Delta_e| \beta_e)^{1/2}$, are confined to the parallel
wavenumbers satisfying $k_{\|} \rho_e \gtrsim 1$. Their growth is outcompeted by 
the EST and whisper instabilites (see sections \ref{negpres_elec_EST} and 
\ref{negpres_subelectron_phantom}, respectively), which have $k_{\|} \rho_e < 1$; this is 
illustrated numerically in figure \ref{Figure_obltransverse}a for a particular 
large, negative value of $\Delta_e \beta_e$. 

As for their analytical characterisation, transverse modes have
identical growth rates to those obtained in the $\Delta_e > 0$ case, given by (\ref{transverse_oblique_growthrate}\textit{a},\textit{b}).
For $\Delta_e < 0$, only the first mode can have positive growth, and such growth is only 
realised if $k_{\bot} > k_{\|}$. Now 
taking the quasi-perpendicular unmagnetised limit $k_{\bot} \rho_e \gg k_{\|} \rho_e \gg 1$, we 
find that this mode has the growth rate
\begin{equation}
\gamma \approx \frac{k_{\bot} v_{\mathrm{th}e}}{\sqrt{\upi}} \left(-\Delta_e - \frac{k_{\bot}^2 \rho_e^2}{\beta_e}\right) \label{tranverseinstab}
\, .
\end{equation}
This expression is mathematically identical to the parallel transverse instability (\ref{partransverse_growthrate}) (section \ref{pospres_electron_trans}), except with 
substitution $k_{\|} \rightarrow k_{\bot}$; the maximum growth rate of the oblique transverse instability is, therefore,
\begin{equation}
\gamma_{\mathrm{max}} = \frac{2}{3 \sqrt{3 \upi}} (|\Delta_e| \beta_e)^{1/2} |\Delta_e| \Omega_e   \label{oblique_transverseinstab_peakgrowth}
\end{equation}
at the (perpendicular) wavenumber 
\begin{equation}
 (k_{\bot} \rho_e)_{\mathrm{peak}} = (\Delta_e \beta_e/3)^{1/2} \, .
\end{equation}
(\ref{tranverseinstab}) is compared with the numerically determined growth rate 
in figure \ref{Figure_obltransverse}b; we find that the approximation is 
excellent provided $k_{\|} \rho_e \gtrsim 1$. 

We note that, based on our analysis, the oblique transverse mode is anticipated always to have a smaller growth rate than the EST instability (\ref{EST_maxgrowth}) when $1 \ll |\Delta_e| \beta_e \lesssim \beta_e^{2/7}$:
\begin{equation}
 \frac{\gamma_{\rm EST}}{\gamma_{\rm trans}} \sim \frac{|\Delta_e| \beta_e}{\sqrt{\log{|\Delta_e| \beta_e}}} 
 \gg 1 \, .
\end{equation}
 
 \subsubsection{Whisper instability} \label{negpres_subelectron_phantom}

When $\Delta_e \lesssim -\beta_e^{-5/7}$ (but $\Delta_e \gg -\beta_e^{-1/3}$), the dominant 
CES microinstability is the CES whisper instability. The 
instability is so named, because it consists in the destablisation of the whisper 
wave, a plasma wave whose existence has not previously been identified: it is therefore of some interest. The likely reason for 
its previous neglect relates to the somewhat esoteric regime in 
which such a wave exists -- a magnetised plasma with $\beta_e \gg 1$ that might naively be expected to support essentially 
unmagnetised perturbations at $k \rho_e \gg 1$. The energetically dominant magnetic component of the wave is
perpendicular to both $\boldsymbol{k}$ and $\boldsymbol{B}_0$ (viz., $\delta B_y$), 
and the wave itself has no electron-number-density perturbation unless $\beta_e$ is extremely 
large. Its operation (and also the operation of its instability in a CE plasma) involves both resonant and non-resonant interactions 
between electrons and the wave. More specifically, it is the non-resonant interaction of electrons at the edge of their Larmor orbits with the parallel electric field associated with the whisper wave that gives rise to the phase-shifted current perturbation necessary for wave propagation, while the primary damping mechanisms (Landau and Barnes' damping, respectively) of whisper waves are mediated by resonant wave-particle interactions.
The physical mechanism of this wave and its instability (which is most clearly 
explored within the quasi-perpendicular limit of gyrokinetics) will be discussed further in a 
future paper.

\begin{figure}
\centerline{\includegraphics[width=0.99\textwidth]{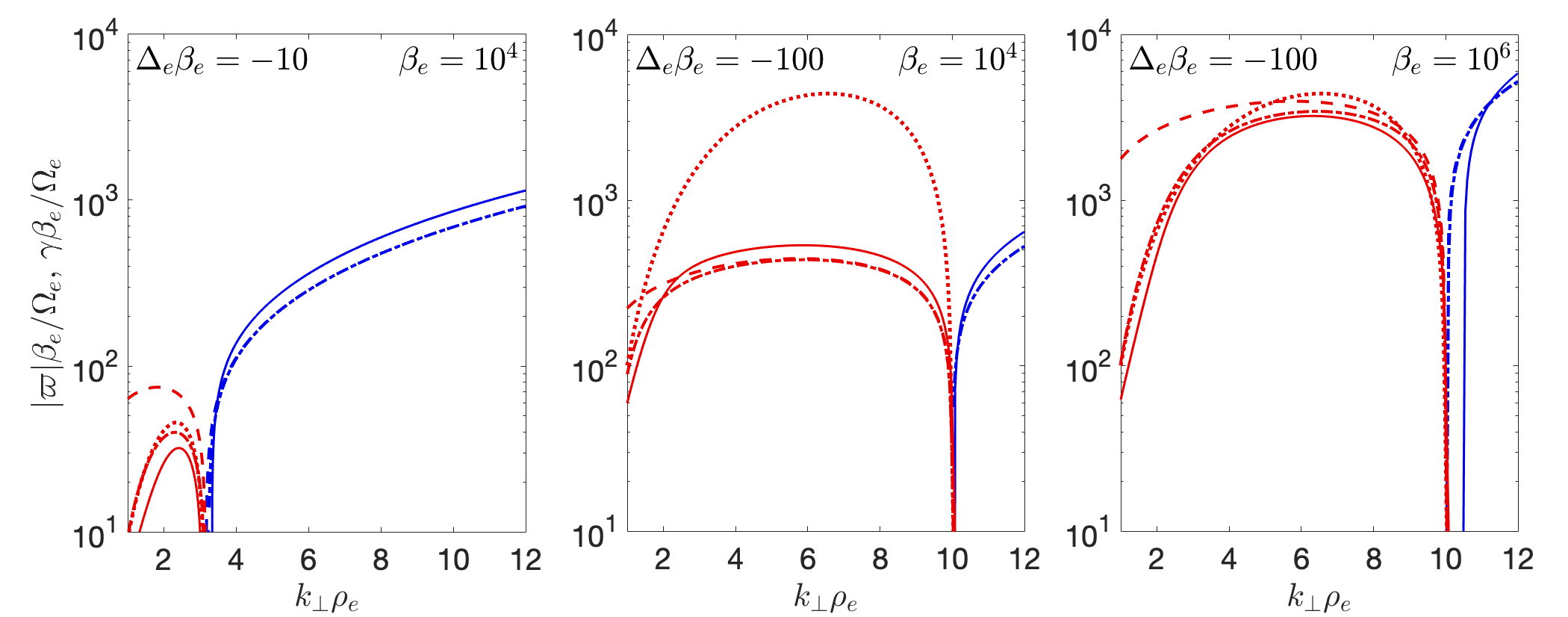}}
\caption{\textit{CES EST vs. whisper instability}. Growth rates of EST and whisper modes whose instability is driven by the CE electron-shear term in the CE distribution function (\ref{CEsheardistfuncexpression}), 
for quasi-perpendicular ($k_{\|} \ll k_{\perp}$) wavevectors with respect to the background 
magnetic field. 
The growth rates (solid lines) of all modes are calculated in the same way as figure \ref{newFig_elecfirehose_marginal} for a selection of different values of $\Delta_e \beta_e$ 
and $\beta_e$, and $k_{\|} \rho_e = 0.35$. 
The approximations (\ref{EST_frequency}), (\ref{fullmagmodedisp}), and (\ref{growthrate_whisperinstab}) for 
the real frequency (dotted, dot-dashed and dashed blue, respectively) and growth rate (dotted, dot-dashed and dashed red, respectively) in the limit $k_{\|} \rho_e \ll 1$, $k_{\perp} \rho_e \gg 1$, are also plotted. 
 \label{newFig_whisperEST}}
\end{figure}

We characterise the whisper instability's growth analytically  
in the limits $\mu_e^{1/2} \ll k_{\|} \rho_e \ll 
1$, $k_{\bot} \rho_e \gg 1$ and $\Delta_e \beta_e \gg 1$ under the orderings 
\begin{equation}
\tilde{\omega}_{e\|} = \frac{\omega}{k_{\|} v_{\mathrm{th}e}}  \sim \frac{1}{\beta_e^{2/7}} \sim \frac{1}{k_{\bot}^2 \rho_e^2} 
\sim \frac{1}{\Delta_e \beta_e} , \quad 
  k_{\|} \rho_e \sim \frac{1}{\sqrt{\log{|\Delta_e| \beta_e}}} \ll 1 \, 
  .
\end{equation}
It can be shown (see appendix \ref{derivation_whisper}) that such modes 
have complex frequencies 
  \begin{eqnarray}
 \frac{\omega}{\Omega_e} & = & -\mathrm{i} \left[\frac{\sqrt{\upi}}{2 k_{\|} \rho_e} \exp{\left(-\frac{1}{k_{\|}^2 \rho_e^2}\right)} + \frac{k_{\|} \rho_e}{8 \sqrt{\upi} k_{\bot}^2 \rho_e^2} 
 \right] \nonumber \\
 &\pm &k_{\|} \rho_e \sqrt{\frac{\sqrt{\upi}}{4} k_{\bot} \rho_e \left(\frac{k_{\bot}^2 \rho_e^2}{\beta_e}+\Delta_e \right) - \left[\frac{\sqrt{\upi}}{2 k_{\|}^2 \rho_e^2} \exp{\left(-\frac{1}{k_{\|}^2 \rho_e^2}\right)} + \frac{1}{8 \sqrt{\upi} k_{\bot}^2 \rho_e^2} 
 \right]^2} 
  . \qquad \,  \label{fullmagmodedisp}
 \end{eqnarray}
It is a simple matter to ascertain that the right-hand-side of (\ref{fullmagmodedisp}) 
is either purely real or purely imaginary, and thus modes are approximately either non-propagating with 
growth rate $\gamma$ or purely oscillating with frequency $\varpi$. The 
dispersion curves $\varpi(k_\perp)$ and $\gamma(k_\perp)$ are plotted in figure 
\ref{newFig_whisperEST}.
To interpret (\ref{fullmagmodedisp}), we take subsidiary limits.

We first consider $1 \ll k_{\bot} \rho_e \sim (|\Delta_e| \beta_e)^{1/2} \ll 
\beta_e^{1/7}$: in this case, the expression for the `$+$' root simplifies to the dispersion relation (\ref{EST_frequency}) of the EST instability. 
However, when $k_\perp \rho_e \gtrsim \beta_e^{1/7}/2^{4/7} \upi^{1/7} \approx 0.57  
\beta_e^{1/7}$, this simplification is no longer justifiable, and 
so when
\begin{equation}
|\Delta_e| \beta_e \gtrsim \frac{5^{6/7}}{2^{10/7} 3^{4/7} \upi^{3/7}} \beta_e^{2/7} \approx 0.79 \beta_e^{2/7} ,  \label{ESTinvaliditybound}
\end{equation}
the perpendicular wavenumber (\ref{EST_maxgrowth_kval}\textit{a}) of the EST instability's peak growth derived from (\ref{EST_frequency}) is so large that (\ref{EST_frequency}) is no 
longer, in fact, a valid description of the EST mode's growth rate. 

Now considering the subsidiary
limit $k_{\perp} \rho_e \sim (|\Delta_e| \beta_e)^{1/2} \gg \beta_e^{1/7}$ and $k_{\|} \rho_e \ll 1/\sqrt{\log{|\Delta_e| 
\beta_e}}$ of (\ref{fullmagmodedisp}),
we find two propagating modes:
\begin{equation}
\frac{\omega}{\Omega_e} \approx \pm \frac{\upi^{1/4}}{2} k_{\|} \rho_e \sqrt{k_{\bot} \rho_e \left(\frac{k_{\bot}^2 \rho_e^2}{\beta_e}+\Delta_e \right)} 
\, . \label{whisperwave_freq}
\end{equation}
If we set $\Delta_e = 0$ in order to identify the underlying Maxwellian mode, this reduces to
\begin{equation}
\frac{\omega}{\Omega_e} \approx \pm \frac{\upi^{1/4}}{2} k_{\|} \rho_e \frac{(k_{\bot} \rho_e)^{3/2}}{\beta_e^{1/2}}
\, ,
\end{equation}
This dispersion relation, which does not coincide with any previously identified plasma 
wave, is that of the whisper wave. 

The presence of this wave in the case of $\Delta_e < 0$ results in 
a purely unstable mode provided $\beta_e^{-1/7} \ll k_{\bot} \rho_e < (|\Delta_e| \beta_e)^{1/2}$ and retaining finite $k_{\|} \rho_e$. 
In this subsidiary limit, the growth rate of the instability is
 \begin{eqnarray}
 \frac{\gamma}{\Omega_e} & = & - \frac{\sqrt{\upi}}{2 k_{\|} \rho_e} \exp{\left(-\frac{1}{k_{\|}^2 \rho_e^2}\right)} \nonumber \\
 && \pm k_{\|} \rho_e \sqrt{\frac{\sqrt{\upi}}{4} k_{\bot} \rho_e \left(|\Delta_e| - \frac{k_{\bot}^2 \rho_e^2}{\beta_e}\right) + \frac{\upi}{2 k_{\|}^4 \rho_e^4} \exp{\left(-\frac{2}{k_{\|}^2 \rho_e^2}\right)}} 
 \, . \qquad \label{growthrate_whisperinstab}
 \end{eqnarray}
 This has the maximum value
 \begin{equation}
  \gamma_{\mathrm{max}} \approx \frac{\upi^{1/4}}{\sqrt{2}}  (k_{\|} \rho_e)_{\mathrm{peak}} \left(|\Delta_e| 
  \beta_e\right)^{1/4}
  |\Delta_e|^{1/2} \Omega_e \, , \label{whiswavinstab_gamma}
 \end{equation}
 at the wavenumbers
 \begin{subeqnarray}
(k_{\bot} \rho_e)_{\mathrm{peak}} & = & \left(\frac{|\Delta_e| \beta_e}{3}\right)^{1/2} , \\
 (k_{\|} \rho_e)_{\mathrm{peak}} & = & \frac{2}{\sqrt{3 \log{|\Delta_e| \beta_e}}} \left[1-\frac{4 \log{3 \left(\log{|\Delta_e| \beta_e}/4\right)}}{3 \log{|\Delta_e| \beta_e}}\right] \, . \label{whiswavinstab_k}
 \end{subeqnarray}
 
\begin{figure}
\centerline{\includegraphics[width=0.99\textwidth]{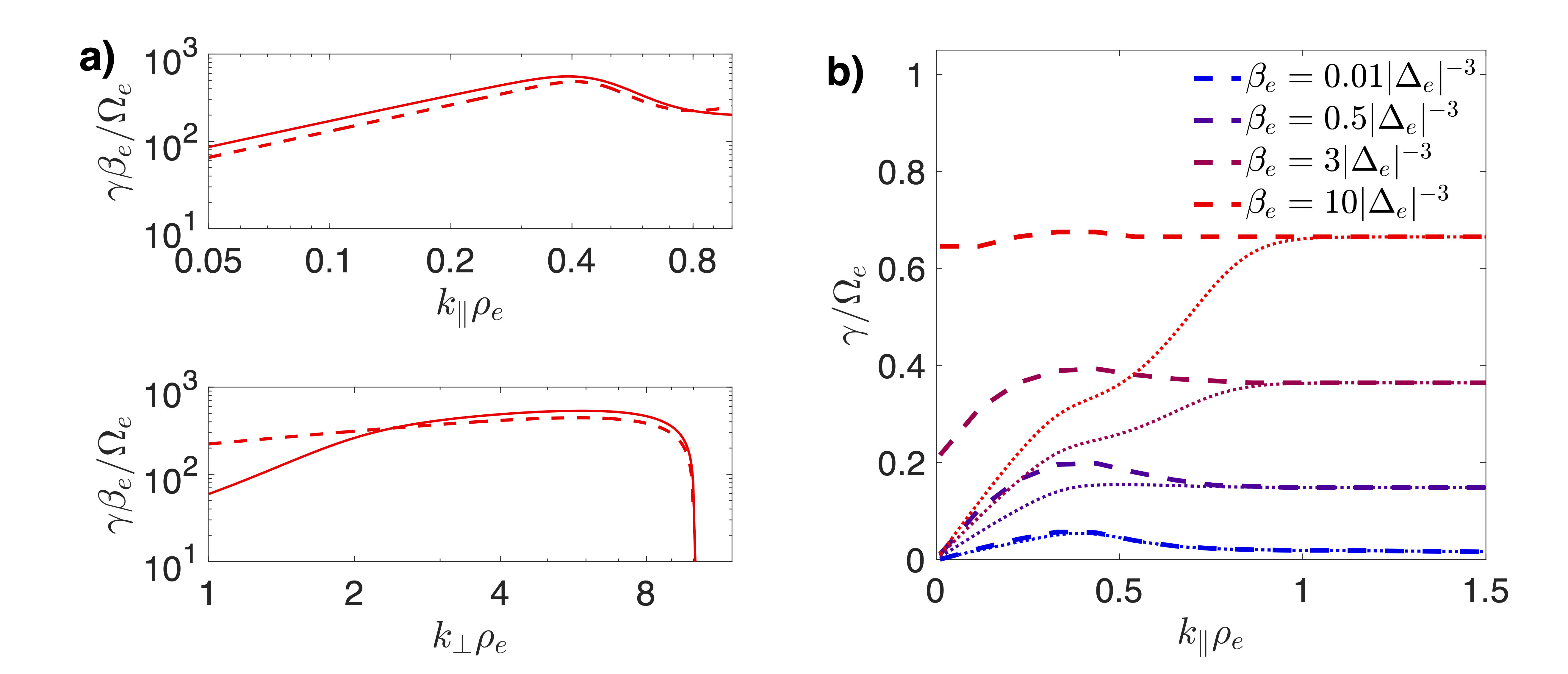}}
\caption{\textit{CES whisper instability}. 
\textbf{a)} Plot of the whisper instability's growth rate as a function of $k_{\|} \rho_e$ with $k_{\bot} \rho_e = (|\Delta_e|\beta_e/3)^{1/2}$ (top), and as a function of $k_{\perp} \rho_e$ with $k_{\|} \rho_e = (k_{\|} \rho_e)_{\rm peak}$ (bottom), where $(k_{\|} \rho_e)_{\rm peak}$ is
given by (\ref{whiswavinstab_k}\textit{b}), for $\Delta_e \beta = -100$ and $\beta_e = 10^4$ [cf. figure (\ref{Figure_obltransverse})]. The numerically determined value 
of this quantity is calculated in the same way as figure \ref{newFig_elecfirehose_marginal}.  The 
dotted and dashed lines show the analytical prediction (\ref{growthrate_whisperinstab}). 
\textbf{b)} Divergence of growth rates (dotted lines) arising from using the approach outlined in appendix \ref{CES_method_append} (which involves solving a simplified `low-frequency' dispersion relation) as compared to the full hot-plasma 
dispersion relation (dashed lines) when $\beta_e \sim |\Delta_e|^{-3}$. 
Here $\Delta_e = -0.01$, $k_{\bot} \rho_e = \left(|\Delta_e| \beta_e/3\right)^{1/2}$, and $\beta_e$ is varied.
For reference, $\beta_e =  0.01 |\Delta_e|^{-3} = 10^{4}$ corresponds to the 
same values of $\beta_e$ and $\Delta_e$ as considered in figure 
\ref{Figure_obltransverse}.
\label{newFig_whisper}}
\end{figure}

 Thus, the maximum growth rate of whisper instability has 
 different scalings with $|\Delta_e|$ and $\beta_e$ than either the EST instability (\ref{EST_maxgrowth}) or the oblique transverse 
 instability (\ref{oblique_transverseinstab_peakgrowth}). 
 When $|\Delta_e| \beta_e \gtrsim \beta_e^{2/7}$, (\ref{fullmagmodedisp})
implies that the growth rate $\gamma$ continues to increase beyond the maximum 
value of $k_\perp \rho_e$ at which the EST modes can exist, and thus the 
whisper instability, if it is operating, is always dominant over the EST 
instability. Whether it is also dominant over the oblique transverse instability depends on the choice of $\beta_e$ and $\Delta_e$. 
 We can quantify this explicitly, by considering the ratio of 
 the oblique transverse instability's growth rate (\ref{oblique_transverseinstab_peakgrowth}) to that of the whisper instability:
 \begin{equation}
   \frac{\gamma_{\rm trans}}{\gamma_{\rm whisper}} \sim \sqrt{\log{(|\Delta_e|\beta_e})} \left(|\Delta_e| 
  \beta_e\right)^{1/4}  |\Delta_e|^{1/2} \, . 
 \end{equation}
 We see that for $|\Delta_e|^3 
  \beta_e \ll 1$, $\gamma_{\rm trans} \ll \gamma_{\rm whisper}$. Thus for $|\Delta_e|^{-7/5} \ll \beta_e \ll 
  |\Delta_e|^{-3}$, the whisper instability dominates. This condition certainly holds for 
  the particular value of $\Delta_e$ considered in figure 
  \ref{Figure_obltransverse}; to support our claim, in figure \ref{newFig_whisper}a we 
  plot the analytical approximation (\ref{growthrate_whisperinstab}) along with the 
  numerically determined growth rate for the fixed values of $k_{\perp} \rho_e$ and $k_{\|} 
  \rho_e$, respectively, at which the whisper instability is predicted to 
  achieve its maximum growth. 
The growth rate of the whisper instability, which is correctly captured by our 
analytic approximation, does indeed exceed that of the transverse instability by 
an appreciable factor. 
  
For $\beta_e \gtrsim |\Delta_e|^{-3}$, (\ref{whiswavinstab_gamma}) 
implies that, in fact, $\gamma/k_{\|} v_{\mathrm{th}e} \sim 1$. This violates the condition of validity of the method that we
have generally used to evaluate CES microinstability growth rates numerically (see section \ref{shortcomings_othermicro}, and also appendix \ref{CES_method_append}). 
The divergence of the true growth rates (calculated by solving the full hot-plasma dispersion relation numerically) 
from those arising from the solution of the low-frequency ($\omega \ll k_{\|} v_{\mathrm{th}e}$) dispersion relation (\ref{sheardisprel_D}) for increasing $\beta_e$ is illustrated in figure \ref{newFig_whisper}b. 
For $\gamma \gtrsim \Omega_e$, we find that the distinction between $k_{\|} \rho_e < 1$ 
modes and $k_{\|} \rho_e > 1$ modes vanishes; futhermore, all modes (including the modes with $k_{\|} = 0$) come to resemble
the transverse instability when $\beta_e \gg |\Delta_e|^{-3}$; this feature, which indicates 
the emergence of yet another distinct CES instability, is discussed in the next section. 
  
   \subsubsection{Ordinary-mode instability} \label{negpres_subelectron_ord}
   
   The final instability we consider in this paper is the CES ordinary-mode (electromagnetic)
   instability: the destabilisation of the ordinary mode at sub-electron-Larmor scales by negative electron 
   pressure anisotropy.   The bi-Maxwellian equivalent of the
   instability was first identified by~\citet{DW70}; for a more recent linear 
   study of the instability, see~\citet{ILS12}. For the characteristically small electron pressure 
   anisotropies that are associated with the CE electron-shear term, this 
   instability can only arise at very large values of $\beta_e$. 
   For purely perpendicular modes ($k_{\|} = 0$) in a magnetised plasma, resonant 
   wave-particle interactions cannot arise, and so the ordinary-mode's 
   instability mechanism is non-resonant.
   
\begin{figure}
\centerline{\includegraphics[width=0.99\textwidth]{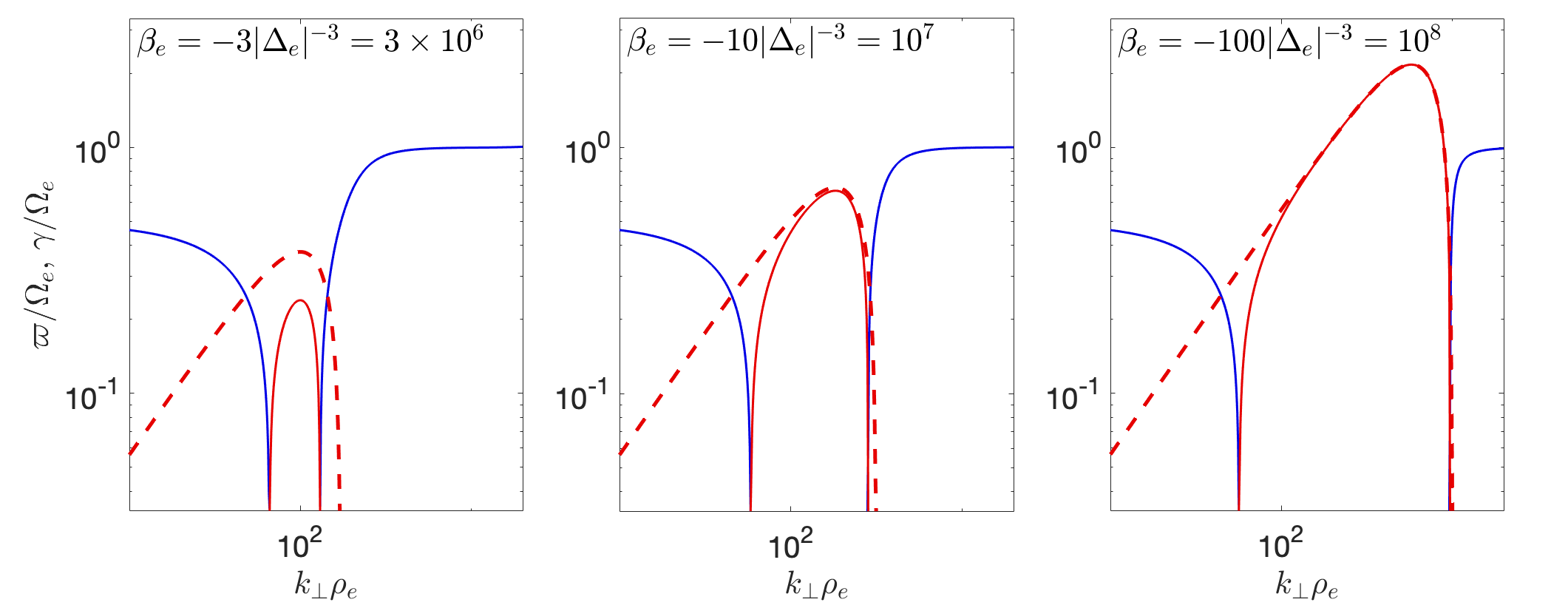}}
\caption{\textit{CES ordinary-mode instability}. Growth rates of ordinary modes whose instability is driven by the CE electron-shear term in the CE distribution function (\ref{CEsheardistfuncexpression}), 
for perpendicular ($k_{\|} = 0$) wavevectors with respect to the background 
magnetic field. 
The growth rates (solid lines) of the modes are calculated using (\ref{disprel_ordmode_D}) and (\ref{disprel_ordmode_stable}).
We show the growth rates for a selection of different values of $\beta_e$ and for $\Delta_e = 
0.01$. 
The approximation (\ref{ordmode_lgekpp}) of the growth rate in the limit $k_{\perp} \rho_e \gg  \gamma/\Omega_e \gg 1$ is also plotted. 
 \label{newFig_ordmode}}
\end{figure}

 The CES ordinary-mode instability is most simply characterised by considering 
  modes that are exactly perpendicular to the guide magnetic field (viz., $k_{\|} = 
  0$). In this case, it can be shown (see appendix \ref{derivation_ordinarymode})
  that, if the ordinary mode is destabilised, its growth rate is given by the equation
 \begin{equation}
\sum_{n = 1}^{\infty} \frac{2 \gamma^2}{\gamma^2+n^2 \Omega_e^2} 
 \exp \left(-\frac{k_{\bot}^2 \rho_{e}^2}{2}\right) I_n\!\left(\frac{k_{\bot}^2 \rho_{e}^2}{2}\right) 
 = -\Delta_e-  k_{\bot}^2 d_e^2  - \exp \left(-\frac{k_{\bot}^2 \rho_{e}^2}{2}\right) I_0\!\left(\frac{k_{\bot}^2 \rho_{e}^2}{2}\right)  \, . \label{disprel_ordmode_D}  
\end{equation}
This dispersion relation is very similar to that derived by~\citet{DW70}
for the ordinary-mode instability in the case of a bi-Maxwellian 
distribution. If the electron pressure anisotropy is insufficient to destabilise 
the ordinary mode, the mode is undamped, and its real frequency satisfies
 \begin{equation}
\sum_{n = 1}^{\infty} \frac{2 \varpi^2}{n^2 \Omega_e^2-\varpi^2} 
 \exp \left(-\frac{k_{\bot}^2 \rho_{e}^2}{2}\right) I_n\!\left(\frac{k_{\bot}^2 \rho_{e}^2}{2}\right) 
 = \Delta_e+  k_{\bot}^2 d_e^2  + \exp \left(-\frac{k_{\bot}^2 \rho_{e}^2}{2}\right) I_0\!\left(\frac{k_{\bot}^2 \rho_{e}^2}{2}\right)  \, . \label{disprel_ordmode_stable}  
\end{equation}
The dispersion curves $\varpi(k_{\perp})$ and $\gamma(k_{\perp})$ for a 
selection of different values of $\beta_e$ and at fixed $\Delta_e$ are shown in 
figure \ref{newFig_ordmode}.

We can use the ordinary-mode dispersion relation (\ref{disprel_ordmode_D}) to derive the threshold for this 
instability at exactly perpendicular wavevectors. We note that the left-hand side of~(\ref{disprel_ordmode_D}) 
is strictly positive; thus for solutions to exist, it is required that there 
exist a range of perpendicular wavenumbers over which the right-hand side of (\ref{disprel_ordmode_D}) 
is also positive. For $k_{\bot} \rho_e \lesssim 1$, the right-hand side is always negative
because $|\Delta_e| \ll 1$. We therefore consider the limit $k_{\bot} \rho_e \gg 
1$ (assuming $\gamma \sim \Omega_e$), for which
\begin{equation}
\frac{1}{\sqrt{\upi} k_{\bot} \rho_e} \sum_{n = 1}^{\infty} \frac{2 \gamma^2}{\gamma^2+n^2 \Omega_e^2} 
 \approx |\Delta_e|-  \frac{k_{\bot}^2 \rho_e^2}{\beta_e}  - \frac{1}{\sqrt{\upi} k_{\bot} \rho_e}  \, . \label{disprel_ordmode_E}  
\end{equation}
The right-hand side of (\ref{disprel_ordmode_E}) is maximal when 
\begin{equation}
k_{\bot} \rho_e = \left(\frac{\beta_e}{2\sqrt{\upi}}\right)^{1/3} \, ,
\end{equation}
and, when maximal, also greater than zero if and only if
\begin{equation}
|\Delta_e|^3 \beta_e > \frac{27}{4 \upi}  \, . \label{ordmodeinstab_thresh}
\end{equation}
Therefore the threshold (\ref{ordmodeinstab_thresh}) is a necessary 
condition for a purely perpendicular instability to exist. It is also a 
sufficient condition, because the left-hand side of (\ref{disprel_ordmode_E}) 
becomes arbitrarily small for small $\gamma$. Comparing the threshold (\ref{ordmodeinstab_thresh})
to figure \ref{newFig_whisper}b, we conclude that the emergence of an instability with a purely perpendicular 
wavenumber at around $\beta_e \sim |\Delta_e|^{-3}$ is consistent with 
numerical expectations. 

One can also show analytically that for $\gamma \gg \Omega_e$, the ordinary-mode 
instability becomes identical to the oblique transverse instability (section \ref{negpres_subelectron_obliquetrans}). Motivated by the 
fact that $\gamma \ll k_{\bot} v_{\mathrm{th}e}$ for the oblique transverse instability, or,
equivalently, $\gamma/\Omega_e \ll k_{\bot} \rho_e$, we first consider (\ref{disprel_ordmode_D}) 
in the limit $k_{\bot} \rho_e \gg  \gamma/\Omega_e \sim 1$; we will subsequently 
take the subsidiary limit $\gamma/\Omega_e \gg 1$. The relevant dispersion 
relation is (\ref{disprel_ordmode_E}), which can be rewritten as
\begin{equation}
\frac{1}{\sqrt{\upi} k_{\bot} \rho_e} \left[\frac{\gamma \upi}{\Omega_e} \coth{\left(\frac{\gamma \upi}{\Omega_e} \right)} - 1\right] 
 \approx -\Delta_e-  \frac{k_{\bot}^2 \rho_e^2}{\beta_e}  - \frac{1}{\sqrt{\upi} k_{\bot} \rho_e} \label{disprel_ordmode_F}  
\end{equation}  
using the summation identity
\begin{equation}
\sum_{n = 1}^{\infty} \frac{2 \gamma^2}{\gamma^2+n^2 \Omega_e^2} = \frac{\gamma \upi}{\Omega_e} \coth{\left(\frac{\gamma \upi}{\Omega_e} \right)} - 1
\, .
\end{equation}
Now assuming $\gamma \gg \Omega_e$ and using $\coth{x} \approx 1$ for any number $x \gg 1$, 
we deduce
\begin{equation}
 \frac{\gamma}{\Omega_e} = -\frac{k_{\bot} \rho_e}{\sqrt{\upi}} \left(\Delta_e +  \frac{k_{\bot}^2 \rho_e^2}{\beta_e}\right)
\, , \label{ordmode_lgekpp}
\end{equation}
which is equivalent to (\ref{tranverseinstab}). Since $|\Delta_e| \ll 1$, our result is consistent with our initial assumption $\gamma/\Omega_e \ll k_{\bot} 
\rho_e$.

Thus, we conclude that, when $\beta_e  \gg |\Delta_e|^{-3}$, the CES ordinary-mode 
instability is the dominant CES microinstability, but that in this limit, the instability is essentially 
identical to the unmagnetised oblique transverse instability already described in section \ref{negpres_subelectron_obliquetrans}. 

\section{Discussion and conclusions} \label{Discussion}

In this paper, we have shown that the Chapman-Enskog description of classical, collisional plasma is valid for 
a wide range of plasma conditions. 
Microinstabilities are stabilised in such plasmas by two effects: collisional 
damping of instabilities, or $\beta$-dependent 
thresholds arising from a non-zero macroscopic magnetic field. 
By identifying the stable region for the leading-order corrections in the Chapman-Enskog 
expansion, we have \textit{de facto} identified the stable region for corrections to arbitrary order: if 
one of the above effects is enough to maintain stability, any perturbations arising from 
smaller corrections will be unable to overcome the same effect.
However, we have also demonstrated that for $\beta \gg 1$ there exists a significant
region of the ($d_e/L$, $\lambda/L$) parameter space in which fast, small-scale instabilities
are both possible and, in fact, generic. Indeed, in the strongly magnetised plasmas (that is, $\rho_s \ll \lambda_s$
for both electrons and ions) on which we have focused our investigation, it transpires that collisional damping is 
never able to prevent the most important kinetic instabilities, and thus strongly magnetised, high-$\beta$ plasmas 
cannot be modelled by standard Chapman-Enskog theory if $\lambda/L \gtrsim 1/\beta$. This 
finding has significant implications for our understanding of various plasma environments, including those found in astrophysical 
contexts and also those created in laser-plasma experiments on high-energy laser facilities. 

When kinetic instabilities do arise in a Chapman-Enskog plasma, we have characterised all of them systematically, 
deriving simple expressions for their thresholds and growth rates in terms of basic parameters
such as $\beta$, $\lambda/L$ and the mass ratio $\mu_e = m_e/m_i$ using a novel 
analytical approach. Three of the instabilities -- the CET whistler instability (section \ref{electron_heatflux_instab_whistl}), the CET slow-wave instability (section \ref{ion_heatflux_instab_slowwave}), and  
the CET long-wavelength kinetic-Alfv\'en wave (KAW) instability (section \ref{CET_KAW_instab}) -- are 
driven by heat fluxes in a Chapman-Enskog plasma, while the remaining ten -- the CES mirror 
instability (section \ref{pospress_ion_mirror}), the CES whistler instability (section 
\ref{pospres_electron_EC}), the CES transverse instability (sections \ref{pospres_electron_trans} and 
\ref{negpres_subelectron_obliquetrans}), the CES electron mirror instability (section 
\ref{pospres_electron_oblique}), the CES firehose instability (sections 
\ref{negpres_fire_par}, \ref{negpres_fire_oblique}, \ref{negpres_fire_critline}, and~\ref{negpres_fire_subion}), the CES parallel and oblique electron firehose instabilities (sections 
\ref{negpress_electronfire_prl} and \ref{negpres_electron_oblique}, respectively), the 
CES electron-scale-transition (EST) instability (section \ref{negpres_elec_EST}), the CES whisper instability (section 
\ref{negpres_subelectron_phantom}), and the CES ordinary-mode instability (section \ref{negpres_subelectron_ord}) 
-- are driven by ion- and/or electron-velocity shears. While many of these 
instabilities, or versions thereof, had been considered previously, four of them (the CET slow-wave, CET long-wavelength KAW, CES EST and CES whisper 
instabilities) are new; the whisper instability in particular seems to be of some interest 
both conceptually and practically, because it is associated with a newly 
discovered plasma wave (the whisper wave), and the instability is much faster than its 
competitors over quite a wide range of values of $\lambda/L$ and $\beta$. 

An important question to address is that of the dominant 
microinstability overall: in a given plasma (with fixed $d_e/L$, $\lambda/L$, and $\beta$), amongst the many instabilities that we have found, 
which is the dominant one? As we explained in section \ref{rel_size_CEterms}, the answer to this question  
depends on assumptions about the relative magnitude of temperature- and velocity-gradient scale lengths $L_T$ and $L_V$. 
Assuming the scalings (\ref{brag_multiscale}) in section \ref{rel_size_CEterms} for a Chapman-Enskog plasma whose largest-scale fluid
motions are sonic (in other words, $\mathrm{Ma} \lesssim 1$), we find that, assuming also $\mathrm{Ma}\, \lambda/L_V$ to be large 
enough to trigger all of the aforementioned instabilities, the three most 
competitive ones are on electron scales: the CET whistler, CES 
whisper, and transverse instabilities. These have growth rates [see (\ref{maxgrowthrate_heatflux}), (\ref{transverse_maxgrowth}) and (\ref{whiswavinstab_gamma}), respectively]
\begin{subeqnarray}
  \gamma_{\mathrm{whistler,T}} & \sim & \eta_e 
\Omega_e \sim \mu_e^{1/4} \mathrm{Ma} \frac{\lambda_{i}}{L_V} \Omega_e \, , \\
\gamma_{\mathrm{whisper}} & \sim & \frac{|\epsilon_e|^{3/4} \beta_e^{1/4}}{\left[\log{|\epsilon_e| \beta_e}\right]^{1/2}}
 \Omega_e  \sim \left(\frac{\lambda_{i}}{L_V}\right)^{3/4} \frac{\mu_e^{3/8} \mathrm{Ma}^{3/4} \beta_e^{1/4}}{\left[\log{\left(\mu_e^{1/2} \beta_e \mathrm{Ma} \, \lambda_i/L_V \right)}\right]^{1/2}} \Omega_e \, , \\
\gamma_{\mathrm{trans}} & \sim & \epsilon_e^{3/2} \beta_e^{1/2} \Omega_e \sim \mu_e^{3/4} \left(\frac{\lambda_{i}}{L_V}\right)^{3/2} \beta_e^{1/2} \Omega_e  \, . 
\end{subeqnarray}
Although the threshold for the CET whistler instability is less 
restrictive than for the whisper instability, at the whisper instability 
threshold $|\epsilon_e| \beta_e \sim \beta_e^{2/7} \sim |\epsilon_e|^{-2/5}$ it follows that
\begin{equation}
 \frac{\gamma_{\mathrm{whistler,T}}}{\gamma_{\mathrm{whisper}}} \sim \frac{\eta_e \left[\log{\epsilon_e \beta_e}\right]^{1/2}}{\epsilon_e^{2/5}}  \sim 
 \mu_e^{1/20} \left(\frac{\lambda_{i}}{L_V}\right)^{3/5} \left[\log{\left(\mu_e^{1/2} \beta_e \mathrm{Ma} \, \lambda_i/L_V \right)}\right]^{1/2} 
 \ll 1 . 
\end{equation}
Thus, the fact that CE plasmas typically support fluid motions on smaller scales 
than temperature gradients (see section \ref{rel_size_CEterms}) implies that CES microinstabilities are more 
potent at sufficiently high plasma $\beta_e$. Yet, for $\beta_e \lesssim \mu_e^{-1/2} \mathrm{Ma}^{-1} \, {L_V}/{\lambda_{i}}$, the CET whistler instability is the most rapidly 
growing microinstability. Finally, for $\beta_e \lesssim \mu_e^{-1/4} \mathrm{Ma}^{-1} L_V/ 
\lambda_i$, none of these electron-scale instabilities is triggered at all, with 
only the ion-scale firehose and mirror instabilities operating. 
In short, the dominant microinstability is a 
complicated function of the parameter regime. For reference, in table \ref{tab:microinstabs_prop} of section \ref{sec:introduction} we show the (approximate) growth rates for 
all of the instabilities considered in this paper if the scalings (\ref{brag_multiscale})
are adopted, and figure \ref{stabilitymapintro}
shows a schematic stability map for the same case
\footnote{A note of caution is warranted: if a Chapman-Enskog plasma is unstable to 
microinstabilities, then the heat fluxes and rate-of-strain tensors will be modified, potentially altering 
both $L_T$ and $L_V$. There is no \textit{a priori} reason to think that such a plasma will 
obey Braginskii-type scalings of the form (\ref{brag_multiscale}) -- and so using this ordering to estimate 
microinstability growth rates is incorrect in kinetically unstable Chapman-Enskog plasmas.}. 

We believe that our study -- which is the first 
systematic investigation of the kinetic stability of a classical, collisional, magnetised plasma -- 
provides a significant step forward towards a comprehensive understanding of this state of matter. 
It is perhaps inevitable, however, given the conceptual extent of the problem, 
that there remain a number of questions concerning the 
stability of the Chapman-Enskog distribution function that we have not addressed here. 
In terms of linear theory, a numerical study using 
a better collision operator to find the exact stability boundaries could be usefully 
carried out -- although we do not anticipate that this would lead to an 
alteration of the basic scalings of those boundaries derived in this paper. 
Another issue not addressed by this work is that of linear coupling between CET and 
CES microinstabilities; it is not immediately obvious to what extent microinstabilities with similar 
growth rates might aid each other's growth. 
The analysis could also be extended to two-species plasmas not in thermal equilibrium,
as well as high-$Z$ plasmas (with important applications in laser-plasma 
physics). 

Perhaps the most interesting future development of this work would be the determination of transport coefficients for plasmas
falling into the unstable regimes. This requires quasi-linear or nonlinear treatment. 
Nonetheless, the results presented here can be seen as both a guide and a warning to those wishing
to address this fundamental question. 
They are a guide in the sense that a correct characterisation of transport coefficients 
requires knowledge of the fastest-growing linear modes, which our study provides. 
But they are also as a warning in that an isolated treatment of one type of microinstability without 
reference to the full range of possible others could lead to a mischaracterisation of transport properties. 
The best hope for a correct calculation of transport in a weakly collisional, high-$\beta$ plasma is, 
therefore, the following programme: for a plasma with particular conditions, 
identify the fastest microinstability, calculate the saturated magnitude of the fluctuations 
produced by it, determine the anomalous transport coefficients with those fluctuations present, 
re-calculate of the stability of this plasma, and so on, until a self-consistent picture emerges. 
It is likely that such a picture will involve a distribution function whose underlying nature depends 
on macroscopic motions, and hence transport coefficients that are themselves properties of flow 
shear, temperature gradients, and large-scale magnetic fields. 
Carrying out such calculations is a non-trivial task, but not impossible.

To carry out this research, AFAB was supported by DOE awards DE-SC0019046 and DE-SC0019047 
through the NSF/DOE Partnership in Basic Plasma Science and Engineering, and also by UKRI (grant number MR/W006723/1). The work of AAS was supported in part by grants from STFC (ST/N000919/1 and ST/W000903/1) and EPSRC (EP/M022331/1 and EP/R034737/1), as well as by the Simons Foundation via a Simons Investigator award. This research was in part funded by Plan S funders (UKRI, STFC and EPSRC); for the purpose of Open Access, the author has applied a CC BY public copyright licence to any Author Accepted Manuscript version arising from this submission. AFAB would like to express his deepest gratitude to Matthew Kunz and Eliot Quataert for many helpful discussions about the paper generally, for highlighting some important considerations pertaining to the linear theory of the firehose instability, and for their ongoing support, without which the paper would never have been completed. All the authors would also like to thank Anatoly Spitkovsky
for bringing several key early papers on the topic of this paper to the authors' attention. 

\begin{appendix}

\section{Glossary of notation used in the paper} \label{Glossary}

As an aid to reading, we provide a glossary of the notation that we use 
in our paper in tables \ref{tab:glossaryI} and \ref{tab:glossaryII} of this appendix. 
\begin{table}
\centering
{\renewcommand{\arraystretch}{1.0}
\renewcommand{\tabcolsep}{0.15cm}
\begin{tabular}[c]{c|c}
Notation & Quantity \\
\hline
$\boldsymbol{r}$ & Spatial position \\
$t$ & Time \\
$e$ & Elementary charge \\
$c$ & Speed of light \\
$Z_s$ ($Z$) & Charge of species $s$ ($s = i$ in two-species plasma) in units of $e$  \\
$m_s$ & Mass of a particle of species $s$ \\
$\mu_e = m_e/m_i$ & Electron-to-ion mass ratio \\ 
$\boldsymbol{E}$ & Electric field \\
$\boldsymbol{B}$ & Magnetic field \\
$\boldsymbol{B}_0$ & Macroscopic magnetic field \\
$\hat{\boldsymbol{z}}$ & Direction vector of the macroscopic magnetic field \\
$\boldsymbol{v}$ ($\boldsymbol{v}_{\perp}$) & Particle velocity (in the direction perpendicular to $\boldsymbol{B}_0$)  \\
$v$ ($v_{\perp}$) & Particle speed (in the direction perpendicular to $\boldsymbol{B}_0$)\\
$v_{\|}$ & Particle velocity in the direction parallel to $\boldsymbol{B}_0$ \\
$\phi$ & Gyrophase angle \\
$f_s(\boldsymbol{r},\boldsymbol{v},t)$ & Distribution function of particles of species $s$ \\
$\mathfrak{C}(f_s,f_{s'})$ & Collision operator for interactions between species $s$ and $s'$ \\
$n_s$ & Density of particles of species $s$ (\ref{dens_vel_temp_def}\textit{a}) \\
$\boldsymbol{V}_s$ & Bulk fluid velocity of particles of species $s$ (\ref{dens_vel_temp_def}\textit{b}) \\
$T_s$ & Temperature of particles of species $s$ (\ref{dens_vel_temp_def}\textit{c}) \\
$p_s$ ($p_{s\|}/p_{s\perp}$)& (Parallel/perpendicular) pressure of particles of species $s$ (\ref{fluxes_fluid}\textit{a}), (\ref{press_perpparr})  \\
$\boldsymbol{\pi}_s$ & Viscosity tensor of particles of species $s$ (\ref{fluxes_fluid}\textit{b}) \\
$\Delta_s = (p_{s\perp}-p_{s\|})/p_s$ & Pressure anisotropy of particles of species $s$ (\ref{pressureanisop}) \\
$\boldsymbol{q}_s$ (${q}_{s\|}$) & (Parallel) heat flux of particles of species $s$ (\ref{fluxes_fluid}\textit{c}), (\ref{par_heat_flux}) \\
$\boldsymbol{R}_s$ (${R}_{s\|}$) & (Parallel) frictional force on species $s$ due to collisions (\ref{fluxes_fluid}\textit{d}) \\
$\mathcal{Q}_s$ & Heating rate due to inter-species collisions (\ref{fluxes_fluid}\textit{e}) \\
$\boldsymbol{u}_{ei}$ (${u}_{ei\|}$) & (Parallel) relative electron-ion drift \\
$v_{\mathrm{th}s}$ & Thermal speed of particles of species $s$ \\
${\boldsymbol{v}}_s' = \boldsymbol{v}-\boldsymbol{V}_s$ & Peculiar velocity of particles of species $s$  \\
${v}_{s\|}'$ (${\boldsymbol{v}}_{s\perp}'$) & Peculiar parallel (perpendicular) velocity of particles of species $s$  \\
$\tilde{\boldsymbol{v}}_s = (\boldsymbol{v}-\boldsymbol{V}_i)/v_{\mathrm{th}s}$ & Non-dimensionalised particle velocity in ion-fluid rest frame  \\
$\tilde{v}_{s\|}$ & Non-dimensionalised parallel particle velocity, ion-fluid rest frame  \\
$\tilde{v}_s$ ($\tilde{v}_{s\perp}$) & Non-dim. (perpendicular) particle speed, ion-fluid rest frame  \\
$\lambda_s$ & Mean free path of species $s$ \\
$\rho_s$ ($\tilde{\rho}_s$) & (Signed) Larmor radius of species $s$ \\
$\tau_s$ & Collision time of species $s$ \\
$\Omega_s$ ($\tilde{\Omega}_s$) & (Signed) Larmor frequency of species $s$ \\
$L$ & Macroscopic length scale of variation in the direction parallel to $\boldsymbol{B}_0$ \\
$L_T$ ($L_{T_i}$) & Electron-(ion)-temperature length scale parallel to $\boldsymbol{B}_0$ (\ref{parallel_vel_scale}\textit{a},\textit{b}) \\
$L_V$ ($L_{V_e}$) & Ion-(electron)-bulk-flow length scale parallel to $\boldsymbol{B}_0$ (\ref{parallel_vel_scale}\textit{c},\textit{d}) \\
$\tau_L$ & Macroscopic time scale over which CE distribution varies \\
$\eta_e = \eta_e^{T}$ & Small parameter (\ref{smallparam}\textit{a}) $\propto$ CE electron-temperature-gradient term \\
$\eta_e^{R}$ & Small parameter (\ref{smallparam}\textit{b}) $\propto$ CE electron-friction term \\
$\eta_e^{u}$ & Small parameter (\ref{smallparam}\textit{c}) $\propto$ CE electron-ion-drift term \\
$\eta_i$ & Small parameter (\ref{smallparam}\textit{d}) $\propto$ CE ion-temperature-gradient term \\
$\epsilon_e$ & Small parameter (\ref{smallparam}\textit{e}) $\propto$ CE electron-shear term \\
$\epsilon_i$ & Small parameter (\ref{smallparam}\textit{f}) $\propto$ CE ion-shear term \\
$A_e^{T}(\tilde{v}_e)$ & Function arising in CE electron-temperature-gradient term \\
$A_e^{R}(\tilde{v}_e)$ & Function arising in CE electron-friction term \\
$A_e^{u}(\tilde{v}_e)$ & Function arising in CE electron-ion-drift term \\
$A_i(\tilde{v}_i)$ & Function arising in CE ion-temperature-gradient term \\
$C_e(\tilde{v}_e)$ & Function arising in CE electron-shear term \\
$C_i(\tilde{v}_i)$ & Function arising in CE ion-shear term \\
\end{tabular}} 
\caption{\textbf{Glossary of notation I.}}
\label{tab:glossaryI}
\end{table}
\begin{table}
\centering
{\renewcommand{\arraystretch}{1.0}
\renewcommand{\tabcolsep}{0.15cm}
\begin{tabular}[c]{c|c}
Notation & Quantity \\
\hline
$\mathsfbi{W}_s$ & Traceless rate-of-strain tensor of species $s$ (\ref{ChapEnsTerms}) \\
$\mathrm{Ma} = V_i/v_{\mathrm{th}i}$ & Mach number \\
$\log{\Lambda_{\rm CL}}$ & Coulomb logarithm \\
$\kappa_e^{\|}$ ($\kappa_i^{\|}$) & Parallel electron (ion) thermal conductivity (\ref{parallel_cond_electrons}) [(\ref{parallel_cond_ions})] \\
$\chi = 2 \kappa_e^{\|}/3 n_e$ & Parallel thermal diffusivity \\
$\mu_{\mathrm{v}s}$ & Dynamic viscosity of particles of species $s$ \\
$\nu = \mu_{\mathrm{v}i}/m_i n_i$ & Kinematic viscosity \\
$\tau_{ie}^{\rm eq}$ & Ion-electron temperature equilibration time \\
$d_s$ & Inertial scale of particles of species $s$ (\ref{skin_depth_def}\textit{b}) \\
$\lambda_{\rm D}$ & Debye length (\ref{skin_depth_def}\textit{c}) \\
$\beta_s$ & Plasma beta of particles of species $s$ (\ref{plasma_beta_def}) \\
$\omega_{\mathrm{p}s}$ & Plasma frequency of particles of species $s$ (\ref{plasmafrequency_def}) \\
$f_{s0}(\tilde{v}_{s\|},\tilde{v}_{s\perp})$ & Equilibrium distribution function of particles of species $s$ \\
$\tilde{f}_{s0}(\tilde{v}_{s\|},\tilde{v}_{s\perp})$ & Renormalised equilibrium dist. func. of particles of species $s$ (\ref{renorm_distfunc_def}) \\
$\delta \boldsymbol{E}$ & Microscale electric-field perturbation \\
$\delta \boldsymbol{B}$ & Microscale magnetic-field perturbation \\
$\widehat{\delta \boldsymbol{E}}$ & Fourier component of $\delta \boldsymbol{E}$ \\
$\widehat{\delta \boldsymbol{B}}$ & Fourier component of $\delta \boldsymbol{B}$ \\
$\boldsymbol{k}$ ($\boldsymbol{k}_{\perp}$) & (Perpendicular) wavevector of electromagnetic perturbation \\
$k$ ($k_{\|}/k_{\perp}$) & (Parallel/perpendicular) wavenumber of electromagnetic perturbation \\
$\hat{\boldsymbol{k}}$ ($\hat{\boldsymbol{x}}$) & (Perpendicular) wavevector direction \\ 
$\hat{\boldsymbol{y}}$ & Direction perpendicular to both $\hat{\boldsymbol{x}}$ and $\hat{\boldsymbol{z}}$ \\
$\theta$ & Wavevector angle \\
$\omega$ & Complex frequency of electromagnetic perturbation \\
$\varpi$ & Real frequency of electromagnetic perturbation \\
$\gamma$ & Growth rate of electromagnetic perturbation \\
$\tilde{\omega}_{s\|} = \omega/|k_{\|}| v_{\mathrm{th}s}$ & Ratio of complex frequency to parallel streaming rate (\ref{normfrequency}) \\
$\hat{\omega}_{s\|} = \tilde{\omega}_{s\|} + \mathrm{i}/|k_{\|}| \lambda_s$ & Modified frequency-to-streaming-rate ratio (\ref{collresconduct}) \\
$\mathsfbi{I}$ & Unit dyadic \\
$\boldsymbol{\mathfrak{E}}$ & Plasma dielectric tensor (\ref{dielecttensfull}) \\
$\boldsymbol{\mathfrak{E}}_s$ & Contribution of species $s$ to $\boldsymbol{\mathfrak{E}}$\\
$\boldsymbol{\sigma}$ & Plasma conductivity tensor (\ref{conductivity}) \\
$\boldsymbol{\sigma}_s$ & Contribution of species $s$ to $\boldsymbol{\sigma}$\\
$\zeta_{sn} = \tilde{\omega}_{s\|}-n/|k_{\|}| \tilde{\rho}_s$  & Resonant non-dimensionalised velocity (\ref{zeta_def}) \\
$\tilde{w}_{s\|} = k_{\|} \tilde{v}_{s\|}/|k_{\|}|$ & Non-dimensionalised parallel particle velocity (\ref{w_pl_def}) \\
$\Lambda_s(\tilde{w}_{s\|},\tilde{v}_{s\bot})$ & Velocity-space anisotropy of distribution function (\ref{anisotropyfunc}) \\
$ \Xi_s(\tilde{w}_{s\|},\tilde{v}_{s\bot})$ & Velocity-space integrand of conductivity tensor (\ref{IntgradCond}) \\
$\boldsymbol{\mathfrak{E}}^{(0)}$ & Leading-order term in expansion of $\boldsymbol{\mathfrak{E}}$ in $\tilde{\omega}_{s\|} \ll 1$   \\
$\boldsymbol{\mathfrak{E}}^{(1)}$ & First-order correction in expansion of $\boldsymbol{\mathfrak{E}}$ in $\tilde{\omega}_{s\|} \ll 1$   \\
$\mathsfbi{M}_s$ & Maxwellian component of $\boldsymbol{\mathfrak{E}}_s$ -- see (\ref{Maxnonmaxsep_s}) \\
$\mathsfbi{P}_s$ & Non-Maxwellian component of $\boldsymbol{\mathfrak{E}}_s$ -- see (\ref{Maxnonmaxsep_s}) \\
$\mathsfbi{M}_s^{(0)}$ ($\mathsfbi{P}_s^{(0)}$) &  Leading-order term in expansion of $\mathsfbi{M}_s$ ($\mathsfbi{P}_s$) in $\tilde{\omega}_{s\|} \ll 1$  \\
$\mathsfbi{M}_s^{(1)}$ &  First-order correction in expansion of $\mathsfbi{M}_s$ in $\tilde{\omega}_{s\|} \ll 1$  \\
$\{\boldsymbol{e}_1,\boldsymbol{e}_2,\boldsymbol{e}_3\}$ & Coordinate basis (\ref{e_coordinatebasis_def}) \\
$\widehat{\delta \boldsymbol{E}}_T = \widehat{\delta \boldsymbol{E}} \bcdot (\mathsfbi{I}-\hat{\boldsymbol{k}}\hat{\boldsymbol{k}})$ & Transverse electric-field perturbation \\
$Z(z)$ & Plasma dispersion function (\ref{Plasma_disp_func_def}) \\
$F(x,y)$, $G(x,y)$, $H(x,y)$ & Special mathematical functions (\ref{specialfuncMax}) \\
$L(x,y)$, $N(x,y)$ & Special mathematical functions (\ref{LandNdef_Append}) \\
$I(x,y)$, $J(x,y)$, $K(x,y)$ & Special mathematical functions (\ref{heatfluxasymptoticfuncs}) \\
$W(x,y)$, $X(x,y)$, $Y(x,y)$ & Special mathematical functions (\ref{shearasymptoticfuncs}) 
\end{tabular}} 
\caption{\textbf{Glossary of notation II.}}
\label{tab:glossaryII}
\end{table}

\section{Derivation of the Chapman-Enskog distribution function} 

\subsection{The Chapman-Enskog expansion in strongly magnetised plasma} \label{ChapEnskogDev_Append_comp}

There exist a number of lucid explanations of how the 
CE distribution functions (\ref{ChapEnskogFunc}) arise in a collisional, strongly magnetised two-species electron-ion plasma ($\rho_s \ll \lambda_s$ for $s = i, e$) -- the monograph 
of~\citet{B65}, but also (for example)~\citet{HS05}, 
Chapter 4. For that reason, we do not provide a full derivation of 
(\ref{ChapEnskogFunc}). However, in this appendix, we describe a calculation 
that allows for a direct derivation of the CE distribution function for a 
strongly magnetised collisional plasma, without first having to perform the CE 
expansion for arbitrary values of $\rho_s/\lambda_s$. 

The first part of the calculation is the same as in~\citet{HS05}, 
pp. 76-78. For the reader's convenience, we present a summarised version.
We consider the Maxwell-Vlasov-Landau equation (\ref{MaxVlasLan}) of 
species $s$ in a frame co-moving with the fluid rest frame of 
that species. Defining the peculiar velocity variable $\boldsymbol{v}_s' = \boldsymbol{v}-\boldsymbol{V}_s$ 
in the fluid rest frame, (\ref{MaxVlasLan}) becomes
\begin{equation}
  \frac{\mathrm{D} f_{s}}{\mathrm{D} t} + \boldsymbol{v}_s' \bcdot \bnabla f_{s} + \left[\frac{Z_s e}{m_s} \left(\boldsymbol{E}'+ \frac{\boldsymbol{v}_s' \times \boldsymbol{B}}{c}\right) -  \frac{\mathrm{D} \boldsymbol{V}_s}{\mathrm{D} t}\right]\bcdot \frac{\partial f_s}{\partial \boldsymbol{v}_s'} - \boldsymbol{v}_s' \bcdot \bnabla \boldsymbol{V}_s \bcdot \frac{\partial f_s}{\partial \boldsymbol{v}_s'} = 
  \sum_{s'} \mathfrak{C}(f_s,f_{s'}) , \label{MaxVlasLan_movframe}
\end{equation}
where $\boldsymbol{E}' \equiv \boldsymbol{E} + \boldsymbol{V}_s \times \boldsymbol{B}/c$ 
is the electric field measured in the moving frame, and 
\begin{equation}
\frac{\mathrm{D}}{\mathrm{D} t} \equiv \frac{\p}{\p t} + \boldsymbol{V}_s \bcdot 
\bnabla
\end{equation}
is the convective derivative. Initially ordering $\lambda_s \sim \rho_s$, and 
assuming the plasma is collisional ($\lambda_{s}/L \ll 1$), we rearrange (\ref{MaxVlasLan_movframe}) 
so that the largest terms are grouped together (on the left-hand side):
\begin{eqnarray}
 \sum_{s'} \mathfrak{C}(f_s,f_{s'}) & - & \frac{Z_s e}{m_s c} \left(\boldsymbol{v}_s' \times \boldsymbol{B} \right) \bcdot \frac{\partial f_s}{\partial \boldsymbol{v}_s'} \nonumber \\ &=& \frac{\mathrm{D} f_{s}}{\mathrm{D} t} + \boldsymbol{v}_s' \bcdot \bnabla f_{s} + \left(\frac{Z_s e}{m_s} \boldsymbol{E}' -  \frac{\mathrm{D} \boldsymbol{V}_s}{\mathrm{D} t}\right)\bcdot \frac{\partial f_s}{\partial \boldsymbol{v}_s'} - \boldsymbol{v}_s' \bcdot \left( \bnabla \boldsymbol{V}_s \right)\bcdot \frac{\partial f_s}{\partial \boldsymbol{v}_s'} 
  . \qquad \label{MaxVlasLan_movframe_group} 
\end{eqnarray}
We then expand the distribution functions $f_s$ in small parameter $\lambda_{s}/L \ll 1$:
\begin{equation}
f_s = f_{s}^{(0)} + f_{s}^{(1)} + \ldots , \label{CE_expand}
\end{equation}
and solve (\ref{MaxVlasLan_movframe_group}) order by order in $\lambda_{s}/L$ for
$f_{s}^{(0)}$ and $f_{s}^{(1)}$. The subsequent treatment of the collision operator for the electron distribution 
function is a little different from the ion distribution function, so we treat 
each case individually. 

\subsubsection{Electrons}

For the electrons, we can rewrite the total collision operator in a convenient form if we assume that $T_i \sim T_e$, and $\boldsymbol{V}_i \sim 
v_{\mathrm{th}i}$:
\begin{equation}
 \sum_{s'} \mathfrak{C}(f_e,f_{s'}) = \mathfrak{C}_{ee}(f_e) + \mathfrak{C}_{ei}^{(0)}(f_e) 
 +  \mathfrak{C}_{ei}^{(1)}(f_e) \, ,  \label{electroncollop_tot}
\end{equation}
where the electron-electron collision operator $\mathfrak{C}_{ee}(f_e)$ and electron-ion collision operators $\mathfrak{C}_{ei}^{(0)}(f_e)$ and $\mathfrak{C}_{ei}^{(1)}(f_e)$ are 
\begin{subeqnarray}
\mathfrak{C}_{ee}(f_e) & \equiv & \mathfrak{C}(f_e,f_e) , \\
\mathfrak{C}_{ei}^{(0)}(f_e) & \equiv & \nu_{ei}(v) v^3 \frac{\p}{\p \boldsymbol{v}} \bcdot \left[\frac{1}{v} \left(\mathsfbi{I}- \hat{\boldsymbol{v}} \hat{\boldsymbol{v}} \right) \bcdot \frac{\p f_e}{\p \boldsymbol{v}} \right] , \\
\mathfrak{C}_{ee}^{(1)}(f_e) & \equiv & \nu_{ei}(v) \frac{m_e \boldsymbol{v}_e' \bcdot \boldsymbol{u}_{ei}}{T_e} \frac{n_{e}}{\upi^{3/2} v_{\mathrm{th}e}^3}  \exp \left(-\tilde{v}_{e}^2\right)
. \label{electroncollop_defs}
\end{subeqnarray}
Here $\nu_{ei}(v)$ is the velocity-dependent collision frequency
\begin{equation}
\nu_{ei}(v) \equiv \frac{3 \sqrt{\upi}}{4 \tau_e} \left(\frac{v_{\mathrm{th}e}}{v}\right)^3 
,
\end{equation}
and the total electron-ion collision operator $\mathfrak{C}(f_e,f_i)$ is given 
by $\mathfrak{C}(f_e,f_i) = \mathfrak{C}_{ei}^{(0)}(f_e) + 
\mathfrak{C}_{ei}^{(1)}(f_e)$. This reformulation of the electron-ion collision 
operator is possible, because the assumptions $T_i \sim T_e$, and $\boldsymbol{V}_i \sim 
v_{\mathrm{th}i}$ mean that, from the perspective of the electrons, the ion distribution is sharply peaked around the ion fluid velocity: in other words, $f_i \approx n_{i} 
\delta(\boldsymbol{v}-\boldsymbol{V}_i)$. Furthermore, the reformulation is 
convenient because the total electron collision operator (\ref{electroncollop_tot}) 
becomes independent of the ion distribution function. Thus, the asymptotic expansion (\ref{CE_expand}) 
for the electron distribution function is decoupled from the ions. 

Substituting (\ref{electroncollop_tot}), the ordered kinetic equation (\ref{MaxVlasLan_movframe_group}) for the electron distribution  
becomes
\begin{subeqnarray}
\mathfrak{C}_{ee}(f_e) & + & \mathfrak{C}_{ei}^{(0)}(f_e) 
 + \frac{e}{m_e c} \left(\boldsymbol{v}_e' \times \boldsymbol{B} \right)\bcdot \frac{\partial f_e}{\partial \boldsymbol{v}_e'} \nonumber \\
  & = & \frac{\mathrm{D} f_{e}}{\mathrm{D} t} + \boldsymbol{v}_e' \bcdot \bnabla f_{e} - \left(\frac{e}{m_e} \boldsymbol{E}' +  \frac{\mathrm{D} \boldsymbol{V}_e}{\mathrm{D} t}\right)\bcdot \frac{\partial f_e}{\partial \boldsymbol{v}_e'} - \boldsymbol{v}_e' \bcdot \left( \bnabla \boldsymbol{V}_e \right) \bcdot \frac{\partial f_e}{\partial \boldsymbol{v}_e'} 
  - \mathfrak{C}_{ei}^{(1)}(f_e)  
  , \qquad \quad \label{MaxVlasLan_movframe_group_elec}   
\end{subeqnarray}
where we note that under assumptions $T_i \sim T_e$, and $\boldsymbol{V}_i \sim 
v_{\mathrm{th}i}$ , $\mathfrak{C}_{ei}^{(1)}(f_e) \sim \mu_e^{1/2} \mathfrak{C}_{ei}^{(0)}(f_e)$ is much smaller than 
$\mathfrak{C}_{ei}^{(0)}(f_e)$.  Then applying expansion (\ref{CE_expand}) with $s = e$ 
gives  
\begin{equation}
\mathfrak{C}_{ee}(f_e^{(0)}) + \mathfrak{C}_{ei}^{(0)}(f_e^{(0)}) 
 + \frac{e}{m_e c} \left(\boldsymbol{v}_e' \times \boldsymbol{B} \right) \bcdot \frac{\partial f_e^{(0)}}{\partial \boldsymbol{v}_e'} = 0 \, .  \label{leadingorderdist}
\end{equation}
It can be shown~\citep{HS05} that the only solution of (\ref{leadingorderdist}) 
is (as expected) a Maxwellian distribution:
\begin{equation}
f_e^{(0)} = \frac{n_{e}}{\upi^{3/2} v_{\mathrm{th}e}^3}  \exp \left(-\frac{|\boldsymbol{v}_e'|^2}{v_{\mathrm{th}e}^2}\right) 
\, .
\end{equation}
After some algebraic manipulation, it can also be shown that the leading-order perturbed electron distribution function $f_{e}^{(1)}(\boldsymbol{v})$ satisfies
\begin{eqnarray}
\mathfrak{C}_{ee}(f_{e}^{(1)}) & + & \mathfrak{C}_{ei}^{(0)}(f_{e}^{(1)}) + \frac{e}{m_e c} \left( \boldsymbol{v}_{e}' \times \boldsymbol{B} \right) \bcdot \frac{\p f_{e}^{(1)}}{\p \boldsymbol{v}_{e}'} \nonumber\\
&=& \Bigg\{\left(\frac{|\boldsymbol{v}_e'|^2}{v_{\mathrm{th}e}^2} -\frac{5}{2}\right)\boldsymbol{v}_{e}' \bcdot \nabla \log T_e \nonumber \\
&& \qquad + \boldsymbol{v}_{e}' \bcdot \left[\frac{\boldsymbol{R}_e}{p_e} + \frac{m_e \boldsymbol{u}_{ei} \nu_{ei}(v)}{T_e} \right]  + \frac{m_e}{2 T_e} \left(\boldsymbol{v}_{e}'  \boldsymbol{v}_{e}' - \frac{|\boldsymbol{v}_e'|^2}{3}\mathsfbi{I} \right) \! \boldsymbol{:} \! \mathsfbi{W}_e \Bigg\} f_{e}^{(0)} , \qquad  \label{cEexpFokkerPlanck}
\end{eqnarray}
where $\boldsymbol{R}_e$ and so on are defined in the main text, in equations 
(\ref{ChapEnsTerms}). 

\subsubsection{Electrons in strongly magnetised limit}

We now solve for $f_{e}^{(1)}$ in a strongly magnetised plasma, i.e., $\rho_e \ll 
\lambda_e$. In this subsidiary limit, both the collision integrals on the 
left-hand-side of (\ref{cEexpFokkerPlanck}) and the terms on its right-hand side are much smaller than the term 
proportional to the magnetic field; in other words, 
\begin{equation}
\boldsymbol{v}_{e}' \times \boldsymbol{B} \bcdot \frac{\p f_{e}^{(1)}}{\p \boldsymbol{v}_{e}'} \approx 0 \, .\label{strongmag_fe1}
\end{equation}
We then define coordinate system $\left\{v_{e\|}',v_{e\bot}',\phi'\right\}$ by $v_{e\|}' \equiv \hat{\boldsymbol{z}} \bcdot 
\boldsymbol{v}_{e}'$, $\boldsymbol{v}_{e\bot}' = \boldsymbol{v}_{e}' - v_{e\|}' 
\hat{\boldsymbol{z}}$, ${v}_{e\bot}' = |\boldsymbol{v}_{e\bot}' |$
and $\phi' = \phi$, where $\hat{\boldsymbol{z}} = \boldsymbol{B}/|\boldsymbol{B}|$ and $\phi$ is the
gyrophase angle. The velocity gradient operator in this system is
\begin{equation}
\frac{\p f_{e}^{(1)}}{\p \boldsymbol{v}_e'}  = \hat{\boldsymbol{z}}\frac{\p f_{e}^{(1)}}{\p v_{e\|}'} + \frac{\boldsymbol{v}_{e\bot}'}{v_{e\bot}'} \frac{\p f_{e}^{(1)}}{\p v_{e\bot}'} + \frac{1}{v_{e\bot}'^2} \boldsymbol{v}_e' \times \hat{\boldsymbol{z}} \frac{\p f_{e}^{(1)}}{\p \phi'} \, .
\end{equation}
This, when combined with (\ref{strongmag_fe1}), implies that $f_{e}^{(1)}$ is approximately gyrotropic: 
\begin{equation}
f_{e}^{(1)}(\boldsymbol{v}') \approx  \langle f_{e}^{(1)}\rangle_{\phi'}(v_{\|}',v_{\bot}') 
,
\end{equation}
where we have defined the gyro-average $\langle f_{e}^{(1)}\rangle_{\phi'}$ of the electron distribution function by
\begin{equation}
\langle f_{e}^{(1)} \rangle_{\phi'} \equiv \frac{1}{2 \upi} \int_0^{2 \upi} \mathrm{d} \phi' \, f_{e}^{(1)} \, .
\end{equation}

Now gyro-averaging (\ref{cEexpFokkerPlanck}), we obtain
\begin{eqnarray}
\mathfrak{C}_{ee}\left(\langle f_{e}^{(1)}\rangle_{\phi'}\right) & + & \mathfrak{C}_{ei}^{(0)}\left(\langle f_{e}^{(1)}\rangle_{\phi'}\right) \nonumber\\
&=& \Bigg\{\left[\left(\frac{|\boldsymbol{v}_e'|^2}{v_{\mathrm{th}e}^2} -\frac{5}{2}\right) \nabla_{\|} \log T_e + \frac{R_{e\|}}{p_e} + \frac{m_e u_{ei\|} \nu_{ei}(v)}{T_e} \right] v_{e\|}' \nonumber \\
&& \qquad \qquad + \left(\hat{\boldsymbol{z}} \hat{\boldsymbol{z}} - \frac{1}{3}\mathsfbi{I} \right) \! \boldsymbol{:} \! \mathsfbi{W}_e \left(\frac{v_{e\|}'^2}{v_{\mathrm{th}e}^2} - \frac{v_{e\bot}'^2}{2 v_{\mathrm{th}e}^2} \right) \Bigg\} f_{e}^{(0)} \, ,\label{cEexpFokkerPlanckgyro}
\end{eqnarray}
where we have used the gyrophase isotropy of the collision operators to commute the order of gyro-averaging on the left-hand side. (\ref{cEexpFokkerPlanckgyro}) is a linear equation for $\langle f_{e}^{(1)}\rangle_{\phi'}$, so by tensor invariance, it must have a solution of the form
\begin{eqnarray}
 \langle f_{e}^{(1)} \rangle_{\phi'} & = & \tau_e \Bigg\{ \bigg[A_e^{T}\!\left(\frac{|\boldsymbol{v}_e'|}{v_{\mathrm{th}e}}\right) \nabla_{\|} \log T_e +A_e^{R}\!\left(\frac{|\boldsymbol{v}_e'|}{v_{\mathrm{th}e}}\right) \frac{R_{e\|}}{p_e}+ \left(A_e^{u}\!\left(\frac{|\boldsymbol{v}_e'|}{v_{\mathrm{th}e}}\right) -1\right)\frac{m_e u_{ei\|}}{T_e \tau_{e}} \bigg] v_{e\|}' 
 \nonumber \\
&& \qquad + C_e\!\left(\frac{|\boldsymbol{v}_e'|}{v_{\mathrm{th}e}}\right) \left(\hat{\boldsymbol{z}} \hat{\boldsymbol{z}} - \frac{1}{3}\mathsfbi{I} \right) \! \boldsymbol{:} \! \mathsfbi{W}_e \left(\frac{v_{e\|}'^2}{v_{\mathrm{th}e}^2} - \frac{v_{e\bot}'^2}{2 v_{\mathrm{th}e}^2} \right) \Bigg\} f_{e}^{(0)} \, ,\label{gyro_CEdist}
\end{eqnarray}
where $\tau_e$ is defined by equation (\ref{colltimes}\textit{a}) in the main text, and the isotropic functions $A_e^{T}\!\left({|\boldsymbol{v}_e'|}/{v_{\mathrm{th}e}}\right)$, $A_e^{R}\!\left({|\boldsymbol{v}_e'|}/{v_{\mathrm{th}e}}\right)$ 
and $C\!\left({|\boldsymbol{v}_e'|}/{v_{\mathrm{th}e}}\right)$ are determined by 
inverting the collision operators (see appendix \ref{ChapEnskogIsoFunc} for an example of how this calculation is done for a simple choice of collision operator). The total electron CE 
distribution function becomes
\begin{eqnarray}
 f_{e}(v_{e\|}',v_{e\bot}') & = & \Bigg\{1 + \tau_e \bigg[A_e^{T}\!\left(\frac{|\boldsymbol{v}_e'|}{v_{\mathrm{th}e}}\right) \nabla_{\|} \log T_e +A_e^{R}\!\left(\frac{|\boldsymbol{v}_e'|}{v_{\mathrm{th}e}}\right) \frac{R_{e\|}}{p_e} \nonumber \\
 && \qquad +\left(A_e^{u}\!\left(\frac{|\boldsymbol{v}_e'|}{v_{\mathrm{th}e}}\right) -1 \right) \frac{m_e u_{ei\|}}{T_e \tau_e} \bigg] v_{e\|}' \nonumber \\
&& \qquad \qquad + C_e\!\left(\frac{|\boldsymbol{v}_e'|}{v_{\mathrm{th}e}}\right) \left(\hat{\boldsymbol{z}} \hat{\boldsymbol{z}} - \frac{1}{3}\mathsfbi{I} \right) \! \boldsymbol{:} \! \mathsfbi{W}_e \left(\frac{v_{e\|}'^2}{v_{\mathrm{th}e}^2} - \frac{v_{e\bot}'^2}{2 v_{\mathrm{th}e}^2} \right) \Bigg\} f_{e}^{(0)} \, .\label{CE_dist_electronframe}
\end{eqnarray}
We emphasize that this quantity is expressed in the rest frame of the electron fluid\footnote{Reintroducing the parameters $\eta_e^{T}$, $\eta_e^{R}$, $\eta_e^{u}$ and $\epsilon_e$ into (\ref{CE_dist_electronframe}) gives the expression (\ref{CE_dist_electronframe_maintext}) that is quoted in section \ref{fluxfluids}.}. 

Finally, we recover (\ref{ChapEnskogFunc}\textit{a}) by transforming (\ref{CE_dist_electronframe}) into the frame 
co-moving with the ion fluid. Since ${u}_{ei\|} \sim \lambda_e v_{\mathrm{th}e}/L \ll 
v_{\mathrm{th}e}$, this transformation applied to the non-Maxwellian component $f_{e}^{(1)}$
of the electron distribution function only produces corrections of magnitude $\sim 
(\lambda_e/L) f_{e}^{(1)}$, and thus any correction terms are negligible. The only important 
contribution is from the shifted Maxwellian:
\begin{equation}
\exp \left(-\frac{|\boldsymbol{v}_e'|^2}{v_{\mathrm{th}e}^2}\right) \approx \exp \left(-\tilde{v}_e^{2}\right) 
\left[1 + 2 \tilde{v}_{e\|} \frac{u_{ei\|}}{v_{\mathrm{th}e}}\right] + \ldots , \label{Maxshift}
\end{equation}
where $\tilde{\boldsymbol{v}}_e = (\boldsymbol{v}- 
\boldsymbol{V}_i)/v_{\mathrm{th}e}$. Combining (\ref{Maxshift}) with 
(\ref{CE_dist_electronframe}), we deduce
\begin{eqnarray}
 f_{e}(\tilde{v}_{e\|},\tilde{v}_{e\bot}) & = & \Bigg\{1 + \left[A_e^{T}\!\left(\tilde{v}_e\right) \lambda_e \nabla_{\|} \log T_e +A_e^{R}\!\left(\tilde{v}_e\right) \lambda_e \frac{R_{e\|}}{p_e} + A_e^{u}\!\left(\tilde{v}_e\right) \lambda_e \frac{m_e u_{ei\|}}{T_e \tau_e} \right] \tilde{v}_{e\|} \nonumber \\
&& \qquad \qquad + \tau_e C_e\!\left(\tilde{v}_e\right) \left(\hat{\boldsymbol{z}} \hat{\boldsymbol{z}} - \frac{1}{3}\mathsfbi{I} \right) \! \boldsymbol{:} \! \mathsfbi{W}_e \left(\tilde{v}_{e\|}^2 - \frac{\tilde{v}_{e\bot}^2}{2} \right) \Bigg\} f_{e}^{(0)} \, .\label{CE_elecdist_ionframe}
\end{eqnarray}
Introducing the parameters $\eta_e^{T}$, $\eta_e^{R}$, $\eta_e^{u}$ and $\epsilon_e$ defined by equations (\ref{smallparam}\textit{a}), (\ref{smallparam}\textit{b}), (\ref{smallparam}\textit{c}) and (\ref{smallparam}\textit{e}) gives the final result 
(\ref{ChapEnskogFunc}\textit{a}).

\subsubsection{Ions}

The derivation of the equivalent result (\ref{ChapEnskogFunc}\textit{b}) for the ion distribution is mostly similar, but with 
one key difference: the total ion collision 
operator is dominated by the ion-ion collision operator $\mathfrak{C}_{ii}(f_i) \equiv  \mathfrak{C}(f_i,f_i)$:
\begin{equation}
 \sum_{s'} \mathfrak{C}(f_i,f_{s'}) = \mathfrak{C}_{ii}(f_i) + \mathfrak{C}(f_i,f_e) \approx \mathfrak{C}_{ii}(f_e) \, .  \label{ioncollop_tot}
\end{equation}  
This is because ion-electron collisions are small in the mass ratio compared to ion-electron 
collisions. After some algebra, it can be shown that the equivalent of
(\ref{cEexpFokkerPlanck}) for the perturbed ion distribution $f_i^{(1)}$ is
\begin{eqnarray}
\mathfrak{C}_{ii}(f_{i}^{(1)}) & - & \frac{Z_i e}{m_i c} \left(\boldsymbol{v}_{i}' \times \boldsymbol{B} \right) \bcdot \frac{\p f_{i}^{(1)}}{\p \boldsymbol{v}_{i}'} \nonumber\\
&=& \Bigg[\left(\frac{|\boldsymbol{v}_i'|^2}{v_{\mathrm{th}i}^2} -\frac{5}{2}\right)\boldsymbol{v}_{i}' \bcdot \nabla \log T_i + \frac{m_i}{2 T_i} \left(\boldsymbol{v}_{i}'  \boldsymbol{v}_{i}' - \frac{|\boldsymbol{v}_i'|^2}{3}\mathsfbi{I} \right) \! \boldsymbol{:} \! \mathsfbi{W}_i \Bigg] f_{i}^{(0)} \, , \label{cEexpFokkerPlanck_ion}
\end{eqnarray}
where the lowest-order distribution is Maxwellian:
\begin{equation}
f_i^{(0)}(\boldsymbol{v}) = \frac{n_{i}}{\upi^{3/2} v_{\mathrm{th}i}^3}  \exp \left(-\frac{|\boldsymbol{v}_i'|^2}{v_{\mathrm{th}i}^2}\right) 
\, .
\end{equation}
We emphasise that the main differences between (\ref{cEexpFokkerPlanck}) and (\ref{cEexpFokkerPlanck_ion}) 
are the presence of only one collision operator on the left-hand side of (\ref{cEexpFokkerPlanck_ion}) and the absence of any term 
proportional to the ion-electron friction force $\boldsymbol{R}_{ie}$ on the 
right-hand-side of (\ref{cEexpFokkerPlanck_ion}). 

Once (\ref{cEexpFokkerPlanck_ion}) 
has been written down, the method for obtaining the ion CE distribution function 
(\ref{ChapEnskogFunc}\textit{b}) in a strongly magnetised plasma is near-identical to 
that of the electron distribution function. Gyro-averaging gives
\begin{equation}
\mathfrak{C}_{ii}(f_{i}^{(1)}) = \Bigg[\left(\frac{|\boldsymbol{v}_i'|^2}{v_{\mathrm{th}i}^2} -\frac{5}{2}\right) v_{i\|}' \nabla_{\|} \log T_i + \left(\hat{\boldsymbol{z}} \hat{\boldsymbol{z}} - \frac{1}{3}\mathsfbi{I} \right) \! \boldsymbol{:} \! \mathsfbi{W}_i \left(\frac{v_{i\|}'^2}{v_{\mathrm{th}i}^2} - \frac{v_{i\bot}'^2}{2 v_{\mathrm{th}i}^2} \right) \Bigg] f_{i}^{(0)} \, , \label{cEexpFokkerPlanck_gyro_ion}
\end{equation} 
from which it follows that
\begin{eqnarray}
 f_{i}(v_{i\|}',v_{i\bot}') & = & \Bigg[1 + \tau_i A_i\!\left(\frac{|\boldsymbol{v}_i'|}{v_{\mathrm{th}i}}\right) v_{i\|}' \nabla_{\|} \log T_i \nonumber \\
 && \qquad + C_i\!\left(\frac{|\boldsymbol{v}_i'|}{v_{\mathrm{th}i}}\right) \left(\hat{\boldsymbol{z}} \hat{\boldsymbol{z}} - \frac{1}{3}\mathsfbi{I} \right) \! \boldsymbol{:} \! \mathsfbi{W}_i \left(\frac{v_{i\|}'^2}{v_{\mathrm{th}i}^2} - \frac{v_{i\bot}'^2}{2 v_{\mathrm{th}i}^2} \right) \Bigg] f_{i}^{(0)} \, .\label{CE_dist_ion_ionframe}
\end{eqnarray}
On substituting for parameters $\eta_i$ and $\epsilon_i$ defined by 
(\ref{smallparam}\textit{d}) and (\ref{smallparam}\textit{f}), respectively, we obtain 
(\ref{ChapEnskogFunc}\textit{b}).

\subsection{Deriving isotropic functions of velocity for the CE solution} \label{ChapEnskogIsoFunc}

In this appendix, we illustrate how to calculate the isotropic functions $A_e^{T}\!\left(\tilde{v}_e\right)$, 
$A_e^{R}\!\left(\tilde{v}_e\right)$, $A_e^{u}\!\left(\tilde{v}_e\right)$,
$A_i\!\left(\tilde{v}_i\right)$, $C_e\!\left(\tilde{v}_e\right)$
and $C_i\!\left(\tilde{v}_i\right)$
arising in the electron and ion CE distribution functions for the particular 
cases of two simplified collision operators: the Krook collision operator and the Lorentz 
collision operator. 

\subsubsection{Krook collision operator} \label{ChapEnskogIsoFunc_Krook}

The Krook collision operator~\citep{BGK54} for species $s$ is given by
\begin{equation}
 \mathfrak{C}_{K}(f_s) \equiv -\frac{1}{\tau_s} \left(f_s - f_{s}^{(0)}\right) 
 , \label{Krook}
\end{equation}
where $\tau_s$ is the collision time of species $s$ (assumed velocity-independent), and 
\begin{equation}
f_{s}^{(0)} = \frac{n_{s}}{\upi^{3/2} v_{\mathrm{th}s}^3}  \exp \left(-\frac{|\boldsymbol{v}_e'|^2}{v_{\mathrm{th}s}^2}\right) 
\end{equation}
is a Maxwellian distribution with density $n_s$, mean velocity $\boldsymbol{V}_e$ and temperature $T_s$ determined from $f_s$ via (\ref{dens_vel_temp_def}). For this 
choice of collision operator, i.e., assuming
\begin{equation}
  \sum_{s'} \mathfrak{C}(f_s,f_{s'})  =  \mathfrak{C}_{K}(f_s)
\end{equation}
for all particle species, calculating the CE distribution function is 
particularly simple. Substituting equation (\ref{CE_dist_electronframe}) for the 
electron CE distribution function into the electron Krook collision operator, we find
\begin{eqnarray}
 \mathfrak{C}_{K}(f_e) & = & - \Bigg\{ \bigg[A_e^{T}\!\left(\frac{|\boldsymbol{v}_e'|}{v_{\mathrm{th}e}}\right) \nabla_{\|} \log T_e +A_e^{R}\!\left(\frac{|\boldsymbol{v}_e'|}{v_{\mathrm{th}e}}\right) \frac{R_{e\|}}{p_e}+ \left(A_e^{u}\!\left(\frac{|\boldsymbol{v}_e'|}{v_{\mathrm{th}e}}\right) -1\right)\frac{m_e u_{ei\|}}{T_e \tau_{e}} \bigg] v_{e\|}' 
 \nonumber \\
&& \qquad + C_e\!\left(\frac{|\boldsymbol{v}_e'|}{v_{\mathrm{th}e}}\right) \left(\hat{\boldsymbol{z}} \hat{\boldsymbol{z}} - \frac{1}{3}\mathsfbi{I} \right) \! \boldsymbol{:} \! \mathsfbi{W}_e \left(\frac{v_{e\|}'^2}{v_{\mathrm{th}e}^2} - \frac{v_{e\bot}'^2}{2 v_{\mathrm{th}e}^2} \right) \Bigg\} f_{e}^{(0)} \, .\label{CEdist_elec_Krook}
\end{eqnarray}
By comparison to (\ref{cEexpFokkerPlanckgyro}), which, on substituting the Krook operator, becomes 
\begin{eqnarray}
 \mathfrak{C}_{K}(f_e^{(1)}) &=& \Bigg\{\left[\left(\frac{|\boldsymbol{v}_e'|^2}{v_{\mathrm{th}e}^2} -\frac{5}{2}\right) \nabla_{\|} \log T_e + \frac{R_{e\|}}{p_e} + \frac{m_e u_{ei\|}}{T_e \tau_{e}} \right] v_{e\|}' \nonumber \\
&& \qquad \qquad + \left(\hat{\boldsymbol{z}} \hat{\boldsymbol{z}} - \frac{1}{3}\mathsfbi{I} \right) \! \boldsymbol{:} \! \mathsfbi{W}_e \left(\frac{v_{e\|}'^2}{v_{\mathrm{th}e}^2} - \frac{v_{e\bot}'^2}{2 v_{\mathrm{th}e}^2} \right) \Bigg\} f_{e}^{(0)} \, ,\label{cEexpFokkerPlanckgyro_B}
\end{eqnarray}
we can immediately deduce that
\begin{subeqnarray}
  A_e^T(\tilde{v}_e) & = & -\left(\tilde{v}_e^2-\frac{5}{2}\right) \, , \\
  A_e^R(\tilde{v}_e) & = & -1 \, , \\
  A_e^u(\tilde{v}_e) & = & 0 \, , \\
  C_e(\tilde{v}_e) & = & -1 \, .
\end{subeqnarray}
The CE electron-ion-drift term vanishes for a Krook operator because the operator neglects inter-species collisions; by the same token, neither $T_i$ and $T_e$ nor $\boldsymbol{V}_i$ and $\boldsymbol{V}_e$ will equilibrate. 

For the ion CE distribution, it follows from (\ref{CE_dist_ion_ionframe}) substituted into (\ref{Krook}) that
\begin{eqnarray}
 \mathfrak{C}_{K}(f_i) = - \Bigg[ A_i\!\left(\frac{|\boldsymbol{v}_i'|}{v_{\mathrm{th}i}}\right) v_{i\|}' \nabla_{\|} \log T_i + C_i\!\left(\frac{|\boldsymbol{v}_i'|}{v_{\mathrm{th}i}}\right) \left(\hat{\boldsymbol{z}} \hat{\boldsymbol{z}} - \frac{1}{3}\mathsfbi{I} \right) \! \boldsymbol{:} \! \mathsfbi{W}_i \left(\frac{v_{i\|}'^2}{v_{\mathrm{th}i}^2} - \frac{v_{i\bot}'^2}{2 v_{\mathrm{th}i}^2} \right) \Bigg] f_{i}^{(0)} , \qquad \, \label{CEdist_ion_Krook}
\end{eqnarray}
which gives, on comparison with (\ref{cEexpFokkerPlanck_gyro_ion}),
that
\begin{subeqnarray}
  A_i(\tilde{v}_i) & = & -\left(\tilde{v}_i^2-\frac{5}{2}\right) \, , \\
  C_i(\tilde{v}_i) & = & -1 \, .
\end{subeqnarray}

\subsubsection{Lorentz collision operator} \label{ChapEnskogIsoFunc_Lorentz}

The Lorentz collision operator for species $s$ is defined by
\begin{equation}
 \mathfrak{C}_{L}(f_s) \equiv \nu_s{(v)} v^3 \frac{\p}{\p \boldsymbol{v}} \bcdot \left[\frac{1}{v}\left(\mathsfbi{I}- \hat{\boldsymbol{v}} \hat{\boldsymbol{v}} \right) \bcdot \frac{\p f_s}{\p \boldsymbol{v}} \right] 
 , \label{Lorentz_coll_op}  
\end{equation}
where $\nu_s(v)$ is a velocity-dependent scattering rate. We emphasise that the Lorentz collision operator
is still simplified and physically complete compared to the full Landau collision operator, as it merely isotropises the 
distribution function over long times. However, such an operator does arise as 
the largest component of the electron-ion collision operator [see (\ref{electroncollop_defs}\textit{b}) in appendix 
\ref{ChapEnskogDev_Append_comp}], and is, in fact, the exact electron collision operator in the limit of highly-charged 
ions: the so called `Lorentz approximation'~\citep{HS05}. 

To calculate the electron CE distribution function, we substitute (\ref{CE_dist_electronframe}) 
into the collision operator (\ref{Lorentz_coll_op}) (with $s = e$). Using the
identities
\begin{subeqnarray}
  \frac{\p}{\p \boldsymbol{v}} \bcdot \left[\frac{1}{v}\left(\mathsfbi{I}- \hat{\boldsymbol{v}} \hat{\boldsymbol{v}} \right) \bcdot \frac{\p}{\p \boldsymbol{v}} \left( \boldsymbol{a} \bcdot \boldsymbol{v}\right) \right] & = & - \frac{2 \boldsymbol{a} \bcdot \boldsymbol{v}}{v^3}  \, , \\
  \frac{\p}{\p \boldsymbol{v}} \bcdot \left[\frac{1}{v}\left(\mathsfbi{I}- \hat{\boldsymbol{v}} \hat{\boldsymbol{v}} \right) \bcdot \frac{\p}{\p \boldsymbol{v}} \left( \boldsymbol{v} \bcdot \mathsfbi{A} \bcdot \boldsymbol{v} \right)\right] & = & -\frac{6 \boldsymbol{v} \bcdot \mathsfbi{A} \bcdot \boldsymbol{v}}{v^3} 
\end{subeqnarray}
for any constant vector $\boldsymbol{a}$ and any symmetric, traceless, constant 
matrix $\mathsfbi{A}$, it follows that
\begin{eqnarray}
  \mathfrak{C}_{L}(f_e) & = & - \hat{\nu}_{e}(\tilde{v}_e) \Bigg\{ \bigg[2 A_e^{T}\!\left(\frac{|\boldsymbol{v}_e'|}{v_{\mathrm{th}e}}\right) \nabla_{\|} \log T_e +2 A_e^{R}\!\left(\frac{|\boldsymbol{v}_e'|}{v_{\mathrm{th}e}}\right) \frac{R_{e\|}}{p_e} \nonumber \\
  && \qquad + 2 \left(A_e^{u}\!\left(\frac{|\boldsymbol{v}_e'|}{v_{\mathrm{th}e}}\right) -1\right)\frac{m_e u_{ei\|}}{T_e \tau_{e}} \bigg] v_{e\|}' 
  \nonumber \\
&& \qquad \qquad + 6 C_e\!\left(\frac{|\boldsymbol{v}_e'|}{v_{\mathrm{th}e}}\right) \left(\hat{\boldsymbol{z}} \hat{\boldsymbol{z}} - \frac{1}{3}\mathsfbi{I} \right) \! \boldsymbol{:} \! \mathsfbi{W}_e \left(\frac{v_{e\|}'^2}{v_{\mathrm{th}e}^2} - \frac{v_{e\bot}'^2}{2 v_{\mathrm{th}e}^2} \right) \Bigg\} f_{e}^{(0)} \, ,\label{CEdist_elec_Lorentz}
\end{eqnarray}
where $\hat{\nu}_{s} \equiv \nu_{s}(\tilde{v}_s) \tau_{s}$ is the non-dimensionalised 
collision rate for species $s$. As with the Krook operator, we compare (\ref{CEdist_elec_Lorentz}) to 
(\ref{cEexpFokkerPlanckgyro}), substituting a Lorentz collision operator for the latter, viz., 
\begin{eqnarray}
 \mathfrak{C}_{L}(f_e^{(1)}) &=& \Bigg\{\left[\left(\frac{|\boldsymbol{v}_e'|^2}{v_{\mathrm{th}e}^2} -\frac{5}{2}\right) \nabla_{\|} \log T_e + \frac{R_{e\|}}{p_e} + \frac{m_e u_{ei\|}\nu_{e}(\tilde{v}_e)}{T_e} \right] v_{e\|}' \nonumber \\
&& \qquad \qquad + \left(\hat{\boldsymbol{z}} \hat{\boldsymbol{z}} - \frac{1}{3}\mathsfbi{I} \right) \! \boldsymbol{:} \! \mathsfbi{W}_e \left(\frac{v_{e\|}'^2}{v_{\mathrm{th}e}^2} - \frac{v_{e\bot}'^2}{2 v_{\mathrm{th}e}^2} \right) \Bigg\} f_{e}^{(0)} \, .\label{cEexpFokkerPlanckgyro_C}
\end{eqnarray}
We deduce from the comparison that
\begin{subeqnarray}
  A_e^T(\tilde{v}_e) & = & -\frac{1}{2 \hat{\nu}_{e}(\tilde{v}_e)} \left(\tilde{v}_e^2-\frac{5}{2}\right) \, , \\
  A_e^R(\tilde{v}_e) & = & -\frac{1}{2 \hat{\nu}_{e}(\tilde{v}_e)} \, , \\
  A_e^u(\tilde{v}_e) & = & \frac{1}{2} \, , \\
  C_e(\tilde{v}_e) & = & -\frac{1}{6 \hat{\nu}_{e}(\tilde{v}_e)} \, .
\end{subeqnarray}
The isotropic functions $A_i(\tilde{v}_i)$ and $C_i(\tilde{v}_i)$, which are given by 
\begin{subeqnarray}
  A_i(\tilde{v}_i) & = & -\frac{1}{2 {\nu}_{i}(\tilde{v}_i) \tau_i} \left(\tilde{v}_i^2-\frac{5}{2}\right) \, , \\
  C_i(\tilde{v}_i) & = & -\frac{1}{6 {\nu}_{i}(\tilde{v}_i) \tau_i} \, .
\end{subeqnarray}
 can be deduced 
in an analogous manner. 

\section{Derivation of hot, magnetised plasma dispersion relation for arbitrary distribution functions} \label{HotPlasmDispDeriva}

In this appendix we re-derive the hot-plasma 
dispersion relation, given by (\ref{hotplasmadisprel}) in section \ref{HotPlasmaDispDis}~\citep[see also][the latter of whose approaches we follow]{D83,P17}. Our derivation also introduces a (simplified) 
collision operator in order to show that substitution (\ref{collresconduct}) stated in section \ref{shortcomings_coll} 
provides a simple technique for including the effect of collisions on linear 
electromagnetic perturbations. 

Consider a kinetic, magnetised plasma in equilibrium composed of one electron species and multiple ions 
species, with (assumed constant) background magnetic field $\boldsymbol{B}_0$. As in section \ref{HotPlasmaDispDis}, we denote the (gyrotropic) equilibrium distribution function of species $s$ as $f_{s0} = f_{s0}(v_{\|},v_{\bot})$.  
and then consider a collisionless, linear perturbation $\delta f_s$ to this equilibrium state, with wavevector $\boldsymbol{k}$ and complex frequency~$\omega$:
\begin{equation}
\delta f_s = \widehat{\delta f}_s \exp\left\{\mathrm{i}\left(\mathbf{k} \bcdot \mathbf{r} - \omega t\right)\right\} \, . \label{Fourierdf}
\end{equation}
The electromagnetic perturbations associated with the perturbed distribution functions 
have the forms given in (\ref{FourierEBfields}), viz., 
\begin{subeqnarray}
   \delta \boldsymbol{E} & = &
   \widehat{\delta \boldsymbol{E}} \exp\left\{\mathrm{i}\left(\mathbf{k} \bcdot \mathbf{r} - \omega t\right)\right\} \label{FourierE} ,\\[3pt]
  \delta \boldsymbol{B} & = &
   \widehat{\delta \boldsymbol{B}} \exp\left\{\mathrm{i}\left(\mathbf{k} \bcdot \mathbf{r} - \omega t\right)\right\} \label{FourierB}  ,
\end{subeqnarray}
and satisfy Faraday's law and the Maxwell-Am\`pere's law:
\begin{subeqnarray}
   \frac{\p \delta \boldsymbol{B}}{\p t} & = &
   - c\bnabla \times \delta \boldsymbol{E} \label{Faraday},\\[3pt]
 \bnabla \times \delta \boldsymbol{B} & = &
   \frac{4 \upi}{c} \delta \boldsymbol{j} + \frac{1}{c} \frac{\p \delta \boldsymbol{E}}{\p t} \label{Ampere} ,
\end{subeqnarray}
where the current perturbation is
\begin{equation}
\delta \boldsymbol{j} = \widehat{\delta \boldsymbol{j}} \exp\left\{\mathrm{i}\left(\mathbf{k} \bcdot \mathbf{r} - \omega t\right)\right\}  = \sum_{s} Z_s e \int \mathrm{d}^3 \boldsymbol{v} \, \boldsymbol{v} \, \delta f_s \, . \label{FourierJ}
\end{equation}  
To close these equations, we relate $\delta f_s$ to the electromagnetic field perturbations by linearising the Maxwell-Vlasov-Landau equation 
(\ref{MaxVlasLan}).
The linearisation $f_s= f_{s0} + \delta f_s$ then gives that the perturbed distribution function of species $s$ satisfies 
\begin{equation}
\frac{\p \delta{f}_{s}}{\p t} + \boldsymbol{v} \bcdot \nabla \delta f_s + \frac{Z_s e}{m_s c} \left(\boldsymbol{v} \times \boldsymbol{B}_0 \right) \bcdot \frac{\p \delta {f}_{s}}{\p \boldsymbol{v}} = -\frac{Z_s e}{m_s} \left(\delta \boldsymbol{E} + \frac{\boldsymbol{v} \times \delta \boldsymbol{B}}{c}\right) \bcdot \frac{\p f_{s0}}{\p \boldsymbol{v}} -\nu_s \delta f_s \, , \label{Vlasov}
\end{equation}
where we have replaced the full linearised collision operator with a
simplified Krook collision operator with constant collision frequency $\nu_{s} = \tau_s^{-1}$ for species $s$. 
For any particular equilibrium distribution function, (\ref{Faraday}\textit{a}),  (\ref{Ampere}\textit{b}),  (\ref{FourierJ}) and (\ref{Vlasov}) 
are a closed set of governing equations.

We now write these equations in terms of $\boldsymbol{k}$ and $\omega$ using (\ref{Fourierdf}),  (\ref{FourierE}\textit{a}),  
and (\ref{FourierB}\textit{b}):
\begin{subeqnarray}
   -\mathrm{i} \omega \widehat{\delta \boldsymbol{B}} & = &
    -\mathrm{i} c \boldsymbol{k} \times \widehat{\delta \boldsymbol{E}} \label{FaradayB} ,\\[3pt]
   \mathrm{i} \boldsymbol{k} \times \widehat{\delta \boldsymbol{B}}  & = &
   \frac{4 \upi}{c} \widehat{\delta \boldsymbol{j}} - \frac{\mathrm{i} \omega}{c} \widehat{\delta \boldsymbol{E}} \label{AmpereB} ,\\[3pt]
   \widehat{\delta \boldsymbol{j}} & = &
   \sum_s Z_s e \int \mathrm{d}^3 \boldsymbol{v} \, \boldsymbol{v} \, \widehat{\delta f_s} \label{currentFour},\\[3pt]
   \left(-\mathrm{i} \hat{\omega}_s + \mathrm{i} \boldsymbol{k} \bcdot \boldsymbol{v} + \tilde{\Omega}_s \frac{\p}{\p \phi}\right) \widehat{\delta f_s} & = &
   -\frac{Z_s e}{m_s} \left(\widehat{\delta \boldsymbol{E}} + \frac{\boldsymbol{v} \times \widehat{\delta \boldsymbol{B}}}{c}\right) \bcdot \frac{\p f_{s0}}{\p \boldsymbol{v}} \, \label{VlasovB} ,
\end{subeqnarray}
where we have defined the (signed) Larmor 
frequency of species $s$ as
\begin{equation}
\tilde{\Omega}_s \equiv \frac{Z_{s} e B_0}{m_s c} = \frac{Z_s}{|Z_s|} \Omega_s ,
\end{equation}
and introduced the modified complex frequency $\hat{\omega}_s \equiv \omega + \mathrm{i} 
\nu_s$. Note that $Z_e = -1$, so that $\tilde{\Omega}_e < 0$. We then eliminate $\widehat{\delta \boldsymbol{B}}$  in (\ref{AmpereB}\textit{b}) and (\ref{VlasovB}\textit{d}) using (\ref{FaradayB}\textit{a}) to give
\begin{subeqnarray}
    \frac{k^2 c^2}{\omega^2} \left[\widehat{\delta \boldsymbol{E}} - \hat{\boldsymbol{k}} \left(\hat{\boldsymbol{k}} \cdot \widehat{\delta \boldsymbol{E}}\right)\right]  & = &
   \frac{4 \upi \mathrm{i}}{\omega}  \widehat{\delta \boldsymbol{j}} - \widehat{\delta \boldsymbol{E}} \label{AmpereC} ,\\[3pt]
       \widehat{\delta \boldsymbol{j}} & = &
   \sum_s Z_s e \int \mathrm{d}^3 \boldsymbol{v} \, \boldsymbol{v} \, \widehat{\delta f_s} \label{currentFourB},\\[3pt]
   \left(-\mathrm{i} \hat{\omega}_s + \mathrm{i} \boldsymbol{k} \bcdot \boldsymbol{v} + \tilde{\Omega}_s \frac{\p}{\p \phi}\right) \widehat{\delta f_s} & = &
   -\frac{Z_s e}{m_s} \left[\widehat{\delta \boldsymbol{E}} + \frac{k}{\omega} \boldsymbol{v} \times \left(\hat{\boldsymbol{k}} \times \widehat{\delta \boldsymbol{E}}\right)\right] \bcdot \frac{\p f_{s0}}{\p \boldsymbol{v}} \, \label{VlasovC}  
   .
\end{subeqnarray}

Next, we derive an expression for $\widehat{\delta f}_s$ in terms of $\widehat{\delta 
\boldsymbol{E}}$. For arbitrary wavelengths compared to the Larmor radius $\rho_s$ of species $s$, expressing  $\widehat{\delta f}_s$ in terms of the equilibrium distribution function and $\widehat{\delta \boldsymbol{E}}$ requires inversion of the gyrophase-angle derivative in (\ref{VlasovC}). This can be done for any $f_{s0}$  
in an orthonormal coordinate system with basis vectors $\left\{\hat{\boldsymbol{x}},\hat{\boldsymbol{y}},\hat{\boldsymbol{z}}\right\}$ defined by equations (\ref{basisdef}). By Fourier transforming $\widehat{\delta f}_s$ in $\phi$, it can then be shown that
\begin{equation}
\widehat{\delta f_s} = -\frac{Z_s e \mathrm{i}}{m_s \omega} \left(\frac{\p f_{s0}}{\p v_{\|}}-\frac{v_{\|}}{v_{\bot}} \frac{\p f_{s0}}{\p v_{\bot}}\right) \hat{\boldsymbol{z}} \bcdot \widehat{\delta \boldsymbol{E}} + \exp \left(-\mathrm{i} k_{\bot} \tilde{\rho}_s \tilde{v}_{s\bot} \sin \phi \right) \sum_{n = -\infty}^{\infty} \widehat{\delta f}_{s,n} \exp \left(\mathrm{i} m \phi\right)\, 
,
\end{equation}
where the series coefficients are given by 
\begin{equation}
\widehat{\delta f}_{s,n} = -\frac{Z_s e \mathrm{i}}{m_s} \frac{1}{\hat{\omega}_s - k_{\|} v_{\|} - n \tilde{\Omega}_s} \left[\frac{\p f_{s0}}{\p v_{\bot}}+\frac{k_{\|}}{\omega}\left(v_{\bot} \frac{\p f_{s0}}{\p v_{\|}}-v_{\|} \frac{\p f_{s0}}{\p v_{\bot}}\right)\right] \boldsymbol{u}_n^{*} \bcdot \widehat{\delta \boldsymbol{E}} \, ,
\end{equation}
and the vector $\boldsymbol{u}_n$ in the basis $\left\{\hat{\boldsymbol{x}},\hat{\boldsymbol{y}},\hat{\boldsymbol{z}}\right\}$ is
\begin{equation}
\boldsymbol{u}_n = \frac{v_{\|}}{v_{\bot}}  \, J_n(k_{\bot} \tilde{\rho}_s \tilde{v}_{s\bot}) \hat{\boldsymbol{z}} + \frac{n J_n(k_{\bot} \tilde{\rho}_s \tilde{v}_{s\bot})}{k_{\bot} \tilde{\rho}_s \tilde{v}_{s\bot}} \hat{\boldsymbol{x}} - \mathrm{i} J_n'( k_{\bot} \tilde{\rho}_s \tilde{v}_{s\bot}) \hat{\boldsymbol{y}} \, ,
\end{equation}
$J_n( k_{\bot} \tilde{\rho}_s \tilde{v}_{s\bot})$ denoting the $n$-th order Bessel function of the first kind. We can then take advantage of the independence of $f_{s0}$ of the gyroangle to show that the current perturbation is
\begin{eqnarray}
\widehat{\delta \boldsymbol{j}} & = & - \sum_s \frac{2 \upi Z_s^2 e^2 \mathrm{i}}{m_s \omega} \int_{-\infty}^{\infty} \mathrm{d} v_{\|} \int_0^{\infty} \mathrm{d} v_{\bot}  \left(v_\bot \frac{\p f_{s0}}{\p v_{\|}}-v_{\|} \frac{\p f_{s0}}{\p v_{\bot}}\right) v_{\|} \hat{\boldsymbol{z}} \left(\hat{\boldsymbol{z}} \bcdot \widehat{\delta \boldsymbol{E}} \right) \nonumber \\
&& \mbox{} + \sum_s 2 \upi Z_s e \int_{C_L} \mathrm{d} v_{\|} \int_0^{\infty} \mathrm{d} v_{\bot} v_\bot^2 \sum_{n = -\infty}^{\infty} \widehat{\delta f}_{s,n} \boldsymbol{u}_n \, , \label{fullmagcurrent}
\end{eqnarray}
where $C_L$ denotes the usual Landau contour. This can be written as Ohm's law:
\begin{eqnarray}
  \widehat{\delta \boldsymbol{j}}  = \boldsymbol{\sigma} \bcdot \widehat{\delta 
  \boldsymbol{E}} , \label{Ohm_law}
\end{eqnarray}
where $\boldsymbol{\sigma}$ is the conductivity tensor. In the absence of 
collisions ($\nu_s = 0$), this is given by~(\ref{conductivity}). If the collision frequency $\nu_s \neq 0$ is non-zero, then 
\begin{equation}
  \frac{\hat{\omega}_s}{|k_{\|}| v_{\mathrm{th}s}} = \tilde{\omega}_{\|s} + \frac{\mathrm{i}}{|k_{\|}| \tau_s v_{\mathrm{th}s}} 
  =  \tilde{\omega}_{\|s} + \frac{\mathrm{i}}{|k_{\|}| \lambda_s} ,
\end{equation}
from which the substitution (\ref{collresconduct}) proposed in section \ref{shortcomings_coll}
follows. 

Substituting Ohm's law (\ref{Ohm_law}) into Amp\`ere's law (\ref{AmpereC}\textit{a}) gives the 
singular nonlinear eigenvalue equation
\begin{equation}
   \left[\frac{c^2 k^2}{\omega^2} \left(\hat{\boldsymbol{k}}\hat{\boldsymbol{k}}-\mathsfbi{I}\right)+ \boldsymbol{\mathfrak{E}} \right] \bcdot \widehat{\delta \boldsymbol{E}} =0, 
   \label{hotplasmadisprel_vecform_Append} 
\end{equation}
where
\begin{equation}
  \boldsymbol{\mathfrak{E}} \equiv \mathsfbi{I} + \frac{4 \upi \mathrm{i}}{\omega} 
  \boldsymbol{\sigma}
\end{equation}
is the plasma dielectric tensor (\ref{dielecttensfull}). 
Taking the determinant of (\ref{hotplasmadisprel_vecform_Append}) gives the 
desired result (\ref{hotplasmadisprel}).

\section{Electrostatic instabilities of CE plasma} \label{Electrostatic}

 In this appendix, we calculate the electrostatic hot-plasma dispersion relation
 for arbitrary distribution functions (appendix \ref{Electrostatic_derive}). We 
 then show (appendix \ref{lowfrequency_electrostatic}) that for frequencies $\omega$ such that 
 $\tilde{\omega}_{s\|} = \omega/k_{\|} v_{\mathrm{th}s} \ll 1$, the dominant contribution to the longitudinal conductivity
 $\hat{\boldsymbol{k}} \bcdot \boldsymbol{\sigma}  \bcdot \hat{\boldsymbol{k}} 
$ is from the Maxwellian component, and strictly positive; 
the small $\textit{O}(\eta_s,\epsilon_s)$ non-Maxwellian distortion associated with the
CE distribution function results in only an $\textit{O}(\eta_s,\epsilon_s)$ distortion to $\hat{\boldsymbol{k}} \bcdot \boldsymbol{\sigma}  \bcdot \hat{\boldsymbol{k}} 
$. We then illustrate the possibility of electrostatic instabilities associated 
with the CE distribution function by calculating the growth rate of the parallel CE 
bump-on-tail instability (appendix \ref{existence_electrostatic}). Finally, 
in appendix \ref{impossibility_electrostatic_slowgrowth}, we show that the only electrostatic instabilities that can 
occur have a growth rate which is exponentially small in dimensionless 
parameters $\textit{O}(\eta_s,\epsilon_s)$, for arbitrary frequencies. Thus, it 
follows that electrostatic instabilities generally have a small growth rate in 
comparison to electromagnetic instabilities for a CE plasma.  

\subsection{The electrostatic hot-plasma dispersion relation} \label{Electrostatic_derive}

Beginning from the singular eigenvalue equation (\ref{singulareigenvaleqnFull}), viz., 
\begin{equation}
   \left[\frac{c^2 k^2}{\omega^2} \left(\hat{\boldsymbol{k}}\hat{\boldsymbol{k}}-\mathsfbi{I}\right)+ \boldsymbol{\mathfrak{E}} \right] \bcdot \widehat{\delta \boldsymbol{E}} =0, 
   \label{hotplasmadisprel_vecform_Append_B} 
\end{equation}
we consider the electrostatic modes, for which $\widehat{\delta \boldsymbol{E}} = (\hat{\boldsymbol{k}} \bcdot \widehat{\delta \boldsymbol{E}}) \hat{\boldsymbol{k}} 
$. For them, the hot-plasma dispersion relation becomes 
\begin{equation}
\mathfrak{E}_{33} = k^2 + \frac{4 \upi \mathrm{i}}{\omega} \hat{\boldsymbol{k}} \bcdot \boldsymbol{\sigma}  \bcdot \hat{\boldsymbol{k}} 
= 0\, . \label{electrostatic_hotplasma}
\end{equation}
Employing the expression (\ref{conductivity}) for the conductivity tensor, we 
calculate the longitudinal conductivity:
\begin{eqnarray}
\hat{\boldsymbol{k}} \bcdot \boldsymbol{\sigma} \bcdot \hat{\boldsymbol{k}} & = & - \frac{\mathrm{i}}{4 \upi \omega} \sum_s \omega_{\mathrm{p}s}^2 \bigg[ \frac{2}{\sqrt{\upi}} \frac{k_{\|}^2}{k^2}  \int_{-\infty}^{\infty} \mathrm{d} \tilde{v}_{s\|} \, \tilde{v}_{s\|} \int_0^{\infty} \mathrm{d} \tilde{v}_{s\bot} \Lambda_s(\tilde{v}_{s\|},\tilde{v}_{s\bot}) \nonumber \\
&& \mbox{} + \tilde{\omega}_{s\|} \frac{2}{\sqrt{\upi}} \int_{C_L} \mathrm{d} \tilde{v}_{s\|} \int_0^{\infty} \mathrm{d} \tilde{v}_{s\bot} \tilde{v}_{s\bot}^2 \Xi_s(\tilde{v}_{s\|},\tilde{v}_{s\bot}) \sum_{n = -\infty}^{\infty} \frac{\hat{\boldsymbol{k}}\bcdot \mathsfbi{R}_{sn} \bcdot \hat{\boldsymbol{k}}}{\zeta_{sn} -\tilde{v}_{s\|}}  \bigg] \, , \label{conductivity_33}
\end{eqnarray}
where 
\begin{eqnarray}
\hat{\boldsymbol{k}}\bcdot \mathsfbi{R}_{sn} \bcdot \hat{\boldsymbol{k}} & = & \frac{J_n(k_{\bot} \tilde{\rho}_s \tilde{v}_{s\bot})^2}{k^2 \tilde{\rho}_s^2 \tilde{v}_{s\bot}^2}  \left(n^2 + 2 n k_{\|} \tilde{\rho}_s \tilde{v}_{s\|} + k_{\|}^2 \tilde{\rho}_s^2 \tilde{v}_{s\|}^2 \right) 
\nonumber \\
& = & \frac{k_{\|}^2 J_n(k_{\bot} \tilde{\rho}_s \tilde{v}_{s\bot})^2}{k^2 \tilde{v}_{s\bot}^2} 
\left(\frac{n}{k_{\|} \tilde{\rho}_s} + \tilde{v}_{s\|}\right)^2 \, .
\end{eqnarray}
By way of the identity
\begin{equation}
\sum_{n = -\infty}^{\infty} J_n(k_{\bot} \tilde{\rho}_s \tilde{v}_{s\bot})^2 \frac{\left(n/{k_{\|} \tilde{\rho}_s} + \tilde{v}_{s\|}\right)^2}{\zeta_{sn} -\tilde{v}_{s\|}}  
= -\tilde{v}_{s\|} + \tilde{\omega}_{s\|} \sum_{n = -\infty}^{\infty} J_n(k_{\bot} \tilde{\rho}_s \tilde{v}_{s\bot})^2 \frac{n/{k_{\|} \tilde{\rho}_s} + \tilde{v}_{s\|}}{\zeta_{sn} -\tilde{v}_{s\|}}
\, ,
\end{equation}
which follows directly from the Bessel function identity
\begin{equation}
  \sum_{n=-\infty}^{\infty} J_n(k_{\bot} \tilde{\rho}_s \tilde{v}_{s\bot})^2 = 1
  ,
\end{equation}
it follows that
\begin{eqnarray}
 & \tilde{\omega}_{s\|} & \frac{2}{\sqrt{\upi}} \int_{C_L} \mathrm{d} \tilde{v}_{s\|} \int_0^{\infty} \mathrm{d} \tilde{v}_{s\bot} \tilde{v}_{s\bot}^2 \Xi_s(\tilde{v}_{s\|},\tilde{v}_{s\bot}) \sum_{n = -\infty}^{\infty} \frac{\hat{\boldsymbol{k}}\bcdot \mathsfbi{R}_{sn} \bcdot \hat{\boldsymbol{k}}}{\zeta_{sn} -\tilde{v}_{s\|}} 
 \nonumber \\
 && = - \tilde{\omega}_{s\|} \frac{2}{\sqrt{\upi}} \frac{k_{\|}^2}{k^2} \int_{C_L} \mathrm{d} \tilde{v}_{s\|} \, \tilde{v}_{s\|} \int_0^{\infty} \mathrm{d} \tilde{v}_{s\bot} \left[\frac{\p \tilde{f}_{s0}}{\p \tilde{v}_{s\bot}}
  + \frac{\Lambda_s(\tilde{v}_{s\|},\tilde{v}_{s\bot})}{\tilde{\omega}_{s\|}} \right] 
  \nonumber \\
  && \quad + \tilde{\omega}_{s\|} \frac{2}{\sqrt{\upi}} \frac{k_{\|}^2}{k^2} \int_{C_L} \mathrm{d} \tilde{v}_{s\|} \int_0^{\infty} \mathrm{d} \tilde{v}_{s\bot} \Lambda_s(\tilde{v}_{s\|},\tilde{v}_{s\bot})
 \sum_{n = -\infty}^{\infty} J_n(k_{\bot} \tilde{\rho}_s \tilde{v}_{s\bot})^2 \frac{n/{k_{\|} \tilde{\rho}_s} + \tilde{v}_{s\|}}{\zeta_{sn} -\tilde{v}_{s\|}} 
  \nonumber \\
  && \quad + \tilde{\omega}_{s\|}^2 \frac{2}{\sqrt{\upi}} \frac{k_{\|}^2}{k^2} \int_{C_L} \mathrm{d} \tilde{v}_{s\|} \int_0^{\infty} \mathrm{d} \tilde{v}_{s\bot} \frac{\p \tilde{f}_{s0}}{\p \tilde{v}_{s\bot}}
\sum_{n = -\infty}^{\infty} J_n(k_{\bot} \tilde{\rho}_s \tilde{v}_{s\bot})^2 \frac{n/{k_{\|} \tilde{\rho}_s} + \tilde{v}_{s\|}}{\zeta_{sn} -\tilde{v}_{s\|}} 
\nonumber \\
&& = -\frac{2}{\sqrt{\upi}} \frac{k_{\|}^2}{k^2}  \int_{-\infty}^{\infty} \mathrm{d} \tilde{v}_{s\|} \, \tilde{v}_{s\|} \int_0^{\infty} \mathrm{d} \tilde{v}_{s\bot} \Lambda_s(\tilde{v}_{s\|},\tilde{v}_{s\bot}) 
\nonumber \\
&& \quad + \frac{2 \tilde{\omega}_{s}^2}{\sqrt{\upi}} \int_{C_L} \mathrm{d} \tilde{v}_{s\|} \int_0^{\infty} \mathrm{d} \tilde{v}_{s\bot}
\sum_{n = -\infty}^{\infty} \left(\tilde{v}_{s\bot} \frac{\p \tilde{f}_{s0}}{\p \tilde{v}_{s\|}} + \frac{n}{k_{\|} \tilde{\rho}_s} \frac{\p \tilde{f}_{s0}}{\p \tilde{v}_{s\bot}}\right)
 \frac{J_n(k_{\bot} \tilde{\rho}_s \tilde{v}_{s\bot})^2}{\zeta_{sn} -\tilde{v}_{s\|}} \, 
 ,
\end{eqnarray}
where $\tilde{\omega}_{s} \equiv \omega/k v_{\mathrm{th}s}$. We conclude that
\begin{equation}
\hat{\boldsymbol{k}} \bcdot \boldsymbol{\sigma} \bcdot \hat{\boldsymbol{k}} = - \frac{\mathrm{i}}{4 \upi \omega} \sum_s \omega_{\mathrm{p}s}^2 \left[ \tilde{\omega}_{s}^2 \frac{2}{\sqrt{\upi}} \int_{C_L} \mathrm{d} \tilde{v}_{s\|} \int_0^{\infty} \mathrm{d} \tilde{v}_{s\bot}
\sum_{n = -\infty}^{\infty} \Pi_n(\tilde{v}_{s\|},\tilde{v}_{s\bot})
 \frac{J_n(k_{\bot} \tilde{\rho}_s \tilde{v}_{s\bot})^2}{\zeta_{sn} -\tilde{v}_{s\|}} \right] \, , \label{conductivity_33_B}
\end{equation}
where
\begin{equation}
\Pi_n(\tilde{v}_{s\|},\tilde{v}_{s\bot}) \equiv  \tilde{v}_{s\bot} \frac{\p \tilde{f}_{s0}}{\p \tilde{v}_{s\|}} + \frac{n}{k_{\|} \tilde{\rho}_s} \frac{\p \tilde{f}_{s0}}{\p \tilde{v}_{s\bot}} 
\, .
\end{equation}
The electrostatic component of the dielectric tensor is then
\begin{equation}
 \mathfrak{E}_{33} = k^2 + \sum_s k_{\mathrm{D}s}^2 \left[ \frac{1}{\sqrt{\upi}} \int_{C_L} \mathrm{d} \tilde{v}_{s\|} \int_0^{\infty} \mathrm{d} \tilde{v}_{s\bot}
\sum_{n = -\infty}^{\infty} \Pi_n(\tilde{v}_{s\|},\tilde{v}_{s\bot})
 \frac{J_n(k_{\bot} \tilde{\rho}_s \tilde{v}_{s\bot})^2}{\zeta_{sn} -\tilde{v}_{s\|}} \right] ,
 \label{electrostatic_dielectric_comp}
\end{equation}
and the electrostatic hot-plasma dispersion relation (\ref{electrostatic_hotplasma}) becomes
\begin{equation}
k^2 + \sum_s k_{\mathrm{D}s}^2 \left[ \frac{1}{\sqrt{\upi}} \int_{C_L} \mathrm{d} \tilde{v}_{s\|} \int_0^{\infty} \mathrm{d} \tilde{v}_{s\bot}
\sum_{n = -\infty}^{\infty} \Pi_n(\tilde{v}_{s\|},\tilde{v}_{s\bot})
 \frac{J_n(k_{\bot} \tilde{\rho}_s \tilde{v}_{s\bot})^2}{\zeta_{sn} -\tilde{v}_{s\|}} \right] = 0 , 
 \label{electrostatic_disprel}
\end{equation}
where the Debye wavenumber $k_{\mathrm{D}s}$ of species $s$ is defined by
\begin{equation}
  k_{\mathrm{D}s} \equiv \frac{\sqrt{2} \omega_{\mathrm{p}s}}{v_{\mathrm{th}s}} \, . \label{Debyewav}
\end{equation}

\subsection{The electrostatic dielectric response at low frequencies} \label{lowfrequency_electrostatic}

In this appendix, we perform a Taylor expansion of the electrostatic component $ \mathfrak{E}_{33}$ 
of the dielectric tensor (\ref{electrostatic_dielectric_comp}) in $\tilde{\omega}_{s\|} \ll 1$. Before carrying out the expansion, we first substitute the identity
\begin{equation}
\Pi_n(\tilde{v}_{s\|},\tilde{v}_{s\bot}) = \tilde{\omega}_{s\|} \Xi_s(\tilde{v}_{s\|},\tilde{v}_{s\bot}) 
+ \left(\tilde{v}_{s\|}-\zeta_{sn}\right) \frac{\p \tilde{f}_{s0}}{\p \tilde{v}_{s\bot}}
\end{equation}
into (\ref{electrostatic_dielectric_comp}), which then becomes
\begin{eqnarray}
\mathfrak{E}_{33} & = & k^2 - \sum_s k_{\mathrm{D}s}^2 \frac{1}{\sqrt{\upi}} \int_{-\infty}^{\infty} \mathrm{d} \tilde{v}_{s\|} \int_0^{\infty} \mathrm{d} \tilde{v}_{s\bot} \frac{\p \tilde{f}_{s0}}{\p \tilde{v}_{s\bot}} \nonumber \\ 
&& + \sum_s k_{\mathrm{D}s}^2 \left[ \frac{\tilde{\omega}_{s\|}}{\sqrt{\upi}} \int_{C_L} \mathrm{d} \tilde{v}_{s\|} \int_0^{\infty} \mathrm{d} \tilde{v}_{s\bot}
\Xi_s(\tilde{v}_{s\|},\tilde{v}_{s\bot}) \sum_{n = -\infty}^{\infty} 
 \frac{J_n(k_{\bot} \tilde{\rho}_s \tilde{v}_{s\bot})^2}{\zeta_{sn} -\tilde{v}_{s\|}} \right] .
 \qquad 
 \label{electrostatic_disprel_alt}
\end{eqnarray}
Now carrying out the Taylor expansion in $\tilde{\omega}_{s\|}  \ll 1$, we see that, to the leading order in this expansion,
\begin{equation}
\mathfrak{E}_{33} \approx k^2 + \sum_s k_{\mathrm{D}s}^2 \frac{1}{\sqrt{\upi}} \int_{-\infty}^{\infty} \mathrm{d} \tilde{v}_{s\|} \tilde{f}_{s0}(\tilde{v}_{s\|},0)
\, . \label{lowfrequency_electrostaticcomponent}
\end{equation} 

For the CE distribution
\begin{equation}
\tilde{f}_{s0}(\tilde{v}_{s\|},0) = \exp \left(-\tilde{v}_{s\|}^2\right) \left\{1+\eta_s A_s(\tilde{v}_{s\|}) \tilde{v}_{s\|} + \epsilon_s C_s(\tilde{v}_{s\|}) \tilde{v}_{s\|}^2\right\} ,  
\end{equation}
we have
\begin{equation}
 \frac{1}{\sqrt{\upi}} \int_{-\infty}^{\infty} \mathrm{d} \tilde{v}_{s\|} \tilde{f}_{s0}(\tilde{v}_{s\|},0) = 1 + \frac{\epsilon_s}{2 \sqrt{\upi}} \int_{0}^{\infty} \mathrm{d} \tilde{v}_{s\|} 
 \tilde{v}_{s\|}^2 C_s(\tilde{v}_{s\|}) \exp \left(-\tilde{v}_{s\|}^2\right) , \label{lowfrequency_electrostaticcomponent_func}
\end{equation}
where the term in the CE distribution function proportional to $\eta_s$ has
vanished on account of having odd parity with respect to $\tilde{v}_{s\|}$.
We conclude that the non-Maxwellian contribution to (\ref{lowfrequency_electrostaticcomponent_func})
is $\textit{O}(\eta_s,\epsilon_s)$ in comparison to the Maxwellian contribution, and so 
the electrostatic component of the dielectric tensor for low-frequency fluctuations is just
\begin{equation}
\mathfrak{E}_{33} \approx k^2 + \sum_s k_{\mathrm{D}s}^2 \label{lowfrequency_electrostaticcomponent_CE}
\, ,
\end{equation} 
or, writing (\ref{lowfrequency_electrostaticcomponent_CE}) explictly in terms of $\tilde{\omega}_{s\|}$ 
and the plasma frequency $\omega_{\mathrm{p}s}$ of species $s$, 
\begin{equation}
\mathfrak{E}_{33} \approx k^2 + \sum_s \frac{\omega_{\mathrm{p}s}^2}{\omega^2} \frac{2 k_{\|}^2}{k^2} \tilde{\omega}_{s\|}^2 \label{lowfrequency_electrostaticcomponent_CE_B}
\, .
\end{equation} 
It follows that $\mathfrak{E}_{33}^{(0)}$ and  
$\mathfrak{E}_{33}^{(1)}$ defined by (\ref{dielectric_expand})
are given by
\begin{subeqnarray}
 \mathfrak{E}_{33}^{(0)} & = & 0 , \\
 \mathfrak{E}_{33}^{(1)} & = & \frac{\omega_{\mathrm{p}e}^2}{\omega^2} \sum_s \frac{Z_s T_e}{T_s} \frac{2 k_{\|}^2}{k^2} . \label{lowfrequency_electrostaticcomponent_CE_C}
\end{subeqnarray}
where we have neglected the displacement current term ($k \ll k_{\mathrm{D}e}$), and the temperature of species $s$ is denoted by $T_s$. 

\subsection{Existence of electrostatic instabilities for a CE plasma} 
\label{existence_electrostatic}

That electrostatic instabilities can exist is most simply shown in the limit of 
purely parallel, high-frequency fluctuations: $k_{\bot} = 0$, $k_{\|} = k$, $\tilde{\omega}_{s} = \tilde{\omega}_{s\|}\gg 
1$, and
\begin{equation}
  \varpi\equiv \Real{\; \omega} \gg  \Imag{\; \omega} \equiv \gamma \, . \label{unstable_electromodes}
\end{equation}
For purely parallel modes, the only non-zero term in the sum of Bessel functions 
in the electrostatic hot-plasma dispersion relation (\ref{electrostatic_disprel}) 
is the $n = 0$ term; thus, (\ref{electrostatic_disprel}) simplifies to
\begin{equation}
\mathfrak{E}_{33}  = k^2 + \sum_s k_{\mathrm{D}s}^2 \left( \frac{1}{\sqrt{\upi}} \int_{C_L} \mathrm{d} \tilde{v}_{s\|} \int_0^{\infty} \mathrm{d} \tilde{v}_{s\bot}
 \tilde{v}_{s\bot} \frac{\p \tilde{f}_{s0}}{\p \tilde{v}_{s\|}} 
 \frac{1}{\tilde{\omega}_{s}  -\tilde{v}_{s\|}} \right) = 0 . \label{electrostatic_disprel_par}
\end{equation}
Next, we expand (\ref{electrostatic_disprel_par}) around the real frequency 
$\varpi$, using (\ref{unstable_electromodes}); this gives
\begin{equation}
\mathfrak{E}_{33}(\omega, k) \approx  \mathfrak{E}_{33}(\varpi, k) + \mathrm{i} 
\gamma \frac{\partial \mathfrak{E}_{33}}{\partial \omega}(\varpi, k) .  \label{electrostatic_disprel_par_B}
\end{equation}
Taking the imaginary part of (\ref{electrostatic_disprel_par_B}) allows for an 
expression for $\gamma$ to be derived in terms of $\varpi$:
\begin{equation}
 \gamma \approx - \left[\frac{\partial \, \Real{ \, \mathfrak{E}_{33}}}{\partial \omega}(\varpi, k)\right]^{-1} \Imag{\, \mathfrak{E}_{33}(\varpi, k)}  
\label{electrostatic_growth} \, . 
\end{equation}
To calculate $\gamma$, we use
\begin{subeqnarray}
 \Real{\, \mathfrak{E}_{33}\!(\varpi, k)} & = & k^2 + \sum_s k_{\mathrm{D}s}^2 \left( \frac{1}{\sqrt{\upi}} \textit{P} \! \int \mathrm{d} \tilde{v}_{s\|} \int_0^{\infty} \mathrm{d} \tilde{v}_{s\bot} \tilde{v}_{s\bot} \frac{\p \tilde{f}_{s0}}{\p \tilde{v}_{s\|}} \frac{1}{\tilde{\omega}_{s}  -\tilde{v}_{s\|}} \right) \, , \quad  \\
 \Imag{\, \mathfrak{E}_{33}\!(\varpi, k)}  & = & -\sqrt{\upi} k^{2} \int_0^{\infty} \mathrm{d} \tilde{v}_{s\bot} \tilde{v}_{s\bot} \frac{\p \tilde{f}_{s0}}{\p \tilde{v}_{s\|}}\!\left(\tilde{\omega}_{s},\tilde{v}_{s\bot}\right)  \, , 
 \label{electrostatic_realimagparts}
\end{subeqnarray}
where, to the leading order, $\tilde{\omega}_{s} \approx \varpi/k v_{\mathrm{th}s}$.
Now expanding (\ref{electrostatic_realimagparts}\textit{a}) in $\tilde{\omega}_{s} \gg 1$, we find that
\begin{equation}
 \Real{\, \mathfrak{E}_{33}\!(\varpi, k)}  \approx  k^2 - \sum_s \frac{k_{\mathrm{D}s}^2}{\tilde{\omega}_{s}^2}  \approx k^2 \left(1-\frac{\omega_{\mathrm{p}e}^2}{\varpi^2}\right) \, , 
 \label{electrostatic_realpart_exp}
\end{equation}
where we have integrated (\ref{electrostatic_realimagparts}\textit{a}) by parts, used 
identity
\begin{equation}
 \int_{-\infty}^{\infty} \mathrm{d} \tilde{v}_{s\|} \int_0^{\infty} \mathrm{d} \tilde{v}_{s\bot} \,  \tilde{v}_{s\bot} \tilde{f}_{s0}\!\left(\tilde{v}_{s\|},\tilde{v}_{s\bot}\right) 
 = \sqrt{\upi} \, ,
\end{equation}
and neglected the small ion contribution to the dielectric tensor. We conclude that -- as expected -- the real frequency of such modes is simply the plasma frequency: $\varpi\approx 
\pm \omega_{\mathrm{p}e}$. This in turn implies that
\begin{equation}
\tilde{\omega}_{e} = \frac{k_{\mathrm{D}e}}{\sqrt{2} k}  \gg 1 \, .
\end{equation}
In other words, electrostatic modes in this limit are simply plasma oscillations 
with wavelengths much greater than the Debye length. 

We immediately deduce that if $\varpi \approx \omega_{\mathrm{p}e}$ (without loss of generality, we can consider the mode with $\varpi > 0$), then
\begin{equation}
 \frac{\partial \, \Real{\, \mathfrak{E}_{33}}}{\partial \omega}(\varpi, k) \approx \frac{2 k^2}{\omega_{\mathrm{p}e}}  
 \, ,
\end{equation}
which in turn implies that $\gamma$ is positive if and only if, for some $k$,
\begin{equation}
 \Imag{\, \mathfrak{E}_{33}\!(\omega_{\mathrm{p}e}, k)} > 0  . 
\end{equation}
For the electron CE distribution function (\ref{ChapEnskogFunc_s}), we have
\begin{eqnarray}
\frac{\p \tilde{f}_{e0}}{\p \tilde{v}_{e\|}} & =&  - \exp \left(-\tilde{v}_{e}^2\right) 
\Bigg\{2 \tilde{v}_{e\|} + \eta_e \left[\left(2 \tilde{v}_{e\|}^2 - 1\right) A_e(\tilde{v}_e) - \frac{\tilde{v}_{e\|}^2}{\tilde{v}_e} A_e'(\tilde{v}_e) \right] 
\nonumber \\
&& \quad + \epsilon_e \left[ 2 \tilde{v}_{e\|} C_e(\tilde{v}_e) \left(\tilde{v}_{e\|}^2-\frac{\tilde{v}_{e\bot}^2}{2}-1\right) 
- \frac{\tilde{v}_{e\|}}{\tilde{v}_e} \left(\tilde{v}_{e\|}^2-\frac{\tilde{v}_{e\bot}^2}{2}\right) 
C_e'(\tilde{v}_e) \right] \Bigg\} . \,  
\end{eqnarray}
As shown in appendix \ref{ChapEnskogIsoFunc_Krook}, for a Krook collision operator it follows that (assuming $\eta_e^R = \eta_e^u = 0$)
\begin{subeqnarray}
  A_e(\tilde{v}_e) & = & -\left(\tilde{v}_e^2 - \frac{5}{2}\right) \, , \\
  C_e(\tilde{v}_e) & = & -1 \, . 
\end{subeqnarray}
We then see that
\begin{eqnarray}
\Imag{\, \mathfrak{E}_{33}\!(\omega_{\mathrm{p}e}, k)}  & = & \sqrt{\upi} k^2 \Bigg[ \frac{k_{\mathrm{D}e}}{\sqrt{2} k} - \eta_e \left(\frac{k_{\mathrm{D}e}^2}{4 k^2} -\frac{3}{4} \right) \left(\frac{k_{\mathrm{D}e}^2}{k^2} - 1\right) \nonumber \\
&& \qquad - \epsilon_e \frac{k_{\mathrm{D}e}}{\sqrt{2} k} \left(\frac{k_{\mathrm{D}e}^2}{2 k^2}-\frac{3}{2}\right)\Bigg] \exp \left(-\frac{k_{\mathrm{D}e}^2}{2 k^2}\right) 
\, . \label{Im_bumpintail_instab}
\end{eqnarray}
This expression changes sign from negative to positive when $k \lesssim \eta_e^{1/3} 
k_{\mathrm{D}e}$, or $k \lesssim \epsilon_e^{1/2} k_{\mathrm{D}e}$; thus, plasma waves with sufficiently 
long wavelengths are driven unstable by the non-Maxwellian component of the 
CE distribution function. Physically, this is the bump-on-tail instability; this arises because the distribution function is no longer monotonically decreasing at (parallel) particle velocities $v_{\|} \gtrsim \eta_e^{-1/3} v_{\mathrm{th}e}$, or  $v_{\|} \gtrsim \eta_e^{-1/3} v_{\mathrm{th}e}$, and so plasma waves can extract energy from particles via the Landau resonance. 

Substituting (\ref{Im_bumpintail_instab}) into (\ref{electrostatic_growth}), the growth rate of instabilities satisfying $k \ll k_{\mathrm{D}e}$
becomes
\begin{equation}
\gamma \approx \omega_{\mathrm{p}e} \frac{\sqrt{\upi}}{2 \sqrt{2}} \frac{k_{\mathrm{D}e}}{k}  \left( 1 - \eta_e \frac{k_{\mathrm{D}e}^3}{2 \sqrt{2} k^3} 
-\epsilon_e \frac{k_{\mathrm{D}e}^2}{2 k^2} \right) \exp \left(-\frac{k_{\mathrm{D}e}^2}{2 k^2}\right) 
.
\end{equation}
Maximising this expression with respect to $k$, it can then be shown that the peak growth rate for CE electron-temperature-gradient-driven 
microinstabilities ($\epsilon_e = 0$) is
\begin{equation}
\gamma_{\mathrm{max}} \approx \frac{3 \sqrt{\upi}}{4} \eta_e^{1/3} \exp \left(-\eta_e^{-2/3}-1\right) \omega_{\mathrm{p}e}  
\end{equation}
at the wavenumber 
\begin{equation}
 k_{\mathrm{peak}} \approx \frac{\eta_e^{1/3}}{\sqrt{2}} \left[1-\frac{\eta_e^{2/3}}{2} \right] 
 k_{\mathrm{D}e} \, ,
\end{equation}
whereas for CE electron-shear-driven microinstabilities ($\eta_e = 0$), 
\begin{equation}
\gamma_{\mathrm{max}} \approx \frac{\sqrt{\upi}}{2} \epsilon_e^{1/2} \exp \left(-\epsilon_e^{-1}-1\right) \omega_{\mathrm{p}e} 
\end{equation}
at the wavenumber 
\begin{equation}
 k_{\mathrm{peak}} \approx \frac{\epsilon_e^{1/2}}{\sqrt{2}} \left[1-\frac{\epsilon_e}{2} \right] 
 k_{\mathrm{D}e} \, .
\end{equation}

\subsection{Impossibility of electrostatic instabilities with `fast' growth rates} 
\label{impossibility_electrostatic_slowgrowth}

The existence of electrostatic instabilities was demonstrated in appendix (\ref{existence_electrostatic}); however, the growth rates of the exemplified instabilities were shown 
to be exponentially small in the 
parameters $\eta_e$ or $\epsilon_e$. In this appendix, 
we provide a proof that there cannot exist electrostatic instabilities whose growth rate scales algebraically with
 $\eta_s$ or $\epsilon_s$.

To substantiate this claim properly, it is necessary to consider perturbations with frequencies $\omega$ satisfying $\omega \ll k_{\|} v_{\mathrm{th}s}$
and $\omega \gtrsim k_{\|} v_{\mathrm{th}s}$ separately. 

\subsubsection{Low-frequency electrostatic modes: $\omega \ll k_{\|} v_{\mathrm{th}s}$}  \label{low_freq_electrostatic_stab}

The impossibility of low-frequency electrostatic instabilities follows 
immediately from equation (\ref{lowfrequency_electrostaticcomponent_CE}), which 
shows that the leading-order term in the $\tilde{\omega}_{s\|} \ll 1$ expansion 
of the electrostatic component of the dielectric tensor is non-zero. It 
follows that the electrostatic component of the dielectric tensor is strictly positive 
at low frequencies. Since the electrostatic component of the dielectric tensor must vanish
in order for the electrostatic dispersion relation (\ref{electrostatic_disprel}) 
to be satisfied, we conclude that there do not exist electrostatic modes with $\omega \ll k_{\|} 
v_{\mathrm{th}s}$, let alone instabilities. 

\subsubsection{Other electrostatic modes: $\omega \gtrsim k_{\|} v_{\mathrm{th}s}$} 
\label{high_freq_electrostatic_stab}

For all other electrostatic perturbations, we suppose that there exist microinstabilities with growth rates which scale algebraically 
with $\eta_s$, $\epsilon_s$, and then prove that that such an supposition is incompatible with the hot-plasma 
electrostatic dispersion relation. 

Consider some unstable perturbation satisfying the electrostatic dispersion relation (\ref{electrostatic_disprel}), with complex frequency $\omega = \varpi + 
\mathrm{i} \gamma$, and $\gamma > 0$. We then define
\begin{subeqnarray}
\tilde{\varpi}_{s\|} & \equiv & \frac{\varpi}{k_{\|} v_{\mathrm{th}s}} , \\
\tilde{\gamma}_{s\|} & \equiv & \frac{\gamma}{k_{\|} v_{\mathrm{th}s}} ,
\end{subeqnarray}
so that $\tilde{\omega}_{s\|} = \tilde{\varpi}_{s\|} + \mathrm{i} \tilde{\gamma}_{s\|}$. 
For unstable perturbations satisfying (\ref{electrostatic_disprel}), it follows 
from the real and imaginary parts of the dispersion relation that
\begin{subeqnarray}
0 & = & k^2 - \sum_s k_{\mathrm{D}s}^2 \Bigg\{ \frac{1}{\sqrt{\upi}} \int_{-\infty}^{\infty} \mathrm{d} \tilde{v}_{s\|} \int_0^{\infty} \mathrm{d} \tilde{v}_{s\bot}
\sum_{n = -\infty}^{\infty} \Bigg[ \Pi_n(\tilde{v}_{s\|},\tilde{v}_{s\bot}) 
\nonumber \\
&& \qquad \qquad \qquad \qquad \times
 \frac{\left(\tilde{v}_{s\|} - \tilde{\varpi}_{s\|}  + n/k_{\|} \tilde{\rho}_s\right)J_n(k_{\bot} \tilde{\rho}_s \tilde{v}_{s\bot})^2}{\left(\tilde{v}_{s\|} - \tilde{\varpi}_{s\|}  + n/k_{\|} \tilde{\rho}_s\right)^2 + \tilde{\gamma}_{s\|}^2} \Bigg] \Bigg\} 
 , \\
0 & = & \gamma \sum_s k_{\mathrm{D}s}^2 \mu_s^{-1/2} \Bigg\{ \frac{1}{\sqrt{\upi}} \int_{-\infty}^{\infty} \mathrm{d} \tilde{v}_{s\|} \int_0^{\infty} \mathrm{d} \tilde{v}_{s\bot}
\sum_{n = -\infty}^{\infty} \Bigg[ \Pi_n(\tilde{v}_{s\|},\tilde{v}_{s\bot}) 
\nonumber \\
&& \qquad \qquad \qquad \qquad \times \frac{J_n(k_{\bot} \tilde{\rho}_s \tilde{v}_{s\bot})^2}{\left(\tilde{v}_{s\|} - \tilde{\varpi}_{s\|}  + n/k_{\|} \tilde{\rho}_s\right)^2 + \tilde{\gamma}_{s\|}^2} \Bigg] \Bigg\} , \label{electrostatic_disp_rel_ReIm}
\end{subeqnarray}
where $\mu_s \equiv m_e/m_s$, and we have utilised the fact that the Landau contour simplifies to the real line
for unstable perturbations. Using (\ref{electrostatic_disp_rel_ReIm}\textit{b}), we can 
eliminate part of (\ref{electrostatic_disp_rel_ReIm}\textit{a}) to give
\begin{eqnarray}
0 & = & k^2 - \sum_s k_{\mathrm{D}s}^2 \Bigg\{ \frac{1}{\sqrt{\upi}} \int_{-\infty}^{\infty} \mathrm{d} \tilde{v}_{s\|} \int_0^{\infty} \mathrm{d} \tilde{v}_{s\bot}
\sum_{n = -\infty}^{\infty} \Bigg[ \Pi_n(\tilde{v}_{s\|},\tilde{v}_{s\bot}) 
\nonumber \\
&& \qquad \qquad \qquad \qquad \times
 \frac{\left(\tilde{v}_{s\|} + n/k_{\|} \tilde{\rho}_s\right)J_n(k_{\bot} \tilde{\rho}_s \tilde{v}_{s\bot})^2}{\left(\tilde{v}_{s\|} - \tilde{\varpi}_{s\|}  + n/k_{\|} \tilde{\rho}_s\right)^2 + \tilde{\gamma}_{s\|}^2} \Bigg] \Bigg\}
 . \label{electrostatic_disp_rel_ReB}
\end{eqnarray}
Next, we substitute for $\Pi_n(\tilde{v}_{s\|},\tilde{v}_{s\bot})$ using
\begin{equation}
\Pi_n(\tilde{v}_{s\|},\tilde{v}_{s\bot}) =  \Lambda_s(\tilde{v}_{s\|},\tilde{v}_{s\bot}) 
+ \left(\tilde{v}_{s\|}+\frac{n}{k_{\|} \tilde{\rho}_s}\right) \frac{\p \tilde{f}_{s0}}{\p \tilde{v}_{s\bot}} 
,
\end{equation}
to give
\begin{eqnarray}
0 & = & k^2 - \sum_s k_{\mathrm{D}s}^2 \Bigg\{ \frac{1}{\sqrt{\upi}} \int_{-\infty}^{\infty} \mathrm{d} \tilde{v}_{s\|} \int_0^{\infty} \mathrm{d} \tilde{v}_{s\bot}
\sum_{n = -\infty}^{\infty} \Bigg[\frac{\p \tilde{f}_{s0}}{\p \tilde{v}_{s\bot}}  \frac{\left(\tilde{v}_{s\|} + n/k_{\|} \tilde{\rho}_s\right)^2 J_n(k_{\bot} \tilde{\rho}_s \tilde{v}_{s\bot})^2}{\left(\tilde{v}_{s\|} - \tilde{\varpi}_{s\|}  + n/k_{\|} \tilde{\rho}_s\right)^2 + \tilde{\gamma}_{s\|}^2} \nonumber \\
&& \qquad \qquad \qquad \qquad + \Lambda_s(\tilde{v}_{s\|},\tilde{v}_{s\bot}) \frac{\left(\tilde{v}_{s\|} + n/k_{\|} \tilde{\rho}_s\right) J_n(k_{\bot} \tilde{\rho}_s \tilde{v}_{s\bot})^2}{\left(\tilde{v}_{s\|} - \tilde{\varpi}_{s\|}  + n/k_{\|} \tilde{\rho}_s\right)^2 + \tilde{\gamma}_{s\|}^2}\Bigg] \Bigg\}
 . \label{electrostatic_disp_rel_ReC}
\end{eqnarray}
This expression is very helpful for contradicting the premise of the existence 
of unstable electrostatic modes. We illustrate this claim with a simple example -- a pure Maxwellian distribution function -- before considering the 
CE distribution. 

For a Maxwellian distribution for which $\Lambda_s(\tilde{v}_{s\|},\tilde{v}_{s\bot}) = 
0$, and 
\begin{equation}
\frac{\p \tilde{f}_{s0}}{\p \tilde{v}_{s\bot}} = - 2 \tilde{v}_{s\bot} \exp 
\left(-\tilde{v}_{s}^2\right), 
\end{equation}
(\ref{electrostatic_disp_rel_ReC}) becomes
\begin{eqnarray}
0 & = & k^2 + \sum_s k_{\mathrm{D}s}^2 \Bigg[ \frac{2}{\sqrt{\upi}} \int_{-\infty}^{\infty} \mathrm{d} \tilde{v}_{s\|} \int_0^{\infty} \mathrm{d} \tilde{v}_{s\bot}
\tilde{v}_{s\bot} \exp 
\left(-\tilde{v}_{s}^2\right) \nonumber \\
&&  \qquad \qquad \qquad \qquad \times \sum_{n = -\infty}^{\infty} 
\frac{\left(\tilde{v}_{s\|} + n/k_{\|} \tilde{\rho}_s\right)^2 J_n(k_{\bot} \tilde{\rho}_s \tilde{v}_{s\bot})^2}{\left(\tilde{v}_{s\|} - \tilde{\varpi}_{s\|}  + n/k_{\|} \tilde{\rho}_s\right)^2 + \tilde{\gamma}_{s\|}^2} \Bigg] 
 . \label{electrostatic_disp_rel_ReD}
\end{eqnarray}
The integrand on the right-hand-side of (\ref{electrostatic_disp_rel_ReD}) is strictly positive 
-- a contradiction. Therefore, we recover the standard result that there cannot 
exist unstable perturbations if the underlying distribution is Maxwellian. 

We now consider the CE distribution (\ref{ChapEnskogFunc_s}). In order for an instability to arise, it is 
clear that the integrand on the right-hand-side of (\ref{electrostatic_disp_rel_ReC}) 
has to be positive -- and further, the contribution of the integrand from that 
interval has to dominate all other (negative) contributions to the total integral. To 
prove that these conditions cannot be satisfied for the CE distribution 
function, we consider the two terms in the integrand on the right-hand-side of (\ref{electrostatic_disp_rel_ReC}) 
separately. 

For the first term, 
\begin{equation}
\frac{\p \tilde{f}_{s0}}{\p \tilde{v}_{s\bot}}  \frac{\left(\tilde{v}_{s\|} + n/k_{\|} \tilde{\rho}_s\right)^2 J_n(k_{\bot} \tilde{\rho}_s \tilde{v}_{s\bot})^2}{\left(\tilde{v}_{s\|} - \tilde{\varpi}_{s\|}  + n/k_{\|} \tilde{\rho}_s\right)^2 + \tilde{\gamma}_{s\|}^2} > 0  
\end{equation}
if and only if
\begin{equation}
\frac{\p \tilde{f}_{s0}}{\p \tilde{v}_{s\bot}}  < 0 .
\end{equation}
For the CE distribution function (\ref{ChapEnskogFunc_s}), 
\begin{eqnarray}
\frac{\p \tilde{f}_{s0}}{\p \tilde{v}_{s\bot}} & =&  - \tilde{v}_{s\bot} \exp \left(-\tilde{v}_{s}^2\right) 
\Bigg\{2  + \eta_s \left[2 \tilde{v}_{s\|} A_s(\tilde{v}_s) - \frac{\tilde{v}_{s\|}}{\tilde{v}_s} A_s'(\tilde{v}_s) \right] 
\nonumber \\
&& \quad + \epsilon_s \left[ 2 C_s(\tilde{v}_s) \left(\tilde{v}_{s\|}^2-\frac{\tilde{v}_{s\bot}^2}{2}+\frac{1}{2}\right) 
- \frac{1}{\tilde{v}_s} \left(\tilde{v}_{s\|}^2-\frac{\tilde{v}_{s\bot}^2}{2}\right) 
C_s'(\tilde{v}_s) \right] \Bigg\} \, . \label{CE_dist_perpder}
\end{eqnarray}
Thus, for $\tilde{v}_{s\bot} \lesssim 1$ and $\tilde{v}_{s\|} \lesssim 1$, we see that  ${\p \tilde{f}_{s0}}/{\p \tilde{v}_{s\bot}}  < 
0$, because $\eta_s, \epsilon_s \ll 1$. The only values of $\tilde{v}_{s}$ where this inequality 
could be reversed are large: $\tilde{v}_{s} \gg 1$. Assuming that $A_s(\tilde{v}_{s}) \sim 
\tilde{v}_{s}^{\iota_\eta}$ and $C_s(\tilde{v}_{s}) \sim
\tilde{v}_{s}^{\iota_\epsilon}$ for $\tilde{v}_{s} \gg 1$, where ${\iota_\eta}$ and ${\iota_\epsilon}$ 
are constants, it follows that for 
\begin{equation}
\tilde{v}_{s} \gtrsim \eta_s^{-{1}/{(\iota_\eta+1)}} , \epsilon_s^{-{1}/{(\iota_\epsilon+2)}} \, , \label{largev_electrostat}
\end{equation}
the non-Maxwellian terms are comparable to the Maxwellian ones. However, for 
such $\tilde{v}_{s}$,
\begin{equation}
\frac{\p \tilde{f}_{s0}}{\p \tilde{v}_{s\bot}} \sim \eta_s^{-1/(\iota_\eta+1)} \exp \left(-\eta_s^{-2/(\iota_\eta+1)}\right) 
, \epsilon_s^{-1/(\iota_\epsilon+1)} \exp \left(-\epsilon_s^{-2/(\iota_\epsilon+1)}\right) ,
\end{equation}
while 
\begin{equation}
\frac{\left(\tilde{v}_{s\|} + n/k_{\|} \tilde{\rho}_s\right)^2 J_n(k_{\bot} \tilde{\rho}_s \tilde{v}_{s\bot})^2}{\left(\tilde{v}_{s\|} - \tilde{\varpi}_{s\|}  + n/k_{\|} \tilde{\rho}_s\right)^2 + \tilde{\gamma}_{s\|}^2} \lesssim \frac{\tilde{\varpi}_{s\|}^2}{\tilde{\gamma}_{s\|}^2} 
\end{equation}
if it is assumed that $|\varpi| \gg |\gamma|$. Since we assumed that $\tilde{\gamma}_{s\|}$ is only algebraically small in $\epsilon_s$ and/or $\eta_s$, we 
conclude that the contribution to the integrand on the right-hand-side of (\ref{electrostatic_disp_rel_ReC}) from 
$\tilde{v}_{s}$ satisfying (\ref{largev_electrostat}) is asymptotically small compared to other 
contributions, and thus cannot change the sign of the total integral. 

For the second term, we consider the $n$th term of the sum independently. Recalling from (\ref{anisopparameter}) that
\begin{equation}
 \Lambda_s(\tilde{v}_{s\|},\tilde{v}_{s\bot})= - \tilde{v}_{s\bot} \exp \left(-\tilde{v}_{s}^2\right) \left[\eta_s A_s(\tilde{v}_s) - 3 \epsilon_s C_s(\tilde{v}_s)  \tilde{v}_{s\|} \right] 
  , \label{anisopparameter_Append}
\end{equation}
it follows that for $\tilde{v}_{s} \sim 1$, 
\begin{equation}
  \frac{ \Lambda_s(\tilde{v}_{s\|},\tilde{v}_{s\bot})}{{\p \tilde{f}_{s0}}/{\p \tilde{v}_{s\bot}} } 
  \sim \frac{\eta_s}{\tilde{v}_{s\|} + n/k_{\|} \tilde{\rho}_s} \, , \frac{\epsilon_s}{\tilde{v}_{s\|} + n/k_{\|} \tilde{\rho}_s} 
  \, .
  \label{aniso_limit_electrostat}
\end{equation}
Thus, for $\tilde{v}_{s} \sim 1$, the non-Maxwellian term is only comparable to 
the Maxwellian one for $|\tilde{v}_{s\|} + n/k_{\|} \tilde{\rho}_s| \lesssim \eta_s, 
\epsilon_s$. However, this non-Maxwellian contribution is in fact always smaller that other 
non-Maxwellian contributions, which by (\ref{aniso_limit_electrostat}) are in 
turn smaller than the equivalent Maxwellian contributions. 

Depending on the magnitude of $|n/k_{\|} \tilde{\rho}_s|$, this claim is justified in two different ways.
\begin{itemize}
  
 \item \underline{$|n/k_{\|} \tilde{\rho}_s| \lesssim 1$}: in this case, let the interval of non-dimensionalised 
 parallel velocities~$\tilde{v}_{s\|}$ satisfying $|\tilde{v}_{s\|} + n/k_{\|} \tilde{\rho}_s| \lesssim \eta_s, 
\epsilon_s$ be denoted by $\mathcal{I}$. Then, there exists another finite interval of $\tilde{v}_{s\|} \sim 1$ 
 such that $|\tilde{v}_{s\|} + n/k_{\|} \tilde{\rho}_s|  \sim 1$. It therefore follows 
 that
 \begin{eqnarray}
 \int_{\mathcal{I}}  & \mathrm{d} \tilde{v}_{s\|} & \; \Lambda_s(\tilde{v}_{s\|},\tilde{v}_{s\bot}) \frac{\left(\tilde{v}_{s\|} + n/k_{\|} \tilde{\rho}_s\right) J_n(k_{\bot} \tilde{\rho}_s \tilde{v}_{s\bot})^2}{\left(\tilde{v}_{s\|} - \tilde{\varpi}_{s\|}  + n/k_{\|} \tilde{\rho}_s\right)^2 + \tilde{\gamma}_{s\|}^2}  \nonumber \\
 &\sim & \eta_s^2 \int_{-\infty}^{\infty} \mathrm{d} \tilde{v}_{s\|}  \Lambda_s(\tilde{v}_{s\|},\tilde{v}_{s\bot}) \frac{\left(\tilde{v}_{s\|} + n/k_{\|} \tilde{\rho}_s\right) J_n(k_{\bot} \tilde{\rho}_s \tilde{v}_{s\bot})^2}{\left(\tilde{v}_{s\|} - \tilde{\varpi}_{s\|}  + n/k_{\|} \tilde{\rho}_s\right)^2 + \tilde{\gamma}_{s\|}^2}  , 
 \end{eqnarray}
  where we have assumed that $|\tilde{\varpi}_{s\|}| \gg 
  |\tilde{\gamma}_{s\|}|$ (and also $|\tilde{\varpi}_{s\|}| \gtrsim 1$). The claim 
  immediately follows. 
  
  \item \underline{$|n/k_{\|} \tilde{\rho}_s| \gg 1$}: in this case, it follows 
  immediately that $|\tilde{v}_{s\|} + n/k_{\|} \tilde{\rho}_s| \lesssim \eta_s, 
\epsilon_s$ if and only if $\tilde{v}_{s\|} \gg 1$. Via a similar argument to 
that presented for large $\tilde{v}_{s\|}$ for the first term in the integrand on the right-hand-side of 
(\ref{electrostatic_disp_rel_ReC}), contributions to the total integral will be 
exponentially small in $\eta_s, \epsilon_s$, and thus are unable to reverse the 
sign of the total integral. 
  
\end{itemize}

Thus, we have confirmed that there cannot exist electrostatic instabilities 
with growth rates which are algebraic in small parameters $\eta_s, \epsilon_s$.

\section{Weak growth of high-frequency perturbations} \label{arg_stab_highfreq}

In this appendix, we present an argument that all perturbations in
a CE plasma with complex frequency $\omega = \varpi+\mathrm{i}\gamma$ satisfying the `high-frequency' 
conditions $|\omega| \gtrsim k_{\|} v_{\mathrm{th}s}$ and $|\varpi| \gg |\gamma|$ for all particle species 
have a growth rate that is at most exponentially small in $\eta_s$, and $\epsilon_s$. This argument does not prove that all perturbations satisfying $|\omega| \gtrsim k_{\|} v_{\mathrm{th}s}$
in a CE plasma are stable, in that it does not apply to perturbations whose damping or growth rate is not small compared to their frequency.  

\subsection{Deriving conditions for stability}

We begin with the result that for any linear electromagnetic perturbation with real frequency $\varpi > 0$,
growth rate $\gamma$, wavevector $\boldsymbol{k}$, and electric-field 
perturbation
\begin{equation}
 \delta \boldsymbol{E} =
   \widehat{\delta \boldsymbol{E}} \exp\left\{\mathrm{i}\left(\mathbf{k} \bcdot \mathbf{r} - \varpi t\right)+\gamma t\right\} \label{FourierE} ,\\[3pt]
  \  
\end{equation}
the dissipation rate $\mathfrak{Q}$ of the 
perturbation is related to the anti-Hermitian part of the plasma dielectric 
tensor evaluated at the perturbation's real frequency~\citep{LL81}:
\begin{equation}
 \mathfrak{Q} = \mathrm{i} \varpi  \widehat{\delta \boldsymbol{E}}^{*} \bcdot \boldsymbol{\mathfrak{E}}^{A}(\boldsymbol{k},\varpi) \bcdot  \widehat{\delta \boldsymbol{E}}
 \, , \label{disprate_linearperturb}
\end{equation}
where the anti-Hermitian part $\boldsymbol{\mathfrak{E}}^{A}$ is defined by
\begin{equation}
  \boldsymbol{\mathfrak{E}}^{A} = 
  \frac{1}{2}\left(\boldsymbol{\mathfrak{E}}-\boldsymbol{\mathfrak{E}}^{\dagger}\right) 
  ,
\end{equation}
with $\boldsymbol{\mathfrak{E}}^{\dagger}$ representing the conjugate transpose of $\boldsymbol{\mathfrak{E}}$.  
If the mode is damped, then the dissipation rate is positive: $\mathfrak{Q} > 
0$. Since $ \boldsymbol{\mathfrak{E}}^{A}$ is anti-Hermitian, it is diagonalisable in some orthonormal basis $\left\{\hat{\boldsymbol{e}}_a,\hat{\boldsymbol{e}}_b,\hat{\boldsymbol{e}}_c\right\}$, 
with imaginary eigenvalues 
$\left(-\mathrm{i}\varsigma_{a},-\mathrm{i}\varsigma_{b},-\mathrm{i}\varsigma_{c}\right)$, 
where $\varsigma_{a}$, $\varsigma_{b}$, and $\varsigma_{c}$ are real numbers. The 
dissipation rate $\mathfrak{Q}$ can be written in terms of these eigenvectors as
\begin{equation}
 \mathfrak{Q} = \varpi \left( \varsigma_a \left|\hat{\boldsymbol{e}}_a \bcdot \widehat{\delta \boldsymbol{E}}\right|^2  + \varsigma_b \left|\hat{\boldsymbol{e}}_b \bcdot \widehat{\delta \boldsymbol{E}}\right|^2 + \varsigma_c \left|\hat{\boldsymbol{e}}_c \bcdot \widehat{\delta \boldsymbol{E}}\right|^2\right) 
 \, .
\end{equation}
Thus, for unstable perturbations to exist, it must be the case that at least one of the 
numbers $\varsigma_{a}$, $\varsigma_{b}$, and $\varsigma_{c}$ has to be negative (without loss of generality, we will assume $\varsigma_{a} < 0$); if this is the case, then the dissipation rate (and hence the growth rate) 
is a linear function of $\varsigma_{a}$. We will show that if $|\omega| \gtrsim k_{\|} v_{\mathrm{th}s}$, 
$\varsigma_{a}$, $\varsigma_{b}$, and $\varsigma_{c}$ can only be negative if 
they are exponentially small in $\eta_s$ and $\epsilon_s$. 

To prove this, consider the characteristic polynomial
\begin{equation}
\varrho(\varsigma) \equiv \mbox{det}\left[\boldsymbol{\mathfrak{E}}^{A}(\boldsymbol{k},\varpi) - \varsigma \mathsfbi{I}\right]
\end{equation}
of $\boldsymbol{\mathfrak{E}}^{A}$ evaluated at the real frequency $\varpi$ and wavevector $\boldsymbol{k}$; it is a cubic, and thus can be written
\begin{equation}
 \varrho(\varsigma) = -\varsigma^3 - \mathrm{i} \varrho_2 \varsigma^2 + \varrho_1 \varsigma 
 + \mathrm{i} \varrho_0 \, ,
\end{equation}
where $\varrho_0$, $\varrho_1$, and $\varrho_2$ depend on 
$\boldsymbol{\mathfrak{E}}^{A}$.
Since $\boldsymbol{\mathfrak{E}}^{A}$ has eigenvalues 
$\left(-\mathrm{i}\varsigma_{a},-\mathrm{i}\varsigma_{b},-\mathrm{i}\varsigma_{c}\right)$, 
it follows that
\begin{eqnarray}
 \varrho(\varsigma) & = & -\left(\varsigma+\mathrm{i}\varsigma_{a}\right) \left(\varsigma+\mathrm{i}\varsigma_{b}\right) \left(\varsigma+\mathrm{i}\varsigma_{c}\right) 
 \nonumber \\
 & = & -\varsigma^3 - \mathrm{i} \varsigma^2 \left(\varsigma_{a}+\varsigma_{b}+\varsigma_{c}\right) 
 + \varsigma \left(\varsigma_{a}\varsigma_{b}+\varsigma_{b}\varsigma_{c}+\varsigma_{c}\varsigma_{a}\right) 
 + \mathrm{i} \varsigma_{a}\varsigma_{b} \varsigma_{c} , \label{char_poly_A}
\end{eqnarray}
and so 
\begin{subeqnarray}
 \varrho_0 & = & \varsigma_{a}\varsigma_{b} \varsigma_{c} , \\
  \varrho_1 & = & \varsigma_{a}\varsigma_{b}+\varsigma_{b}\varsigma_{c}+\varsigma_{c}\varsigma_{a} , \\
   \varrho_2 & = & \varsigma_{a}+\varsigma_{b}+\varsigma_{c} . 
\end{subeqnarray}
This implies that $\varsigma_{a}$, $\varsigma_{b}$, and $\varsigma_{c}$ are 
positive if $\varrho_0$, $\varrho_1$, and $\varrho_2$ are positive. Furthermore, $\varrho_0$, $\varrho_1$, and $\varrho_2$ 
can be used to provide bounds for $\varsigma_{a}$, $\varsigma_{b}$, and $\varsigma_{c}$ 
using an inequality discovered by~\citet{L80}: 
\begin{equation}
\varsigma_{-} \leq \varsigma_a, \varsigma_b, \varsigma_c \leq \varsigma_{+},  \label{dielectric_antiHerm_eigenval_bound}
\end{equation}
where
\begin{equation}
\varsigma_{\pm} = -\frac{\varrho_2}{3} \pm \frac{2}{3} \sqrt{\varrho_2^2 - 3 \varrho_1^2} 
.  \label{dielectric_antiHerm_eigenval_bound_vals}
\end{equation}
In particular, the expression (\ref{dielectric_antiHerm_eigenval_bound_vals}) for the root bounds implies that if $\varrho_1$ and $\varrho_2$ are 
exponentially small in $\eta_s$ and $\epsilon_s$, then so are $\varsigma_{a}$, $\varsigma_{b}$, and 
$\varsigma_{c}$.

We can also evaluate $ \varrho(\varsigma)$ in terms of the components of $\boldsymbol{\mathfrak{E}}^{A}$ 
in the coordinate basis 
$\left\{\hat{\boldsymbol{x}},\hat{\boldsymbol{y}},\hat{\boldsymbol{z}}\right\}$:
\begin{eqnarray}
 \varrho(\varsigma) & = & -\varsigma^3 + \varsigma^2 \left(\mathfrak{E}_{xx}^{A}+\mathfrak{E}_{yy}^{A}+\mathfrak{E}_{zz}^{A}\right) \nonumber \\
 && - \varsigma \left(\mathfrak{E}_{xx}^{A} \mathfrak{E}_{yy}^{A} + \mathfrak{E}_{yy}^{A} \mathfrak{E}_{zz}^{A}+ \mathfrak{E}_{zz}^{A} \mathfrak{E}_{xx}^{A} + (\mathfrak{E}_{xy}^{A})^2 + (\mathfrak{E}_{yz}^{A})^2 + (\mathfrak{E}_{xz}^{A})^2\right) + \mbox{det}\,\boldsymbol{\mathfrak{E}}^{A} 
 , \qquad \quad \label{char_poly_B}
\end{eqnarray}
where we have used the symmetries (\ref{dielectricsimsgen}) of the dielectric tensor to give $ \varrho(\varsigma)$ in terms of only the (six) independent components
of $\boldsymbol{\mathfrak{E}}^{A}$. (\ref{char_poly_B}) gives
\begin{subeqnarray}
 \varrho_0 & = & - \mathrm{i} \mbox{det} \, \boldsymbol{\mathfrak{E}}^{A} , \\
  \varrho_1 & = & -\mathfrak{E}_{xx}^{A} \mathfrak{E}_{yy}^{A} - \mathfrak{E}_{yy}^{A} \mathfrak{E}_{zz}^{A}- \mathfrak{E}_{zz}^{A} \mathfrak{E}_{xx}^{A} - (\mathfrak{E}_{xy}^{A})^2-(\mathfrak{E}_{yz}^{A})^2 - (\mathfrak{E}_{xz}^{A})^2 , \\
   \varrho_2 & = & -\mathrm{i} \left(\mathfrak{E}_{xx}^{A}+\mathfrak{E}_{yy}^{A}+\mathfrak{E}_{zz}^{A}\right) . \label{char_coeffs_B}
\end{subeqnarray}
The anti-Hermiticity of $\boldsymbol{\mathfrak{E}}^{A}$ implies 
that $\Imag{\, \mathfrak{E}_{xx}^{A}} = - \mathrm{i} \mathfrak{E}_{xx}^{A}$, $\Imag{\, \mathfrak{E}_{yy}^{A}}  = 
- \mathrm{i} \mathfrak{E}_{yy}^{A}$, $\Imag{\, \mathfrak{E}_{zz}^{A}}  = - \mathrm{i} \mathfrak{E}_{zz}^{A}$, 
and $\Imag{\, \mathfrak{E}_{xz}^{A}}  = - \mathrm{i} \mathfrak{E}_{xz}^{A}$, while $\Real{\, \mathfrak{E}_{xy}^{A}} = \mathfrak{E}_{xy}^{A}$
and $\Real{\, \mathfrak{E}_{yz}^{A}} = \mathfrak{E}_{yz}^{A}$, as is indeed necessary for
$\varrho_0$, $\varrho_1$, and $\varrho_2$ to be real numbers. Thus, in order to establish stability it is 
sufficient for our purposes to show that
\begin{subeqnarray}
  \mathrm{i} \mbox{det}\, \boldsymbol{\mathfrak{E}}^{A} & < & 0, \\
\mathfrak{E}_{xx}^{A} \mathfrak{E}_{yy}^{A} + \mathfrak{E}_{yy}^{A} \mathfrak{E}_{zz}^{A}+ \mathfrak{E}_{zz}^{A} \mathfrak{E}_{xx}^{A} + (\mathfrak{E}_{xy}^{A})^2+(\mathfrak{E}_{yz}^{A})^2 + (\mathfrak{E}_{xz}^{A})^2 & < & 0, \\
 \mathrm{i} \left(\mathfrak{E}_{xx}^{A}+\mathfrak{E}_{yy}^{A}+\mathfrak{E}_{zz}^{A}\right) & < & 0 . \label{gen_stab_conditions}
\end{subeqnarray}
When these inequalities are not strictly satisfied, 
then we can instead estimate the magnitude of (\ref{char_coeffs_B}\textit{b}) and 
(\ref{char_coeffs_B}\textit{c}) to determine bounds for $\varsigma_{a}$, $\varsigma_{b}$, and 
$\varsigma_{c}$.

\subsection{Evaluating conditions for stability}

Combining equations (\ref{dielecttensfull}) with (\ref{conductivity}) gives an
expression for the general plasma dielectric tensor (assuming $k_{\|} > 0$ without loss of generality):
\begin{eqnarray}
\boldsymbol{\mathfrak{E}} & = & \mathsfbi{I} + \sum_s \frac{\omega_{\mathrm{p}s}^2}{\omega^2} \bigg[ \frac{2}{\sqrt{\upi}} \int_{-\infty}^{\infty} \mathrm{d} \tilde{v}_{s\|} \, \tilde{v}_{s\|} \int_0^{\infty} \mathrm{d} \tilde{v}_{s\bot} \Lambda_s(\tilde{v}_{s\|},\tilde{v}_{s\bot}) \hat{\boldsymbol{z}} \hat{\boldsymbol{z}} \nonumber \\
&& \mbox{} + \tilde{\omega}_{s\|} \frac{2}{\sqrt{\upi}} \int_{C_L} \mathrm{d} \tilde{v}_{s\|} \int_0^{\infty} \mathrm{d} \tilde{v}_{s\bot} \tilde{v}_{s\bot}^2 \Xi_s(\tilde{v}_{s\|},\tilde{v}_{s\bot}) \sum_{n = -\infty}^{\infty} \frac{\mathsfbi{R}_{sn}}{\zeta_{sn} -\tilde{v}_{s\|}}  \bigg] \, , \label{dielectric_full_Append}
\end{eqnarray}
where all salient quantities are defined in section \ref{HotPlasmaDispDis}. Now 
evaluating the anti-Hermitian part of (\ref{dielectric_full_Append}) for $\omega = \varpi$, $\tilde{\omega}_{s\|} = 
\tilde{\varpi}_{s\|}$, we find 
\begin{eqnarray}
\boldsymbol{\mathfrak{E}}^A & = & -\mathrm{i} \sum_s \frac{\omega_{\mathrm{p}s}^2}{\varpi^2} \bigg[2\sqrt{\upi} \tilde{\varpi}_{s\|} \int_0^{\infty} \mathrm{d} \tilde{v}_{s\bot} \tilde{v}_{s\bot}^2 \sum_{n = -\infty}^{\infty} \Xi_s(\zeta_{sn},\tilde{v}_{s\bot}) \mathsfbi{R}_{sn}(\zeta_{sn},\tilde{v}_{s\bot}) \bigg] \, . \label{dielectric_anitHerm_full_Append}
\end{eqnarray}
We now consider stability conditions (\ref{gen_stab_conditions}) in turn.

First evaluating (\ref{gen_stab_conditions}\textit{c}), it can be shown that
\begin{eqnarray}
 \mathrm{i} \left(\mathfrak{E}_{xx}^{A} \right. &+& \left. \mathfrak{E}_{yy}^{A} \, + \, \mathfrak{E}_{zz}^{A}\right) = 2\sqrt{\upi} \sum_s \frac{\omega_{\mathrm{p}s}^2}{\varpi^2} \tilde{\varpi}_{s\|} \sum_{n = -\infty}^{\infty} \bigg\{ \int_0^{\infty} \mathrm{d} \tilde{v}_{s\bot} \tilde{v}_{s\bot}^2 \Xi_s(\zeta_{sn},\tilde{v}_{s\bot}) \nonumber \\
 & \times &\bigg[\frac{n^2 J_n(k_{\bot} \tilde{\rho}_s \tilde{v}_{s\bot})^2}{k_{\bot}^2 \tilde{\rho}_s^2 \tilde{v}_{s\bot}^2} + J_n'(k_{\bot} \tilde{\rho}_s \tilde{v}_{s\bot})^2+ \frac{\zeta_{sn}^2}{\tilde{v}_{s\bot}^2} J_n(k_{\bot} \tilde{\rho}_s \tilde{v}_{s\bot})^2\bigg] \bigg\} \, . \label{dielectric_anitHerm_trace_Append}
\end{eqnarray}
It is clear that the right-hand-side (\ref{dielectric_anitHerm_trace_Append}) is 
negative if 
\begin{equation}
\Xi_s(\zeta_{sn},\tilde{v}_{s\bot}) < 0 \, .
\end{equation}
For a Maxwellian distribution, 
\begin{equation}
 \Xi_s(\zeta_{sn},\tilde{v}_{s\bot}) = \frac{\p \tilde{f}_{s0}}{\p \tilde{v}_{s\bot}}(\zeta_{sn},\tilde{v}_{s\bot}) = - 2 \tilde{v}_{s\bot} \exp 
\left(-\tilde{v}_{s\bot}^2\right) \exp 
\left(-\zeta_{sn}^2\right) < 0 \, ,
\end{equation}
and thus $\mathrm{i} \left(\mathfrak{E}_{xx}^{A}+\mathfrak{E}_{yy}^{A}+\mathfrak{E}_{zz}^{A}\right) < 
0$, as required. For the CE distribution (\ref{ChapEnskogFunc_s}), 
\begin{eqnarray}
 \Xi_s(\tilde{v}_{s\|},\tilde{v}_{s\bot}) & = &  - \tilde{v}_{s\bot} \exp \left(-\tilde{v}_{s}^2\right) 
\Bigg\{2  + \eta_s \left[2 \tilde{v}_{s\|} A_s(\tilde{v}_s) - \frac{\tilde{v}_{s\|}}{\tilde{v}_s} A_s'(\tilde{v}_s) \right] 
\nonumber \\
&& + \epsilon_s \left[ 2 C_s(\tilde{v}_s) \left(\tilde{v}_{s\|}^2-\frac{\tilde{v}_{s\bot}^2}{2}+\frac{1}{2}\right) 
- \frac{1}{\tilde{v}_s} \left(\tilde{v}_{s\|}^2-\frac{\tilde{v}_{s\bot}^2}{2}\right) 
C_s'(\tilde{v}_s) \right] \Bigg\} \nonumber \\
&& - \frac{\tilde{v}_{s\bot}}{ \tilde{\omega}_{s\|}} \exp \left(-\tilde{v}_{s}^2\right) \left[\eta_s A_s(\tilde{v}_s) - 3 \epsilon_s C_s(\tilde{v}_s)  \tilde{v}_{s\|} \right] 
\, . \label{CE_dist_Xi}
\end{eqnarray}
For $|\tilde{\omega}_{s\|}| \gtrsim 1$, it is clear for $\tilde{v}_s \lesssim 
1$ that the largest contribution to $\Xi_s(\tilde{v}_{s\|},\tilde{v}_{s\bot})$ comes 
from the Maxwellian term; the non-Maxwellian terms are 
$\textit{O}(\eta_s,\epsilon_s)$. Thus, for $\zeta_{sn}, \tilde{v}_{s\bot} \lesssim 
1$, $\Xi_s(\zeta_{sn},\tilde{v}_{s\bot}) < 0$. As discussed in 
appendix (\ref{high_freq_electrostatic_stab}), for $\zeta_{sn} \gg 1$, 
the sign of $\Xi_s(\zeta_{sn},\tilde{v}_{s\bot}) < 0$ can in principle be 
reversed. However, the magnitude of $\Xi_s(\zeta_{sn},\tilde{v}_{s\bot})$ is exponentially small for
such $\zeta_{sn}$, and thus so is $\varrho_2$. 

The remaining conditions (\ref{gen_stab_conditions}\textit{a}) and (\ref{gen_stab_conditions}\textit{b}) are much more tedious to 
treat; thus for simplicity, we explicitly consider only the case when a single particle species provides the dominant contribution to the dielectric tensor.
Under this assumption, it can be shown that
\begin{eqnarray}
 \mathfrak{E}_{xx}^{A} \mathfrak{E}_{yy}^{A} &+& \mathfrak{E}_{yy}^{A} \mathfrak{E}_{zz}^{A} \, + \, \mathfrak{E}_{zz}^{A} \mathfrak{E}_{xx}^{A} + (\mathfrak{E}_{xy}^{A})^2+(\mathfrak{E}_{yz}^{A})^2 + (\mathfrak{E}_{xz}^{A})^2 \nonumber \\
 & = & 2 \upi \frac{\omega_{\mathrm{p}s}^4}{\varpi^4} \tilde{\varpi}_{s\|}^2 \sum_{m = -\infty}^{\infty} \sum_{n = -\infty}^{\infty} \bigg\{ \int_0^{\infty} \mathrm{d} \tilde{v}_{s\bot}^{(1)} \int_0^{\infty} \mathrm{d} \tilde{v}_{s\bot}^{(2)} \,  \tilde{v}_{s\bot}^{(1)} \tilde{v}_{s\bot}^{(2)} \nonumber \\
 && \qquad \qquad \times \bigg[ \Xi_s(\zeta_{sm},\tilde{v}_{s\bot}^{(1)}) \Xi_s(\zeta_{sn},\tilde{v}_{s\bot}^{(2)}) \mathfrak{A} {\,}_{mn} \, (\alpha_s, \tilde{v}_{s\bot}^{(1)},\tilde{v}_{s\bot}^{(2)}) \bigg] \bigg\} \, , \qquad \label{dielectric_anitHerm_rho1_Append}
\end{eqnarray}
where $\alpha_s \equiv k_{\bot} \tilde{\rho}_s$ and
\begin{eqnarray}
\mathfrak{A}& {\!}_{mn}&\!(\alpha_s,\tilde{v}_{s\bot}^{(1)},\tilde{v}_{s\bot}^{(2)})
   \nonumber \\
&  \equiv & \frac{1}{\alpha_s^2}  \left[m \tilde{v}_{s\bot}^{(2)} J_m(\alpha_s \tilde{v}_{s\bot}^{(1)}) J_n'(\alpha_s \tilde{v}_{s\bot}^{(2)}) - n \tilde{v}_{s\bot}^{(1)} J_m'(\alpha_s \tilde{v}_{s\bot}^{(1)}) J_n(\alpha_s \tilde{v}_{s\bot}^{(2)}) \right]^2 
 \nonumber \\
&&    + \frac{1}{\alpha_s^2}  \left[m \zeta_{sn} \tilde{v}_{s\bot}^{(2)} J_m(\alpha_s \tilde{v}_{s\bot}^{(1)}) J_n'(\alpha_s \tilde{v}_{s\bot}^{(2)}) - n \zeta_{sm} \tilde{v}_{s\bot}^{(1)} J_m'(\alpha_s \tilde{v}_{s\bot}^{(1)}) J_n(\alpha_s \tilde{v}_{s\bot}^{(2)}) \right]^2 
\nonumber \\
&&   +  \left[\zeta_{sn} \tilde{v}_{s\bot}^{(2)} J_m(\alpha_s \tilde{v}_{s\bot}^{(1)}) J_n'(\alpha_s \tilde{v}_{s\bot}^{(2)}) - \zeta_{sm} \tilde{v}_{s\bot}^{(1)} J_m'(\alpha_s \tilde{v}_{s\bot}^{(1)}) J_n(\alpha_s \tilde{v}_{s\bot}^{(2)}) \right]^2 
 \, . \quad 
\end{eqnarray}
Being a sum of positive terms, $\mathfrak{A} {\,}_{mn}$ is positive for all $n$ 
and $m$, and thus we again conclude that the integrand on the right-hand side of (\ref{dielectric_anitHerm_rho1_Append}) 
is negative if $\Xi_s(\zeta_{sm},\tilde{v}_{s\bot}) < 0$ and $\Xi_s(\zeta_{sn},\tilde{v}_{s\bot}) < 
0$. Via similar reasoning to that applied to $\varrho_2$ in the previous paragraph, it follows that for 
the CE distribution function, the only way in which this condition can be 
violated is for either $\zeta_{sm} \gg 1$ or $\zeta_{sn} \gg 1$ -- both of 
which give rise to exponentially small terms. Thus, either $\varrho_1 > 0$ or $\varrho_1$ 
is exponentially small in $\eta_s$ and $\epsilon_s$. 

Finally, for (\ref{gen_stab_conditions}\textit{a}), it is necessary to evaluate 
$\mbox{det}\, \boldsymbol{\mathfrak{E}}^{A}$; this becomes (after much tedious algebra)
\begin{eqnarray}
\mbox{det} &\boldsymbol{\mathfrak{E}}^{A}& = -\frac{4}{3} \mathrm{i} \upi^{3/2} \frac{\omega_{\mathrm{p}s}^6}{\varpi^6} \tilde{\varpi}_{s\|}^3 \nonumber \\
&\times &  \sum_{m = -\infty}^{\infty} \sum_{n = -\infty}^{\infty}  \sum_{l = -\infty}^{\infty}  \bigg\{ \int_0^{\infty} \mathrm{d} \tilde{v}_{s\bot}^{(1)} \int_0^{\infty} \mathrm{d} \tilde{v}_{s\bot}^{(2)} \int_0^{\infty} \mathrm{d} \tilde{v}_{s\bot}^{(3)} \,  \tilde{v}_{s\bot}^{(1)} \tilde{v}_{s\bot}^{(2)} \tilde{v}_{s\bot}^{(3)} \nonumber \\
 &\times &  \bigg[ \Xi_s(\zeta_{sm},\tilde{v}_{s\bot}^{(1)}) \Xi_s(\zeta_{sn},\tilde{v}_{s\bot}^{(2)}) \Xi_s(\zeta_{sl},\tilde{v}_{s\bot}^{(3)}) \mathfrak{B} {\,}_{mnl} \, (\alpha_s, \tilde{v}_{s\bot}^{(1)},\tilde{v}_{s\bot}^{(2)},\tilde{v}_{s\bot}^{(3)}) \bigg] \bigg\} \, , \qquad \label{dielectric_anitHerm_rho0_Append}
\end{eqnarray}
where
\begin{eqnarray}
\mathfrak{B}& {\!}_{mnl}&\!(\alpha_s,\tilde{v}_{s\bot}^{(1)},\tilde{v}_{s\bot}^{(2)},\tilde{v}_{s\bot}^{(3)})
   \nonumber \\
&  \equiv & \bigg\{m J_m(\alpha_s \tilde{v}_{s\bot}^{(1)})\left[\tilde{v}_{s\bot}^{(1)} \zeta_{sn}  J_n(\alpha_s \tilde{v}_{s\bot}^{(2)}) J_l'(\alpha_s \tilde{v}_{s\bot}^{(3)}) - \tilde{v}_{s\bot}^{(3)} \zeta_{sl}  J_n'(\alpha_s \tilde{v}_{s\bot}^{(2)}) J_l(\alpha_s \tilde{v}_{s\bot}^{(3)}) 
\right]
 \nonumber \\
 & &  + n J_n(\alpha_s \tilde{v}_{s\bot}^{(1)})\left[\tilde{v}_{s\bot}^{(2)} \zeta_{sl}  J_l(\alpha_s \tilde{v}_{s\bot}^{(2)}) J_m'(\alpha_s \tilde{v}_{s\bot}^{(3)}) - \tilde{v}_{s\bot}^{(1)} \zeta_{sm}  J_l'(\alpha_s \tilde{v}_{s\bot}^{(2)}) J_m(\alpha_s \tilde{v}_{s\bot}^{(3)}) 
\right]
 \nonumber \\
&& \qquad \qquad \qquad + l J_l(\alpha_s \tilde{v}_{s\bot}^{(1)})\left[\tilde{v}_{s\bot}^{(3)} \zeta_{sm}  J_m(\alpha_s \tilde{v}_{s\bot}^{(2)}) J_n'(\alpha_s \tilde{v}_{s\bot}^{(3)}) 
\right. \nonumber \\
&&  \qquad \qquad \qquad \qquad \qquad  \left. - \tilde{v}_{s\bot}^{(2)} \zeta_{sn}  J_m'(\alpha_s \tilde{v}_{s\bot}^{(2)}) J_n(\alpha_s \tilde{v}_{s\bot}^{(3)}) \right] \bigg\}^2 
 \, . \quad 
\end{eqnarray}
Similarly to $\mathfrak{A} {\,}_{mn}$, $\mathfrak{B} {\,}_{mnl}$ is strictly 
positive for all $m$, $n$ and $l$, meaning that the integrand  on the right-hand side of (\ref{dielectric_anitHerm_rho0_Append}) 
is negative if $\Xi_s(\zeta_{sm},\tilde{v}_{s\bot}) < 0$, $\Xi_s(\zeta_{sn},\tilde{v}_{s\bot}) < 
0$, and $\Xi_s(\zeta_{sl},\tilde{v}_{s\bot}) < 0$. For the CE distribution, 
exactly the same argument as before applies to show that either $\varrho_0 > 
0$ or it is exponentially small. 

In summary, we have now verified that the only situation in which the stability 
conditions (\ref{gen_stab_conditions}) are not satisfied are those for which 
$\varrho_0$, $\varrho_1$ and $\varrho_2$ are exponentially small in $\eta_s$ and $\epsilon_s$. In the latter case, considerations of 
bounds (\ref{dielectric_antiHerm_eigenval_bound}) and (\ref{dielectric_antiHerm_eigenval_bound_vals}) 
implies that $\varsigma_a$, $\varsigma_b$, and $\varsigma_c$ are also all 
exponentially small in $\eta_s$ and $\epsilon_s$. The claim of the appendix 
follows. 

\section{Properties of leading-order expansion $\boldsymbol{\mathfrak{E}}^{(0)}$ of dielectric tensor (\ref{dielecttensfull}) in $\tilde{\omega}_{s\|} \ll 1$ for a weakly anisotropic distribution function} 
\label{DispLowFreq}

\subsection{Symmetries of $\boldsymbol{\mathfrak{E}}_s^{(0)}$ in coordinate basis $\left\{\hat{\boldsymbol{x}},\hat{\boldsymbol{y}},\hat{\boldsymbol{z}}\right\}$}

In this appendix, we show that the leading-order expansion $\boldsymbol{\mathfrak{E}}_s^{(0)}$ [cf. (\ref{Dielectric_0}\textit{a})] of the dielectric tensor $\boldsymbol{\mathfrak{E}}_s$ of species $s$ [cf. (\ref{dielectric_species_s})] in $\tilde{\omega}_{s\|} \ll 1$ arising in a non-relativistic plasma with only weak anisotropy
of its particle distribution function obeys additional symmetries 
(\ref{dielectricsymsB}), viz., 
\begin{subeqnarray}
  (\boldsymbol{\mathfrak{E}}_s^{(0)})_{xz} & = & - \frac{k_{\bot}}{k_{\|}} (\boldsymbol{\mathfrak{E}}_s^{(0)})_{xx} \, , \\
   (\boldsymbol{\mathfrak{E}}_s^{(0)})_{yz} & = & \frac{k_{\bot}}{k_{\|}} (\boldsymbol{\mathfrak{E}}_s^{(0)})_{xy} \, , \\
   (\boldsymbol{\mathfrak{E}}_s^{(0)})_{zz} & = & \frac{k_{\bot}^2}{k_{\|}^2} (\boldsymbol{\mathfrak{E}}_s^{(0)})_{xx} \, . 
  \label{dielectricsymsB_Append}
\end{subeqnarray}
when $k \rho_s \sim 1$. 
The term `weak anisotropy' means that the magnitude of angular anisotropy -- mathematically represented by the function $\Lambda_s$ defined by~(\ref{anisotropyfunc}) -- satisfies 
$\Lambda_s \lesssim \tilde{\omega}_{s\|}$ for all particle species when $\tilde{v}_s \sim 
1$. 

We begin the proof by substituting (\ref{conductivity}) into (\ref{dielectric_species_s}) to give 
\begin{eqnarray}
\boldsymbol{\mathfrak{E}}_s & \equiv & \frac{\omega_{\mathrm{p}s}^2}{\omega^2} \bigg[ \frac{2}{\sqrt{\upi}} \frac{k_{\|}}{|k_{\|}|}  \int_{-\infty}^{\infty} \mathrm{d} \tilde{v}_{s\|} \, \tilde{v}_{s\|} \int_0^{\infty} \mathrm{d} \tilde{v}_{s\bot} \Lambda_s(\tilde{v}_{s\|},\tilde{v}_{s\bot}) \hat{\boldsymbol{z}} \hat{\boldsymbol{z}} \nonumber \\
&& \mbox{} + \tilde{\omega}_{s\|} \frac{2}{\sqrt{\upi}} \int_{C_L} \mathrm{d} \tilde{v}_{s\|} \int_0^{\infty} \mathrm{d} \tilde{v}_{s\bot} \tilde{v}_{s\bot}^2 \Xi_s(\tilde{v}_{s\|},\tilde{v}_{s\bot}) \sum_{n = -\infty}^{\infty} \frac{\mathsfbi{R}_{sn}}{\zeta_{sn} -\tilde{v}_{s\|}}  \bigg] \, . \label{conductivity_S}
\end{eqnarray}
Then, under the assumed ordering $\tilde{\omega}_{s\|} \sim \Lambda_s$,  the function $\Xi_s$ defined by (\ref{IntgradCond}) satisfies $\Xi_s \sim 1$ for $\tilde{v}_s \sim 1$; therefore, $\boldsymbol{\mathfrak{E}}_s$
has order-unity elements as $\tilde{\omega}_{s\|} \rightarrow 0$. Let us expand $\boldsymbol{\mathfrak{E}}_s$ in a Taylor series around $\tilde{\omega}_{s\|} = 0$:
\begin{equation}
 \boldsymbol{\mathfrak{E}}_s = \tilde{\omega}_{s\|} \boldsymbol{\mathfrak{E}}_s^{(0)} 
  + \delta \boldsymbol{\mathfrak{E}}_s ,
\end{equation}
where $\delta \boldsymbol{\mathfrak{E}}_s = \textit{O}(\tilde{\omega}_{s\|}^2)$, and the matrix elements of $\boldsymbol{\mathfrak{E}}_s^{(0)}$ 
are given below: 
\begin{subeqnarray}
 (\boldsymbol{\mathfrak{E}}_s^{(0)} )_{xx} & \equiv &  -\frac{2 \omega_{\mathrm{p}s}^2}{\sqrt{\upi} \omega^2} \sum_{n=-\infty}^{\infty} \Bigg[ \frac{n^2}{k_{\bot}^2 \tilde{\rho}_s^2}\int_{C_L} \frac{\mathrm{d} \tilde{v}_{s\|}}{\tilde{v}_{s\|}+n/|k_{\|}| \tilde{\rho}_s} \nonumber \\
 && \qquad \qquad \qquad \times \int_0^{\infty} \mathrm{d} \tilde{v}_{s\bot} \Xi_s(\tilde{v}_{s\|},\tilde{v}_{s\bot}) J_n(k_{\bot} \tilde{\rho}_s \tilde{v}_{s\bot})^2 \Bigg] ,  \\
 (\boldsymbol{\mathfrak{E}}_s^{(0)} )_{xy} & \equiv & -\frac{2 \mathrm{i} \omega_{\mathrm{p}s}^2}{\sqrt{\upi}  \omega^2} \sum_{n=-\infty}^{\infty} \left[ \frac{n}{k_{\bot} \tilde{\rho}_s} \int_{C_L} \frac{\mathrm{d} \tilde{v}_{s\|}}{\tilde{v}_{s\|}+n/|k_{\|}| \tilde{\rho}_s} \right. \nonumber \\
 && \left. \qquad \qquad \quad \times \int_0^{\infty} \mathrm{d} \tilde{v}_{s\bot} \, \tilde{v}_{s\bot} \Xi_s(\tilde{v}_{s\|},\tilde{v}_{s\bot}) J_n(k_{\bot} \tilde{\rho}_s \tilde{v}_{s\bot}) J_n'(k_{\bot} \tilde{\rho}_s \tilde{v}_{s\bot}) \right] , \, \\
 (\boldsymbol{\mathfrak{E}}_s^{(0)} )_{xz} & \equiv & -\frac{2 \omega_{\mathrm{p}s}^2}{\sqrt{\upi}  \omega^2} \sum_{n=-\infty}^{\infty} \Bigg[ \frac{n}{k_{\bot} \tilde{\rho}_s}\int_{C_L} \frac{\tilde{v}_{s\|} \mathrm{d} \tilde{v}_{s\|}}{\tilde{v}_{s\|}+n/|k_{\|}| \tilde{\rho}_s} \nonumber \\ 
 && \qquad \qquad \qquad \times \int_0^{\infty} \mathrm{d} \tilde{v}_{s\bot} \Xi_s(\tilde{v}_{s\|},\tilde{v}_{s\bot}) J_n(k_{\bot} \tilde{\rho}_s \tilde{v}_{s\bot})^2 \Bigg] , \\
 (\boldsymbol{\mathfrak{E}}_s^{(0)} )_{yx} & \equiv & - (\boldsymbol{\mathfrak{E}}_s^{(0)} )_{xy} ,\\
 (\boldsymbol{\mathfrak{E}}_s^{(0)} )_{yy} & \equiv & -\frac{2 \omega_{\mathrm{p}s}^2}{\sqrt{\upi}  \omega^2} \sum_{n=-\infty}^{\infty} \Bigg[ \int_{C_L} \frac{\mathrm{d} \tilde{v}_{s\|}}{\tilde{v}_{s\|}+n/|k_{\|}| \tilde{\rho}_s} \nonumber \\
 && \qquad \qquad \qquad \times \int_0^{\infty} \mathrm{d} \tilde{v}_{s\bot} \, \tilde{v}_{s\bot}^2 \Xi_s(\tilde{v}_{s\|},\tilde{v}_{s\bot}) J_n'(k_{\bot} \tilde{\rho}_s \tilde{v}_{s\bot})^2 \Bigg] , \\
 (\boldsymbol{\mathfrak{E}}_s^{(0)} )_{yz} & \equiv & -\frac{2 \mathrm{i} \omega_{\mathrm{p}s}^2}{\sqrt{\upi}  \omega^2} \sum_{n=-\infty}^{\infty}  \left[  \int_{C_L} \frac{\tilde{v}_{s\|} \mathrm{d} \tilde{v}_{s\|}}{\tilde{v}_{s\|}+n/|k_{\|}| \tilde{\rho}_s} \right. \nonumber \\
 && \left. \qquad \qquad \quad \times \int_0^{\infty} \mathrm{d} \tilde{v}_{s\bot} \, \tilde{v}_{s\bot} \Xi_s(\tilde{v}_{s\|},\tilde{v}_{s\bot}) J_n(k_{\bot} \tilde{\rho}_s \tilde{v}_{s\bot}) J_n'(k_{\bot} \tilde{\rho}_s \tilde{v}_{s\bot}) \right] , \\
 (\boldsymbol{\mathfrak{E}}_s^{(0)} )_{zx} & \equiv & (\boldsymbol{\mathfrak{E}}_s^{(0)} )_{xz} ,\\
 (\boldsymbol{\mathfrak{E}}_s^{(0)} )_{zy} & \equiv & -(\boldsymbol{\mathfrak{E}} _s^{(0)} )_{yz} ,\\
 (\boldsymbol{\mathfrak{E}}_s^{(0)} )_{zz} & \equiv & \frac{2 \omega_{\mathrm{p}s}^2}{\sqrt{\upi} \tilde{\omega}_{s\|} \omega^2} \int_{-\infty}^{\infty} \mathrm{d} \tilde{v}_{s\|} \tilde{v}_{s\|} \int_0^{\infty} \mathrm{d} \tilde{v}_{s\bot} \Lambda_s(\tilde{v}_{s\|},\tilde{v}_{s\bot}) \nonumber \\
 & & - \frac{2 \omega_{\mathrm{p}s}^2}{\sqrt{\upi}  \omega^2} \sum_{n=-\infty}^{\infty} \int_{C_L} \frac{\tilde{v}_{s\|}^2 \mathrm{d} \tilde{v}_{s\|}}{\tilde{v}_{s\|}+n/|k_{\|}| \tilde{\rho}_s} \int_0^{\infty} \mathrm{d} \tilde{v}_{s\bot} \Xi_s(\tilde{v}_{s\|},\tilde{v}_{s\bot}) J_n(k_{\bot} \tilde{\rho}_s \tilde{v}_{s\bot})^2 . \qquad
 \label{dielectric0elements}
\end{subeqnarray}

Next, noting that
\begin{equation}
  \frac{\tilde{v}_{s\|}}{\tilde{v}_{s\|}+n/|k_{\|}| \tilde{\rho}_s}  
  = 1 - \frac{n}{|k_{\|}| \tilde{\rho}_s} \frac{\tilde{v}_{s\|}}{\tilde{v}_{s\|}+n/|k_{\|}| \tilde{\rho}_s}   
  ,
\end{equation}
as well as 
\begin{equation}
 \sum_{n=-\infty}^{\infty} \frac{n}{k_{\bot} \tilde{\rho}_s} \int_{C_L} \mathrm{d} \tilde{v}_{s\|} \int_0^{\infty} \mathrm{d} \tilde{v}_{s\bot} \Xi_s(\tilde{v}_{s\|},\tilde{v}_{s\bot}) J_n(k_{\bot} \tilde{\rho}_s \tilde{v}_{s\bot})^2   
  = 0,
\end{equation}
we see that the double integral in (\ref{dielectric0elements}\textit{c})
can be rearranged to give
\begin{eqnarray}
   (\boldsymbol{\mathfrak{E}}_s^{(0)})_{xz} & = & \frac{2 \omega_{\mathrm{p}s}^2}{\sqrt{\upi} \omega^2} \sum_{n=-\infty}^{\infty} \Bigg[ \frac{n^2}{ |k_{\|}| k_{\bot} \tilde{\rho}_s^2}\int_{C_L} \frac{\mathrm{d} \tilde{v}_{s\|}}{\tilde{v}_{s\|}+n/|k_{\|}| \tilde{\rho}_s} \nonumber \\
   && \qquad \qquad \quad \times \int_0^{\infty} \mathrm{d} \tilde{v}_{s\bot} \Xi_s(\tilde{v}_{s\|},\tilde{v}_{s\bot}) J_n(k_{\bot} \tilde{\rho}_s \tilde{v}_{s\bot})^2 \Bigg] , 
  \nonumber\\
  & = & -\frac{k_{\bot}}{|k_{\|}|}  (\boldsymbol{\mathfrak{E}}_s^{(0)})_{xx} .
 \end{eqnarray}
Similarly, it can be shown that
\begin{eqnarray}
  (\boldsymbol{\mathfrak{E}}_s^{(0)})_{yz} & = & \frac{2 \mathrm{i} \omega_{\mathrm{p}s}^2}{\sqrt{\upi} \omega^2} \sum_{n=-\infty}^{\infty} \left[ \frac{n}{|k_{\|}| \tilde{\rho}_s} \int_{C_L} \frac{\mathrm{d} \tilde{v}_{s\|}}{\tilde{v}_{s\|}+n/|k_{\|}| \tilde{\rho}_s} \right. \nonumber \\
  && \left. \qquad \qquad \qquad \times \int_0^{\infty} \mathrm{d} \tilde{v}_{s\bot} \, \tilde{v}_{s\bot} \Xi_s(\tilde{v}_{s\|},\tilde{v}_{s\bot}) J_n(k_{\bot} \tilde{\rho}_s \tilde{v}_{s\bot}) J_n'(k_{\bot} \tilde{\rho}_s \tilde{v}_{s\bot}) \right] , \nonumber \\
  & = & \frac{k_{\bot}}{|k_{\|}|} (\boldsymbol{\mathfrak{E}}_s^{(0)})_{xy} . 
 \end{eqnarray}
Finally, $(\boldsymbol{\mathfrak{E}}_s^{(0)})_{zz}$ can also be written in terms of $(\boldsymbol{\mathfrak{E}}_s^{(0)})_{xx}$: because
 \begin{equation}
  \frac{\tilde{v}_{s\|}^2}{\tilde{v}_{s\|}+n/|k_{\|}| \tilde{\rho}_s}  
  = \tilde{v}_{s\|} - \frac{n}{|k_{\|}| \tilde{\rho}_s} + \frac{n^2}{|k_{\|}| ^2\tilde{\rho}_s^2}  \frac{1}{\tilde{v}_{s\|}+n/|k_{\|}| \tilde{\rho}_s}   
  ,
\end{equation}
it follows that
\begin{eqnarray}
 (\boldsymbol{\mathfrak{E}}_s^{(0)})_{zz} & = & \frac{k_{\bot}^2}{k_{\|}^2}  (\boldsymbol{\mathfrak{E}}_s^{(0)})_{xx} +\frac{2 \omega_{\mathrm{p}s}^2}{\sqrt{\upi} \tilde{\omega}_{s\|} \omega^2} \int_{-\infty}^{\infty} \mathrm{d} \tilde{v}_{s\|} \tilde{v}_{s\|} \int_0^{\infty} \mathrm{d} \tilde{v}_{s\bot} \Lambda_s(\tilde{v}_{s\|},\tilde{v}_{s\bot}) \nonumber \\
 & & - \frac{2 \omega_{\mathrm{p}s}^2}{\sqrt{\upi} \omega^2} \sum_{n=-\infty}^{\infty} \int_{-\infty}^{\infty} \tilde{v}_{s\|} \mathrm{d} \tilde{v}_{s\|} \int_0^{\infty} \mathrm{d} \tilde{v}_{s\bot} \Xi_s(\tilde{v}_{s\|},\tilde{v}_{s\bot}) J_n(k_{\bot} \tilde{\rho}_s \tilde{v}_{s\bot})^2  \qquad  
\nonumber \\
& & + \frac{2 \omega_{\mathrm{p}s}^2}{\sqrt{\upi} \omega^2} \sum_{n=-\infty}^{\infty} \frac{n}{k_{\bot} \tilde{\rho}_s} \int_{-\infty}^{\infty} \mathrm{d} \tilde{v}_{s\|} \int_0^{\infty} \mathrm{d} \tilde{v}_{s\bot} \Xi_s(\tilde{v}_{s\|},\tilde{v}_{s\bot}) J_n(k_{\bot} \tilde{\rho}_s \tilde{v}_{s\bot})^2 
\nonumber \\
& = & \frac{k_{\bot}^2}{k_{\|}^2}  (\boldsymbol{\mathfrak{E}}_s^{(0)})_{xx} - \frac{2 \omega_{\mathrm{p}s}^2}{\sqrt{\upi} \omega^2} \int_{-\infty}^{\infty} \mathrm{d} \tilde{v}_{s\|} \int_0^{\infty} \mathrm{d} \tilde{v}_{s\bot} \, \tilde{v}_{s\|} \frac{\partial \tilde{f}_{s0}}{\partial \tilde{v}_{s\bot}} \qquad   \nonumber 
\\
& & + \frac{2 \omega_{\mathrm{p}s}^2}{\sqrt{\upi} \tilde{\omega}_{s\|} \omega^2} \int_{-\infty}^{\infty} \mathrm{d} \tilde{v}_{s\|} \tilde{v}_{s\|} \int_0^{\infty} \mathrm{d} \tilde{v}_{s\bot} \Lambda_s(\tilde{v}_{s\|},\tilde{v}_{s\bot}) \left[1-\sum_{n=-\infty}^{\infty} J_n(k_{\bot} \tilde{\rho}_s \tilde{v}_{s\bot})^2 \right] \nonumber \\
& = & \frac{k_{\bot}^2}{k_{\|}^2}  (\boldsymbol{\mathfrak{E}}_s^{(0)})_{xx} + \frac{2 \omega_{\mathrm{p}s}^2}{\sqrt{\upi} \omega^2} \int_{-\infty}^{\infty} \mathrm{d} \tilde{v}_{s\|} \int_0^{\infty} \mathrm{d} \tilde{v}_{s\bot} 
\Lambda_s(\tilde{v}_{s\|},\tilde{v}_{s\bot}) ,
\end{eqnarray}
where we have used the identity
\begin{equation}
  \sum_{n=-\infty}^{\infty} J_n(k_{\bot} \tilde{\rho}_s \tilde{v}_{s\bot})^2 = 1
  .
\end{equation}
Thus, we conclude that since the anisotropy is assumed small, 
\begin{equation}
  (\boldsymbol{\mathfrak{E}}_s^{(0)})_{zz} = \frac{k_{\bot}^2}{k_{\|}^2}  (\boldsymbol{\mathfrak{E}}_s^{(0)})_{xx} 
  + \textit{O}(\tilde{\omega}_{s\|})  ,
\end{equation}
completing the proof.

\subsection{Evaluating the dielectric tensor in coordinate basis $\{\boldsymbol{e}_1,\boldsymbol{e}_2,\boldsymbol{e}_3\}$}

To demonstrate that the components of the dielectric tensor $\boldsymbol{\mathfrak{E}}_s^{(0)}$ are given by 
(\ref{dielectric123}), viz., 
\begin{subeqnarray}
  (\boldsymbol{\mathfrak{E}}_s^{(0)})_{11} & =  & \frac{k^2}{k_{\|}^2} (\boldsymbol{\mathfrak{E}}_s^{(0)})_{xx}  \, , \\
   (\boldsymbol{\mathfrak{E}}_s^{(0)})_{12} & = & -  (\boldsymbol{\mathfrak{E}}_s^{(0)})_{21} = \frac{k}{k_{\|}}  (\boldsymbol{\mathfrak{E}}_s^{(0)})_{xy}  \, , \\
  (\boldsymbol{\mathfrak{E}}_s^{(0)})_{22} & = &  (\boldsymbol{\mathfrak{E}}_s^{(0)})_{yy} \, , 
     \label{dielectric123_append}
\end{subeqnarray}
we use (\ref{dielectricsymsB_Append}) to express $\boldsymbol{\mathfrak{E}_s}^{(0)}$ 
in the form
\begin{eqnarray}
 \boldsymbol{\mathfrak{E}}_s^{(0)} & = & (\boldsymbol{\mathfrak{E}}_s^{(0)})_{xx} \hat{\boldsymbol{x}} \hat{\boldsymbol{x}} 
  + (\boldsymbol{\mathfrak{E}}_s^{(0)})_{xy} \left(\hat{\boldsymbol{x}} \hat{\boldsymbol{y}} 
  -\hat{\boldsymbol{y}} \hat{\boldsymbol{x}}\right) + (\boldsymbol{\mathfrak{E}}_s^{(0)})_{yy} \hat{\boldsymbol{y}} \hat{\boldsymbol{y}} 
 \nonumber \\
  && - \frac{k_{\bot}}{|k_{\|}|} (\boldsymbol{\mathfrak{E}}_s^{(0)})_{xx} \left(\hat{\boldsymbol{x}} \hat{\boldsymbol{z}} 
  +\hat{\boldsymbol{z}} \hat{\boldsymbol{x}}\right) + \frac{k_{\bot}}{|k_{\|}|} (\boldsymbol{\mathfrak{E}}_s^{(0)})_{xy} \left(\hat{\boldsymbol{y}} \hat{\boldsymbol{z}} 
  -\hat{\boldsymbol{z}} \hat{\boldsymbol{y}}\right) + \frac{k_{\bot}^2}{k_{\|}^2} (\boldsymbol{\mathfrak{E}}_s^{(0)})_{xx}  \hat{\boldsymbol{z}} \hat{\boldsymbol{z}}  
 \,  . \qquad \label{dielectric_0_tensorexpress}
\end{eqnarray}
Noting that 
\begin{subeqnarray}
  \hat{\boldsymbol{k}} & = & \frac{k_{\bot}}{k} \hat{\boldsymbol{x}} +\frac{k_{\|}}{k} \hat{\boldsymbol{z}}  
  , \\
  \hat{\boldsymbol{y}} \times \hat{\boldsymbol{k}} & = & \frac{k_{\|}}{k} \hat{\boldsymbol{x}} -\frac{k_{\bot}}{k} \hat{\boldsymbol{z}} 
  ,
\end{subeqnarray}
we can rewrite (\ref{dielectric_0_tensorexpress}) as
\begin{eqnarray}
  \boldsymbol{\mathfrak{E}}_s^{(0)} & = & \frac{k^2}{k_{\|}^2} (\boldsymbol{\mathfrak{E}}_s^{(0)})_{xx}  \left(\hat{\boldsymbol{y}} \times \hat{\boldsymbol{k}}\right) 
   \left(\hat{\boldsymbol{y}} \times \hat{\boldsymbol{k}}\right) \nonumber \\
  && + \frac{k}{|k_{\|}|} (\boldsymbol{\mathfrak{E}}_s^{(0)})_{xy}  \left[ \left(\hat{\boldsymbol{y}} \times \hat{\boldsymbol{k}}\right)\hat{\boldsymbol{y}} 
  -\hat{\boldsymbol{y}}  \left(\hat{\boldsymbol{y}} \times \hat{\boldsymbol{k}}\right)\right] + (\boldsymbol{\mathfrak{E}}_s^{(0)})_{yy}\hat{\boldsymbol{y}} \hat{\boldsymbol{y}}  
   \\
  & = & \frac{k^2}{k_{\|}^2} (\boldsymbol{\mathfrak{E}}_s^{(0)})_{xx} \boldsymbol{e}_1 \boldsymbol{e}_1 + \frac{k}{|k_{\|}|} (\boldsymbol{\mathfrak{E}}_s^{(0)})_{xy}  \left(\boldsymbol{e}_1 
  \boldsymbol{e}_2 - \boldsymbol{e}_2 \boldsymbol{e}_1 \right) + (\boldsymbol{\mathfrak{E}}_s^{(0)})_{yy} 
  \boldsymbol{e}_2 \boldsymbol{e}_2 \, ,
\end{eqnarray}
leading to the desired results (\ref{dielectric123_append}). In addition, we see that $\boldsymbol{\mathfrak{E}}_s^{(0)} \cdot \hat{\boldsymbol{k}} = 0$; thus, the results (\ref{dielectric3_smallterm}) claiming 
that certain components of  $\boldsymbol{\mathfrak{E}}_s$ are small in $\tilde{\omega}_{s\|}$ 
are justified. 

\section{Dielectric tensor components for the CE distribution function (\ref{ChapEnskogFunc})} \label{ChapEnskogFunc_Append_comp} 

In this appendix, we calculate the components of the dielectric tensor arising 
from the CE distribution function (\ref{ChapEnskogFunc}), with isotropic functions $A_e^T(\tilde{v}_e)$, $A_e^R(\tilde{v}_e)$, $A_e^u(\tilde{v}_e)$, $C_e(\tilde{v}_e)$, $A_i(\tilde{v}_i)$ and 
$C_i(\tilde{v}_i)$ chosen as appropriate for a Krook collision operator (see appendix \ref{ChapEnskogIsoFunc_Krook}), viz., 
\begin{subeqnarray}
  A_e^T(\tilde{v}_e) & = & -\left(\tilde{v}_e^2 - \frac{5}{2}\right) \, , \\
  A_e^R(\tilde{v}_e) & = & -1 \, , \\
  A_e^u(\tilde{v}_e) & = & 0 \, , \\
  A_i(\tilde{v}_i) & = & -\left(\tilde{v}_i^2 - \frac{5}{2}\right) \, , \\
  C_e(\tilde{v}_e) & = & -1 \, , \\
  C_i(\tilde{v}_i) & = & -1 \, . 
\end{subeqnarray}
This, via (\ref{dielectric_0_multispecies}), allows for the dielectric 
tensor $\boldsymbol{\mathfrak{E}}_s$ 
to be calculated order by order in $\tilde{\omega}_{s\|}$. 
We carry out these calculations in the case of non-relativistic fluctuations, 
and so
\begin{equation}
\boldsymbol{\mathfrak{E}} \approx \frac{4 \upi \mathrm{i}}{\omega} \boldsymbol{\sigma} = \sum_s \boldsymbol{\mathfrak{E}}_s ,
\end{equation}
where we remind the reader that [cf. (\ref{conductivity_S})]
\begin{eqnarray}
\boldsymbol{\mathfrak{E}}_s & = & \frac{\omega_{\mathrm{p}s}^2}{\omega^2} \bigg[\frac{2}{\sqrt{\upi}} \frac{k_{\|}}{|k_{\|}|}  \int_{-\infty}^{\infty} \mathrm{d} \tilde{v}_{s\|} \, \tilde{v}_{s\|} \int_0^{\infty} \mathrm{d} \tilde{v}_{s\bot} \Lambda_s(\tilde{v}_{s\|},\tilde{v}_{s\bot}) \hat{\boldsymbol{z}} \hat{\boldsymbol{z}} \nonumber \\
&& \mbox{} + \tilde{\omega}_{s\|} \frac{2}{\sqrt{\upi}} \int_{C_L} \mathrm{d} \tilde{v}_{s\|} \int_0^{\infty} \mathrm{d} \tilde{v}_{s\bot} \tilde{v}_{s\bot}^2 \Xi_s(\tilde{v}_{s\|},\tilde{v}_{s\bot}) \sum_{n = -\infty}^{\infty} \frac{\mathsfbi{R}_{sn}}{\zeta_{sn} -\tilde{v}_{s\|}} \bigg] \, , \label{dielectricnonrel_AppendB}
\end{eqnarray}
\begin{equation}
  \zeta_{sn} \equiv \tilde{\omega}_{s\|} - \frac{n}{|k_{\|}| \tilde{\rho}_s} ,
\end{equation}
\begin{equation}
  \tilde{f}_{s0}(\tilde{v}_{s\|},\tilde{v}_{s\bot}) \equiv \frac{\upi^{3/2} v_{\mathrm{th}s}^3}{n_{s0}} f_{s0}\left(\frac{k_{\|}}{|k_{\|}|} v_{\mathrm{th}s} \tilde{v}_{s\|},v_{\mathrm{th}s} \tilde{v}_{s\bot}\right)
  ,
\end{equation}
\begin{equation}
  \Lambda_s(\tilde{v}_{s\|},\tilde{v}_{s\bot}) \equiv \tilde{v}_{s\bot} \frac{\p \tilde{f}_{s0}}{\p \tilde{v}_{s\|}}-\tilde{v}_{s\|} \frac{\p \tilde{f}_{s0}}{\p \tilde{v}_{s\bot}} 
  , \label{anisotropyfunc_Append}
\end{equation}
\begin{equation}
  \Xi_s(\tilde{v}_{s\|},\tilde{v}_{s\bot}) \equiv \frac{\p \tilde{f}_{s0}}{\p \tilde{v}_{s\bot}}
  + \frac{\Lambda_s(\tilde{v}_{s\|},\tilde{v}_{s\bot})}{\tilde{\omega}_{s\|}} , \label{IntgradCond_Append}
\end{equation}
and
\normalsize
\begin{subeqnarray}
 (\mathsfbi{R}_{sn} )_{xx} & \equiv & \frac{n^2 J_n(k_{\bot} \tilde{\rho}_s \tilde{v}_{s\bot})^2}{k_{\bot}^2 \tilde{\rho}_s^2 \tilde{v}_{s\bot}^2} , \\
 (\mathsfbi{R}_{sn} )_{xy} & \equiv & \frac{\mathrm{i} n J_n(k_{\bot} \tilde{\rho}_s \tilde{v}_{s\bot}) J_n'(k_{\bot} \tilde{\rho}_s \tilde{v}_{s\bot})}{k_{\bot} \tilde{\rho}_s \tilde{v}_{s\bot}} , \\
 (\mathsfbi{R}_{sn} )_{xz} & \equiv & \frac{n J_n(k_{\bot} \tilde{\rho}_s \tilde{v}_{s\bot})^2}{k_{\bot} \tilde{\rho}_s \tilde{v}_{s\bot}} \frac{k_{\|} \tilde{v}_{s\|}}{|k_{\|}| \tilde{v}_{s\bot}}  ,\\
 (\mathsfbi{R}_{sn} )_{yx} & \equiv & - (\mathsfbi{R}_{sn} )_{xy}\\
 (\mathsfbi{R}_{sn} )_{yy} & \equiv & J_n'(k_{\bot} \tilde{\rho}_s \tilde{v}_{s\bot})^2 ,\\
 (\mathsfbi{R}_{sn} )_{yz} & \equiv & \mathrm{i} n J_n(k_{\bot} \tilde{\rho}_s \tilde{v}_{s\bot}) J_n'(k_{\bot} \tilde{\rho}_s \tilde{v}_{s\bot}) \frac{k_{\|} \tilde{v}_{s\|}}{|k_{\|}| \tilde{v}_{s\bot}} , \\
 (\mathsfbi{R}_{sn} )_{zx} & \equiv & (\mathsfbi{R}_{sn} )_{xz}\\
 (\mathsfbi{R}_{sn} )_{zy} & \equiv & -(\mathsfbi{R}_{sn} )_{yz}\\
 (\mathsfbi{R}_{sn} )_{zz} & \equiv & \frac{\tilde{v}_{s\|}^2}{\tilde{v}_{s\bot}^2} J_n(k_{\bot} \tilde{\rho}_s \tilde{v}_{s\bot})^2 . \label{defMsn_Append}
\end{subeqnarray}
The components of the dielectric tensor $\boldsymbol{\mathfrak{E}}_s$ in coordinate basis $\{\boldsymbol{e}_1,\boldsymbol{e}_2,\boldsymbol{e}_3\}$ 
are related to the components in coordinate basis $\left\{\hat{\boldsymbol{x}},\hat{\boldsymbol{y}},\hat{\boldsymbol{z}}\right\}$ 
by
\begin{subeqnarray}
 (\boldsymbol{\mathfrak{E}}_s)_{11} & = & \frac{k_{\|}^2}{k^2} (\boldsymbol{\mathfrak{E}}_s)_{xx} - \frac{2 k_{\|} k_{\bot}}{k^2} (\boldsymbol{\mathfrak{E}}_s)_{xz} + \frac{k_{\bot}^2}{k^2} (\boldsymbol{\mathfrak{E}}_s)_{zz}\, , \\
 (\boldsymbol{\mathfrak{E}}_s)_{12} & = & \frac{k_{\|}}{k}(\boldsymbol{\mathfrak{E}}_s)_{xy} + \frac{k_{\bot}}{k}(\boldsymbol{\mathfrak{E}}_s)_{yz} \,  ,\\
 (\boldsymbol{\mathfrak{E}}_s)_{13} & = & \frac{k_{\|} k_{\bot}}{k^2} \left[(\boldsymbol{\mathfrak{E}}_s)_{xx} - (\boldsymbol{\mathfrak{E}}_s)_{zz}\right] + \left(\frac{k_{\|}^2}{k^2}- \frac{k_{\bot}^2}{k^2}\right) (\boldsymbol{\mathfrak{E}}_s)_{xz} \, , \\
 (\boldsymbol{\mathfrak{E}}_s)_{21} & = & -(\boldsymbol{\mathfrak{E}}_s)_{12} \, , \\
 (\boldsymbol{\mathfrak{E}}_s)_{22} & = & (\boldsymbol{\mathfrak{E}}_s)_{yy} \, , \\
 (\boldsymbol{\mathfrak{E}}_s)_{23} & = & -\frac{k_{\bot}}{k}(\boldsymbol{\mathfrak{E}}_s)_{xy} + \frac{k_{\|}}{k}(\boldsymbol{\mathfrak{E}}_s)_{yz} \, , \\
 (\boldsymbol{\mathfrak{E}}_s)_{31} & = & (\boldsymbol{\mathfrak{E}}_s)_{13} \, , \\
 (\boldsymbol{\mathfrak{E}}_s)_{32} & = & -(\boldsymbol{\mathfrak{E}}_s)_{23} \, , \\
 (\boldsymbol{\mathfrak{E}}_s)_{33} & = & \frac{k_{\bot}^2}{k^2} (\boldsymbol{\mathfrak{E}}_s)_{xx} + \frac{2 k_{\|} k_{\bot}}{k^2} (\boldsymbol{\mathfrak{E}}_s)_{xz} + \frac{k_{\|}^2}{k^2} (\boldsymbol{\mathfrak{E}}_s)_{zz} \, 
 . \label{dielectric123trans}
 \end{subeqnarray}
For clarity, we calculate separately the Maxwellian 
contribution $\mathsfbi{M}_s$ of the total CE distribution function and the non-Maxwellian contribution $\mathsfbi{P}_s$ associated with the CE 
electron friction, temperature-gradient, and shear terms to $\boldsymbol{\mathfrak{E}}_s$ -- viz., we decompose $\boldsymbol{\mathfrak{E}}_s$ as follows [cf. (\ref{Maxnonmaxsep_s})]:
\begin{equation}
\boldsymbol{\mathfrak{E}}_s = \frac{\omega_{\mathrm{p}s}^2}{\omega^2} \left(\mathsfbi{M}_s + \mathsfbi{P}_s \right) 
\label{Maxnonmaxsep_s_Append}
\, .
\end{equation}

\subsection{Maxwellian distribution} \label{Maxwellresponse}

\subsubsection{General dielectric tensor} \label{Maxwellresponse_gen}

Consider a non-dimensionalised Maxwellian distribution function:
\begin{equation}
 \tilde{f}_s(\tilde{v}_{s\|},\tilde{v}_{s\bot}) = \exp \left(-\tilde{v}_{s}^2\right) .  
\end{equation}
The Maxwellian is isotropic, so (\ref{anisotropyfunc_Append}) gives
\begin{equation}
 \Lambda_s(\tilde{v}_{s\|},\tilde{v}_{s\bot}) = 0,  
\end{equation}
while (\ref{IntgradCond_Append}) becomes
\begin{equation}
  \Xi_s(\tilde{v}_{s\|},\tilde{v}_{s\bot}) = - 2 \tilde{v}_{s\bot} \exp \left(-\tilde{v}_{s}^2\right) . 
  \label{Xi_maxwellian}
\end{equation}
Substituting this into (\ref{dielectricnonrel_AppendB}) gives 
\begin{subeqnarray}
 (\mathsfbi{M}_s)_{xx} & = &  \frac{4}{\sqrt{\upi}} \tilde{\omega}_{s\|} \sum_{n=-\infty}^{\infty} \left[ \frac{n^2}{k_{\bot}^2 \tilde{\rho}_s^2}\int_{C_L} \frac{\exp \left(-\tilde{v}_{s\|}^2\right) \mathrm{d} \tilde{v}_{s\|}}{\tilde{v}_{s\|}-\zeta_{sn}} \right. \nonumber \\
 && \left. \qquad \qquad \qquad \times \int_0^{\infty} \mathrm{d} \tilde{v}_{s\bot} \, \tilde{v}_{s\bot} J_n(k_{\bot} \tilde{\rho}_s \tilde{v}_{s\bot})^2 \exp \left(-\tilde{v}_{s\bot}^2\right) \right] , \qquad\\
 (\mathsfbi{M}_s)_{xy} & = & \frac{4 \mathrm{i}}{\sqrt{\upi}} \tilde{\omega}_{s\|} \sum_{n=-\infty}^{\infty} \left[ \frac{n}{k_{\bot} \tilde{\rho}_s} \int_{C_L} \frac{\exp \left(-\tilde{v}_{s\|}^2\right) \mathrm{d} \tilde{v}_{s\|}}{\tilde{v}_{s\|}-\zeta_{sn}} \right. \nonumber \\
 && \left. \qquad \times \int_0^{\infty} \mathrm{d} \tilde{v}_{s\bot} \, \tilde{v}_{s\bot}^2 J_n(k_{\bot} \tilde{\rho}_s \tilde{v}_{s\bot}) J_n'(k_{\bot} \tilde{\rho}_s \tilde{v}_{s\bot}) \exp \left(-\tilde{v}_{s\bot}^2\right) \right] \, , \\
 (\mathsfbi{M}_s)_{xz} & = & \frac{4}{\sqrt{\upi}} \tilde{\omega}_{s\|} \sum_{n=-\infty}^{\infty} \left[ \frac{n}{k_{\bot} \tilde{\rho}_s}\int_{C_L} \frac{\tilde{v}_{s\|} \exp \left(-\tilde{v}_{s\|}^2\right) \mathrm{d} \tilde{v}_{s\|}}{\tilde{v}_{s\|}-\zeta_{sn}} \right. \\
 && \left. \qquad \qquad \qquad \times \int_0^{\infty} \mathrm{d} \tilde{v}_{s\bot} \, \tilde{v}_{s\bot} J_n(k_{\bot} \tilde{\rho}_s \tilde{v}_{s\bot})^2 \exp \left(-\tilde{v}_{s\bot}^2\right) \right] , \qquad\\
 (\mathsfbi{M}_s)_{yx} & = & (\mathsfbi{M}_s)_{xy} ,\\
 (\mathsfbi{M}_s)_{yy} & = & \frac{4}{\sqrt{\upi}} \tilde{\omega}_{s\|} \sum_{n=-\infty}^{\infty} \left[ \int_{C_L} \frac{\exp \left(-\tilde{v}_{s\|}^2\right) \mathrm{d} \tilde{v}_{s\|}}{\tilde{v}_{s\|}-\zeta_{sn}} \right. \nonumber \\
 && \left. \qquad \qquad \qquad \times \int_0^{\infty} \mathrm{d} \tilde{v}_{s\bot} \, \tilde{v}_{s\bot}^3 J_n'(k_{\bot} \tilde{\rho}_s \tilde{v}_{s\bot})^2 \exp \left(-\tilde{v}_{s\bot}^2\right)\right] , \qquad\\
 (\mathsfbi{M}_s)_{yz} & = & -\frac{4 \mathrm{i}}{\sqrt{\upi}} \tilde{\omega}_{s\|} \sum_{n=-\infty}^{\infty} \left[ \int_{C_L} \frac{\tilde{v}_{s\|} \exp \left(-\tilde{v}_{s\|}^2\right) \mathrm{d} \tilde{v}_{s\|}}{\tilde{v}_{s\|}-\zeta_{sn}} \right. \nonumber \\
 && \left. \qquad \times \int_0^{\infty} \mathrm{d} \tilde{v}_{s\bot} \, \tilde{v}_{s\bot}^2 J_n(k_{\bot} \tilde{\rho}_s \tilde{v}_{s\bot}) J_n'(k_{\bot} \tilde{\rho}_s \tilde{v}_{s\bot}) \exp \left(-\tilde{v}_{s\bot}^2\right)\right] \, ,\\
 (\mathsfbi{M}_s)_{zx} & = & (\mathsfbi{M}_s)_{xz} \, ,\\
 (\mathsfbi{M}_s)_{zy} & = & -(\mathsfbi{M}_s)_{yz} \, , \\
 (\mathsfbi{M}_s)_{zz} & = & \frac{4}{\sqrt{\upi}} \tilde{\omega}_{s\|} \sum_{n=-\infty}^{\infty} \left[ \int_{C_L} \frac{\tilde{v}_{s\|}^2 \exp \left(-\tilde{v}_{s\|}^2\right) \mathrm{d} \tilde{v}_{s\|}}{\tilde{v}_{s\|}-\zeta_{sn}} \right. \nonumber \\
 && \left. \qquad \qquad \qquad \times \int_0^{\infty} \mathrm{d} \tilde{v}_{s\bot} \, \tilde{v}_{s\bot} J_n(k_{\bot} \tilde{\rho}_s \tilde{v}_{s\bot})^2 \exp \left(-\tilde{v}_{s\bot}^2\right) \right] . \qquad
 \label{dielectricelements_maxA}
\end{subeqnarray}
Using the integral identities
\begin{subeqnarray}
\frac{1}{\sqrt{\upi}}\int_{C_L} \frac{u \exp{\left(-u^2\right)} \mathrm{d}u}{u-z}  & = & 1 + z Z\!\left(z\right) 
\, , \\
\frac{1}{\sqrt{\upi}}\int_{C_L} \frac{u^2 \exp{\left(-u^2\right)} \mathrm{d}u}{u-z}  & = & z\left[1 + z Z\!\left(z\right) 
\right] \, , \label{plasmadispfunc_iden}
\end{subeqnarray}
involving the plasma dispersion function, and the identities
\begin{subeqnarray}
\int_0^{\infty} \mathrm{d} t \, t \, J_n(\alpha t)^2 \exp \left(-t^2\right)& = & \frac{1}{2} 
\exp \left(-\frac{\alpha^2}{2}\right) I_n\!\left(\frac{\alpha^2}{2}\right) \, , \\
\int_0^{\infty} \mathrm{d} t \, t^2 J_n(\alpha t) J_n'(\alpha t) \exp \left(-t^2\right)  & = & \frac{\alpha}{4} 
\exp \left(-\frac{\alpha^2}{2}\right) \left[I_n'\!\left(\frac{\alpha^2}{2}\right)-I_n\!\left(\frac{\alpha^2}{2}\right)\right] \, , \qquad \\
\int_0^{\infty} \mathrm{d} t \, t^3 J_n'(\alpha t)^2 \exp \left(-t^2\right)  & = & \frac{1}{4} 
\exp \left(-\frac{\alpha^2}{2}\right)\left\{\frac{2n^2}{\alpha^2} I_n\!\left(\frac{\alpha^2}{2}\right) \right. \nonumber \\
&& \left. - \alpha^2 \left[I_n'\!\left(\frac{\alpha^2}{2}\right)-I_n\!\left(\frac{\alpha^2}{2}\right)\right] \right\} \, , \qquad  
\label{Besselfunc_iden}
\end{subeqnarray}
involving Bessel functions (here $\alpha$ a real number), we obtain expressions for the dielectric components (\ref{dielectricelements_maxA}) 
in terms of special functions: 
\begin{subeqnarray}
 (\mathsfbi{M}_s)_{xx} & = &  2\tilde{\omega}_{s\|} \sum_{n=-\infty}^{\infty} \frac{n^2}{k_{\bot}^2 \tilde{\rho}_s^2} Z\!\left(\zeta_{sn}\right) \exp \left(-\frac{k_{\bot}^2 \tilde{\rho}_s^2}{2}\right) I_n\!\left(\frac{k_{\bot}^2 \tilde{\rho}_s^2}{2}\right) , \qquad\\
 (\mathsfbi{M}_s)_{xy} & = & \mathrm{i} \tilde{\omega}_{s\|} \sum_{n=-\infty}^{\infty} n Z\!\left(\zeta_{sn}\right) \exp \left(-\frac{k_{\bot}^2 \tilde{\rho}_s^2}{2}\right) \left[I_n'\!\left(\frac{k_{\bot}^2 \tilde{\rho}_s^2}{2}\right)-I_n\!\left(\frac{k_{\bot}^2 \tilde{\rho}_s^2}{2}\right) \right]\, , \\
 (\mathsfbi{M}_s)_{xz} & = & 2 \tilde{\omega}_{s\|} \sum_{n=-\infty}^{\infty} \frac{n}{k_{\bot} \tilde{\rho}_s} \left[1+\zeta_{sn} Z\!\left(\zeta_{sn}\right)\right] \exp \left(-\frac{k_{\bot}^2 \tilde{\rho}_s^2}{2}\right) I_n\!\left(\frac{k_{\bot}^2 \tilde{\rho}_s^2}{2}\right) , \qquad \\
 (\mathsfbi{M}_s)_{yx} & = & (\mathsfbi{M}_s)_{xy}\, ,\\
 (\mathsfbi{M}_s)_{yy} & = & \tilde{\omega}_{s\|} \sum_{n=-\infty}^{\infty} Z\!\left(\zeta_{sn}\right) \nonumber \\
 && \times \exp \left(-\frac{k_{\bot}^2 \tilde{\rho}_s^2}{2}\right) \left[\left( \frac{2 n^2}{k_{\bot}^2 \tilde{\rho}_s^2}+k_{\bot}^2 \tilde{\rho}_s^2\right) I_n\!\left(\frac{k_{\bot}^2 \tilde{\rho}_s^2}{2}\right) -  k_{\bot}^2 \tilde{\rho}_s^2 I_n'\!\left(\frac{k_{\bot}^2 \tilde{\rho}_s^2}{2}\right) \right] , \qquad\\
 (\mathsfbi{M}_s)_{yz} & = & -\mathrm{i} \tilde{\omega}_{s\|} \sum_{n=-\infty}^{\infty} k_{\bot} \tilde{\rho}_s \left[1+\zeta_{sn} Z\!\left(\zeta_{sn}\right)\right] \nonumber \\
 && \qquad \qquad \times \exp \left(-\frac{k_{\bot}^2 \tilde{\rho}_s^2}{2}\right)  \left[I_n'\!\left(\frac{k_{\bot}^2 \tilde{\rho}_s^2}{2}\right)-I_n\!\left(\frac{k_{\bot}^2 \tilde{\rho}_s^2}{2}\right) \right]\, , 
 \, \\
 (\mathsfbi{M}_s)_{zx} & = & (\mathsfbi{M}_s)_{xz} \, ,\\
 (\mathsfbi{M}_s)_{zy} & = & -(\mathsfbi{M}_s)_{yz} \, , \\
 (\mathsfbi{M}_s)_{zz} & = & 2 \tilde{\omega}_{s\|} \sum_{n=-\infty}^{\infty} \zeta_{sn} \left[1+\zeta_{sn} Z\!\left(\zeta_{sn}\right)\right] \exp \left(-\frac{k_{\bot}^2 \tilde{\rho}_s^2}{2}\right) I_n\!\left(\frac{k_{\bot}^2 \tilde{\rho}_s^2}{2}\right) 
 \, .
 \label{dielectricelements_maxB}
\end{subeqnarray}
The components of the dielectric tensor (\ref{Maxnonmaxsep_s_Append}) in coordinate basis $\{\boldsymbol{e}_1,\boldsymbol{e}_2,\boldsymbol{e}_3\}$ then follow from 
(\ref{dielectric123trans}), though we do not write these out 
explicitly. 

\subsubsection{Dielectric tensor in low-frequency limit, $\left\{\hat{\boldsymbol{x}},\hat{\boldsymbol{y}},\hat{\boldsymbol{z}}\right\}$ coordinate frame} 
\label{Maxwellresponse_lowxyz}

Now, to consider the low-frequency limit $\tilde{\omega}_{s\|} \ll 1$, we Taylor expand 
(\ref{dielectricelements_maxB}) in $\tilde{\omega}_{s\|}$. Noting that $\tilde{\omega}_{s\|}$ 
only appears via the argument $\zeta_{sn} = \tilde{\omega}_{s\|} - n/|k_{\|}| 
\tilde{\rho}_s$, we use the differential identity $Z'\!\left(z\right) = -2\left[1 + z Z\!\left(z\right)\right]$
to obtain the expansions
\begin{subeqnarray}
 Z\!\left(\zeta_{sn}\right) & = & Z\!\left(-\frac{n}{|k_{\|}| 
\tilde{\rho}_s}\right) - 2 \tilde{\omega}_{s\|} \left[1 - \frac{n}{|k_{\|}| \tilde{\rho}_s} Z\!\left(-\frac{n}{|k_{\|}| \tilde{\rho}_s}\right)\right] 
+ \textit{O}(\tilde{\omega}_{s\|}^2) , \qquad \quad \\
 1+ \zeta_{sn} Z\!\left(\zeta_{sn}\right) & = & 1 -\frac{n}{|k_{\|}| \tilde{\rho}_s} Z\!\left(-\frac{n}{|k_{\|}| \tilde{\rho}_s}\right) \nonumber \\
 && + \tilde{\omega}_{s\|} \left[\left(1- \frac{2 n^2}{|k_{\|}|^2 \tilde{\rho}_s^2} \right)Z\!\left(-\frac{n}{|k_{\|}| \tilde{\rho}_s}\right)+ \frac{2 n}{|k_{\|}| \tilde{\rho}_s} \right] + \textit{O}(\tilde{\omega}_{s\|}^2) \, , \qquad \\
\zeta_{sn} \left[1+\zeta_{sn} Z\!\left(\zeta_{sn}\right)\right] & = &- \frac{n}{|k_{\|}| \tilde{\rho}_s} \left[1 - \frac{n}{|k_{\|}| \tilde{\rho}_s} Z\!\left(-\frac{n}{|k_{\|}| \tilde{\rho}_s}\right)\right] 
+ \tilde{\omega}_{s\|} \left[1- \frac{2 n^2}{|k_{\|}|^2 \tilde{\rho}_s^2} \right. \nonumber \\
&& \left. \qquad \qquad - \frac{2 n}{|k_{\|}| \tilde{\rho}_s} \left(1- \frac{n^2}{|k_{\|}|^2 \tilde{\rho}_s^2} \right)Z\!\left(-\frac{n}{|k_{\|}| \tilde{\rho}_s}\right) \right] + \textit{O}(\tilde{\omega}_{s\|}^2) 
 . \qquad 
\end{subeqnarray} 
Then, expanding the dielectric tensor as
\begin{equation}
\mathsfbi{M}_s = \tilde{\omega}_{s\|} \mathsfbi{M}_s^{(0)} + \tilde{\omega}_{s\|}^2 
\mathsfbi{M}_s^{(1)} + \textit{O}(\tilde{\omega}_{s\|}^3) \, ,
\end{equation}
we have
\begin{subeqnarray}
 (\mathsfbi{M}_s^{(0)})_{xx} & = & 2 \sum_{n=-\infty}^{\infty} \frac{n^2}{k_{\bot}^2 \tilde{\rho}_s^2} Z\!\left(-\frac{n}{|k_{\|}| 
\tilde{\rho}_s}\right) \exp \left(-\frac{k_{\bot}^2 \tilde{\rho}_s^2}{2}\right) I_n\!\left(\frac{k_{\bot}^2 \tilde{\rho}_s^2}{2}\right) , \qquad\\
 (\mathsfbi{M}_s^{(0)})_{xy} & = & \mathrm{i} \sum_{n=-\infty}^{\infty} n Z\!\left(-\frac{n}{|k_{\|}| \tilde{\rho}_s}\right) \nonumber \\
 && \qquad \qquad \times  \exp \left(-\frac{k_{\bot}^2 \tilde{\rho}_s^2}{2}\right) \left[I_n'\!\left(\frac{k_{\bot}^2 \tilde{\rho}_s^2}{2}\right)-I_n\!\left(\frac{k_{\bot}^2 \tilde{\rho}_s^2}{2}\right) \right]\, , \qquad \\
 (\mathsfbi{M}_s^{(0)})_{xz} & = & 2 \sum_{n=-\infty}^{\infty} \frac{n}{k_{\bot} \tilde{\rho}_s} \left[1 -\frac{n}{|k_{\|}| \tilde{\rho}_s} Z\!\left(-\frac{n}{|k_{\|}| \tilde{\rho}_s}\right)\right] \nonumber \\
 && \qquad \qquad \qquad \times \exp \left(-\frac{k_{\bot}^2 \tilde{\rho}_s^2}{2}\right) I_n\!\left(\frac{k_{\bot}^2 \tilde{\rho}_s^2}{2}\right) , \qquad \\
 (\mathsfbi{M}_s^{(0)})_{yy} & = & \sum_{n=-\infty}^{\infty} Z\!\left(-\frac{n}{|k_{\|}| \tilde{\rho}_s}\right) \nonumber \\
 && \times \exp \left(-\frac{k_{\bot}^2 \tilde{\rho}_s^2}{2}\right) \left[\left( \frac{2 n^2}{k_{\bot}^2 \tilde{\rho}_s^2}+k_{\bot}^2 \tilde{\rho}_s^2\right) I_n\!\left(\frac{k_{\bot}^2 \tilde{\rho}_s^2}{2}\right) -  k_{\bot}^2 \tilde{\rho}_s^2 I_n'\!\left(\frac{k_{\bot}^2 \tilde{\rho}_s^2}{2}\right) \right] , \qquad \qquad \\
 (\mathsfbi{M}_s^{(0)})_{yz} & = & \mathrm{i} \sum_{n=-\infty}^{\infty} k_{\bot} \tilde{\rho}_s \left[1 -\frac{n}{|k_{\|}| \tilde{\rho}_s} Z\!\left(-\frac{n}{|k_{\|}| \tilde{\rho}_s}\right)\right] \nonumber \\
 && \qquad \qquad \times \exp \left(-\frac{k_{\bot}^2 \tilde{\rho}_s^2}{2}\right)  \left[I_n'\!\left(\frac{k_{\bot}^2 \tilde{\rho}_s^2}{2}\right)-I_n\!\left(\frac{k_{\bot}^2 \tilde{\rho}_s^2}{2}\right) \right]\, , 
 \, \\
 (\mathsfbi{M}_s^{(0)})_{zz} & = & -2 \sum_{n=-\infty}^{\infty} \frac{n}{|k_{\|}| \tilde{\rho}_s} \left[1 - \frac{n}{|k_{\|}| \tilde{\rho}_s} Z\!\left(-\frac{n}{|k_{\|}| \tilde{\rho}_s}\right)\right] \nonumber \\
 && \qquad \qquad \qquad \times \exp \left(-\frac{k_{\bot}^2 \tilde{\rho}_s^2}{2}\right) I_n\!\left(\frac{k_{\bot}^2 \tilde{\rho}_s^2}{2}\right) 
 \, ,
 \label{dielectricelements_max0}
\end{subeqnarray}
and
\begin{subeqnarray}
 (\mathsfbi{M}_s^{(1)})_{xx} & = &  -4 \sum_{n=-\infty}^{\infty} \frac{n^2}{k_{\bot}^2 \tilde{\rho}_s^2} \left[1 - \frac{n}{|k_{\|}| \tilde{\rho}_s} Z\!\left(-\frac{n}{|k_{\|}| \tilde{\rho}_s}\right)\right] \nonumber \\
 && \qquad \qquad \qquad \times \exp \left(-\frac{k_{\bot}^2 \tilde{\rho}_s^2}{2}\right) I_n\!\left(\frac{k_{\bot}^2 \tilde{\rho}_s^2}{2}\right) , \qquad\\
 (\mathsfbi{M}_s^{(1)})_{xy} & = & -2\mathrm{i} \sum_{n=-\infty}^{\infty} n \left[1 - \frac{n}{|k_{\|}| \tilde{\rho}_s} Z\!\left(-\frac{n}{|k_{\|}| \tilde{\rho}_s}\right)\right] \nonumber \\
 && \qquad \qquad \times \exp \left(-\frac{k_{\bot}^2 \tilde{\rho}_s^2}{2}\right) \left[I_n'\!\left(\frac{k_{\bot}^2 \tilde{\rho}_s^2}{2}\right)-I_n\!\left(\frac{k_{\bot}^2 \tilde{\rho}_s^2}{2}\right) \right]\, , \\
 (\mathsfbi{M}_s^{(1)})_{xz} & = & 2 \sum_{n=-\infty}^{\infty} \frac{n}{k_{\bot} \tilde{\rho}_s} \left[\left(1- \frac{2 n^2}{|k_{\|}|^2 \tilde{\rho}_s^2} \right)Z\!\left(-\frac{n}{|k_{\|}| \tilde{\rho}_s}\right) + \frac{2 n}{|k_{\|}| \tilde{\rho}_s} \right] \nonumber \\
 && \qquad \qquad \times \exp \left(-\frac{k_{\bot}^2 \tilde{\rho}_s^2}{2}\right) I_n\!\left(\frac{k_{\bot}^2 \tilde{\rho}_s^2}{2}\right) , \qquad \\
 (\mathsfbi{M}_s^{(1)})_{yy} & = & -2 \sum_{n=-\infty}^{\infty} \left[1 - \frac{n}{|k_{\|}| \tilde{\rho}_s} Z\!\left(-\frac{n}{|k_{\|}| \tilde{\rho}_s}\right)\right] \nonumber \\
 && \times \exp \left(-\frac{k_{\bot}^2 \tilde{\rho}_s^2}{2}\right) \left[\left( \frac{2 n^2}{k_{\bot}^2 \tilde{\rho}_s^2}+k_{\bot}^2 \tilde{\rho}_s^2\right) I_n\!\left(\frac{k_{\bot}^2 \tilde{\rho}_s^2}{2}\right) -  k_{\bot}^2 \tilde{\rho}_s^2 I_n'\!\left(\frac{k_{\bot}^2 \tilde{\rho}_s^2}{2}\right) \right] , \qquad \quad\\
 (\mathsfbi{M}_s^{(1)})_{yz} & = & -\mathrm{i} \sum_{n=-\infty}^{\infty} k_{\bot} \tilde{\rho}_s\left[\left(1- \frac{2 n^2}{|k_{\|}|^2 \tilde{\rho}_s^2} \right)Z\!\left(-\frac{n}{|k_{\|}| \tilde{\rho}_s}\right) + \frac{2 n}{|k_{\|}| \tilde{\rho}_s} \right] \nonumber \\
 && \qquad \qquad \times \exp \left(-\frac{k_{\bot}^2 \tilde{\rho}_s^2}{2}\right)  \left[I_n'\!\left(\frac{k_{\bot}^2 \tilde{\rho}_s^2}{2}\right)-I_n\!\left(\frac{k_{\bot}^2 \tilde{\rho}_s^2}{2}\right) \right]\, , 
 \, \\
 (\mathsfbi{M}_s^{(1)})_{zz} & = & 2 \sum_{n=-\infty}^{\infty} \left[1- \frac{2 n^2}{|k_{\|}|^2 \tilde{\rho}_s^2} - \frac{2 n}{|k_{\|}| \tilde{\rho}_s} \left(1- \frac{n^2}{|k_{\|}|^2 \tilde{\rho}_s^2} \right)Z\!\left(-\frac{n}{|k_{\|}| \tilde{\rho}_s}\right) \right] \nonumber \\
 && \qquad \qquad \qquad \times \exp \left(-\frac{k_{\bot}^2 \tilde{\rho}_s^2}{2}\right) I_n\!\left(\frac{k_{\bot}^2 \tilde{\rho}_s^2}{2}\right) 
 \, .
 \label{dielectricelements_max1}
\end{subeqnarray}

These expressions can be simplified somewhat using two further types of algebraic manipulation. 
First, for $z$ a real number, we can split the plasma dispersion into real and imaginary parts 
as
\begin{eqnarray}
Z\!\left(z\right) & = & \frac{1}{\sqrt{\upi}} \mathcal{P} \int_{-\infty}^{\infty} \frac{\exp{\left(-u^2\right)} \mathrm{d}u}{u-z} 
+ \mathrm{i} \sqrt{\upi} \exp{\left(-z^2\right)} \nonumber \\
&& = \Real{\; Z\!\left(z\right)} +  \mathrm{i} \sqrt{\upi} \exp{\left(-z^2\right)} 
\, .
\end{eqnarray}
Thus, we see that the real part of $Z\!\left(z\right)$ is an odd function
for real $z$, while the imaginary part is an even function. As a consequence, 
only one of the real or imaginary parts of the plasma dispersion function will 
enter into the summations in (\ref{dielectricelements_max0}) and (\ref{dielectricelements_max1}). Secondly, we utilise the generating function of the 
modified Bessel function, viz., 
\begin{equation}
  \sum_{n = -\infty}^{\infty} I_n\!\left(\alpha\right) t^{n} = \exp{\left[\frac{\alpha}{2}\left(t+\frac{1}{t}\right)\right]} 
  \, ,
\end{equation}
to deduce the following identities:
\begin{subeqnarray}
  \sum_{n = -\infty}^{\infty} I_n\!\left(\alpha\right) & = & \exp{\left(\alpha\right)} 
  \, , \\
  \sum_{n = -\infty}^{\infty} n^2 I_n\!\left(\alpha\right) & = & \alpha \exp{\left(\alpha\right)} 
  \, , \\
    \sum_{n = -\infty}^{\infty} \left[ I_n'\!\left(\alpha\right) - I_n\!\left(\alpha\right) \right] & = & 
    0
  \, , \\
    \sum_{n = -\infty}^{\infty} n^2 \left[I_n'\!\left(\alpha\right)- I_n\!\left(\alpha\right)\right] & = & \exp{\left(\alpha\right)}  
  \, .  \label{Besselfuncsums}
\end{subeqnarray}
Combining these results, we obtain from (\ref{dielectricelements_max0}) and (\ref{dielectricelements_max1}) the following expressions for the components 
of $\mathsfbi{M}_s^{(0)}$ and $\mathsfbi{M}_s^{(1)}$:
\begin{subeqnarray}
 (\mathsfbi{M}_s^{(0)})_{xx} & = &  4 \mathrm{i} \sqrt{\upi} \sum_{m=1}^{\infty} \frac{m^2}{k_{\bot}^2 \tilde{\rho}_s^2} \exp{\left(-\frac{m^2}{k_{\|}^2 \tilde{\rho}_s^2}\right)} \exp \left(-\frac{k_{\bot}^2 \tilde{\rho}_s^2}{2}\right) I_m\!\left(\frac{k_{\bot}^2 \tilde{\rho}_s^2}{2}\right) \nonumber \\
 & = & \mathrm{i} F\!\left(k_{\|} \tilde{\rho}_s,k_{\bot} \tilde{\rho}_s\right)  \, , \\
 (\mathsfbi{M}_s^{(0)})_{xy} & = & -\mathrm{i} \sum_{m=-\infty}^{\infty} m \, \Real{\left[Z\left(\frac{m}{|k_{\|}| \tilde{\rho}_s}\right)\right]} \exp \left(-\frac{k_{\bot}^2 \tilde{\rho}_s^2}{2}\right) \left[I_m'\!\left(\frac{k_{\bot}^2 \tilde{\rho}_s^2}{2}\right)-I_m\!\left(\frac{k_{\bot}^2 \tilde{\rho}_s^2}{2}\right) \right] \nonumber \\
 & = & -\mathrm{i} G\!\left(k_{\|} \tilde{\rho}_s,k_{\bot} \tilde{\rho}_s\right)  \, , \\
 (\mathsfbi{M}_s^{(0)})_{xz} & = & -4 \mathrm{i} \sqrt{\upi} \sum_{m=-\infty}^{\infty} \frac{m^2}{k_{\bot} |k_{\|}| \tilde{\rho}_s^2} \exp{\left(-\frac{m^2}{k_{\|}^2 \tilde{\rho}_s^2}\right)} \exp \left(-\frac{k_{\bot}^2 \tilde{\rho}_s^2}{2}\right) I_n\!\left(\frac{k_{\bot}^2 \tilde{\rho}_s^2}{2}\right) \nonumber \\
  & = & -\frac{\mathrm{i} k_{\bot}}{|k_{\|}|} F\!\left(k_{\|} \tilde{\rho}_s,k_{\bot} \tilde{\rho}_s\right)  \, , \\
 (\mathsfbi{M}_s^{(0)})_{yy} & = & \mathrm{i} \sqrt{\upi} \sum_{m=-\infty}^{\infty}  \exp{\left(-\frac{m^2}{k_{\|}^2 \tilde{\rho}_s^2}\right)} \nonumber \\
 && \times \exp \left(-\frac{k_{\bot}^2 \tilde{\rho}_s^2}{2}\right) \left[\left( \frac{2 m^2}{k_{\bot}^2 \tilde{\rho}_s^2}+k_{\bot}^2 \tilde{\rho}_s^2\right) I_m\!\left(\frac{k_{\bot}^2 \tilde{\rho}_s^2}{2}\right) -  k_{\bot}^2 \tilde{\rho}_s^2 I_m'\!\left(\frac{k_{\bot}^2 \tilde{\rho}_s^2}{2}\right) \right] \nonumber \\
 & = & \mathrm{i}H\!\left(k_{\|} \tilde{\rho}_s,k_{\bot} \tilde{\rho}_s\right)  \, , \\
 (\mathsfbi{M}_s^{(0)})_{yz} & = & -\mathrm{i} \sum_{m=-\infty}^{\infty} \frac{m k_{\bot}}{|k_{\|}|} \, \Real{\left[Z\left(\frac{m}{|k_{\|}| \tilde{\rho}_s}\right)\right]} \exp \left(-\frac{k_{\bot}^2 \tilde{\rho}_s^2}{2}\right) \left[I_m'\!\left(\frac{k_{\bot}^2 \tilde{\rho}_s^2}{2}\right)-I_m\!\left(\frac{k_{\bot}^2 \tilde{\rho}_s^2}{2}\right) \right] \nonumber \\
 & = & -\frac{\mathrm{i} k_{\bot}}{|k_{\|}|} G\!\left(k_{\|} \tilde{\rho}_s,k_{\bot} \tilde{\rho}_s\right)  \, , \\
 (\mathsfbi{M}_s^{(0)})_{zz} & = &  4 \mathrm{i} \sqrt{\upi} \sum_{m=1}^{\infty} \frac{m^2}{k_{\|}^2 \tilde{\rho}_s^2} \exp{\left(-\frac{m^2}{k_{\|}^2 \tilde{\rho}_s^2}\right)} \exp \left(-\frac{k_{\bot}^2 \tilde{\rho}_s^2}{2}\right) I_m\!\left(\frac{k_{\bot}^2 \tilde{\rho}_s^2}{2}\right) \nonumber \\
 & = & \frac{\mathrm{i} k_{\bot}^2 }{k_{\|}^2} F\!\left(k_{\|} \tilde{\rho}_s,k_{\bot} \tilde{\rho}_s\right)  \, ,
 \label{dielectricelements_max0_B}
\end{subeqnarray}
and
\begin{subeqnarray}
 (\mathsfbi{M}_s^{(1)})_{xx} & = &  -2 \left\{1 + \sum_{m=-\infty}^{\infty} \frac{2 m^3}{|k_{\|}| k_{\bot}^2 \tilde{\rho}_s^3} \Real{\left[Z\left(\frac{m}{|k_{\|}| \tilde{\rho}_s}\right)\right]} \right. \nonumber \\
 && \left. \qquad \qquad \qquad \times \exp \left(-\frac{k_{\bot}^2 \tilde{\rho}_s^2}{2}\right) I_m\!\left(\frac{k_{\bot}^2 \tilde{\rho}_s^2}{2}\right) \right\} \nonumber \\
 &=& -\frac{4}{3} W\!\left(k_{\|} \tilde{\rho}_s,k_{\bot} \tilde{\rho}_s\right) , \qquad\\
 (\mathsfbi{M}_s^{(1)})_{xy} & = & 4 \sqrt{\upi} \sum_{m=1}^{\infty} \frac{m^2}{|k_{\|}| \tilde{\rho}_s} \exp{\left(-\frac{m^2}{k_{\|}^2 \tilde{\rho}_s^2}\right)} \nonumber \\
 && \qquad \qquad \times \exp \left(-\frac{k_{\bot}^2 \tilde{\rho}_s^2}{2}\right) \left[I_m'\!\left(\frac{k_{\bot}^2 \tilde{\rho}_s^2}{2}\right)-I_m\!\left(\frac{k_{\bot}^2 \tilde{\rho}_s^2}{2}\right) \right]\, , \\
 (\mathsfbi{M}_s^{(1)})_{xz} & = & 2 \left\{\frac{k_{\bot}}{|k_{\|}|} + \sum_{m=-\infty}^{\infty} \left(\frac{2 m^3}{|k_{\|}|^2 k_{\bot} \tilde{\rho}_s^3} -\frac{m}{k_{\bot} \tilde{\rho}_s}\right)\Real{\left[Z\left(\frac{m}{|k_{\|}| \tilde{\rho}_s}\right)\right]} \right. \nonumber \\
 && \left. \qquad \qquad \qquad \times \exp \left(-\frac{k_{\bot}^2 \tilde{\rho}_s^2}{2}\right) I_m\!\left(\frac{k_{\bot}^2 \tilde{\rho}_s^2}{2}\right) \right\} , \qquad\\
 (\mathsfbi{M}_s^{(1)})_{yy} & = & -2 \left\{ 1 + \sum_{m=-\infty}^{\infty} \frac{m}{|k_{\|}| \tilde{\rho}_s} \Real{\left[Z\left(\frac{m}{|k_{\|}| \tilde{\rho}_s}\right)\right]} \right. \nonumber \\
 &\times & \left. \exp \left(-\frac{k_{\bot}^2 \tilde{\rho}_s^2}{2}\right) \left[\left( \frac{2 m^2}{k_{\bot}^2 \tilde{\rho}_s^2}+k_{\bot}^2 \tilde{\rho}_s^2\right) I_m\!\left(\frac{k_{\bot}^2 \tilde{\rho}_s^2}{2}\right) -  k_{\bot}^2 \tilde{\rho}_s^2 I_m'\!\left(\frac{k_{\bot}^2 \tilde{\rho}_s^2}{2}\right) \right] \right\} \nonumber \\
  &=& -\frac{4}{3} Y\!\left(k_{\|} \tilde{\rho}_s,k_{\bot} \tilde{\rho}_s\right) , \qquad\\
 (\mathsfbi{M}_s^{(1)})_{yz} & = & -\sqrt{\upi} \sum_{m=-\infty}^{\infty} k_{\bot} \tilde{\rho}_s \left(1-\frac{2 m^2}{|k_{\|}|^2 \tilde{\rho}_s^2} \right) \exp{\left(-\frac{m^2}{k_{\|}^2 \tilde{\rho}_s^2}\right)} \nonumber \\
 && \qquad \qquad \times \exp \left(-\frac{k_{\bot}^2 \tilde{\rho}_s^2}{2}\right)  \left[I_m'\!\left(\frac{k_{\bot}^2 \tilde{\rho}_s^2}{2}\right)-I_m\!\left(\frac{k_{\bot}^2 \tilde{\rho}_s^2}{2}\right) \right]\, , 
 \, \\
 (\mathsfbi{M}_s^{(1)})_{zz} & = & 2 \left\{1 - \frac{k_{\bot}^2}{k_{\|}^2} + \sum_{m=-\infty}^{\infty} \frac{2 m}{|k_{\|}| \tilde{\rho}_s} \left(1- \frac{m^2}{|k_{\|}|^2 \tilde{\rho}_s^2} \right)\Real{\left[Z\left(\frac{m}{|k_{\|}| \tilde{\rho}_s}\right)\right]} \right. \nonumber \\
 && \left. \qquad \qquad \qquad \times \exp \left(-\frac{k_{\bot}^2 \tilde{\rho}_s^2}{2}\right) I_m\!\left(\frac{k_{\bot}^2 \tilde{\rho}_s^2}{2}\right) 
 \right\} \, ,
 \label{dielectricelements_max1_B}
\end{subeqnarray}
where we have reintroduced the special functions $F\!\left(x,y\right)$, $G\!\left(x,y\right)$ and 
$H\!\left(x,y\right)$ defined by (\ref{specialfuncMax}), as well as $W\!\left(x,y\right)$ and $Y\!\left(x,y\right)$ defined 
by (\ref{shearasymptoticfuncs}). As anticipated 
from the arguments presented in appendix \ref{DispLowFreq}, $\mathsfbi{M}_s^{(0)}$ 
obeys the symmetries
\begin{subeqnarray}
  (\mathsfbi{M}_s^{(0)})_{xz} & = & - \frac{k_{\bot}}{k_{\|}}  (\mathsfbi{M}_s^{(0)})_{xx} \, , \\
  (\mathsfbi{M}_s^{(0)})_{yz} & = & \frac{k_{\bot}}{k_{\|}}  (\mathsfbi{M}_s^{(0)})_{xy} \, , \\
  (\mathsfbi{M}_s^{(0)})_{zz} & = & \frac{k_{\bot}^2}{k_{\|}^2} (\mathsfbi{M}_s^{(0)})_{xx}  \, 
  .
\end{subeqnarray}

\subsubsection{Dielectric tensor in low-frequency limit, $\{\boldsymbol{e}_1,\boldsymbol{e}_2,\boldsymbol{e}_3\}$ coordinate frame} 
\label{Maxwellresponse_low123}

Having evaluated the first- and second-order terms in the expansion for components of the dielectric tensor in the
coordinate basis $\left\{\hat{\boldsymbol{x}},\hat{\boldsymbol{y}},\hat{\boldsymbol{z}}\right\}$, 
we can use (\ref{dielectric123trans}) to find equivalent expressions 
in the coordinate basis $\{\boldsymbol{e}_1,\boldsymbol{e}_2,\boldsymbol{e}_3\}$. 
Explicitly, we have the following transformations for 
$\mathsfbi{M}_s^{(0)}$:
\begin{subeqnarray}
 (\mathsfbi{M}_s^{(0)})_{11} & = & \frac{k_{\|}^2}{k^2} (\mathsfbi{M}_s^{(0)})_{xx} - \frac{2 k_{\|} k_{\bot}}{k^2} (\mathsfbi{M}_s^{(0)})_{xz} + \frac{k_{\bot}^2}{k^2} (\mathsfbi{M}_s^{(0)})_{zz}\, , \\
 (\mathsfbi{M}_s^{(0)})_{12} & = & \frac{k_{\|}}{k}(\mathsfbi{M}_s^{(0)})_{xy} + \frac{k_{\bot}}{k}(\mathsfbi{M}_s^{(0)})_{yz} \,  ,\\
 (\mathsfbi{M}_s^{(0)})_{13} & = & \frac{k_{\|} k_{\bot}}{k^2} \left[(\mathsfbi{M}_s^{(0)})_{xx} - (\mathsfbi{M}_s^{(0)})_{zz}\right] + \left(\frac{k_{\|}^2}{k^2}- \frac{k_{\bot}^2}{k^2}\right) (\mathsfbi{M}_s^{(0)})_{xz} \, , \\
 (\mathsfbi{M}_s^{(0)})_{22} & = & (\mathsfbi{M}_s^{(0)})_{yy} \, , \\
 (\mathsfbi{M}_s^{(0)})_{23} & = & -\frac{k_{\bot}}{k}(\mathsfbi{M}_s^{(0)})_{xy} + \frac{k_{\|}}{k}(\mathsfbi{M}_s^{(0)})_{yz} \, , \\
 (\mathsfbi{M}_s^{(0)})_{33} & = & \frac{k_{\bot}^2}{k^2} (\mathsfbi{M}_s^{(0)})_{xx} + \frac{2 k_{\|} k_{\bot}}{k^2} (\mathsfbi{M}_s^{(0)})_{xz} + \frac{k_{\|}^2}{k^2} (\mathsfbi{M}_s^{(0)})_{zz} \, 
 , \label{dielectric123trans_Append_0}
 \end{subeqnarray}
and similiarly for $\mathsfbi{M}_s^{(1)}$. 

On account of the symmetries derived in appendix \ref{Maxwellresponse_lowxyz}, we find for $\mathsfbi{M}_s^{(0)}$ that
\begin{subeqnarray}
 (\mathsfbi{M}_s^{(0)})_{11} & = & \frac{k^2}{k_{\|}^2}(\mathsfbi{M}_s^{(0)})_{xx} \, , \\
 (\mathsfbi{M}_s^{(0)})_{12} & = & \frac{k}{k_{\|}}(\mathsfbi{M}_s^{(0)})_{xy} \,  ,\\
 (\mathsfbi{M}_s^{(0)})_{21} & = & -(\mathsfbi{M}_s^{(0)})_{12} \, , \\
 (\mathsfbi{M}_s^{(0)})_{22} & = & (\mathsfbi{M}_s^{(0)})_{yy} \, , 
 \label{dielectric123trans_mat0}
 \end{subeqnarray}
with all other components vanishing. This agrees with (\ref{dielectric123}) 
stated in the main text. On substitution of 
identities (\ref{dielectricelements_max0_B}), (\ref{Mmaxcomp}) are 
recovered.

As for $\mathsfbi{M}_s^{(1)}$, from the results (\ref{dielectricelements_max1_B}) derived in appendix \ref{Maxwellresponse_lowxyz}, we have the following identities:
\begin{subeqnarray}
(\mathsfbi{M}_s^{(1)})_{xz} + \frac{k_{\bot}}{k_{\|}} (\mathsfbi{M}_s^{(1)})_{xx} & = & -2 \sum_{m=-\infty}^{\infty} \frac{m}{k_{\bot} \tilde{\rho}_s} \Real{\left[Z\left(\frac{m}{|k_{\|}| \tilde{\rho}_s}\right)\right]} \nonumber \\
 && \qquad \qquad \qquad \times \exp \left(-\frac{k_{\bot}^2 \tilde{\rho}_s^2}{2}\right) I_m\!\left(\frac{k_{\bot}^2 \tilde{\rho}_s^2}{2}\right) , \\ 
(\mathsfbi{M}_s^{(1)})_{yz} -  \frac{k_{\bot}}{k_{\|}} (\mathsfbi{M}_s^{(1)})_{xy} & = & -\sqrt{\upi} \sum_{m=-\infty}^{\infty} k_{\bot} \tilde{\rho}_s \exp{\left(-\frac{m^2}{k_{\|}^2 \tilde{\rho}_s^2}\right)} \nonumber \\
 && \qquad \times \exp \left(-\frac{k_{\bot}^2 \tilde{\rho}_s^2}{2}\right) \left[I_m'\!\left(\frac{k_{\bot}^2 \tilde{\rho}_s^2}{2}\right)-I_m\!\left(\frac{k_{\bot}^2 \tilde{\rho}_s^2}{2}\right) \right] , \,  \qquad\\ 
  (\mathsfbi{M}_s^{(1)})_{zz} + \frac{k_{\bot}}{k_{\|}} (\mathsfbi{M}_s^{(1)})_{xz} & = & 2 \left\{1 + \sum_{m=-\infty}^{\infty} \frac{m}{|k_{\|}| \tilde{\rho}_s} \Real{\left[Z\left(\frac{m}{|k_{\|}| \tilde{\rho}_s}\right)\right]} \right. \nonumber \\
 && \left. \qquad \qquad \qquad \times \exp \left(-\frac{k_{\bot}^2 \tilde{\rho}_s^2}{2}\right) I_m\!\left(\frac{k_{\bot}^2 \tilde{\rho}_s^2}{2}\right) \right\} \, .
\end{subeqnarray}
Thus, we can decompose the dielectric components $(\mathsfbi{M}_s^{(1)})_{xz}$, $(\mathsfbi{M}_s^{(1)})_{yz}$ and $(\mathsfbi{M}_s^{(1)})_{zz}$ in terms of the remaining components of $\mathsfbi{M}_s^{(1)}$ as follows:
\begin{subeqnarray}
(\mathsfbi{M}_s^{(1)})_{xz} & = & -\frac{k_{\bot}}{k_{\|}} (\mathsfbi{M}_s^{(1)})_{xx} -  L\!\left(k_{\|} \tilde{\rho}_s,k_{\bot} \tilde{\rho}_s \right) \, , \\
(\mathsfbi{M}_s^{(1)})_{yz} & = &  \frac{k_{\bot}}{k_{\|}} (\mathsfbi{M}_s^{(1)})_{xy} -  N\!\left(k_{\|} \tilde{\rho}_s,k_{\bot} \tilde{\rho}_s \right) \, , \\
(\mathsfbi{M}_s^{(1)})_{zz} & = & -\frac{k_{\bot}}{k_{\|}} (\mathsfbi{M}_s^{(1)})_{xz} + \left[2 + \frac{k_{\bot}}{k_{\|}} L\!\left(k_{\|} \tilde{\rho}_s,k_{\bot} \tilde{\rho}_s \right)\right] 
\nonumber \\
& = & \frac{k_{\bot}^2}{k_{\|}^2} (\mathsfbi{M}_s^{(1)})_{xx} + 2\left[1 + \frac{k_{\bot}}{k_{\|}} L\!\left(k_{\|} \tilde{\rho}_s,k_{\bot} \tilde{\rho}_s \right)\right] 
\, ,
\end{subeqnarray}
where the special functions $L(x,y)$ and $N(x,y)$ are defined by
\begin{subeqnarray}
 L\!\left(x,y\right) & \equiv & \sum_{m=-\infty}^{\infty} \frac{2 m}{y} \Real{\; Z\!\left(\frac{m}{x}\right)} \exp \left(-\frac{y^2}{2}\right) I_m\!\left(\frac{y^2}{2}\right) , \\ 
 N\!\left(x,y\right) & \equiv & \sqrt{\upi} \sum_{m=-\infty}^{\infty} y \exp{\left(-\frac{m^2}{x^2}\right)} \exp \left(-\frac{y^2}{2}\right) \left[I_m'\!\left(\frac{y^2}{2}\right)-I_m\!\left(\frac{y^2}{2}\right) \right] 
 . \, \label{LandNdef_Append}
\end{subeqnarray} 
This leads to the following expressions:
\begin{subeqnarray}
 (\mathsfbi{M}_s^{(1)})_{11} & = & \frac{k^2} {k_{\|}^2} (\mathsfbi{M}_s^{(1)})_{xx} + 2 \left[\frac{k_{\bot}^2}{k^2} + \frac{k_{\bot}}{k_{\|}} L\!\left(k_{\|} \tilde{\rho}_s,k_{\bot} \tilde{\rho}_s \right)\right] \, , \\
 (\mathsfbi{M}_s^{(1)})_{12} & = & \frac{k}{k_{\|}} (\mathsfbi{M}_s^{(1)})_{xy} -\frac{k_{\bot}}{k} N\!\left(k_{\|} \tilde{\rho}_s,k_{\bot} \tilde{\rho}_s \right) \,  ,\\
 (\mathsfbi{M}_s^{(1)})_{13} & = & - \frac{2 k_{\bot} k_{\|}}{k^2} - L\!\left(k_{\|} \tilde{\rho}_s,k_{\bot} \tilde{\rho}_s \right) \, , \\
 (\mathsfbi{M}_s^{(1)})_{22} & = & (\mathsfbi{M}_s^{(1)})_{yy} \, , \\
 (\mathsfbi{M}_s^{(1)})_{23} & = & -\frac{k_{\|}}{k} N\!\left(k_{\|} \tilde{\rho}_s,k_{\bot} \tilde{\rho}_s \right) \, , \\
 (\mathsfbi{M}_s^{(1)})_{33} & = & \frac{2 k_{\|}^2}{k^2} \, 
 . \label{dielectric123trans_1}
 \end{subeqnarray}
 We note that $\mathsfbi{M}_s^{(1)}$ does not possess the same symmetry 
 properties as $\mathsfbi{M}_s^{(0)}$. 

\subsubsection{Asymptotic forms of $\mathsfbi{M}_s^{(0)}$ and $\mathsfbi{M}_s^{(1)}$} 
\label{asympspecialfuncsappend}

In this appendix, we write down asymptotic forms at small 
and large $x$ and $y$ for the special functions 
$F\!\left(x,y\right)$, $G\!\left(x,y\right)$, $H\!\left(x,y\right)$, $L(x,y)$ and $N(x,y)$ defined 
by (\ref{specialfuncMax}) and (\ref{LandNdef_Append}), respectively. Physically, this corresponds via (\ref{Mmaxcomp}) to considering the 
dielectric response associated with $\mathsfbi{M}_s^{(0)}$ and $\mathsfbi{M}_s^{(1)}$ for modes with parallel and perpendicular wavenumbers very 
small (or very large) with respect to the inverse Larmor radius of species $s$. 
Detailed derivations are left as an exercise to keen readers (and can be verified 
numerically). 

Proceeding systematically through various limits, we have the following results:
\begin{itemize} 
  \item  $x \sim 1$, $y \ll 1$:
\begin{subeqnarray}
  F\!\left(x,y\right) & = & \sqrt{\upi} \exp{\left(-\frac{1}{x^2}\right)} \left[1+\textit{O}\!\left(y^2\right) \right]  \, , \\
  G\!\left(x,y\right) & = & \Real{\left[Z\left(\frac{1}{x}\right)\right]} \left[1+\textit{O}\!\left(y^2\right) \right] \, , \\
  H\!\left(x,y\right) & = & \sqrt{\upi} \exp{\left(-\frac{1}{x^2}\right)} \left[1+\textit{O}\!\left(y^2\right) \right]  \, , \\
  L\!\left(x,y\right) & = & y \Real{\left[Z\left(\frac{1}{x}\right)\right]} \left[1+\textit{O}\!\left(y^2\right) \right]  \, ,  \\ 
  N\!\left(x,y\right) & = & \sqrt{\upi} y \left[2 \exp{\left(-\frac{1}{x^2}\right)} - 1 \right] \left[1+\textit{O}\!\left(y^2\right) \right]  \, .   
 \label{specialfuncMax_x1_ysmall}
\end{subeqnarray}
  \item  $x, y \gg 1$
\begin{subeqnarray}
  F\!\left(x,y\right) & = & \frac{\sqrt{\upi} x^3}{\left(x^2+y^2\right)^{3/2}} \left[1+\textit{O}\!\left(\frac{1}{x^2+y^2}\right) \right] \, , \\
  G\!\left(x,y\right) & = & -\frac{2 x^3}{\left(x^2+y^2\right)^{2}} \left[1+\textit{O}\!\left(\frac{1}{x^2+y^2}\right) \right] \, , \\
  H\!\left(x,y\right) & = & \frac{\sqrt{\upi} x}{\left(x^2+y^2\right)^{1/2}} \left[1+\textit{O}\!\left(\frac{1}{x^2+y^2}\right) \right] \, , \\
  L\!\left(x,y\right) & = & -\frac{2 x y}{x^2+y^2} \left[1+\textit{O}\!\left(\frac{1}{x^2+y^2}\right) \right]  \, ,  \\
  N\!\left(x,y\right) & = & \frac{\sqrt{\upi} x}{y \left(x^2+y^2\right)^{1/2}} \left[1+\textit{O}\!\left(\frac{1}{x^2+y^2}\right) \right]  \, .   
 \label{specialfuncMax_xylarge}
\end{subeqnarray}
 We observe that the asymptotic forms (\ref{specialfuncMax_xylarge}) are in fact valid even for $y \lesssim 1$.
  \item  $x \ll 1$, $y \sim 1$:
\begin{subeqnarray}
  F\!\left(x,y\right) & = & \frac{4 \sqrt{\upi}}{y^2} \exp{\left(-\frac{y^2}{2}\right)} I_1\!\left(\frac{y^2}{2}\right)\exp{\left(-\frac{1}{x^2}\right)} \left\{1+\textit{O}\!\left[\exp{\left(-\frac{3}{x^2}\right)} \right] \right\}  \, , \qquad \\
  G\!\left(x,y\right) & = & -x \exp{\left(-\frac{y^2}{2}\right)} \left[I_0\!\left(\frac{y^2}{2}\right)- I_1\!\left(\frac{y^2}{2}\right)\right] \left[1+\textit{O}\!\left(x^2\right) \right] \, , \\
  H\!\left(x,y\right) & = & \sqrt{\upi} y^2 \exp{\left(-\frac{y^2}{2}\right)} \left[I_0\!\left(\frac{y^2}{2}\right)- I_1\!\left(\frac{y^2}{2}\right)\right] \left[1+\textit{O}\!\left(x^2\right) \right]  \, , \\  
  L\!\left(x,y\right) & = & -\frac{2 x}{y} \left[1-\exp \left(-\frac{y^2}{2}\right) I_0\!\left(\frac{y^2}{2}\right)\right] \left[1+\textit{O}\!\left(x^2\right) \right]  \, , \\  
  N\!\left(x,y\right) & = & -\sqrt{\upi} y \exp{\left(-\frac{y^2}{2}\right)} \left[I_0\!\left(\frac{y^2}{2}\right)- I_1\!\left(\frac{y^2}{2}\right)\right] \left[1+\textit{O}\!\left(x^2\right) \right]  \, .   
 \label{specialfuncMax_xsmall_y1}
 \end{subeqnarray}
   \item  $x, y \ll 1$:
\begin{subeqnarray}
  F\!\left(x,y\right) & = & \sqrt{\upi} \exp{\left(-\frac{1}{x^2}\right)} \left\{1+\textit{O}\!\left[\exp{\left(-\frac{3}{x^2}\right)},y^2 \right] \right\}  \, , \\
  G\!\left(x,y\right) & = & -x \left[1 - \left(\frac{3}{4}y^2 - \frac{1}{2} x^2\right) \right. \nonumber \\
  && \left. \qquad + \left(\frac{3}{4} x^4-\frac{15}{32} x^2 y^2 + \frac{5}{16}y^4\right) \right] \left[1+\textit{O}\!\left(x^6, x^4 y^2, x^2 y^4, y^6\right) \right] \, , \\
  H\!\left(x,y\right) & = & \sqrt{\upi} y^2 \left[ 1 - \left(\frac{3}{4}y^2 - \frac{1}{2} x^2\right) \right. \nonumber \\
  && \left.  \qquad + \left(\frac{3}{4} x^4-\frac{15}{32} x^2 y^2 + \frac{5}{16}y^4\right) \right] \left[1+\textit{O}\!\left(x^6, x^4 y^2, x^2 y^4, y^6\right) \right]  \, , \\
  L\!\left(x,y\right) & = & -x y \left[1+\textit{O}\!\left(x^2, y^2\right) \right]  \, , \\  
  N\!\left(x,y\right) & = & -\sqrt{\upi} y \left[1+\textit{O}\!\left(x^2\right) \right] \left[1+\textit{O}\!\left(x^2, y^2\right) \right]  \, .     
 \label{specialfuncMax_xysmall}
\end{subeqnarray}
   \item  $x \ll 1$, $y \gg 1$:
\begin{subeqnarray}
  F\!\left(x,y\right) & = &  \frac{4}{y^3} \exp{\left(-\frac{1}{x^2}\right)} \left\{1+\textit{O}\!\left[\exp{\left(-\frac{3}{x^2}\right)}, \frac{1}{y^2} \right] \right\}   \, , \\
  G\!\left(x,y\right) & = & -\frac{x}{\sqrt{\upi} y^3} \left[1+\textit{O}\!\left(\frac{1}{y^2}\right) \right] \, , \\
  H\!\left(x,y\right) & = & \frac{1}{y} \left[1+\textit{O}\!\left(\frac{1}{y^2}\right) \right]  \, , \\
  L\!\left(x,y\right) & = & -\frac{2 x}{y} \left[1-\frac{1}{\sqrt{\upi} y}\right]\left[1+\textit{O}\!\left(x^2,\frac{1}{y^3}\right) \right]  \, , \\  
  N\!\left(x,y\right) & = & -\frac{1}{y^2} \left[1+\textit{O}\!\left(\frac{1}{y^2}\right) \right] \, .    
 \label{specialfuncMax_xsmall_ylarge}
\end{subeqnarray}
\end{itemize}

\subsubsection{Unmagnetised Maxwellian dielectric response} \label{unmagresponse}

In this paper, we consider microinstabilities over a wide range of scales, from $k \rho_i \ll 1$ 
to sub-electron-scale microinstabilities with $k \rho_e \gg 1$. Therefore, the 
ordering $k \rho_s \sim 1$ assumed in section \ref{disprel_simps_II} for the derivation of the low-frequency 
dielectric tensor in a magnetised plasma cannot hold for both ions and electrons (as was noted in section \ref{consequences} and discussed in section \ref{shortcomings_twospecies}). While the 
derivation of the dielectric tensor in a strongly magnetised plasma ($k \rho_s \ll 1$) 
is straightforwardly performed by asymptotic analysis applied directly to the hot, magnetised plasma conductivity tensor (\ref{conductivity}), 
the equivalent calculation for $k \rho_s \gg 1$ is most easily done by direct 
analysis of the Vlasov equation with $\boldsymbol{B}_0 = 0$. In this appendix, we 
present such a calculation. 

We begin from (\ref{VlasovC}), but 
with  $\tilde{\Omega}_s = 0$ (and ignoring the displacement current):
\begin{subeqnarray}
    \frac{k^2 c^2}{\omega^2} \left[\widehat{\delta \boldsymbol{E}} - \hat{\boldsymbol{k}} \left(\hat{\boldsymbol{k}} \cdot \widehat{\delta \boldsymbol{E}}\right)\right]  & = &
   \frac{4 \upi \mathrm{i}}{\omega}  \widehat{\delta \boldsymbol{j}} ,\\[3pt]
       \widehat{\delta \boldsymbol{j}} & = &
   \sum_s Z_s e \int \mathrm{d}^3 \boldsymbol{v} \, \boldsymbol{v} \, \widehat{\delta f_s} , \\[3pt]
   \left(-\mathrm{i} \omega + \mathrm{i} \boldsymbol{k} \bcdot \boldsymbol{v} \right) \widehat{\delta f_s} & = &
   -\frac{Z_s e}{m_s} \left[\widehat{\delta \boldsymbol{E}} + \frac{k}{\omega} \boldsymbol{v} \times \left(\hat{\boldsymbol{k}} \times \widehat{\delta \boldsymbol{E}}\right)\right] \bcdot \frac{\p f_{s0}}{\p \boldsymbol{v}} \, \label{Vlasov_unmag} 
   .
\end{subeqnarray}
As with the magnetised case, we substitute the perturbed distribution 
function (\ref{Vlasov_unmag}\textit{c}) into the current (\ref{Vlasov_unmag}\textit{b}) :
\begin{equation}
   \widehat{\delta \boldsymbol{j}} = 
   -\mathrm{i} \sum_s \frac{Z_s^2 e^2}{m_s} \int \mathrm{d}^3 \boldsymbol{v} \, \frac{\boldsymbol{v} }{\omega - \boldsymbol{k} \bcdot \boldsymbol{v}} \left[\widehat{\delta \boldsymbol{E}} + \frac{k}{\omega} \boldsymbol{v} \times \left(\hat{\boldsymbol{k}} \times \widehat{\delta \boldsymbol{E}}\right)\right] \bcdot \frac{\p f_{s0}}{\p \boldsymbol{v}} \, \label{current_unmag_A} 
   .
\end{equation}
Non-dimensionalising the distribution function via
\begin{equation}
  \tilde{f}_{s0}(\tilde{\boldsymbol{v}}_{s}) \equiv \frac{\upi^{3/2} v_{\mathrm{th}s}^3}{n_{s0}} f_{s0}\left(v_{\mathrm{th}s} \tilde{\boldsymbol{v}}_{s}\right)
  ,
\end{equation}
we obtain 
\begin{equation}
   \widehat{\delta \boldsymbol{j}} = 
   -\frac{\mathrm{i}}{4 \upi \omega} \sum_s \omega_{\mathrm{p}s}^2 \frac{\tilde{\omega}_s}{\upi^{3/2}} \int \mathrm{d}^3 \tilde{\boldsymbol{v}_s} \, \frac{ \tilde{\boldsymbol{v}_s}}{\tilde{\omega}_s - \hat{ \boldsymbol{k}} \bcdot \tilde{\boldsymbol{v}}_s} \left[\widehat{\delta \boldsymbol{E}} + \frac{1}{\tilde{\omega}_s} \tilde{\boldsymbol{v}}_s \times \left(\hat{\boldsymbol{k}} \times \widehat{\delta \boldsymbol{E}}\right)\right] \bcdot \frac{\p \tilde{f}_{s0}}{\p \tilde{\boldsymbol{v}}_s} \, \label{current_unmag_B} 
   ,
\end{equation}
where $\tilde{\omega}_s = \omega/k v_{\mathrm{th}s}$. For a Maxwellian distribution, 
with
\begin{equation}
 \tilde{f}_{s0}(\tilde{\boldsymbol{v}}_{s}) = \exp{\left(-\tilde{v}_s^2\right)} \, ,  
\end{equation}
the second term in (\ref{current_unmag_B}) vanishes, leaving
\begin{equation}
   \widehat{\delta \boldsymbol{j}} = 
   \boldsymbol{\sigma} \bcdot \widehat{\delta \boldsymbol{E}} \, \label{current_unmag_B} 
   ,
\end{equation}
where the conductivity tensor is
\begin{equation}
\boldsymbol{\sigma} =  \frac{\mathrm{i}}{4 \upi \omega} \sum_s \omega_{\mathrm{p}s}^2 \frac{2 \tilde{\omega}_s}{\upi^{3/2}} \int \mathrm{d}^3 \tilde{\boldsymbol{v}_s} \, \frac{ \tilde{\boldsymbol{v}_s} \tilde{\boldsymbol{v}_s}}{\tilde{\omega}_s - \hat{ \boldsymbol{k}} \bcdot \tilde{\boldsymbol{v}}_s} \exp{\left(-\tilde{v}_s^2\right)} 
\, .
\end{equation}
The integral can be evaluated to give
\begin{equation}
\boldsymbol{\sigma} = -\frac{\mathrm{i}}{4 \upi \omega} \sum_s \omega_{\mathrm{p}s}^2 \tilde{\omega}_s 
\left\{Z\!\left(\tilde{\omega}_s\right)\left(\mathsfbi{I}-\hat{\boldsymbol{k}}\hat{\boldsymbol{k}}\right) + 2 \left[\tilde{\omega}_s + \tilde{\omega}_s^2 Z\!\left(\tilde{\omega}_s\right) \right]\hat{\boldsymbol{k}}\hat{\boldsymbol{k}}\right\} 
\, .
\end{equation}
The dielectric tensor in an unmagnetised Maxwellian plasma for general $\tilde{\omega}_s$ is, therefore,
\begin{equation}
\boldsymbol{\mathfrak{E}}^{\rm (UM)} = \sum_s \frac{\omega_{\mathrm{p}s}^2}{\omega^2} \tilde{\omega}_s 
\left\{Z\!\left(\tilde{\omega}_s\right) \left(\mathsfbi{I}-\hat{\boldsymbol{k}}\hat{\boldsymbol{k}}\right) + 2 \left[\tilde{\omega}_s + \tilde{\omega}_s^2 Z\!\left(\tilde{\omega}_s\right) \right]\hat{\boldsymbol{k}}\hat{\boldsymbol{k}}\right\} 
\, .
\end{equation}
Note that it follows from (\ref{Vlasov_unmag}) that $\boldsymbol{\mathfrak{E}} \bcdot \hat{\boldsymbol{k}} = 
0$, so we conclude that for non-zero fluctuations, either $\hat{\boldsymbol{k}}  \bcdot \widehat{\delta \boldsymbol{E}} = 
0$ or $1 + \tilde{\omega}_s Z\!\left(\tilde{\omega}_s\right) = 
0$. We do not find the conventional longitudinal plasma waves because we have neglected the 
displacement current in Maxwell's equations. The only modes that satisfy $1 + \tilde{\omega}_s Z\!\left(\tilde{\omega}_s\right) = 
0$ are 
strongly damped, with $\tilde{\omega}_s \sim 1$. Thus, all modes satisfying 
$\tilde{\omega}_s \ll 1$ must be purely transverse. 

For $\tilde{\omega}_s \ll 1$, the unmagnetised dielectric response therefore simplifies to
\begin{equation}
\boldsymbol{\mathfrak{E}}^{\rm (UM)} = \mathrm{i} \sqrt{\upi} \left(\mathsfbi{I}-\hat{\boldsymbol{k}}\hat{\boldsymbol{k}}\right) \sum_s \frac{\omega_{\mathrm{p}s}^2}{\omega^2} \tilde{\omega}_s 
\left[1+\textit{O}\!\left(\tilde{\omega}_s \right) \right]
\, . \label{dielectricresponselowfreq_unmag}
\end{equation}

\subsubsection{Validity of approximation $\mathsfbi{M}_s \approx \mathsfbi{M}_s^{(0)}$ for large or small $k_{\|} \rho_s $ and $k_{\bot} \rho_s $} \label{ordering}

In carrying out the expansion of the Maxwellian dielectric tensor (\ref{dielectricelements_maxB}) in $\tilde{\omega}_{s\|}$,
we assumed that $k \rho_s \sim 1$; however, in general, we will wish to consider microinstabilities that exist at 
typical wavenumbers $k \rho_s \ll 1$ or $k \rho_s \gg 1$. Indeed, since the 
mass ratio $\mu_e = m_e/m_i$ is very small, if we wish to 
consider the combined response of both species, it is inevitable that for one of them, $k \rho_s \ll 1$ or $k \rho_s \gg 1$.
Thus, it remains to assess when the approximation $\mathsfbi{M}_s \approx \mathsfbi{M}_s^{(0)}$ 
is valid in these limits. We show in this appendix that this approximation is appropriate in the 
limit $k_{\|} \rho_s \gg 1$, for arbitrary $k_{\bot} \rho_s$; however, for $k_{\|} \rho_s \ll 1$, the 
approximation breaks down for some dielectric components -- indeed, in the limit $k_\| \rho_s, k_\bot \rho_s \ll 1$, it breaks down for all but two components. 
For these instances, an alternative expression for the dielectric tensor is derived below. 

The validity of the $k_{\|} \rho_s \gg 1$ limit is most simply demonstrated by comparing the components of 
$\mathsfbi{M}_s^{(0)}$ to the unmagnetised dielectric response 
(\ref{dielectricresponselowfreq_unmag}). 
Recalling that
\begin{subeqnarray}
 (\mathsfbi{M}_s^{(0)})_{11} & = & \mathrm{i} \tilde{\omega}_{s\|} \frac{k^2}{k_{\|}^2} F\!\left(k_{\|} \tilde{\rho}_s,k_{\bot} \tilde{\rho}_s\right) \, , \\
 (\mathsfbi{M}_s^{(0)})_{12} & = & \mathrm{i} \tilde{\omega}_{s\|} \frac{k}{k_{\|}} G\!\left(k_{\|} \tilde{\rho}_s,k_{\bot} \tilde{\rho}_s\right) \,  ,\\
 (\mathsfbi{M}_s^{(0)})_{21} & = & -(\mathsfbi{M}_s^{(0)})_{12} \, , \\
 (\mathsfbi{M}_s^{(0)})_{22} & = & \mathrm{i} \tilde{\omega}_{s\|} H\!\left(k_{\|} \tilde{\rho}_s,k_{\bot} \tilde{\rho}_s\right) \, , 
 \label{dielectric123trans_mat_Append_B}
 \end{subeqnarray}
and applying the asymptotic results (\ref{specialfuncMax_xylarge}), we find 
\begin{subeqnarray}
 (\mathsfbi{M}_s^{(0)})_{11} & \approx & \mathrm{i} \sqrt{\upi} \frac{\tilde{\omega}_{s\|} k_{\|}}{k} \, , \\
 (\mathsfbi{M}_s^{(0)})_{12} & \approx & -2 \mathrm{i} \frac{\tilde{\omega}_{s\|} k_{\|}^2}{k^2} \frac{1}{k \rho_s} \,  ,\\
 (\mathsfbi{M}_s^{(0)})_{22} & \approx & \mathrm{i} \sqrt{\upi} \frac{\tilde{\omega}_{s\|} k_{\|}}{k} \, . 
 \label{dielectric123trans_mat_unmaglim}
 \end{subeqnarray}
 We note these expressions are valid for arbitrary $k_{\bot} \rho_s$. The equivalent components of the unmagnetised (normalised) dielectric tensor $\mathsfbi{M}_s \approx \omega^2 \boldsymbol{\mathfrak{E}}_s^{\rm (UM)}/\omega_{\mathrm{p}s}^2$ are
\begin{subeqnarray}
 (\mathsfbi{M}_s)_{11} & = & \mathrm{i} \sqrt{\upi} \tilde{\omega}_{s} \, , \\
 (\mathsfbi{M}_s)_{12} & = &  (\mathsfbi{M}_s^{(0)})_{21} = 0 \,  , \\
 (\mathsfbi{M}_s)_{22} & = & \mathrm{i} \sqrt{\upi}  \tilde{\omega}_{s}  \, . 
 \label{dielectric123trans_mat_unmag}
 \end{subeqnarray}
Noting that $\tilde{\omega}_{s} = \tilde{\omega}_{s\|} k_{\|}/k$, we see that the 
diagonal terms are identical, while the non-zero $\boldsymbol{e}_1 \boldsymbol{e}_2$ term present in the $k \rho_s \gg 1$ 
limit of $\mathsfbi{M}_s^{(0)}$ becomes asymptotically small in $1/k \rho_s \ll 1$. 

To demonstrate that the approximation $\mathsfbi{M}_s \approx \mathsfbi{M}_s^{(0)}$  
is not accurate in the limit $k_{\|} \rho_s \ll 1$, we consider the full 
Maxwellian dielectric tensor assuming $\tilde{\omega}_{s\|} \lesssim 1$ and $k_{\|} \rho_s \ll 1$. 
If this long-wavenumber dielectric tensor subsequently evaluated at low frequencies $\tilde{\omega}_{s\|} \ll 1$ gives the same 
result as $\mathsfbi{M}_s^{(0)}$ for any particular component of $\mathsfbi{M}_s$, then 
the approximation for that component is reasonable; otherwise, the approximation 
has to be modified at sufficiently small $k_{\|} \rho_s \ll 1$.

If $k_{\|} \rho_s \ll 1$ and $\tilde{\omega}_{s\|} \lesssim 1$, it follows that 
for $n \neq 0$, 
\begin{equation}
  |\zeta_{sn}| \equiv \left|\tilde{\omega}_{s\|} - \frac{n}{k_{\|} \tilde{\rho}_s}\right| \gg 1 \, .
\end{equation}
In this case, we can simplify the plasma dispersion function via a large-argument 
expansion:
\begin{equation}
 Z\!\left(\zeta_{sn}\right) \approx -\frac{1}{\zeta_{sn}}-\frac{1}{2 \zeta_{sn}^3} + 
 \ldots 
\end{equation}
The long-wavelength dielectric tensor is then
\begin{subeqnarray}
 (\mathsfbi{M}_s)_{xx} & \approx &  -2 \tilde{\omega}_{s\|} \sum_{n=-\infty}^{\infty} \frac{n^2}{ \zeta_{sn} k_{\bot}^2 \tilde{\rho}_s^2} \exp \left(-\frac{k_{\bot}^2 \tilde{\rho}_s^2}{2}\right) I_n\!\left(\frac{k_{\bot}^2 \tilde{\rho}_s^2}{2}\right) , \qquad\\
 (\mathsfbi{M}_s)_{xy} & \approx & -\mathrm{i} \tilde{\omega}_{s\|} \sum_{n=-\infty}^{\infty} \frac{n}{\zeta_{sn}} \exp \left(-\frac{k_{\bot}^2 \tilde{\rho}_s^2}{2}\right) \left[I_n'\!\left(\frac{k_{\bot}^2 \tilde{\rho}_s^2}{2}\right)-I_n\!\left(\frac{k_{\bot}^2 \tilde{\rho}_s^2}{2}\right) \right]\, , \\
 (\mathsfbi{M}_s)_{xz} & \approx & - \tilde{\omega}_{s\|} \sum_{n=-\infty}^{\infty} \frac{n}{\zeta_{sn}^2 k_{\bot} \tilde{\rho}_s} \exp \left(-\frac{k_{\bot}^2 \tilde{\rho}_s^2}{2}\right) I_n\!\left(\frac{k_{\bot}^2 \tilde{\rho}_s^2}{2}\right) , \qquad \\
 (\mathsfbi{M}_s)_{yx} & = & -(\mathsfbi{M}_s)_{xy} , \\
 (\mathsfbi{M}_s)_{yy} & \approx & - \tilde{\omega}_{s\|} \Bigg[ \sum_{n \in \mathbb{Z}^{\neq}} \bigg\{\frac{1}{\zeta_{sn}} \exp \left(-\frac{k_{\bot}^2 \tilde{\rho}_s^2}{2}\right)  \nonumber \\
 && \times \left[\left( \frac{2 n^2}{k_{\bot}^2 \tilde{\rho}_s^2}+k_{\bot}^2 \tilde{\rho}_s^2\right) I_n\!\left(\frac{k_{\bot}^2 \tilde{\rho}_s^2}{2}\right) -  k_{\bot}^2 \tilde{\rho}_s^2 I_n'\!\left(\frac{k_{\bot}^2 \tilde{\rho}_s^2}{2}\right) \right] \bigg\} \qquad \nonumber\\
 && - Z\!\left(\tilde{\omega}_{s\|}\right) k_{\bot}^2 \tilde{\rho}_s^2 \exp \left(-\frac{k_{\bot}^2 \tilde{\rho}_s^2}{2}\right) \left\{I_0\!\left(\frac{k_{\bot}^2 \tilde{\rho}_s^2}{2}\right)-I_1\!\left(\frac{k_{\bot}^2 \tilde{\rho}_s^2}{2}\right) \right\} 
 \Bigg] \, , \\
 (\mathsfbi{M}_s)_{yz} & \approx & \mathrm{i} \tilde{\omega}_{s\|} \Bigg[ \sum_{n \in \mathbb{Z}^{\neq}} \bigg\{ \frac{1}{2 \zeta_{sn}^2} k_{\bot} \tilde{\rho}_s \exp \left(-\frac{k_{\bot}^2 \tilde{\rho}_s^2}{2}\right)  \left[I_n'\!\left(\frac{k_{\bot}^2 \tilde{\rho}_s^2}{2}\right)-I_n\!\left(\frac{k_{\bot}^2 \tilde{\rho}_s^2}{2}\right) \right] \bigg\} \nonumber \\
 &&+ \left[1+ \tilde{\omega}_{s\|} Z\!\left(\tilde{\omega}_{s\|}\right) \right] k_{\bot} \tilde{\rho}_s \exp \left(-\frac{k_{\bot}^2 \tilde{\rho}_s^2}{2}\right) \left\{I_0\!\left(\frac{k_{\bot}^2 \tilde{\rho}_s^2}{2}\right)-I_1\!\left(\frac{k_{\bot}^2 \tilde{\rho}_s^2}{2}\right) \right\} 
 \Bigg] , \qquad \quad  \\
 (\mathsfbi{M}_s)_{zx} & = & (\mathsfbi{M}_s)_{xz} \, ,\\
 (\mathsfbi{M}_s)_{zy} & = & -(\mathsfbi{M}_s)_{yz} \, , \\
 (\mathsfbi{M}_s)_{zz} & \approx & - \tilde{\omega}_{s\|} \Bigg[ \sum_{n \in \mathbb{Z}^{\neq}} \bigg\{ \frac{1}{\zeta_{sn}} \exp \left(-\frac{k_{\bot}^2 \tilde{\rho}_s^2}{2}\right) I_n\!\left(\frac{k_{\bot}^2 \tilde{\rho}_s^2}{2}\right) 
 \bigg\} \nonumber \\
&& \qquad - 2 \tilde{\omega}_{s\|} \left[1+ \tilde{\omega}_{s\|} Z\!\left(\tilde{\omega}_{s\|}\right) \right] \exp \left(-\frac{k_{\bot}^2 \tilde{\rho}_s^2}{2}\right) I_0\!\left(\frac{k_{\bot}^2 \tilde{\rho}_s^2}{2}\right) 
\Bigg] \, ,
 \label{dielectricelements_maxC}
\end{subeqnarray}
where $\mathbb{Z}^{\neq}$ denotes non-zero integers. 
We note that the error associated with neglecting higher-order terms in $\zeta_{sn}$ 
is $\textit{O}(k_{\|}^2 \rho_s^2)$. Next, using
\begin{equation}
  -\frac{1}{\zeta_{sn}} = \frac{1}{n/k_{\|} \tilde{\rho}_s-\tilde{\omega}_{s\|}} \approx \frac{k_{\|} \tilde{\rho}_s}{n} 
  \left[1+\frac{\tilde{\omega}_{s\|}k_{\|} \tilde{\rho}_s}{n} + \textit{O}\!\left(\frac{\omega^2}{\Omega_e^2}\right) \right] , 
\end{equation}
we can isolate the dependence of each dielectric tensor component on 
$\tilde{\omega}_{s\|}$. It is clear that any sum involving an odd power of $n$ vanishes, meaning that 
the leading-order contributions in $k_{\|} \tilde{\rho}_s$ from the summation terms arise from the highest power of $\tilde{\omega}_{s\|}$ 
gives an even power of $n$. The resulting approximate expressions are
\begin{subeqnarray}
 (\mathsfbi{M}_s)_{xx} & \approx &   \frac{2 k_{\|}^2}{k_{\bot}^2} \tilde{\omega}_{s\|}^2 \left[1-\exp \left(-\frac{k_{\bot}^2 \tilde{\rho}_s^2}{2}\right) I_0\!\left(\frac{k_{\bot}^2 \tilde{\rho}_s^2}{2}\right)\right] , \qquad\\
 (\mathsfbi{M}_s)_{xy} & \approx &  \mathrm{i} \tilde{\omega}_{s\|} k_{\|} \tilde{\rho}_s \exp \left(-\frac{k_{\bot}^2 \tilde{\rho}_s^2}{2}\right) \left[I_0\!\left(\frac{k_{\bot}^2 \tilde{\rho}_s^2}{2}\right)-I_1\!\left(\frac{k_{\bot}^2 \tilde{\rho}_s^2}{2}\right) \right]\, , \\
 (\mathsfbi{M}_s)_{xz} & \approx & -4 k_{\|}^2 \tilde{\rho}_s^2  \frac{k_{\|}}{k_{\bot}} \tilde{\omega}_{s\|}^2 \sum_{n =1}^{\infty} \frac{1}{n^2} \exp \left(-\frac{k_{\bot}^2 \tilde{\rho}_s^2}{2}\right) I_n\!\left(\frac{k_{\bot}^2 \tilde{\rho}_s^2}{2}\right) , \qquad \\
 (\mathsfbi{M}_s)_{yy} & \approx & \tilde{\omega}_{s\|} \exp \left(-\frac{k_{\bot}^2 \tilde{\rho}_s^2}{2}\right) \Bigg\{ Z\!\left(\tilde{\omega}_{s\|}\right) k_{\bot}^2 \tilde{\rho}_s^2  \left[I_0\!\left(\frac{k_{\bot}^2 \tilde{\rho}_s^2}{2}\right)-I_1\!\left(\frac{k_{\bot}^2 \tilde{\rho}_s^2}{2}\right) \right] \nonumber \\
 &&+ 2 \tilde{\omega}_{s\|} k_{\|}^2 \tilde{\rho}_s^2 \sum_{n =1}^{\infty} \left[\left( \frac{2}{k_{\bot}^2 \tilde{\rho}_s^2}+\frac{k_{\bot}^2 \tilde{\rho}_s^2}{n^2}\right) I_n\!\left(\frac{k_{\bot}^2 \tilde{\rho}_s^2}{2}\right) -  \frac{k_{\bot}^2 \tilde{\rho}_s^2}{n^2} I_n'\!\left(\frac{k_{\bot}^2 \tilde{\rho}_s^2}{2}\right) \right] \Bigg\} , \qquad \quad  \\
 (\mathsfbi{M}_s)_{yz} & \approx & \mathrm{i} \tilde{\omega}_{s\|} \left[1+ \tilde{\omega}_{s\|} Z\!\left(\tilde{\omega}_{s\|}\right) \right] \nonumber \\
 && \qquad \qquad \times k_{\bot} \tilde{\rho}_s \exp \left(-\frac{k_{\bot}^2 \tilde{\rho}_s^2}{2}\right)  \left[I_0\!\left(\frac{k_{\bot}^2 \tilde{\rho}_s^2}{2}\right)-I_1\!\left(\frac{k_{\bot}^2 \tilde{\rho}_s^2}{2}\right) \right]\, , \\
 (\mathsfbi{M}_s)_{zz} & \approx & 2 \tilde{\omega}_{s\|}^2 \left[1+ \tilde{\omega}_{s\|} Z\!\left(\tilde{\omega}_{s\|}\right) \right] \exp \left(-\frac{k_{\bot}^2 \tilde{\rho}_s^2}{2}\right) I_0\!\left(\frac{k_{\bot}^2 \tilde{\rho}_s^2}{2}\right) 
  \, ,
 \label{dielectricelements_maxD}
\end{subeqnarray}
where we have again used the sum identities (\ref{Besselfuncsums}). 
Note that we have retained a term in $ (\mathsfbi{M}_s)_{yy}$ which is quadratic in $k_{\|} 
\tilde{\rho}_s$, even though there exists another term which is independent of $k_{\|} 
\tilde{\rho}_s$. This is because the latter term becomes arbitrarily small in the limit $k_{\bot} 
\rho_s \ll 1$, whereas the former is independent of $k_{\bot} 
\rho_s$; hence, if $k_{\bot} \rho_s \ll k_{\|} \rho_s$, the latter term can become dominant. 

Now considering 
the limit $\tilde{\omega}_{s\|} \ll 1$, while holding $k_{\|} \rho_s \ll 1$ at some fixed value , the plasma 
dispersion function can now be approximated by its small-argument expansion
\begin{equation}
 Z\!\left(\tilde{\omega}_{s\|}\right) \approx i \sqrt{\upi} \, ,
\end{equation}
to give 
\begin{subeqnarray}
 (\mathsfbi{M}_s)_{xx} & \approx &  \frac{2 k_{\|}^2}{k_{\bot}^2} \tilde{\omega}_{s\|}^2 \left[1-\exp \left(-\frac{k_{\bot}^2 \tilde{\rho}_s^2}{2}\right) I_0\!\left(\frac{k_{\bot}^2 \tilde{\rho}_s^2}{2}\right)\right] , \qquad\\
 (\mathsfbi{M}_s)_{xy} & \approx & \mathrm{i} \tilde{\omega}_{s\|} k_{\|} \tilde{\rho}_s \exp \left(-\frac{k_{\bot}^2 \tilde{\rho}_s^2}{2}\right) \left[I_0\!\left(\frac{k_{\bot}^2 \tilde{\rho}_s^2}{2}\right)-I_1\!\left(\frac{k_{\bot}^2 \tilde{\rho}_s^2}{2}\right) \right]\, , \\
 (\mathsfbi{M}_s)_{xz} & \approx & -4 k_{\|}^2 \tilde{\rho}_s^2  \frac{k_{\|}}{k_{\bot}} \tilde{\omega}_{s\|}^2 \sum_{n =1}^{\infty} \frac{1}{n^2} \exp \left(-\frac{k_{\bot}^2 \tilde{\rho}_s^2}{2}\right) I_n\!\left(\frac{k_{\bot}^2 \tilde{\rho}_s^2}{2}\right) , \qquad \\
 (\mathsfbi{M}_s)_{yy} & \approx & \tilde{\omega}_{s\|} \exp \left(-\frac{k_{\bot}^2 \tilde{\rho}_s^2}{2}\right) \Bigg\{ \mathrm{i} \sqrt{\upi} k_{\bot}^2 \tilde{\rho}_s^2 \left[I_0\!\left(\frac{k_{\bot}^2 \tilde{\rho}_s^2}{2}\right)-I_1\!\left(\frac{k_{\bot}^2 \tilde{\rho}_s^2}{2}\right) \right] \nonumber \\
 && + 2 \tilde{\omega}_{s\|}^2 k_{\|}^2 \tilde{\rho}_s^2 \sum_{n =1}^{\infty} \left[\left( \frac{2}{k_{\bot}^2 \tilde{\rho}_s^2}+\frac{k_{\bot}^2 \tilde{\rho}_s^2}{n^2}\right) I_n\!\left(\frac{k_{\bot}^2 \tilde{\rho}_s^2}{2}\right) -  \frac{k_{\bot}^2 \tilde{\rho}_s^2}{n^2} I_n'\!\left(\frac{k_{\bot}^2 \tilde{\rho}_s^2}{2}\right) \right] \Bigg\} , \qquad \quad \\
 (\mathsfbi{M}_s)_{yz} & \approx & \mathrm{i} \tilde{\omega}_{s\|} \left[1+ \mathrm{i} \sqrt{\upi} \tilde{\omega}_{s\|} \right] \nonumber \\
 && \qquad \qquad \times k_{\bot} \tilde{\rho}_s \exp \left(-\frac{k_{\bot}^2 \tilde{\rho}_s^2}{2}\right)  \left[I_0\!\left(\frac{k_{\bot}^2 \tilde{\rho}_s^2}{2}\right)-I_1\!\left(\frac{k_{\bot}^2 \tilde{\rho}_s^2}{2}\right) \right]\, , \\
 (\mathsfbi{M}_s)_{zz} & \approx & 2 \tilde{\omega}_{s\|}^2 \left[1+ \mathrm{i} \sqrt{\upi} \tilde{\omega}_{s\|} \right] \exp \left(-\frac{k_{\bot}^2 \tilde{\rho}_s^2}{2}\right) I_0\!\left(\frac{k_{\bot}^2 \tilde{\rho}_s^2}{2}\right) 
  \, .
 \label{dielectricelements_maxD}
\end{subeqnarray}
For comparison, we state below the long-wavelength limit of $\mathsfbi{M}_s^{(0)}$ 
using asymptotic expressions (\ref{specialfuncMax_xsmall_y1}):
\begin{subeqnarray}
 (\mathsfbi{M}_s^{(0)})_{xx} & = &  4 \mathrm{i} \sqrt{\upi} \frac{\tilde{\omega}_{s\|} }{k_{\bot}^2 \tilde{\rho}_s^2} \exp{\left(-\frac{1}{k_{\|}^2 \tilde{\rho}_s^2}\right)} \exp \left(-\frac{k_{\bot}^2 \tilde{\rho}_s^2}{2}\right) I_1\!\left(\frac{k_{\bot}^2 \tilde{\rho}_s^2}{2}\right) \, , \\
 (\mathsfbi{M}_s^{(0)})_{xy} & = & \mathrm{i} \tilde{\omega}_{s\|} |k_{\|}| \tilde{\rho}_s \exp \left(-\frac{k_{\bot}^2 \tilde{\rho}_s^2}{2}\right) \left[I_0\!\left(\frac{k_{\bot}^2 \tilde{\rho}_s^2}{2}\right)-I_1\!\left(\frac{k_{\bot}^2 \tilde{\rho}_s^2}{2}\right) \right] \, , \\
 (\mathsfbi{M}_s^{(0)})_{xz} & = & -4 \mathrm{i} \sqrt{\upi} \frac{\tilde{\omega}_{s\|} }{k_{\bot} k_{\|} \tilde{\rho}_s^2} \exp{\left(-\frac{1}{k_{\|}^2 \tilde{\rho}_s^2}\right)} \exp \left(-\frac{k_{\bot}^2 \tilde{\rho}_s^2}{2}\right) I_1\!\left(\frac{k_{\bot}^2 \tilde{\rho}_s^2}{2}\right)\, , \\
 (\mathsfbi{M}_s^{(0)})_{yy} & = & \mathrm{i} \sqrt{\upi} \tilde{\omega}_{s\|} k_{\bot}^2 \tilde{\rho}_s^2 \exp \left(-\frac{k_{\bot}^2 \tilde{\rho}_s^2}{2}\right) \left[I_0\!\left(\frac{k_{\bot}^2 \tilde{\rho}_s^2}{2}\right)-I_1\!\left(\frac{k_{\bot}^2 \tilde{\rho}_s^2}{2}\right) \right] , \qquad \\
(\mathsfbi{M}_s^{(0)})_{yz} & = & \mathrm{i} \tilde{\omega}_{s\|} k_{\bot} \tilde{\rho}_s \exp \left(-\frac{k_{\bot}^2 \tilde{\rho}_s^2}{2}\right)  \left[I_0\!\left(\frac{k_{\bot}^2 \tilde{\rho}_s^2}{2}\right)-I_1\!\left(\frac{k_{\bot}^2 \tilde{\rho}_s^2}{2}\right) \right]\, , \\
 (\mathsfbi{M}_s^{(0)})_{zz} & = &  4 \mathrm{i} \sqrt{\upi} \frac{\tilde{\omega}_{s\|}}{k_{\|}^2 \tilde{\rho}_s^2} \exp{\left(-\frac{1}{k_{\|}^2 \tilde{\rho}_s^2}\right)} \exp \left(-\frac{k_{\bot}^2 \tilde{\rho}_s^2}{2}\right) I_1\!\left(\frac{k_{\bot}^2 \tilde{\rho}_s^2}{2}\right)  \, 
 .
 \label{dielectricelements_max0_smallxlim}
\end{subeqnarray}
Assuming $k_{\bot} \rho_s \sim 1$, we observe that while three of the six unique dielectric tensor components are 
identical for both $\tilde{\omega}_{s\|} \rightarrow 0$, $k_{\|} \rho_s \ll 1$ 
fixed, and $k_{\|} \rho_s \rightarrow 0$, $\tilde{\omega}_{s\|} \ll 1$ fixed [$(\mathsfbi{M}_s)_{xy}$, $(\mathsfbi{M}_s)_{yy}$, and $(\mathsfbi{M}_s)_{yz}$], 
the other three [$(\mathsfbi{M}_s)_{xx}$, $(\mathsfbi{M}_s)_{xz}$, and $(\mathsfbi{M}_s)_{zz}$] are not.
Instead, the dominant terms are the quadratic terms $(\mathsfbi{M}_s^{(1)})_{xx}$, $(\mathsfbi{M}_s^{(1)})_{xz}$, and $(\mathsfbi{M}_s^{(1)})_{zz}$ in the $\tilde{\omega}_{s\|} \ll 1$ 
expansion. In the limit $k_{\bot} \rho_s \ll 1$, $(\mathsfbi{M}_s)_{yy}$ 
also departs from the approximation $(\mathsfbi{M}_s^{(0)})_{yy}$ 
for sufficiently small $k_{\bot} \rho_s$ as compared to $k_{\|} \rho_s$,
instead being accurately described by $(\mathsfbi{M}_s^{(1)})_{yy}$. 

As a consequence, we must assess the conditions under 
which one approximation or the other is valid. This is most simply answered by 
observing that the expressions for $(\mathsfbi{M}_s^{(0)})_{xx}$, $(\mathsfbi{M}_s^{(0)})_{xz}$, and $(\mathsfbi{M}_s^{(0)})_{zz}$ 
from (\ref{dielectricelements_max0_smallxlim}\textit{a}), (\ref{dielectricelements_max0_smallxlim}\textit{c}) 
and (\ref{dielectricelements_max0_smallxlim}\textit{f}) are exponentially small; thus, 
for $k_{\|} \rho_s \ll 1/\log{(1/\tilde{\omega}_{s\|})}$, we must use 
approximations (\ref{dielectricelements_maxD}\textit{a}), 
(\ref{dielectricelements_maxD}\textit{c}), (\ref{dielectricelements_maxD}\textit{e}) for $(\mathsfbi{M}_s)_{xx}$, $(\mathsfbi{M}_s)_{xz}$, and 
$(\mathsfbi{M}_s)_{zz}$. In addition, if $k_{\bot}^2 \rho_s^2 \ll \tilde{\omega}_{s\|} k_{\|}^2 
\rho_s^2 \ll 1$, then 
\begin{equation}
 (\mathsfbi{M}_s)_{yy} \approx \frac{2\omega_{\mathrm{p}s}^2}{\omega^2} \tilde{\omega}_{s\|}^2 k_{\|}^2 \tilde{\rho}_s^2
\end{equation}
becomes the appropriate approximation for $(\mathsfbi{M}_s)_{yy}$. 

\subsubsection{Calculation of secord-order corrections to dispersion relation} \label{secondordercorrect}

In this appendix, we justify the relations (\ref{secondordercorrect_orderings}) used in appendix \ref{CES_method_append} -- 
that is, for $k_{\|} \rho_s \ll 1$, 
\begin{subeqnarray}
 \frac{\left[(\mathsfbi{M}_{s})_{13}\right]^2}{(\mathsfbi{M}_s^{(1)})_{33}} & \lesssim & (\mathsfbi{M}_{s})_{11} \, , \\
 \frac{(\mathsfbi{M}_{s})_{13}(\mathsfbi{M}_{s})_{23}}{(\mathsfbi{M}_s^{(1)})_{33}} & \lesssim & \tilde{\omega}_{e\|} (\mathsfbi{M}_{s})_{12} \ll (\mathsfbi{M}_{s})_{12} \, , \\
 \frac{\left[(\mathsfbi{M}_{s})_{23}\right]^2}{(\mathsfbi{M}_s^{(1)})_{33}} & \lesssim & \tilde{\omega}_{e\|} (\mathsfbi{M}_{s})_{22} \ll (\mathsfbi{M}_{s})_{22}  \, . 
  \label{secondordercorrect_orderings_Append}
\end{subeqnarray}
We also prove the identity (\ref{secondorderident_special}), or 
    \begin{equation}
  (\mathsfbi{M}_{e}^{(1)} + \mathsfbi{M}_{i}^{(1)})_{11} - \frac{\left[(\mathsfbi{M}_{e}^{(1)} + \mathsfbi{M}_{i}^{(1)})_{13}\right]^2}{2 (\mathsfbi{M}_{e}^{(1)})_{33}} =   -\frac{4}{3} W_e - \frac{4}{3} W_i - \frac{1}{4} \left(L_e + L_i\right)^2 
  \, \label{secondorderident_special_Append}
  \end{equation}
  used to derive the dispersion relation (\ref{sheardisprel_D}). 
  
To complete the first task, we begin with the expressions (\ref{dielectric123trans_smallkpl}) for the 
dielectric components, and substitute (\ref{dielectricelements_max1_B}\textit{a}), (\ref{dielectricelements_max0_B}\textit{b}), (\ref{dielectricelements_max0_B}\textit{d})
 and (\ref{dielectricelements_max1_B}\textit{d}) for $(\mathsfbi{M}_{s}^{(1)})_{xx}$, 
 $(\mathsfbi{M}_{s}^{(0)})_{xy}$, $(\mathsfbi{M}_{s}^{(0)})_{xy}$, $(\mathsfbi{M}_{s}^{(0)})_{yy}$ 
 and $(\mathsfbi{M}_{s}^{(1)})_{xy}$, respectively. This gives (\ref{dielectric123trans_smallkpl})
 directly in terms of special functions $G(x,y)$, $H(x,y)$, $L(x,y)$, $N(x,y)$, $W(x,y)$ and $Y(x,y)$:
 \begin{subeqnarray}
 (\mathsfbi{M}_{s})_{11} & \approx & -\frac{4 k^2} {3 k_{\|}^2} \tilde{\omega}_{s\|}^2 W\!\left(k_{\|} \tilde{\rho}_s,k_{\bot} \tilde{\rho}_s \right) + 2 \tilde{\omega}_{s\|}^2 \left[\frac{k_{\bot}^2}{k^2} + \frac{k_{\bot}}{k_{\|}} L\!\left(k_{\|} \tilde{\rho}_s,k_{\bot} \tilde{\rho}_s \right)\right] \, , \\
 (\mathsfbi{M}_{s})_{12} & \approx &  - \mathrm{i} \frac{k}{k_{\|}} \tilde{\omega}_{s\|} G\!\left(k_{\|} \tilde{\rho}_s,k_{\bot} \tilde{\rho}_s \right) \,  ,\\
 (\mathsfbi{M}_{s})_{13} & \approx & -\tilde{\omega}_{s\|}^2 \left[\frac{2 k_{\bot} k_{\|}}{k^2} + L\!\left(k_{\|} \tilde{\rho}_s,k_{\bot} \tilde{\rho}_s \right)\right] \, , \\
 (\mathsfbi{M}_{s})_{22} & \approx & \mathrm{i} \tilde{\omega}_{s\|} H\!\left(k_{\|} \tilde{\rho}_s,k_{\bot} \tilde{\rho}_s \right) -\frac{4}{3} \tilde{\omega}_{s\|}^2 Y\!\left(k_{\|} \tilde{\rho}_s,k_{\bot} \tilde{\rho}_s \right) \, , \\
 (\mathsfbi{M}_{s})_{23} & \approx & -\frac{k_{\|}}{k} \tilde{\omega}_{s\|}^2 N\!\left(k_{\|} \tilde{\rho}_s,k_{\bot} \tilde{\rho}_s \right) \, , \\
 (\mathsfbi{M}_{s})_{33} & \approx & \frac{2 k_{\|}^2}{k^2} \tilde{\omega}_{s\|}^2 \, 
 . \label{dielectric123trans_smallkpl_Append_secondorder_specfunc}
 \end{subeqnarray}
 We then apply the $k_{\|} \rho_s \ll 1$ limits of the aforementioned special functions  
 using Appendices \ref{asympspecialfuncsappend} and \ref{asympspecialfuncsappendB} -- in 
 particular, (\ref{specialfuncMax_xsmall_y1}\textit{b}), 
 (\ref{specialfuncMax_xsmall_y1}\textit{c}), (\ref{specialfuncMax_xsmall_y1}\textit{d}), 
 (\ref{specialfuncMax_xsmall_y1}\textit{e}), (\ref{shearasymptoticfuncs_xsmall_y1}\textit{a}), and 
 (\ref{shearasymptoticfuncs_xysmall}\textit{c}):
  \begin{subeqnarray}
 (\mathsfbi{M}_{s})_{11} & \approx & 2 \tilde{\omega}_{s\|}^2 \exp \left(-\frac{k_{\bot}^2 \tilde{\rho}_s^2}{2}\right) I_0\!\left(\frac{k_{\bot}^2 \tilde{\rho}_s^2}{2}\right) \, , \\
 (\mathsfbi{M}_{s})_{12} & \approx & \mathrm{i} \tilde{\omega}_{s\|} k \tilde{\rho}_s \exp{\left(-\frac{k_{\bot}^2 \tilde{\rho}_s^2}{2}\right)} \left[I_0\!\left(\frac{k_{\bot}^2 \tilde{\rho}_s^2}{2}\right)- I_1\!\left(\frac{k_{\bot}^2 \tilde{\rho}_s^2}{2}\right)\right] \,  ,\\
 (\mathsfbi{M}_{s})_{13} & \approx & -\tilde{\omega}_{s\|}^2 \frac{2 k_{\|}}{k} \exp \left(-\frac{k_{\bot}^2 \tilde{\rho}_s^2}{2}\right) I_0\!\left(\frac{k_{\bot}^2 \tilde{\rho}_s^2}{2}\right) \, , \\
 (\mathsfbi{M}_{s})_{22} & \approx &   \mathrm{i} \sqrt{\upi} \tilde{\omega}_{s\|} k_{\bot}^2 \tilde{\rho}_s^2 \exp{\left(-\frac{k_{\bot}^2 \tilde{\rho}_s^2}{2}\right)} \left[I_0\!\left(\frac{k_{\bot}^2 \tilde{\rho}_s^2}{2}\right)- I_1\!\left(\frac{k_{\bot}^2 \tilde{\rho}_s^2}{2}\right)\right] + \tilde{\omega}_{s\|}^2 k_{\|}^2 \tilde{\rho}_s^2  \, , \qquad \\
 (\mathsfbi{M}_{s})_{23} & \approx & \sqrt{\upi} \tilde{\omega}_{s\|}^2 k_{\|} \tilde{\rho}_s \exp{\left(-\frac{k_{\bot}^2 \tilde{\rho}_s^2}{2}\right)} \left[I_0\!\left(\frac{k_{\bot}^2 \tilde{\rho}_s^2}{2}\right)- I_1\!\left(\frac{k_{\bot}^2 \tilde{\rho}_s^2}{2}\right)\right] \, , \\
 (\mathsfbi{M}_{s})_{33} & \approx & \frac{2 k_{\|}^2}{k^2} \tilde{\omega}_{s\|}^2 \, 
 . \label{dielectric123trans_smallkpl_Append_secondorder_specfunc_sub}
 \end{subeqnarray}
We can now make the relevant comparisons presented in 
(\ref{secondordercorrect_orderings_Append}), and obtain the desired results:
\begin{subeqnarray}
\frac{\left[(\mathsfbi{M}_{s})_{13}\right]^2}{(\mathsfbi{M}_{s})_{11} (\mathsfbi{M}_{s})_{33}} & \approx 
& \exp \left(-\frac{k_{\bot}^2 \tilde{\rho}_s^2}{2}\right) I_0\!\left(\frac{k_{\bot}^2 \tilde{\rho}_s^2}{2}\right) \lesssim 1, \\
\frac{(\mathsfbi{M}_{s})_{13}(\mathsfbi{M}_{s})_{23}}{(\mathsfbi{M}_{s})_{12} \mathsfbi{M}_{s})_{33}} 
& \approx & \mathrm{i} \tilde{\omega}_{s\|} \exp \left(-\frac{k_{\bot}^2 \tilde{\rho}_s^2}{2}\right) I_0\!\left(\frac{k_{\bot}^2 \tilde{\rho}_s^2}{2}\right) \lesssim \tilde{\omega}_{s\|} , \\
\frac{\left[(\mathsfbi{M}_{s})_{23}\right]^2}{(\mathsfbi{M}_{s})_{22} (\mathsfbi{M}_{s})_{33}} 
& \approx & - \mathrm{i} \frac{\sqrt{\upi}}{2} \tilde{\omega}_{s\|} \exp{\left(-\frac{k_{\bot}^2 \tilde{\rho}_s^2}{2}\right)} \left[I_0\!\left(\frac{k_{\bot}^2 \tilde{\rho}_s^2}{2}\right)- I_1\!\left(\frac{k_{\bot}^2 \tilde{\rho}_s^2}{2}\right)\right] \lesssim 
\tilde{\omega}_{s\|} \, , \qquad
\end{subeqnarray}
where we used the inequalities 
\begin{subeqnarray}
\exp \left(-\frac{k_{\bot}^2 \tilde{\rho}_s^2}{2}\right) I_0\!\left(\frac{k_{\bot}^2 \tilde{\rho}_s^2}{2}\right) &  
\leq & 1 , \\
\exp{\left(-\frac{k_{\bot}^2 \tilde{\rho}_s^2}{2}\right)} \left[I_0\!\left(\frac{k_{\bot}^2 \tilde{\rho}_s^2}{2}\right)- I_1\!\left(\frac{k_{\bot}^2 \tilde{\rho}_s^2}{2}\right)\right] &  
\leq & 1 ,
\end{subeqnarray}
valid for arbitrary values of $k_{\bot} \tilde{\rho}_s$. 
 
 To derive (\ref{secondorderident_special_Append}), we use  
 (\ref{dielectric123trans_smallkpl_Append_secondorder_specfunc}\textit{a}),  (\ref{dielectric123trans_smallkpl_Append_secondorder_specfunc}\textit{c})
 and (\ref{dielectric123trans_smallkpl_Append_secondorder_specfunc}\textit{f}) to derive the 
 following expressions:
 \begin{subeqnarray}
   (\mathsfbi{M}_{e}^{(1)} + \mathsfbi{M}_{i}^{(1)})_{11} & = &  \frac{k^2} {k_{\|}^2}\left[(\mathsfbi{M}_{e}^{(1)})_{xx} + (\mathsfbi{M}_{i}^{(1)})_{xx} \right] + 2 \left[\frac{2 k_{\bot}^2}{k^2} + \frac{k_{\bot}}{k_{\|}} \left( L_e + L_i \right)\right] 
   , \\
    (\mathsfbi{M}_{e}^{(1)} + \mathsfbi{M}_{i}^{(1)})_{13} & = &  \frac{4 k_{\bot} k_{\|}}{k^2} + L_e + L_i 
   , \\
  2 (\mathsfbi{M}_{e}^{(1)})_{33} & = & \frac{4 k_{\|}^2}{k^2} ,
 \end{subeqnarray}
 where we have introduced the notation $L_e = L\!\left(k_{\|} \tilde{\rho}_e,k_{\bot} \tilde{\rho}_e 
 \right)$, $L_i = L\!\left(k_{\|} \rho_i,k_{\bot} \rho_i \right)$. Then,
 \begin{equation}
\frac{\left[(\mathsfbi{M}_{e}^{(1)} + \mathsfbi{M}_{i}^{(1)})_{13}\right]^2}{2 (\mathsfbi{M}_{e}^{(1)})_{33}} 
= \left[\frac{2 k_{\bot}}{k} + \frac{k}{2 k_{\|}} \left( L_e + L_i\right) \right]^2 ,
 \end{equation}
 which in turn gives
 \begin{equation}
   (\mathsfbi{M}_{e}^{(1)} + \mathsfbi{M}_{i}^{(1)})_{11} - \frac{\left[(\mathsfbi{M}_{e}^{(1)} + \mathsfbi{M}_{i}^{(1)})_{13}\right]^2}{2 (\mathsfbi{M}_{e}^{(1)})_{33}} =  \frac{k^2} {k_{\|}^2}\left[(\mathsfbi{M}_{e}^{(1)})_{xx} + (\mathsfbi{M}_{i}^{(1)})_{xx} - \frac{1}{4} \left(L_e + L_i\right)^2\right] .
 \end{equation}
The identities (\ref{secondordercorrect_identities}) give 
(\ref{secondorderident_special_Append}), completing the proof. 

\subsection{CE electron-friction term} \label{resistiveterm}

For an electron distribution of the form 
\begin{equation}
 \tilde{f}_e(\tilde{v}_{e\|},\tilde{v}_{e\bot}) = - \eta_e^{R} \tilde{v}_{e\|} \exp \left(-\tilde{v}_{e}^2\right)   
\end{equation}
with $\eta_s^{R} \ll 1$ a constant, it follows that
\begin{equation}
 \Lambda_e(\tilde{v}_{e\|},\tilde{v}_{e\bot}) = - \eta_e^{R} \tilde{v}_{e\bot} \exp \left(-\tilde{v}_{e}^2\right),  \label{Lambda_func_friction} 
\end{equation}
while
\begin{equation}
  \Xi_e(\tilde{v}_{e\|},\tilde{v}_{e\bot}) = - \frac{\eta_e^R}{\tilde{\omega}_{e\|}} \tilde{v}_{e\bot} \exp \left(-\tilde{v}_{e}^2\right) + \textit{O}(\eta_e) .
\end{equation}
Since
\begin{equation}
\int_{-\infty}^{\infty} \mathrm{d} \tilde{v}_{e\|} \, \tilde{v}_{e\|} \int_0^{\infty} \mathrm{d} \tilde{v}_{e\bot} \Lambda_e(\tilde{v}_{e\|},\tilde{v}_{e\bot}) = 0  
\end{equation}
when $\Lambda_e(\tilde{v}_{e\|},\tilde{v}_{e\bot})$ is given by (\ref{Lambda_func_friction}), 
the function $\Xi_e(\tilde{v}_{e\|},\tilde{v}_{e\bot})$ is just proportional to that arising for a Maxwellian distribution [cf. (\ref{Xi_maxwellian})],
and so the dielectric response associated with the CE electron-friction 
term is too:
\begin{equation}
\mathsfbi{P}_e = \frac{\eta_e^R}{2} \mathsfbi{M}_e 
\, . \label{Presismat_Append}
\end{equation}

\subsection{CE temperature-gradient-driven terms} \label{heatfluxterm}

For the CE temperature-gradient-driven term arising from a Krook operator, which takes the form
\begin{equation}
 \tilde{f}_s(\tilde{v}_{s\|},\tilde{v}_{s\bot}) = - \eta_s \tilde{v}_{s\|} \left(\tilde{v}_{s}^2 - \frac{5}{2}\right) \exp \left(-\tilde{v}_{s}^2\right) ,      
\end{equation}
it follows (assuming $\eta_e^R = 0$) that
\begin{equation}
 \Lambda_s(\tilde{v}_{s\|},\tilde{v}_{s\bot}) = - \eta_s \tilde{v}_{s\bot}  \left(\tilde{v}_{s}^2 - \frac{5}{2}\right)\exp \left(-\tilde{v}_{s}^2\right),  
\end{equation}
and
\begin{equation}
  \Xi_s(\tilde{v}_{s\|},\tilde{v}_{s\bot}) = - \frac{\eta_s}{\tilde{\omega}_{s\|}} \tilde{v}_{s\bot} \left(\tilde{v}_{s}^2 - \frac{5}{2}\right) \exp \left(-\tilde{v}_{s}^2\right) + \textit{O}(\eta_s) .
\end{equation}
Then, to leading order in $\eta_s$,
\begin{subeqnarray}
 (\mathsfbi{P}_s)_{xx} & = &  \frac{2}{\sqrt{\upi}} \eta_s \sum_{n=-\infty}^{\infty} \left[ \frac{n^2}{k_{\bot}^2 \tilde{\rho}_s^2}\int_{C_L} \frac{\exp \left(-\tilde{v}_{s\|}^2\right) \mathrm{d} \tilde{v}_{s\|}}{\tilde{v}_{s\|}-\zeta_{sn}} \right. \nonumber \\
 && \left. \qquad \qquad  \times \int_0^{\infty} \mathrm{d} \tilde{v}_{s\bot} \, \tilde{v}_{s\bot} J_n(k_{\bot} \tilde{\rho}_s \tilde{v}_{s\bot})^2 \exp \left(-\tilde{v}_{s\bot}^2\right) \left(\tilde{v}_{s}^2 - \frac{5}{2}\right) \right] , \qquad\\
 (\mathsfbi{P}_s)_{xy} & = & \frac{2 \mathrm{i}}{\sqrt{\upi}} \eta_s\sum_{n=-\infty}^{\infty} \left[ \frac{n}{k_{\bot} \tilde{\rho}_s} \int_{C_L} \frac{\exp \left(-\tilde{v}_{s\|}^2\right) \mathrm{d} \tilde{v}_{s\|}}{\tilde{v}_{s\|}-\zeta_{sn}} \right. \nonumber \\
 && \left. \, \times \int_0^{\infty} \mathrm{d} \tilde{v}_{s\bot} \, \tilde{v}_{s\bot}^2 J_n(k_{\bot} \tilde{\rho}_s \tilde{v}_{s\bot}) J_n'(k_{\bot} \tilde{\rho}_s \tilde{v}_{s\bot}) \exp \left(-\tilde{v}_{s\bot}^2\right)\left(\tilde{v}_{s}^2 - \frac{5}{2}\right) \right] \, , \qquad \\
 (\mathsfbi{P}_s)_{xz} & = & \frac{2}{\sqrt{\upi}} \eta_s \sum_{n=-\infty}^{\infty} \left[ \frac{n}{k_{\bot} \tilde{\rho}_s}\int_{C_L} \frac{\tilde{v}_{s\|} \exp \left(-\tilde{v}_{s\|}^2\right) \mathrm{d} \tilde{v}_{s\|}}{\tilde{v}_{s\|}-\zeta_{sn}} \right. \\
 && \left.  \qquad \qquad \times \int_0^{\infty} \mathrm{d} \tilde{v}_{s\bot} \, \tilde{v}_{s\bot} J_n(k_{\bot} \tilde{\rho}_s \tilde{v}_{s\bot})^2 \exp \left(-\tilde{v}_{s\bot}^2\right) \left(\tilde{v}_{s}^2 - \frac{5}{2}\right) \right] , \qquad\\
 (\mathsfbi{P}_s)_{yx} & = & -(\mathsfbi{P}_s)_{xy} \, , \\
 (\mathsfbi{P}_s)_{yy} & = & \frac{2}{\sqrt{\upi}} \eta_s \sum_{n=-\infty}^{\infty} \left[ \int_{C_L} \frac{\exp \left(-\tilde{v}_{s\|}^2\right) \mathrm{d} \tilde{v}_{s\|}}{\tilde{v}_{s\|}-\zeta_{sn}} \right. \nonumber \\
 && \left. \qquad \qquad \times \int_0^{\infty} \mathrm{d} \tilde{v}_{s\bot} \, \tilde{v}_{s\bot}^3 J_n'(k_{\bot} \tilde{\rho}_s \tilde{v}_{s\bot})^2 \exp \left(-\tilde{v}_{s\bot}^2\right) \left(\tilde{v}_{s}^2 - \frac{5}{2}\right) \right] , \qquad\\
 (\mathsfbi{P}_s)_{yz} & = & -\frac{2 \mathrm{i} }{\sqrt{\upi}} \eta_s \sum_{n=-\infty}^{\infty} \left[ \int_{C_L} \frac{\tilde{v}_{s\|} \exp \left(-\tilde{v}_{s\|}^2\right) \mathrm{d} \tilde{v}_{s\|}}{\tilde{v}_{s\|}-\zeta_{sn}} \right. \nonumber \\
 && \left. \, \times \int_0^{\infty} \mathrm{d} \tilde{v}_{s\bot} \, \tilde{v}_{s\bot}^2 J_n(k_{\bot} \tilde{\rho}_s \tilde{v}_{s\bot}) J_n'(k_{\bot} \tilde{\rho}_s \tilde{v}_{s\bot}) \exp \left(-\tilde{v}_{s\bot}^2\right) \left(\tilde{v}_{s}^2 - \frac{5}{2}\right) \right] \, , \qquad \\
 (\mathsfbi{P}_s)_{zx} & = & (\mathsfbi{P}_s)_{xz} \, ,\\
 (\mathsfbi{P}_s)_{zy} & = & -(\mathsfbi{P}_s)_{yz} \, , \\
 (\mathsfbi{P}_s)_{zz} & = & \frac{2 }{\sqrt{\upi}} \eta_s \sum_{n=-\infty}^{\infty} \left[ \int_{C_L} \frac{\tilde{v}_{s\|}^2 \exp \left(-\tilde{v}_{s\|}^2\right) \mathrm{d} \tilde{v}_{s\|}}{\tilde{v}_{s\|}-\zeta_{sn}} \right. \nonumber \\
 && \left. \qquad \qquad \times \int_0^{\infty} \mathrm{d} \tilde{v}_{s\bot} \, \tilde{v}_{s\bot} J_n(k_{\bot} \tilde{\rho}_s \tilde{v}_{s\bot})^2 \exp \left(-\tilde{v}_{s\bot}^2\right) \left(\tilde{v}_{s}^2 - \frac{5}{2}\right) \right] . \qquad
 \label{dielectricelements_heatfluxA}
\end{subeqnarray}
In addition to the plasma-dispersion-function identities (\ref{plasmadispfunc_iden}) and Bessel-function identities (\ref{Besselfunc_iden}), we use
\begin{subeqnarray}
\frac{1}{\sqrt{\upi}}\int_{C_L} \frac{u^3 \exp{\left(-u^2\right)} \mathrm{d}u}{u-z}  & = & \frac{1}{2} + z^2\left[1 + z Z\!\left(z\right) 
\right] \, , \\
\frac{1}{\sqrt{\upi}}\int_{C_L} \frac{u^4 \exp{\left(-u^2\right)} \mathrm{d}u}{u-z}  & = & z\left\{\frac{1}{2} + z^2\left[1 + z Z\!\left(z\right) , 
\right] \right\} \label{plasmadispfunc_iden_B}
\end{subeqnarray}
and
\begin{subeqnarray}
\int_0^{\infty} \mathrm{d} t \, t^3 \, J_n(\alpha t)^2 \exp \left(-t^2\right)& = & \frac{1}{2}  \exp \left(-\frac{\alpha^2}{2}\right)
 \bigg\{I_n\!\left(\frac{\alpha^2}{2}\right) \nonumber \\
 && \qquad + \frac{\alpha^2}{2} \left[I_n'\!\left(\frac{\alpha^2}{2}\right) - I_n\!\left(\frac{\alpha^2}{2}\right)\right]\bigg\} \, , \\
\int_0^{\infty} \mathrm{d} t^4 \, t^2 J_n(\alpha t) J_n'(\alpha t) \exp \left(-t^2\right)  & = & \frac{\alpha}{4} 
\exp \left(-\frac{\alpha^2}{2}\right) \bigg[ \left(\alpha^2-2 + \frac{2 n^2}{\alpha^2} \right) I_n\!\left(\frac{\alpha^2}{2}\right) \nonumber \\
&& \qquad + \left(1-\alpha^2 \right) I_n'\!\left(\frac{\alpha^2}{2}\right) \bigg] \, , \qquad \\
\int_0^{\infty} \mathrm{d} t^5 \, t^3 J_n'(\alpha t)^2 \exp \left(-t^2\right)  & = & \frac{1}{2} 
\exp \left(-\frac{\alpha^2}{2}\right) \nonumber \\
&& \times \bigg\{ \left[\frac{3 \alpha^2 }{2} - \frac{\alpha^4}{2} + n^2 \left(\frac{1}{\alpha^2} -\frac{3}{2}\right) \right] I_n\!\left(\frac{\alpha^2}{2}\right) \nonumber \\
&& \qquad + \left( \frac{\alpha^4}{2} + \frac{n^2}{2} - \alpha^2 \right) I_n'\!\left(\frac{\alpha^2}{2}\right) \bigg\} \, ,  
\label{Besselfunc_iden_B}
\end{subeqnarray}
to obtain again the expressions for the dielectric components (\ref{dielectricelements_heatfluxA}) 
in terms of special mathematical functions (a tedious, but elementary calculation): 
\begin{subeqnarray}
 (\mathsfbi{P}_s)_{xx} & = & \eta_s \sum_{n=-\infty}^{\infty} \frac{n^2}{k_{\bot}^2 \tilde{\rho}_s^2} \exp \left(-\frac{k_{\bot}^2 \tilde{\rho}_s^2}{2}\right) \Bigg\{ \frac{k_{\bot}^2 \tilde{\rho}_s^2}{2} Z\!\left(\zeta_{sn}\right) I_n'\!\left(\frac{k_{\bot}^2 \tilde{\rho}_s^2}{2}\right) \nonumber \\
&& \qquad \qquad + \left[\zeta_{sn} + Z\!\left(\zeta_{sn}\right) \left(\zeta_{sn}^2 - \frac{3}{2} -\frac{k_{\bot}^2 \tilde{\rho}_s^2}{2}\right) \right] I_n\!\left(\frac{k_{\bot}^2 \tilde{\rho}_s^2}{2}\right) \Bigg\} \, ,\\
 (\mathsfbi{P}_s)_{xy} & = & \frac{\mathrm{i} \eta_s}{2} \sum_{n=-\infty}^{\infty} n \exp \left(-\frac{k_{\bot}^2 \tilde{\rho}_s^2}{2}\right) \Bigg\{ \left[ \zeta_{sn} + Z\!\left(\zeta_{sn}\right) \left(\zeta_{sn}^2 - \frac{3}{2} -\frac{k_{\bot}^2 \tilde{\rho}_s^2}{2}\right) \right] I_n'\!\left(\frac{k_{\bot}^2 \tilde{\rho}_s^2}{2}\right) \nonumber \\
 && \qquad + \left[Z\!\left(\zeta_{sn}\right) \left(\frac{1}{2} +\frac{k_{\bot}^2 \tilde{\rho}_s^2}{2} + \frac{2 n^2}{k_{\bot}^2 \tilde{\rho}_s^2}-\zeta_{sn}^2 \right) -\zeta_{sn} \right] I_n\!\left(\frac{k_{\bot}^2 \tilde{\rho}_s^2}{2}\right) \Bigg\} \, ,\\
 (\mathsfbi{P}_s)_{xz} & = & \eta_s \sum_{n=-\infty}^{\infty} \frac{n}{k_{\bot} \tilde{\rho}_s} \exp \left(-\frac{k_{\bot}^2 \tilde{\rho}_s^2}{2}\right) \Bigg\{ \frac{k_{\bot}^2 \tilde{\rho}_s^2}{2} \left[1 + \zeta_{sn} Z\!\left(\zeta_{sn}\right) \right] I_n'\!\left(\frac{k_{\bot}^2 \tilde{\rho}_s^2}{2}\right) \nonumber \\
 && + \left[\zeta_{sn}^2 - 1 -\frac{k_{\bot}^2 \tilde{\rho}_s^2}{2} + \zeta_{sn} Z\!\left(\zeta_{sn}\right) \left(\zeta_{sn}^2 - \frac{3}{2} -\frac{k_{\bot}^2 \tilde{\rho}_s^2}{2}\right) \right] I_n\!\left(\frac{k_{\bot}^2 \tilde{\rho}_s^2}{2}\right)\Bigg\} \, , \\
 (\mathsfbi{P}_s)_{yx} & = & (\mathsfbi{P}_s)_{xy} \, , \\
 (\mathsfbi{P}_s)_{yy} & = & \eta_s \sum_{n=-\infty}^{\infty} \exp \left(-\frac{k_{\bot}^2 \tilde{\rho}_s^2}{2}\right) \Bigg\{ \bigg[ \left(\frac{n^2}{k_{\bot}^2 \tilde{\rho}_s^2} + \frac{k_{\bot}^2 \tilde{\rho}_s^2}{2} \right)\zeta_{sn} \nonumber \\
  && + Z\!\left(\zeta_{sn}\right)\left(\frac{n^2 \zeta_{sn}^2}{k_{\bot}^2 \tilde{\rho}_s^2} + \frac{k_{\bot}^2 \tilde{\rho}_s^2 \zeta_{sn}^2}{2} + \frac{k_{\bot}^2 \tilde{\rho}_s^2}{4} - \frac{k_{\bot}^4 \tilde{\rho}_s^4}{2} - \frac{3 n^2}{2} - \frac{3 n^2}{2 k_{\bot}^2 \tilde{\rho}_s^2} \right) \bigg] I_n\!\left(\frac{k_{\bot}^2 \tilde{\rho}_s^2}{2}\right) \Bigg\} \nonumber \\
&& + \left[Z\!\left(\zeta_{sn}\right) \left(\frac{1}{2} + k_{\bot}^2 \tilde{\rho}_s^2 + \frac{n^2}{k_{\bot}^2 \tilde{\rho}_s^2} - \zeta_{sn}^2 \right) - \zeta_{sn} \right] \frac{k_{\bot}^2 \tilde{\rho}_s^2}{2} I_n'\!\left(\frac{k_{\bot}^2 \tilde{\rho}_s^2}{2}\right) \, , \\
 (\mathsfbi{P}_s)_{yz} & = & -\frac{\mathrm{i} \eta_s}{2} \sum_{n=-\infty}^{\infty} k_{\bot} \tilde{\rho}_s \exp \left(-\frac{k_{\bot}^2 \tilde{\rho}_s^2}{2}\right) \nonumber\\
 && \times \Bigg\{ \left[k_{\bot}^2 \tilde{\rho}_s^2 + \frac{2 n^2}{k_{\bot}^2 \tilde{\rho}_s^2}  - \zeta_{sn}^2 + \zeta_{sn} Z\!\left(\zeta_{sn}\right) \left(k_{\bot}^2 \tilde{\rho}_s^2 + \frac{1}{2} + \frac{2 n^2}{k_{\bot}^2 \tilde{\rho}_s^2}  - \zeta_{sn}^2\right)\right] I_n\!\left(\frac{k_{\bot}^2 \tilde{\rho}_s^2}{2}\right) \nonumber \\
&& + \bigg[\zeta_{sn}^2  -1  - k_{\bot}^2 \tilde{\rho}_s^2 + \zeta_{sn} Z\!\left(\zeta_{sn}\right) \left(\zeta_{sn}^2 - \frac{3}{2} - k_{\bot}^2 \tilde{\rho}_s^2\right) \bigg]  I_n'\!\left(\frac{k_{\bot}^2 \tilde{\rho}_s^2}{2}\right) \Bigg\} \, ,\\
 (\mathsfbi{P}_s)_{zx} & = & (\mathsfbi{P}_s)_{xz} \, ,\\
 (\mathsfbi{P}_s)_{zy} & = & -(\mathsfbi{P}_s)_{yz} \, , \\
 (\mathsfbi{P}_s)_{zz} & = & \eta_s \sum_{n=-\infty}^{\infty} \exp \left(-\frac{k_{\bot}^2 \tilde{\rho}_s^2}{2}\right) \Bigg\{ \frac{k_{\bot}^2 \tilde{\rho}_s^2}{2} \zeta_{sn} \left[1 + \zeta_{sn} Z\!\left(\zeta_{sn}\right) \right] I_n'\!\left(\frac{k_{\bot}^2 \tilde{\rho}_s^2}{2}\right) \nonumber \\
 &+& \left[\zeta_{sn}^3 - \zeta_{sn} -\frac{k_{\bot}^2 \tilde{\rho}_s^2 \zeta_{sn}}{2} + \zeta_{sn}^2 Z\!\left(\zeta_{sn}\right) \left(\zeta_{sn}^2 - \frac{3}{2} -\frac{k_{\bot}^2 \tilde{\rho}_s^2}{2}\right) \right] I_n\!\left(\frac{k_{\bot}^2 \tilde{\rho}_s^2}{2}\right)\Bigg\} . \, 
\quad
 \label{dielectricelements_heatfluxB}
\end{subeqnarray}

\subsubsection{Dielectric tensor in low-frequency limit} \label{heatflux_dieelectrc_lowfreq}

In the low-frequency limit $\tilde{\omega}_{s\|} \ll 1$ under the ordering $k_{\|} \rho_s \sim k_{\bot} \rho_s \sim 1$,
the expressions~(\ref{dielectricelements_heatfluxB}) can be approximated by the 
leading-order term of the expansion of $\mathsfbi{P}_s$, that is 
\begin{equation}
\mathsfbi{P}_s \approx \mathsfbi{P}_s^{(0)} + \textit{O}(\tilde{\omega}_{s\|}^2) \, 
,
\end{equation}
where
\begin{subeqnarray}
 (\mathsfbi{P}_s^{(0)})_{xx} & = & \eta_s \sum_{n=-\infty}^{\infty} \frac{n^2}{k_{\bot}^2 \tilde{\rho}_s^2} \exp \left(-\frac{k_{\bot}^2 \tilde{\rho}_s^2}{2}\right) \Bigg\{ \frac{k_{\bot}^2 \tilde{\rho}_s^2}{2} Z\!\left(-\frac{n}{|k_{\|}| \tilde{\rho}_s}\right) I_n'\!\left(\frac{k_{\bot}^2 \tilde{\rho}_s^2}{2}\right) \nonumber \\
&& + \left[-\frac{n}{|k_{\|}| \tilde{\rho}_s} + Z\!\left(-\frac{n}{|k_{\|}| \tilde{\rho}_s}\right) \left(\frac{n^2}{|k_{\|}|^2 \tilde{\rho}_s^2} - \frac{3}{2} -\frac{k_{\bot}^2 \tilde{\rho}_s^2}{2}\right) \right] I_n\!\left(\frac{k_{\bot}^2 \tilde{\rho}_s^2}{2}\right) \Bigg\} \, ,\\
 (\mathsfbi{P}_s^{(0)})_{xy} & = & \frac{\mathrm{i} \eta_s}{2} \sum_{n=-\infty}^{\infty} n \exp \left(-\frac{k_{\bot}^2 \tilde{\rho}_s^2}{2}\right) \nonumber \\
 && \times \Bigg\{ \left[Z\!\left(-\frac{n}{|k_{\|}| \tilde{\rho}_s}\right) \left(\frac{1}{2} +\frac{k_{\bot}^2 \tilde{\rho}_s^2}{2} + \frac{2 n^2}{k_{\bot}^2 \tilde{\rho}_s^2}-\frac{n^2}{|k_{\|}|^2 \tilde{\rho}_s^2} \right) + \frac{n}{|k_{\|}| \tilde{\rho}_s}\right] I_n\!\left(\frac{k_{\bot}^2 \tilde{\rho}_s^2}{2}\right) \nonumber \\
 && + \left[ -\frac{n}{|k_{\|}| \tilde{\rho}_s} + Z\!\left(-\frac{n}{|k_{\|}| \tilde{\rho}_s}\right) \left(\frac{n^2}{|k_{\|}|^2 \tilde{\rho}_s^2} - \frac{3}{2} -\frac{k_{\bot}^2 \tilde{\rho}_s^2}{2}\right) \right] I_n'\!\left(\frac{k_{\bot}^2 \tilde{\rho}_s^2}{2}\right) \Bigg\} \, , \qquad \\
 (\mathsfbi{P}_s^{(0)})_{xz} & = & \eta_s \sum_{n=-\infty}^{\infty} \frac{n}{k_{\bot} \tilde{\rho}_s} \exp \left(-\frac{k_{\bot}^2 \tilde{\rho}_s^2}{2}\right) \Bigg\{ \frac{k_{\bot}^2 \tilde{\rho}_s^2}{2} \left[1 -\frac{n}{|k_{\|}| \tilde{\rho}_s} Z\!\left(-\frac{n}{|k_{\|}| \tilde{\rho}_s}\right) \right] I_n'\!\left(\frac{k_{\bot}^2 \tilde{\rho}_s^2}{2}\right) \nonumber \\
 && \qquad \qquad +  I_n\!\left(\frac{k_{\bot}^2 \tilde{\rho}_s^2}{2}\right) \bigg[\frac{n^2}{|k_{\|}|^2 \tilde{\rho}_s^2} - 1 -\frac{k_{\bot}^2 \tilde{\rho}_s^2}{2} \nonumber \\
 && \qquad \qquad \quad -\frac{n}{|k_{\|}| \tilde{\rho}_s} Z\!\left(-\frac{n}{|k_{\|}| \tilde{\rho}_s}\right) \left(\frac{n^2}{|k_{\|}|^2 \tilde{\rho}_s^2} - \frac{3}{2} -\frac{k_{\bot}^2 \tilde{\rho}_s^2}{2}\right) \bigg] \Bigg\} \, , \\
 (\mathsfbi{P}_s^{(0)})_{yx} & = & (\mathsfbi{P}_s^{(0)})_{xy} \, , \\
 (\mathsfbi{P}_s^{(0)})_{yy} & = & \eta_s \sum_{n=-\infty}^{\infty} \exp \left(-\frac{k_{\bot}^2 \tilde{\rho}_s^2}{2}\right) \Bigg\{ \Bigg[ -\left(\frac{n^2}{k_{\bot}^2 \tilde{\rho}_s^2} + \frac{k_{\bot}^2 \tilde{\rho}_s^2}{2} \right)\frac{n}{|k_{\|}| \tilde{\rho}_s} \nonumber \\
 && \qquad + Z\!\left(-\frac{n}{|k_{\|}| \tilde{\rho}_s}\right)\left(\frac{n^4}{k_{\bot}^2 k_{\|}^2 \tilde{\rho}_s^4} + \frac{n^2 k_{\bot}^2}{k_{\|}^2} + \frac{k_{\bot}^2 \tilde{\rho}_s^2}{4}
 - \frac{k_{\bot}^4 \tilde{\rho}_s^4}{2} \right. \nonumber \\ 
 && \qquad \left. - \frac{3 n^2}{2} - \frac{3 n^2}{2 k_{\bot}^2 \tilde{\rho}_s^2} \right) \Bigg] I_n\!\left(\frac{k_{\bot}^2 \tilde{\rho}_s^2}{2}\right) + \frac{k_{\bot}^2 \tilde{\rho}_s^2}{2} I_n'\!\left(\frac{k_{\bot}^2 \tilde{\rho}_s^2}{2}\right) \nonumber \\
&& \qquad \times \left[Z\!\left(-\frac{n}{|k_{\|}| \tilde{\rho}_s}\right) \left(\frac{1}{2} + k_{\bot}^2 \tilde{\rho}_s^2 + \frac{n^2}{k_{\bot}^2 \tilde{\rho}_s^2} -\frac{n^2}{k_{\|}^2 \tilde{\rho}_s^2} \right) +\frac{n}{|k_{\|}| \tilde{\rho}_s} \right] \Bigg\} \, , \\
 (\mathsfbi{P}_s^{(0)})_{yz} & = & -\frac{\mathrm{i} \eta_s}{2} \sum_{n=-\infty}^{\infty} k_{\bot} \tilde{\rho}_s \exp \left(-\frac{k_{\bot}^2 \tilde{\rho}_s^2}{2}\right) \nonumber\\
 && \qquad \times \Bigg\{ I_n\!\left(\frac{k_{\bot}^2 \tilde{\rho}_s^2}{2}\right) \Bigg[k_{\bot}^2 \tilde{\rho}_s^2 + \frac{2 n^2}{k_{\bot}^2 \tilde{\rho}_s^2}  -\frac{n^2}{k_{\|}^2 \tilde{\rho}_s^2} \nonumber\\
 && \qquad \qquad -\frac{n}{|k_{\|}| \tilde{\rho}_s} Z\!\left(-\frac{n}{|k_{\|}| \tilde{\rho}_s}\right) \left(k_{\bot}^2 \tilde{\rho}_s^2 + \frac{1}{2} + \frac{2 n^2}{k_{\bot}^2 \tilde{\rho}_s^2}  - \frac{n^2}{k_{\|}^2 \tilde{\rho}_s^2}\right)\Bigg] \nonumber \\
&& \qquad + I_n'\!\left(\frac{k_{\bot}^2 \tilde{\rho}_s^2}{2}\right) \Bigg[\frac{n^2}{k_{\|}^2 \tilde{\rho}_s^2}  -1  - k_{\bot}^2 \tilde{\rho}_s^2 \nonumber \\
&& \qquad \qquad -\frac{n}{|k_{\|}| \tilde{\rho}_s} Z\!\left(-\frac{n}{|k_{\|}| \tilde{\rho}_s}\right) \left(\frac{n^2}{k_{\|}^2 \tilde{\rho}_s^2} - \frac{3}{2} - k_{\bot}^2 \tilde{\rho}_s^2\right) \Bigg] \Bigg\} \, ,\\
 (\mathsfbi{P}_s^{(0)})_{zx} & = & (\mathsfbi{P}_s^{(0)})_{xz} \, ,\\
 (\mathsfbi{P}_s^{(0)})_{zy} & = & -(\mathsfbi{P}_s^{(0)})_{yz} \, , \\
 (\mathsfbi{P}_s^{(0)})_{zz} & = & \eta_s \sum_{n=-\infty}^{\infty} \exp \left(-\frac{k_{\bot}^2 \tilde{\rho}_s^2}{2}\right) \Bigg\{ -\frac{n k_{\bot}^2 \tilde{\rho}_s}{2 |k_{\|}|} \left[1 - \frac{n}{|k_{\|}| \tilde{\rho}_s} Z\!\left(-\frac{n}{|k_{\|}| \tilde{\rho}_s}\right) \right] I_n'\!\left(\frac{k_{\bot}^2 \tilde{\rho}_s^2}{2}\right) \nonumber \\
 && + I_n\!\left(\frac{k_{\bot}^2 \tilde{\rho}_s^2}{2}\right) \Bigg[\frac{n}{|k_{\|}| \tilde{\rho}_s} -\frac{n^3}{|k_{\|}|^3 \tilde{\rho}_s^3} +\frac{n k_{\bot}^2 \tilde{\rho}_s}{2 |k_{\|}|} \nonumber \\
 && + \frac{n^2}{k_{\|}^2 \tilde{\rho}_s^2} Z\!\left( - \frac{n}{|k_{\|}| \tilde{\rho}_s}\right) \left(\frac{n^2}{k_{\|}^2 \tilde{\rho}_s^2} - \frac{3}{2} -\frac{k_{\bot}^2 \tilde{\rho}_s^2}{2}\right) \Bigg] \Bigg\} \, .
 \label{dielectricelements_heatflux0_A}
\end{subeqnarray}
In this limit, we have utilised the approximation $\zeta_{sn} \approx - n/|k_{\|}| 
\tilde{\rho}_s$. Similarly to the Maxwellian case, we can use the Bessel-function-summation identities 
(\ref{Besselfuncsums}) and the symmetry properties of the plasma dispersion 
function with a real argument to show that
\begin{subeqnarray}
 (\mathsfbi{P}_s^{(0)})_{xx} & = &  2 \mathrm{i} \sqrt{\upi} \eta_s \exp \left(-\frac{k_{\bot}^2 \tilde{\rho}_s^2}{2}\right) \sum_{n=1}^{\infty} \frac{n^2}{k_{\bot}^2 \tilde{\rho}_s^2} \exp{\left(-\frac{n^2}{k_{\|}^2 \tilde{\rho}_s^2}\right)} \nonumber \\
 && \qquad \qquad \times \left[\left(\frac{n^2}{k_{\|}^2 \tilde{\rho}_s^2} - \frac{3}{2} -\frac{k_{\bot}^2 \tilde{\rho}_s^2}{2}\right)I_n\!\left(\frac{k_{\bot}^2 \tilde{\rho}_s^2}{2}\right) + \frac{k_{\bot}^2 \tilde{\rho}_s^2}{2} I_n'\!\left(\frac{k_{\bot}^2 \tilde{\rho}_s^2}{2}\right)\right] \nonumber \\
 & = & \mathrm{i} \eta_s I\!\left(k_{\|} \tilde{\rho}_s,k_{\bot} \tilde{\rho}_s\right)  \, , \\
 (\mathsfbi{P}_s^{(0)})_{xy} & = & -\mathrm{i} \eta_s \Bigg\{\frac{1}{2 |k_{\|}| \tilde{\rho}_s} + \frac{1}{2} \sum_{n=-\infty}^{\infty} n \, \Real{\left[Z\left(\frac{n}{|k_{\|}| \tilde{\rho}_s}\right)\right]} \exp \left(-\frac{k_{\bot}^2 \tilde{\rho}_s^2}{2}\right) \nonumber \\
 && \qquad \times \Bigg\{\left(\frac{n^2}{k_{\|}^2 \tilde{\rho}_s^2} - \frac{3}{2} - k_{\bot}^2 \tilde{\rho}_s^2 \right) I_n'\!\left(\frac{k_{\bot}^2 \tilde{\rho}_s^2}{2}\right) \nonumber \\
 && \qquad +\left(\frac{1}{2} + k_{\bot}^2 \tilde{\rho}_s^2 +  \frac{2 n^2}{k_{\bot}^2 \tilde{\rho}_s^2} -  \frac{n^2}{k_{\|}^2 \tilde{\rho}_s^2} \right) I_n\!\left(\frac{k_{\bot}^2 \tilde{\rho}_s^2}{2}\right) \Bigg\} \nonumber \\
 & = & -\mathrm{i} \eta_s J\!\left(k_{\|} \tilde{\rho}_s,k_{\bot} \tilde{\rho}_s\right)  \, , \\
 (\mathsfbi{P}_s^{(0)})_{xz} & = &  -2 \mathrm{i} \sqrt{\upi} \eta_s \exp \left(-\frac{k_{\bot}^2 \tilde{\rho}_s^2}{2}\right) \sum_{n=1}^{\infty} \frac{n^2}{k_{\bot}^2 \tilde{\rho}_s^2} \exp{\left(-\frac{n^2}{|k_{\|}| k_{\bot} \tilde{\rho}_s^2}\right)} \nonumber \\
 && \qquad \qquad \times \left[\left(\frac{n^2}{k_{\|}^2 \tilde{\rho}_s^2} - \frac{3}{2} -\frac{k_{\bot}^2 \tilde{\rho}_s^2}{2}\right)I_n\!\left(\frac{k_{\bot}^2 \tilde{\rho}_s^2}{2}\right) + \frac{k_{\bot}^2 \tilde{\rho}_s^2}{2} I_n'\!\left(\frac{k_{\bot}^2 \tilde{\rho}_s^2}{2}\right)\right] \nonumber \\
 & = & -\frac{\mathrm{i} k_{\bot}}{|k_{\|}|} \eta_s I\!\left(k_{\|} \tilde{\rho}_s,k_{\bot} \tilde{\rho}_s\right)  \, , \\ 
 (\mathsfbi{P}_s^{(0)})_{yy} & = & \frac{\mathrm{i} \sqrt{\upi}}{2} \eta_s \exp \left(-\frac{k_{\bot}^2 \tilde{\rho}_s^2}{2}\right) \sum_{n=-\infty}^{\infty} \exp{\left(-\frac{n^2}{k_{\|}^2 \tilde{\rho}_s^2}\right)} \nonumber \\
&& \qquad \times \Bigg\{\left(n^2 + \frac{1}{2} k_{\bot}^2 \tilde{\rho}_s^2 + k_{\bot}^4 \tilde{\rho}_s^4 - \frac{n^2 k_{\bot}^2}{k_{\|}^2}\right) I_n'\!\left(\frac{k_{\bot}^2 \tilde{\rho}_s^2}{2}\right) \nonumber \\
&& \qquad +\left(\frac{2 n^4}{k_{\|}^2  k_{\bot}^2 \tilde{\rho}_s^4} - \frac{3 n^2}{k_{\bot}^2 \tilde{\rho}_s^2} - 3 n^2 + \frac{1}{2} k_{\bot}^2 \tilde{\rho}_s^2 - k_{\bot}^4 \tilde{\rho}_s^4 + \frac{n^2 k_{\bot}^2}{k_{\|}^2}\right) I_n\!\left(\frac{k_{\bot}^2 \tilde{\rho}_s^2}{2}\right) \Bigg\} \nonumber \\
 & = & \mathrm{i} \eta_s K\!\left(k_{\|} \tilde{\rho}_s,k_{\bot} \tilde{\rho}_s\right)  \, , \\
 (\mathsfbi{P}_s^{(0)})_{yz} & = & -\mathrm{i} \eta_s \Bigg\{\frac{k_{\bot}}{2 k_{\|}^2 \tilde{\rho}_s} + \frac{1}{2} \sum_{n=-\infty}^{\infty} \frac{n k_{\bot}}{|k_{\|}|} \, \Real{\left[Z\left(\frac{n}{|k_{\|}| \tilde{\rho}_s}\right)\right]} \exp \left(-\frac{k_{\bot}^2 \tilde{\rho}_s^2}{2}\right) \nonumber \\
 && \qquad \times \Bigg\{\left(\frac{n^2}{k_{\|}^2 \tilde{\rho}_s^2} - \frac{3}{2} - k_{\bot}^2 \tilde{\rho}_s^2 \right) I_n'\!\left(\frac{k_{\bot}^2 \tilde{\rho}_s^2}{2}\right) \nonumber \\
 && \qquad +\left(\frac{1}{2} + k_{\bot}^2 \tilde{\rho}_s^2 +  \frac{2 n^2}{k_{\bot}^2 \tilde{\rho}_s^2} -  \frac{n^2}{k_{\|}^2 \tilde{\rho}_s^2} \right) I_n\!\left(\frac{k_{\bot}^2 \tilde{\rho}_s^2}{2}\right) \Bigg\} \nonumber \\
 & = & -\frac{\mathrm{i} k_{\bot}}{|k_{\|}|} \eta_s J\!\left(k_{\|} \tilde{\rho}_s,k_{\bot} \tilde{\rho}_s\right)  \, , \\
 (\mathsfbi{P}_s^{(0)})_{zz} & = &  2 \mathrm{i} \sqrt{\upi} \eta_s \exp \left(-\frac{k_{\bot}^2 \tilde{\rho}_s^2}{2}\right) \sum_{n=1}^{\infty} \frac{n^2}{k_{\|}^2 \tilde{\rho}_s^2} \exp{\left(-\frac{n^2}{k_{\|}^2 \tilde{\rho}_s^2}\right)} \nonumber \\
 && \qquad \qquad \times \left[\left(\frac{n^2}{k_{\|}^2 \tilde{\rho}_s^2} - \frac{3}{2} -\frac{k_{\bot}^2 \tilde{\rho}_s^2}{2}\right)I_n\!\left(\frac{k_{\bot}^2 \tilde{\rho}_s^2}{2}\right) + \frac{k_{\bot}^2 \tilde{\rho}_s^2}{2} I_n'\!\left(\frac{k_{\bot}^2 \tilde{\rho}_s^2}{2}\right)\right] \nonumber \\
 & = & \frac{\mathrm{i} k_{\bot}^2}{k_{\|}^2} \eta_s I\!\left(k_{\|} \tilde{\rho}_s,k_{\bot} \tilde{\rho}_s\right)  \, 
 ,
 \label{dielectricelements_heatflux0_B}
\end{subeqnarray}
where the functions $I\!\left(x,y\right)$, $J\!\left(x,y\right)$ and $K\!\left(x,y\right)$ are 
defined by 
\begin{subeqnarray}
  I\!\left(x,y\right) & \equiv & \frac{2 \sqrt{\upi}}{y^2} \exp{\left(-\frac{y^2}{2}\right)} \nonumber \\
  & & \times \sum_{m=1}^{\infty} m^2 \exp{\left(-\frac{m^2}{x^2}\right)} \left[\frac{y^2}{2} I_m'\!\left(\frac{y^2}{2}\right)+ \left(\frac{m^2}{x^2}-\frac{3+y^2}{2}\right)I_m\!\left(\frac{y^2}{2}\right)\right] , \qquad \\
  J\!\left(x,y\right) & \equiv & \frac{1}{2x} + \frac{1}{2} \exp{\left(-\frac{y^2}{2}\right)}  \sum_{m = -\infty}^{\infty} \left\{ m \, \Real{\; Z\!\left(\frac{m}{x}\right)} \left[\left(\frac{m^2}{x^2} - \frac{3}{2} - y^2\right) I_m'\!\left(\frac{y^2}{2}\right) \right. \right. \nonumber \\
  && \left. \left. + \left(\frac{1}{2}+y^2 +\frac{2 m^2}{y^2}-\frac{m^2}{x^2}\right)I_m\!\left(\frac{y^2}{2}\right)\right] \right\} , \\
  K\!\left(x,y\right) & \equiv & \frac{\sqrt{\upi}}{2} \exp{\left(-\frac{y^2}{2}\right)}  \sum_{m = -\infty}^{\infty} \left\{\exp{\left(-\frac{m^2}{x^2}\right)} \left[\left(m^2 + \frac{1}{2}y^2 +y^4 - \frac{m^2 y^2}{x^2}\right) I_m'\!\left(\frac{y^2}{2}\right) \right. \right. \nonumber \\ \,  
 && \left. \left. + \left(\frac{2 m^4}{x^2 y^2} - \frac{3 m^2}{y^2}-3m^2+\frac{1}{2}y^2 - y^4 + \frac{m^2 y^2}{x^2}\right)I_m\!\left(\frac{y^2}{2}\right)\right]  \right\} . 
 \,  \qquad \quad \label{heatfluxasymptoticfuncs}
\end{subeqnarray}

\subsubsection{Asymptotic limits of $\mathsfbi{P}_s^{(0)}$} 
\label{asympspecialfuncsappend_heatflux}

In this appendix, we give simplified expressions in the limits of small 
and large $x$ and $y$ for the special functions 
$I\!\left(x,y\right)$, $J\!\left(x,y\right)$ and $K\!\left(x,y\right)$ defined 
by (\ref{heatfluxasymptoticfuncs}). Physically, this correspond,s via (\ref{Pelecheatflux}), to considering the 
dielectric response associated with $\mathsfbi{P}_s^{(0)}$ for modes with parallel and perpendicular wavenumbers that are very 
small or very large with respect to the inverse Larmor radius of species $s$.

Proceeding systematically through various limits, we have the following results:
\begin{itemize} 
  \item  $x \sim 1$, $y \ll 1$:
\begin{subeqnarray}
  I\!\left(x,y\right) & = & \frac{\sqrt{\upi}}{2} \left(\frac{1}{x^2}-\frac{1}{2}\right)\exp{\left(-\frac{1}{x^2}\right)} \left[1+\textit{O}\!\left(y^2\right) \right]  \, , \\
  J\!\left(x,y\right) & = & \left[\left(\frac{1}{4}-\frac{1}{2 x^2}\right)\Real{\; Z\left(\frac{1}{x}\right)} + \frac{1}{2 x} \right] \left[1+\textit{O}\!\left(y^2\right) \right] \, , \\
  K\!\left(x,y\right) & = & \frac{\sqrt{\upi}}{2} \left(\frac{1}{x^2}-\frac{1}{2}\right)\exp{\left(-\frac{1}{x^2}\right)} \left[1+\textit{O}\!\left(y^2\right) \right]   \, .  
 \label{heatfluxasymptoticfuncs_x1_ysmall}
\end{subeqnarray}
  \item  $x, y \gg 1$:
\begin{subeqnarray}
  I\!\left(x,y\right) & = & -\frac{\sqrt{\upi} x^3}{4 \left(x^2+y^2\right)^{3/2}} \left[1+\textit{O}\!\left(\frac{1}{x^2+y^2}\right) \right] \, , \\
  J\!\left(x,y\right) & = & -\frac{x^3}{\left(x^2+y^2\right)^{2}} \left[1+\textit{O}\!\left(\frac{1}{x^2+y^2}\right) \right] \, , \\
  K\!\left(x,y\right) & = & -\frac{\sqrt{\upi} x}{4 \left(x^2+y^2\right)^{1/2}} \left[1+\textit{O}\!\left(\frac{1}{x^2+y^2}\right) \right] \, .  
 \label{heatfluxasymptoticfuncs_xylarge}
\end{subeqnarray}
  \item  $x \ll 1, y \sim 1$:
\begin{subeqnarray}
  I\!\left(x,y\right) & = & \frac{2 \sqrt{\upi}}{x^2 y^2} \exp{\left(-\frac{y^2}{2}-\frac{1 }{x^2}\right)} I_1\!\left(\frac{y^2}{2}\right) \left[1+\textit{O}\!\left(x^2\right) \right] \, , \\
  J\!\left(x,y\right) & = & -\frac{x}{2} \exp{\left(-\frac{y^2}{2}\right)} \nonumber \\
  && \quad \times \left[y^2\left(I_0\!\left(\frac{y^2}{2}\right)-I_1\!\left(\frac{y^2}{2}\right)\right) - I_1\!\left(\frac{y^2}{2}\right)\right] \left[1+\textit{O}\!\left(x^2\right) \right] \, , \\
  K\!\left(x,y\right) & = & \frac{\sqrt{\upi}}{2} \exp{\left(-\frac{y^2}{2}\right)} \bigg[\left(\frac{1}{2}y^2 -y^4 \right) I_0\!\left(\frac{y^2}{2}\right) \nonumber \\
  && \qquad \qquad +\left(\frac{1}{2}y^2 +y^4 \right) I_1\!\left(\frac{y^2}{2}\right)\bigg] \left[1+\textit{O}\!\left(x^2\right) \right] \, .  
 \label{heatfluxasymptoticfuncs_xsmall_y1}
\end{subeqnarray}
   \item  $x, y \ll 1$:
\begin{subeqnarray}
  I\!\left(x,y\right) & = & \frac{\sqrt{\upi}}{2 x^2} \exp{\left(-\frac{1}{x^2}\right)} \left[1+\textit{O}\!\left(\exp{\left(-\frac{3}{x^2}\right)},y^2 \right) \right]  \, , \\
  J\!\left(x,y\right) & = & -x \left(\frac{3}{8}y^2 - \frac{1}{4} x^2 \right) \left[1+\textit{O}\!\left(x^4, x^2 y^2, y^4\right) \right] \, , \\
  K\!\left(x,y\right) & = & \frac{\sqrt{\upi}}{4} y^2 \left[1+\textit{O}\!\left(x^2, y^2\right) \right]  \, .  
 \label{heatfluxasymptoticfuncs_xysmall}
\end{subeqnarray}
\end{itemize}

\subsection{CE shear terms} \label{nonzeroshear}

For a CE shear term of the form
\begin{equation}
 \tilde{f}_s(\tilde{v}_{s\|},\tilde{v}_{s\bot}) = - \epsilon_s \left(\tilde{v}_{s\|}^2 - \frac{\tilde{v}_{s\bot}^2}{2} \right) \exp \left(-\tilde{v}_{s}^2\right) ,   
\end{equation}
we have
\begin{subeqnarray}
  \Lambda_s(\tilde{v}_{s\|},\tilde{v}_{s\bot}) & = & -3 \epsilon_s \tilde{v}_{s\|} \tilde{v}_{s\bot} \exp \left(-\tilde{v}_{s}^2\right) 
  , \\
  \Xi_s(\tilde{v}_{s\|},\tilde{v}_{s\bot}) & = & -\frac{3 \epsilon_s}{\tilde{\omega}_{s\|}} \tilde{v}_{s\|} \tilde{v}_{s\bot} \exp \left(-\tilde{v}_{s}^2\right)  
  + \textit{O}(\epsilon_s) .
\end{subeqnarray}
This gives
\begin{subeqnarray}
 (\mathsfbi{P}_s)_{xx} & = &  \frac{6}{\sqrt{\upi}} \epsilon_s \sum_{n=-\infty}^{\infty} \left[ \frac{n^2}{k_{\bot}^2 \tilde{\rho}_s^2}\int_{C_L} \frac{\tilde{v}_{s\|} \exp \left(-\tilde{v}_{s\|}^2\right) \mathrm{d} \tilde{v}_{s\|}}{\tilde{v}_{s\|}-\zeta_{sn}} \right. \nonumber \\
 && \left. \qquad \qquad \qquad \times \int_0^{\infty} \mathrm{d} \tilde{v}_{s\bot} \, \tilde{v}_{s\bot} J_n(k_{\bot} \tilde{\rho}_s \tilde{v}_{s\bot})^2 \exp \left(-\tilde{v}_{s\bot}^2\right) \right] , \qquad\\
 (\mathsfbi{P}_s)_{xy} & = & \frac{6 \mathrm{i}}{\sqrt{\upi}} \epsilon_s \sum_{n=-\infty}^{\infty} \left[ \frac{n}{k_{\bot} \tilde{\rho}_s} \int_{C_L} \frac{\tilde{v}_{s\|} \exp \left(-\tilde{v}_{s\|}^2\right) \mathrm{d} \tilde{v}_{s\|}}{\tilde{v}_{s\|}-\zeta_{sn}} \right. \nonumber \\
 && \left. \qquad \times \int_0^{\infty} \mathrm{d} \tilde{v}_{s\bot} \, \tilde{v}_{s\bot}^2 J_n(k_{\bot} \tilde{\rho}_s \tilde{v}_{s\bot}) J_n'(k_{\bot} \tilde{\rho}_s \tilde{v}_{s\bot}) \exp \left(-\tilde{v}_{s\bot}^2\right) \right] \, , \\
 (\mathsfbi{P}_s)_{xz} & = & \frac{6}{\sqrt{\upi}} \epsilon_s \sum_{n=-\infty}^{\infty} \left[ \frac{n}{k_{\bot} \tilde{\rho}_s}\int_{C_L} \frac{\tilde{v}_{s\|}^2 \exp \left(-\tilde{v}_{s\|}^2\right) \mathrm{d} \tilde{v}_{s\|}}{\tilde{v}_{s\|}-\zeta_{sn}} \right. \\
 && \left. \qquad \qquad \qquad \times \int_0^{\infty} \mathrm{d} \tilde{v}_{s\bot} \, \tilde{v}_{s\bot} J_n(k_{\bot} \tilde{\rho}_s \tilde{v}_{s\bot})^2 \exp \left(-\tilde{v}_{s\bot}^2\right) \right] , \qquad\\
 (\mathsfbi{P}_s)_{yx} & = & (\mathsfbi{P}_s)_{xy}\\
 (\mathsfbi{P}_s)_{yy} & = & \frac{6}{\sqrt{\upi}} \epsilon_s \sum_{n=-\infty}^{\infty} \left[ \int_{C_L} \frac{\tilde{v}_{s\|} \exp \left(-\tilde{v}_{s\|}^2\right) \mathrm{d} \tilde{v}_{s\|}}{\tilde{v}_{s\|}-\zeta_{sn}} \right. \nonumber \\
 && \left. \qquad \qquad \qquad \times \int_0^{\infty} \mathrm{d} \tilde{v}_{s\bot} \, \tilde{v}_{s\bot}^3 J_n'(k_{\bot} \tilde{\rho}_s \tilde{v}_{s\bot})^2 \exp \left(-\tilde{v}_{s\bot}^2\right)\right] , \qquad\\
 (\mathsfbi{P}_s)_{yz} & = & -\frac{6 \mathrm{i}}{\sqrt{\upi}} \epsilon_s \sum_{n=-\infty}^{\infty} \left[ \int_{C_L} \frac{\tilde{v}_{s\|}^2 \exp \left(-\tilde{v}_{s\|}^2\right) \mathrm{d} \tilde{v}_{s\|}}{\tilde{v}_{s\|}-\zeta_{sn}} \right. \nonumber \\
 && \left. \qquad \times \int_0^{\infty} \mathrm{d} \tilde{v}_{s\bot} \, \tilde{v}_{s\bot}^2 J_n(k_{\bot} \tilde{\rho}_s \tilde{v}_{s\bot}) J_n'(k_{\bot} \tilde{\rho}_s \tilde{v}_{s\bot}) \exp \left(-\tilde{v}_{s\bot}^2\right)\right] \, ,\\
 (\mathsfbi{P}_s)_{zx} & = & (\mathsfbi{P}_s)_{xz} \, ,\\
 (\mathsfbi{P}_s)_{zy} & = & -(\mathsfbi{P}_s)_{yz} \, , \\
 (\mathsfbi{P}_s)_{zz} & = & \frac{6}{\sqrt{\upi}} \epsilon_s \Bigg\{ \sum_{n=-\infty}^{\infty} \left[ \int_{C_L} \frac{\tilde{v}_{s\|}^3 \exp \left(-\tilde{v}_{s\|}^2\right) \mathrm{d} \tilde{v}_{s\|}}{\tilde{v}_{s\|}-\zeta_{sn}} \right. \nonumber \\
 && \left. \qquad \qquad \qquad \times \int_0^{\infty} \mathrm{d} \tilde{v}_{s\bot} \, \tilde{v}_{s\bot} J_n(k_{\bot} \tilde{\rho}_s \tilde{v}_{s\bot})^2 \exp \left(-\tilde{v}_{s\bot}^2\right) \right] \nonumber \\
 && \qquad \qquad - \int_{-\infty}^{\infty} \mathrm{d} \tilde{v}_{s\|} \, \tilde{v}_{s\|}^2 \int_0^{\infty} \mathrm{d} \tilde{v}_{s\bot} \tilde{v}_{s\bot} \exp \left(-\tilde{v}_{s}^2\right) \Bigg\} \, . 
 \label{dielectricelements_shearA}
\end{subeqnarray}
Again using the Bessel-function identities (\ref{Besselfunc_iden}), and the
identities (\ref{plasmadispfunc_iden}) and (\ref{plasmadispfunc_iden_B}\textit{a}) 
applicable to the plasma dispersion function, the dielectric tensor's elements become
\begin{subeqnarray}
 (\mathsfbi{P}_s)_{xx} & = & 3\epsilon_s \sum_{n=-\infty}^{\infty} \frac{n^2}{k_{\bot}^2 \tilde{\rho}_s^2}  \left[1+\zeta_{sn} Z\!\left(\zeta_{sn}\right)\right] \exp \left(-\frac{k_{\bot}^2 \tilde{\rho}_s^2}{2}\right) I_n\!\left(\frac{k_{\bot}^2 \tilde{\rho}_s^2}{2}\right) , \qquad\\
 (\mathsfbi{P}_s)_{xy} & = & \frac{3 \mathrm{i}\epsilon_s}{2}   \sum_{n=-\infty}^{\infty} n  \left[1+\zeta_{sn} Z\!\left(\zeta_{sn}\right)\right] \nonumber \\
 && \qquad \qquad \times \exp \left(-\frac{k_{\bot}^2 \tilde{\rho}_s^2}{2}\right) \left[I_n'\!\left(\frac{k_{\bot}^2 \tilde{\rho}_s^2}{2}\right)-I_n\!\left(\frac{k_{\bot}^2 \tilde{\rho}_s^2}{2}\right) \right]\, , \\
 (\mathsfbi{P}_s)_{xz} & = & 3 \epsilon_s \sum_{n=-\infty}^{\infty} \frac{n}{k_{\bot} \tilde{\rho}_s} \zeta_{sn} \left[1+\zeta_{sn} Z\!\left(\zeta_{sn}\right)\right] \exp \left(-\frac{k_{\bot}^2 \tilde{\rho}_s^2}{2}\right) I_n\!\left(\frac{k_{\bot}^2 \tilde{\rho}_s^2}{2}\right) , \qquad \\
 (\mathsfbi{P}_s)_{yx} & = & (\mathsfbi{P}_s)_{xy} , \\
 (\mathsfbi{P}_s)_{yy} & = & \frac{3}{2} \epsilon_s  \sum_{n=-\infty}^{\infty} \left[ 1+ \zeta_{sn} Z\!\left(\zeta_{sn}\right) \right] \nonumber \\
 && \times \exp \left(-\frac{k_{\bot}^2 \tilde{\rho}_s^2}{2}\right) \left[\left( \frac{2 n^2}{k_{\bot}^2 \tilde{\rho}_s^2}+k_{\bot}^2 \tilde{\rho}_s^2\right) I_n\!\left(\frac{k_{\bot}^2 \tilde{\rho}_s^2}{2}\right) -  k_{\bot}^2 \tilde{\rho}_s^2 I_n'\!\left(\frac{k_{\bot}^2 \tilde{\rho}_s^2}{2}\right) \right] , \qquad\\
 (\mathsfbi{P}_s)_{yz} & = & -\frac{3 \mathrm{i}\epsilon_s }{2} \sum_{n=-\infty}^{\infty} k_{\bot} \tilde{\rho}_s \zeta_{sn} \left[1+\zeta_{sn} Z\!\left(\zeta_{sn}\right)\right] \nonumber \\
 && \qquad \qquad \times \exp \left(-\frac{k_{\bot}^2 \tilde{\rho}_s^2}{2}\right)  \left[I_n'\!\left(\frac{k_{\bot}^2 \tilde{\rho}_s^2}{2}\right)-I_n\!\left(\frac{k_{\bot}^2 \tilde{\rho}_s^2}{2}\right) \right]\, , 
 \, \\
 (\mathsfbi{P}_s)_{zx} & = & (\mathsfbi{P}_s)_{xz} \, ,\\
 (\mathsfbi{P}_s)_{zy} & = & -(\mathsfbi{P}_s)_{yz} \, , \\
 (\mathsfbi{P}_s)_{zz} & = & 3 \epsilon_s \sum_{n=-\infty}^{\infty} \zeta_{sn}^2 \left[1+\zeta_{sn} Z\!\left(\zeta_{sn}\right)\right] \exp \left(-\frac{k_{\bot}^2 \tilde{\rho}_s^2}{2}\right) I_n\!\left(\frac{k_{\bot}^2 \tilde{\rho}_s^2}{2}\right) 
 \, .
 \label{dielectricelements_shearB}
\end{subeqnarray}

\subsubsection{Dielectric tensor in low-frequency limit} \label{appendix_dielectric_lowfreq_CEshear}

As with the CE temperature-gradient term, under the ordering $k_{\|} \rho_s \sim k_{\bot} \rho_s \sim 1$,
the expressions (\ref{dielectricelements_heatfluxB}) can be approximated by the 
leading-order term of the expansion of~$\mathsfbi{P}_s$ in the low-frequency limit $\tilde{\omega}_{s\|} \ll 
1$. Namely, we have
\begin{equation}
\mathsfbi{P}_s \approx \mathsfbi{P}_s^{(0)} + \textit{O}(\tilde{\omega}_{s\|}^2) \, 
,
\end{equation}
where
\begin{subeqnarray}
 (\mathsfbi{P}_s^{(0)})_{xx} & = &  3 \epsilon_s  \sum_{n=-\infty}^{\infty} \frac{n^2}{k_{\bot}^2 \tilde{\rho}_s^2}  \left[1-\frac{n}{|k_{\|}| \tilde{\rho}_s} Z\!\left(-\frac{n}{|k_{\|}| \tilde{\rho}_s}\right)\right] \nonumber \\
 && \qquad \qquad \times \exp \left(-\frac{k_{\bot}^2 \tilde{\rho}_s^2}{2}\right) I_n\!\left(\frac{k_{\bot}^2 \tilde{\rho}_s^2}{2}\right) \, , \\
 (\mathsfbi{P}_s^{(0)})_{xy} & = & \frac{3 \mathrm{i} \epsilon_s }{2} \sum_{n=-\infty}^{\infty} n  \left[1-\frac{n}{|k_{\|}| \tilde{\rho}_s} Z\!\left(-\frac{n}{|k_{\|}| \tilde{\rho}_s}\right)\right] \nonumber \\
 && \qquad \qquad \times \exp \left(-\frac{k_{\bot}^2 \tilde{\rho}_s^2}{2}\right) \left[I_n'\!\left(\frac{k_{\bot}^2 \tilde{\rho}_s^2}{2}\right)-I_n\!\left(\frac{k_{\bot}^2 \tilde{\rho}_s^2}{2}\right) \right]\, , \\
 (\mathsfbi{P}_s^{(0)})_{xz} & = & -3 \epsilon_s  \sum_{n=-\infty}^{\infty} \frac{n^2}{k_{\bot} |k_{\|}| \tilde{\rho}_s^2} \left[1-\frac{n}{|k_{\|}| \tilde{\rho}_s}Z\!\left(-\frac{n}{|k_{\|}| \tilde{\rho}_s}\right)\right] \nonumber \\
 && \qquad \qquad \times \exp \left(-\frac{k_{\bot}^2 \tilde{\rho}_s^2}{2}\right) I_n\!\left(\frac{k_{\bot}^2 \tilde{\rho}_s^2}{2}\right) , \qquad \\
 (\mathsfbi{P}_s^{(0)})_{yx} & = & (\mathsfbi{P}_s^{(0)})_{xy} , \\
 (\mathsfbi{P}_s^{(0)})_{yy} & = & \frac{3}{2} \epsilon_s  \sum_{n=-\infty}^{\infty} \left[1-\frac{n}{|k_{\|}| \tilde{\rho}_s} Z\!\left(-\frac{n}{|k_{\|}| \tilde{\rho}_s}\right) \right] \nonumber \\
 && \times \exp \left(-\frac{k_{\bot}^2 \tilde{\rho}_s^2}{2}\right) \left[\left( \frac{2 n^2}{k_{\bot}^2 \tilde{\rho}_s^2}+k_{\bot}^2 \tilde{\rho}_s^2\right) I_n\!\left(\frac{k_{\bot}^2 \tilde{\rho}_s^2}{2}\right) -  k_{\bot}^2 \tilde{\rho}_s^2 I_n'\!\left(\frac{k_{\bot}^2 \tilde{\rho}_s^2}{2}\right) \right] , \qquad \quad \\
 (\mathsfbi{P}_s^{(0)})_{yz} & = & \frac{3 \mathrm{i} \epsilon_s}{2} \sum_{n=-\infty}^{\infty} \frac{n k_{\bot}}{|k_{\|}|} \left[1-\frac{n}{|k_{\|}| \tilde{\rho}_s} Z\!\left(-\frac{n}{|k_{\|}| \tilde{\rho}_s}\right)\right] \nonumber \\
 && \qquad \qquad \times \exp \left(-\frac{k_{\bot}^2 \tilde{\rho}_s^2}{2}\right)  \left[I_n'\!\left(\frac{k_{\bot}^2 \tilde{\rho}_s^2}{2}\right)-I_n\!\left(\frac{k_{\bot}^2 \tilde{\rho}_s^2}{2}\right) \right]\, , 
 \, \\
 (\mathsfbi{P}_s^{(0)})_{zx} & = & (\mathsfbi{P}_s^{(0)})_{xz} \, ,\\
 (\mathsfbi{P}_s^{(0)})_{zy} & = & -(\mathsfbi{P}_s^{(0)})_{yz} \, , \\
 (\mathsfbi{P}_s^{(0)})_{zz} & = & 3 \sum_{n=-\infty}^{\infty} \frac{n^2}{k_{\|}^2 \tilde{\rho}_s^2} \left[1-\frac{n}{|k_{\|}| \tilde{\rho}_s} Z\!\left(-\frac{n}{|k_{\|}| \tilde{\rho}_s}\right)\right] \exp \left(-\frac{k_{\bot}^2 \tilde{\rho}_s^2}{2}\right) I_n\!\left(\frac{k_{\bot}^2 \tilde{\rho}_s^2}{2}\right) 
 \, .
 \label{dielectricelements0_shear}
\end{subeqnarray}
In this calculation, we have utilised the approximation $\zeta_{sn} \approx - n/|k_{\|}| 
\tilde{\rho}_s$. Similarly to the Maxwellian case, we can use the Bessel-function-summation identities 
(\ref{Besselfuncsums}) and the symmetry properties of the plasma dispersion 
function with a real argument to show that
\begin{subeqnarray}
 (\mathsfbi{P}_s^{(0)})_{xx} & = &  3 \epsilon_s  \Bigg\{ \frac{1}{2} +  \exp \left(-\frac{k_{\bot}^2 \tilde{\rho}_s^2}{2}\right) \sum_{n=-\infty}^{\infty} \frac{n^3}{|k_{\|}| k_{\bot}^2 \tilde{\rho}_s^3} \Real{\left[ Z\!\left(\frac{n}{|k_{\|}| \tilde{\rho}_s}\right) \right] } I_n\!\left(\frac{k_{\bot}^2 \tilde{\rho}_s^2}{2}\right) \Bigg\} \nonumber \\
 & = & \epsilon_s W(|k_{\|}| \tilde{\rho}_s,k_{\bot} \tilde{\rho}_s) \, , \\
 (\mathsfbi{P}_s^{(0)})_{xy} & = & 3  \sqrt{\upi} \epsilon_s  \exp \left(-\frac{k_{\bot}^2 \tilde{\rho}_s^2}{2}\right) \nonumber \\
 && \qquad \times \sum_{n=1}^{\infty}\frac{n^2}{|k_{\|}| \tilde{\rho}_s}  \exp{\left(-\frac{n^2}{k_{\|}^2 \tilde{\rho}_s^2}\right)} \left[I_n'\!\left(\frac{k_{\bot}^2 \tilde{\rho}_s^2}{2}\right)-I_n\!\left(\frac{k_{\bot}^2 \tilde{\rho}_s^2}{2}\right) \right] \nonumber \\
  & = & -\epsilon_s X(|k_{\|}| \tilde{\rho}_s,k_{\bot} \tilde{\rho}_s) \, , \\
 (\mathsfbi{P}_s^{(0)})_{xz} & = & -3 \epsilon_s  \Bigg\{ \frac{k_{\bot}}{2 |k_{\|}|} +  \exp \left(-\frac{k_{\bot}^2 \tilde{\rho}_s^2}{2}\right) \nonumber \\
 && \qquad \qquad \times \sum_{n=-\infty}^{\infty} \frac{n^3}{k_{\bot} k_{\|}^2 \tilde{\rho}_s^3} \Real{\left[ Z\!\left(\frac{n}{|k_{\|}| \tilde{\rho}_s}\right) \right] } I_n\!\left(\frac{k_{\bot}^2 \tilde{\rho}_s^2}{2}\right) \Bigg\} \nonumber \\
 & = & -\frac{k_{\bot}}{|k_{\|}|} \epsilon_s  W(|k_{\|}| \tilde{\rho}_s,k_{\bot} \tilde{\rho}_s) \, , \\
 (\mathsfbi{P}_s^{(0)})_{yx} & = & (\mathsfbi{P}_s^{(0)})_{xy} , \\
 (\mathsfbi{P}_s^{(0)})_{yy} & = & \frac{3}{2} \epsilon_s \Bigg\{ 1 + \exp \left(-\frac{k_{\bot}^2 \tilde{\rho}_s^2}{2}\right) \sum_{n=-\infty}^{\infty} \frac{2 n^3}{|k_{\|}| k_{\bot}^2 \tilde{\rho}_s^3} \Real{\left[ Z\!\left(\frac{n}{|k_{\|}| \tilde{\rho}_s}\right) \right] } I_n\!\left(\frac{k_{\bot}^2 \tilde{\rho}_s^2}{2}\right) \nonumber \\
 && \qquad \, + k_{\bot}^2 \tilde{\rho}_s^2  \exp \left(-\frac{k_{\bot}^2 \tilde{\rho}_s^2}{2}\right) \sum_{n=-\infty}^{\infty} \frac{n}{|k_{\|}| \tilde{\rho}_s} \nonumber \\
 && \qquad \qquad \quad \times \Real{\left[ Z\!\left(\frac{n}{|k_{\|}| \tilde{\rho}_s}\right) \right] } \left[ I_n\!\left(\frac{k_{\bot}^2 \tilde{\rho}_s^2}{2}\right) - I_n'\!\left(\frac{k_{\bot}^2 \tilde{\rho}_s^2}{2}\right) \right] \Bigg\} , \nonumber \\
 & = & \epsilon_s Y(|k_{\|}| \tilde{\rho}_s,k_{\bot} \tilde{\rho}_s) \, , \\
 (\mathsfbi{P}_s^{(0)})_{yz} & = & 3  \sqrt{\upi} \epsilon_s \exp \left(-\frac{k_{\bot}^2 \tilde{\rho}_s^2}{2}\right) \nonumber \\
 && \qquad \times \sum_{n=1}^{\infty}\frac{k_{\bot} n^2}{k_{\|}^2 \tilde{\rho}_s}  \exp{\left(-\frac{n^2}{k_{\|}^2 \tilde{\rho}_s^2}\right)} \left[I_n'\!\left(\frac{k_{\bot}^2 \tilde{\rho}_s^2}{2}\right)-I_n\!\left(\frac{k_{\bot}^2 \tilde{\rho}_s^2}{2}\right) \right] \nonumber \\
  & = & -\frac{k_{\bot}}{|k_{\|}|} \epsilon_s X(|k_{\|}| \tilde{\rho}_s,k_{\bot} \tilde{\rho}_s) \, , \\
 (\mathsfbi{P}_s^{(0)})_{zx} & = & (\mathsfbi{P}_s^{(0)})_{xz} \, ,\\
 (\mathsfbi{P}_s^{(0)})_{zy} & = & -(\mathsfbi{P}_s^{(0)})_{yz} \, , \\
 (\mathsfbi{P}_s^{(0)})_{zz} & = & 3 \epsilon_s \Bigg\{ \frac{k_{\bot}^2}{2 k_{\|}^2} +  \exp \left(-\frac{k_{\bot}^2 \tilde{\rho}_s^2}{2}\right) \sum_{n=-\infty}^{\infty} \frac{n^3}{|k_{\|}|^3 \tilde{\rho}_s^3} \Real{\left[ Z\!\left(\frac{n}{|k_{\|}| \tilde{\rho}_s}\right) \right] } I_n\!\left(\frac{k_{\bot}^2 \tilde{\rho}_s^2}{2}\right) \Bigg\} \nonumber \\
 & = & \frac{k_{\bot}^2}{k_{\|}^2} \epsilon_s W(|k_{\|}| \tilde{\rho}_s,k_{\bot} \tilde{\rho}_s) \, 
 ,
 \label{dielectricelements0_shearB}
\end{subeqnarray}
where the functions $W\!\left(x,y\right)$, $Y\!\left(x,y\right)$ and $X\!\left(x,y\right)$ are 
defined by
\begin{subeqnarray}
  W\!\left(x,y\right) & \equiv & \frac{3}{2} + \frac{3}{xy^2} \exp{\left(-\frac{y^2}{2}\right)} \sum_{m = -\infty}^{\infty} m^3 \, \Real{\; Z\!\left(\frac{m}{x}\right)} I_m\!\left(\frac{y^2}{2}\right) \, , \\
  X\!\left(x,y\right) & \equiv & \frac{3 \sqrt{\upi}}{x} \exp{\left(-\frac{y^2}{2}\right)} \sum_{m=1}^{\infty} m^2 \left[ I_m\!\left(\frac{y^2}{2}\right)- I_m'\!\left(\frac{y^2}{2}\right)\right] \exp{\left(-\frac{m^2}{x^2}\right)} \, , \qquad \\
  Y\!\left(x,y\right) & \equiv & W\!\left(x,y\right) - \frac{3}{2} \frac{y^2 G\!\left(x,y\right)}{x}  \,  
  .
  \qquad \qquad \label{shearasymptoticfuncs}
\end{subeqnarray}

\subsubsection{Asymptotic limits of $\mathsfbi{P}_s^{(0)}$} 
\label{asympspecialfuncsappendB}

As we have done for the other special functions defined in this paper, in this appendix we 
provide asymptotic expressions in the limits where $x$ and $y$ 
are very small or large for the special functions $W(x,y)$, $X(x,y)$ and $Y(x,y)$ 
defined in (\ref{shearasymptoticfuncs}). These limits again correspond to parallel and perpendicular 
wavenumbers that are very small or very large with respect to the inverse Larmor radius of species $s$. 

Considering various asymptotic limits in a systematic fashion, we find
\begin{itemize} 
  \item  $x \sim 1$, $y \ll 1$:
\begin{subeqnarray}
  W\!\left(x,y\right) & = & \left[\frac{3}{2} +\frac{3}{2 x} \Real{\; Z\left(\frac{1}{x}\right)} \right] \left[1+\textit{O}\!\left(y^2\right) \right]  \, , \\
  X\!\left(x,y\right) & = & -\frac{3 \sqrt{\upi}}{2 x} \exp{\left(-\frac{1}{x^2}\right)} \left[1+\textit{O}\!\left(y^2\right) \right] \, , \\
  Y\!\left(x,y\right) & = &  \left[\frac{3}{2} +\frac{3}{2 x} \Real{\; Z\left(\frac{1}{x}\right)} \right] \left[1+\textit{O}\!\left(y^2\right) \right]  \, .  
 \label{shearasymptoticfuncs_x1_ysmall}
\end{subeqnarray}
  \item  $x, y \gg 1$:
\begin{subeqnarray}
  W\!\left(x,y\right) & = & \frac{3 x^2 \left(x^2-y^2\right)}{2 \left(x^2+y^2\right)^2} \left[1+\textit{O}\!\left(\frac{1}{x^2+y^2}\right) \right] \, , \\
  X\!\left(x,y\right) & = & \frac{3 \sqrt{\upi} x^2 \left(y^2-2 x^2\right)}{4 \left(x^2+y^2\right)^{5/2}} \left[1+\textit{O}\!\left(\frac{1}{x^2+y^2}\right) \right] \, , \\
  Y\!\left(x,y\right) & = & \frac{3 x^2}{2\left(x^2 + y^2\right)} \left[1+\textit{O}\!\left(\frac{1}{x^2+y^2}\right) \right] \, .  
 \label{shearasymptoticfuncs_xylarge}
\end{subeqnarray}
 \item  $x \ll 1$, $y \sim 1$:
\begin{subeqnarray}
  W\!\left(x,y\right) & = & -\frac{3 x^2}{2 y^2} \left[1-\exp \left(-\frac{y^2}{2}\right) I_0\!\left(\frac{y^2}{2}\right)\right] \left[1+\textit{O}\!\left(x^2\right) \right] \, , \qquad \\
  X\!\left(x,y\right) & = & \frac{3 \sqrt{\upi}}{x} \exp{\left(-\frac{y^2}{2}\right)} \left[I_0\!\left(\frac{y^2}{2}\right)-I_1\!\left(\frac{y^2}{2}\right)\right] \nonumber \\
  && \qquad \qquad \times \exp{\left(-\frac{1}{x^2}\right)} \left\{1+\textit{O}\!\left[\exp{\left(-\frac{3}{x^2}\right)} \right] \right\} \, , \\
  Y\!\left(x,y\right) & = & \frac{3}{2} y^2 \exp{\left(-\frac{y^2}{2}\right)} \left[I_0\!\left(\frac{y^2}{2}\right)- I_1\!\left(\frac{y^2}{2}\right)\right] \left[1+\textit{O}\!\left(x^2\right) \right]  \, .  
 \label{shearasymptoticfuncs_xsmall_y1}
 \end{subeqnarray}
   \item  $x, y \ll 1$:
\begin{subeqnarray}
  W\!\left(x,y\right) & = & -\frac{3}{4}x^2 \left[1+\textit{O}\!\left(x^2,y^2 \right) \right]  \, , \\
  X\!\left(x,y\right) & = & \frac{3 \sqrt{\upi}}{x} \exp{\left(-\frac{1}{x^2}\right)} \left\{1+\textit{O}\!\left[\exp{\left(-\frac{3}{x^2}\right)},y^2 \right] \right\} \, , \\
  Y\!\left(x,y\right) & = & \left[\frac{3}{2} y^2 -\frac{3}{4} x^2 -\frac{9}{8}\left(x^4- \frac{2}{3} x^2 y^2 + y^4\right) \right] \nonumber \\
  && \qquad \qquad \qquad \times \left[1+\textit{O}\!\left(x^6, x^4 y^2, x^2 y^4, y^6 \right) \right]  \, .  
 \label{shearasymptoticfuncs_xysmall}
\end{subeqnarray}
   \item  $x \ll 1$, $y \gg 1$:
\begin{subeqnarray}
  W\!\left(x,y\right) & = &  -\frac{3 x^2}{2 y^2} \left[1+\textit{O}\!\left(x^2,\frac{1}{y^2}\right) \right]   \, , \\
  X\!\left(x,y\right) & = &  \frac{3}{x y^3} \exp{\left(-\frac{1}{x^2}\right)} \left\{1+\textit{O}\!\left[\exp{\left(-\frac{3}{x^2}\right)},\frac{1}{y^2} \right] \right\} \, , \\
  Y\!\left(x,y\right) & = & \frac{3}{2 \sqrt{\upi} y} \left[1+\textit{O}\!\left(x^2,\frac{1}{y^2}\right) \right]  \, .  
 \label{shearasymptoticfuncs_xsmall_ylarge}
\end{subeqnarray}

\end{itemize}

\section{Density perturbations for low-frequency modes} \label{CES_density_modes}

In this appendix, we derive an expression for the (Fourier-transformed) perturbation of number density 
$\widehat{\delta n}_s$ of species $s$ associated with a low-frequency mode, in terms of the expanded terms of the dielectric tensor $\boldsymbol{\mathfrak{E}}_s = \tilde{\omega}_{s\|}\boldsymbol{\mathfrak{E}}_s^{(0)} + \tilde{\omega}_{s\|}^2 \boldsymbol{\mathfrak{E}}_s^{(1)} + \ldots$
of species $s$ and the perturbed electric field, $\widehat{\delta 
\boldsymbol{E}}$; we will show that $\widehat{\delta n}_s$ is, in fact, independent 
of $\boldsymbol{\mathfrak{E}}_s^{(0)}$.
We then derive an expression for the perturbed density of all sub-ion-Larmor 
scale ($k \rho_i \gg 1$), low-frequency modes. 

\subsection{Derivation of general expressions}

We begin with the continuity equation (\ref{fluideqns}\textit{a}), which describes the 
time evolution of the density of species $s$ in terms of itself and the bulk 
velocity of the same species. For any small-amplitude perturbation (with perturbed density $\delta n_s$ and bulk velocity $\delta \boldsymbol{V}_s$)
of some (much more slowly evolving) quasi-equilibrium state (with mean density $n_{s0} \gg \delta n_s$ and bulk velocity $\boldsymbol{V}_{s0} \gg \delta \boldsymbol{V}_s$), viz., 
\begin{equation}
  n_s = n_{s0} + \delta n_s , \quad \boldsymbol{V}_s = \boldsymbol{V}_{s0} + \delta \boldsymbol{V}_s 
  \, ,
\end{equation}
the continuity equation governing that perturbation then becomes
\begin{equation}
  \frac{\partial \delta n_s}{\partial t} + n_{s0} \bnabla \bcdot \delta \boldsymbol{V}_s = 0  \, 
  . \label{densityeqnlin}
\end{equation}
Assuming the perturbation has the form 
\begin{subeqnarray}
\delta n_s & = & \widehat{\delta n}_s \exp\left\{\mathrm{i}\left(\boldsymbol{k} \bcdot \boldsymbol{r} - \omega t\right)\right\} , 
\\
\delta \boldsymbol{V}_s& = & \widehat{\delta \boldsymbol{V}}_s \exp\left\{\mathrm{i}\left(\boldsymbol{k} \bcdot \boldsymbol{r} - \omega t\right)\right\} , 
\, 
\end{subeqnarray}
we deduce from (\ref{densityeqnlin}) that
\begin{equation}
\widehat{\delta n}_s = \frac{n_{s0} \boldsymbol{k} \bcdot \delta \boldsymbol{V}_s}{\omega} 
\, . \label{densityperturb_fourier_vel}
\end{equation}
The perturbed velocity $\widehat{\delta \boldsymbol{V}}_s$ can be written in terms of 
the dielectric tensor of species $s$ using Ohm's law (\ref{Ohm_law}) and 
(\ref{dielectric_species_s}):
\begin{equation}
\delta \boldsymbol{V}_s = -\frac{\mathrm{i} \omega}{4 \upi Z_s e n_{s0}} \boldsymbol{\mathfrak{E}}_s \bcdot \widehat{\delta \boldsymbol{E}} 
\, ,
\end{equation}
whence, by way of (\ref{densityperturb_fourier_vel}), 
\begin{equation}
\widehat{\delta n}_s = -\frac{\mathrm{i}}{4 \upi Z_s e} \boldsymbol{k} \bcdot \boldsymbol{\mathfrak{E}}_s \bcdot \widehat{\delta \boldsymbol{E}}
\, . \label{densityperturb_fourier}
\end{equation}
Finally, we note that the symmetries (\ref{dielectric_zeroth_order}) of $\boldsymbol{\mathfrak{E}}_s^{(0)}$ 
imply that it does not contribute to the right-hand side of 
(\ref{densityperturb_fourier}), which implies in turn that
\begin{equation}
\widehat{\delta n}_s \approx -\frac{\mathrm{i} \tilde{\omega}_{s\|}^2}{4 \upi Z_s e} \boldsymbol{k} \bcdot \boldsymbol{\mathfrak{E}}_s^{(1)} \bcdot \widehat{\delta \boldsymbol{E}}
\, . \label{densityperturb_fourier_B}
\end{equation}
Thus, for low-frequency modes, $\widehat{\delta n}_s$ is a function of the electric 
field and $\boldsymbol{\mathfrak{E}}_s^{(1)}$, but not 
of~$\boldsymbol{\mathfrak{E}}_s^{(0)}$.

We note that the condition (\ref{singulareigenvaleqnFull_33}) implies that, for 
low-frequency modes, quasi-neutrality is maintained: 
\begin{equation}
  \sum_s Z_s \widehat{\delta n}_s = -\frac{\mathrm{i}}{4 \upi e} \boldsymbol{k} \bcdot \boldsymbol{\mathfrak{E}}_s \bcdot \widehat{\delta \boldsymbol{E}} 
  = 0 \, .
\end{equation}
Thus, in a two-species plasma, the ion number density
associated with a perturbation can be calculated if the electron number density 
is known, and visa versa. 

\subsection{Special case: sub-ion-Larmor scale modes in a two-species plasma} \label{KAWs_densityperturb}

In the special case of a two-species plasma whose characteristic parallel wavenumber satisfies $k_{\|} \rho_i \gg 1$,
 a particularly simple expression for the perturbed number densities of ions 
(and electrons) can be derived: the Boltzmann response. This arises because the ion dielectric tensor $\boldsymbol{\mathfrak{E}}_i$
is unmagnetised, and so takes the simple form (valid for arbitrary $\tilde{\omega}_i = \omega/k v_{\mathrm{th}i}$) that was derived in appendix \ref{unmagresponse}:
\begin{equation}
\boldsymbol{\mathfrak{E}}_i \approx \boldsymbol{\mathfrak{E}}_i^{\rm (UM)} = \frac{\omega_{\mathrm{p}i}^2}{\omega^2} \tilde{\omega}_i 
\left\{\left(\mathsfbi{I}-\hat{\boldsymbol{k}}\hat{\boldsymbol{k}}\right) Z\!\left(\tilde{\omega}_i\right) + 2 \left[\tilde{\omega}_i + \tilde{\omega}_i^2 Z\!\left(\tilde{\omega}_i\right) \right]\hat{\boldsymbol{k}}\hat{\boldsymbol{k}}\right\} 
\, .
\end{equation}
It follows that
\begin{equation}
 \boldsymbol{k} \bcdot \boldsymbol{\mathfrak{E}}_i \bcdot \widehat{\delta \boldsymbol{E}} \approx \frac{\omega_{\mathrm{p}i}^2}{\omega^2} 
  2 \tilde{\omega}_i^2 \left[1+ \tilde{\omega}_i Z\!\left(\tilde{\omega}_i\right) \right] {\boldsymbol{k}} \bcdot \widehat{\delta \boldsymbol{E}}
\, .
\end{equation}
Now assuming that $\tilde{\omega}_i \ll 1$, it follows that
\begin{equation}
 \boldsymbol{k} \bcdot \boldsymbol{\mathfrak{E}}_i^{(1)} \bcdot \widehat{\delta \boldsymbol{E}} \approx \frac{2 \omega_{\mathrm{p}i}^2}{\omega^2} \frac{k_{\|}^2}{k^2}
   {\boldsymbol{k}} \bcdot \widehat{\delta \boldsymbol{E}}
\, .
\end{equation}
Expression (\ref{densityperturb_fourier_B}) with $s = i$ then gives
\begin{equation}
 \widehat{\delta n}_i \approx -\frac{Z e \mathrm{i} n_{i0}}{T_i} 
 \frac{\hat{\boldsymbol{k}} \bcdot \widehat{\delta \boldsymbol{E}}}{k} \, .
\end{equation}
Finally, introducing the electrostatic potential $\varphi$, whose Fourier 
transform is related to the electrostatic component of the electric field via
\begin{equation}
\hat{\varphi} =  \frac{\mathrm{i} \hat{\boldsymbol{k}} \bcdot \widehat{\delta \boldsymbol{E}}}{k} 
\, ,
\end{equation}
we deduce that 
\begin{equation}
 \widehat{\delta n}_i \approx -\frac{Z e \mathrm{i} n_{i0}}{T_i} 
\hat{\varphi}  \, , 
\end{equation}
and 
\begin{equation}
 \widehat{\delta n}_e \approx -\frac{Z e \mathrm{i} n_{e0}}{T_i} 
\hat{\varphi}  \, ,
\end{equation}
where we have used the quasi-neutrality relation $n_{e0} = Z n_{i0}$ for the 
equilibrium state.

\section{Calculating the electrostatic field from the transverse electric field} 
\label{electrostaticcompcalc}

In appendix \ref{Maxwellresponse_low123}, it was shown that for any function 
with a small anisotropy,
\begin{equation}
\boldsymbol{\mathfrak{E}}_s^{(0)} \bcdot \hat{\boldsymbol{k}} = 0 \, ,
\end{equation}
which implies that the leading-order terms (in $\tilde{\omega}_{s\|} \ll 1$) of the dielectric tensor are 
insufficient to determine the electrostatic field. To 
do this, we must go to the next order in $\tilde{\omega}_{s\|} \ll 1$. 
To illustrate how such a calculation is done, 
in this appendix, we derive an expression for the electrostatic field component $\hat{\boldsymbol{k}} \bcdot \widehat{\delta \boldsymbol{E}} $ in terms of the 
transverse electric field $\widehat{\delta \boldsymbol{E}}_{T}$ and special functions when the underlying particle distribution function is Maxwellian. 

To achieve this aim, we first derive a relation between the components of the electric 
field in the coordinate basis 
$\left\{\hat{\boldsymbol{x}},\hat{\boldsymbol{y}},\hat{\boldsymbol{z}}\right\}$. We begin 
with the consistency condition (\ref{consistencycond}) appropriate for non-relativistic 
electromagnetic fluctuations:
\begin{equation}
  \boldsymbol{k} \bcdot \boldsymbol{\mathfrak{E}} \bcdot \widehat{\delta \boldsymbol{E}}  
  = 0 \, . \label{consistencycond_Append}
\end{equation}
Writing $\hat{\boldsymbol{k}}$, $\boldsymbol{\mathfrak{E}}$ and $\widehat{\delta \boldsymbol{E}}$ 
in the basis 
$\left\{\hat{\boldsymbol{x}},\hat{\boldsymbol{y}},\hat{\boldsymbol{z}}\right\}$, 
this becomes
\begin{equation}
  \left(k_{\bot} \mathfrak{E}_{xx} + k_{\|} \mathfrak{E}_{xz} \right) \widehat{\delta 
  E}_{x} + \left(k_{\bot} \mathfrak{E}_{xy} - k_{\|} \mathfrak{E}_{yz} \right) \widehat{\delta 
  E}_{y} + \left(k_{\bot} \mathfrak{E}_{xz} + k_{\|} \mathfrak{E}_{zz} \right) \widehat{\delta 
  E}_{z} = 0 \, .
\end{equation}
Now considering the case of fluctuations that satisfy $\tilde{\omega}_{s\|} \ll 1$ for all particle species 
$s$, and expanding the components of the dielectric in $\tilde{\omega}_{s\|} \ll 1$,
we find
\begin{equation}
  \left(k_{\bot} \mathfrak{E}_{xx}^{(1)} + k_{\|} \mathfrak{E}_{xz}^{(1)} \right) \widehat{\delta 
  E}_{x} + \left(k_{\bot} \mathfrak{E}_{xy}^{(1)} - k_{\|} \mathfrak{E}_{yz}^{(1)} \right) \widehat{\delta 
  E}_{y} + \left(k_{\bot} \mathfrak{E}_{xz}^{(1)} + k_{\|} \mathfrak{E}_{zz}^{(1)} \right) \widehat{\delta 
  E}_{z} = \textit{O}(\tilde{\omega}_{s\|}^3) \, ,
\end{equation}
where 
\begin{equation}
\boldsymbol{\mathfrak{E}}^{(1)} = \sum_{s} \tilde{\omega}_{s\|}^2 \boldsymbol{\mathfrak{E}}_s^{(1)} 
\, .
\end{equation}
From (\ref{dielectricelements_max1_B}), we have
\begin{subeqnarray}
 k_{\bot} \mathfrak{E}_{xx}^{(1)} + k_{\|} \mathfrak{E}_{xz}^{(1)}  & = & - \sum_s \frac{2 k_{\|} \omega_{\mathrm{p}s}^2 \tilde{\omega}_{s\|}^2}{\omega^2} \sum_{m=-\infty}^{\infty} \frac{m}{k_{\bot} \tilde{\rho}_s} \Real{\; Z\left(\frac{m}{|k_{\|}| \tilde{\rho}_s}\right)} \nonumber \\
 && \qquad \qquad \qquad \times \exp \left(-\frac{k_{\bot}^2 \tilde{\rho}_s^2}{2}\right) I_m\!\left(\frac{k_{\bot}^2 \tilde{\rho}_s^2}{2}\right) \, , \\ 
 k_{\bot} \mathfrak{E}_{xy}^{(1)} - k_{\|} \mathfrak{E}_{yz}^{(1)}  & = & \sum_s \frac{\sqrt{\upi} k_{\|} \omega_{\mathrm{p}s}^2 \tilde{\omega}_{s\|}^2}{\omega^2}  \sum_{m=-\infty}^{\infty} k_{\bot} \tilde{\rho}_s \exp{\left(-\frac{m^2}{k_{\|}^2 \tilde{\rho}_s^2}\right)} \nonumber \\
 && \qquad \qquad \times \exp \left(-\frac{k_{\bot}^2 \tilde{\rho}_s^2}{2}\right) \left[I_m'\!\left(\frac{k_{\bot}^2 \tilde{\rho}_s^2}{2}\right)-I_m\!\left(\frac{k_{\bot}^2 \tilde{\rho}_s^2}{2}\right) \right]\, , \\ 
 k_{\bot} \mathfrak{E}_{xz}^{(1)} + k_{\|} \mathfrak{E}_{zz}^{(1)} & = & \sum_s \frac{2 k_{\|} \omega_{\mathrm{p}s}^2 \tilde{\omega}_{s\|}^2}{\omega^2} \left[1 + \sum_{m=-\infty}^{\infty} \frac{m}{|k_{\|}| \tilde{\rho}_s} \Real{\; Z\left(\frac{m}{|k_{\|}| \tilde{\rho}_s}\right)} \right. \nonumber \\
 && \left. \qquad \qquad \qquad \times \exp \left(-\frac{k_{\bot}^2 \tilde{\rho}_s^2}{2}\right) I_m\!\left(\frac{k_{\bot}^2 \tilde{\rho}_s^2}{2}\right) \right] \, .
\end{subeqnarray}
Thus, we have the following relationship between $\widehat{\delta E}_{x}$, $\widehat{\delta E}_{y}$ 
and $\widehat{\delta E}_{z}$:
\begin{eqnarray}
  \sum_s \frac{k_{\mathrm{D}s}^2}{2 k_{\|}^2} \Bigg\{-L\!\left(|k_{\|}| \tilde{\rho}_s,k_{\bot} \tilde{\rho}_s\right) \widehat{\delta E}_{x} & + & N\!\left(|k_{\|}| \tilde{\rho}_s,k_{\bot} \tilde{\rho}_s\right) \widehat{\delta E}_{y} \nonumber \\
  & + & \left[2+\frac{k_{\bot}}{k_{\|}}  L\!\left(|k_{\|}| \tilde{\rho}_s,k_{\bot} \tilde{\rho}_s\right)\right] \widehat{\delta E}_{z} \Bigg\} = 0 \, , 
   \label{consistentrel_asymp}
\end{eqnarray}
where $k_{\mathrm{D}s}$ is the Debye wavenumber (\ref{Debyewav}), and $L(x,y)$ and $N(x,y)$ were defined previously by (\ref{LandNdef_Append}).
Using the identities
\begin{subeqnarray}
\widehat{\delta E}_{x} & = & \frac{k_{\|}}{k} \widehat{\delta E}_{1} + \frac{k_{\bot}}{k} 
\widehat{\delta E}_{3} \, , \\
\widehat{\delta E}_{y} & = & \widehat{\delta E}_{2} \, , \\
\widehat{\delta E}_{z} & = & -\frac{k_{\bot}}{k} \widehat{\delta E}_{1} + \frac{k_{\|}}{k} 
\widehat{\delta E}_{3} \, ,
\end{subeqnarray}
we can rearrange (\ref{consistentrel_asymp}) to give
\begin{eqnarray}
\frac{1}{k_{\|} k}\left(\sum_s k_{\mathrm{D}s}^2 \right) \widehat{\delta E}_{3} & = & \sum_s \frac{k_{\mathrm{D}s}^2}{2 k_{\|}^2} \Bigg\{ \left[\frac{k}{k_{\|}} L\!\left(|k_{\|}| \tilde{\rho}_s,k_{\bot} \tilde{\rho}_s\right) + 2 \frac{k_{\bot}}{k} \right] \widehat{\delta E}_{1} \nonumber \\ && \qquad - N\!\left(|k_{\|}| \tilde{\rho}_s,k_{\bot} \tilde{\rho}_s\right) \widehat{\delta E}_{2} \Bigg\} \, .
   \label{consistentrel_asympB}
\end{eqnarray}
Thus, the electrostatic field is related to the transverse field by
\begin{eqnarray}
  \hat{\boldsymbol{k}} \bcdot \widehat{\delta \boldsymbol{E}} & = & \left(\sum_s \frac{Z_s T_e}{T_s} \right)^{-1} \sum_s \frac{Z_s T_e}{T_s} \Bigg\{ \left[\frac{k^2}{2 k_{\|}^2} L\!\left(|k_{\|}| \tilde{\rho}_s,k_{\bot} \tilde{\rho}_s\right) + \frac{k_{\bot}}{k_{\|}} \right] \widehat{\delta E}_{1} \nonumber \\ && \qquad \qquad \qquad \qquad \qquad \qquad - \frac{k}{2 k_{\|}} N\!\left(|k_{\|}| \tilde{\rho}_s,k_{\bot} \tilde{\rho}_s\right) \widehat{\delta E}_{2} \Bigg\}
  \, . \label{electrostaticcomp_Append}
\end{eqnarray}

\section{Methodology for characterising CET microinstabilities} \label{CET_method}

In this appendix, we describe our method for calculating the real frequencies and growth rates
of microinstabilities driven by the CE electron- and ion-temperature-gradient, 
and electron-friction terms when the Krook collision 
operator is assumed. The method follows that outlined in section 
\ref{linear_stab_method}: that is, motivated by the considerations of section \ref{sec:CharacMicroQual},
we assume that all significant CET microinstabilities are low frequency ($\omega \ll k_{\|} v_{\mathrm{th}s}$ for at least one particle species), 
and derive algebraic dispersion relations of such microinstabilities [a particular example 
of which is given by (\ref{disprel_simp_2})]. 
The growth rate of CET microinstabilities [and, therefore, the stability of the electron and ion CE distribution functions 
(\ref{CEheatflux}\textit{a}) and (\ref{CEheatflux}\textit{b})] as a function of their parallel and perpendicular 
wavenumbers $k_{\|}$ and $k_{\perp}$ is assessed by solving this dispersion relation 
for the complex frequency $\omega$, and then evaluating its imaginary part. 

As we explained in section \ref{linear_stab_method}, to construct the 
algebraic, low-frequency dispersion relation for particular forms
of CE distribution function for each particle species $s$, we must evaluate its (leading-order) non-Maxwellian 
contribution to the dielectric tensor, $\mathsfbi{P}_{s} \approx \mathsfbi{P}_{s}^{(0)}$ [see (\ref{Maxnonmaxsep_s}) and 
(\ref{Maxnonmaxsep_s_Append}) for the precise relation of this quantity to the dielectric tensor $\boldsymbol{\mathfrak{E}}_s$]. 
This is done for the CE electron-friction term
in appendix \ref{heatfluxrterm_di_res}, and for the CE temperature-gradient terms in 
appendix \ref{heatfluxrterm_di_heatflux}. We then deduce the algebraic dispersion 
relations of CE electron-temperature-gradient-driven microinstabilities in appendix 
\ref{electron_heatflux_instab}, and of CE ion-temperature-gradient-driven 
microinstabilities in appendix \ref{ion_heatflux_instab}. Within these two
appendices, respectively, we also present derivations of the (further) simplified 
dispersion relations for the parallel CET whistler instability (appendix 
\ref{derivation_parwhistlerheatflux}), the parallel CET slow-hydromagnetic-wave 
instability (appendix \ref{derivation_slowwave}), and the CET long-wavelength KAW instability 
(appendix \ref{derivation_CETKAW}), from which the frequencies and growth rates 
of these instabilities that are stated in section \ref{CETclass} are calculated. 
 
\subsection{Dielectric response of CE electron-friction term} \label{heatfluxrterm_di_res}

We first consider the CE electron-friction 
term when evaluating $\mathsfbi{P}_{e}^{(0)}$, defined in (\ref{Maxnonmaxsep_s}).  
We showed in  
appendix \ref{resistiveterm} that, when a Krook collision operator was assumed, if $\eta_e^T = \eta_i = 
0$, then [see (\ref{Presismat_Append})]
\begin{subeqnarray}
  (\mathsfbi{P}_{e}^{(0)})_{11} & = & \frac{\eta_e^R}{2} (\mathsfbi{M}_{e}^{(0)})_{11} \, , \\
  (\mathsfbi{P}_{e}^{(0)})_{12} & = & \frac{\eta_e^R}{2} (\mathsfbi{M}_{e}^{(0)})_{12} \, , \\
  (\mathsfbi{P}_{e}^{(0)})_{21} & = & \frac{\eta_e^R}{2} (\mathsfbi{M}_{e}^{(0)})_{21} \, , \\  
  (\mathsfbi{P}_{e}^{(0)})_{22} & = & \frac{\eta_e^R}{2} (\mathsfbi{M}_{e}^{(0)})_{22} \, 
  . \label{Presismat}
\end{subeqnarray}
It follows that the dispersion relation of all plasma modes is identical to that in a
Maxwellian plasma, only with shifted complex 
frequencies $\tilde{\omega}_{e\|}^{*} \equiv \tilde{\omega}_{e\|} + \eta_e^R/2 
$. Since $\Imag{(\tilde{\omega}_{e\|})} < 0$ for all modes in a Maxwellian plasma, we conclude 
that $\Imag{(\tilde{\omega}_{e\|}^{*})} < 0$ also, and hence the 
CE electron-friction term cannot drive any microinstabilities when a Krook collision operator is employed: instead, it merely modifies the
real frequency of the waves. Thus, when characteristing CET microinstabilities, 
we henceforth ignore the CE electron-friction term, as well as the electron-ion-drift 
term (viz., $\eta_e^R = \eta_e^u = 0$). 

\subsection{Dielectric response of CE temperature-gradient terms} \label{heatfluxrterm_di_heatflux}

Now consider the CE temperature-gradient terms. It is shown in appendix \ref{heatfluxterm} 
that $\mathsfbi{P}_{s}^{(0)}$ is given by
\begin{subeqnarray}
  (\mathsfbi{P}_{e}^{(0)})_{11} & = & \mathrm{i} \eta_e^T \frac{k^2}{k_{\|}^2} I\!\left(k_{\|} \tilde{\rho}_e,k_{\bot} \tilde{\rho}_e\right) \, , \\
  (\mathsfbi{P}_{e}^{(0)})_{12} & = & - \mathrm{i} \eta_e^T \frac{k}{k_{\|}} J\!\left(k_{\|} \tilde{\rho}_e,k_{\bot} \tilde{\rho}_e\right) \, , \\
  (\mathsfbi{P}_{e}^{(0)})_{21} & = & \mathrm{i} \eta_e^T \frac{k}{k_{\|}}  J\!\left(k_{\|} \tilde{\rho}_e,k_{\bot} \tilde{\rho}_e\right) \, , \\  
  (\mathsfbi{P}_{e}^{(0)})_{22} & = & \mathrm{i} \eta_e^T K\!\left(k_{\|} \tilde{\rho}_e,k_{\bot} \tilde{\rho}_e\right) \, 
  , \label{Pelecheatflux}
\end{subeqnarray}
where the special functions $I(x,y)$, $J(x,y)$ and $K(x,y)$ are defined by (\ref{heatfluxasymptoticfuncs}). Note that $\tilde{\rho}_e < 0$, by definition. The contribution $\mathsfbi{P}_{i}^{(0)}$ associated with the CE 
ion-temperature-gradient terms is given by
\begin{subeqnarray}
  (\mathsfbi{P}_{i}^{(0)})_{11} & = & \mathrm{i} \eta_i \frac{k^2}{k_{\|}^2} I\!\left(k_{\|} \rho_i,k_{\bot} \rho_i\right) \, , \\
  (\mathsfbi{P}_{i}^{(0)})_{12} & = & - \mathrm{i} \eta_i \frac{k}{k_{\|}} J\!\left(k_{\|} \rho_i,k_{\bot} \rho_i\right) \, , \\
  (\mathsfbi{P}_{i}^{(0)})_{21} & = & \mathrm{i} \eta_i \frac{k}{k_{\|}}  J\!\left(k_{\|} \rho_i,k_{\bot} \rho_i\right) \, , \\  
  (\mathsfbi{P}_{i}^{(0)})_{22} & = & \mathrm{i} \eta_i K\!\left(k_{\|} \rho_i,k_{\bot} \rho_i\right) \, 
  . \label{Pionheatflux}
\end{subeqnarray}

\subsection{Approximate dispersion relation of CE electron-temperature-gradient-driven microinstabilities} \label{electron_heatflux_instab}

We first consider microinstabilities for which $\tilde{\omega}_{e\|} = \omega/k_{\|} v_{\mathrm{th}e} \sim \eta_e^T$. It follows that $\tilde{\omega}_{i\|} = \omega/k_{\|} v_{\mathrm{th}i} \sim \eta_e^T \mu_e^{-1/2} \gg  
\eta_i$. Therefore, the CE ion-temperature-gradient term is irrelevant for such instabilities, and we need consider only the electron-temperature-gradient term.   
We also assume that the Maxwellian contribution to the dielectric tensor, $\mathsfbi{M}_{i}$,  can be ignored for such microinstabilities 
-- the validity of this assumption is discussed at the end of this section. 

The dispersion relation for microinstabilities under the ordering $\tilde{\omega}_{e\|} \sim \eta_e^T \sim 1/\beta_e$ 
is then given by (\ref{disprel_simp_2}), with $\mathsfbi{M}_{e}^{(0)}$ and $\mathsfbi{P}_{e}^{(0)}$ 
substituted for by (\ref{Mmaxcomp}) and (\ref{Pelecheatflux}), 
respectively: 
 \begin{eqnarray}
   \left[\tilde{\omega}_{e\|} F\!\left(k_{\|} \tilde{\rho}_e,k_{\bot} \tilde{\rho}_e\right) \right. & + & \left. \eta_e^T I\!\left(k_{\|} \tilde{\rho}_e,k_{\bot} \tilde{\rho}_e\right)+ \mathrm{i} k_{\|}^2 d_e^2\right] \nonumber \\
   &\times& \left[\tilde{\omega}_{e\|} H\!\left(k_{\|} \tilde{\rho}_e,k_{\bot} \tilde{\rho}_e\right) + \eta_e^T K\!\left(k_{\|} \tilde{\rho}_e,k_{\bot} \tilde{\rho}_e\right) 
+\mathrm{i} k^2 d_e^2\right] 
   \nonumber \\
  & + &\left[\tilde{\omega}_{e\|} G\!\left(k_{\|} \tilde{\rho}_e,k_{\bot} \tilde{\rho}_e\right) + \eta_e^T J\!\left(k_{\|} \tilde{\rho}_e,k_{\bot} \tilde{\rho}_e\right)\right]^2 = 0 \, 
  .
  \qquad \label{dispheatfluxA}
 \end{eqnarray}
 We remind the reader that we have ordered $k^2 d_e^2 \sim \eta_e^T$ and $k \rho_e \sim 1$. Noting that $\beta_e = \rho_e^2/d_e^2$, we can rewrite the skin-depth terms as follows:
 \begin{equation}
   k_{\|}^2 d_e^2 = \frac{k_{\|}^2 \rho_e^2}{\beta_e} \, , \quad k^2 d_e^2 = \frac{k^2 \rho_e^2}{\beta_e} \, .
 \end{equation}
 This allows for the dispersion relation (\ref{dispheatfluxA}) to be arranged as a 
 quadratic in the complex variable $\tilde{\omega}_{e\|} \beta_e$:
 \begin{equation}
   A_{\mathrm{T}}\!\left(k_{\|} \rho_e,k_{\bot} \rho_e\right) \tilde{\omega}_{e\|}^2 \beta_e^2 +B_{\mathrm{T}}\!\left(k_{\|} \rho_e,k_{\bot} \rho_e\right) \tilde{\omega}_{e\|} \beta_e + C_{\mathrm{T}}\!\left(k_{\|} \rho_e,k_{\bot} \rho_e\right) = 0 
   \, , \label{heatflux_elec_frequency_disprel_Append}
 \end{equation}
 where 
 \begin{eqnarray}
   A_{\mathrm{T}}\!\left(k_{\|} \rho_e,k_{\bot} \rho_e\right) & = & F_e H_e + G_e^2 \, , \\
   B_{\mathrm{T}}\!\left(k_{\|} \rho_e,k_{\bot} \rho_e\right) & = & \eta_e^T \beta_e \left(F_e K_e +H_e I_e + 2G_e J_e\right) + \mathrm{i} \left(F_e k^2 \rho_e^2 + H_e k_{\|}^2 \rho_e^2\right) \, , \\
   C_{\mathrm{T}}\!\left(k_{\|} \rho_e,k_{\bot} \rho_e\right) & = & \left(\eta_e^T \beta_e\right)^2 \left(I_e K_e + J_e^2\right) - k^2 k_{\|}^2 \rho_e^4 +\mathrm{i} \eta_e^T \beta_e  \left(I_e k^2 \rho_e^2 + K_e k_{\|}^2 \rho_e^2\right) \, ,   \label{heatflux_elec_frequency_coeffs_Append}
 \end{eqnarray}
 and $F_e \equiv F\!\left(k_{\|} \tilde{\rho}_e,k_{\bot} \tilde{\rho}_e\right)$, $G_e \equiv G\!\left(k_{\|} \tilde{\rho}_e,k_{\bot} 
 \tilde{\rho}_e\right)$, etc. Solving (\ref{heatflux_elec_frequency_disprel_Append}) gives two roots; restoring dimensions to the complex frequency, they 
 are
\begin{equation}
  \omega = \frac{\Omega_e}{\beta_e} k_{\|} \rho_e \frac{-B_{\mathrm{T}} \pm \sqrt{B_{\mathrm{T}}^2 + 4A_{\mathrm{T}}C_{\mathrm{T}}}}{2 A_{\mathrm{T}}} 
  \, , \label{heatflux_elec_frequency_Append}
\end{equation}
recovering (\ref{heatflux_elec_frequency}). 
For a given wavenumber, we use (\ref{heatflux_elec_frequency_Append}) to calculate the growth rates 
of the perturbations -- and, in particular, to see if positive growth rates 
are present. If they are, it is anticipated that they will have typical size 
$\gamma \sim \Omega_e/\beta_e \sim \eta_e^T \Omega_e $ (or $\tilde{\omega}_{e\|} \sim 1/\beta_e \sim \eta_e^T$).

When deriving (\ref{heatflux_elec_frequency_Append}), we assumed that neglecting the Maxwellian ion response was legitimate. It is clear that 
if $\tilde{\omega}_{i\|} \gg 1$, then thermal ions are effectively static to 
electromagnetic perturbations, and so their contribution $\mathsfbi{M}_{i}$ to the dielectric tensor can be ignored. 
In terms of a condition on 
$\eta_e^{T}$, the scaling $\eta_e^{T} \sim \tilde{\omega}_{e\|}$ gives $\eta_e^{T} \gg \mu_e^{1/2}$, so this regime is valid for sufficiently 
large $\eta_e^{T}$. For $\tilde{\omega}_{i\|} \lesssim 1$, it is not immediately 
clear in the same way that the ion contribution to the dielectric tensor is 
small. However, having deduced the typical magnitude of the complex frequency of perturbations 
whilst ignoring ion contributions, we are now able to confirm that our 
neglect of $\mathsfbi{M}_{i}$ was justified. 

Since $k \rho_e \sim 1$ under the ordering assumed when deriving (\ref{dispheatfluxA}), we conclude that the Maxwellian ion response is 
unmagnetised: $k \rho_i \gg 1$. As a consequence, it can be 
shown (see appendix \ref{unmagresponse}) that the transverse components of $\mathsfbi{M}_{i}$ 
are given by
\begin{equation}
   \left(\mathsfbi{M}_{i}\right)_{11} = \left(\mathsfbi{M}_{i}\right)_{22} = \tilde{\omega}_i Z\!\left(\tilde{\omega}_i\right) \, , 
   \quad \left(\mathsfbi{M}_{i}\right)_{12} = \left(\mathsfbi{M}_{i}\right)_{21} = 0 \, , 
   \label{ion_unmagresponse}
\end{equation}
where $ \tilde{\omega}_i  \equiv \omega/k v_{\mathrm{th}i} = k_{\|} 
\tilde{\omega}_{i\|}/k$.
Then, estimating the size of the neglected Maxwellian ion contribution to the dielectric tensor (assuming $k_{\|} \sim k$) as compared with the equivalent electron contribution, we find
\begin{equation}
  \frac{(\boldsymbol{\mathfrak{E}}_i)_{11}}{(\boldsymbol{\mathfrak{E}}_e^{(0)})_{11}} \sim  \frac{(\boldsymbol{\mathfrak{E}}_i)_{22}}{(\boldsymbol{\mathfrak{E}}_e^{(0)})_{22}} \sim \frac{\mu_e \tilde{\omega}_i}{\tilde{\omega}_{e\|}} 
  |Z\!\left(\tilde{\omega}_i\right)| \sim \mu_e^{1/2} |Z\!\left(\tilde{\omega}_i\right)| 
  ,
\end{equation}
where we have used $\boldsymbol{\mathfrak{E}}_i = \mu_e \mathsfbi{M}_{i}$ and $\boldsymbol{\mathfrak{E}}_e^{(0)} = \tilde{\omega}_{e\|} \mathsfbi{M}_{e}^{(0)} + 
\mathsfbi{P}_{e}^{(0)}$ (see section \ref{disprel_simps_II}). Since $|Z\!\left(z\right)| \lesssim 1$ for all $z$ with positive imaginary part~\citep{F61}, we conclude that the ion contribution to the dielectric tensor 
is indeed small for unstable perturbations, irrespective of the value of $\tilde{\omega}_{i\|}$, and so its 
neglect was valid. 

\subsubsection{Derivation of frequency and growth rate of the parallel CET whistler instability} \label{derivation_parwhistlerheatflux}

The dispersion relation of 
unstable whistler waves with their wavevector parallel to $\boldsymbol{B}_0$ is obtained 
by taking the subsidiary limit $k_\bot \rho_e \rightarrow 0$ in (\ref{dispheatfluxA}), and substituting $\tilde{\rho}_e = - \rho_e$:
\begin{eqnarray}
\left[\tilde{\omega}_{e\|} \beta_e \sqrt{\upi} \exp{\left(-\frac{1}{k_\|^2 \rho_e^2}\right)} + \eta_e^T \beta_e \frac{\sqrt{\upi}}{2} \left(\frac{1}{k_\|^2 \rho_e^2}-\frac{1}{2}\right) \exp{\left(-\frac{1}{k_\|^2 \rho_e^2}\right)} + \mathrm{i} k_\|^2 \rho_e^2 
\right]^2 \qquad \qquad \nonumber \\
+\left\{\tilde{\omega}_{e\|} \beta_e \, \Real{\; Z\!\left(\frac{1}{k_{\|} \rho_e}\right)} + \eta_e^T \beta_e \left[\frac{1}{2 k_{\|} \rho_e} + \left(\frac{1}{2 k_\|^2 \rho_e^2} - \frac{1}{4}\right)\Real{\; Z\!\left(\frac{1}{k_\| \rho_e}\right)}\right]\right\}^2 = 0 \, . \qquad \label{disprel_parallel_elecheatflux}
\end{eqnarray}
This can be factorised to give two roots; separating the complex frequency into real and imaginary parts via
$\omega = \varpi + \mathrm{i} \gamma$, and defining 
\begin{equation}
 \tilde{\varpi}_{e\|} \equiv \frac{\varpi}{k_{\|} v_{\mathrm{th}e}} \, , \quad \tilde{\gamma}_{e\|} \equiv \frac{\gamma}{k_{\|} v_{\mathrm{th}e}} \, , \label{real_imag_frequency_defs}
\end{equation}
we have
\begin{subeqnarray}
\tilde{\varpi}_{e\|} \beta_e & = &  \eta_e^T \beta_e \left(\frac{1}{2 k_\|^2 \rho_e^2}-\frac{1}{4}\right) 
+ \frac{\left({\eta_e^T \beta_e}/{2 k_\| \rho_e} - k_\|^2 \rho_e^2\right) \Real{\; Z\!\left({1}/{k_\| \rho_e}\right)} }{\left[\Real{\; Z\!\left({1}/{k_\| \rho_e}\right)}\right]^2 + \upi \exp{\left(-{2}/{k_\|^2 \rho_e^2}\right)}} 
\, , \\
\tilde{\gamma}_{e\|} \beta_e & = & 
\frac{\sqrt{\upi}\left({\eta_e^T \beta_e}/{2 k_\| \rho_e} - k_\|^2 \rho_e^2\right) }{\left[\Real{\; Z\!\left({1}/{k_\| \rho_e}\right)}\right]^2 \exp{\left({1}/{k_\|^2 \rho_e^2}\right)}+ \upi \exp{\left(-{1}/{k_\|^2 \rho_e^2}\right)}} 
\, , \label{whistlerwave_growthrate_allkpl_Append}
\end{subeqnarray}
whence (\ref{whistlerwave_growthrate_allkpl}) follows immediately. 

\subsection{Approximate dispersion relation of CE ion-temperature-gradient-driven microinstabilities}  \label{ion_heatflux_instab}

We now explain the method used to characterise microinstabilities driven by the ion-temperature-gradient term. For these, we set the electron-temperature-gradient terms to zero, $\eta_e^T = 0$, assume 
the ordering $\tilde{\omega}_{i\|} \sim \eta_i$, and anticipate that
such microinstabilities will occur on ion rather than electron scales, i.e., $k \rho_i \sim 1$. Under the ordering $\tilde{\omega}_{i\|} \sim \eta_i \ll 1$, 
it follows that $\tilde{\omega}_{e\|} \sim \mu_e^{1/2} \tilde{\omega}_{i\|}  \ll 1$; 
therefore, we can use (\ref{Mmaxcomp}) 
to quantity the contribution of Maxwellian electrons to the total dielectric tensor. 
However, since $k \rho_i \sim 1$, we must consider the matrix $\mathsfbi{M}_{e}^{(0)}$
in the limit $k_{\|} \rho_e \sim k_{\bot} \rho_e \sim \mu_e^{1/2} \ll 1$. 
Asymptotic forms of (\ref{Mmaxcomp}) appropriate for this limit are given by
(\ref{specialfuncMax_xysmall}), and lead to\footnote{As noted in section \ref{shortcomings_twospecies}, for $k_{\|} \rho_e \ll 1$, the approximation $(\mathsfbi{M}_{e})_{11} \approx \tilde{\omega}_{e\|} (\mathsfbi{M}_{e}^{(0)})_{11}$ 
in fact breaks down, on account of $(\mathsfbi{M}_{e}^{(0)})_{11}$ becoming 
exponentially small in $k_{\|} \rho_e \ll 1$. However, it turns out that when $k_{\|} \rho_i \sim k_{\bot} \rho_i \sim 
1$, $(\mathsfbi{M}_{e})_{11} \ll (\mathsfbi{M}_{i})_{11}$, and so this subtlety
can be ignored for the CE ion-temperature-gradient-driven instabilities.}
\begin{subeqnarray}
  (\mathsfbi{M}_{e}^{(0)})_{11} & = & \textit{O}\left[\exp{\left(-\frac{1}{k_{\|}^2 \rho_e^2}\right)}\right] \, , \\
  (\mathsfbi{M}_{e}^{(0)})_{12} & \approx & -\mathrm{i} \frac{k}{k_{\|}} \left[k_{\|} \rho_e + \textit{O}(k^3 \rho_e^3)\right]  \, , \\
  (\mathsfbi{M}_{e}^{(0)})_{21} & = & \mathrm{i} \frac{k}{k_{\|}}  \left[k_{\|} \rho_e + \textit{O}(k^3 \rho_e^3)\right] \, , \\  
  (\mathsfbi{M}_{e}^{(0)})_{22} & = & \mathrm{i} \left[\sqrt{\upi} k_{\bot}^2 \rho_e^2 + \textit{O}(k_{\bot}^4 \rho_e^4)\right] \, . \label{Mmaxcomp_smallk_e}
\end{subeqnarray}

We now combine (\ref{Mmaxcomp_smallk_e}) with (\ref{Mmaxcomp}) for $\mathsfbi{M}_i^{(0)}$ 
and (\ref{Pionheatflux}) for $\mathsfbi{P}_i^{(0)}$, and find
 the dispersion relation for CE ion-temperature-gradient-driven microinstabilities by 
 substituting the dielectric tensor (\ref{dielectric_0_multispecies}) into 
 (\ref{disprel_simp_1}):
 \begin{eqnarray}
   \left[\tilde{\omega}_{i\|} F\!\left(k_{\|} \rho_i,k_{\bot} \rho_i\right) \right. & + & \left. \eta_i I\!\left(k_{\|} \rho_i,k_{\bot} \rho_i\right)+ \mathrm{i} k_{\|}^2 d_i^2\right] \nonumber \\
   &\times& \left[\tilde{\omega}_{i\|} H\!\left(k_{\|} \rho_i,k_{\bot} \rho_i\right) + \eta_i K\!\left(k_{\|} \rho_i,k_{\bot} \rho_i\right) 
+\mathrm{i} k^2 d_i^2\right] 
   \nonumber \\
  & + &\left[\tilde{\omega}_{i\|} \left[G\!\left(k_{\|} \rho_i,k_{\bot} \rho_i\right) + k_{\|} \rho_i\right]+ \eta_i J\!\left(k_{\|} \rho_i,k_{\bot} \rho_i\right)\right]^2 = 0 \, 
  ,
  \qquad \label{dispheatflux_ion}
 \end{eqnarray}
where $d_i = c/\omega_{pi}$ is the ion inertial scale, and we have ordered $\eta_i \sim 1/\beta_i \sim k^2 d_i^2$. 
This dispersion relation is 
very similar to (\ref{dispheatfluxA}), save for the addition of one term [the middle term in the third line of (\ref{dispheatflux_ion})] providing a linear coupling between the $\widehat{\delta \boldsymbol{E}}_1$ and  $\widehat{\delta \boldsymbol{E}}_2$ components of the electric field 
perturbation. Similarly to (\ref{heatflux_elec_frequency_Append}), the dispersion relation (\ref{dispheatflux_ion}) can be written as a quadratic in 
$\tilde{\omega}_{i\|} \beta_i$, which is then solved to give the following 
expression for the complex frequency:
\begin{equation}
  \omega = \frac{\Omega_i}{\beta_i} k_{\|} \rho_i \frac{-\tilde{B}_{\mathrm{T}} \pm \sqrt{\tilde{B}_{\mathrm{T}}^2 + 4\tilde{A}_{\mathrm{T}}\tilde{C}_{\mathrm{T}}}}{2 \tilde{A}_{\mathrm{T}}} 
  \, , \label{heatflux_ion_frequency_Append}
\end{equation}
where
 \begin{eqnarray}
   \tilde{A}_{\mathrm{T}} & = & F_i H_i + \left[G_i+ k_{\|} \rho_i\right]^2 \, , \\
   \tilde{B}_{\mathrm{T}} & = & \eta_i \beta_i \left[F_i K_i +H_i I_i + 2J_i \left(G_i+ k_{\|} \rho_i\right)\right] + \mathrm{i} \left(F_i k^2 \rho_e^2 + H_i k_{\|}^2 \rho_e^2\right) \, , \\
   \tilde{C}_{\mathrm{T}} & = & \left(\eta_i \beta_i\right)^2 \left(I_i K_i + J_i^2\right) - k^2 k_{\|}^2 \rho_e^4 +\mathrm{i} \eta_i \beta_i  \left(I_i k^2 \rho_e^2 + K_i k_{\|}^2 \rho_e^2\right) \, 
   .  \label{CEiontemp_quadcoeff}
 \end{eqnarray}
This expression is the one that is used to evaluate the real frequencies and growth rates of ion-scale CET microinstabilities in sections \ref{ion_heatflux_instab_slowwave}.

\subsubsection{Derivation of frequency and growth rate of the parallel CET slow-hydromagnetic-wave instability} \label{derivation_slowwave}

We obtain the dispersion relation of the parallel slow-wave instability by considering the general dispersion 
relation (\ref{dispheatflux_ion}) of CE ion-temperature-gradient-driven instabilities in the limit $k_{\perp} \rightarrow 0$:
\begin{eqnarray}
\left[\tilde{\omega}_{i\|} \beta_i \sqrt{\upi} \exp{\left(-\frac{1}{k_\|^2 \rho_i^2}\right)} \right. & + & \left.\eta_i \beta_i \frac{\sqrt{\upi}}{2} \left(\frac{1}{k_\|^2 \rho_i^2}-\frac{1}{2}\right) \exp{\left(-\frac{1}{k_\|^2 \rho_i^2}\right)} + \mathrm{i} k_\|^2 \rho_i^2 
\right]^2 \qquad \qquad \nonumber \\
&+& \left\{\tilde{\omega}_{i\|} \beta_i \, \left[\Real{\; Z\!\left(\frac{1}{k_\| \rho_i}\right)} + k_{\|} \rho_i\right] \right. \nonumber \\
&+& \left. \eta_i \beta_i \left[\frac{1}{2 k_{\|} \rho_i} + \left(\frac{1}{2 k_\|^2 \rho_i^2} - \frac{1}{4}\right)\Real{\; Z\!\left(\frac{1}{k_{\|} 
\rho_i}\right)}\right]\right\}^2 = 0 \, . \qquad \label{disprel_parallel_ionheatflux}
\end{eqnarray}
As before, this can be factorised to give two roots; for $\tilde{\omega}_{i\|} = \tilde{\varpi}_{i\|} + \mathrm{i} 
\tilde{\gamma}_{i\|}$ [cf. (\ref{real_imag_frequency_defs})], it follows that
\begin{subeqnarray}
\tilde{\varpi}_{i\|} \beta_i & = &  \eta_i \beta_i \left(\frac{1}{2 k_\|^2 \rho_i^2}-\frac{1}{4}\right) 
+ \frac{k_{\|} \rho_i \left[\Real{\; Z\!\left(\frac{1}{k_\| \rho_i}\right)} + k_{\|} \rho_i\right] \left({\eta_i \beta_i}/{4} - k_\| \rho_i\right) }{\left[\Real{\; Z\!\left(\frac{1}{k_\| \rho_i}\right)} + k_{\|} \rho_i\right]^2 + \upi \exp{\left(-\frac{2}{k_\|^2 \rho_i^2}\right)}} 
\, , \\
\tilde{\gamma}_{i\|} \beta_i & = & 
\frac{\sqrt{\upi} k_{\|} \rho_i \left({\eta_i \beta_i}/{4} - k_\| \rho_i\right) }{\left[\Real{\; Z\!\left(\frac{1}{k_\| \rho_i}\right)} + k_{\|} \rho_i\right]^2 \exp{\left(\frac{1}{k_\|^2 \rho_i^2}\right)}+ \upi \exp{\left(-\frac{1}{k_\|^2 \rho_i^2}\right)}} 
\, . 
\end{subeqnarray}
These can be rearranged to give (\ref{slowwavegen}). 

\subsubsection{Derivation of frequency and growth rate of the CET long-wavelength KAW instability} \label{derivation_CETKAW}

In the limit $k_{\|} \rho_i \ll 
1$, $k_\perp \rho_i \sim 1$, the general dispersion relation (\ref{dispheatflux_ion}) 
of CE ion-temperature-gradient-driven instabilities
becomes
\begin{eqnarray}
  \Bigg[\tilde{\omega}_{i\|} (1 & - & \mathcal{F}_i ) - \frac{\eta_i}{2} \mathcal{G}_i \Bigg ]^2 
\nonumber  \\
& + & \frac{k_{\perp}^2 \rho_i^2}{\beta_i} \Bigg[ \mathrm{i} \sqrt{\upi} \left(\mathcal{F}_i  + \sqrt{\frac{\mu_e Z^2}{\tau}}\right) \tilde{\omega}_{i\|} - \frac{1}{\beta_i} + \frac{\mathrm{i} \sqrt{\upi} \eta_i}{2} \left(
\mathcal{G}_i -\frac{1}{2} \mathcal{F}_i  \right) \Bigg]  = 0  , \qquad \quad \label{CETKAW_disprel}
\end{eqnarray}
where we remind the reader that $\mathcal{F}_i = \mathcal{F}(k_{\perp} \rho_i)$, $\mathcal{G}_i = \mathcal{G}(k_{\perp} 
\rho_i)$, with the functions $\mathcal{F}(\alpha)$ and $\mathcal{G}(\alpha)$ being defined by (\ref{specialfunction_quasiperp}). 
Equation (\ref{heatfluxKAWfreq}) for the complex frequency of the CET KAW 
modes in the main text is then derived by solving (\ref{CETKAW_disprel}) for $\tilde{\omega}_{i\|} = {\omega}/{k_{\|} 
v_{\mathrm{th}i}}$.

\section{Methodology for characterising CES microinstabilities} \label{CES_method_append}

This appendix outlines the method used to determine the growth rates 
of microinstabilities driven by the CE electron- and ion-shear terms. Once 
again (cf. appendix \ref{CET_method}), section \ref{linear_stab_method} presents the general 
framework of our approach: determine a simplified algebraic dispersion relation
satisfied by the (complex) frequencies $\omega$ of CES microinstabilities with
parallel and perpendicular wavenumber $k_{\|}$ and $k_{\perp}$ under the assumption that 
they are low frequency [viz., $\omega \ll k_{\|} v_{\mathrm{th}s}$; cf. (\ref{omegascale})], solve for $\omega$, then 
calculate the growth rate $\gamma$ from its imaginary part (and the real frequency 
$\varpi$ from its real part). To construct the dispersion relation, we first 
need to know the tensor $\mathsfbi{P}_{s}^{(0)}$ for a CE distribution function of the form (\ref{CEsheardistfuncexpression}); 
this result is given in appendix \ref{dielectricterms_shear}. Then, in 
appendix \ref{formfordispshear_derivation}, we determine 
an approximate quadratic dispersion relation for CES microinstabilities, show in appendix \ref{thresholdcal_quadratic}
how that dispersion relation can be used in certain cases to evaluate the CES instability thresholds 
semi-analytically, then demonstrate
the significant shortcomings of the quadratic approximation in appendix \ref{shortcomings_quadratic}.  
In appendix \ref{CES_method_append_quartic_derivation}, we address these shortcomings 
by constructing a revised quartic dispersion relation for CES microinstabilities. This quartic dispersion relation 
is then used to derive simplified dispersion relations for the various different 
CES microinstabilities discussed in the main text: the mirror instability in 
appendix \ref{derivation_mirror}, the parallel (CES) whistler instability in 
appendix \ref{derivation_parwhistler}, the transverse instability in appendix \ref{derivation_transverse},
the electron mirror instability in appendix \ref{derivation_elecmirror}, the 
parallel, oblique and critical-line firehose instabilities in Appendicies 
\ref{derivation_firehose_par}, \ref{derivation_firehose_oblique}, and 
\ref{derivation_firehose_critline}, the parallel and oblique electron firehose instabilities in 
Appendices \ref{derivation_parelecfirehose} and 
\ref{derivation_obliqueelecfirehose}, the EST instability in appendix 
\ref{derivation_EST}, and the whisper instability in appendix 
\ref{derivation_whisper}. 
 Finally, in appendix \ref{derivation_ordinarymode}, we derive the dispersion 
 relation of the CET ordinary-mode instabilty 
-- the one CES (or CET) microinstability that does not satisfy $\omega \ll k_{\|} v_{\mathrm{th}s}$ 
for either electrons or ions (see section \ref{shortcomings_othermicro}) -- directly from the hot-plasma dispersion 
relation. 

\subsection{Dielectric response of CE shear terms} \label{dielectricterms_shear}

First, we evaluate the elements of $\mathsfbi{P}_{s}^{(0)}$:
\begin{subeqnarray}
  (\mathsfbi{P}_{s}^{(0)})_{11} & = & \epsilon_s \frac{k^2}{k_{\|}^2} W\!\left(k_{\|} \tilde{\rho}_s,k_{\bot} \tilde{\rho}_s\right) \, , \\
  (\mathsfbi{P}_{s}^{(0)})_{12} & = & - \epsilon_s \frac{k}{k_{\|}} X\!\left(k_{\|} \tilde{\rho}_s,k_{\bot} \tilde{\rho}_s\right) \, , \\
  (\mathsfbi{P}_{s}^{(0)})_{21} & = & \epsilon_s \frac{k}{k_{\|}}  X\!\left(k_{\|} \tilde{\rho}_s,k_{\bot} \tilde{\rho}_s\right) \, , \\  
  (\mathsfbi{P}_{s}^{(0)})_{22} & = & \epsilon_s Y\!\left(k_{\|} \tilde{\rho}_s,k_{\bot} \tilde{\rho}_s\right) \, 
  , \label{Pshearmat}
\end{subeqnarray}
where the special functions $W\!\left(x,y\right)$, $Y\!\left(x,y\right)$ and 
$X\!\left(x,y\right)$ are defined by (\ref{shearasymptoticfuncs}). These results are derived in appendix 
\ref{nonzeroshear}. 

\subsection{Quadratic approximation to dispersion relation of CE shear-driven microinstabilities} 
\label{formfordispshear}

\subsubsection{Derivation} \label{formfordispshear_derivation}

Considering the relative magnitude of $\tilde{\omega}_{i\|} = \omega/k_{\|} v_{\mathrm{th}i}$ and 
$\tilde{\omega}_{e\|} = \omega/k_{\|} v_{\mathrm{th}e} \ll \tilde{\omega}_{i\|}$, we observe that, unlike CET microinstabilities, CES 
microinstabilities satisfy the low-frequency condition (\ref{omegascale}) for both 
electrons and ions. This claim holds because
any microinstability involving the CE electron-shear term must satisfy $\tilde{\omega}_{e\|} \sim \epsilon_e \ll (m_e/m_i)^{1/2}$, 
where the last inequality arises from the scaling relation $\epsilon_e \sim (m_e/m_i)^{1/2} \epsilon_i$ 
given by (\ref{smallparamscalings}\textit{d}); thus, from the scaling relation (\ref{ratio_tildeomegas}) with $T_e = T_i$, it follows that $\tilde{\omega}_{i\|} \sim \epsilon_e (m_i/m_e)^{1/2} \sim \epsilon_i \ll 1$. 
Therefore, it is consistent to expand both the Maxwellian electron and ion terms 
in $\tilde{\omega}_{s\|} \ll 1$. 

We therefore initially approximate $\boldsymbol{\mathfrak{E}}$ as follows:
  \begin{equation}
    \boldsymbol{\mathfrak{E}} \approx \tilde{\omega}_{e\|} \boldsymbol{\mathfrak{E}}^{(0)}  = \frac{\omega_{\mathrm{p}e}^2}{\omega^2} \left(\sum_s \tilde{\omega}_{s\|} \mu_s \mathsfbi{M}_s^{(0)} + \sum_s \mu_s \mathsfbi{P}_s^{(0)} \right)
    \, , \label{dielectricapproximationshear_zerothorder}
  \end{equation}
  where the expansion of $\mathsfbi{M}_s$ and $\mathsfbi{P}_s$ in $\tilde{\omega}_{s\|}$, i.e., 
 \begin{equation}
   \mathsfbi{M}_s\!\left(\tilde{\omega}_{s\|},\boldsymbol{k}\right) \approx \tilde{\omega}_{s\|} \mathsfbi{M}_s^{(0)}\!\left(\boldsymbol{k}\right) 
   \, , \quad
   \mathsfbi{P}_s\!\left(\tilde{\omega}_{s\|},\boldsymbol{k}\right) \approx \mathsfbi{P}_s^{(0)}\!\left(\boldsymbol{k}\right) \, 
   , \label{leadingorder_MandP}
  \end{equation}
  applies to both ion and electron species. By analogy to the derivation presented in section \ref{consequences}, this approximation 
  gives rise to a simplified dispersion relation [cf. (\ref{disprel_simp_1})]
\begin{equation}
\left(\tilde{\omega}_{e\|} \mathfrak{E}_{11}^{(0)}-\frac{k^2 c^2}{\omega^2}\right)\left(\tilde{\omega}_{e\|} \mathfrak{E}_{22}^{(0)}-\frac{k^2 c^2}{\omega^2}\right)+\left(\tilde{\omega}_{e\|} \mathfrak{E}_{12}^{(0)}\right)^2
 = 0 \, . \label{disprel_simp_shear}
\end{equation}
We emphasise that here each component of $\boldsymbol{\mathfrak{E}}^{(0)}$ has both electron and 
 ion contributions. Expressing $\tilde{\omega}_{i\|} = \tilde{\omega}_{e\|} \mu_e^{-1/2}$ in (\ref{dielectricapproximationshear_zerothorder}), (\ref{disprel_simp_shear}) can be written as
 \begin{eqnarray}
   && \left[\tilde{\omega}_{e\|} (\mathsfbi{M}_{e}^{(0)} + \mu_e^{1/2} \mathsfbi{M}_{i}^{(0)})_{11}  + (\mathsfbi{P}_{e}^{(0)} + \mu_e^{1/2} \mathsfbi{P}_{i}^{(0)})_{11} - k^2 d_e^2\right] \nonumber \\
   & \quad & \times \left[\tilde{\omega}_{e\|} (\mathsfbi{M}_{e}^{(0)} + \mu_e^{1/2} \mathsfbi{M}_{i}^{(0)})_{22} + (\mathsfbi{P}_{e}^{(0)} + \mu_e^{1/2} \mathsfbi{P}_{i}^{(0)})_{22} - k^2 d_e^2\right] 
   \nonumber \\
  & \quad &  \qquad \qquad \qquad + \left[ \tilde{\omega}_{e\|} (\mathsfbi{M}_{e}^{(0)} + \mu_e^{1/2} \mathsfbi{M}_{i}^{(0)})_{12} + (\mathsfbi{P}_{e}^{(0)} + \mu_e^{1/2} \mathsfbi{P}_{i}^{(0)})_{12} \right]^2 = 0 \, 
  . \label{disprel_simp_shear_A}
 \end{eqnarray} 
 Combining the expressions (\ref{Pshearmat}) for $\mathsfbi{P}_s^{(0)}$ with (\ref{Mmaxcomp}) for $\mathsfbi{M}_s^{(0)}$
  and substituting $\mathsfbi{M}_s^{(0)}$ and $\mathsfbi{P}_s^{(0)}$ into (\ref{disprel_simp_shear_A}) gives 
 \begin{eqnarray}
   && \left[\mathrm{i} \tilde{\omega}_{e\|} \left( F_e + \mu_e^{1/2} F_i \right) +\epsilon_e \left( W_e + \mu_e^{1/2} W_i \right)- k_{\|}^2 d_e^2\right] \qquad \qquad \nonumber \\
  & \qquad & \times \left[ \mathrm{i}  \tilde{\omega}_{e\|} \left( H_e + \mu_e^{1/2} H_i \right) + \epsilon_e \left( Y_e + \mu_e^{1/2} Y_i \right)- k^2 d_e^2\right] 
   \nonumber \\
  & \qquad & \qquad \qquad \qquad +\left[\mathrm{i} \tilde{\omega}_{e\|} \left( G_e + \mu_e^{1/2} G_i \right) +  \epsilon_e \left( X_e + \mu_e^{1/2} X_i \right) \right]^2 = 0 \, 
  , \label{sheardisprel_A}
  \qquad
 \end{eqnarray}
 where we have used $\epsilon_i = \epsilon_e \mu_e^{-1/2}$. For brevity of notation, 
 we have also defined $F_s \equiv F\!\left(k_{\|} \tilde{\rho}_s,k_{\bot} \tilde{\rho}_s\right)$, $G_s \equiv G\!\left(k_{\|} \tilde{\rho}_s,k_{\bot} 
\tilde{\rho}_s\right)$, and so on. 

Using (\ref{skin_depth_def}\textit{b}) for the terms $\propto d_e^2$ explicitly introduces a $\beta_e$ dependence into (\ref{sheardisprel_A}). 
After some elementary manipulations, we obtain the quadratic
 \begin{equation}
   A_{\mathrm{S}} \tilde{\omega}_{e\|}^2 \beta_e^2 + \mathrm{i} B_{\mathrm{S}} \tilde{\omega}_{e\|} \beta_e - C_{\mathrm{S}} = 0  \label{sheardisprel_B}
   \, ,
 \end{equation}
 where 
 \begin{subeqnarray}
   A_{\mathrm{S}} & = & \left(F_e+\mu_e^{1/2} F_i\right)\left(H_e+\mu_e^{1/2} H_i\right) +\left(G_e+\mu_e^{1/2} G_i\right)^2 \, , \\
   B_{\mathrm{S}} & = & \left(H_e+\mu_e^{1/2} H_i\right)\left[k_{\|}^2 \rho_e^2 - \epsilon_e \beta_e \left(W_e+\mu_e^{1/2} W_i\right)\right] - 2  \epsilon_e \beta_e \left(G_e+\mu_e^{1/2} G_i\right) \left(X_e+\mu_e^{1/2} X_i\right) \nonumber \\
   && +\left(F_e+\mu_e^{1/2} F_i\right)\left[k^2 \rho_e^2 - \epsilon_e \beta_e \left(Y_e+\mu_e^{1/2} Y_i\right)\right] \, , \\
   C_{\mathrm{S}} & = & \left[k_{\|}^2 \rho_e^2 - \epsilon_e \beta_e \left(W_e+\mu_e^{1/2} W_i\right)\right] \left[k^2 \rho_e^2 - \epsilon_e \beta_e \left(Y_e+\mu_e^{1/2} Y_i\right)\right] \nonumber \\
   && + \epsilon_e^2 \beta_e^2 \left(X_e+\mu_e^{1/2} X_i\right)^2\, .
 \end{subeqnarray}
As before, this can be solved explicitly for the complex frequency:
\begin{equation}
  \omega = \frac{\Omega_e}{\beta_e} k_{\|} \rho_e \frac{- \mathrm{i} B_{\mathrm{S}} \pm \sqrt{-B_{\mathrm{S}}^2 + 4A_{\mathrm{S}}C_{\mathrm{S}}}}{2 A_{\mathrm{S}}} 
  \, . \label{shear_frequency_Append}
\end{equation}
From this expression, we can extract the real frequency $\varpi$ and the growth 
rate $\gamma$ explicitly. In the case when $4A_{\mathrm{S}}C_{\mathrm{S}} > B_{\mathrm{S}}^2$, we have two oppositely 
propagating modes with the same growth rate:
\begin{subeqnarray}
 \varpi& = & \pm \frac{\Omega_e}{\beta_e} k_{\|} \rho_e \frac{\sqrt{-B_{\mathrm{S}}^2 + 4A_{\mathrm{S}}C_{\mathrm{S}}}}{2 A_{\mathrm{S}}}  
 \, , 
\\
 \gamma & = & \frac{\Omega_e}{\beta_e} k_{\|} \rho_e \frac{B_S}{2 A_{\mathrm{S}}}  
 \, .
\end{subeqnarray}
 For $4A_{\mathrm{S}}C_{\mathrm{S}} < B_{\mathrm{S}}^2$, both modes are non-propagating, with distinct 
 growth rates:
 \begin{equation}
  \gamma = \frac{\Omega_e}{\beta_e} k_{\|} \rho_e \frac{B_{\mathrm{S}} \pm \sqrt{B_{\mathrm{S}}^2 - 4A_{\mathrm{S}}C_{\mathrm{S}}}}{2 A_{\mathrm{S}}} 
  \, . \label{shear_growthrateB}
\end{equation}

\subsubsection{Semi-analytic estimates of CES instability thresholds using quadratic approximation} 
\label{thresholdcal_quadratic}

In the case of non-propagating modes whose growth rate is given by (\ref{shear_growthrateB}), 
we can determine semi-analytic formulae for the thresholds 
of any instabilities. This is done by noting that, at 
marginal stability, $\tilde{\omega}_{e\|} = 0$. Therefore, it follows from (\ref{sheardisprel_B}) that $C_{\rm S} = 
0$, or, equivalently, 
\begin{equation}
  \left[k_{\|}^2 \rho_e^2 - \epsilon_e \beta_e \left(W_e+\mu_e^{1/2} W_i\right)\right] \left[k^2 \rho_e^2 - \epsilon_e \beta_e \left(Y_e+\mu_e^{1/2} Y_i\right)\right] + \epsilon_e^2 \beta_e^2 \left(X_e+\mu_e^{1/2} X_i\right)^2\, 
  = 0 \, .
\end{equation}
This is a quadratic in $\epsilon_e \beta_e$ which can be solved exactly 
to give the threshold value of $\epsilon_e \beta_e$ as a function of 
perpendicular and parallel wavenumber:
\begin{eqnarray}
\epsilon_e \beta_e & = & \frac{1}{2}\left[\left(W_e+\mu_e^{1/2} W_i\right) \left(Y_e+\mu_e^{1/2} Y_i\right) + \left(X_e+\mu_e^{1/2} X_i\right)^2\right]^{-1} \nonumber \\
&& \times \Bigg(k^2 \rho_e^2 \left(W_e+\mu_e^{1/2} W_i\right) + k_{\|}^2 \rho_e^2 \left(Y_e+\mu_e^{1/2} Y_i\right) \nonumber \\
&& \pm \Bigg\{\left[k^2 \rho_e^2 \left(W_e+\mu_e^{1/2} W_i\right) + k_{\|}^2 \rho_e^2 \left(Y_e+\mu_e^{1/2} Y_i\right)\right]^2  \nonumber \\&& - 4 k_{\|}^2 k^2 \rho_e^4 \left[\left(W_e+\mu_e^{1/2} W_i\right) \left(Y_e+\mu_e^{1/2} Y_i\right) + \left(X_e+\mu_e^{1/2} X_i\right)^2\right]\Bigg\}^{1/2} \Bigg) . \qquad \label{thresholdval_semianalyticform}
\end{eqnarray}
Expression (\ref{thresholdval_semianalyticform}) is used in sections \ref{negpres_fire} and \ref{negpres_electron_oblique} to evaluate the wavevector-dependent thresholds of the CES ion and electron firehose instabilities, respectively. 

\subsubsection{Shortcomings of quadratic approximation} \label{shortcomings_quadratic}

In contrast to quadratic approximations to the dispersion relations of 
CET microinstabilities being sufficient to characterise all instabilities of note (see, e.g., appendix \ref{electron_heatflux_instab}),
not all CES microinstabilities are captured by the quadratic dispersion relation (\ref{sheardisprel_B}), because there are important
microinstabilities whose correct description requires keeping
higher-order terms in the $\tilde{\omega}_{s\|} \ll 1$ expansion. The 
mathematical reason for this is that some microinstabilities occur in  
wavenumber regimes where either $k_{\|} \rho_i \ll 1$ and/or $k_{\|} \rho_e \ll 1$. 
As a result, the issues raised in section~\ref{shortcomings_twospecies} regarding the 
commutability of the ${\omega}_{s\|} \ll 1$ and $k_{\|} \rho_s \ll 1$ limits must be carefully 
resolved. In appendix \ref{ordering}, it is shown that, if $k_{\|} \rho_s \ll 
1/\log{(1/\tilde{\omega}_{s\|})}$, then the dominant contributions to  
$(\mathsfbi{M}_s)_{xx}$, $(\mathsfbi{M}_s)_{xz}$, and $(\mathsfbi{M}_s)_{zz}$ 
arise from the quadratic term in $\tilde{\omega}_{s\|} \ll 1$ expansion, namely 
\begin{subeqnarray}
  (\mathsfbi{M}_{s})_{xx} & \approx & \tilde{\omega}_{s\|}^2 (\mathsfbi{M}_{s}^{(1)})_{xx} \, , \\
  (\mathsfbi{M}_{s})_{xz} &\approx & \tilde{\omega}_{s\|}^2 (\mathsfbi{M}_{s}^{(1)})_{xz} \, , \\
  (\mathsfbi{M}_{s})_{zz} & \approx & \tilde{\omega}_{s\|}^2 (\mathsfbi{M}_{s}^{(1)})_{zz} \, . 
  \label{Mmaxcomp_ordering}
\end{subeqnarray}
If $k_{\bot} \rho_s \ll k_{\|} \rho_s \tilde{\omega}_{s\|}^{1/2}$, then
\begin{equation}
(\mathsfbi{M}_{s})_{yy}  \approx  \tilde{\omega}_{s\|}^2 (\mathsfbi{M}_{s}^{(1)})_{yy} \, . 
\end{equation}
In the $\left\{\boldsymbol{e}_1,\boldsymbol{e}_2,\boldsymbol{e}_3\right\}$ 
coordinate frame, this means that the dominant contributions to each component 
of $\mathsfbi{M}_{s}$ are (see appendix \ref{Maxwellresponse_low123})
\begin{subeqnarray}
 (\mathsfbi{M}_{s})_{11} & \approx & \tilde{\omega}_{s\|}^2 (\mathsfbi{M}_{s}^{(1)})_{11} = \frac{k^2} {k_{\|}^2} {\omega}_{s\|}^2 (\mathsfbi{M}_{s}^{(1)})_{xx} + 2 \tilde{\omega}_{s\|}^2 \left[\frac{k_{\bot}^2}{k^2} + \frac{k_{\bot}}{k_{\|}} L\!\left(k_{\|} \tilde{\rho}_s,k_{\bot} \tilde{\rho}_s \right)\right] \, , \quad \\
 (\mathsfbi{M}_{s})_{12} & \approx & \tilde{\omega}_{s\|} (\mathsfbi{M}_{s}^{(0)})_{12} = \frac{k}{k_{\|}} \tilde{\omega}_{s\|} (\mathsfbi{M}_{s}^{(0)})_{xy} \,  ,\\
 (\mathsfbi{M}_{s})_{13} & \approx & \tilde{\omega}_{s\|}^2 (\mathsfbi{M}_{s}^{(1)})_{13}= -\tilde{\omega}_{s\|}^2 \left[\frac{2 k_{\bot} k_{\|}}{k^2} + L\!\left(k_{\|} \tilde{\rho}_s,k_{\bot} \tilde{\rho}_s \right)\right] \, , \\
 (\mathsfbi{M}_{s})_{22} & \approx & \tilde{\omega}_{s\|} (\mathsfbi{M}_{s}^{(0)})_{22} + \tilde{\omega}_{s\|}^2 (\mathsfbi{M}_{s}^{(1)})_{22} = \tilde{\omega}_{s\|} (\mathsfbi{M}_{s}^{(0)})_{yy} + \tilde{\omega}_{s\|}^2 (\mathsfbi{M}_{s}^{(1)})_{yy} \, , \\
 (\mathsfbi{M}_{s})_{23} & \approx & \tilde{\omega}_{s\|}^2 (\mathsfbi{M}_{s}^{(1)})_{23} =  -\frac{k_{\|}}{k} \tilde{\omega}_{s\|}^2 N\!\left(k_{\|} \tilde{\rho}_s,k_{\bot} \tilde{\rho}_s \right) \, , \\
 (\mathsfbi{M}_{s})_{33} & \approx & \tilde{\omega}_{s\|}^2 (\mathsfbi{M}_{s}^{(1)})_{33} = \frac{2 k_{\|}^2}{k^2} \tilde{\omega}_{s\|}^2 \, 
 , \label{dielectric123trans_smallkpl}
 \end{subeqnarray}
where the special functions $L(x,y)$ and $N(x,y)$ are given by (\ref{LandNdef_Append}).
The quadratic dispersion relation (\ref{sheardisprel_B}) must, therefore, be 
revised to capture correctly all relevant microinstabilities. 

\subsection{Quartic approximation to dispersion relation of CE shear-driven microinstabilities} \label{CES_method_append_quartic}

\subsubsection{Derivation of general quartic CES dispersion relation} \label{CES_method_append_quartic_derivation}

To assess how the new terms identified in section \ref{shortcomings_quadratic} change the dispersion relation (\ref{sheardisprel_A}), we now return to the full hot-plasma dispersion relation~(\ref{hotplasmadisprel}), which 
we write in the form
\begin{equation}
\left(\mathfrak{E}_{11}-\frac{k^2 c^2}{\omega^2} - \frac{\mathfrak{E}_{13}^2}{\mathfrak{E}_{33}}\right)\left(\mathfrak{E}_{22}-\frac{k^2 c^2}{\omega^2}+ \frac{\mathfrak{E}_{23}^2}{\mathfrak{E}_{33}}\right)+\left(\mathfrak{E}_{12}-\frac{\mathfrak{E}_{13}\mathfrak{E}_{23}}{\mathfrak{E}_{33}}\right)^2 = 0 \, 
. \label{hotplasmadisprel_C}
\end{equation}
Reminding the reader that, for a two-species plasma,
  \begin{equation}
    \boldsymbol{\mathfrak{E}} = \sum_s \boldsymbol{\mathfrak{E}}_s = \frac{\omega_{\mathrm{p}e}^2}{\omega^2} \sum_s \mu_s \left(\mathsfbi{M}_s + \mathsfbi{P}_s \right) 
    \, , \label{dielectric_expand_shear}
  \end{equation}
  and also that the electrostatic component of the dielectric tensor is 
  determined by the Maxwellian components only (which in turn are equal for electrons and ions when $T_i = T_e$ -- see appendix 
  \ref{lowfrequency_electrostatic}), viz., 
    \begin{equation}
    \mathfrak{E}_{33} \approx  \tilde{\omega}_{e\|}^2 \mathfrak{E}_{33}^{(1)} = \frac{\omega_{\mathrm{p}e}^2}{\omega^2} \sum_s  \mu_s \tilde{\omega}_{s\|}^2
    (\mathsfbi{M}_s^{(1)})_{33} = \frac{4 \omega_{\mathrm{p}e}^2}{\omega^2}  \tilde{\omega}_{e\|}^2 \frac{k_{\|}^2}{k^2}
    \, ,
  \end{equation}
we show in appendix \ref{secondordercorrect} that, in the limit $k_{\|} \rho_s \ll 1$,
\begin{subeqnarray}
 \frac{\left[(\mathsfbi{M}_{s})_{13}\right]^2}{(\mathsfbi{M}_s^{(1)})_{33}} & \lesssim & (\mathsfbi{M}_{s})_{11} \, , \\
 \frac{(\mathsfbi{M}_{s})_{13}(\mathsfbi{M}_{s})_{23}}{(\mathsfbi{M}_s^{(1)})_{33}} & \lesssim & \tilde{\omega}_{e\|} (\mathsfbi{M}_{s})_{12} \ll (\mathsfbi{M}_{s})_{12} \, , \\
 \frac{\left[(\mathsfbi{M}_{s})_{23}\right]^2}{(\mathsfbi{M}_s^{(1)})_{33}} & \lesssim & \tilde{\omega}_{e\|} (\mathsfbi{M}_{s})_{22} \ll (\mathsfbi{M}_{s})_{22}  \, . 
  \label{secondordercorrect_orderings}
\end{subeqnarray}
On the other hand, the shear-perturbation components $\mathsfbi{P}_s$ satisfy
\begin{equation}
(\mathsfbi{P}_{s})_{11} \sim (\mathsfbi{P}_{s})_{22} \gg 
(\mathsfbi{P}_{s})_{12}  
\, .
\end{equation}
Substituting for $\mathsfbi{M}_{s}$ and $\mathsfbi{P}_{s}$ in (\ref{dielectric_expand_shear}) using (\ref{dielectric123trans_smallkpl}) 
and (\ref{leadingorder_MandP}\textit{b}), respectively, and then substituting (\ref{dielectric_expand_shear}) into 
(\ref{hotplasmadisprel_C}), we obtain the following quartic dispersion relation:
 \begin{eqnarray}
   && \left\{\tilde{\omega}_{e\|}^2 \left[(\mathsfbi{M}_{e}^{(1)} + \mathsfbi{M}_{i}^{(1)})_{11} - \frac{(\mathsfbi{M}_{e}^{(1)} + \mathsfbi{M}_{i}^{(1)})_{13}^{2}}{2 (\mathsfbi{M}_{e}^{(1)})_{33}} \right]  + (\mathsfbi{P}_{e}^{(0)} + \mu_e^{1/2} \mathsfbi{P}_{i}^{(0)})_{11} - k^2 d_e^2\right\} \nonumber \\
   & \quad & \times \left\{\tilde{\omega}_{e\|}^2 \left[(\mathsfbi{M}_{e}^{(1)} + \mathsfbi{M}_{i}^{(1)})_{22} \right] + \tilde{\omega}_{e\|} \left[(\mathsfbi{M}_{e}^{(0)} + \mu_e^{1/2} \mathsfbi{M}_{i}^{(0)})_{22} \right]+ (\mathsfbi{P}_{e}^{(0)} + \mu_e^{1/2} \mathsfbi{P}_{i}^{(0)})_{22} - k^2 d_e^2\right\} 
   \nonumber \\
  & \quad &  \qquad \qquad \qquad + \tilde{\omega}_{e\|}^2 \left[(\mathsfbi{M}_{e}^{(0)} + \mu_e^{1/2} \mathsfbi{M}_{i}^{(0)})_{12} \right]^2 = 0 \, 
  . \label{sheardisprel_C}
 \end{eqnarray}
We have assumed $k \rho_e \ll k \rho_i \ll 1$ and so we now have additional quadratic terms for both electrons and 
ions, as explained in section \ref{shortcomings_quadratic}.

We note that the dispersion relation (\ref{sheardisprel_C}) is similar to (\ref{sheardisprel_A}) except for the 
addition of two quadratic terms in $\tilde{\omega}_{e\|}$, and the absence of the linear terms
$\tilde{\omega}_{e\|} (\mathsfbi{M}_{s}^{(0)})_{11}$
and $(\mathsfbi{P}_{s}^{(0)})_{12}$. This motivates our approach to finding 
modes at arbitrary wavevectors: we solve a quartic dispersion relation that 
includes all the terms in (\ref{sheardisprel_C}) and also those linear terms 
which were present in (\ref{sheardisprel_A}), but absent in 
(\ref{sheardisprel_C}). Explicitly, this dispersion relation is
 \begin{eqnarray}
   && \left\{-\tilde{\omega}_{e\|}^2 \left[\frac{4}{3} W_e + \frac{4}{3} W_i + \frac{1}{4} \left(L_e + L_i\right)^2\right] + \mathrm{i} \tilde{\omega}_{e\|} \left( F_e + \mu_e^{1/2} F_i \right) + \epsilon_e \left( W_e + \mu_e^{1/2} W_i \right)- k_{\|}^2 d_e^2\right\} \nonumber \\
  & \, & \times \left[-\tilde{\omega}_{e\|}^2 \left(\frac{4}{3} Y_i + \frac{4}{3} Y_e \right)+ \mathrm{i}  \tilde{\omega}_{e\|} \left( H_e + \mu_e^{1/2} H_i \right) + \epsilon_e \left( Y_e + \mu_e^{1/2} Y_i \right)- k^2 d_e^2\right] 
   \nonumber \\
 & \, & \qquad \qquad +\left[\mathrm{i} \tilde{\omega}_{e\|} \left( G_e + \mu_e^{1/2} G_i \right) +  \epsilon_e \left( X_e + \mu_e^{1/2} X_i \right) \right]^2 = 0 \, 
  , \label{sheardisprel_D}
  \qquad
 \end{eqnarray}
where $L_s \equiv L\!\left(k_{\|} \tilde{\rho}_s,k_{\bot} \tilde{\rho}_s\right)$. 
The special functions $W(x,y)$ and $Y(x,y)$, defined in (\ref{shearasymptoticfuncs}), appear due to their relationship to the matrix $(\mathsfbi{M}_{s}^{(1)})$ (derived in appendix \ref{Maxwellresponse_lowxyz}): 
\begin{subeqnarray}
 W\!\left(k_{\|} \tilde{\rho}_s,k_{\bot} \tilde{\rho}_s\right) & = &  -\frac{3}{4} (\mathsfbi{M}_{s}^{(1)})_{xx}
 \, , \\
 Y\!\left(k_{\|} \tilde{\rho}_s,k_{\bot} \tilde{\rho}_s\right) & = & -\frac{3}{4} (\mathsfbi{M}_{e}^{(1)})_{yy} \, 
 ,
 \label{secondordercorrect_identities}
  \end{subeqnarray}
  combined with the identity
  \begin{equation}
  (\mathsfbi{M}_{e}^{(1)} + \mathsfbi{M}_{i}^{(1)})_{11} - \frac{(\mathsfbi{M}_{e}^{(1)} + \mathsfbi{M}_{i}^{(1)})_{13}^2}{2 (\mathsfbi{M}_{e}^{(1)})_{33}} =   - \frac{k^2}{k_{\|}^2} \left[\frac{4}{3} W_e + \frac{4}{3} W_i + \frac{1}{4} \left(L_e + L_i\right)^2 \right]
  \, , \label{secondorderident_special}
  \end{equation}
  proven in appendix \ref{secondordercorrect}. 
  
The dispersion relation (\ref{sheardisprel_D}) recovers all the roots of interest because 
it captures approximate values for 
all of the roots of the dispersion relations (\ref{sheardisprel_B}) and (\ref{sheardisprel_C}) 
in their respective wavenumber regions of validity. 
We note that, in situations when there are 
fewer than four physical modes (e.g., in the $k_{\|} \rho_e \gtrsim 1$ regime), solving (\ref{sheardisprel_D})
will also return non-physical modes that are the result of the addition of 
higher-order terms in a regime where such terms are illegitimate. However, 
by construction, such modes can be distinguished by their large magnitude ($\tilde{\omega}_{e\|} \sim 1$) as 
compared to the others. We acknowledge that our approach does not maintain consistent orderings: indeed, 
depending on the scale of a particular instability, there may be terms retained 
that are, in fact, smaller than other 
terms we have neglected when carrying out the $\tilde{\omega}_{i\|} \ll 1$ 
expansion. However, unlike the quadratic dispersion relation 
(\ref{sheardisprel_B}), the quartic dispersion relation (\ref{sheardisprel_D}) 
always captures the leading order terms for arbitrary wavevectors, and so  
provides reasonable approximations to the complex frequency of all possible CES microinstabilities. 

\subsubsection{Derivation of frequency and growth rate of the CES mirror instability} \label{derivation_mirror}

To derive the CES mirror instability's growth rate when it is close to marginality, we consider the
dispersion relation (\ref{sheardisprel_D}) under the orderings 
(\ref{mirror_order_marg}), viz., 
\begin{equation}
  k_{\|} \rho_i  \sim k_{\bot}^2 \rho_i^2 \sim \Gamma_i \ll 1 \, , \quad \tilde{\omega}_{i\|} = \mu_e^{-1/2} \tilde{\omega}_{e\|} \sim \frac{\Gamma_i}{\beta_i} 
  \, ,
  \label{mirror_order_Append}
\end{equation}
where $\Gamma_i = \Delta \beta_i-1$, 
and $\Delta = \Delta_i + \Delta_e = 3(\epsilon_i+\epsilon_e)/2$.  
Using the asymptotic identities (\ref{specialfuncMax_xysmall}) for the special functions $F_s$, $G_s$, $H_s$, $L_s$, and $N_s$, 
and (\ref{shearasymptoticfuncs_xysmall}) for $W_s$, $X_s$, and $Y_s$, 
(\ref{sheardisprel_D}) becomes, after dropping terms that are asymptotically small under the ordering (\ref{mirror_order_Append}), 
 \begin{equation}
\mathrm{i} \sqrt{\upi} k_{\bot}^2 \rho_i^2 \tilde{\omega}_{i\|} + \Delta \left(k_{\bot}^2 \rho_i^2 - \frac{1}{2} k_{\|}^2 \rho_i^2  - \frac{3}{4} k_{\bot}^4 \rho_i^4\right)- \frac{k^2 \rho_i^2}{\beta_i}
= 0 \, , \label{mirror_disp_Append}
 \end{equation}
 which in turn can be rearranged to give (\ref{mirrorgrowthrate_marg}) in section \ref{pospress_ion_mirror} and the subsequent results. We note that, save for the term $\propto G_e$, which cancels to 
 leading order with its ion equivalent, and the term $\propto Y_e$, 
 which we retain in order to capture correctly the mirror 
 instability's exact stability threshold, the electron terms in (\ref{sheardisprel_D}) 
 are negligibly small under the ordering (\ref{mirror_order_Append}). We also observe that by assuming frequency ordering
 (\ref{mirror_order_Append}), we have removed the shear Alfv\'en 
 wave from the dispersion relation. As we demonstrate 
 when characterising the growth rate of
 firehose-unstable shear Alfv\'en waves (see section \ref{negpres_fire_oblique} and appendix \ref{derivation_firehose_oblique}),
 a different ordering is 
 required to extract this mode (which is, in any case, stable for $\Delta_i > 
 0$). 
 
 To derive the growth rate of long-wavelength ($k_{\|} \rho_i \sim k_{\perp} \rho_i \ll1$) mirror modes away from marginality, when 
 $\Gamma_i \gtrsim 1$, we adopt the alternative ordering (\ref{mirror_order}), which is 
 equivalent to
 \begin{equation}
 \tilde{\omega}_{i\|} \sim \frac{1}{\beta_i} \sim {\Delta} \ll 1
  \, .
  \label{mirror_order_unity_Append}
\end{equation}
Again using the identities (\ref{specialfuncMax_xysmall}) and (\ref{shearasymptoticfuncs_xysmall}) to evaludate the special functions, the dispersion relation (\ref{sheardisprel_D}) is then
 \begin{equation}
\mathrm{i} \sqrt{\upi} k_{\bot}^2 \rho_i^2 \tilde{\omega}_{i\|} + \Delta \left(k_{\bot}^2 \rho_i^2 - \frac{1}{2} k_{\|}^2 \rho_i^2 \right)- \frac{k^2 \rho_i^2}{\beta_i}
= 0 \, , \label{mirror_disp_unity_Append}
 \end{equation}
 which, after some algebraic manipulation, gives (\ref{mirrorgrowthrate}) in section \ref{pospress_ion_mirror} and the subsequent results. 
 
 Finally, the expression (\ref{ionmirror_subionscale}) for the growth rate of 
 sub-ion-Larmor scale mirror modes is derived 
 by adopting the orderings (\ref{order_subionmirror}):
 \begin{equation}
  k_{\|} \rho_i \sim k_{\perp} \rho_i \sim (\Delta_i \beta_i)^{1/2} \gg 1 , \quad \tilde{\omega}_{i\|} \sim \frac{\Delta_i^{1/2}}{\beta_i^{1/2}} 
  , 
\label{order_subionmirror_append}
\end{equation}
and then using the asymptotic identities (\ref{specialfuncMax_xylarge}) 
for evaluating $F_i$, $G_i$, $H_i$, $L_i$, and $N_i$, (\ref{specialfuncMax_xysmall})
for $F_e$, $G_e$, $H_e$, $L_e$, and $N_e$, (\ref{shearasymptoticfuncs_xylarge})
for $W_i$, $X_i$, and $Y_i$, and (\ref{shearasymptoticfuncs_xysmall}) for 
$W_e$, $X_e$ and $Y_e$. Once again neglecting small terms under the assumed ordering, the 
dispersion relation (\ref{sheardisprel_D}) simpifies to a quadratic of the 
form (\ref{sheardisprel_A}):
 \begin{equation}
\left[-\frac{\Delta_i}{2} \frac{2 k_{\|}^2 \left(k_{\|}^2- k_{\bot}^2\right)}{k^4} + \frac{k_{\|}^2 \rho_i^2}{\beta_i}\right] \left(-\Delta_i \frac{k_{\|}^2}{k^2}+ \frac{k^2 \rho_i^2}{\beta_i}\right) -\tilde{\omega}_{i\|}^2 k_{\|}^2 \rho_i^2 = 0 \, 
  , \label{sheardisprel_oblique}
 \end{equation}
from which follow (\ref{ionmirror_subionscale}) and the subsequent results in \ref{pospress_ion_mirror}. 

\subsubsection{Derivation of frequency and growth rate of the parallel CES whistler instability} \label{derivation_parwhistler}

We derive the expressions (\ref{electronweibelgrowthrate}) for the real frequency and growth rate of the parallel 
CES whistler instability by adopting the ordering (\ref{electronWeibel_order}), 
\begin{equation}
  \tilde{\omega}_{e\|} \sim \Delta_e \sim \frac{1}{\beta_e} , \: k_{\|} \rho_e \sim 1 ,  
  \label{electronWeibel_order_Append}
\end{equation}
and evaluating $F_s$, $G_s$, $H_s$, $L_s$, and $N_s$ via 
(\ref{specialfuncMax_x1_ysmall}), and 
$W_s$, $X_s$, and $Y_s$ via (\ref{shearasymptoticfuncs_x1_ysmall}). 
The special functions with $s = i$ are simplified further by assuming additionally
that $k_{\|} \rho_i \gg 1$.  
Under these assumptions and simplifications, the dispersion relation (\ref{sheardisprel_D}) becomes
 \begin{eqnarray}
   \left\{\mathrm{i} \tilde{\omega}_{e\|} \sqrt{\upi} \left[\exp{\left(-\frac{1}{k_\|^2 \rho_e^2}\right)} + \mu_e^{1/2} \right] +\Delta_e \left[1+\frac{1}{k_{\|} \rho_e} \Real{\; Z\!\left(\frac{1}{k_{\|} \rho_e}\right)} + \mu_e^{1/2}\right] - \frac{k_\|^2 \rho_e^2}{\beta_e}\right\}^2 \nonumber \\
   \qquad +\left\{\mathrm{i} \tilde{\omega}_{e\|} \Real{\; Z\!\left(\frac{1}{k_{\|} \rho_e}\right)}  -  \frac{\Delta_e}{k_\| \rho_e} \left[\sqrt{\upi} \exp{\left(-\frac{1}{k_\|^2 \rho_e^2}\right)} + \mu_e \right] \right\}^2 = 0 \, 
  , \label{sheardisprel_electron_parallel}
  \qquad
 \end{eqnarray}
where we have substituted $\tilde{\rho}_e = - \rho_e$, and the only ion terms that we retain -- the terms proportional to $\mu_e^{1/2}$ or $\mu_e$ -- 
are those that we find to affect the dispersion relation qualitatively (as explained in the main text,
these terms are formally small under the assumed ordering, but cannot be neglected 
in certain subsidiary limits, e.g. $k_{\|} \rho_e \ll 1$, which we will subsequently wish to explore). 
(\ref{sheardisprel_electron_parallel}) can then be factorised to give two complex roots,
the real and imaginary parts of which become (\ref{electronweibelgrowthrate}\textit{a})
and (\ref{electronweibelgrowthrate}\textit{b}), respectively. 

\subsubsection{Derivation of frequency and growth rate of the CES transverse instability} \label{derivation_transverse}

To obtain the growth rate (\ref{transverse_oblique_growthrate}) of the two CES transverse modes, 
we take directly the unmagnetised limit of the full CES dispersion relation 
(\ref{sheardisprel_D}) under the orderings
\begin{equation}
k_{\bot} \rho_e \sim k_{\|} \rho_e \sim \left(\Delta_e \beta_e\right)^{1/2} \gg 1 
, \qquad \tilde{\omega}_{e\|}  \sim \Delta_e \ll 1  ,
\end{equation}
and then employ asymptotic identities (\ref{specialfuncMax_xylarge})
for $F_s$, $G_s$, $H_s$, $L_s$, and $N_s$, and (\ref{shearasymptoticfuncs_xylarge})
for $W_s$, $X_s$, and $Y_s$.
We then obtain a dispersion relation similar to 
(\ref{sheardisprel_A}), but with two separable roots:
 \begin{equation}
\left[\mathrm{i} \tilde{\omega}_{e\|} \sqrt{\upi} \frac{k_{\|}^3}{k^3} +\Delta_e \frac{k_{\|}^2 (k_{\|}^2- k_{\bot}^2)}{k^4} - \frac{k_{\|}^2 \rho_e^2}{\beta_e}\right] \left(\mathrm{i} \tilde{\omega}_{e\|} \sqrt{\upi} \frac{k_{\|}}{k} +\Delta_e \frac{k_{\|}^2}{k^2} - \frac{k^2 \rho_e^2}{\beta_e}\right) = 0 \, 
  . \label{sheardisprel_electron_klarge}
 \end{equation}
When rearranged, the first bracket gives expression (\ref{transverse_oblique_growthrate}\textit{a}), 
and the second bracket gives~(\ref{transverse_oblique_growthrate}\textit{b}). 

\subsubsection{Derivation of frequency and growth rate of the CES electron mirror instability} \label{derivation_elecmirror}

When its marginality parameter $\Gamma_e = \Delta_e \beta_e -1$ is small, the growth rate (\ref{electronmirror_growthrate}) (and zero real frequency) of the CES electron mirror instability's can be derived from the dispersion relation (\ref{sheardisprel_D}) by adopting the ordering (\ref{KAW_ords_marg}), viz., 
\begin{equation}
k_{\perp}^2 \rho_e^2 \sim  k_{\|} \rho_e \sim \tilde{\omega}_{e\|} \beta_e \sim \Gamma_e  
\ll 1 ,
\label{KAW_ords_marg_Append}
\end{equation}
and assuming that $\Gamma_e \gg \mu_e^{1/2}$. This latter inequality implies that $1 \ll k_{\|} \rho_i \ll k_{\perp} \rho_i$, so 
we use the asymptotic identities (\ref{specialfuncMax_xylarge})
to simplify $F_i$, $G_i$, $H_i$, $L_i$, and $N_i$, (\ref{shearasymptoticfuncs_xylarge})
to simplify $W_i$, $X_i$, and $Y_i$, (\ref{specialfuncMax_xysmall})
for $F_e$, $G_e$, $H_e$, $L_e$, and $N_e$, and (\ref{shearasymptoticfuncs_xysmall}) for 
$W_e$, $X_e$ and $Y_e$. Collecting terms, using the identity $\Delta_e = (1+\Gamma_e)/\beta_e$, and keeping only leading-order ones, the dispersion relation simplifies to
 \begin{equation}
 \frac{3}{2 \beta_e} k_{\|}^2 \rho_e^2 \left( -\frac{\Gamma_e}{\beta_e} k_{\bot}^2 \rho_e^2 + \frac{3}{2 \beta_e} k_{\|}^2 \rho_e^2 + \frac{3}{4 \beta_e} k_{\bot}^4 \rho_e^4 + \mathrm{i} \sqrt{\upi} k_{\bot}^2 \rho_e^2 \tilde{\omega}_{e\|}  \right) -\tilde{\omega}_{e\|}^2 k_{\|}^2 \rho_e^2 = 0 \, 
  . \label{sheardisprel_oblique_marginal}
 \end{equation}
Because the discriminant of the quadratic (\ref{sheardisprel_oblique_marginal}) is negative, it follows that its solution satisfies $\omega = \mathrm{i} \gamma$, with $\gamma$ being given by (\ref{electronmirror_growthrate}). 

To derive the expression (\ref{obliqueinstab_freq_gen}) for the complex
frequency of long-wavelength electron mirror modes, we adopt the ordering (\ref{KAW_ords}), 
\begin{equation}
\tilde{\omega}_{e\|}  \sim \frac{k \rho_e}{\beta_e}  \sim \Delta_e k \rho_e  ,
\label{KAW_ords_append}
\end{equation}
and then consider the subsidiary limit $k_{\|} \rho_e \sim k_{\perp} \rho_e \sim \mu_e^{1/4} \ll 1$ of the 
dispersion relation~(\ref{sheardisprel_D}). Using the asymptotic identities (\ref{specialfuncMax_xylarge}) 
for $F_i$, $G_i$, $H_i$, $L_i$, and $N_i$, (\ref{specialfuncMax_xysmall})
for $F_e$, $G_e$, $H_e$, $L_e$, and $N_e$, (\ref{shearasymptoticfuncs_xylarge})
for $W_i$, $X_i$, and $Y_i$, and (\ref{shearasymptoticfuncs_xysmall}) for 
$W_e$, $X_e$ and $Y_e$, we find that
 \begin{eqnarray}
   && \left\{\frac{\Delta_e}{2} \left[ k_{\|}^2 \rho_e^2 - \mu_e^{1/2} \frac{2 k_{\|}^2 \left(k_{\|}^2- k_{\bot}^2\right)}{k^4} \right]+ \frac{k_{\|}^2 \rho_e^2}{\beta_e}\right\} \qquad \qquad \nonumber \\
  & \qquad & \times \left[ \frac{\Delta_e}{2} \left( k_{\|}^2 \rho_e^2 - 2 k_{\bot}^2 \rho_e^2  - \mu_e^{1/2} \frac{2 k_{\|}^2}{k^2}  \right)+ \frac{k^2 \rho_e^2}{\beta_e}\right] -\tilde{\omega}_{e\|}^2 k_{\|}^2 \rho_e^2 = 0 \, 
  , \label{sheardisprel_oblique}
  \qquad
 \end{eqnarray}
 where both the CE ion- and electron-shear terms are kept on account of their 
 equal size under the assumed ordering. Solving for $\omega$ gives (\ref{obliqueinstab_freq_gen}).
 
 \subsubsection{Derivation of frequency and growth rate of the parallel CES firehose instability} \label{derivation_firehose_par}

The relevant orderings of parameters to adopt in order to derive 
the complex frequency~(\ref{growthrate_firehose_parallel}) of the parallel CES firehose instability is (\ref{quasiparallelord}), viz., 
   \begin{equation}
     \tilde{\omega}_{i\|} \sim \frac{1}{\beta_i^{1/2}} \sim |\Delta_i|^{1/2} \sim 
     k_{\|} \rho_i \ll 1 \, , \label{quasiparallelord_Append}
   \end{equation}
   with an additional small wavenumber-angle condition $k_{\bot} \rho_i \ll \beta_i^{-3/4}$ (which 
   we shall justify \emph{a posteriori}). Under this ordering, 
   the special functions $F_s$, $G_s$, $H_s$, $L_s$, and $N_s$ can be simplified 
   using  (\ref{specialfuncMax_xysmall}), and 
$W_s$, $X_s$, and $Y_s$ using (\ref{shearasymptoticfuncs_xysmall}), and 
so the dispersion relation (\ref{sheardisprel_D}) 
   reduces to
    \begin{equation}
\left(\tilde{\omega}_{i\|}^2 - \frac{\Delta_i}{2} - \frac{1}{\beta_i}\right)^2 - \frac{\tilde{\omega}_{i\|}^2}{4} k_{\|}^2 \rho_i^2 = 0 \, 
  , \label{sheardisprel_firehose_parallel}
 \end{equation}
 where the only non-negligible electron term is the one $\propto \tilde{\omega}_{e\|} G_e$. Similarly to the CES mirror instability (see appendix \ref{derivation_mirror}), 
 this term cancels to leading order with its ion equivalent, and the next-order electron term is much 
 smaller than the equivalent ion term. This dispersion relation can be rearranged to give 
 (\ref{growthrate_firehose_parallel}). 
 
 We also note that, in deriving (\ref{sheardisprel_firehose_parallel}) from (\ref{sheardisprel_D}), 
 we have assumed that the linear term $\propto \tilde{\omega}_{e\|} \mu_e^{1/2} H_i$ is  
much smaller than the quadratic term $\propto \tilde{\omega}_{e\|}^2 Y_i$; their 
relative magnitude is given by
 \begin{equation}
   \frac{\tilde{\omega}_{e\|} \mu_e^{1/2} H_i}{\tilde{\omega}_{e\|}^2 Y_i} 
   \sim \frac{k_{\bot}^2 \rho_i^2}{ \tilde{\omega}_{i\|} k_{\|}^2 \rho_i^2} \sim \beta_i^{3/2} k_{\bot}^2 
   \rho_i^2 . 
 \end{equation}
 Thus, this assumption (which it is necessary to make in order for there to be both left-handed 
 and right-handed Alfv\'en modes in high-$\beta$ plasma) is only justified if the small-angle condition
$k_{\bot} \rho_i \ll \beta_i^{-3/4} \ll 1$ holds true. 

\subsubsection{Derivation of frequency and growth rate of the oblique CES firehose instability} \label{derivation_firehose_oblique}

To derive the oblique firehose's growth rate (\ref{growthrate_firehose_oblique}), we use the ordering (\ref{obliqueord}), viz., 
   \begin{equation}
     \tilde{\omega}_{i\|} \sim \frac{1}{\beta_i^{1/2}} \sim |\Delta_i|^{1/2} \sim 
     k_{\|}^2 \rho_i^2 \sim k_{\bot}^2 \rho_i^2 \ll 1 . \label{obliqueord_Append}
   \end{equation}
Simplifying the special functions $F_s$, $G_s$, $H_s$, $L_s$, and $N_s$
   via (\ref{specialfuncMax_xysmall}), and 
$W_s$, $X_s$, and $Y_s$ via (\ref{shearasymptoticfuncs_xysmall}), 
the dispersion relation (\ref{sheardisprel_D}) becomes
    \begin{equation}
   \mathrm{i} \sqrt{\upi} \left(\tilde{\omega}_{i\|}^2 - \frac{\Delta_i}{2} - \frac{1}{\beta_i}\right) k_{\bot}^2 \rho_i^2 \tilde{\omega}_{i\|} 
- \frac{\tilde{\omega}_{i\|}^2}{4} \left(k_{\|}^2 \rho_i^2  - \frac{3}{2} k_{\bot}^2 \rho_i^2 \right)^2 = 0 \, 
  , \label{sheardisprel_firehose_oblique}
 \end{equation}
 where, in contrast to the quasi-parallel firehose, the linear term $\propto \tilde{\omega}_{e\|} \mu_e^{1/2} H_i$ in (\ref{sheardisprel_D}) 
 is larger than the quadratic term $\propto \tilde{\omega}_{e\|}^2 Y_i$. 
 (\ref{sheardisprel_firehose_oblique}) can be solved to give two roots: 
 $\omega \approx 0$, corresponding to the stable slow mode 
 (whose damping rate is asymptotically small under the assumed ordering), 
 and the expression (\ref{growthrate_firehose_oblique}) for the complex frequency of the (sometimes firehose-unstable) 
 shear Alfv\'en mode. 
 
 \subsubsection{Derivation of frequency and growth rate of the critical-line CES firehose instability} \label{derivation_firehose_critline}

To characterise the growth of the critical-line firehose when $\beta_i \gg 10^{6}$, we set $k_\perp = 2 
k_{\|}/3$, and 
order 
\begin{equation}
  \tilde{\omega}_{i\|} \sim \beta_i^{-3/5} \sim  k_{\|}^6 \rho_i^6 \sim \left|\Delta_i + \frac{2}{\beta_i}\right|^{1/2} \, . \label{firehose_critline_orderingB}
\end{equation}
The dispersion relation (\ref{sheardisprel_D}) transforms similarly to 
(\ref{sheardisprel_firehose_oblique}) in this case, with two important exceptions: first, the 
term in (\ref{sheardisprel_D}) $\propto \tilde{\omega}_{e\|} G_e + \mu_e^{1/2} \tilde{\omega}_{e\|} G_i$ is $\textit{O}(k_\|^5 
\rho_i^5)$ on the critical line, rather than $\textit{O}(k_{\|}^3
\rho_i^3)$; secondly, our choice of ordering requires that we retain $\textit{O}(k_\|^4 
\rho_i^4)$. This gives
    \begin{equation}
  \mathrm{i} \sqrt{\upi} \left(\tilde{\omega}_{i\|}^2 - \frac{1}{2} \Delta_i- \frac{1}{\beta_i} - \frac{5}{8} \Delta_i k_{\|}^{2} \rho_i^{2}\right) \tilde{\omega}_{i\|} 
- \frac{6889}{13824} \tilde{\omega}_{i\|}^2 k_{\|}^{6} \rho_i^{6} = 0 \, 
  . \label{sheardisprel_firehose_critline_Barnes_gen}
 \end{equation}
 
To obtain the expression (\ref{growthrate_firehose_critline_Barnes})
 for the critical-line firehose's growth rate in the limit $\beta_i \gg 
 10^6$ that is valid under the ordering (\ref{ordering_firehose_critline_Barnes}), we consider the subsidiary limit
\begin{equation}
  \left|\Delta_i+\frac{2}{\beta_i}\right| \gg \beta_i^{-6/5}  , \label{sublim_firehose_critline_Barnes}
\end{equation}
in which case (\ref{sheardisprel_firehose_critline_Barnes_gen}) becomes 
\begin{equation}
  \mathrm{i} \sqrt{\upi} \left(\tilde{\omega}_{i\|}^2 - \frac{\Delta_i}{2} - \frac{1}{\beta_i}\right) \tilde{\omega}_{i\|} 
- \frac{6889}{13824} \tilde{\omega}_{i\|}^2 k_{\|}^{6} \rho_i^{6} = 0 \, 
  . \label{sheardisprel_firehose_critline_Barnes}
 \end{equation}
  The expression (\ref{growthrate_firehose_critline_Barnes}) follows from solving (\ref{sheardisprel_firehose_critline_Barnes}) 
  for $\omega$ (and once again neglecting the $\omega \approx 0$ solution).
  
  The expression (\ref{growthrate_firehose_margcritline_Barnes}) for the growth of critical-line firehose modes when $\beta_i \simeq -2/\Delta_i \gg 10^6$, 
  can be deduced by considering the opposite subsidiary limit to (\ref{sublim_firehose_critline_Barnes}), viz., 
  \begin{equation}
  \left|\Delta_i+\frac{2}{\beta_i}\right| \ll \beta_i^{-6/5}. \label{sublim_firehose_critline_Barnes_alt}
\end{equation} 
In this limit, (\ref{sheardisprel_firehose_critline_Barnes_gen}) simplifies to
    \begin{equation}
  \mathrm{i} \sqrt{\upi} \left(\tilde{\omega}_{i\|}^2 + \frac{5}{4 \beta_i} k_{\|}^2 \rho_i^2  \right) \tilde{\omega}_{i\|} 
- \frac{6889}{13824} \tilde{\omega}_{i\|}^2 k_{\|}^{6} \rho_i^{6} = 0 \, 
  . \label{sheardisprel_firehose_critline_Barnes_marg}
 \end{equation}
 Noting that the quadratic (\ref{sheardisprel_firehose_critline_Barnes_marg}) has a negative discriminant, we deduce that $\omega = \mathrm{i} \gamma$; 
then solving (\ref{sheardisprel_firehose_critline_Barnes_marg}) for $\gamma$ gives (\ref{growthrate_firehose_margcritline_Barnes}). 
  
  When $\beta_i \ll 10^6$, the appropriate ordering to adopt in order to simplify the dispersion 
  relation of critical-line is no longer (\ref{firehose_critline_orderingB}), but instead
  \begin{equation}
  \tilde{\omega}_{i\|} \sim  \frac{1}{\sqrt{\beta_i \log{\beta_i}}} \sim \left|\Delta_i + \frac{2}{\beta_i}\right|^{1/2} , \quad k_{\|} \rho_i \sim \frac{1}{\sqrt{\log{\beta_i}}} . \label{firehose_critline_orderingB}
\end{equation}
Under this ordering, the term $\propto \mu_e^{1/2} \tilde{\omega}_{e\|} 
F_i$ in (\ref{sheardisprel_D}) is retained, while the term $\propto \tilde{\omega}_{e\|} G_e + \mu_e^{1/2} \tilde{\omega}_{e\|} G_i$ 
is neglected. This gives
      \begin{equation}
\left[\tilde{\omega}_{i\|}^2 + \mathrm{i} \frac{\sqrt{\upi}}{k_{\|}^2 \rho_i^2}  \exp{\left(-\frac{1}{k_{\|}^2 \rho_i^2}\right)} \tilde{\omega}_{i\|} - \frac{1}{2} \Delta_i- \frac{1}{\beta_i} - \frac{5}{8} \Delta_i k_{\|}^{2} \rho_i^{2}\right] \tilde{\omega}_{i\|} = 0 \, 
  . \label{sheardisprel_firehose_critline_cyclo_gen}
 \end{equation}
 
 To obtain the expression (\ref{growthrate_firehose_critline_cyclotron})
  for the critical-line firehose instability's growth rate in the case when ordering (\ref{ordering_firehose_critline_cyclo}) holds -- that is, when $\Delta_i \beta_i  +2| \sim 1$, 
  we consider the appropriate subsidiary limit of (\ref{sheardisprel_firehose_critline_cyclo_gen}): 
  \begin{equation}
  \left|\Delta_i+\frac{2}{\beta_i}\right| \gg  \frac{1}{\beta_i \log{\beta_i}}. \label{sublim_firehose_critline_cyclo}
\end{equation} 
In this case, the last term in the square brackets on the LHS of (\ref{sheardisprel_firehose_critline_cyclo_gen}) can be neglected, 
leaving the only non-trivial roots to satisfy
    \begin{equation}
 \tilde{\omega}_{i\|}^2 + \mathrm{i} \frac{\sqrt{\upi}}{k_{\|}^2 \rho_i^2} \exp{\left(-\frac{1}{k_{\|}^2 \rho_i^2}\right)} \tilde{\omega}_{i\|} - \frac{\Delta_i}{2} - \frac{1}{\beta_i}
 = 0 \, 
  , \label{sheardisprel_firehose_critline_cyclo}
 \end{equation}
whence (\ref{growthrate_firehose_critline_cyclotron}) follows immediately. The case of growth when $\Delta_i \simeq -2/\beta_i$ can be recovered from the opposite subsidiary limit, 
  \begin{equation}
  \left|\Delta_i+\frac{2}{\beta_i}\right| \ll  \frac{1}{\beta_i \log{\beta_i}}. \label{sublim_firehose_critline_cyclo_alt}
\end{equation} 
In this case, the dispersion relation of the critical-line firehose modes is 
    \begin{equation}
 \tilde{\omega}_{i\|}^2 + \mathrm{i} \frac{\sqrt{\upi}}{k_{\|}^2 \rho_i^2} \exp{\left(-\frac{1}{k_{\|}^2 \rho_i^2}\right)} \tilde{\omega}_{i\|} + \frac{5}{4 \beta_i} k_{\|}^2 \rho_i^2 = 0 \, 
  , \label{sheardisprel_firehose_critline_cyclo_marg}
 \end{equation}
 which, when solved for the growth rate $\gamma = -i \omega$, gives (\ref{growthrate_firehose_critline_cyclo_marg}). 

 \subsubsection{Derivation of frequency and growth rate of the CES parallel electron firehose instability} \label{derivation_parelecfirehose}

This derivation is identical to that given in appendix 
\ref{derivation_parwhistler} for the frequency and growth rate of the parallel CES whistler 
instability, and the same expressions (\ref{electronweibelgrowthrate}) are used in section \ref{negpres_electron_oblique}. 

\subsubsection{Derivation of frequency and growth rate of the CES oblique electron firehose instability} \label{derivation_obliqueelecfirehose}

The complex frequency (\ref{ESTmodefreq}) 
of the electron-firehose modes with $\mu_e^{1/2} \ll k_{\|} \rho_e \ll k_{\perp} \rho_e \sim 1$
is derived by applying the ordering
\begin{equation}
\tilde{\omega}_{e\|} \sim |\Delta_e| \sim \frac{1}{\beta_e}   
\end{equation}
to (\ref{sheardisprel_D}) and using the asymptotic identities (\ref{specialfuncMax_xylarge}) 
for $F_i$, $G_i$, $H_i$, $L_i$, and $N_i$, (\ref{specialfuncMax_xsmall_y1})
for $F_e$, $G_e$, $H_e$, $L_e$, and $N_e$, (\ref{shearasymptoticfuncs_xylarge})
for $W_i$, $X_i$, and $Y_i$, and (\ref{shearasymptoticfuncs_xsmall_y1}) for 
$W_e$, $X_e$ and $Y_e$. We obtain the simplified dispersion relation
 \begin{eqnarray}
   && \left\{-\Delta_e \frac{k_{\|}^2}{k_{\bot}^2} \left[1 - \exp \left(-\frac{k_{\bot}^2 \rho_e^2}{2}\right) I_0\!\left(\frac{k_{\bot}^2 \rho_e^2}{2}\right)\right]- \frac{k_{\|}^2 \rho_e^2}{\beta_e}\right\} \nonumber \\
  & \, & \times \left\{\left(\mathrm{i} \sqrt{\upi} \tilde{\omega}_{e\|} + \Delta_e \right) k_{\bot}^2 \rho_e^2 \exp \left(-\frac{k_{\bot}^2 \rho_e^2}{2}\right) \left[I_0\!\left(\frac{k_{\bot}^2 \rho_e^2}{2}\right) - I_1\!\left(\frac{k_{\bot}^2 \rho_e^2}{2}\right)  \right] - \frac{k^2 \rho_e^2}{\beta_e}\right\} 
   \nonumber \\
 & \, & \qquad \qquad - k_{\|}^2 \rho_e^2 \tilde{\omega}_{e\|}^2 \exp \left(-k_{\bot}^2 \rho_e^2\right) \left[I_0\!\left(\frac{k_{\bot}^2 \rho_e^2}{2}\right) - I_1\!\left(\frac{k_{\bot}^2 \rho_e^2}{2}\right)\right]^2 = 0 \, 
  . \label{sheardisprel_ESTs}
  \qquad
 \end{eqnarray}
Introducing the special functions $\mathcal{F}(k_{\perp} \rho_e)$ and $\mathcal{H}(k_{\perp} \rho_e)$ given by 
(\ref{EST_specfunc}), and then rearranging (\ref{sheardisprel_ESTs}), 
leads to (\ref{ESTmodefreq}).

\subsubsection{Derivation of frequency and growth rate of the CES EST instability} \label{derivation_EST}

To derive the expression (\ref{EST_frequency}) for the growth rate of the EST instability in the limits $\mu_e^{1/2} \ll k_{\|} \rho_e \ll 
1 \ll k_{\bot} \rho_e \ll \beta_e^{1/7}$, and $\Delta_e \beta_e \gg 1$, we apply the orderings 
(\ref{ESTorder}), viz., 
\begin{equation}
k_{\bot} \rho_e \sim (\Delta_e \beta_e)^{1/2} , \quad \tilde{\omega}_{e\|}  \sim \Delta_e^{5/2} \beta_e^{3/2} , \quad
  k_{\|} \rho_e \sim \frac{1}{\sqrt{\log{|\Delta_e| \beta_e}}} \ll 1 \, 
  . \label{ESTorder_append}
\end{equation}
to (\ref{sheardisprel_D}). We then use the asymptotic identities (\ref{specialfuncMax_xylarge}) 
for $F_i$, $G_i$, $H_i$, $L_i$, and $N_i$, (\ref{specialfuncMax_xsmall_ylarge})
for $F_e$, $G_e$, $H_e$, $L_e$, and $N_e$, (\ref{shearasymptoticfuncs_xylarge})
for $W_i$, $X_i$, and $Y_i$, and (\ref{shearasymptoticfuncs_xsmall_ylarge}) for 
$W_e$, $X_e$ and $Y_e$ to give
   \begin{equation}
   \mathrm{i} \frac{\tilde{\omega}_{e\|}}{k_{\bot} \rho_e } \left\{\mathrm{i} \frac{\tilde{\omega}_{e\|}}{k_{\bot}^3 \rho_e^3} \left[4 \exp{\left(-\frac{1}{k_{\|}^2 \rho_e^2}\right)} +\sqrt{\upi} \mu_e^{1/2} k_{\|}^3 \rho_e^{3} \right]- \Delta_e  \frac{k_{\|}^2 \rho_e^2}{k_{\bot}^2 \rho_e^2} - \frac{k_{\|}^2 \rho_e^2}{\beta_e}\right\}
- \frac{k_{\|}^2 \rho_e^2 \tilde{\omega}_{e\|}^2}{\upi k_{\bot}^6 \rho_i^6 } = 0 
  , \label{sheardisprel_EST}
  \quad
 \end{equation} 
 where the only ion contribution that is not always small, and thus cannot be 
 neglected, is the term proportional to $\mu_e^{1/2}$. 
 Solving for the frequency gives $\omega \approx 0$ -- 
 corresponding to a damped mode whose frequency is asymptotically small under the assumed ordering (\ref{ESTorder_append}) -- and the EST mode, whose
 growth rate is given by (\ref{EST_frequency}).

\subsubsection{Derivation of frequency and growth rate of the CES whisper instability} \label{derivation_whisper}

In the limits $\mu_e^{1/2} \ll k_{\|} \rho_e \ll 
1 \gg k_{\bot} \rho_e$ and $\Delta_e \beta_e \gg 1$ under the orderings 
\begin{equation}
\tilde{\omega}_{e\|}  \sim \frac{1}{\beta_e^{2/7}} \sim \frac{1}{k_{\bot}^2 \rho_e^2} 
\sim \frac{1}{\Delta_e \beta_e} , \quad
  k_{\|} \rho_e \sim \frac{1}{\sqrt{\log{|\Delta_e| \beta_e}}} \ll 1 \, 
  , \label{whisperorder_append}
\end{equation}
the dispersion relation (\ref{sheardisprel_D}) becomes 
   \begin{equation}
   \mathrm{i} \frac{\tilde{\omega}_{e\|}}{k_{\bot} \rho_e } \Bigg\{ \frac{k_{\|}^2 \rho_e^2}{k_{\bot}^2 \rho_e^2} \frac{4 \tilde{\omega}_{e\|}^2}{\sqrt{\upi} k_{\bot} \rho_e} + \mathrm{i} \frac{4 \tilde{\omega}_{e\|}}{k_{\bot}^3 \rho_e^3} \exp{\left(-\frac{1}{k_{\|}^2 \rho_e^2}\right)} - \Delta_e  \frac{k_{\|}^2 \rho_e^2}{k_{\bot}^2 \rho_e^2} - \frac{k_{\|}^2 \rho_e^2}{\beta_e}\Bigg\} 
 - \frac{k_{\|}^2 \rho_e^2 \tilde{\omega}_{e\|}^2}{\upi k_{\bot}^6 \rho_e^6 } = 0 \, 
  , \label{sheardisprel_electronmag_oblique}
  \quad
 \end{equation}
 where we have once again evaluated 
$F_i$, $G_i$, $H_i$, $L_i$, and $N_i$ using (\ref{specialfuncMax_xylarge}), 
$F_e$, $G_e$, $H_e$, $L_e$, and $N_e$ using (\ref{specialfuncMax_xsmall_ylarge}), 
$W_i$, $X_i$, and $Y_i$ using (\ref{shearasymptoticfuncs_xylarge}), and 
$W_e$, $X_e$ and $Y_e$ using (\ref{shearasymptoticfuncs_xsmall_ylarge}), and 
neglected all terms that are small under the ordering (\ref{whisperorder_append}). 
Solving for the non-trivial root of (\ref{sheardisprel_electronmag_oblique})
gives the expression (\ref{fullmagmodedisp}) for the complex frequency of whisper 
waves. 

\subsubsection{Derivation of frequency and growth rate of the CES ordinary-mode instability} \label{derivation_ordinarymode}

  Because the low-frequency assumption $\tilde{\omega}_{e\|} \ll 1$ is broken in the regime 
  of relevance to the CES ordinary-mode instability, the dispersion relation (\ref{sheardisprel_D}) is not valid; 
  to characterise these modes, we must instead return to considering the full 
  hot-plasma dispersion relation. 
  
  We choose to categorise the ordinary-mode instability for modes with $k_{\|} = 0$. 
  In this special case, the plasma dielectric tensor simplifies considerably, and has 
  the convenient property that 
  \begin{equation}
 \hat{\boldsymbol{z}} \bcdot \boldsymbol{\mathfrak{E}} = \left(\hat{\boldsymbol{z}} \bcdot \boldsymbol{\mathfrak{E}} \bcdot \hat{\boldsymbol{z}} \right) \hat{\boldsymbol{z}}  
 \, ,
  \end{equation}
 if the particle distribution functions have even parity with respect to the 
 parallel velocity $v_{\|}$~\citep{D83} -- a condition satisfied by the CE distribution functions (\ref{CEsheardistfuncexpression}). 
 Thus, perturbations whose associated eigenmode satisfies $\widehat{\delta \boldsymbol{E}} = \widehat{\delta 
  E}_z \hat{\boldsymbol{z}}$ decouple from other modes in the plasma. The dispersion 
  relation for such modes follows from (\ref{HotPlasmaDispDis}):
  \begin{equation}
  \mathfrak{E}_{zz} - \frac{c^2 k_{\bot}^2}{\omega^2}= 0  \, .
  \end{equation}
  In terms of matrices $\mathsfbi{M}_{s}$ and $\mathsfbi{P}_{s}$ defined by (\ref{Maxnonmaxsep_s}), this can be 
  written
  \begin{equation}
 \sum_s (\mathsfbi{M}_{s})_{zz}  + \sum_s (\mathsfbi{P}_{s})_{zz} - k_{\bot}^2 d_e^2 = 
 0 \, . \label{disprel_ordmode_A}
  \end{equation}
  
For $k_{\|} = 0$, the matrix components $(\mathsfbi{M}_{s})_{zz}$ and $(\mathsfbi{P}_{s})_{zz}$ 
are given by [see (\ref{dielectricelements_maxB}\textit{i}) and (\ref{dielectricelements_shearB}\textit{i})]
\begin{subeqnarray}
 (\mathsfbi{M}_{s})_{zz} & = & - \sum_{n = -\infty}^{\infty} \frac{\omega}{\omega-n \tilde{\Omega}_s} 
 \exp \left(-\frac{k_{\bot}^2 \tilde{\rho}_s^2}{2}\right) I_n\!\left(\frac{k_{\bot}^2 \tilde{\rho}_s^2}{2}\right) 
 \, , \\
  (\mathsfbi{P}_{s})_{zz} & = & - \frac{3 \epsilon_s}{2} \sum_{n = -\infty}^{\infty}
 \exp \left(-\frac{k_{\bot}^2 \tilde{\rho}_s^2}{2}\right) I_n\!\left(\frac{k_{\bot}^2 \tilde{\rho}_s^2}{2}\right) = -\Delta_s \, .
\end{subeqnarray}
Therefore, the dispersion relation (\ref{disprel_ordmode_A}) becomes
\begin{eqnarray}
  k_{\bot}^2 d_e^2 & =&  -\sum_s \frac{m_e}{m_s} \left[\Delta_s + \sum_{n = -\infty}^{\infty} \frac{\omega}{\omega-n \tilde{\Omega}_s} 
 \exp \left(-\frac{k_{\bot}^2 \tilde{\rho}_s^2}{2}\right) I_n\!\left(\frac{k_{\bot}^2 \tilde{\rho}_s^2}{2}\right) 
\right] \nonumber \\
& =&  -\sum_s \frac{m_e}{m_s} \left[\Delta_s + \sum_{n = 1}^{\infty} \frac{2 \omega^2}{\omega^2-n^2 \tilde{\Omega}_s^2} 
 \exp \left(-\frac{k_{\bot}^2 \tilde{\rho}_s^2}{2}\right) I_n\!\left(\frac{k_{\bot}^2 \tilde{\rho}_s^2}{2}\right) 
\right]\, . \label{disprel_ordmode_B}
\end{eqnarray}
Since the left-hand side of (\ref{disprel_ordmode_B}) is real, and the imaginary 
part of the right-hand side is non-zero if and only if the complex frequency $\omega$ has 
non-zero real and imaginary parts, we conclude that all solutions must be either purely propagating, or purely growing modes.  
Looking for purely growing roots, we substitute $\omega = \mathrm{i} \gamma$ 
into (\ref{disprel_ordmode_B}), and deduce that
\begin{eqnarray}
\sum_s \frac{m_e}{m_s} \left[\sum_{n = 1}^{\infty} \frac{2 \gamma^2}{\gamma^2+n^2 \tilde{\Omega}_s^2} 
 \exp \left(-\frac{k_{\bot}^2 \tilde{\rho}_s^2}{2}\right) I_n\!\left(\frac{k_{\bot}^2 \tilde{\rho}_s^2}{2}\right) 
\right] \qquad \qquad \nonumber \\
\qquad \qquad = -  k_{\bot}^2 d_e^2 - \sum_s \frac{m_e}{m_s} \left[ \Delta_s + \exp \left(-\frac{k_{\bot}^2 \tilde{\rho}_s^2}{2}\right) I_0\!\left(\frac{k_{\bot}^2 \tilde{\rho}_s^2}{2}\right) \right] \, . \label{disprel_ordmode_C}  
\end{eqnarray}
Neglecting the ion contributions (which are smaller than the electron ones by a $(m_e/m_i)^{1/2}$ factor) and considering $\Delta_e < 0$, we arrive at 
(\ref{disprel_ordmode_D}).
\end{appendix}

\bibliographystyle{jpp}

\bibliography{Linear_stability_of_CE_dist_paper_Oct23_vsub}

\begin{thebibliography}{149}
\expandafter\ifx\csname natexlab\endcsname\relax\def\natexlab#1{#1}\fi
\def\au#1{#1} \def\ed#1{#1} \def\yr#1{#1}\def\at#1{#1}\def\jt#1{\textit{#1}}
  \def\bt#1{#1}\def\bvol#1{\textbf{#1}} \def\vol#1{#1} \def\pg#1{#1}
  \def\publ#1{#1}\def\arxiv#1{#1}\def\org#1{#1}\def\st#1{\textit{#1}}

\bibitem[Achterberg(2013)]{A13}
{\sc \au{Achterberg, A.}} \yr{2013}  \at{Mirror, firehose and cosmic-ray-driven
  instabilities in a high-beta plasma}.  \jt{Mon. Not. R. Astron. Soc.}
  \bvol{436},  \pg{705}.

\bibitem[{Albright}(1970{\natexlab{{\em a\/}}})]{A70B}
{\sc \au{{Albright}, N.~W.}} \yr{1970{\natexlab{{\em a\/}}}}  \at{{Quasilinear
  stabilization of the transverse instability}}.  \jt{Phys. Fluids}  \bvol{13},
   \pg{1021}.

\bibitem[{Albright}(1970{\natexlab{{\em b\/}}})]{A70}
{\sc \au{{Albright}, N.~W.}} \yr{1970{\natexlab{{\em b\/}}}}  \at{Transverse
  wave instability driven by shear flow in an unmagnetized plasma}.  \jt{Phys.
  Plasmas}  \bvol{13},  \pg{2728}.

\bibitem[Astfalk \& Jenko(2016)]{AF16}
{\sc \au{Astfalk, P.} \& \au{Jenko, F.}} \yr{2016}  \at{Parallel and oblique
  firehose instability thresholds for bi-kappa distributed protons}.  \jt{J.
  Geophys. Res. Space Phys.}  \bvol{121},  \pg{2842}.

\bibitem[Balbus(2000)]{Balb00}
{\sc \au{Balbus, Steven~A.}} \yr{2000}  \at{Stability, instability, and
  ``backward'' transport in stratified fluids}.  \jt{Astrophys. J.}
  \bvol{534},  \pg{420}.

\bibitem[Balbus(2001)]{Balb01}
{\sc \au{Balbus, Steven~A.}} \yr{2001}  \at{Convective and rotational stability
  of a dilute plasma}.  \jt{Astrophys. J.}  \bvol{562},  \pg{909}.

\bibitem[Balbus(2004)]{Balb2004}
{\sc \au{Balbus, Steven~A.}} \yr{2004}  \at{Viscous shear instability in weakly
  magnetized, dilute plasmas}.  \jt{Astrophys. J.}  \bvol{616},  \pg{857}.

\bibitem[{Balbus} \& {Hawley}(1991)]{BH91a}
{\sc \au{{Balbus}, Steven~A.} \& \au{{Hawley}, John~F.}} \yr{1991}  \at{{A
  Powerful Local Shear Instability in Weakly Magnetized Disks. I. Linear
  Analysis}}.  \jt{Astrophys. J.}  \bvol{376},  \pg{214}.

\bibitem[Barkana \& Loeb(2001)]{BL01}
{\sc \au{Barkana, R.} \& \au{Loeb, A.}} \yr{2001}  \at{In the beginning: the
  first sources of light and the reionization of the universe}.  \jt{Phys.
  Rept.}  \bvol{349},  \pg{125}.

\bibitem[Barnes(1966)]{B66}
{\sc \au{Barnes, Aaron}} \yr{1966}  \at{Collisionless damping of hydromagnetic
  waves}.  \jt{Phys. Fluids}  \bvol{9},  \pg{1483}.

\bibitem[{Basu} \& {Coppi}(1984)]{BC84}
{\sc \au{{Basu}, B.} \& \au{{Coppi}, B.}} \yr{1984}  \at{{Theory of
  field-swelling instability in anisotropic plasmas}}.  \jt{Phys. Fluids}
  \bvol{27},  \pg{1187}.

\bibitem[Bell {\em et~al.\/}(1981)Bell, Evans \& Nicholas]{BEN81}
{\sc \au{Bell, A.~R.}, \au{Evans, R.~G.} \& \au{Nicholas, D.~J.}} \yr{1981}
  \at{Electron energy transport in steep temperature gradients in
  laser-produced plasmas}.  \jt{Phys. Rev. Lett.}  \bvol{46},  \pg{243}.

\bibitem[Bell {\em et~al.\/}(2020)Bell, Kingham, Watkins \& Matthews]{BKWM20}
{\sc \au{Bell, A~R}, \au{Kingham, R~J}, \au{Watkins, H~C} \& \au{Matthews,
  J~H}} \yr{2020}  \at{Instability in a magnetised collisional plasma driven by
  a heat flow or a current}.  \jt{Plasma Phys. Controlled Fusion}  \bvol{62},
  \pg{095026}.

\bibitem[Bernstein(1958)]{B58}
{\sc \au{Bernstein, I.~B.}} \yr{1958}  \at{Waves in a plasma in a magnetic
  field}.  \jt{Phys. Rev.}  \bvol{109},  \pg{10}.

\bibitem[Bhatnagar {\em et~al.\/}(1954)Bhatnagar, Gross \& Krook]{BGK54}
{\sc \au{Bhatnagar, P.~L.}, \au{Gross, E.~P.} \& \au{Krook, M.}} \yr{1954}
  \at{A model for collision processes in gases. {I.} {S}mall amplitude
  processes in charged and neutral one-component systems}.  \jt{Phys. Rev.}
  \bvol{94},  \pg{511}.

\bibitem[{Bobylev}(1982)]{B82}
{\sc \au{{Bobylev}, A.~V.}} \yr{1982}  \at{{The Chapman-Enskog and Grad methods
  for solving the Boltzmann equation}}.  \jt{Sov. Phys. Dokl.}  \bvol{27},
  \pg{29}.

\bibitem[{Boldyrev} {\em et~al.\/}(2013){Boldyrev}, {Horaites}, {Xia} \&
  {Perez}]{BHXP13}
{\sc \au{{Boldyrev}, S.}, \au{{Horaites}, K.}, \au{{Xia}, Q.} \& \au{{Perez},
  J.~C.}} \yr{2013}  \at{{Toward a theory of astrophysical plasma turbulence at
  subproton scales}}.  \jt{Astrophys. J.}  \bvol{777},  \pg{41}.

\bibitem[Bott {\em et~al.\/}(2021{\natexlab{{\em a\/}}})Bott, Arzamasskiy,
  Kunz, Quataert \& Squire]{BAKQS21}
{\sc \au{Bott, A. F.~A.}, \au{Arzamasskiy, L.}, \au{Kunz, M.~W.}, \au{Quataert,
  E.} \& \au{Squire, J.}} \yr{2021{\natexlab{{\em a\/}}}}  \at{Adaptive
  critical balance and firehose instability in an expanding, turbulent,
  collisionless plasma}.  \jt{The Astrophysical Journal Letters}  \bvol{922},
  \pg{L35}.

\bibitem[Bott {\em et~al.\/}(2021{\natexlab{{\em b\/}}})Bott, Chen, Boutoux,
  Caillaud, Duval, Koenig, Khiar, Lantu\'ejoul, Le-Deroff, Reville, Rosch, Ryu,
  Spindloe, Vauzour, Villette, Schekochihin, Lamb, Tzeferacos, Gregori \&
  Casner]{B21b}
{\sc \au{Bott, A. F.~A.}, \au{Chen, L.}, \au{Boutoux, G.}, \au{Caillaud, T.},
  \au{Duval, A.}, \au{Koenig, M.}, \au{Khiar, B.}, \au{Lantu\'ejoul, I.},
  \au{Le-Deroff, L.}, \au{Reville, B.}, \au{Rosch, R.}, \au{Ryu, D.},
  \au{Spindloe, C.}, \au{Vauzour, B.}, \au{Villette, B.}, \au{Schekochihin,
  A.~A.}, \au{Lamb, D.~Q.}, \au{Tzeferacos, P.}, \au{Gregori, G.} \&
  \au{Casner, A.}} \yr{2021{\natexlab{{\em b\/}}}}  \at{Inefficient
  magnetic-field amplification in supersonic laser-plasma turbulence}.
  \jt{Phys. Rev. Lett.}  \bvol{127},  \pg{175002}.

\bibitem[Bott {\em et~al.\/}(2022)Bott, Chen, Tzeferacos, Palmer, Bell,
  Bingham, Birkel, Froula, Katz, Kunz, Li, Park, Petrasso, Ross, Reville, Ryu,
  S{\'e}guin, White, Schekochihin, Lamb \& Gregori]{B23}
{\sc \au{Bott, A. F.~A.}, \au{Chen, L.}, \au{Tzeferacos, P.}, \au{Palmer, C.
  A.~J.}, \au{Bell, A.~R.}, \au{Bingham, R.}, \au{Birkel, A.}, \au{Froula,
  D.~H.}, \au{Katz, J.}, \au{Kunz, M.~W.}, \au{Li, C.~K.}, \au{Park, H-S.},
  \au{Petrasso, R.}, \au{Ross, J.~S.}, \au{Reville, B.}, \au{Ryu, D.},
  \au{S{\'e}guin, F.~H.}, \au{White, T.~G.}, \au{Schekochihin, A.~A.},
  \au{Lamb, D.~Q.} \& \au{Gregori, G.}} \yr{2022}  \at{Insensitivity of a
  turbulent laser-plasma dynamo to initial conditions}.  \jt{Matter and
  Radiation at Extremes}  \bvol{7},  \pg{046901}.

\bibitem[{Bott} {\em et~al.\/}(2021){Bott}, {Tzeferacos}, {Chen}, {Palmer},
  {Rigby}, {Bell}, {Bingham}, {Birkel}, {Graziani}, {Froula}, {Katz}, {Koenig},
  {Kunz}, {Li}, {Meinecke}, {Miniati}, {Petrasso}, {Park}, {Remington},
  {Reville}, {Ross}, {Ryu}, {Ryutov}, {S{\'e}guin}, {White}, {Schekochihin},
  {Lamb} \& {Gregori}]{B21a}
{\sc \au{{Bott}, A.~F.~A.}, \au{{Tzeferacos}, P.}, \au{{Chen}, L.},
  \au{{Palmer}, C. A.~J.}, \au{{Rigby}, A.}, \au{{Bell}, A.~R.}, \au{{Bingham},
  R.}, \au{{Birkel}, A.}, \au{{Graziani}, C.}, \au{{Froula}, D.~H.},
  \au{{Katz}, J.}, \au{{Koenig}, M.}, \au{{Kunz}, M.~W.}, \au{{Li}, C.},
  \au{{Meinecke}, J.}, \au{{Miniati}, F.}, \au{{Petrasso}, R.}, \au{{Park},
  H.~S.}, \au{{Remington}, B.~A.}, \au{{Reville}, B.}, \au{{Ross}, J.~S.},
  \au{{Ryu}, D.}, \au{{Ryutov}, D.}, \au{{S{\'e}guin}, F.~H.}, \au{{White},
  T.~G.}, \au{{Schekochihin}, A.~A.}, \au{{Lamb}, D.~Q.} \& \au{{Gregori}, G.}}
  \yr{2021}  \at{{Time-resolved turbulent dynamo in a laser plasma}}.
  \jt{Proc. Nat. Acad. Sci. USA}  \bvol{118},  \pg{e2015729118}.

\bibitem[{Braginskii}(1965)]{B65}
{\sc \au{{Braginskii}, S.~I.}} \yr{1965}  \at{{Transport processes in a
  plasma}}.  \jt{Rev. Plasma Phys.}  \bvol{1},  \pg{205}.

\bibitem[Califano {\em et~al.\/}(2002)Califano, Cecchi \& Chiuderi]{CCC02}
{\sc \au{Califano, F.}, \au{Cecchi, T.} \& \au{Chiuderi, C.}} \yr{2002}
  \at{Nonlinear kinetic regime of the {W}eibel instability in an electron--ion
  plasma}.  \jt{Phys. Plasmas}  \bvol{9},  \pg{451}.

\bibitem[Califano {\em et~al.\/}(1998)Califano, Pegoraro, Bulanov \&
  Mangeney]{CPBM98}
{\sc \au{Califano, F.}, \au{Pegoraro, F.}, \au{Bulanov, S.~V.} \& \au{Mangeney,
  A.}} \yr{1998}  \at{Kinetic saturation of the {W}eibel instability in a
  collisionless plasma}.  \jt{Phys. Rev. E}  \bvol{57},  \pg{7048}.

\bibitem[Camporeale \& Burgess(2008)]{CB08}
{\sc \au{Camporeale, E.} \& \au{Burgess, D.}} \yr{2008}  \at{Electron firehose
  instability: Kinetic linear theory and two-dimensional particle-in-cell
  simulations}.  \jt{J. Geophys. Res. Space Phys.}  \bvol{113}.

\bibitem[{Camporeale} \& {Burgess}(2010)]{CB10}
{\sc \au{{Camporeale}, E.} \& \au{{Burgess}, D.}} \yr{2010}  \at{{Electron
  Temperature Anisotropy in an Expanding Plasma: Particle-in-Cell
  Simulations}}.  \jt{Astrophys. J.}  \bvol{710},  \pg{1848--1856}.

\bibitem[Cercignani(1988)]{C88}
{\sc \au{Cercignani, C.}} \yr{1988} {\em The Boltzmann Equation and Its
  Applications\/}.  \publ{New York: Springer New York}.

\bibitem[Chandrasekhar {\em et~al.\/}(1958)Chandrasekhar, Kaufman \&
  Watson]{CKW58}
{\sc \au{Chandrasekhar, S.}, \au{Kaufman, A.~N.} \& \au{Watson, K.~M.}}
  \yr{1958}  \at{The stability of the pinch}.  \jt{Proc. R. Soc Lond. A}
  \bvol{245},  \pg{435}.

\bibitem[Chapman \& Cowling(1970)]{CC70}
{\sc \au{Chapman, S.} \& \au{Cowling, T.G.}} \yr{1970} {\em The Mathematical
  Theory of Non-Uniform Gases\/}, 2nd edn.  \publ{Cambridge: Cambridge
  University Press}.

\bibitem[Chew {\em et~al.\/}(1956)Chew, Goldberger, Low \&
  Chandrasekhar]{CGL56}
{\sc \au{Chew, G.~F.}, \au{Goldberger, M.~L.}, \au{Low, F.~E.} \&
  \au{Chandrasekhar, Subrahmanyan}} \yr{1956}  \at{The {B}oltzmann equation and
  the one-fluid hydromagnetic equations in the absence of particle collisions}.
   \jt{Proc. R. Soc Lond. A}  \bvol{236},  \pg{112}.

\bibitem[Corless {\em et~al.\/}(1996)Corless, Gonnet, Hare, Jeffrey \&
  Knuth]{CGHJK96}
{\sc \au{Corless, R.~M.}, \au{Gonnet, G.~H.}, \au{Hare, D. E.~G.}, \au{Jeffrey,
  D.~J.} \& \au{Knuth, D.~E.}} \yr{1996}  \at{On the lambert w function}.
  \jt{Advances in Computational Mathematics}  \bvol{5},  \pg{329--359}.

\bibitem[{Davidson}(1983)]{D83}
{\sc \au{{Davidson}, R.~C.}} \yr{1983} {Kinetic waves and instabilities in a
  uniform plasma}.  \bt{In {\em Basic Plasma Physics: Selected Chapters,
  Handbook of Plasma Physics, Volume 1\/} (ed. \ed{A.~A. {Galeev} \& R.~N.
  {Sudan}})},  \pg{p. 229}.

\bibitem[Davidson {\em et~al.\/}(1972)Davidson, Hammer, Haber \&
  Wagner]{DHHW72}
{\sc \au{Davidson, R.~C.}, \au{Hammer, D.~A.}, \au{Haber, I.} \& \au{Wagner,
  C.~E.}} \yr{1972}  \at{Nonlinear development of electromagnetic instabilities
  in anisotropic plasmas}.  \jt{Phys. Fluids}  \bvol{15},  \pg{317}.

\bibitem[Davidson \& V{\"o}lk(1968)]{DV68}
{\sc \au{Davidson, Ronald~C.} \& \au{V{\"o}lk, Heinrich~J.}} \yr{1968}
  \at{Macroscopic quasilinear theory of the garden‐hose instability}.
  \jt{Phys. Fluids}  \bvol{11},  \pg{2259}.

\bibitem[Davidson \& Wu(1970)]{DW70}
{\sc \au{Davidson, R.~C.} \& \au{Wu, C.~S.}} \yr{1970}  \at{Ordinary‐mode
  electromagnetic instability in high‐{$\beta$} plasmas}.  \jt{Phys. Fluids}
  \bvol{13},  \pg{1407}.

\bibitem[Drake {\em et~al.\/}(2021)Drake, Pfrommer, Reynolds, Ruszkowski,
  Swisdak, Einarsson, Thomas, Hassam \& Roberg-Clark]{DPRR20}
{\sc \au{Drake, J.~F.}, \au{Pfrommer, C.}, \au{Reynolds, C.~S.},
  \au{Ruszkowski, M.}, \au{Swisdak, M.}, \au{Einarsson, A.}, \au{Thomas, T.},
  \au{Hassam, A.~B.} \& \au{Roberg-Clark, G.~T.}} \yr{2021}
  \at{Whistler-regulated magnetohydrodynamics: Transport equations for electron
  thermal conduction in the high-$\beta$ intracluster medium of galaxy
  clusters}.  \jt{Astrophys. J.}  \bvol{923},  \pg{245}.

\bibitem[Enskog(1917)]{E1917}
{\sc \au{Enskog, D.}} \yr{1917} {\em Kinetische Theorie der Vorg{\"a}nge in
  m{\"a}ssig verd{\"u}nnten Gasen. 1, Allgemeiner Teil.\/}.  \publ{Uppsala:
  Almqvist {\&} Wiksell}.

\bibitem[Epperlein(1984)]{E84}
{\sc \au{Epperlein, E.~M.}} \yr{1984}  \at{The accuracy of {B}raginskii's
  transport coefficients for a {L}orentz plasma}.  \jt{J. Phys. D}  \bvol{17},
  \pg{1823}.

\bibitem[Epperlein \& Bell(1987)]{EB87}
{\sc \au{Epperlein, E.~M.} \& \au{Bell, A.~R.}} \yr{1987}  \at{Non-local
  analysis of the collisional weibel instability in planar laser-ablated
  targets}.  \jt{Plasma Phys. Controlled Fusion}  \bvol{29},  \pg{85}.

\bibitem[Epperlein \& Haines(1986)]{EH86}
{\sc \au{Epperlein, E.~M.} \& \au{Haines, M.~G.}} \yr{1986}  \at{Plasma
  transport coefficients in a magnetic field by direct numerical solution of
  the {F}okker--{P}lanck equation}.  \jt{Phys. Fluids}  \bvol{29},  \pg{1029}.

\bibitem[Fabian(1994)]{F94}
{\sc \au{Fabian, A.~C.}} \yr{1994}  \at{Cooling flows in clusters of galaxies}.
   \jt{Annu. Rev. Astron. Astrophys.}  \bvol{32},  \pg{277}.

\bibitem[{Foote} \& {Kulsrud}(1979)]{FK79}
{\sc \au{{Foote}, E.~A.} \& \au{{Kulsrud}, R.~M.}} \yr{1979}
  \at{{Hydromagnetic waves in high beta plasmas}}.  \jt{Astrophys. J.}
  \bvol{233},  \pg{302}.

\bibitem[Fried \& Conte(1961)]{F61}
{\sc \au{Fried, B.D.} \& \au{Conte, S.D.}} \yr{1961} {\em The Plasma Dispersion
  Function\/}.  \publ{New York: Academic Press}.

\bibitem[Fried(1959)]{F59}
{\sc \au{Fried, B.~D.}} \yr{1959}  \at{Mechanism for instability of transverse
  plasma waves}.  \jt{Phys. Fluids}  \bvol{2},  \pg{337}.

\bibitem[Furth(1963)]{F62}
{\sc \au{Furth, H.~P.}} \yr{1963}  \at{Prevalent instability of nonthermal
  plasmas}.  \jt{Phys. Fluids}  \bvol{6},  \pg{48}.

\bibitem[Galtier \& Meyrand(2015)]{GM15}
{\sc \au{Galtier, S.} \& \au{Meyrand, R.}} \yr{2015}  \at{Entanglement of
  helicity and energy in kinetic {A}lfv{\'e}n wave/whistler turbulence}.
  \jt{J. Plasma Phys.}  \bvol{81},  \pg{325810106}.

\bibitem[Garc{\'\i}a-Col{\'\i}n {\em et~al.\/}(2008)Garc{\'\i}a-Col{\'\i}n,
  Velasco \& Uribe]{GVU08}
{\sc \au{Garc{\'\i}a-Col{\'\i}n, L.~S.}, \au{Velasco, R.~M.} \& \au{Uribe,
  F.~J.}} \yr{2008}  \at{Beyond the {N}avier--{S}tokes equations: {B}urnett
  hydrodynamics}.  \jt{Phys. Reports}  \bvol{465},  \pg{149}.

\bibitem[Gary(1993)]{G93}
{\sc \au{Gary, S.~P.}} \yr{1993} {\em Theory of Space Plasma
  Microinstabilities\/}.  \publ{Cambridge: Cambridge University Press}.

\bibitem[{Gary} \& {Li}(2000)]{GL00}
{\sc \au{{Gary}, S.~P.} \& \au{{Li}, H.}} \yr{2000}  \at{Whistler heat flux
  instability at high beta}.  \jt{Astrophys. J.}  \bvol{529},  \pg{1131}.

\bibitem[Gary \& Madland(1985)]{GM85}
{\sc \au{Gary, S.~P.} \& \au{Madland, C.~D.}} \yr{1985}  \at{Electromagnetic
  electron temperature anisotropy instabilities}.  \jt{J. Geophys. Res. Space
  Phys.}  \bvol{90},  \pg{7607}.

\bibitem[Gary \& Nishimura(2003)]{GN03}
{\sc \au{Gary, S.~P.} \& \au{Nishimura, K.}} \yr{2003}  \at{Resonant electron
  firehose instability: Particle-in-cell simulations}.  \jt{Phys. Plasmas}
  \bvol{10},  \pg{3571}.

\bibitem[Gary \& Wang(1996)]{GW96}
{\sc \au{Gary, S.~Peter} \& \au{Wang, Joseph}} \yr{1996}  \at{Whistler
  instability: Electron anisotropy upper bound}.  \jt{Journal of Geophysical
  Research: Space Physics}  \bvol{101},  \pg{10749--10754}.

\bibitem[{Guo} {\em et~al.\/}(2014){Guo}, {Sironi} \& {Narayan}]{GSN14}
{\sc \au{{Guo}, X.}, \au{{Sironi}, L.} \& \au{{Narayan}, R.}} \yr{2014}
  \at{Non-thermal electron acceleration in low mach number collisionless
  shocks. {I}. {P}article energy spectra and acceleration mechanism}.
  \jt{Astrophys. J.}  \bvol{794},  \pg{153}.

\bibitem[{Guo} {\em et~al.\/}(2018){Guo}, {Sironi} \& {Narayan}]{GSN18}
{\sc \au{{Guo}, X.}, \au{{Sironi}, L.} \& \au{{Narayan}, R.}} \yr{2018}
  \at{Electron heating in low mach number perpendicular shocks. {II}.
  {D}ependence on the pre-shock conditions}.  \jt{Astrophys. J.}  \bvol{858},
  \pg{95}.

\bibitem[Hall(1981)]{H81}
{\sc \au{Hall, A.~N.}} \yr{1981}  \at{The firehose instability in interstellar
  space}.  \jt{Mon. Not. R. Astron. Soc.}  \bvol{195},  \pg{685}.

\bibitem[Hall {\em et~al.\/}(1964)Hall, Heckrotte \& Kammash]{HHK64}
{\sc \au{Hall, L.~S.}, \au{Heckrotte, W.} \& \au{Kammash, T.}} \yr{1964}
  \at{Electrostatic instabilities near cyclotron frequency in a plasma with
  anisotropic velocity distribution}.  \jt{Phys. Rev. Lett.}  \bvol{13},
  \pg{603}.

\bibitem[Harris(1959)]{H59}
{\sc \au{Harris, E.~G.}} \yr{1959}  \at{Unstable plasma oscillations in a
  magnetic field}.  \jt{Phys. Rev. Lett.}  \bvol{2},  \pg{34}.

\bibitem[Hasegawa(1969)]{H69}
{\sc \au{Hasegawa, A.}} \yr{1969}  \at{Drift mirror instability in the
  magnetosphere}.  \jt{The Physics of Fluids}  \bvol{12},  \pg{2642}.

\bibitem[Hasegawa(2012)]{H12}
{\sc \au{Hasegawa, A.}} \yr{2012} {\em Plasma Instabilities and Nonlinear
  Effects\/}.  \publ{Berlin: Springer Berlin}.

\bibitem[{Hawley} \& {Balbus}(1991)]{BH91b}
{\sc \au{{Hawley}, John~F.} \& \au{{Balbus}, Steven~A.}} \yr{1991}  \at{{A
  Powerful Local Shear Instability in Weakly Magnetized Disks. II. Nonlinear
  Evolution}}.  \jt{Astrophys. J.}  \bvol{376},  \pg{223}.

\bibitem[Helander {\em et~al.\/}(1994)Helander, Krasheninnikov \& Catto]{HKC94}
{\sc \au{Helander, P.}, \au{Krasheninnikov, S.~I.} \& \au{Catto, P.~J.}}
  \yr{1994}  \at{Fluid equations for a partially ionized plasma}.  \jt{Phys.
  Plasmas}  \bvol{1},  \pg{3174}.

\bibitem[Helander \& Sigmar(2005)]{HS05}
{\sc \au{Helander, P.} \& \au{Sigmar, D.J.}} \yr{2005} {\em Collisional
  Transport in Magnetized Plasmas\/}.  \publ{Cambridge: Cambridge University
  Press}.

\bibitem[Hellinger(2007)]{H07}
{\sc \au{Hellinger, P.}} \yr{2007}  \at{Comment on the linear mirror
  instability near the threshold}.  \jt{Phys. Plasmas}  \bvol{14},
  \pg{082105}.

\bibitem[{Hellinger} {\em et~al.\/}(2009){Hellinger}, {Kuznetsov}, {Passot},
  {Sulem} \& {Tr{\'a}vn{\'\i}{\v{c}}ek}]{HKPS09}
{\sc \au{{Hellinger}, P.}, \au{{Kuznetsov}, E.~A.}, \au{{Passot}, T.},
  \au{{Sulem}, P.~L.} \& \au{{Tr{\'a}vn{\'\i}{\v{c}}ek}, P.~M.}} \yr{2009}
  \at{{Mirror instability: From quasi-linear diffusion to coherent
  structures}}.  \jt{Geophys. Res. Lett.}  \bvol{36},  \pg{L06103}.

\bibitem[Hellinger \& Matsumoto(2000)]{HM00}
{\sc \au{Hellinger, P.} \& \au{Matsumoto, H.}} \yr{2000}  \at{New kinetic
  instability: Oblique {A}lfv{\'e}n fire hose}.  \jt{J. Geophys. Res.}
  \bvol{105},  \pg{10519}.

\bibitem[Hellinger \& {\v S}tver{\'a}k(2018)]{HS18}
{\sc \au{Hellinger, P.} \& \au{{\v S}tver{\'a}k, S.}} \yr{2018}  \at{Electron
  mirror instability: particle-in-cell simulations}.  \jt{J. Plasma Phys.}
  \bvol{84},  \pg{905840402}.

\bibitem[Hellinger \& Tr{\'a}vn{\'\i}{\v c}ek(2008)]{HT08}
{\sc \au{Hellinger, P.} \& \au{Tr{\'a}vn{\'\i}{\v c}ek, P.~M.}} \yr{2008}
  \at{Oblique proton fire hose instability in the expanding solar wind: Hybrid
  simulations}.  \jt{J. Geophys. Res. Space Phys.}  \bvol{113}.

\bibitem[Hellinger \& Tr{\'a}vn{\'\i}{\v c}ek(2015)]{HT15}
{\sc \au{Hellinger, P.} \& \au{Tr{\'a}vn{\'\i}{\v c}ek, P.~M.}} \yr{2015}
  \at{Proton temperature-anisotropy-driven instabilities in weakly collisional
  plasmas: Hybrid simulations}.  \jt{J. Plasma Phys.}  \bvol{81},
  \pg{305810103}.

\bibitem[Hellinger {\em et~al.\/}(2014)Hellinger, Tr{\'a}vn{\'\i}{\v c}ek,
  Decyk \& Schriver]{HTDS14}
{\sc \au{Hellinger, P.}, \au{Tr{\'a}vn{\'\i}{\v c}ek, P.~M.}, \au{Decyk, V.~K.}
  \& \au{Schriver, D.}} \yr{2014}  \at{Oblique electron fire hose instability:
  {P}article-in-cell simulations}.  \jt{J. Geophys. Res. Space Phys.}
  \bvol{119},  \pg{59}.

\bibitem[Hollweg \& V{\"o}lk(1970)]{HV70}
{\sc \au{Hollweg, J.~V.} \& \au{V{\"o}lk, H.~J.}} \yr{1970}  \at{New plasma
  instabilities in the solar wind}.  \jt{J. Geophys. Res.}  \bvol{75},
  \pg{5297}.

\bibitem[Huba(1994)]{H94}
{\sc \au{Huba, J.~D.}} \yr{1994} {\em NRL Plasma Formulary\/}.
  \publ{Washington, DC: Naval Research Laboratory}.

\bibitem[Hurricane {\em et~al.\/}(2014)Hurricane, Callahan, Casey, Celliers,
  Cerjan, Dewald, Dittrich, D{\"o}ppner, Hinkel, Hopkins, Kline, Le~Pape, Ma,
  MacPhee, Milovich, Pak, Park, Patel, Remington, Salmonson, Springer \&
  Tommasini]{H14}
{\sc \au{Hurricane, O.~A.}, \au{Callahan, D.~A.}, \au{Casey, D.~T.},
  \au{Celliers, P.~M.}, \au{Cerjan, C.}, \au{Dewald, E.~L.}, \au{Dittrich,
  T.~R.}, \au{D{\"o}ppner, T.}, \au{Hinkel, D.~E.}, \au{Hopkins, L. F.~Berzak},
  \au{Kline, J.~L.}, \au{Le~Pape, S.}, \au{Ma, T.}, \au{MacPhee, A.~G.},
  \au{Milovich, J.~L.}, \au{Pak, A.}, \au{Park, H.~S.}, \au{Patel, P.~K.},
  \au{Remington, B.~A.}, \au{Salmonson, J.~D.}, \au{Springer, P.~T.} \&
  \au{Tommasini, R.}} \yr{2014}  \at{Fuel gain exceeding unity in an inertially
  confined fusion implosion}.  \jt{Nature}  \bvol{506},  \pg{343}.

\bibitem[Ibscher {\em et~al.\/}(2012)Ibscher, Lazar \& Schlickeiser]{ILS12}
{\sc \au{Ibscher, D.}, \au{Lazar, M.} \& \au{Schlickeiser, R.}} \yr{2012}
  \at{On the existence of {W}eibel instability in a magnetized plasma. {II}.
  {P}erpendicular wave propagation: the ordinary mode}.  \jt{Phys. Plasmas}
  \bvol{19},  \pg{072116}.

\bibitem[Islam \& Balbus(2005)]{IB05}
{\sc \au{Islam, Tanim} \& \au{Balbus, Steven}} \yr{2005}  \at{Dynamics of the
  magnetoviscous instability}.  \jt{Astrophys. J.}  \bvol{633},  \pg{328}.

\bibitem[Kahn(1962)]{K62}
{\sc \au{Kahn, F.~D.}} \yr{1962}  \at{Transverse plasma waves and their
  instability}.  \jt{J. Fluid Mech.}  \bvol{14},  \pg{321}.

\bibitem[Kahn(1964)]{K64}
{\sc \au{Kahn, F.~D.}} \yr{1964}  \at{Transverse plasma waves and their
  instability. {P}art 2}.  \jt{J. Fluid Mech.}  \bvol{19},  \pg{210}.

\bibitem[Kalman {\em et~al.\/}(1968)Kalman, Montes \& Quemada]{KMQ68}
{\sc \au{Kalman, G.}, \au{Montes, C.} \& \au{Quemada, D.}} \yr{1968}
  \at{Anisotropic temperature plasma instabilities}.  \jt{Phys. Fluids}
  \bvol{11},  \pg{1797}.

\bibitem[Kato(2005)]{K05}
{\sc \au{Kato, T.~N.}} \yr{2005}  \at{Saturation mechanism of the {W}eibel
  instability in weakly magnetized plasmas}.  \jt{Phys. Plasmas}  \bvol{12},
  \pg{080705}.

\bibitem[{Kennel} \& {Petschek}(1966)]{KP66}
{\sc \au{{Kennel}, C.~F.} \& \au{{Petschek}, H.~E.}} \yr{1966}  \at{{Limit on
  stably trapped particle fluxes}}.  \jt{J. Geophys. Res.}  \bvol{71},  \pg{1}.

\bibitem[Kennel \& Sagdeev(1967)]{KS67}
{\sc \au{Kennel, C.~F.} \& \au{Sagdeev, R.~Z.}} \yr{1967}  \at{Collisionless
  shock waves in high-$\beta$ plasmas}.  \jt{J. Geophys. Res.}  \bvol{72},
  \pg{3303}.

\bibitem[Komarov {\em et~al.\/}(2018)Komarov, Schekochihin, Churazov \&
  Spitkovsky]{KSCS17}
{\sc \au{Komarov, S.}, \au{Schekochihin, A.~A.}, \au{Churazov, E.} \&
  \au{Spitkovsky, A.}} \yr{2018}  \at{Self-inhibiting thermal conduction in a
  high-beta, whistler-unstable plasma}.  \jt{J. Plasma Phys.}  \bvol{84},
  \pg{905840305}.

\bibitem[Komarov {\em et~al.\/}(2016)Komarov, Churazov, Kunz \&
  Schekochihin]{KCKS16}
{\sc \au{Komarov, S.~V.}, \au{Churazov, E.~M.}, \au{Kunz, M.~W.} \&
  \au{Schekochihin, A.~A.}} \yr{2016}  \at{Thermal conduction in a
  mirror-unstable plasma}.  \jt{Mon. Not. R. Astron. Soc.}  \bvol{460},
  \pg{467}.

\bibitem[Krall \& Trivelpiece(1973)]{CL14}
{\sc \au{Krall, N.~A.} \& \au{Trivelpiece, A.~W.}} \yr{1973} {\em Principles of
  plasma physics\/}.  \publ{McGraw-Hill}.

\bibitem[Kull(1991)]{Kull91}
{\sc \au{Kull, H.J.}} \yr{1991}  \at{Theory of the rayleigh-taylor
  instability}.  \jt{Phys. Rept.}  \bvol{206},  \pg{197--325}.

\bibitem[Kunz(2011)]{K11}
{\sc \au{Kunz, Matthew~W.}} \yr{2011}  \at{{Dynamical stability of a thermally
  stratified intracluster medium with anisotropic momentum and heat
  transport}}.  \jt{Mon. Not. R. Astron. Soc.}  \bvol{417},  \pg{602}.

\bibitem[Kunz {\em et~al.\/}(2018)Kunz, Abel, Klein \& Schekochihin]{KAKS18}
{\sc \au{Kunz, M.~W.}, \au{Abel, I.~G.}, \au{Klein, K.~G.} \& \au{Schekochihin,
  A.~A.}} \yr{2018}  \at{Astrophysical gyrokinetics: turbulence in
  pressure-anisotropic plasmas at ion scales and beyond}.  \jt{J. Plasma Phys.}
   \bvol{84},  \pg{715840201}.

\bibitem[Kunz {\em et~al.\/}(2015)Kunz, Schekochihin, Chen, Abel \&
  Cowley]{KSCAC15}
{\sc \au{Kunz, M.~W.}, \au{Schekochihin, A.~A.}, \au{Chen, C. H.~K.}, \au{Abel,
  I.~G.} \& \au{Cowley, S.~C.}} \yr{2015}  \at{Inertial-range kinetic
  turbulence in pressure-anisotropic astrophysical plasmas}.  \jt{J. Plasma
  Phys.}  \bvol{81},  \pg{325810501}.

\bibitem[Kunz {\em et~al.\/}(2014)Kunz, Schekochihin \& Stone]{KSS14}
{\sc \au{Kunz, M.~W.}, \au{Schekochihin, A.~A.} \& \au{Stone, J.~M.}} \yr{2014}
   \at{Firehose and mirror instabilities in a collisionless shearing plasma}.
  \jt{Phys. Rev. Lett.}  \bvol{112},  \pg{205003}.

\bibitem[Kunz {\em et~al.\/}(2020)Kunz, Squire, Schekochihin \&
  Quataert]{KSSQ20}
{\sc \au{Kunz, M.~W.}, \au{Squire, J.}, \au{Schekochihin, A.~A.} \&
  \au{Quataert, E.}} \yr{2020}  \at{Self-sustaining sound in collisionless,
  high-$\beta$ plasma}.  \jt{J. Plasma Phys.}  \bvol{86},  \pg{905860603}.

\bibitem[Kuzichev {\em et~al.\/}(2019)Kuzichev, Vasko, Soto-Chavez, Tong,
  Artemyev, Bale \& Spitkovsky]{KVC19}
{\sc \au{Kuzichev, Ilya~V.}, \au{Vasko, Ivan~Y.}, \au{Soto-Chavez,
  Angel~Rualdo}, \au{Tong, Yuguang}, \au{Artemyev, Anton~V.}, \au{Bale,
  Stuart~D.} \& \au{Spitkovsky, Anatoly}} \yr{2019}  \at{Nonlinear evolution of
  the whistler heat flux instability}.  \jt{Astrophys. J.}  \bvol{882},
  \pg{81}.

\bibitem[Kuznetsov {\em et~al.\/}(2007)Kuznetsov, Passot \& Sulem]{KPS07}
{\sc \au{Kuznetsov, E.~A.}, \au{Passot, T.} \& \au{Sulem, P.~L.}} \yr{2007}
  \at{Dynamical model for nonlinear mirror modes near threshold}.  \jt{Phys.
  Rev. Lett.}  \bvol{98},  \pg{235003}.

\bibitem[Laguerre(1880)]{L80}
{\sc \au{Laguerre, E.}} \yr{1880}  \at{Sur une m{\'e}thode pour obtenir par
  approximation les racines d'une {\'e}quation alg{\'e}brique qui a toutes ses
  racines r{\'e}elles}.  \jt{Nouvelles Annales de Math{\'e}matiques.}
  \bvol{19},  \pg{161}.

\bibitem[Landau(1946)]{L46}
{\sc \au{Landau, L.~D.}} \yr{1946}  \at{{On the vibrations of the electronic
  plasma}}.  \jt{J. Phys. USSR}  \bvol{10},  \pg{25}, [\it{Zh. Eksp. Teor.
  Fiz.} \bf{16}\normalfont, 574 (1946)].

\bibitem[Lazar {\em et~al.\/}(2009)Lazar, Schlickeiser \& Poedts]{LSP09}
{\sc \au{Lazar, M.}, \au{Schlickeiser, R.} \& \au{Poedts, S.}} \yr{2009}
  \at{On the existence of {W}eibel instability in a magnetized plasma. {I}.
  {P}arallel wave propagation}.  \jt{Phys. Plasmas}  \bvol{16},  \pg{012106}.

\bibitem[Lemons {\em et~al.\/}(1979)Lemons, Winske \& Gary]{LWG79}
{\sc \au{Lemons, D.S.}, \au{Winske, D.} \& \au{Gary, S.P.}} \yr{1979}
  \at{Nonlinear theory of the {W}eibel instability}.  \jt{J. Plasma Phys.}
  \bvol{21},  \pg{287}.

\bibitem[{Levinson} \& {Eichler}(1992)]{LE92}
{\sc \au{{Levinson}, A.} \& \au{{Eichler}, D.}} \yr{1992}  \at{{Inhibition of
  electron thermal conduction by electromagnetic instabilities}}.
  \jt{Astrophys. J.}  \bvol{387},  \pg{212}.

\bibitem[Li {\em et~al.\/}(2007)Li, S\'eguin, Frenje, Rygg, Petrasso, Town,
  Landen, Knauer \& Smalyuk]{L06}
{\sc \au{Li, C.~K.}, \au{S\'eguin, F.~H.}, \au{Frenje, J.~A.}, \au{Rygg,
  J.~R.}, \au{Petrasso, R.~D.}, \au{Town, R. P.~J.}, \au{Landen, O.~L.},
  \au{Knauer, J.~P.} \& \au{Smalyuk, V.~A.}} \yr{2007}  \at{Observation of
  megagauss-field topology changes due to magnetic reconnection in
  laser-produced plasmas}.  \jt{Phys. Rev. Lett.}  \bvol{99},  \pg{055001}.

\bibitem[Li \& Habbal(2000)]{LH00}
{\sc \au{Li, X.} \& \au{Habbal, S.R.}} \yr{2000}  \at{Electron kinetic firehose
  instability}.  \jt{J. Geophys. Res. Space Phys.}  \bvol{105},  \pg{27377}.

\bibitem[{Meinecke} {\em et~al.\/}(2022){Meinecke}, {Tzeferacos}, {Ross},
  {Bott}, {Feister}, {Park}, {Bell}, {Blandford}, {Berger}, {Bingham},
  {Casner}, {Chen}, {Foster}, {Froula}, {Goyon}, {Kalantar}, {Koenig},
  {Lahmann}, {Li}, {Lu}, {Palmer}, {Petrasso}, {Poole}, {Remington}, {Reville},
  {Reyes}, {Rigby}, {Ryu}, {Swadling}, {Zylstra}, {Miniati}, {Sarkar},
  {Schekochihin}, {Lamb} \& {Gregori}]{M21}
{\sc \au{{Meinecke}, J.}, \au{{Tzeferacos}, P.}, \au{{Ross}, J.~S.},
  \au{{Bott}, A.~F.~A.}, \au{{Feister}, S.}, \au{{Park}, H.~S.}, \au{{Bell},
  A.~R.}, \au{{Blandford}, R.}, \au{{Berger}, R.~L.}, \au{{Bingham}, R.},
  \au{{Casner}, A.}, \au{{Chen}, L.~E.}, \au{{Foster}, J.}, \au{{Froula},
  D.~H.}, \au{{Goyon}, C.}, \au{{Kalantar}, D.}, \au{{Koenig}, M.},
  \au{{Lahmann}, B.}, \au{{Li}, C.~K.}, \au{{Lu}, Y.}, \au{{Palmer}, C.~A.~J.},
  \au{{Petrasso}, R.}, \au{{Poole}, H.}, \au{{Remington}, B.}, \au{{Reville},
  B.}, \au{{Reyes}, A.}, \au{{Rigby}, A.}, \au{{Ryu}, D.}, \au{{Swadling}, G.},
  \au{{Zylstra}, A.}, \au{{Miniati}, F.}, \au{{Sarkar}, S.},
  \au{{Schekochihin}, A.~A.}, \au{{Lamb}, D.~Q.} \& \au{{Gregori}, G.}}
  \yr{2022}  \at{Strong suppression of heat conduction in a laboratory replica
  of galaxy-cluster turbulent plasmas}.  \jt{Science Advances}  \bvol{8},
  \pg{eabj6799}.

\bibitem[Melville {\em et~al.\/}(2016)Melville, Schekochihin \& Kunz]{MSK16}
{\sc \au{Melville, S.}, \au{Schekochihin, A.A.} \& \au{Kunz, M.W.}} \yr{2016}
  \at{Pressure-anisotropy-driven microturbulence and magnetic-field evolution
  in shearing, collisionless plasma}.  \jt{Mon. Not. R. Astron. Soc.}
  \bvol{459},  \pg{2701}.

\bibitem[{Messmer}(2002)]{M02}
{\sc \au{{Messmer}, P.}} \yr{2002}  \at{{Temperature isotropization in solar
  flare plasmas due to the electron firehose instability}}.  \jt{Astron.
  Astrophys.}  \bvol{382},  \pg{301}.

\bibitem[Mikhailovskii \& Tsypin(1984)]{MT84}
{\sc \au{Mikhailovskii, A.~B.} \& \au{Tsypin, V.~S.}} \yr{1984}  \at{Transport
  equations of plasma in a curvilinear magnetic field}.  \jt{Beitr.
  Plasmaphys.}  \bvol{24},  \pg{335}.

\bibitem[Mogavero \& Schekochihin(2014)]{MS14}
{\sc \au{Mogavero, F.} \& \au{Schekochihin, A.A.}} \yr{2014}  \at{Models of
  magnetic field evolution and effective viscosity in weakly collisional
  extragalactic plasmas}.  \jt{Mon. Not. R. Astron. Soc.}  \bvol{440},
  \pg{3226}.

\bibitem[Nicastro {\em et~al.\/}(2008)Nicastro, Mathur \& Elvis]{NME08}
{\sc \au{Nicastro, F.}, \au{Mathur, S.} \& \au{Elvis, M.}} \yr{2008}
  \at{Missing baryons and the warm-hot intergalactic medium}.  \jt{Science}
  \bvol{319},  \pg{55}.

\bibitem[Okabe \& Hattori(2003)]{OH03}
{\sc \au{Okabe, N.} \& \au{Hattori, M.}} \yr{2003}  \at{Spontaneous generation
  of the magnetic field and suppression of the heat conduction in cold fronts}.
   \jt{Astrophys. J.}  \bvol{599},  \pg{964}.

\bibitem[{Paesold} \& {Benz}(1999)]{PB99}
{\sc \au{{Paesold}, G.} \& \au{{Benz}, A.~O.}} \yr{1999}  \at{{Electron
  Firehose instability and acceleration of electrons in solar flares}}.
  \jt{Astron. Astrophys.}  \bvol{351},  \pg{741}.

\bibitem[Parker(1958)]{P58}
{\sc \au{Parker, E.~N.}} \yr{1958}  \at{Dynamical instability in an anisotropic
  ionized gas of low density}.  \jt{Phys. Rev.}  \bvol{109},  \pg{1874}.

\bibitem[Parra(2017)]{P17}
{\sc \au{Parra, F.~I.}} \yr{2017}  \at{Collisionless plasma physics} .

\bibitem[Pistinner \& Eichler(1998)]{PE98}
{\sc \au{Pistinner, S.~L.} \& \au{Eichler, D.}} \yr{1998}  \at{Self-inhibiting
  heat flux}.  \jt{Mon. Not. R. Astron. Soc.}  \bvol{301},  \pg{49}.

\bibitem[Pitaevskii \& Lifshitz(1981)]{LL81}
{\sc \au{Pitaevskii, L.P.} \& \au{Lifshitz, E.M.}} \yr{1981} {\em Physical
  Kinetics\/}.  \publ{Elsevier Science}.

\bibitem[{Pokhotelov} \& {Amariutei}(2011)]{PA11}
{\sc \au{{Pokhotelov}, O.~A.} \& \au{{Amariutei}, O.~A.}} \yr{2011}
  \at{{Quasi-linear dynamics of Weibel instability}}.  \jt{Ann. Geophys}
  \bvol{29},  \pg{1997}.

\bibitem[Pokhotelov {\em et~al.\/}(2008)Pokhotelov, Sagdeev, Balikhin,
  Onishchenko \& Fedun]{PSBO08}
{\sc \au{Pokhotelov, O.~A.}, \au{Sagdeev, R.~Z.}, \au{Balikhin, M.~A.},
  \au{Onishchenko, O.~G.} \& \au{Fedun, V.~N.}} \yr{2008}  \at{Nonlinear mirror
  waves in non-maxwellian space plasmas}.  \jt{J. Geophys. Res. Space Phys.}
  \bvol{113}.

\bibitem[Quataert(2008)]{Q08}
{\sc \au{Quataert, Eliot}} \yr{2008}  \at{Buoyancy instabilities in weakly
  magnetized low-collisionality plasmas}.  \jt{Astrophys. J.}  \bvol{673},
  \pg{758}.

\bibitem[Quataert {\em et~al.\/}(2002)Quataert, Dorland \& Hammett]{QDH02}
{\sc \au{Quataert, Eliot}, \au{Dorland, William} \& \au{Hammett, Gregory~W.}}
  \yr{2002}  \at{The magnetorotational instability in a collisionless plasma}.
  \jt{Astrophys. J.}  \bvol{577},  \pg{524}.

\bibitem[Ramani \& Laval(1978)]{RL78}
{\sc \au{Ramani, A.} \& \au{Laval, G.}} \yr{1978}  \at{Heat flux reduction by
  electromagnetic instabilities}.  \jt{Phys. Fluids}  \bvol{21},  \pg{980}.

\bibitem[Rayleigh(1883)]{Ray1883}
{\sc \au{Rayleigh, Lord}} \yr{1883}  \at{Investigation of the character of the
  equilibrium of an incompressible heavy fluid of variable density}.  \jt{Proc.
  R. Soc Lond. A}  \bvol{14},  \pg{170}.

\bibitem[Rincon {\em et~al.\/}(2015)Rincon, Schekochihin \& Cowley]{RSC15}
{\sc \au{Rincon, F.}, \au{Schekochihin, A.~A.} \& \au{Cowley, S.~C.}} \yr{2015}
   \at{Non-linear mirror instability}.  \jt{Mon. Not. R. Astron. Soc.}
  \bvol{447},  \pg{L45}.

\bibitem[Riquelme {\em et~al.\/}(2018)Riquelme, Quataert \& Verscharen]{RQV18}
{\sc \au{Riquelme, M.}, \au{Quataert, E.} \& \au{Verscharen, D.}} \yr{2018}
  \at{{PIC} simulations of velocity-space instabilities in a decreasing
  magnetic field: Viscosity and thermal conduction}.  \jt{Astrophys. J.}
  \bvol{854},  \pg{132}.

\bibitem[Riquelme {\em et~al.\/}(2015)Riquelme, Quataert \& Verscharen]{RQV15}
{\sc \au{Riquelme, M.~A.}, \au{Quataert, E.} \& \au{Verscharen, D.}} \yr{2015}
  \at{Particle-in-cell simulations of continuously driven mirror and ion
  cyclotron instabilities in high beta astrophysical and heliospheric plasmas}.
   \jt{Astrophys. J.}  \bvol{800},  \pg{27}.

\bibitem[Riquelme {\em et~al.\/}(2016)Riquelme, Quataert \& Verscharen]{RQV16}
{\sc \au{Riquelme, M.~A.}, \au{Quataert, E.} \& \au{Verscharen, D.}} \yr{2016}
  \at{{PIC} simulations of the effect of velocity space instabilities on
  electron viscosity and thermal conduction}.  \jt{Astrophys. J.}  \bvol{824},
  \pg{123}.

\bibitem[{Roberg-Clark} {\em et~al.\/}(2016){Roberg-Clark}, {Drake}, {Reynolds}
  \& {Swisdak}]{RDRS16}
{\sc \au{{Roberg-Clark}, G.~T.}, \au{{Drake}, J.~F.}, \au{{Reynolds}, C.~S.} \&
  \au{{Swisdak}, M.}} \yr{2016}  \at{{Suppression of electron thermal
  conduction in the high {$\beta$} intracluster medium of galaxy clusters}}.
  \jt{Astrophys. J. Lett.}  \bvol{830},  \pg{L9}.

\bibitem[Roberg-Clark {\em et~al.\/}(2018{\natexlab{{\em a\/}}})Roberg-Clark,
  Drake, Reynolds \& Swisdak]{RDRS18}
{\sc \au{Roberg-Clark, G.~T.}, \au{Drake, J.~F.}, \au{Reynolds, C.~S.} \&
  \au{Swisdak, M.}} \yr{2018{\natexlab{{\em a\/}}}}  \at{Suppression of
  electron thermal conduction by whistler turbulence in a sustained thermal
  gradient}.  \jt{Phys. Rev. Lett.}  \bvol{120},  \pg{035101}.

\bibitem[Roberg-Clark {\em et~al.\/}(2018{\natexlab{{\em b\/}}})Roberg-Clark,
  Drake, Swisdak \& Reynolds]{R18b}
{\sc \au{Roberg-Clark, G.~T.}, \au{Drake, J.~F.}, \au{Swisdak, M.} \&
  \au{Reynolds, C.~S.}} \yr{2018{\natexlab{{\em b\/}}}}  \at{Wave generation
  and heat flux suppression in astrophysical plasma systems}.  \jt{Astrophys.
  J.}  \bvol{867},  \pg{154}.

\bibitem[Rogister(1971)]{R71}
{\sc \au{Rogister, A.}} \yr{1971}  \at{Parallel propagation of nonlinear low
  frequency waves in high beta plasma}.  \jt{Phys. Fluids}  \bvol{14},
  \pg{2733}.

\bibitem[Rosenbluth(1956)]{R56}
{\sc \au{Rosenbluth, M.N.}} \yr{1956}  \at{The stability of the pinch}.
  \jt{Los Alamos Sci. Lab. Rep.}  \bvol{LA-2030}.

\bibitem[Rosin {\em et~al.\/}(2011)Rosin, Schekochihin, Rincon \&
  Cowley]{RSRC11}
{\sc \au{Rosin, M.~S.}, \au{Schekochihin, A.~A.}, \au{Rincon, F.} \&
  \au{Cowley, S.~C.}} \yr{2011}  \at{A non-linear theory of the parallel
  firehose and gyrothermal instabilities in a weakly collisional plasma}.
  \jt{Mon. Not. R. Astron. Soc.}  \bvol{413},  \pg{7}.

\bibitem[Ruyer {\em et~al.\/}(2015)Ruyer, Gremillet, Debayle \&
  Bonnaud]{RGDB15}
{\sc \au{Ruyer, C.}, \au{Gremillet, L.}, \au{Debayle, A.} \& \au{Bonnaud, G.}}
  \yr{2015}  \at{Nonlinear dynamics of the ion {W}eibel-filamentation
  instability: An analytical model for the evolution of the plasma and spectral
  properties}.  \jt{Phys. Plasmas}  \bvol{22},  \pg{032102}.

\bibitem[Ryu {\em et~al.\/}(2008)Ryu, Kang, Cho \& Das]{RKCD08}
{\sc \au{Ryu, D.}, \au{Kang, H.}, \au{Cho, J.} \& \au{Das, S.}} \yr{2008}
  \at{Turbulence and magnetic fields in the large-scale structure of the
  universe}.  \jt{Science}  \bvol{320},  \pg{909}.

\bibitem[Schekochihin \& Cowley(2006)]{SC06}
{\sc \au{Schekochihin, A.~A.} \& \au{Cowley, S.~C.}} \yr{2006}  \at{Turbulence,
  magnetic fields, and plasma physics in clusters of galaxies}.  \jt{Phys.
  Plasmas}  \bvol{13},  \pg{056501}.

\bibitem[Schekochihin {\em et~al.\/}(2009)Schekochihin, Cowley, Dorland,
  Hammett, Howes, Quataert \& Tatsuno]{SCDHHQ09}
{\sc \au{Schekochihin, A.~A.}, \au{Cowley, S.~C.}, \au{Dorland, W.},
  \au{Hammett, G.~W.}, \au{Howes, G.~G.}, \au{Quataert, E.} \& \au{Tatsuno,
  T.}} \yr{2009}  \at{Astrophysical gyrokinetics: Kinetic and fluid turbulent
  cascades in magnetized weakly collisional plasmas}.  \jt{The Astrophysical
  Journal Supplement Series}  \bvol{182},  \pg{310}.

\bibitem[Schekochihin {\em et~al.\/}(2005)Schekochihin, Cowley, Kulsrud,
  Hammett \& Sharma]{SCKHS05}
{\sc \au{Schekochihin, A.~A.}, \au{Cowley, S.~C.}, \au{Kulsrud, R.~M.},
  \au{Hammett, G.~W.} \& \au{Sharma, P.}} \yr{2005}  \at{Plasma instabilities
  and magnetic field growth in clusters of galaxies}.  \jt{Astrophys. J.}
  \bvol{629},  \pg{139}.

\bibitem[Schekochihin {\em et~al.\/}(2008)Schekochihin, Cowley, Kulsrud, Rosin
  \& Heinemann]{SCKR08}
{\sc \au{Schekochihin, A.~A.}, \au{Cowley, S.~C.}, \au{Kulsrud, R.~M.},
  \au{Rosin, M.~S.} \& \au{Heinemann, T.}} \yr{2008}  \at{Nonlinear growth of
  firehose and mirror fluctuations in astrophysical plasmas}.  \jt{Phys. Rev.
  Lett.}  \bvol{100},  \pg{081301}.

\bibitem[{Schekochihin} {\em et~al.\/}(2010){Schekochihin}, {Cowley}, {Rincon}
  \& {Rosin}]{ACRR10}
{\sc \au{{Schekochihin}, A.~A.}, \au{{Cowley}, S.~C.}, \au{{Rincon}, F.} \&
  \au{{Rosin}, M.~S.}} \yr{2010}  \at{{Magnetofluid dynamics of magnetized
  cosmic plasma: firehose and gyrothermal instabilities}}.  \jt{Mon. Not. R.
  Astron. Soc.}  \bvol{405},  \pg{291}.

\bibitem[Shaaban {\em et~al.\/}(2019)Shaaban, Lazar, Yoon, Poedts \&
  L{\'o}pez]{SLYP19}
{\sc \au{Shaaban, S~M}, \au{Lazar, M}, \au{Yoon, P~H}, \au{Poedts, S} \&
  \au{L{\'o}pez, R~A}} \yr{2019}  \at{{Quasi-linear approach of the whistler
  heat-flux instability in the solar wind}}.  \jt{Mon. Not. R. Astron. Soc.}
  \bvol{486},  \pg{4498--4507}.

\bibitem[Shima \& Hall(1965)]{SH65}
{\sc \au{Shima, Y.} \& \au{Hall, L.~S.}} \yr{1965}  \at{Electrostatic
  instabilities in a plasma with anisotropic velocity distribution}.  \jt{Phys.
  Rev.}  \bvol{139},  \pg{A1115}.

\bibitem[Simakov \& Catto(2004)]{SC04}
{\sc \au{Simakov, A.~N.} \& \au{Catto, P.~J.}} \yr{2004}  \at{Drift-ordered
  fluid equations for modelling collisional edge plasma}.  \jt{Contrib. Plasma
  Phys.}  \bvol{44},  \pg{83}.

\bibitem[Southwood \& Kivelson(1993)]{SK93}
{\sc \au{Southwood, D.~J.} \& \au{Kivelson, M.~G.}} \yr{1993}  \at{Mirror
  instability: 1. physical mechanism of linear instability}.  \jt{J. Geophys.
  Res. Space Phys.}  \bvol{98},  \pg{9181}.

\bibitem[Squire {\em et~al.\/}(2017)Squire, Kunz, Quataert \&
  Schekochihin]{SKQS17}
{\sc \au{Squire, J.}, \au{Kunz, M.~W.}, \au{Quataert, E.} \& \au{Schekochihin,
  A.~A.}} \yr{2017}  \at{Kinetic simulations of the interruption of
  large-amplitude shear-alfv\'en waves in a high-$\ensuremath{\beta}$ plasma}.
  \jt{Phys. Rev. Lett.}  \bvol{119},  \pg{155101}.

\bibitem[Stix(2012)]{S12}
{\sc \au{Stix, M.}} \yr{2012} {\em The Sun: An Introduction\/}.  \publ{Springer
  Berlin}.

\bibitem[{Stix}(1962)]{S62}
{\sc \au{{Stix}, T.~H.}} \yr{1962} {\em {The Theory of Plasma Waves}\/}.
  \publ{New York: McGraw-Hill}.

\bibitem[Takabe {\em et~al.\/}(1985)Takabe, Mima, Montierth \& Morse]{TMMM85}
{\sc \au{Takabe, H.}, \au{Mima, K.}, \au{Montierth, L.} \& \au{Morse, R.~L.}}
  \yr{1985}  \at{{Self‐consistent growth rate of the Rayleigh--Taylor
  instability in an ablatively accelerating plasma}}.  \jt{Phys. Fluids}
  \bvol{28},  \pg{3676--3682}.

\bibitem[Taylor(1950)]{Tay50}
{\sc \au{Taylor, Geoffrey~Ingram}} \yr{1950}  \at{The instability of liquid
  surfaces when accelerated in a direction perpendicular to their planes. i}.
  \jt{Proc. R. Soc Lond. A}  \bvol{201},  \pg{192}.

\bibitem[Tzeferacos {\em et~al.\/}(2018)Tzeferacos, Rigby, Bott, Bell, Bingham,
  Casner, Cattaneo, Churazov, Emig, Fiuza, Forest, Foster, Graziani, Katz,
  Koenig, Li, Meinecke, Petrasso, Park, Remington, Ross, Ryu, Ryutov, White,
  Reville, Miniati, Schekochihin, Lamb, Froula \& Gregori]{T18}
{\sc \au{Tzeferacos, P.}, \au{Rigby, A.}, \au{Bott, A. F.~A.}, \au{Bell,
  A.~R.}, \au{Bingham, R.}, \au{Casner, A.}, \au{Cattaneo, F.}, \au{Churazov,
  E.~M.}, \au{Emig, J.}, \au{Fiuza, F.}, \au{Forest, C.~B.}, \au{Foster, J.},
  \au{Graziani, C.}, \au{Katz, J.}, \au{Koenig, M.}, \au{Li, C.~K.},
  \au{Meinecke, J.}, \au{Petrasso, R.}, \au{Park, H.~S.}, \au{Remington,
  B.~A.}, \au{Ross, J.~S.}, \au{Ryu, D.}, \au{Ryutov, D.}, \au{White, T.~G.},
  \au{Reville, B.}, \au{Miniati, F.}, \au{Schekochihin, A.~A.}, \au{Lamb,
  D.~Q.}, \au{Froula, D.~H.} \& \au{Gregori, G.}} \yr{2018}  \at{Laboratory
  evidence of dynamo amplification of magnetic fields in a turbulent plasma}.
  \jt{Nat. Comm.}  \bvol{9},  \pg{591}.

\bibitem[{Vedenov} \& {Sagdeev}(1958)]{VS61}
{\sc \au{{Vedenov}, A.~A.} \& \au{{Sagdeev}, R.~Z.}} \yr{1958}  \at{{Some
  properties of a plasma with an anisotropic ion velocity distribution in a
  magnetic field}}.  \jt{Sov. Phys. Dokl.}  \bvol{3},  \pg{278}.

\bibitem[Walsh {\em et~al.\/}(2017)Walsh, Chittenden, McGlinchey, Niasse \&
  Appelbe]{W17}
{\sc \au{Walsh, C.~A.}, \au{Chittenden, J.~P.}, \au{McGlinchey, K.},
  \au{Niasse, N. P.~L.} \& \au{Appelbe, B.~D.}} \yr{2017}  \at{Self-generated
  magnetic fields in the stagnation phase of indirect-drive implosions on the
  national ignition facility}.  \jt{Phys. Rev. Lett.}  \bvol{118},
  \pg{155001}.

\bibitem[Weibel(1959)]{W59}
{\sc \au{Weibel, E.~S.}} \yr{1959}  \at{Spontaneously growing transverse waves
  in a plasma due to an anisotropic velocity distribution}.  \jt{Phys. Rev.
  Lett.}  \bvol{2},  \pg{83}.

\bibitem[Wiegelmann {\em et~al.\/}(2014)Wiegelmann, Thalmann \& Solanki]{WTS14}
{\sc \au{Wiegelmann, T.}, \au{Thalmann, J.~K.} \& \au{Solanki, S.~K.}}
  \yr{2014}  \at{The magnetic field in the solar atmosphere}.  \jt{Astron.
  Astrophys. Rev.}  \bvol{22},  \pg{78}.

\bibitem[Yoon {\em et~al.\/}(1993)Yoon, Wu \& de~Assis]{YWA93}
{\sc \au{Yoon, P.~H.}, \au{Wu, C.~S.} \& \au{de~Assis, A.~S.}} \yr{1993}
  \at{Effect of finite ion gyroradius on the fire‐hose instability in a high
  beta plasma}.  \jt{Phys. Fluids B}  \bvol{5},  \pg{1971}.

\bibitem[Yoshida {\em et~al.\/}(2005)Yoshida, Furlanetto \& Hernquist]{YFH05}
{\sc \au{Yoshida, N.}, \au{Furlanetto, S.R.} \& \au{Hernquist, L.}} \yr{2005}
  \at{The temperature structure of the warm-hot intergalactic medium}.
  \jt{Astrophys. J. Lett.}  \bvol{618},  \pg{L91}.

\end{thebibliography}

\end{document}